\documentclass[12pt]{article}
\makeatletter \@addtoreset{equation}{section} \makeatother
\renewcommand{\theequation}{\thesection.\arabic{equation}}
\newcommand{\ba}{\begin{array}} \newcommand{\ea}{\end{array}}
\newcommand{\beq}{\begin{equation}} \newcommand{\eeq}{\end{equation}}
\newcommand{\bea}{\begin{eqnarray}} \newcommand{\eea}{\end{eqnarray}}
 
 \def\bce{\begin{center}} \def\ece{\end{center}}
 \def\nonu{\nonumber} 
\def\pa{\partial}  \def\be{\beta} \def\ga{\gamma}
 \def\de{\delta} \def\De{\Delta} \def\ep{\epsilon}

\def\la{\lambda}

 \newcommand{\tf}{\widetilde{f}}

     \def\eps6{{\displaystyle
    \mathop{\epsilon}^{6}}{}} \def\g6{{\displaystyle
    \mathop{g}^{6}}{}}  \def\nab6{{\displaystyle
    \mathop{\nabla}^{6}}{}}

\def\ft#1#2{{\textstyle{\frac{\scriptstyle #1}{\scriptstyle #2}}}}
\def\fft#1#2{\frac{#1}{#2}} \def\0{{\sst{(0)}}} \def\1{{\sst{(1)}}}
\def\2{{\sst{(2)}}} \def\3{{\sst{(3)}}} \def\4{{\sst{(4)}}}
\def\5{{\sst{(5)}}} \def\6{{\sst{(6)}}} \def\7{{\sst{(7)}}}
\def\8{{\sst{(8)}}}

\def\tf#1#2{\tfrac{#1}{#2}}

\def\ba{\begin{array}} \def\ea{\end{array}} \def\beq{\begin{equation}}
\def\eeq{\end{equation}} \def\be{\begin{equation}}
\def\ee{\end{equation}}

 \def\la{\lambda} \def\eps{\epsilon}

\def\ba{\begin{array}} \def\ea{\end{array}} \def\beq{\begin{equation}}
\def\eeq{\end{equation}} \def\be{\begin{equation}}
\def\ee{\end{equation}}

 \def\la{\lambda} \def\eps{\epsilon}

      \def\eps6{{\displaystyle
    \mathop{\epsilon}^{6}}{}}  \def\nab6{{\displaystyle
    \mathop{\nabla}^{6}}{}}

\newcommand{\bean}{\begin{eqnarray*}}
\newcommand{\eean}{\end{eqnarray*}}

\usepackage{amsmath}

\usepackage{amsmath} \usepackage{graphicx} \usepackage{amssymb}
\usepackage{amsthm} \usepackage{grffile} \usepackage{simplewick}
\usepackage{tikz}

\usepackage{nccmath} \usepackage{physics} \usepackage{wrapfig}
\usepackage{mathtools} \usepackage{stackengine}

\usetikzlibrary{intersections,decorations.markings}

\allowdisplaybreaks

\usepackage{mathrsfs}

\begin{document}
\thispagestyle{empty} \addtocounter{page}{-1}
   \begin{flushright}
\end{flushright}

  
\centerline{ \Large \bf
More on the Celestial
Soft Symmetry
Algebra
}
\vspace*{0.3cm}
\centerline{ \Large \bf  
in  the ${\cal N}=8$ Supergravity
} 
\vspace*{1.5cm}
\centerline {\bf
Changhyun Ahn$^{\dagger,\star,}$\footnote{CA is a visitor to
Seoul
National University of Science and Technology
and a professor emeritus at
Kyungpook National University.  }
}
\vspace*{1.0cm} 
\centerline{\it
$\dagger$
Institute for Convergent Fundamental Studies,
}
\centerline{\it
Seoul
National University of Science and Technology, Seoul
01811, Korea} 
\centerline{\it
$\star$
Department of Physics, Kyungpook National University, Daegu
41566, Korea} 
\vspace*{0.5cm}
\vspace*{0.5cm}
\vspace*{0.8cm} 
\centerline{\tt ahn@knu.ac.kr
} 
\vskip1cm

\centerline{\bf Abstract}
\vspace*{0.5cm}

We construct the operator product expansions (OPEs)
of the ${\cal N}=8$ ${\rm w}_{1+\infty}$ currents
(obtained from the antiholomorphic light transform) with
the matter fields in the ${\cal N}=8$ supergravity. 
The corresponding (anti)commutators
are obtained from the double contour integrals
using the above OPEs.
By computing the triple (anti)commutators
consisting of current-current-matter (i.e., the Jacobi identity
for fixed last element), the previously known
${\cal N}=8$ celestial soft symmetry algebra between these currents
(in the antiholomorphic sector) is reproduced.
We repeat the above procedure by taking the
holomorphic light transform and the OPEs
of the ${\cal N}=8$ ${\rm \bar{w}}_{1+\infty}$ currents
with the matter fields and the corresponding (anti)commutators
are obtained. We also obtain another type
of ${\cal N}=8$
celestial soft symmetry algebra between these currents
(in the holomorphic sector). As a subalgebra in the antiholomorphic
and holomorphic sectors, there exists
the ${\cal N}=8$ extended BMS (Bondi, van der Burg,
Metzner, and Sachs
in 1962)
algebra.
We analyze the above results from the ${\cal N}=8$ conformally
{\it soft}
${\rm w}_{1+\infty}$ (or ${\rm \bar{w}}_{1+\infty}$) currents in the
antiholomorphic (or holomorphic) sector. The nontrivial
soft theorems
for the graviton, the gravitinos and the graviphotons are described.
Finally, using the shadow transform
(the consecutive composition of holomorphic and antiholomorphic light
transforms), we
obtain the OPEs
of the ${\cal N}=8$ {\it shadow} ${\rm w}_{1+\infty}$
(or ${\rm \bar{w}}_{1+\infty}$) currents with
the matter fields where the stress-energy tensor, the supercurrents
and the supertranslation generator of the ${\cal N}=8$
extended BMS algebra are identified. 


\baselineskip=18pt
\newpage


\tableofcontents


\section{
Introduction}

The asymptotic symmetry group of general relativity in four
dimensional asymptotically
flat spacetimes
does not reduce to Poincare group of general relativity. Instead,
the Poincare group is infinitely enhanced to the BMS group \cite{BvM,Sachs}
which is the semidirect product of the Lorentz and
the supertranslations groups.
At null infinity, the supertranslations act as angle-dependent
(of two sphere)
translations in retarded time.
In \cite{HLMS}, it was shown that
the quantum S matrix Ward identity corresponding to 
the above BMS supertranslations is equivalent to the Weinberg's
{\it leading} soft graviton theorem \cite{Weinberg}.
In \cite{BT},
the BMS group was further extended to the group of all local conformal
transformation (superrotation) and the extended BMS group is the semidirect
product of the left Virasoro, the right Virasoro
and the supertranslations groups.
In \cite{KLPS}, it was proven that
the quantum S matrix Ward identity corresponding to 
the superrotation generators of this extended BMS group is equivalent to the
{\it subleading} soft graviton theorem \cite{CS}.
In \cite{CL},
it was shown that the {\it subsubleading} soft graviton theorem
\cite{CS} is equivalent to the conservation of asymptotic charges
from large diffeomorphism.

The (bosonic) extended BMS algebra  in the
Einstein-Yang-Mills theory was described in \cite{FSTZ1912}.
Among the three kinds of generators,
there exist {\it four} nonzero commutator relations in addition to
two trivial zero commutator relations.
There are two kinds of the subalgebras
consisting of i) the left Virasoro and the supertranslations
and ii) the right Virasoro and the supertranslations.
See also \cite{Barnich}.
On the other hand,
the ${\cal N}=1$ supersymmetric extended BMS algebra in the
${\cal N}=1$ supersymmetric Einstein-Yang-Mills theory
was studied in \cite{FSTZ}.
The main observation of this work is the fact that
the infinite dimensional extension of (local) supersymmetry
in the ${\cal N}=1$ supersymmetric extended BMS algebra is given by
the ``square root'' of BMS supertranslations.
In other words, the supertranslation operator can be written
in terms of the composite operators of two supercharges.
The anticommutator relation between these two supersymmetry generators
doesn't vanish.
There are also the commutator between the
left Virasoro generator and one supersymmetry generator
and the commutator
between the
right Virasoro generator and another supersymmetry generator.
See also 
\cite{AGS,TZ,Ahn2202,Ahn2203,AK2309,AK2407,AK2501}
for relevant works.
Moreover, the ${\cal N}=8$
supersymmetric extended BMS algebra was studied in \cite{2212-2,Tropper}
where the $SU(8)$ indices for the supersymmetry generators appear
and consists of {\it seven} nontrivial (anti)commutators as above.

The right Lorentz algebra (with vanishing central term)
between the $SL(2,\mathbb{R})_R$ generators with the right-weight $2$
can be seen from the wedge subalgebra of ${\rm w}_{1+\infty}$ algebra
described in \cite{GHPS,Strominger,HPS}.
The commutator between the modes of supertranslation generators
with right-weight $\frac{3}{2}$
having the fixed holomorphic modes (in order not to mix
$SL(2,\mathbb{R})_L$ primaries and descendants) contains
the mode with the right-weight $1(=\frac{3}{2}+\frac{3}{2}-2)$
on the right-hand side \footnote{
When the two right modes on the left-hand side are equal to each other,
then the coefficient of mode dependent term on the right-hand side
becomes zero. When the two right modes
on the left-hand side are different from each other,
the sum of these is equal to zero from the above wedge condition.
This implies that the zero mode with right-weight $1$ appears on the
right-hand side.
See also (\ref{wwcomm}).}.
Then the zero mode with the right-weight $1$ appears on the
right-hand side but this commutes with all other modes of
generators and can be decoupled completely.
Then there exists the vanishing commutator between
the supertranslation generators. 
Next, the commutator between the mode of
the right Lorentz generators and 
the modes of supertranslation generators
having the fixed holomorphic modes provides
the mode of
supertranslation generators with the right-weight $\frac{3}{2}$
on the right-hand side. The mode dependent term is
the same as the one from the primary condition under the
right Lorentz mode of the left-hand side.
We can see only half of the commutators between
the half of Poincare generators (i.e., three Lorentz and
two supertranslations) described as ii)
in the previous paragraph in this way.

For the celestial holography, see the review papers of
\cite{Strominger1703,Raclariu,Pasterski,PPR,Donnay,Ruzziconi,Zhu}.

\begin{itemize}
\item[]
We would like to construct
the celestial soft symmetry algebra
in ${\cal N}=8$ supergravity \cite{dF} where
the ${\cal N}=8$ supersymmetric (extended) BMS algebra
\cite{2212-2,Tropper},
as a subalgebra,
is manifest \footnote{Sometimes we do not use the word 
{\it extended} for simplicity but we assume that
our ${\cal N}=8$ supersymmetric version is the
supersymmetric
extension of the bosonic extended BMS algebra described in the first
paragraph.}. In \cite{AK2509},
the nontrivial twenty-five (anti)commutator relations,
in the {\it antiholomorphic} sector, 
between the graviton, the gravitinos, the graviphotons,
the graviphotinos and the scalars in the ${\cal N}=8$ supergravity
were found explicitly and the twenty-five couplings
can be written in terms of eight independent couplings.
That is, there exists the $SL(2,\mathbb{R})_R$ transformation
(described in the previous paragraph)
of each mode of the primaries with given right-weights
appearing in the first nine equations of $(4.1)$ in \cite{AK2509}.
Among them, the first {\it two} commutator relations are related to
the {\it three} commutators in the ${\cal N}=8$ BMS algebra
but the modes for the supertranslation generators are {\it restricted}
as in the bosonic pure gravity theory in the previous paragraph.  

Moreover, we should construct
the similar nontrivial twenty-five (anti)commutator relations,
in the {\it holomorphic} sector, 
between the graviton, the gravitinos, the graviphotons,
the graviphotinos and the scalars in the ${\cal N}=8$ supergravity.
This is a necessary step in order to
have
i) the left Virasoro algebra (with vanishing central term)
between the $SL(2,\mathbb{R})_L$ generators
with the left-weight $2$,
ii) the commutator between the mode of
the left Virasoro generators and 
the modes of supertranslation generators
having the fixed antiholomorphic modes,
and
iii) the commutator between the mode of
the left Virasoro generators and 
the modes of supercurrent generators \footnote{
\label{vanishing}
There are also  trivial eight vanishing (anti)commutator relations
in the ${\cal N}=8$ BMS algebra including the vanishing commutator
between the supertranslation generators.
}.
Note that the two kinds of modes of
supertranslation generators (i.e., fixed holomorphic
and antiholomorphic ones) do {\it not} cover the whole range
for the holomorphic and antiholomorphic
modes of supertranslation generators as in \cite{FSTZ1912,FSTZ}.
We should fill this gap completely and observe the important role
of supercurrents (or supersymmetry).

For the construction of bosonic ${\rm w}_{1+\infty}$ currents
(mainly positive helicity graviton currents)
in the antiholomorphic sector
\cite{HPS}, the antiholomorphic integral light transform \cite{KS} of the
first celestial operator in the OPE of two celestial operators
was taken. Then the ${\rm w}_{1+\infty}$ currents
transform as a primary under the $SL(2,\mathbb{R})_R$
and the right-weight is positive half-integer
greater than or equal to $1$.
Then the OPE of the  ${\rm w}_{1+\infty}$ currents
with the matter fields takes the simple form and the
right-hand side of this OPE
depends on i) dummy variable, ii) the right-weight
of the currents, iii) the left- and right-weights of the matter fields,
iv) the complex coordinates $z$ and $\bar{z}$ and
v) the antiholomorphic derivative acting on the matter fields
on the right-hand side.
Similarly,
for the construction of bosonic ${\rm \bar{w}}_{1+\infty}$ currents
\footnote{We are using the barred notation for the currents
in the holomorphic sector in order not to confuse with
the ones in the antiholomorphic sector.}
(for example, the negative helicity graviton currents)
in the holomorphic sector,
we should perform the holomorphic integral light transform of the
first celestial operator in the OPE of two celestial operators.
The ${\rm {\bar w}}_{1+\infty}$ currents
should transform as a primary under the $SL(2,\mathbb{R})_L$
and the left-weight is positive half-integer
greater than or equal to $1$
\footnote{We expect that the OPE of the  ${\rm {\bar w}}_{1+\infty}$ currents
with the matter fields has the simple form and the
right-hand side depends on i) dummy variable, ii) the left-weight
of the currents, iii) the left- and right-weights of the matter fields,
iv) the complex coordinates ${z}$ and $\bar{z}$ and
v) the holomorphic derivative acting on the matter fields
on the right-hand side.}. See also the recent work of
\cite{CP} for the holomorphic descendants appearing on the
right-hand side.

The antiholomorphic and holomorphic
integral light transforms
in the previous paragraph do not affect
the $SU(8)$ indices
from ${\cal N}=8$ supersymmetry appearing
in the celestial operators of both sides in the OPE
and we can take Appendix C of \cite{AK2509} for our starting point.
Because we act the above light transforms on the
first celestial operator in the OPE and the right-weight
for the first celestial operator
(or the left-weight for the holomorphic
light transform) for the antiholomorphic light transform
is specified, other remaining parameter, the left-weight
(or the right-weight)
dependence should appear on the right-hand side of the
OPE. By defining the charges and taking
their algebraic (anti)commutators described in \cite{HPS}, we can avoid
the non-associative ambiguities \cite{BHP}
of the collinear spacetime limits.
The (anti)commutators obey the Jacobi identity making
a well defined infinite dimensional symmetry algebra.
For given OPEs of the ${\cal N}=8$ ${\rm w}_{1+\infty}$
(or ${\rm \bar{w}}_{1+\infty}$)
currents with the matter fields \footnote{
What we mean by the matter fields here is the hard operators
which act as physical primary 256 states of ${\cal N}=8$
supergravity and get transformed by the
symmetries, compared to the soft operators which act as symmetry
currents.},
the corresponding (anti)commutators can be obtained
by performing the double contour integrals. 

In order to check the action of ${\cal N}=8$
${\rm w}_{1+\infty}$ (or ${\rm \bar{w}}_{1+\infty}$)
symmetry, we should consider the action of
(anti)commutators between
the ${\cal N}=8$
${\rm w}_{1+\infty}$ (or ${\rm \bar{w}}_{1+\infty}$) currents
on any matter fields as in \cite{HPS}.
It is nontrivial to check how the product of
two antiholomorphic (or holomorphic) differential operators 
can be written in terms of the single
antiholomorphic (or holomorphic) differential operators
with the linear mode dependent factors
for generic right-weights and modes.
As a subalgebra, there exist, in the antiholomorphic sector,
i) the right Virasoro algebra (with vanishing central term)
between the $SL(2,\mathbb{R})_R$ generators
with the right-weight $2$,
ii) the commutator between the mode of
the right Virasoro generators and 
the modes of supertranslation generators
having the fixed holomorphic modes,
and
iii) the commutator between the mode of
the right Virasoro generators and 
the modes of ${\cal N}=8$ supercurrent generators.
Compared to \cite{FSTZ},
the last commutator is the generalization of
${\cal N}=1$ supercurrents to the ${\cal N}=8$ ones
and see also \cite{2212-2}
\footnote{For the holomorphic sector, the corresponding
subalgebra was described in the previous paragraph
containing the footnote \ref{vanishing}.
Then the ${\cal N}=8$ BMS algebra consists of two kinds of
three commutators given by (\ref{BMS})
and (\ref{BMS1}) generating {\it all} supertranslations
and one anticommutator (\ref{BMS2}) as well as the trivial ones
(\ref{trivialeight}).}. 

Along the line of \cite{FSTZ,2212-2},
the nonvanishing anticommutator 
between the two supercurrents \footnote{This leads to generate
{\it all} supertranslations (i.e., cover the whole ranges
for the left- and right-modes).}
of the ${\cal N}=8$
BMS algebra (which should produce the
supertranslation generators) can be checked explicitly 
because we take the mode of one supercurrent
of the $SU(8)$ fundamental ${\bf 8}$  with weights
$(0,\frac{3}{2})$
in the antiholomorphic sector
and the mode of
other supercurrent
of the $SU(8)$ antifundamental $\bar{\bf 8}$
with weights $(\frac{3}{2},0)$
in the holomorphic sector.
Due to the lowest values of these right or left-weights,
the action of these charges on the matter fields do {\it not} contain
any antiholomorphic (or holomorphic) derivatives.
This is the simplest example where
one of the currents in the (anti)commutator
is from the ${\rm w}_{1+\infty}$ currents
and the other current is from the
${\rm \bar{w}}_{1+\infty}$ currents.
We leave the systematic analysis of the extension of
${\cal N}=8$  ${\rm w}_{1+\infty}$ (or ${\rm \bar{w}}_{1+\infty}$)
algebra where
the nontrivial (anti)commutators
between the ${\rm w}_{1+\infty}$ currents
and the ${\rm \bar{w}}_{1+\infty}$ currents
appear
to future work. 

In \cite{GHPS,Strominger,HPS},
the OPE between the soft gravitons was obtained
by absorbing the two singular factors in the
Euler beta function appearing on the right-hand side
in the OPE between the two gravitons.
On the other hand, we can absorb the single singular factor
related to the first graviton on the left-hand side
\cite{HPS,Jiang}.
Then we can apply their prescription to the
gravitinos, the graviphotons, the graviphotinos and the
scalars in the antiholomorphic sector (or holomorphic sector)
in the ${\cal N}=8$ supergravity.
Furthermore, the commutator relation
between the soft graviton and the matter fields was obtained
in \cite{HPS}. By identifying the ${\rm w}_{1+\infty}$ currents
with the above soft gravitons, they concluded that 
the actions of the soft gravitons respect the ${\rm w}_{1+\infty}$
symmetry algebra. By following the procedure of \cite{HPS},
we can also check that the actions of the 
soft gravitons, (the gravitinos, and so on) respect
the ${\cal N}=8$ ${\rm w}_{1+\infty}$ symmetry algebra.
The soft graviton theorem in pure gravity theory
described in \cite{Jiang}
can be easily applied to
the soft gravitinos theorem and the soft graviphotons theorem
in the ${\cal N}=8$ supergravity.

Finally, it is known that
from the shadow (integral) transform which is given by
the consecutive composition of holomorphic light transform and
antiholomorphic light transform, the holomorphic (or antiholomorphic)
coordinate dependence {\it inside} the integral
can be removed from the celestial operator
which depends on both complex coordinates. See also the work of
\cite{Jiang}. In this way, it is straightforward to
observe that the previous supercurrent of the fundamental
(or antifundamental) ${\bf 8}$ (or $\bar{\bf 8}$)
of $SU(8)$ depends on the complex coordinates $\bar{z}$ and $z$.
Similarly, the graviton currents with weights
$(0,2)$ (or $(2,0)$) from the shadow transform
in the antiholomorphic (or holomorphic) sector,
at the subleading order, depend on the complex coordinate
$z$ and $\bar{z}$. We will observe that the supertranslation
generators can be written in terms of the product of two supercurrents.

\end{itemize}  
  
%
%

In section 2, by following the work of \cite{HPS},
we present the OPEs of the ${\cal N}=8$ ${\rm w}_{1+\infty}$
currents with matter fields in the ${\cal N}=8$ supergravity.
In section 3,
the (anti)commutators from the results of section 2 between
the modes of  the ${\cal N}=8$ ${\rm w}_{1+\infty}$
currents and the matter fields are given explicitly.
In section 4,
we apply further action of
the modes of  the ${\cal N}=8$ ${\rm w}_{1+\infty}$
currents to the (anti)commutators in section 3 and
the ${\cal N}=8$ celestial soft symmetry algebra found in \cite{AK2509}
is reproduced.
In section 5, we continue to calculate
the OPEs and the (anti)commutators
of the ${\cal N}=8$ ${\rm \bar{w}}_{1+\infty}$
currents with matter fields, and
further action of
the modes of  these
currents to the (anti)commutators
as done
in the previous sections 2, 3, and 4.
In  section 6,  the ${\cal N}=8$ ${\rm {w}}_{1+\infty}$ algebra
(or  ${\rm \bar{w}}_{1+\infty}$ algebra) is found by following
the description of \cite{HPS}.
In section 7, the summary of this paper is given and the future
directions related to this work are presented.
In Appendices A, B, C, D, E and F,
the various details used in the previous sections
are presented also.


 \section{The OPEs of ${\cal N}=8$ ${\rm w}_{1+\infty}$
currents with matter fields
\label{OPE}
}

In \cite{HPS},
the (bosonic) positive graviton ${\rm w}_{1+\infty}$
currents were obtained by taking
the integral of a conformal celestial primary operator
(known as antiholomorphic light transform),
which appears in the first operator of the OPE between two primaries,
and
taking a limit for the helicity with a normalization factor
\footnote{We are using the Roman `${\rm w}$' in the current in order not to
confuse this with the complex coordinate $w$ appearing in the
coordinates of the OPE usually.}.

We would like to construct the 
${\cal N}=8$ (supersymmetric) ${\rm w}_{1+\infty}$ currents
by following the work of \cite{HPS}.
The celestial OPEs between the graviton, the gravitinos, the
graviphotons, the graviphotinos and the scalars are
described explicitly in \cite{2212-1,2212-2,AK2509} where the combination
of three helicities is fixed by two. For the time being,
we consider the coordinates for the first and second operators
in the above OPE as $(z_1,\bar{z}_1)=(z,\bar{z})$ and $(z_2,\bar{z}_2)=(0,0)$
in order to calculate the integral of light transform easily
and at the final stage we return to the original coordinates.

Our starting point is the relation $(3.2)$ of \cite{HPS} where
the $p$, which is given by the bulk dimension
of the three-point vertex minus four, is fixed by one
\footnote{The complete expression for the OPE is given by
\bea
{\cal O}_{h_1,\bar{h}_1}(z,\bar{z}) \, {\cal O}_{h_2,\bar{h}_2}(0,0)
& = &
\frac{\bar{z}}{z}\, \ga_{s_3}^{s_1,s_2} \, \sum_{m=0}^{\infty}\,
\frac{1}{m!}\,
B(2\bar{h}_1+1+m, 2\bar{h}_2+1) \, \bar{z}^{m}\,
{\bar{\pa}}^m \, {\cal O}_{h_1+h_2-1,\bar{h}_1+\bar{h}_2+1}(0,0)
\nonu \\
&+& \cdots.
\label{OOOPE}
\eea
The $p$ dependence appearing in the various places is fixed by $1$.
We present only the singular term $\frac{1}{z}$ in the holomorphic
coordinate and ignore the $O(z^0)$ terms. 
In the original coordinates, the derivative on the right-hand side
of this OPE is the derivative with respect to
the antiholomorphic coordinate $\bar{z}_2$.
The structure constant depends on the two right-weights,
$\bar{h}_1$ and $\bar{h}_2$. The reason why we shouldn't consider
the equation $(5.4)$ of \cite{HPS} is that this equation
holds for the gravitons in the pure gravity theory.
In other words, some expressions hold only for $s_1=+2$.
All the $SU(8)$ indices of ${\cal N}=8$ supergravity theory
should appear on
both sides of this OPE and we will see them later.}.
Now we apply the (antiholomorphic) light transform for the
first celestial operator in the OPE
(\ref{OOOPE}).
Then the right-hand side of the OPE of ${\rm w}_{1+\infty}$ currents
with matter fields contains 
the original Euler beta function and the ratio of the
Gamma functions, which comes from the integration of
light transform.
By considering the left- and right-weights for the
two celestial operators
\bea
(h_1, \bar{h}_1) = (s_1+1-q, 1-q+\varepsilon), \qquad
(h_2, \bar{h}_2) = (h, \bar{h}),
\label{fourh}
\eea
where $q$ is the right-weight of the ${\rm w}_{1+\infty}$
currents ($q=1, \frac{3}{2}, 2, \frac{5}{2}, \cdots $.)
\cite{Strominger},
the ratio of two Gamma functions (after
writing down the Euler beta function in terms of
Gamma functions) with non-positive arguments 
can be simplified with finite values.
The $\varepsilon$ (introduced in the
helicity) is a small parameter which will be taken to zero
at the final stage of the calculation.
Note that there exists the helicity $s_1$ dependence on the
left-weight explicitly. Of course, when we fix $s_1=+2$, then
we reproduce the weights for the positive gravitons given in \cite{HPS}.
Furthermore, after taking the antiholomorphic light transform,
the original right-weight $\bar{h}_1$ for the first operator
in the OPE is changed into the one $(1-\bar{h}_1)=q$ for the
${\rm w}_{1+\infty}$ currents
\footnote{We can count the right-weight explicitly.
That is the sum of $-1$
from $d \bar{w}$, $(-2\bar{h}_1+2)$
from the integrand and $\bar{h}_1$ from the right-weight of
corresponding operator. }.
See \cite{HPS} for details
\footnote{We will describe
another type of light transform (known as holomorphic light transform)
in section \ref{holomorphic}.}.

Finally, we obtain the OPE of the
${\rm w}_{1+\infty}$ currents with the matter fields,
by returning to the original complex coordinates,
\bea
{\rm w}^q(z_1,\bar{z}_1) \, {\cal O}_{h,\bar{h}}(z_2,\bar{z}_2)
& = & \frac{1}{z_{12}}\, \frac{1}{2}\,
\sum_{m=0}^{2q-3}\, \frac{(m+1)\, \Gamma(2\bar{h}+1)}{\Gamma(2\bar{h}+4
-2q +m)}\, \bar{z}_{12}^{2-2q+m}\, \bar{\pa}^m \,
{\cal O}_{h+s_1-q,\bar{h}+2-q}(z_2,\bar{z}_2) \nonu \\
&+ & \cdots.
\label{OPEwO}
\eea
We take the same numerical normalization factor as in
\cite{HPS} and ignore the $O(z_{12}^0)$ terms
\footnote{The differential operator appearing in the coefficient
on the right-hand side of (\ref{OPEwO}) becomes
\bea
\frac{1}{z_{12}} \, \frac{1}{2}\,
\sum_{m=0}^{2q-3}\, \frac{(m+1)\, \Gamma(2\bar{h}+1)}{\Gamma(2\bar{h}+4
-2q +m)}\, \bar{z}_{12}^{2-2q+m}\, \bar{\pa}^m & = & \frac{1}{2} \,
\frac{1}{z_{12} \, \bar{z}_{12}}
\quad \mbox{for} \quad q=\frac{3}{2},
\nonu \\
\frac{1}{z_{12}} \, \frac{1}{2}\,
\sum_{m=0}^{2q-3}\, \frac{(m+1)\, \Gamma(2\bar{h}+1)}{\Gamma(2\bar{h}+4
-2q +m)}\, \bar{z}_{12}^{2-2q+m}\, \bar{\pa}^m & = &
\frac{1}{z_{12}} \bigg[
\frac{1}{\bar{z}_{12}^2}\bar{h} + \frac{1}{\bar{z}_{12}} \,
\bar{\pa}_{\bar{z}_2} \bigg] \quad
\mbox{for}  \quad q=2.
\label{twovaluesofq}
\eea
For higher $q \geq \frac{5}{2}$, the higher singularities
in the antiholomorphic coordinates $\bar{z}_{12}$ occur.}.
Note that the maximum value of the dummy variable
in (\ref{OPEwO}) comes from the condition
of nonzero value of the ratio of two Gamma functions
described in the previous paragraph.
This implies that the singular terms in the antiholomorphic
coordinates $\bar{z}_{12}$ appear on the right-hand side of
(\ref{OPEwO}).
The above result is a generalization of $(5.8)$ of \cite{HPS}
although the right-hand side of this OPE is exactly the same as
the one in \cite{HPS} except the left-weight $(h+s_1-q)$ of the matter field
on the right-hand side.
It is obvious to see that
the left- and right-weights for the operator
on the right-hand side in (\ref{OPEwO})
can be determined from (\ref{OOOPE}) by using (\ref{fourh})
with $\varepsilon \rightarrow 0$.
If we take the helicity of the first
operator $s_1=+2$, then the above result becomes
the equation $(5.8)$ of \cite{HPS}.
It is easy to check that the
left- and right-weights of both sides
are equal to each other.
The left-weight of the left-hand side is given by
$(s_1+1-q)+h$ which is equal to $1+(h+s_1-q)$ and this
is the left-weight of the right-hand side.
The right-weight of the left-hand side is given by
$(q+\bar{h})$ which is equal to $(-2+2q-m)+m+(\bar{h}+2-q)$
and 
this is the right-weight of the right-hand side
\footnote{
\label{weightsofw}
Note that the left- and right-weights for the ${\rm w}_{1+\infty}$
current are
given by $(s_1+1-q,q)$.}.

The three operators in (\ref{OPEwO})
can be either the graviton, the gravitinos, the graviphotons,
the graviphotinos or the scalars in the ${\cal N}=8$ supergravity
and we have the following correspondences for the
currents and the matter fields 
\bea
{\rm w}^q & \longleftrightarrow &
{\Phi}_{+2}^{(h)}, \, {\Phi}_{+\frac{3}{2}}^{(h),A}, \, {\Phi}_{+1}^{(h),AB},
\, {\Phi}_{+\frac{1}{2}}^{(h),ABC}, \, {\Phi}_{0}^{(h),ABCD},
\, {\Phi}_{ABC,-\frac{1}{2}}^{(h)}, \, {\Phi}_{AB,-1}^{(h)}, \,
{\Phi}_{A,-\frac{3}{2}}^{(h)},
\, {\Phi}_{-2}^{(h)},
\label{wO} \\
{\cal O}_{h,\bar{h}}
&\longleftrightarrow &
{\cal O}_{+2}^{(h)}, \, {\cal O}_{+\frac{3}{2}}^{(h),A}, \, {\cal O}_{+1}^{(h),AB},
\, {\cal O}_{+\frac{1}{2}}^{(h),ABC}, \, {\cal O}_{0}^{(h),ABCD},
\, {\cal O}_{ABC,-\frac{1}{2}}^{(h)}, \, {\cal O}_{AB,-1}^{(h)},
\, {\cal O}_{A,-\frac{3}{2}}^{(h)},
\, {\cal O}_{-2}^{(h)}.
\nonu 
\eea
According to the analysis of \cite{AK2509},
there exist the forty-five OPEs of the currents with
the matter fields.
It is straightforward to classify
$q$, $h$, $\bar{h}$, $s_1$ (i.e., the parameters of
two weights of the left-hand side)
and three operators
in (\ref{OPEwO}) explicitly.
The only nontrivial part is to write down the complete
$SU(8)$ indices with the matter fields on the right-hand sides.
This can be read off from the work of \cite{AK2509}
\footnote{For the currents from the holomorphic light transform
in section \ref{holomorphic}, the corresponding 
structures of $SU(8)$ indices on the right-hand sides
of the OPEs are slightly
different.}.
We summarize them in Tables \ref{table1} and \ref{table2}
for the helicities $s_1 \geq s_2$ and Tables \ref{table9}
and \ref{table10} (of Appendix A) for the helicities $ s_1 < s_2$.
The former is for twenty-five OPEs
while the latter is for twenty OPEs \footnote{
The right-weights $q$ for the ${\rm w}_{1+\infty}$ currents
are the sum of the upper and lower indices
of each component operator in the ${\cal N}=8$ multiplet \cite{AK2509}.
The left-weights $h$ of the matter fields are common
while the right-weights $\bar{h}$ of those fields
contain the different numerical values. See also Tables \ref{table1} and
\ref{table2}.}.
For fixed $s_1$ and $s_2$ with the constraint for the
$s_3$, the matter fields on the right-hand sides are
determined uniquely.
In other words, the upper indices of the matter fields
on the right-hand sides are the sum of
each upper index on the left-hand side.
Their helicities $s_3$ can be obtained from the
values of helicities $s_1$ and $s_2$.
These identifications of the matter fields
can be checked by analyzing the left- and right-weights
explicitly as in the above Tables.
Although the $SU(8)$ indices for the currents
and the matter fields on the left-hand sides
of the OPEs are complete
in those Tables, they for the matter fields
on the right-hand sides of the OPEs are not complete
due to the various $SU(8)$ tensors appearing on the right-hand
sides of the OPEs. 

\begin{table}[tbp]
\centering
\renewcommand{\arraystretch}{1.7}
\begin{tabular}{|c|c||c|c|c|||c|c|c|c|| }
\hline
$ q$ &
$ {\rm w^q}$ &
$ h$ &
$ \bar{h}$ &
$ {\cal O}_{h,\bar{h}}$ &
$ s_1$ &
$ h+s_1-q$ &
$ \bar{h}+2-q$ &
${\cal O}_{h+s_1-q,\bar{h}+2-q}$  
\\
\hline
\hline
${\scriptstyle h_1+2}$ &
$  \Phi_{+2}^{(h_1)}$ &
${\scriptstyle 1-h_2}$ &
${\scriptstyle -1-h_2}$ &
$ {\cal O}_{+2}^{(h_2)}$ &
${\scriptstyle +2}$ &
${\scriptstyle 1-(h_1+h_2)}$ &
${\scriptstyle -1-(h_1+h_2)} $ &
$ {\cal O}_{+2}^{(h_1+h_2)}$
\\
\hline
${\scriptstyle h_1+2}$ &
$\Phi_{+2}^{(h_1)}$&
${\scriptstyle 1-h_2}$ &
${\scriptstyle -\frac{1}{2}-h_2}$ &
${\cal O}_{+\frac{3}{2}}^{(h_2),A}$ &
${\scriptstyle +2}$ &
${\scriptstyle 1-(h_1+h_2)}$ &
${\scriptstyle -\frac{1}{2}-(h_1+h_2)} $ &
${\cal O}_{+\frac{3}{2}}^{(h_1+h_2),A}$
\\
\hline
${\scriptstyle h_1+2}$ &
$\Phi_{+2}^{(h_1)}$&
${\scriptstyle 1-h_2}$ &
${\scriptstyle -h_2}$ &
${\cal O}_{+1}^{(h_2),AB}$ &
${\scriptstyle +2}$ &
${\scriptstyle 1-(h_1+h_2)}$ &
${\scriptstyle -(h_1+h_2)} $ &
${\cal O}_{+1}^{(h_1+h_2),AB}$
\\
\hline
${\scriptstyle h_1+2}$ &
$\Phi_{+2}^{(h_1)}$&
${\scriptstyle 1-h_2}$ &
${\scriptstyle \frac{1}{2}-h_2}$ &
${\cal O}_{+\frac{1}{2}}^{(h_2),ABC}$ &
${\scriptstyle +2}$ &
${\scriptstyle 1-(h_1+h_2)}$ &
${\scriptstyle \frac{1}{2}-(h_1+h_2)} $ &
${\cal O}_{+\frac{1}{2}}^{(h_1+h_2),ABC}$
\\
\hline
${\scriptstyle h_1+2}$ &
$\Phi_{+2}^{(h_1)}$&
${\scriptstyle 1-h_2}$ &
${\scriptstyle 1-h_2}$ &
${\cal O}_{0}^{(h_2),ABCD}$ &
${\scriptstyle +2}$ &
${\scriptstyle 1-(h_1+h_2)}$ &
${\scriptstyle 1-(h_1+h_2)} $ &
${\cal O}_{0}^{(h_1+h_2),ABCD}$
\\
\hline
${\scriptstyle h_1+2}$ &
$\Phi_{+2}^{(h_1)}$&
${\scriptstyle 1-h_2}$ &
${\scriptstyle \frac{3}{2}-h_2}$ &
${\cal O}_{ABC,-\frac{1}{2}}^{(h_2)}$ &
${\scriptstyle +2}$ &
${\scriptstyle 1-(h_1+h_2)}$ &
${\scriptstyle \frac{3}{2}-(h_1+h_2)} $ &
${\cal O}_{ABC,-\frac{1}{2}}^{(h_1+h_2)}$
\\
\hline
${\scriptstyle h_1+2}$ &
$\Phi_{+2}^{(h_1)}$&
${\scriptstyle 1-h_2}$ &
${\scriptstyle 2-h_2}$ &
${\cal O}_{AB,-1}^{(h_2)}$ &
${\scriptstyle +2}$ &
${\scriptstyle 1-(h_1+h_2)}$ &
${\scriptstyle 2-(h_1+h_2)} $ &
${\cal O}_{AB,-1}^{(h_1+h_2)}$
\\
\hline
${\scriptstyle h_1+2}$ &
$\Phi_{+2}^{(h_1)}$&
${\scriptstyle 1-h_2}$ &
${\scriptstyle \frac{5}{2}-h_2}$ &
${\cal O}_{A,-\frac{3}{2}}^{(h_2)}$ &
${\scriptstyle +2}$ &
${\scriptstyle 1-(h_1+h_2)}$ &
${\scriptstyle \frac{5}{2}-(h_1+h_2)} $ &
${\cal O}_{A,-\frac{3}{2}}^{(h_1+h_2)}$
\\
\hline
${\scriptstyle h_1+2}$ &
$\Phi_{+2}^{(h_1)}$&
${\scriptstyle 1-h_2}$ &
${\scriptstyle 3-h_2}$ &
${\cal O}_{-2}^{(h_2)}$ &
${\scriptstyle +2}$ &
${\scriptstyle 1-(h_1+h_2)}$ &
${\scriptstyle 3-(h_1+h_2)} $ &
${\cal O}_{-2}^{(h_1+h_2)}$
\\
\hline
\hline
${\scriptstyle h_1+\frac{3}{2}}$ &
$\Phi_{+\frac{3}{2}}^{(h_1),A}$&
${\scriptstyle 1-h_2}$ &
${\scriptstyle -\frac{1}{2}-h_2}$ &
${\cal O}_{+\frac{3}{2}}^{(h_2),B}$ &
${\scriptstyle +\frac{3}{2}}$ &
${\scriptstyle 1-(h_1+h_2)}$ &
${\scriptstyle -(h_1+h_2)} $ &
${\cal O}_{+1}^{(h_1+h_2),AB}$
\\
\hline
${\scriptstyle h_1+\frac{3}{2}}$ &
$\Phi_{+\frac{3}{2}}^{(h_1),A}$&
${\scriptstyle 1-h_2}$ &
${\scriptstyle -h_2}$ &
${\cal O}_{+1}^{(h_2),BC}$ &
${\scriptstyle +\frac{3}{2}}$ &
${\scriptstyle 1-(h_1+h_2)}$ &
${\scriptstyle \frac{1}{2}-(h_1+h_2)} $ &
${\cal O}_{+\frac{1}{2}}^{(h_1+h_2),ABC}$
\\
\hline
${\scriptstyle h_1+\frac{3}{2}}$ &
$\Phi_{+\frac{3}{2}}^{(h_1),A}$&
${\scriptstyle 1-h_2}$ &
${\scriptstyle \frac{1}{2}-h_2}$ &
${\cal O}_{+\frac{1}{2}}^{(h_2),BCD}$ &
${\scriptstyle +\frac{3}{2}}$ &
${\scriptstyle 1-(h_1+h_2)}$ &
${\scriptstyle 1-(h_1+h_2)} $ &
${\cal O}_{0}^{(h_1+h_2),ABCD}$
\\
\hline
${\scriptstyle h_1+\frac{3}{2}}$ &
$\Phi_{+\frac{3}{2}}^{(h_1),A}$&
${\scriptstyle 1-h_2}$ &
${\scriptstyle 1-h_2}$ &
${\cal O}_{0}^{(h_2),BCDE}$ &
${\scriptstyle +\frac{3}{2}}$ &
${\scriptstyle 1-(h_1+h_2)}$ &
${\scriptstyle \frac{3}{2}-(h_1+h_2)} $ &
${\cal O}_{FGH,-\frac{1}{2}}^{(h_1+h_2)}$
\\
\hline
${\scriptstyle h_1+\frac{3}{2}}$ &
$\Phi_{+\frac{3}{2}}^{(h_1),A}$&
${\scriptstyle 1-h_2}$ &
${\scriptstyle \frac{3}{2}-h_2}$ &
${\cal O}_{BCD,-\frac{1}{2}}^{(h_2)}$ &
${\scriptstyle +\frac{3}{2}}$ &
${\scriptstyle 1-(h_1+h_2)}$ &
${\scriptstyle 2-(h_1+h_2)} $ &
${\cal O}_{CD,-1}^{(h_1+h_2)} +\cdots$
\\
\hline
${\scriptstyle h_1+\frac{3}{2}}$ &
$\Phi_{+\frac{3}{2}}^{(h_1),A}$&
${\scriptstyle 1-h_2}$ &
${\scriptstyle 2-h_2}$ &
${\cal O}_{BC,-1}^{(h_2)}$ &
${\scriptstyle +\frac{3}{2}}$ &
${\scriptstyle 1-(h_1+h_2)}$ &
${\scriptstyle \frac{5}{2}-(h_1+h_2)} $ &
${\cal O}_{C,-\frac{3}{2}}^{(h_1+h_2)}+\cdots$
\\
\hline
${\scriptstyle h_1+\frac{3}{2}}$ &
$\Phi_{+\frac{3}{2}}^{(h_1),A}$&
${\scriptstyle 1-h_2}$ &
${\scriptstyle \frac{5}{2}-h_2}$ &
${\cal O}_{B,-\frac{3}{2}}^{(h_2)}$ &
${\scriptstyle +\frac{3}{2}}$ &
${\scriptstyle 1-(h_1+h_2)}$ &
${\scriptstyle 3-(h_1+h_2)} $ &
${\cal O}_{-2}^{(h_1+h_2)}$
\\
\hline
\hline
\end{tabular}
\caption{
  The right-weights $q$ (and the helicities $s_1$) for the
  ${\rm w}_{1+\infty}$ currents,
the left- and right-weights
$(h, \bar{h})$ for the matter fields on the left-hand sides
and those $(h+s_1-q, \bar{h}+2-q)$ for the matter fields
on the right-hand sides in (\ref{OPEwO}).
From the weights in the seventh and eighth columns
of this Table,
we can write down the matter fields on the right-hand sides completely.
We list these weights (and helicities) for the 
first sixteen equations of (\ref{25OPEs}) in the section \ref{OPE}.
For the second and the third cases from the below,
some of the terms are abbreviated in the corresponding last columns.}
\label{table1}
\end{table}

\begin{table}[tbp]
\centering
\renewcommand{\arraystretch}{1.7}
\begin{tabular}{|c|c||c|c|c|||c|c|c|c|| }
\hline
$ q$ &
$ {\rm w^q}$ &
$ h$ &
$ \bar{h}$ &
$ {\cal O}_{h,\bar{h}}$ &
$ s_1$ &
$ h+s_1-q$ &
$ \bar{h}+2-q$ &
${\cal O}_{h+s_1-q,\bar{h}+2-q}$  
\\
\hline
\hline
${\scriptstyle h_1+1}$ &
$\Phi_{+1}^{(h_1),AB}$&
${\scriptstyle 1-h_2}$ &
${\scriptstyle -h_2}$ &
${\cal O}_{+1}^{(h_2),CD}$ &
${\scriptstyle +1}$ &
${\scriptstyle 1-(h_1+h_2)}$ &
${\scriptstyle 1-(h_1+h_2)} $ &
${\cal O}_{0}^{(h_1+h_2),ABCD}$
\\
\hline
${\scriptstyle h_1+1}$ &
$  \Phi_{+1}^{(h_1),AB}$ &
${\scriptstyle 1-h_2}$ &
${\scriptstyle \frac{1}{2}-h_2}$ &
$ {\cal O}_{+\frac{1}{2}}^{(h_2),CDE}$ &
${\scriptstyle +1}$ &
${\scriptstyle 1-(h_1+h_2)}$ &
${\scriptstyle \frac{3}{2}-(h_1+h_2)} $ &
$ {\cal O}_{FGH,-\frac{1}{2}}^{(h_1+h_2)}$
\\
\hline
${\scriptstyle h_1+1}$ &
$\Phi_{+1}^{(h_1),AB}$&
${\scriptstyle 1-h_2}$ &
${\scriptstyle 1-h_2}$ &
${\cal O}_{0}^{(h_2),CDEF}$ &
${\scriptstyle +1}$ &
${\scriptstyle 1-(h_1+h_2)}$ &
${\scriptstyle 2-(h_1+h_2)} $ &
${\cal O}_{GH,-1}^{(h_1+h_2)}$
\\
\hline
${\scriptstyle h_1+1}$ &
$\Phi_{+1}^{(h_1),AB}$&
${\scriptstyle 1-h_2}$ &
${\scriptstyle \frac{3}{2}-h_2}$ &
${\cal O}_{CDE,-\frac{1}{2}}^{(h_2)}$ &
${\scriptstyle +1}$ &
${\scriptstyle 1-(h_1+h_2)}$ &
${\scriptstyle \frac{5}{2}-(h_1+h_2)} $ &
${\cal O}_{D,-\frac{3}{2}}^{(h_1+h_2)}+\cdots$
\\
\hline
${\scriptstyle h_1+1}$ &
$\Phi_{+1}^{(h_1),AB}$&
${\scriptstyle 1-h_2}$ &
${\scriptstyle 2-h_2}$ &
${\cal O}_{CD,-1}^{(h_2)}$ &
${\scriptstyle +1}$ &
${\scriptstyle 1-(h_1+h_2)}$ &
${\scriptstyle 3-(h_1+h_2)} $ &
${\cal O}_{-2}^{(h_1+h_2)}$
\\
\hline
\hline
${\scriptstyle h_1+\frac{1}{2}}$ &
$\Phi_{+\frac{1}{2}}^{(h_1),ABC}$&
${\scriptstyle 1-h_2}$ &
${\scriptstyle \frac{1}{2}-h_2}$ &
${\cal O}_{+\frac{1}{2}}^{(h_2),DEF}$ &
${\scriptstyle +\frac{1}{2}}$ &
${\scriptstyle 1-(h_1+h_2)}$ &
${\scriptstyle 2-(h_1+h_2)} $ &
${\cal O}_{GH,-1}^{(h_1+h_2)}$
\\
\hline
${\scriptstyle h_1+\frac{1}{2}}$ &
$\Phi_{+\frac{1}{2}}^{(h_1),ABC}$&
${\scriptstyle 1-h_2}$ &
${\scriptstyle 1-h_2}$ &
${\cal O}_{0}^{(h_2),DEFG}$ &
${\scriptstyle +\frac{1}{2}}$ &
${\scriptstyle 1-(h_1+h_2)}$ &
${\scriptstyle \frac{5}{2}-(h_1+h_2)} $ &
${\cal O}_{H,-\frac{3}{2}}^{(h_1+h_2)}$
\\
\hline
${\scriptstyle h_1+\frac{1}{2}}$ &
$\Phi_{+\frac{1}{2}}^{(h_1),ABC}$&
${\scriptstyle 1-h_2}$ &
${\scriptstyle \frac{3}{2}-h_2}$ &
${\cal O}_{DEF,-\frac{1}{2}}^{(h_2)}$ &
${\scriptstyle +\frac{1}{2}}$ &
${\scriptstyle 1-(h_1+h_2)}$ &
${\scriptstyle 3-(h_1+h_2)} $ &
${\cal O}_{-2}^{(h_1+h_2)}$
\\
\hline
\hline
${\scriptstyle h_1}$ &
$\Phi_{0}^{(h_1),ABCD}$&
${\scriptstyle 1-h_2}$ &
${\scriptstyle 1-h_2}$ &
${\cal O}_{0}^{(h_2),EFGH}$ &
${\scriptstyle 0}$ &
${\scriptstyle 1-(h_1+h_2)}$ &
${\scriptstyle 3-(h_1+h_2)} $ &
${\cal O}_{-2}^{(h_1+h_2)}$
\\
\hline
\hline
\end{tabular}
\caption{This is a continuation of previous Table \ref{table1}.
We list the corresponding weights (and helicities) for  
the remaining nine equations of (\ref{25OPEs}) in the section \ref{OPE}. }
\label{table2}
\end{table}

Finally, we obtain the following twenty-five OPEs
for the case $s_1 \geq s_2$ of two helicities
by i) substituting the values of $q$ and $\bar{h}$
from Tables \ref{table1} and \ref{table2} into
the factors in front of the differential operators
on the right-hand sides of the OPE (\ref{OPEwO}),
ii) identifying the ${\rm w}_{1+\infty}$ currents
and the matter fields appearing on the left-hand sides
(as well as the matter fields on the right-hand sides)
and iii) incorporating the $SU(8)$ indices \cite{AK2509}
with matter fields 
on the right-hand sides of the OPEs appropriately
\footnote{
\label{superBMS}
For $h_1=0$ in the current
$\Phi_{+2}^{(h_1)}(z_1,\bar{z}_1)$, the current
$\Phi_{+2}^{(0)}(z_1,\bar{z}_1)$ plays the role of
$SL(2,\mathbb{R})_R$
stress energy tensor
\cite{HPS}.
The right-hand side of the first equation of (\ref{25OPEs}) consists of
two terms, from (\ref{twovaluesofq}),
\bea
{\Phi}_{+2}^{(0)}(z,\bar{z}) \,
{\cal O}_{+2}^{(h_2)}(0,0)
=  \frac{1}{z}\bigg[ 
\frac{1}{\bar{z}^2}\, (-1-h_2)
+ \frac{1}{\bar{z}}\, \bar{\pa} \bigg]
{\cal O}_{+2}^{(h_2)}(0,0)
+  \cdots,
\label{PhiO}
\eea
where $(-1-h_2)$ at the second order pole in the antiholomorphic
coordinates is the right-weight of the matter field
${\cal O}_{+2}^{(h_2)}(0,0)$
according to Table \ref{table1}.
We can expand $ {\Phi}_{+2}^{(0)}(z,\bar{z})$
with respect to the holomorphic coordinate $z$
via the standard mode expansion.
Each coefficient operator denoted by ${\Phi}_{+2}^{(0)}(\bar{z})$
depends on the antiholomorphic
coordinate $\bar{z}$. See also (\ref{wmode}) where the right-hand side
contains $\frac{1}{z}$ factor.
We take the contour integral over $z$ on both sides of (\ref{PhiO})
and obtain the OPE of the (holomorphic) zero mode of  the current
with the matter field
\bea
{\Phi}_{+2}^{(0)}(\bar{z}) \,
{\cal O}_{+2}^{(h_2)}(0,0)
= \bigg[ 
\frac{1}{\bar{z}^2}\, (-1-h_2)
+ \frac{1}{\bar{z}}\, \bar{\pa} \bigg]
{\cal O}_{+2}^{(h_2)}(0,0)
+  \cdots.
\label{primary}
\eea
We can check other eight operators (the graviton, the gravitinos,
the graviphotons, the graviphotinos and the
scalars) in (\ref{wO}) by using the corresponding equations
which are the second to the ninth relations in (\ref{25OPEs}).
They satisfy their primary conditions as in (\ref{primary})
with their own right-weights given in Table \ref{table1}.

On the other hand, for $h_1=-\frac{1}{2}$, we have
the OPE together with (\ref{twovaluesofq})
\bea
{\Phi}_{+2}^{(-\frac{1}{2})}(z,\bar{z}) \,
{\cal O}_{+2}^{(h_2)}(0,0)
=\frac{1}{z\, \bar{z}} \, \frac{1}{2}\,
{\cal O}_{+2}^{(-\frac{1}{2}+h_2)}(0,0)
+  \cdots.
\label{translation}
\eea
The left- and right-weights of the matter field
${\cal O}_{+2}^{(h_2)}(0,0)$, $(1-h_2)$ and $(-1-h_2)$
are increased by $\frac{1}{2}$, $\frac{1}{2}$ respectively
in the matter field ${\cal O}_{+2}^{(-\frac{1}{2}+h_2)}(0,0)$
on the right-hand side, according to Table \ref{table1}.
Note that ${\Phi}_{+2}^{(-\frac{1}{2})}(z,\bar{z})$
has the left- and right-weights $(3-q,q)=(\frac{3}{2},\frac{3}{2})$
and is identified with the supertranslation generator \cite{HPS}.
See also (\ref{wmode}) where the powers of $z$ and $\bar{z}$
in the denominator
are given by $(\frac{3}{2}+m)$ and $(\frac{3}{2}+n)$
respectively.
As we did before in the previous paragraph,
other eight operators (the graviton, the gravitinos,
the graviphotons, the graviphotinos and the
scalars) in (\ref{wO}) can be analyzed similarly to obtain
the corresponding relations like as (\ref{translation}).

In the notation of \cite{FSTZ}, the right Virasoro generator
and supertranslation generator of extended BMS algebra are given by
${\Phi}_{+2}^{(0)}(\bar{z}) \equiv \bar{T}(\bar{z})$ and
${\Phi}_{+2}^{(-\frac{1}{2})}(z,\bar{z}) \equiv {\cal P}(z,\bar{z})$.
We will comment on these further at the section \ref{twocurrents}.
}
\bea
&& {\Phi}_{+2}^{(h_1)}(z_1,\bar{z}_1) \,
{\cal O}_{+2}^{(h_2)}(z_2,\bar{z}_2)
=  \frac{1}{z_{12}}\, \frac{1}{2}\,
\sum_{m=0}^{2(h_1+2)-3}\,
\frac{(m+1)\, \Gamma(2(-1-h_2)+1)}{
  \Gamma(-2-2(h_1+h_2) +m)}
\nonu \\
&& \times  \bar{z}_{12}^{2-2(h_1+2)+m}
\, \bar{\pa}^m \,
{\cal O}_{+2}^{(h_1+h_2)}(z_2,\bar{z}_2)
+  \cdots,
\nonu \\
&& {\Phi}_{+2}^{(h_1)}(z_1,\bar{z}_1) \,
{\cal O}_{+\frac{3}{2}}^{(h_2),A}(z_2,\bar{z}_2)
=  \frac{1}{z_{12}}\, \frac{1}{2}\,
\sum_{m=0}^{2(h_1+2)-3}\,
\frac{(m+1)\, \Gamma(2(-\frac{1}{2}-h_2)+1)}{
  \Gamma(-1-2(h_1+h_2) +m)}\nonu \\
&& \times  \bar{z}_{12}^{2-2(h_1+2)+m}
\, \bar{\pa}^m \,
{\cal O}_{+\frac{3}{2}}^{(h_1+h_2),A}(z_2,\bar{z}_2)
+  \cdots,
\nonu \\
&& {\Phi}_{+2}^{(h_1)}(z_1,\bar{z}_1) \,
{\cal O}_{+1}^{(h_2),AB}(z_2,\bar{z}_2)
=  \frac{1}{z_{12}}\, \frac{1}{2}\,
\sum_{m=0}^{2(h_1+2)-3}\,
\frac{(m+1)\, \Gamma(2(-h_2)+1)}{
  \Gamma(-2(h_1+h_2) +m)}\, \bar{z}_{12}^{2-2(h_1+2)+m}
\nonu \\
&& \times  \bar{\pa}^m \,
{\cal O}_{+1}^{(h_1+h_2),AB}(z_2,\bar{z}_2)
+  \cdots,
\nonu \\
&&
{\Phi}_{+2}^{(h_1)}(z_1,\bar{z}_1) \,
{\cal O}_{+\frac{1}{2}}^{(h_2),ABC}(z_2,\bar{z}_2)
=  \frac{1}{z_{12}}\, \frac{1}{2}\,
\sum_{m=0}^{2(h_1+2)-3}\,
\frac{(m+1)\, \Gamma(2(\frac{1}{2}-h_2)+1)}{
  \Gamma(1-2(h_1+h_2) +m)}\, \bar{z}_{12}^{2-2(h_1+2)+m}
\nonu \\
&& \times   \bar{\pa}^m \,
{\cal O}_{+\frac{1}{2}}^{(h_1+h_2),ABC}(z_2,\bar{z}_2)
+  \cdots,
\nonu \\
&&
{\Phi}_{+2}^{(h_1)}(z_1,\bar{z}_1) \,
{\cal O}_{0}^{(h_2),ABCD}(z_2,\bar{z}_2)
=  \frac{1}{z_{12}}\, \frac{1}{2}\,
\sum_{m=0}^{2(h_1+2)-3}\,
\frac{(m+1)\, \Gamma(2(1-h_2)+1)}{
  \Gamma(2-2(h_1+h_2) +m)}\, \bar{z}_{12}^{2-2(h_1+2)+m}
\nonu \\
&& \times  \bar{\pa}^m \,
{\cal O}_{0}^{(h_1+h_2),ABCD}(z_2,\bar{z}_2)
+  \cdots,
\nonu \\
&&
{\Phi}_{+2}^{(h_1)}(z_1,\bar{z}_1) \,
{\cal O}_{ABC,-\frac{1}{2}}^{(h_2)}(z_2,\bar{z}_2)
=  \frac{1}{z_{12}}\, \frac{1}{2}\,
\sum_{m=0}^{2(h_1+2)-3}\,
\frac{(m+1)\, \Gamma(2(\frac{3}{2}-h_2)+1)}{
  \Gamma(3-2(h_1+h_2) +m)}\, \bar{z}_{12}^{2-2(h_1+2)+m} 
\nonu \\
&& \times  \bar{\pa}^m \,
{\cal O}_{ABC,-\frac{1}{2}}^{(h_1+h_2)}(z_2,\bar{z}_2)
+  \cdots,
\nonu \\
&& {\Phi}_{+2}^{(h_1)}(z_1,\bar{z}_1) \,
{\cal O}_{AB,-1}^{(h_2)}(z_2,\bar{z}_2)
=  \frac{1}{z_{12}}\, \frac{1}{2}\,
\sum_{m=0}^{2(h_1+2)-3}\,
\frac{(m+1)\, \Gamma(2(2-h_2)+1)}{
  \Gamma(4-2(h_1+h_2) +m)}\, \bar{z}_{12}^{2-2(h_1+2)+m}
\nonu \\
&& \times   \bar{\pa}^m \,
{\cal O}_{AB,-1}^{(h_1+h_2)}(z_2,\bar{z}_2)
+  \cdots,
\nonu \\
&&
{\Phi}_{+2}^{(h_1)}(z_1,\bar{z}_1) \,
{\cal O}_{A,-\frac{3}{2}}^{(h_2)}(z_2,\bar{z}_2)
=  \frac{1}{z_{12}}\, \frac{1}{2}\,
\sum_{m=0}^{2(h_1+2)-3}\,
\frac{(m+1)\, \Gamma(2(\frac{5}{2}-h_2)+1)}{
  \Gamma(5-2(h_1+h_2) +m)}\, \bar{z}_{12}^{2-2(h_1+2)+m}
\nonu \\
&& \times   \bar{\pa}^m \,
{\cal O}_{A,-\frac{3}{2}}^{(h_1+h_2)}(z_2,\bar{z}_2)
+  \cdots,
\nonu \\
&&
{\Phi}_{+2}^{(h_1)}(z_1,\bar{z}_1) \,
{\cal O}_{-2}^{(h_2)}(z_2,\bar{z}_2)
=  \frac{1}{z_{12}}\, \frac{1}{2}\,
\sum_{m=0}^{2(h_1+2)-3}\,
\frac{(m+1)\, \Gamma(2(3-h_2)+1)}{
  \Gamma(6-2(h_1+h_2) +m)}\, \bar{z}_{12}^{2-2(h_1+2)+m}
\nonu \\
&& \times   \bar{\pa}^m \,
{\cal O}_{-2}^{(h_1+h_2)}(z_2,\bar{z}_2)
+  \cdots,
\nonu \\
&&
{\Phi}_{+\frac{3}{2}}^{(h_1),A}(z_1,\bar{z}_1) 
{\cal O}_{+\frac{3}{2}}^{(h_2),B}(z_2,\bar{z}_2)
=  \frac{1}{z_{12}} \frac{1}{2}
\sum_{m=0}^{2(h_1+\frac{3}{2})-3}
\frac{(m+1) \Gamma(2(-\frac{1}{2}-h_2)+1)}{
\Gamma(-2(h_1+h_2) +m)}  \bar{z}_{12}^{2-2(h_1+\frac{3}{2})+m}
\nonu \\
&& \times  \bar{\pa}^m \,
{\cal O}_{+1}^{(h_1+h_2),AB}(z_2,\bar{z}_2)
+  \cdots,
\nonu \\
&&
{\Phi}_{+\frac{3}{2}}^{(h_1),A}(z_1,\bar{z}_1) \,
{\cal O}_{+1}^{(h_2),BC}(z_2,\bar{z}_2)
=  \frac{1}{z_{12}}\, \frac{1}{2}\,
\sum_{m=0}^{2(h_1+\frac{3}{2})-3}\,
\frac{(m+1)\, \Gamma(2(-h_2)+1)}{
  \Gamma(1-2(h_1+h_2) +m)}\, \bar{z}_{12}^{2-2(h_1+\frac{3}{2})+m}
\nonu \\
&& \times   \bar{\pa}^m \,
{\cal O}_{+\frac{1}{2}}^{(h_1+h_2),ABC}(z_2,\bar{z}_2)
+  \cdots,
\nonu \\
&&
{\Phi}_{+\frac{3}{2}}^{(h_1),A}(z_1,\bar{z}_1) 
{\cal O}_{+\frac{1}{2}}^{(h_2),BCD}(z_2,\bar{z}_2)
=  \frac{1}{z_{12}} \frac{1}{2}
\sum_{m=0}^{2(h_1+\frac{3}{2})-3}
\frac{(m+1) \Gamma(2(\frac{1}{2}-h_2)+1)}{
\Gamma(2-2(h_1+h_2) +m)} \bar{z}_{12}^{2-2(h_1+\frac{3}{2})+m}
\nonu \\
&& \times   \bar{\pa}^m \,
{\cal O}_{0}^{(h_1+h_2),ABCD}(z_2,\bar{z}_2)
+  \cdots,
\nonu \\
&&
{\Phi}_{+\frac{3}{2}}^{(h_1),A}(z_1,\bar{z}_1) \,
{\cal O}_{0}^{(h_2),BCDE}(z_2,\bar{z}_2)
=  \frac{1}{z_{12}}\, \frac{1}{2}\,
\sum_{m=0}^{2(h_1+\frac{3}{2})-3}\,
\frac{(m+1)\, \Gamma(2(1-h_2)+1)}{
  \Gamma(3-2(h_1+h_2) +m)}\, \bar{z}_{12}^{2-2(h_1+\frac{3}{2})+m}
\nonu \\
&& \times  
\frac{1}{3!} \, \ep^{ABCDEFGH}\,
 \bar{\pa}^m \, {\cal O}_{FGH,-\frac{1}{2}}^{(h_1+h_2)}(z_2,\bar{z}_2)
+  \cdots,
\nonu \\
&&
{\Phi}_{+\frac{3}{2}}^{(h_1),A}(z_1,\bar{z}_1) \,
{\cal O}_{BCD,-\frac{1}{2}}^{(h_2)}(z_2,\bar{z}_2)
=  \frac{1}{z_{12}}\, \frac{1}{2}\,
\sum_{m=0}^{2(h_1+\frac{3}{2})-3}\,
\frac{(m+1)\, \Gamma(2(\frac{3}{2}-h_2)+1)}{
  \Gamma(4-2(h_1+h_2) +m)}\, \bar{z}_{12}^{2-2(h_1+\frac{3}{2})+m}
\nonu \\
&& \times   
3 \, \de^{A}_{[B} \,\bar{\pa}^m \,
{\cal O}_{CD],-1}^{(h_1+h_2)}(z_2,\bar{z}_2)
+  \cdots,
\nonu \\
&&
{\Phi}_{+\frac{3}{2}}^{(h_1),A}(z_1,\bar{z}_1) \,
{\cal O}_{BC,-1}^{(h_2)}(z_2,\bar{z}_2)
=  \frac{1}{z_{12}}\, \frac{1}{2}\,
\sum_{m=0}^{2(h_1+\frac{3}{2})-3}\,
\frac{(m+1)\, \Gamma(2(2-h_2)+1)}{
  \Gamma(5-2(h_1+h_2) +m)}\, \bar{z}_{12}^{2-2(h_1+\frac{3}{2})+m}
\nonu \\
&& \times  
2! \, \de^{A}_{[B} \, \bar{\pa}^m \,
{\cal O}_{C],-\frac{3}{2}}^{(h_1+h_2)}(z_2,\bar{z}_2)
+  \cdots,
\nonu \\
&&
{\Phi}_{+\frac{3}{2}}^{(h_1),A}(z_1,\bar{z}_1) \,
{\cal O}_{B,-\frac{3}{2}}^{(h_2)}(z_2,\bar{z}_2)
=  \frac{1}{z_{12}}\, \frac{1}{2}\,
\sum_{m=0}^{2(h_1+\frac{3}{2})-3}\,
\frac{(m+1)\, \Gamma(2(\frac{5}{2}-h_2)+1)}{
  \Gamma(6-2(h_1+h_2) +m)}\, \bar{z}_{12}^{2-2(h_1+\frac{3}{2})+m}
\nonu \\
&& \times 
\de^{A}_{B} \, \bar{\pa}^m \,
{\cal O}_{-2}^{(h_1+h_2)}(z_2,\bar{z}_2)
+  \cdots,
\nonu \\
&&
{\Phi}_{+1}^{(h_1),AB}(z_1,\bar{z}_1) \,
{\cal O}_{+1}^{(h_2),CD}(z_2,\bar{z}_2)
=  \frac{1}{z_{12}}\, \frac{1}{2}\,
\sum_{m=0}^{2(h_1+1)-3}\,
\frac{(m+1)\, \Gamma(2(-h_2)+1)}{
  \Gamma(2-2(h_1+h_2) +m)}\, \bar{z}_{12}^{2-2(h_1+1)+m}
\nonu \\
&& \times   \bar{\pa}^m \,
{\cal O}_{0}^{(h_1+h_2),ABCD}(z_2,\bar{z}_2)
+  \cdots,
\nonu \\
&&
{\Phi}_{+1}^{(h_1),AB}(z_1,\bar{z}_1) \,
{\cal O}_{+\frac{1}{2}}^{(h_2),CDE}(z_2,\bar{z}_2)
=  \frac{1}{z_{12}}\, \frac{1}{2}\,
\sum_{m=0}^{2(h_1+1)-3}\,
\frac{(m+1)\, \Gamma(2(\frac{1}{2}-h_2)+1)}{
  \Gamma(3-2(h_1+h_2) +m)}\, \bar{z}_{12}^{2-2(h_1+1)+m}\, 
\nonu \\
&& \times 
\frac{1}{3!}\, \ep^{ABCDEFGH}\, \bar{\pa}^m \,
{\cal O}_{FGH,-\frac{1}{2}}^{(h_1+h_2)}(z_2,\bar{z}_2)
+  \cdots,
\nonu \\
&&
{\Phi}_{+1}^{(h_1),AB}(z_1,\bar{z}_1) \,
{\cal O}_{0}^{(h_2),CDEF}(z_2,\bar{z}_2)
=  \frac{1}{z_{12}}\, \frac{1}{2}\,
\sum_{m=0}^{2(h_1+1)-3}\,
\frac{(m+1)\, \Gamma(2(1-h_2)+1)}{
  \Gamma(4-2(h_1+h_2) +m)}\, \bar{z}_{12}^{2-2(h_1+1)+m}
\nonu \\
&& \times   
\frac{1}{2!}\, \ep^{ABCDEFGH}\,\bar{\pa}^m \,
{\cal O}_{GH,-1}^{(h_1+h_2)}(z_2,\bar{z}_2)
+  \cdots,
\nonu \\
&&
{\Phi}_{+1}^{(h_1),AB}(z_1,\bar{z}_1) \,
{\cal O}_{CDE,-\frac{1}{2}}^{(h_2)}(z_2,\bar{z}_2)
=  \frac{1}{z_{12}}\, \frac{1}{2}\,
\sum_{m=0}^{2(h_1+1)-3}\,
\frac{(m+1)\, \Gamma(2(\frac{3}{2}-h_2)+1)}{
  \Gamma(5-2(h_1+h_2) +m)}\, \bar{z}_{12}^{2-2(h_1+1)+m}
\nonu \\
&& \times  3!\,
\de^{A}_{[C} \, \bar{\pa}^m \,
{\cal O}_{D,-\frac{3}{2}}^{(h_1+h_2)}(z_2,\bar{z}_2) \, \de^{B}_{E]}
+  \cdots,
\nonu \\
&&
{\Phi}_{+1}^{(h_1),AB}(z_1,\bar{z}_1) \,
{\cal O}_{CD,-1}^{(h_2)}(z_2,\bar{z}_2)
=  \frac{1}{z_{12}}\, \frac{1}{2}\,
\sum_{m=0}^{2(h_1+1)-3}\,
\frac{(m+1)\, \Gamma(2(2-h_2)+1)}{
  \Gamma(6-2(h_1+h_2) +m)}\, \bar{z}_{12}^{2-2(h_1+1)+m}
\nonu \\
&& \times  
\de^{AB}_{CD} \, \bar{\pa}^m \,
{\cal O}_{-2}^{(h_1+h_2)}(z_2,\bar{z}_2)
+  \cdots,
\nonu \\
&&
{\Phi}_{+\frac{1}{2}}^{(h_1),ABC}(z_1,\bar{z}_1) \,
{\cal O}_{+\frac{1}{2}}^{(h_2),DEF}(z_2,\bar{z}_2)
=  \frac{1}{z_{12}}\, \frac{1}{2}\,
\sum_{m=0}^{2(h_1+\frac{1}{2})-3}\,
\frac{(m+1)\, \Gamma(2(\frac{1}{2}-h_2)+1)}{
  \Gamma(4-2(h_1+h_2) +m)}\, \bar{z}_{12}^{2-2(h_1+\frac{1}{2})+m}
\nonu \\
&& \times  
\frac{1}{2!} \, \ep^{ABCDEFGH} \, \bar{\pa}^m \,
{\cal O}_{GH,-1}^{(h_1+h_2)}(z_2,\bar{z}_2)
+  \cdots,
\nonu \\
&&
{\Phi}_{+\frac{1}{2}}^{(h_1),ABC}(z_1,\bar{z}_1) \,
{\cal O}_{0}^{(h_2),DEFG}(z_2,\bar{z}_2)
=  \frac{1}{z_{12}}\, \frac{1}{2}\,
\sum_{m=0}^{2(h_1+\frac{1}{2})-3}\,
\frac{(m+1)\, \Gamma(2(1-h_2)+1)}{
  \Gamma(5-2(h_1+h_2) +m)}\, \bar{z}_{12}^{2-2(h_1+\frac{1}{2})+m}
\nonu \\
&& \times  
\ep^{ABCDEFGH} \, \bar{\pa}^m \,
{\cal O}_{H,-\frac{3}{2}}^{(h_1+h_2)}(z_2,\bar{z}_2)
+  \cdots,
\nonu \\
&&
{\Phi}_{+\frac{1}{2}}^{(h_1),ABC}(z_1,\bar{z}_1) \,
{\cal O}_{DEF,-\frac{1}{2}}^{(h_2)}(z_2,\bar{z}_2)
=  \frac{1}{z_{12}}\, \frac{1}{2}\,
\sum_{m=0}^{2(h_1+\frac{1}{2})-3}\,
\frac{(m+1)\, \Gamma(2(\frac{3}{2}-h_2)+1)}{
  \Gamma(6-2(h_1+h_2) +m)}\, \bar{z}_{12}^{2-2(h_1+\frac{1}{2})+m} 
\nonu \\
&& \times 
\de^{ABC}_{DEF} \, \bar{\pa}^m \,
{\cal O}_{-2}^{(h_1+h_2)}(z_2,\bar{z}_2)
+  \cdots,
\nonu \\
&&
{\Phi}_{0}^{(h_1),ABCD}(z_1,\bar{z}_1) \,
{\cal O}_{0}^{(h_2),EFGH}(z_2,\bar{z}_2)
=  \frac{1}{z_{12}}\, \frac{1}{2}\,
\sum_{m=0}^{2(h_1)-3}\,
\frac{(m+1)\, \Gamma(2(1-h_2)+1)}{
  \Gamma(6-2(h_1+h_2) +m)}\, \bar{z}_{12}^{2-2(h_1)+m}
\nonu \\
&& \times   
\ep^{ABCDEFGH} \,\bar{\pa}^m \,
{\cal O}_{-2}^{(h_1+h_2)}(z_2,\bar{z}_2)
+  \cdots.
\label{25OPEs}
\eea
Let us focus on 
the tenth relation of (\ref{25OPEs})
\footnote{ We are using the same normalization of \cite{AK2509}
for the numerical factors appearing on the right-hand sides of
(\ref{25OPEs}). See also the footnotes $9$ and $10$ of \cite{AK2509}.
For the $n$-repeated $SU(8)$ indices, we multiply by $\frac{1}{n!}$
and for the antisymmetric $n$ $SU(8)$ indices, we also
assume $\frac{1}{n!}$. This implies that
for fixed $SU(8)$ indices on the left-hand sides, the corresponding
operators appear once (with the numerical coefficients $1$) up to signs.
This applies to
all the OPEs and all the (anti)commutators in this paper.}.
The current $\Phi_{+\frac{3}{2}}^{(h_1=0),A}$ has the following
OPE with the gravitino matter fields
\bea
{\Phi}_{+\frac{3}{2}}^{(0),A}(z,\bar{z}) 
{\cal O}_{+\frac{3}{2}}^{(h_2),B}(0,0)= \frac{1}{z \,\bar{z} }
\, \frac{1}{2}\,
{\cal O}_{+1}^{(h_2),AB}(0,0)
+  \cdots.
\label{tenth}
\eea
As described in the footnote \ref{superBMS},
the contour integral over $z$ on both sides of (\ref{tenth})
can be applied together with the mode expansion of
${\Phi}_{+\frac{3}{2}}^{(0),A}(z,\bar{z})$ in the holomorphic coordinate
$z$.
This leads to the OPE of the (holomorphic) zero mode
of the current with the matter field
\bea
{\Phi}_{+\frac{3}{2}}^{(0),A}(\bar{z}) 
{\cal O}_{+\frac{3}{2}}^{(h_2),B}(0,0)= \frac{1}{\bar{z} }\, \frac{1}{2}\,
{\cal O}_{+1}^{(h_2),AB}(0,0)
+  \cdots.
\label{simpletenth}
\eea
The left-weight of the left-hand side of
(\ref{simpletenth}) is given by $0+(1-h_2)$
which can be written as $(1-h_2)$ which is the left-weight
of the right-hand side of (\ref{simpletenth}).
Note that the left-weight $1$ of the current
${\Phi}_{+\frac{3}{2}}^{(0),A}(z,\bar{z})$
from Table \ref{table1}
becomes $0$ for the current ${\Phi}_{+\frac{3}{2}}^{(0),A}(\bar{z})$
after the contour integration over $z$.
The right-weight of the left-hand side
is $\frac{3}{2}+(-\frac{1}{2}-h_2)$
which is equal to $1+(-h_2)$.
This is the right-weight of the right-hand side.
That is, the sum of left- and right-weights
of the matter field on the left-hand side,
$(\frac{1}{2}-2h_2)$, is transformed into
$(1-2h_2)$ of the matter field on the right-hand side
(or in the notation of $\Delta$, this shows $\Delta \rightarrow
\Delta+\frac{1}{2}$).
Or the right-weight of the matter field
is increased by $\frac{1}{2}$.

Therefore, we conclude that
the current ${\Phi}_{+\frac{3}{2}}^{(0),A}(\bar{z})$
with weights $(0,\frac{3}{2})$
is the supercurrent of ${\cal N}=8$ BMS algebra
\cite{2212-2,Tropper}
by recognizing (\ref{simpletenth}).
We will also describe the behavior of this
supercurrent in the context of shadow transform
for the currents in section \ref{lastsection}
\footnote{
\label{Sbar}
In the notation of \cite{FSTZ,2212-2},
this current corresponds to $\bar{S}(\bar{z})$ or
$\bar{S}^A(\bar{z})$. Under the action of
the supercurrent, the matter field
${\cal O}^{(h_2)}_{+2}(z,\bar{z})$ is transformed as
${\cal O}^{(h_2),A}_{+\frac{3}{2}}(z,\bar{z})$.
Similarly, we observe that there are
changes of the matter fields 
${\cal O}^{(h_2),B}_{+\frac{3}{2}}(z,\bar{z}) \rightarrow 
{\cal O}^{(h_2),AB}_{+1}(z,\bar{z}), \cdots,
{\cal O}^{(h_2)}_{B,-\frac{3}{2}}(z,\bar{z}) \rightarrow
\de^A_{B}\,
{\cal O}^{(h_2)}_{-2}(z,\bar{z})$.
We can see that  each helicity of the matter field
is decreased by $\frac{1}{2}$: $s_2 \rightarrow s_2-\frac{1}{2}$.
This implies that
under the supercurrent, the left-weights (or $\frac{(\De+s_2)}{2}$)
of the matter fields
remain the same while the right-weights (or $\frac{(\De-s_2)}{2}$)
are increased by $\frac{1}{2}$.}.

We can analyze from the eleventh equation to the sixteenth
equation of (\ref{25OPEs}) similarly.
We can easily observe that
by comparing the fourth column with the eighth
column
in Table \ref{table1} for fixed $h_1=0$,
the right-weights are increased by $\frac{1}{2}$.
Equivalently, the sum of left- and right-weights (i.e., the
conformal dimension $\Delta$)
of the matter fields is increased by $\frac{1}{2}$
because the left-weights for the matter fields on
both sides are the same.
For the three positive helicities of the matter fields on the
left-hand sides, each $SU(8)$ index
on the left-hand sides appears as the product
of $SU(8)$ indices of the matter fields on the right-hand sides.
For the remaining four non-positive helicities of the matter fields on the
left-hand sides, the $SU(8)$ invariant tensors $\ep^{ABCDEFGH}$ or
$\de^A_B$ appear on the right-hand sides of the OPEs.

For the remaining
twenty OPEs with $s_1 < s_2$ of the ${\cal N}=8$ ${\rm w}_{1+\infty}$
currents with matter fields, we present
them in Appendix A with the help of Tables \ref{table9}
and \ref{table10}.
These can be obtained from (\ref{OOOPE}) (or (\ref{25OPEs}))
by interchanging the two operators on the left-hand sides, in principle.

\section{The action of the charges on matter fields
\label{anticommutators}
}

In order to understand the structure of celestial soft symmetry
algebra \cite{AK2509} in ${\cal N}=8$ supergravity fully,
sometimes we need to describe also (anti)commutators in addition to
the previous OPEs.
In the mode expansion of the current ${\rm w}^{q}(z,\bar{z})$
where the left- and right-weights are
given by $(h,\bar{h})=(s_1+1-q,q)$
(see also footnote
\ref{weightsofw}), we have the following
standard decomposition in terms of
the complex coordinates $(z,\bar{z})$
\bea
{\rm w}^q(z, \bar{z}) =\sum_{m,n} \,
\frac{{\rm w}_{m,n}^q}{z^{s_1+1-q+m} \, \bar{z}^{q+n}}.
\label{wmode}
\eea
Of course, each mode ${\rm w}^q_{m,n}$ can be extracted by performing 
the contour integrals over $z$ and $\bar{z}$ on
both sides of (\ref{wmode}).
Then the modes ${\rm w}_{m,n}^q$ have the weights
$(-m,-n)$.

We can generalize
the definition for the action of the modes on the matter fields
in the bosonic theory \cite{HPS} by realizing that
there exists a minus sign when the ordering is
reversed in the radial ordering between two
fermionic operators.
Let us introduce the following (anti)commutators \footnote{On the right-hand
side of (\ref{wOdef}), the OPE of the currents with the matter fields
is the radial ordering and consists of two parts depending on either
$\sqrt{w\,\bar{w}} > \sqrt{z\,\bar{z}}$ or  $\sqrt{w\,\bar{w}} <
\sqrt{z\,\bar{z}}$. For the former region, there is no change in
the currents and the matter fields and for the latter region,
they are reversed with minus sign. Moreover, for fermionic currents
and fermionic matter fields, there exists an extra minus sign
from the definition of radial ordering \cite{CFT}.
Therefore, there exist the commutators for boson-boson, boson-fermion,
fermion-boson cases and the anticommutator for fermion-fermion case
on the left-hand side of (\ref{wOdef}). We are using the graded commutator
notation $[A, B \}$ for two operators $A$ and $B$.}
\bea
\bigg[
{\rm w}^q_{m,n},
{\cal O}_{h,\bar{h}}(z,\bar{z})   \bigg\}
=\oint_z \, \frac{d w}{2 \pi i} \, w^{s_1-q+m} \, \oint_{\bar{z}} \,
\frac{d \bar{w}}{2 \pi i}\, \bar{w}^{q+n-1}\, {\rm w}^q(w,\bar{w})\,
{\cal O}_{h,\bar{h}}(z,\bar{z}).
\label{wOdef}
\eea
The left-hand side of (\ref{wOdef}) becomes the anticommutator
for the two fermionic operators (the current and the
matter field) on the right-hand side.
Otherwise, the left-hand side becomes the usual commutator notation.
The OPE of the ${\rm w}_{1+\infty}$ currents with the matter fields
appearing on the right-hand side of (\ref{wOdef})
is described in (\ref{OPEwO}).
We can check the left- and right-weights of both sides of (\ref{wOdef}).
The left-weight of the right-hand side is given by
$-1-(s_1-q+m)+(s_1+1-q)+h=-m+h$ while
the right-weight of the right-hand side is given by
$-1-(q+n-1)+q+\bar{h}=-n+\bar{h}$. They are exactly
the weights of the left-hand side of (\ref{wOdef}).

For $m=q-s_1$, we observe that there is only $\frac{1}{z}$ term
in the holomorphic coordinate from (\ref{wmode}).
Moreover, the contour integral over $w$ coordinate in (\ref{wOdef})
becomes unity
because the power of $w$ is equal to zero.
After integrating over $\bar{w}$ in (\ref{wOdef})
and using the change of variables, finally we obtain
\footnote{
\label{hat}
We are using the hatted notation $\hat{}$
for the modes of these restricted currents in the (anti)commutators.}
\bea
\bigg[
\hat{\rm w}^q_m,
{\cal O}_{h,\bar{h}}(z,\bar{z})   \bigg\}  & = &
{\mathscr D}_m^{q}(\bar{h}) \, {\cal O}_{h+s_1-q,\bar{h}+2-q}(z,\bar{z}),
\qquad \hat{\rm w}^q_n \equiv {\rm w}^q_{q-s_1,n},
\label{anticomm}
\eea
where the differential operator \cite{HPS,HP} appearing on the
right-hand side of (\ref{anticomm}) is given by
\footnote{We have the explicit forms for these differential operators
for $q=2, \frac{3}{2}$
\bea
{\mathscr D}_m^{q=2}(\bar{h})= \bar{z}^m \bigg( (1+m)\, \bar{h}+\bar{z}
\, \bar{\pa}_{\bar{z}}\bigg), \qquad
{\mathscr D}_m^{q=\frac{3}{2}}(\bar{h})= \frac{1}{2}\, \bar{z}^{m+\frac{1}{2}}.
\nonu
\eea
For $q \geq \frac{5}{2}$, there are positive powers of antiholomorphic
derivative.
\label{Dvalues}}
\bea
{\mathscr D}_m^{q}(\bar{h}) \equiv \frac{1}{2}\, \sum_{l=0}^{2q-3}\,
\left(\begin{array}{c}
q+m-1 \\  l \\
\end{array}\right)    
\, \frac{(2q-2-l)\, \Gamma(2 \bar{h}+1)}{\Gamma(2\bar{h}+1-l)} \,
\bar{z}^{q+m-1-l} \, \bar{\pa}_{\bar{z}}^{2q-3-l}. 
\label{Diff}
\eea
Note that the left-weight of the right-hand side of
(\ref{anticomm}) is given by $h+(s_1-q)$
which is the weight of the matter field on the right-hand side of
(\ref{anticomm}).
This should be equal to the left-weight of the left-hand side.
Note that the left-weight of $\hat{\rm w}^q_m$ is given by
$-(q-s_1)=s_1-q$. Then we have the correct left-weight of the left-hand
side.
The right-weight of the right-hand side 
is given by the sum of $-(q+m-1-l)+2q-3-l=q-m-2$
which is the weight of (\ref{Diff}) and
$(\bar{h}+2-q)$ which
is the weight of the matter field on the right-hand side
of (\ref{anticomm}).
This sum is given by $(\bar{h}-m)$ which is the right-weight of the
left-hand side.
Obviously, the power of the (antiholomorphic) derivative in
(\ref{Diff}) is greater
than or equal to zero.

Finally, we obtain the following twenty-five (anti)commutators
(corresponding to (\ref{25OPEs}))
for the case $s_1 \geq s_2$ as before
by i) substituting the right-weights
$q$ and $\bar{h}$ on the left-hand sides from Tables \ref{table1}
and \ref{table2} into the differential operators (\ref{Diff})
on the right-hand sides,
ii) identifying the ${\rm w}_{1+\infty}$ currents with the hatted notation
and the two kinds of matter fields correctly and iii)
putting the $SU(8)$ indices together with
the matter fields on the right-hand sides  
\footnote{
\label{q2and3half}
For $q=2$, we can check the following
result \cite{HPS} corresponding to (\ref{primary})
\bea
\bigg[
\hat{\Phi}_{+2,m}^{(0)},
{\cal O}_{+2}^{(h_2)}(z,\bar{z}) \bigg]  & = &
{\mathscr D}_m^{h_1+2}(-1-h_2) \, {\cal O}_{+2}^{(h_1+h_2)}(z,\bar{z})=
\bar{z}^m \bigg( (m+1)\, (-1-h_2)+ \bar{z}\, \bar{\pa}
\bigg) \, {\cal O}_{+2}^{(h_2)}(z,\bar{z}).
\nonu
\eea
In the notation of \cite{FSTZ}, $\hat{\Phi}_{+2,m}^{(0)}=\bar{L}_m$
which is the mode of the right Virasoro generator
$\bar{T}(\bar{z})$ appearing in the footnote
\ref{superBMS}. 
For $q=\frac{3}{2}$, the following result \cite{HPS}
corresponding to (\ref{translation})
holds
\bea
\bigg[
{\Phi}_{+2,m,n}^{(-\frac{1}{2})},
{\cal O}_{+2}^{(h_2)}(z,\bar{z}) \bigg]  & = &
\frac{1}{2}\, z^{m+\frac{1}{2}}\,
\bar{z}^{n+\frac{1}{2}}\,
    {\cal O}_{+2}^{(-\frac{1}{2}+h_2)}(z,\bar{z}),
\nonu \\
\bigg[
\hat{\Phi}_{+2,n}^{(-\frac{1}{2})},
    {\cal O}_{+2}^{(h_2)}(z,\bar{z}) \bigg]  & = &
{\mathscr D}_m^{\frac{3}{2}}(-1-h_2) \, {\cal O}_{+2}^{(-\frac{1}{2}+h_2)}(z,\bar{z})=
\frac{1}{2}\, \bar{z}^{n+\frac{1}{2}}\,
{\cal O}_{+2}^{(-\frac{1}{2}+h_2)}(z,\bar{z}).
\label{q3halftwocases}
\eea
In the first case of (\ref{q3halftwocases}),
we can use (\ref{twovaluesofq})
and (\ref{wOdef}).
In the second case of (\ref{q3halftwocases}),
the mode is given by $m=q-s_1=-\frac{1}{2}$.
Then the second case can be obtained from the first case
by fixing $m=-\frac{1}{2}$.
We can express ${\Phi}_{+2,m,n}^{(-\frac{1}{2})}=P_{m,n}$
which is the mode of ${\cal P}(z,\bar{z})$
appearing in the footnote \ref{superBMS}
and $\hat{\Phi}_{+2,n}^{(-\frac{1}{2})}=P_{-\frac{1}{2},n}$
using  the notation of
\cite{FSTZ}. The latter is the subset of the former.
We will describe the similar analysis in section \ref{holomorphic}.}
\bea
\bigg[
\hat{\Phi}_{+2,m}^{(h_1)},
{\cal O}_{+2}^{(h_2)}(z,\bar{z}) \bigg]  & = &
{\mathscr D}_m^{h_1+2}(-1-h_2) \, {\cal O}_{+2}^{(h_1+h_2)}(z,\bar{z}),
\nonu \\
\bigg[
\hat{\Phi}_{+2,m}^{(h_1)},
{\cal O}_{+\frac{3}{2}}^{(h_2),A}(z,\bar{z}) \bigg]  & = &
{\mathscr D}_m^{h_1+2}(-\tfrac{1}{2}-h_2) \,
{\cal O}_{+\frac{3}{2}}^{(h_1+h_2),A}(z,\bar{z}),
\nonu \\
\bigg[
\hat{\Phi}_{+2,m}^{(h_1)},
{\cal O}_{+1}^{(h_2),AB}(z,\bar{z}) \bigg]  & = &
{\mathscr D}_m^{h_1+2}(-h_2) \,
{\cal O}_{+1}^{(h_1+h_2),AB}(z,\bar{z}),
\nonu \\
\bigg[
\hat{\Phi}_{+2,m}^{(h_1)},
{\cal O}_{+\frac{1}{2}}^{(h_2),ABC}(z,\bar{z}) \bigg]  & = &
{\mathscr D}_m^{h_1+2}(\tfrac{1}{2}-h_2) \,
{\cal O}_{+\frac{1}{2}}^{(h_1+h_2),ABC}(z,\bar{z}),
\nonu \\
\bigg[
\hat{\Phi}_{+2,m}^{(h_1)},
{\cal O}_{0}^{(h_2),ABCD}(z,\bar{z}) \bigg]  & = &
{\mathscr D}_m^{h_1+2}(1-h_2) \,
{\cal O}_{0}^{(h_1+h_2),ABCD}(z,\bar{z}),
\nonu \\
\bigg[
\hat{\Phi}_{+2,m}^{(h_1)},
{\cal O}_{ABC,-\frac{1}{2}}^{(h_2)}(z,\bar{z}) \bigg]  & = &
{\mathscr D}_m^{h_1+2}(\tfrac{3}{2}-h_2) \,
{\cal O}_{ABC,-\frac{1}{2}}^{(h_1+h_2)}(z,\bar{z}),
\nonu \\
\bigg[
\hat{\Phi}_{+2,m}^{(h_1)},
{\cal O}_{AB,-1}^{(h_2)}(z,\bar{z}) \bigg]  & = &
{\mathscr D}_m^{h_1+2}(2-h_2) \,
{\cal O}_{AB,-1}^{(h_1+h_2)}(z,\bar{z}),
\nonu \\
\bigg[
\hat{\Phi}_{+2,m}^{(h_1)},
{\cal O}_{A,-\frac{3}{2}}^{(h_2)}(z,\bar{z}) \bigg]  & = &
{\mathscr D}_m^{h_1+2}(\tfrac{5}{2}-h_2) \,
{\cal O}_{A,-\frac{3}{2}}^{(h_1+h_2)}(z,\bar{z}),
\nonu \\
\bigg[
\hat{\Phi}_{+2,m}^{(h_1)},
{\cal O}_{-2}^{(h_2)}(z,\bar{z}) \bigg]  & = &
{\mathscr D}_m^{h_1+2}(3-h_2) \,
{\cal O}_{-2}^{(h_1+h_2)}(z,\bar{z}),
\nonu \\
\bigg\{
\hat{\Phi}_{+\frac{3}{2},m}^{(h_1),A},
{\cal O}_{+\frac{3}{2}}^{(h_2),B}(z,\bar{z}) \bigg\}  & = &
{\mathscr D}_m^{h_1+\frac{3}{2}}(-\tfrac{1}{2}-h_2) \,
{\cal O}_{+1}^{(h_1+h_2),AB}(z,\bar{z}),
\nonu \\
\bigg[
\hat{\Phi}_{+\frac{3}{2},m}^{(h_1),A},
{\cal O}_{+1    }^{(h_2),BC}(z,\bar{z}) \bigg]  & = &
{\mathscr D}_m^{h_1+\frac{3}{2}}(-h_2) \,
{\cal O}_{+\frac{1}{2}}^{(h_1+h_2),ABC}(z,\bar{z}),
\nonu \\
\bigg\{
\hat{\Phi}_{+\frac{3}{2},m}^{(h_1),A},
{\cal O}_{+\frac{1}{2}}^{(h_2),BCD}(z,\bar{z}) \bigg\}  & = &
{\mathscr D}_m^{h_1+\frac{3}{2}}(\tfrac{1}{2}-h_2) \,
{\cal O}_{0}^{(h_1+h_2),ABCD}(z,\bar{z}),
\nonu \\
\bigg[
\hat{\Phi}_{+\frac{3}{2},m}^{(h_1),A},
{\cal O}_{0}^{(h_2),BCDE}(z,\bar{z}) \bigg]  & = &
\frac{1}{3!} \, \ep^{ABCDEFGH} \, {\mathscr D}_m^{h_1+\frac{3}{2}}(1-h_2) \,
{\cal O}_{FGH,-\frac{1}{2}}^{(h_1+h_2)}(z,\bar{z}),
\nonu \\
\bigg\{
\hat{\Phi}_{+\frac{3}{2},m}^{(h_1),A},
{\cal O}_{BCD,-\frac{1}{2}}^{(h_2)}(z,\bar{z}) \bigg\}  & = &
3 \, \de^A_{[B} \, {\mathscr D}_m^{h_1+\frac{3}{2}}(\tfrac{3}{2}-h_2) \,
{\cal O}_{CD],-1}^{(h_1+h_2)}(z,\bar{z}),
\nonu \\
\bigg[
\hat{\Phi}_{+\frac{3}{2},m}^{(h_1),A},
{\cal O}_{BC,-1}^{(h_2)}(z,\bar{z}) \bigg]  & = &
2! \, \de^A_{[B} \, {\mathscr D}_m^{h_1+\frac{3}{2}}(2-h_2) \,
{\cal O}_{C],-\frac{3}{2}}^{(h_1+h_2)}(z,\bar{z}),
\nonu \\
\bigg\{
\hat{\Phi}_{+\frac{3}{2},m}^{(h_1),A},
{\cal O}_{B,-\frac{3}{2}}^{(h_2)}(z,\bar{z}) \bigg\}  & = &
\de^A_{B} \, {\mathscr D}_m^{h_1+\frac{3}{2}}(\tfrac{5}{2}-h_2) \,
{\cal O}_{-2}^{(h_1+h_2)}(z,\bar{z}),
\nonu \\
\bigg[
\hat{\Phi}_{+1,m}^{(h_1),AB},
{\cal O}_{+1}^{(h_2),CD}(z,\bar{z}) \bigg]  & = &
{\mathscr D}_m^{h_1+1}(-h_2) \,
{\cal O}_{0}^{(h_1+h_2),ABCD}(z,\bar{z}),
\nonu \\
\bigg[
\hat{\Phi}_{+1,m}^{(h_1),AB},
{\cal O}_{+\frac{1}{2}}^{(h_2),CDE}(z,\bar{z}) \bigg]  & = &
\frac{1}{3!} \, \ep^{ABCDEFGH}\,
{\mathscr D}_m^{h_1+1}(\tfrac{1}{2}-h_2) \,
{\cal O}_{FGH,-\frac{1}{2}}^{(h_1+h_2)}(z,\bar{z}),
\nonu \\
\bigg[
\hat{\Phi}_{+1,m}^{(h_1),AB},
{\cal O}_{0}^{(h_2),CDEF}(z,\bar{z}) \bigg]  & = &
\frac{1}{2!} \, \ep^{ABCDEFGH}\,
{\mathscr D}_m^{h_1+1}(1-h_2) \,
{\cal O}_{GH,-1}^{(h_1+h_2)}(z,\bar{z}),
\nonu \\
\bigg[
\hat{\Phi}_{+1,m}^{(h_1),AB},
{\cal O}_{CDE,-\frac{1}{2}}^{(h_2)}(z,\bar{z}) \bigg]  & = &
3!\, \de^A_{[C} \,
{\mathscr D}_m^{h_1+1}(\tf32-h_2) \,
{\cal O}_{D,-\frac{3}{2}}^{(h_1+h_2)}(z,\bar{z}) \, \de^{B}_{E]},
\nonu \\
\bigg[
\hat{\Phi}_{+1,m}^{(h_1),AB},
{\cal O}_{CD,-1}^{(h_2)}(z,\bar{z}) \bigg]  & = &
\de^{AB}_{CD} \,
{\mathscr D}_m^{h_1+1}(2-h_2) \,
{\cal O}_{-2}^{(h_1+h_2)}(z,\bar{z}),
\nonu \\
\bigg\{
\hat{\Phi}_{+\frac{1}{2},m}^{(h_1),ABC},
{\cal O}_{+\frac{1}{2}}^{(h_2),DEF}(z,\bar{z}) \bigg\}  & = &
\frac{1}{2!} \, \ep^{ABCDEFGH} \,
{\mathscr D}_m^{h_1+\frac{1}{2}}(\tfrac{1}{2}-h_2) \,
{\cal O}_{GH,-1}^{(h_1+h_2)}(z,\bar{z}),
\nonu \\
\bigg[
\hat{\Phi}_{+\frac{1}{2},m}^{(h_1),ABC},
{\cal O}_{0}^{(h_2),DEFG}(z,\bar{z}) \bigg]  & = &
\ep^{ABCDEFGH} \,
{\mathscr D}_m^{h_1+\frac{1}{2}}(1-h_2) \,
{\cal O}_{H,-\frac{3}{2}}^{(h_1+h_2)}(z,\bar{z}),
\nonu \\
\bigg\{
\hat{\Phi}_{+\frac{1}{2},m}^{(h_1),ABC},
{\cal O}_{DEF,-\frac{1}{2}}^{(h_2)}(z,\bar{z}) \bigg\}  & = &
\de^{ABC}_{DEF} \,
{\mathscr D}_m^{h_1+\frac{1}{2}}(\tf32-h_2) \,
{\cal O}_{-2}^{(h_1+h_2)}(z,\bar{z}),
\nonu \\
\bigg[
\hat{\Phi}_{0,m}^{(h_1),ABCD},
{\cal O}_{0}^{(h_2),EFGH}(z,\bar{z}) \bigg]  & = &
\ep^{ABCDEFGH} \,
{\mathscr D}_m^{h_1}(1-h_2) \,
{\cal O}_{-2}^{(h_1+h_2)}(z,\bar{z}).
\label{comm25OPEs}
\eea
We can write down the anticommutator (corresponding to
(\ref{simpletenth}) with some normalization)
by fixing $h_1=0$ in the tenth equation of
(\ref{comm25OPEs}) together with the footnote \ref{Dvalues} as follows
\footnote{In the notations of \cite{FSTZ,2212-2},
the mode $ \hat{\Phi}_{+\frac{3}{2},m}^{(0),A}$ of the current corresponds to
$\bar{G}_m$ or $\bar{G}_m^A$ which is the mode of
the supercurrents
$\bar{S}(\bar{z})$ or $\bar{S}^A(\bar{z})$ in the footnote \ref{Sbar}.
}:
\bea
\bigg\{
\hat{\Phi}_{+\frac{3}{2},m}^{(0),A},
{\cal O}_{+\frac{3}{2}}^{(h_2),B}(z,\bar{z}) \bigg\}   = 
\frac{1}{2}\, \bar{z}^{m+\frac{1}{2}}\,
{\cal O}_{+1}^{(h_2),AB}(z,\bar{z}).
\label{Two3half}
\eea
From Table \ref{table1}, the left- and right-weights
of the matter field with helicity $+\frac{3}{2}$
on the left-hand side are given by $(1-h_2,-\frac{1}{2}-h_2)$ while
those for the matter field with helicity $+1$
on the right-hand side
are given by $(1-h_2, -h_2)$.
This implies that the sum of these weights,
$(\frac{1}{2}-2h_2)$ on the left-hand side
is changed into $(1-2h_2)$ on the right-hand side.
In the previous notation, this implies the shift of the conformal
dimension $\De \rightarrow \De+\frac{1}{2}$ in the description of
(\ref{simpletenth})
\footnote{The left-weight of the
left-hand side of (\ref{Two3half}) is $0+(1-h_2)$ which is
equal to the left-weight of the right-hand side.
Note that the left-weight of $\hat{\Phi}_{+\frac{3}{2},m}^{(0),A}$
is $s_1-q=\frac{3}{2}-\frac{3}{2}=0$ and the right-weight of this
mode of the current is $-m$.
The right-weight of the left-hand side of (\ref{Two3half})
is given by $-m+(-\frac{1}{2}-h_2)= (-m-\frac{1}{2})-h_2$
which is the right-weight of the right-hand side of (\ref{Two3half}).}.

We can easily check that
this behavior for the two matter fields
satisfies for the last six cases of Table \ref{table1}
by summing over i) the third and the fourth columns and ii)
the seventh and the eighth and taking their differences as done
in the previous section.
It is useful to describe all these (anti)commutators
between the modes of supercurrents and the matter fields explicitly
as follows
\footnote{This holds for the first equation of (\ref{20OPEs}).
For the commutator $\bigg[
\hat{\Phi}_{+\frac{3}{2},m}^{(0),A},
{\cal O}_{-2}^{(h_2)}(z,\bar{z}) \bigg]$ where the matter field
with helicity $-2$ has the left- and right-weights
$(1-h_2,3-h_2)$, there is no matter field
due to the fact that $s_3 \neq s_1+s_2-2=-\frac{5}{2}$
on the right-hand side.
According to Table \ref{table1}, the above combination
$(h+\bar{h})$ for the helicity $-2$ is the maximum value
and there is no room for the matter field having $(h+\bar{h})=
\frac{9}{2}-2h_2$ where $\frac{9}{2}=\frac{3}{2}+3$
is the numerical value of sum of right-weights of this commutator.

As in the footnote \ref{q2and3half}, the following relations
hold for the actions on the primary operators
\bea
\bigg[
\hat{\Phi}_{+2,m}^{(0)},
{\cal O}_{+2}^{(h)}(z,\bar{z}) \bigg]  & = &
\bar{z}^m \bigg( (m+1)\, (-1-h)+ \bar{z}\, \bar{\pa}
\bigg) \, {\cal O}_{+2}^{(h)}(z,\bar{z}),
\nonu \\
\bigg[
\hat{\Phi}_{+2,m}^{(0)},
{\cal O}_{+\frac{3}{2}}^{(h),A}(z,\bar{z}) \bigg]  & = &
\bar{z}^m \bigg( (m+1)\, (-\tfrac{1}{2}-h)+ \bar{z}\, \bar{\pa}
\bigg) \, {\cal O}_{+\frac{3}{2}}^{(h),A}(z,\bar{z}),
\nonu \\
\bigg[
\hat{\Phi}_{+2,m}^{(0)},
{\cal O}_{+1}^{(h),AB}(z,\bar{z}) \bigg]  & = &
\bar{z}^m \bigg( (m+1)\, (-h)+ \bar{z}\, \bar{\pa}
\bigg) \, {\cal O}_{+1}^{(h),AB}(z,\bar{z}),
\nonu \\
\bigg[
\hat{\Phi}_{+2,m}^{(0)},
{\cal O}_{+\frac{1}{2}}^{(h),ABC}(z,\bar{z}) \bigg]  & = &
\bar{z}^m \bigg( (m+1)\, (\tfrac{1}{2}-h)+ \bar{z}\, \bar{\pa}
\bigg) \, {\cal O}_{+\frac{1}{2}}^{(h),ABC}(z,\bar{z}),
\nonu \\
\bigg[
\hat{\Phi}_{+2,m}^{(0)},
{\cal O}_{0}^{(h),ABCD}(z,\bar{z}) \bigg]  & = &
\bar{z}^m \bigg( (m+1)\, (1-h)+ \bar{z}\, \bar{\pa}
\bigg) \, {\cal O}_{0}^{(h),ABCD}(z,\bar{z}),
\nonu \\
\bigg[
\hat{\Phi}_{+2,m}^{(0)},
{\cal O}_{ABC,-\frac{1}{2}}^{(h)}(z,\bar{z}) \bigg]  & = &
\bar{z}^m \bigg( (m+1)\, (\tfrac{3}{2}-h)+ \bar{z}\, \bar{\pa}
\bigg) \, {\cal O}_{ABC,-\frac{1}{2}}^{(h)}(z,\bar{z}),
\nonu \\
\bigg[
\hat{\Phi}_{+2,m}^{(0)},
{\cal O}_{AB,-1}^{(h)}(z,\bar{z}) \bigg]  & = &
\bar{z}^m \bigg( (m+1)\, (2-h)+ \bar{z}\, \bar{\pa}
\bigg) \, {\cal O}_{AB,-1}^{(h)}(z,\bar{z}),
\nonu \\
\bigg[
\hat{\Phi}_{+2,m}^{(0)},
{\cal O}_{A,-\frac{3}{2}}^{(h)}(z,\bar{z}) \bigg]  & = &
\bar{z}^m \bigg( (m+1)\, (\tfrac{5}{2}-h)+ \bar{z}\, \bar{\pa}
\bigg) \, {\cal O}_{A,-\frac{3}{2}}^{(h)}(z,\bar{z}),
\nonu \\
\bigg[
\hat{\Phi}_{+2,m}^{(0)},
{\cal O}_{-2}^{(h)}(z,\bar{z}) \bigg]  & = &
\bar{z}^m \bigg( (m+1)\, (3-h)+ \bar{z}\, \bar{\pa}
\bigg) \, {\cal O}_{-2}^{(h)}(z,\bar{z}).
\label{pri}
\eea
The right-weights from Table \ref{table1} are placed properly.
The corresponding OPEs of (\ref{pri}) can be obtained similarly as in
(\ref{primary}).
Note that the left- and right-weights of the current
are given by
$(0,-m)$ which is also the weights of complex
coordinate dependent factors on the
right-hand side of (\ref{pri}).}
\bea
\bigg[
\hat{\Phi}_{+\frac{3}{2},m}^{(0),A},
{\cal O}_{+2}^{(h)}(z,\bar{z}) \bigg]  & =& 
\frac{1}{2}\, \bar{z}^{m+\frac{1}{2}}\,
{\cal O}_{+\frac{3}{2}}^{(h),A}(z,\bar{z}),
\nonu \\
\bigg\{
\hat{\Phi}_{+\frac{3}{2},m}^{(0),A},
{\cal O}_{+\frac{3}{2}}^{(h),B}(z,\bar{z}) \bigg\} &  = & 
\frac{1}{2}\, \bar{z}^{m+\frac{1}{2}}\,
{\cal O}_{+1}^{(h),AB}(z,\bar{z}),
\nonu \\
\bigg[
\hat{\Phi}_{+\frac{3}{2},m}^{(0),A},
{\cal O}_{+1}^{(h),BC}(z,\bar{z}) \bigg]  & =& 
\frac{1}{2}\, \bar{z}^{m+\frac{1}{2}}\,
{\cal O}_{+\frac{1}{2}}^{(h),ABC}(z,\bar{z}),
\nonu \\
\bigg\{
\hat{\Phi}_{+\frac{3}{2},m}^{(0),A},
{\cal O}_{+\frac{1}{2}}^{(h),BCD}(z,\bar{z}) \bigg\}  & = &
\frac{1}{2}\, \bar{z}^{m+\frac{1}{2}}\,
{\cal O}_{0}^{(h),ABCD}(z,\bar{z}),
\nonu \\
\bigg[
\hat{\Phi}_{+\frac{3}{2},m}^{(0),A},
{\cal O}_{0}^{(h),BCDE}(z,\bar{z}) \bigg] &  = & 
\frac{1}{2}\, \bar{z}^{m+\frac{1}{2}}\,
\frac{1}{3!} \, \ep^{ABCDEFGH}\,
{\cal O}_{FGH,-\frac{1}{2}}^{(h)}(z,\bar{z}),
\nonu \\
\bigg\{
\hat{\Phi}_{+\frac{3}{2},m}^{(0),A},
{\cal O}_{BCD,-\frac{1}{2}}^{(h)}(z,\bar{z}) \bigg\} &  = &
\frac{1}{2}\, \bar{z}^{m+\frac{1}{2}}\,
3 \, \de^{A}_{[B}\, {\cal O}_{CD],-1}^{(h)}(z,\bar{z}),
\nonu \\
\bigg[
\hat{\Phi}_{+\frac{3}{2},m}^{(0),A},
{\cal O}_{BC,-1}^{(h)}(z,\bar{z}) \bigg]  & = & 
\frac{1}{2}\, \bar{z}^{m+\frac{1}{2}}\,
2! \, \de^{A}_{[B}\, {\cal O}_{C],-\frac{3}{2}}^{(h)}(z,\bar{z}),
\nonu \\
\bigg\{
\hat{\Phi}_{+\frac{3}{2},m}^{(0),A},
{\cal O}_{B,-\frac{3}{2}}^{(h)}(z,\bar{z}) \bigg\}  & = & 
\frac{1}{2}\, \bar{z}^{m+\frac{1}{2}}\,
\de^{A}_{B}\, {\cal O}_{-2}^{(h)}(z,\bar{z}).
\label{red25anticomm}
\eea
As described before, from Table \ref{table1},
the left-weights $h$ in the third column
are the same as the left-weights
$(h+s_1-q)$ in the seventh column due to the fact $q=\frac{3}{2}=s_1$.
On the other hand, the right-weights $\bar{h}$
in the fourth column are increased by $\frac{1}{2}$
because the eighth column has $(\bar{h}+2-q)$
which is equal to $(\bar{h}+\frac{1}{2})$ for $q=\frac{3}{2}$.
Moreover, the holomorphic mode of
$\hat{\Phi}_{+\frac{3}{2},m}^{(0),A}$ is fixed by zero mode from
$s_1-q=0$ in (\ref{anticomm}). It is obvious to write down
the corresponding OPEs for the above (anti)commutators
(\ref{red25anticomm}) by
considering the previous OPE (\ref{simpletenth})
\footnote{
For the supertranslation operators, the following identities satisfy
\bea
\bigg[
{\Phi}_{+2,m,n}^{(-\frac{1}{2})},
{\cal O}_{+2}^{(h)}(z,\bar{z}) \bigg]   &=& 
\frac{1}{2}\, z^{m+\frac{1}{2}}\,
\bar{z}^{n+\frac{1}{2}}\,
{\cal O}_{+2}^{(-\frac{1}{2}+h)}(z,\bar{z}),
\nonu \\
\bigg[
{\Phi}_{+2,m,n}^{(-\frac{1}{2})},
{\cal O}_{+\frac{3}{2}}^{(h),A}(z,\bar{z}) \bigg]   &=& 
\frac{1}{2}\, z^{m+\frac{1}{2}}\,
\bar{z}^{n+\frac{1}{2}}\,
{\cal O}_{+\frac{3}{2}}^{(-\frac{1}{2}+h),A}(z,\bar{z}),
\nonu \\
\bigg[
{\Phi}_{+2,m,n}^{(-\frac{1}{2})},
{\cal O}_{+1}^{(h),AB}(z,\bar{z}) \bigg]   & = & 
\frac{1}{2}\, z^{m+\frac{1}{2}}\,
\bar{z}^{n+\frac{1}{2}}\,
{\cal O}_{+1}^{(-\frac{1}{2}+h),AB}(z,\bar{z}),
\nonu \\
\bigg[
{\Phi}_{+2,m,n}^{(-\frac{1}{2})},
{\cal O}_{+\frac{1}{2}}^{(h),ABC}(z,\bar{z}) \bigg]  & = & 
\frac{1}{2}\, z^{m+\frac{1}{2}}\,
\bar{z}^{n+\frac{1}{2}}\,
{\cal O}_{+\frac{1}{2}}^{(-\frac{1}{2}+h),ABC}(z,\bar{z}),
\nonu \\
\bigg[
{\Phi}_{+2,m,n}^{(-\frac{1}{2})},
{\cal O}_{0}^{(h),ABCD}(z,\bar{z}) \bigg]  & = & 
\frac{1}{2}\, z^{m+\frac{1}{2}}\,
\bar{z}^{n+\frac{1}{2}}\,
{\cal O}_{0}^{(-\frac{1}{2}+h),ABCD}(z,\bar{z}),
\nonu \\
\bigg[
{\Phi}_{+2,m,n}^{(-\frac{1}{2})},
{\cal O}_{ABC,-\frac{1}{2}}^{(h)}(z,\bar{z}) \bigg]  & = & 
\frac{1}{2}\, z^{m+\frac{1}{2}}\,
\bar{z}^{n+\frac{1}{2}}\,
{\cal O}_{ABC,-\frac{1}{2}}^{(-\frac{1}{2}+h)}(z,\bar{z}),
\nonu \\
\bigg[
{\Phi}_{+2,m,n}^{(-\frac{1}{2})},
{\cal O}_{AB,-1}^{(h)}(z,\bar{z}) \bigg]  & = & 
\frac{1}{2}\, z^{m+\frac{1}{2}}\,
\bar{z}^{n+\frac{1}{2}}\,
{\cal O}_{AB,-1}^{(-\frac{1}{2}+h)}(z,\bar{z}),
\nonu \\
\bigg[
{\Phi}_{+2,m,n}^{(-\frac{1}{2})},
{\cal O}_{A,-\frac{3}{2}}^{(h)}(z,\bar{z}) \bigg]  & = &
\frac{1}{2}\, z^{m+\frac{1}{2}}\,
\bar{z}^{n+\frac{1}{2}}\,
{\cal O}_{A,-\frac{3}{2}}^{(-\frac{1}{2}+h)}(z,\bar{z}),
\nonu \\
\bigg[
{\Phi}_{+2,m,n}^{(-\frac{1}{2})},
{\cal O}_{-2}^{(h)}(z,\bar{z}) \bigg]  & = & 
\frac{1}{2}\, z^{m+\frac{1}{2}}\,
\bar{z}^{n+\frac{1}{2}}\,
{\cal O}_{-2}^{(-\frac{1}{2}+h)}(z,\bar{z}).
\label{transla}
\eea
The left- and right-weights for matter fields on the left-hand sides
are increased by $(\frac{1}{2},\frac{1}{2})$ respectively.
On the right-hand sides, the factors $z^{\frac{1}{2}}\,
\bar{z}^{\frac{1}{2}}$ have the weights $(-\frac{1}{2},-\frac{1}{2})$
which cancel the above weights when we consider the weights of
both sides.
Moreover, the weights $(-m,-n)$ for the currents on the left-hand sides
can be described as the factors $z^m \, \bar{z}^n$
which has also the same weights $(-m,-n)$ on
the right-hand sides. 
Of course,
the corresponding OPEs of (\ref{transla}) can be obtained similarly as in
(\ref{translation}).}.

In Appendix B,
the remaining twenty (anti)commutators with the helicity
condition $s_1 < s_2$
between the ${\cal N}=8$ ${\rm w}_{1+\infty}$
currents and the matter fields
are presented explicitly based on (\ref{20OPEs}).
Then the full forty-five (anti)commutators
given by (\ref{red25anticomm}) and (\ref{comm20OPEs}) will be used
in the next section \ref{twocurrents}.

\section{The action of ${\cal N}=8$
${\rm w}_{1+\infty}$ symmetry
\label{twocurrents}
}

In order to see the ${\cal N}=8$
${\rm w}_{1+\infty}$
symmetry action on the massless celestial
primaries ${\cal O}_{h,\bar{h}}(z,\bar{z})$,
we should act the ${\cal N}=8$ ${\rm w}_{1+\infty}$
currents on both sides of (\ref{anticomm}) further
and interchange the two ${\cal N}=8$ ${\rm w}_{1+\infty}$
currents with minus (or plus) sign.
We arrive at the final relation
\bea
\bigg[
\hat{\rm w}^p_{m},
\bigg[\hat{\rm w}^q_{n}, {\cal O}_{h,\bar{h}}(z,\bar{z}) \bigg\} \bigg\}
\mp \bigg[
\hat{\rm w}^q_{n},
\bigg[\hat{\rm w}^p_{m}, {\cal O}_{h,\bar{h}}(z,\bar{z}) \bigg\} \bigg\}
= \nonu \\ 
\bigg( (q-1)m-(p-1)n \bigg) \,
\bigg[\hat{\rm w}^{p+q-2}_{m+n}, {\cal O}_{h,\bar{h}}(z,\bar{z}) \bigg\},
\label{wwO}
\eea
where the (anti)commutator for the ${\cal N}=8$ ${\rm w}_{1+\infty}$
currents satisfies
\bea
\bigg[ \hat{\rm w}_m^p, \hat{\rm w}^q_n \bigg\}=
\bigg( (q-1)m-(p-1)n \bigg) \,
\hat{\rm w}^{p+q-2}_{m+n}.
\label{wwcomm}
\eea
The notation for the (anti)commutator
is described in (\ref{wOdef}).
For the two fermionic currents,
this (anti)commutator becomes the anticommutator.
The other three cases become the usual commutator relations.
We can easily see that the left-weight of the right-hand side
of (\ref{wwcomm}) is given by
$s_3-(p+q-2)=s_1+s_2-2-(p+q-2)=(s_1-p)+(s_2-q)$
which is the left-weight of the left-hand side.
The right-weight of the right-hand side
is given by $-(m+n)=-m-n$
which is the right-weight of the left-hand side.

In Tables \ref{table3}, \ref{table4} and \ref{table5},
we classify the possible $p$, $q$ (the right-weights
of the ${\cal N}=8$ ${\rm w}_{1+\infty}$ currents),
$h$, and $\bar{h}$ (the weights of the matter field)
with the currents and the matter fields for the first forty-five
cases with $s_1 \geq s_2$.
For the remaining fifty cases with $s_1 \geq s_2$
we present them 
in Tables \ref{table11}, \ref{table12}, \ref{table13} and \ref{table14}
(of Appendix C)
\footnote{We can also analyze the other cases $s_1 < s_2$ based on
Appendix B.}.
For convenience, we list some combinations of these
right-weights,
i) the right-weights appearing on the last differential operator
in the first term of (\ref{Iden}),
ii) the right-weights appearing on the last differential operator
in the second term of (\ref{Iden}),
and iii) the superscript on the differential operator
in the third term of (\ref{Iden}),
in the last three columns of these Tables
\footnote{The condition for the helicities
$s_1+s_2+s_3 \geq 2$ \cite{Ahn2607} coming from the fact that
the helicity $s_1+s_2+s_4-4$ should be greater than or equal to
$-2$
should be satisfied. See also the last element of
the coupling $\la_{s_1,s_2+s_3-2,-s_1-s_2-s_3+4}$ appearing
in the first term (or other couplings appearing in the second or third
terms which have the same last element) of
(\ref{Iden}). This is the reason why the case of
$s_1=+2$, $s_2=+\frac{3}{2}$ and $s_3=-2$, for example, is not allowed
in Table \ref{table3} although the case of
$s_1=+2$, $s_2=+2$ and $s_3=-2$ is allowed.
Note that the left-weights, $(s_1+1-p)$ and $(s_2+1-q)$,
of the two ${\cal N}=8$ ${\rm w}_{1+\infty}$ currents do not participate in
these Tables.}.

\begin{table}[tbp]
\centering
\renewcommand{\arraystretch}{1.7}
\begin{tabular}{|c|c||c|c||c|c|c||c|c|c| }
\hline
$ p$ &
$ {\rm w^p}$ &
$ q$ &
$ {\rm w^q}$ &
$ h $ &
$ \bar{ h}$ &
$ {\cal O}_{{ h},\bar{ h}}$ &
${\scriptstyle \bar{h}+2-q}$ &
${\scriptstyle \bar{h}+2-p}$ &
${\scriptstyle p+q-2}$
\\
\hline
\hline
${\scriptstyle  h_1+2}$ &
$\Phi_{+2}^{(h_1)}$ &
${\scriptstyle h_2+2}$ &
$\Phi_{+2}^{(h_2)}$ &
${\scriptstyle 1-h_3}$ &
${\scriptstyle -1-h_3}$ &
${\cal O}_{+2}^{(h_3)}$ &
${\scriptstyle -1-h_2-h_3}$ &
${\scriptstyle -1-h_1-h_3}$ &
${\scriptstyle h_1+h_2+2}$
\\
\hline
${\scriptstyle  h_1+2}$ &
$\Phi_{+2}^{(h_1)}$ &
${\scriptstyle h_2+2}$ &
$\Phi_{+2}^{(h_2)}$ &
${\scriptstyle 1-h_3}$ &
${\scriptstyle -\frac{1}{2}-h_3}$ &
${\cal O}_{+\frac{3}{2}}^{(h_3),A}$  &
${\scriptstyle -\frac{1}{2}-h_2-h_3}$ &
${\scriptstyle -\frac{1}{2}-h_1-h_3}$ &
${\scriptstyle h_1+h_2+2}$
\\
\hline
${\scriptstyle  h_1+2}$ &
$\Phi_{+2}^{(h_1)}$ &
${\scriptstyle h_2+2}$ &
$\Phi_{+2}^{(h_2)}$ &
${\scriptstyle 1-h_3}$ &
${\scriptstyle -h_3}$ &
${\cal O}_{+1}^{(h_3),AB}$  &
${\scriptstyle -h_2-h_3} $ &
${\scriptstyle -h_1-h_3}$ &
${\scriptstyle h_1+h_2+2}$
\\
\hline
${\scriptstyle  h_1+2}$ &
$\Phi_{+2}^{(h_1)}$ &
${\scriptstyle h_2+2}$ &
$\Phi_{+2}^{(h_2)}$ &
${\scriptstyle 1-h_3}$ &
${\scriptstyle \frac{1}{2}-h_3}$ &
${\cal O}_{+\frac{1}{2}}^{(h_3),ABC}$  &
${\scriptstyle \frac{1}{2}-h_2-h_3}$ &
${\scriptstyle \frac{1}{2}-h_1-h_3}$ &
${\scriptstyle h_1+h_2+2}$
\\
\hline
${\scriptstyle  h_1+2}$ &
$\Phi_{+2}^{(h_1)}$ &
${\scriptstyle h_2+2}$ &
$\Phi_{+2}^{(h_2)}$ &
${\scriptstyle 1-h_3}$ &
${\scriptstyle 1-h_3}$ &
${\cal O}_{0}^{(h_3),ABCD}$  &
${\scriptstyle 1-h_2-h_3}$ &
${\scriptstyle 1-h_1-h_3}$ &
${\scriptstyle h_1+h_2+2}$
\\
\hline
${\scriptstyle  h_1+2}$ &
$\Phi_{+2}^{(h_1)}$ &
${\scriptstyle h_2+2}$ &
$\Phi_{+2}^{(h_2)}$ &
${\scriptstyle 1-h_3}$ &
${\scriptstyle \frac{3}{2}-h_3}$ &
${\cal O}_{ABC,-\frac{1}{2}}^{(h_3)}$  &
${\scriptstyle \frac{3}{2}-h_2-h_3}$ &
${\scriptstyle \frac{3}{2}-h_1-h_3}$ &
${\scriptstyle h_1+h_2+2}$
\\
\hline
${\scriptstyle  h_1+2}$ &
$\Phi_{+2}^{(h_1)}$ &
${\scriptstyle h_2+2}$ &
$\Phi_{+2}^{(h_2)}$ &
${\scriptstyle 1-h_3}$ &
${\scriptstyle 2-h_3}$ &
${\cal O}_{AB,-1}^{(h_3)}$  &
${\scriptstyle 2-h_2-h_3}$ &
${\scriptstyle 2-h_1-h_3}$ &
${\scriptstyle h_1+h_2+2}$
\\
\hline
${\scriptstyle  h_1+2}$ &
$\Phi_{+2}^{(h_1)}$ &
${\scriptstyle h_2+2}$ &
$\Phi_{+2}^{(h_2)}$ &
${\scriptstyle 1-h_3}$ &
${\scriptstyle \frac{5}{2}-h_3}$ &
${\cal O}_{A,-\frac{3}{2}}^{(h_3)}$  &
${\scriptstyle \frac{5}{2}-h_2-h_3}$ &
${\scriptstyle \frac{5}{2}-h_1-h_3}$ &
${\scriptstyle h_1+h_2+2}$
\\
\hline
${\scriptstyle  h_1+2}$ &
$\Phi_{+2}^{(h_1)}$ &
${\scriptstyle h_2+2}$ &
$\Phi_{+2}^{(h_2)}$ &
${\scriptstyle 1-h_3}$ &
${\scriptstyle 3-h_3}$ &
${\cal O}_{-2}^{(h_3)}$  &
${\scriptstyle 3-h_2-h_3}$ &
${\scriptstyle 3-h_1-h_3}$ &
${\scriptstyle h_1+h_2+2}$
\\
\hline
\hline
${\scriptstyle  h_1+2}$ &
$\Phi_{+2}^{(h_1)}$ &
${\scriptstyle h_2+\frac{3}{2}}$ &
$\Phi_{+\frac{3}{2}}^{(h_2),A}$ &
${\scriptstyle 1-h_3}$ &
${\scriptstyle -1-h_3}$ &
${\cal O}_{+2}^{(h_3)}$  &
${\scriptstyle -\frac{1}{2}-h_2-h_3}$ &
${\scriptstyle -1-h_1-h_3}$ &
${\scriptstyle h_1+h_2+\frac{3}{2}}$
\\
\hline
${\scriptstyle  h_1+2}$ &
$\Phi_{+2}^{(h_1)}$ &
${\scriptstyle h_2+\frac{3}{2}}$ &
$\Phi_{+\frac{3}{2}}^{(h_2),A}$ &
${\scriptstyle 1-h_3}$ &
${\scriptstyle -\frac{1}{2}-h_3}$ &
${\cal O}_{+\frac{3}{2}}^{(h_3),B}$  &
${\scriptstyle -h_2-h_3}$ &
${\scriptstyle -\frac{1}{2}-h_1-h_3}$ &
${\scriptstyle h_1+h_2+\frac{3}{2}}$
\\
\hline
${\scriptstyle  h_1+2}$ &
$\Phi_{+2}^{(h_1)}$ &
${\scriptstyle h_2+\frac{3}{2}}$ &
$\Phi_{+\frac{3}{2}}^{(h_2),A}$ &
${\scriptstyle 1-h_3}$ &
${\scriptstyle -h_3}$ &
${\cal O}_{+1}^{(h_3),BC}$  &
${\scriptstyle \frac{1}{2}-h_2-h_3}$ &
${\scriptstyle -h_1-h_3}$ &
${\scriptstyle h_1+h_2+\frac{3}{2}}$
\\
\hline
${\scriptstyle  h_1+2}$ &
$\Phi_{+2}^{(h_1)}$ &
${\scriptstyle h_2+\frac{3}{2}}$ &
$\Phi_{+\frac{3}{2}}^{(h_2),A}$ &
${\scriptstyle 1-h_3}$ &
${\scriptstyle \frac{1}{2}-h_3}$ &
${\cal O}_{+\frac{1}{2}}^{(h_3),BCD}$  &
${\scriptstyle 1-h_2-h_3}$ &
${\scriptstyle \frac{1}{2}-h_1-h_3}$ &
${\scriptstyle h_1+h_2+\frac{3}{2}}$
\\
\hline
${\scriptstyle  h_1+2}$ &
$\Phi_{+2}^{(h_1)}$ &
${\scriptstyle h_2+\frac{3}{2}}$ &
$\Phi_{+\frac{3}{2}}^{(h_2),A}$ &
${\scriptstyle 1-h_3}$ &
${\scriptstyle 1-h_3}$ &
${\cal O}_{0}^{(h_3),BCDE}$  &
${\scriptstyle \frac{3}{2}-h_2-h_3}$ &
${\scriptstyle 1-h_1-h_3}$ &
${\scriptstyle h_1+h_2+\frac{3}{2}}$
\\
\hline
${\scriptstyle  h_1+2}$ &
$\Phi_{+2}^{(h_1)}$ &
${\scriptstyle h_2+\frac{3}{2}}$ &
$\Phi_{+\frac{3}{2}}^{(h_2),A}$ &
${\scriptstyle 1-h_3}$ &
${\scriptstyle \frac{3}{2}-h_3}$ &
${\cal O}_{BCD,-\frac{1}{2}}^{(h_3)}$  &
${\scriptstyle 2-h_2-h_3}$ &
${\scriptstyle \frac{3}{2}-h_1-h_3}$ &
${\scriptstyle h_1+h_2+\frac{3}{2}}$
\\
\hline
${\scriptstyle  h_1+2}$ &
$\Phi_{+2}^{(h_1)}$ &
${\scriptstyle h_2+\frac{3}{2}}$ &
$\Phi_{+\frac{3}{2}}^{(h_2),A}$ &
${\scriptstyle 1-h_3}$ &
${\scriptstyle 2-h_3}$ &
${\cal O}_{BC,-1}^{(h_3)}$  &
${\scriptstyle \frac{5}{2}-h_2-h_3}$ &
${\scriptstyle 2-h_1-h_3}$ &
${\scriptstyle h_1+h_2+\frac{3}{2}}$
\\
\hline
${\scriptstyle  h_1+2}$ &
$\Phi_{+2}^{(h_1)}$ &
${\scriptstyle h_2+\frac{3}{2}}$ &
$\Phi_{+\frac{3}{2}}^{(h_2),A}$ &
${\scriptstyle 1-h_3}$ &
${\scriptstyle \frac{5}{2}-h_3}$ &
${\cal O}_{B,-\frac{3}{2}}^{(h_3)}$  &
${\scriptstyle 3-h_2-h_3}$ &
${\scriptstyle \frac{5}{2}-h_1-h_3}$ &
${\scriptstyle h_1+h_2+\frac{3}{2}}$
\\
\hline
\hline
\end{tabular}
\caption{
The right-weights for the two ${\cal N}=8$
${\rm w}_{1+\infty}$ currents, the left- and right-weights
for the matter fields in the first
seventeen relations of (\ref{wwO})
with the condition $s_1+s_2+s_3 \geq 2$ in the section \ref{twocurrents}.
Moreover, the combinations of these weights appearing in (\ref{Iden})
are also presented in the last three columns.}
\label{table3}
\end{table}
\begin{table}[tbp]
\centering
\renewcommand{\arraystretch}{1.7}
\begin{tabular}{|c|c||c|c||c|c|c||c|c|c| }
\hline
$ p$ &
$ {\rm w^p}$ &
$ q$ &
$ {\rm w^q}$ &
$ h $ &
$ \bar{ h}$ &
$ {\cal O}_{{ h},\bar{ h}}$ &
${\scriptstyle \bar{h}+2-q}$ &
${\scriptstyle \bar{h}+2-p}$ &
${\scriptstyle p+q-2}$
\\
\hline
\hline
${\scriptstyle  h_1+2}$ &
$\Phi_{+2}^{(h_1)}$ &
${\scriptstyle h_2+1}$ &
$\Phi_{+1}^{(h_2),AB}$ &
${\scriptstyle 1-h_3}$ &
${\scriptstyle -1-h_3}$ &
${\cal O}_{+2}^{(h_3)}$ &
${\scriptstyle -h_2-h_3}$ &
${\scriptstyle -1-h_1-h_3}$ & 
${\scriptstyle h_1+h_2+1}$
\\
\hline
${\scriptstyle  h_1+2}$ &
$\Phi_{+2}^{(h_1)}$ &
${\scriptstyle h_2+1}$ &
$\Phi_{+1}^{(h_2),AB}$ &
${\scriptstyle 1-h_3}$ &
${\scriptstyle -\frac{1}{2}-h_3}$ &
${\cal O}_{+\frac{3}{2}}^{(h_3),C}$  &
${\scriptstyle \frac{1}{2}-h_2-h_3}$ &
${\scriptstyle -\frac{1}{2}-h_1-h_3}$ &
${\scriptstyle h_1+h_2+1}$
\\
\hline
${\scriptstyle  h_1+2}$ &
$\Phi_{+2}^{(h_1)}$ &
${\scriptstyle h_2+1}$ &
$\Phi_{+1}^{(h_2),AB}$ &
${\scriptstyle 1-h_3}$ &
${\scriptstyle -h_3}$ &
${\cal O}_{+1}^{(h_3),CD}$  &
${\scriptstyle 1-h_2-h_3}$ &
${\scriptstyle -h_1-h_3}$ &
${\scriptstyle h_1+h_2+1}$
\\
\hline
${\scriptstyle  h_1+2}$ &
$\Phi_{+2}^{(h_1)}$ &
${\scriptstyle h_2+1}$ &
$\Phi_{+1}^{(h_2),AB}$ &
${\scriptstyle 1-h_3}$ &
${\scriptstyle \frac{1}{2}-h_3}$ &
${\cal O}_{+\frac{1}{2}}^{(h_3),CDE}$  &
${\scriptstyle \frac{3}{2}-h_2-h_3}$ &
${\scriptstyle \frac{1}{2}-h_1-h_3}$ &
${\scriptstyle h_1+h_2+1}$
\\
\hline
${\scriptstyle  h_1+2}$ &
$\Phi_{+2}^{(h_1)}$ &
${\scriptstyle h_2+1}$ &
$\Phi_{+1}^{(h_2),AB}$ &
${\scriptstyle 1-h_3}$ &
${\scriptstyle 1-h_3}$ &
${\cal O}_{0}^{(h_3),CDEF}$  &
${\scriptstyle 2-h_2-h_3}$ &
${\scriptstyle 1-h_1-h_3}$ &
${\scriptstyle h_1+h_2+1}$
\\
\hline
${\scriptstyle  h_1+2}$ &
$\Phi_{+2}^{(h_1)}$ &
${\scriptstyle h_2+1}$ &
$\Phi_{+1}^{(h_2),AB}$ &
${\scriptstyle 1-h_3}$ &
${\scriptstyle \frac{3}{2}-h_3}$ &
${\cal O}_{CDE,-\frac{1}{2}}^{(h_3)}$  &
${\scriptstyle \frac{5}{2}-h_2-h_3}$ &
${\scriptstyle \frac{3}{2}-h_1-h_3}$ &
${\scriptstyle h_1+h_2+1}$
\\
\hline
${\scriptstyle  h_1+2}$ &
$\Phi_{+2}^{(h_1)}$ &
${\scriptstyle h_2+1}$ &
$\Phi_{+1}^{(h_2),AB}$ &
${\scriptstyle 1-h_3}$ &
${\scriptstyle 2-h_3}$ &
${\cal O}_{CD,-1}^{(h_3)}$  &
${\scriptstyle 3-h_2-h_3}$ &
${\scriptstyle 2-h_1-h_3}$ &
${\scriptstyle h_1+h_2+1}$
\\
\hline
\hline
${\scriptstyle  h_1+2}$ &
$\Phi_{+2}^{(h_1)}$ &
${\scriptstyle h_2+\frac{1}{2}}$ &
$\Phi_{+\frac{1}{2}}^{(h_2),ABC}$ &
${\scriptstyle 1-h_3}$ &
${\scriptstyle -1-h_3}$ &
${\cal O}_{+2}^{(h_3)}$  &
${\scriptstyle \frac{1}{2}-h_2-h_3}$ &
${\scriptstyle -1-h_1-h_3}$ &
${\scriptstyle h_1+h_2+\frac{1}{2}}$
\\
\hline
${\scriptstyle  h_1+2}$ &
$\Phi_{+2}^{(h_1)}$ &
${\scriptstyle h_2+\frac{1}{2}}$ &
$\Phi_{+\frac{1}{2}}^{(h_2),ABC}$ &
${\scriptstyle 1-h_3}$ &
${\scriptstyle -\frac{1}{2}-h_3}$ &
${\cal O}_{+\frac{3}{2}}^{(h_3),D}$  &
${\scriptstyle 1-h_2-h_3}$ &
${\scriptstyle -\frac{1}{2}-h_1-h_3}$ &
${\scriptstyle h_1+h_2+\frac{1}{2}}$
\\
\hline
${\scriptstyle  h_1+2}$ &
$\Phi_{+2}^{(h_1)}$ &
${\scriptstyle h_2+\frac{1}{2}}$ &
$\Phi_{+\frac{1}{2}}^{(h_2),ABC}$ &
${\scriptstyle 1-h_3}$ &
${\scriptstyle -h_3}$ &
${\cal O}_{+1}^{(h_3),DE}$  &
${\scriptstyle \frac{3}{2}-h_2-h_3}$ &
${\scriptstyle -h_1-h_3}$ &
${\scriptstyle h_1+h_2+\frac{1}{2}}$
\\
\hline
${\scriptstyle  h_1+2}$ &
$\Phi_{+2}^{(h_1)}$ &
${\scriptstyle h_2+\frac{1}{2}}$ &
$\Phi_{+\frac{1}{2}}^{(h_2),ABC}$ &
${\scriptstyle 1-h_3}$ &
${\scriptstyle \frac{1}{2}-h_3}$ &
${\cal O}_{+\frac{1}{2}}^{(h_3),DEF}$  &
${\scriptstyle 2-h_2-h_3}$ &
${\scriptstyle \frac{1}{2}-h_1-h_3}$ &
${\scriptstyle h_1+h_2+\frac{1}{2}}$
\\
\hline
${\scriptstyle  h_1+2}$ &
$\Phi_{+2}^{(h_1)}$ &
${\scriptstyle h_2+\frac{1}{2}}$ &
$\Phi_{+\frac{1}{2}}^{(h_2),ABC}$ &
${\scriptstyle 1-h_3}$ &
${\scriptstyle 1-h_3}$ &
${\cal O}_{0}^{(h_3),DEFG}$  &
${\scriptstyle \frac{5}{2}-h_2-h_3}$ &
${\scriptstyle 1-h_1-h_3}$ &
${\scriptstyle h_1+h_2+\frac{1}{2}}$
\\
\hline
${\scriptstyle  h_1+2}$ &
$\Phi_{+2}^{(h_1)}$ &
${\scriptstyle h_2+\frac{1}{2}}$ &
$\Phi_{+\frac{1}{2}}^{(h_2),ABC}$ &
${\scriptstyle 1-h_3}$ &
${\scriptstyle \frac{3}{2}-h_3}$ &
${\cal O}_{DEF,-\frac{1}{2}}^{(h_3)}$  &
${\scriptstyle 3-h_2-h_3}$ &
${\scriptstyle \frac{3}{2}-h_1-h_3}$ &
${\scriptstyle h_1+h_2+\frac{1}{2}}$
\\
\hline
\hline
${\scriptstyle  h_1+2}$ &
$\Phi_{+2}^{(h_1)}$ &
${\scriptstyle h_2}$ &
$\Phi_{0}^{(h_2),ABCD}$ &
${\scriptstyle 1-h_3}$ &
${\scriptstyle -1-h_3}$ &
${\cal O}_{+2}^{(h_3)}$  &
${\scriptstyle 1-h_2-h_3}$ &
${\scriptstyle -1-h_1-h_3}$ &
${\scriptstyle h_1+h_2}$
\\
\hline
${\scriptstyle  h_1+2}$ &
$\Phi_{+2}^{(h_1)}$ &
${\scriptstyle h_2}$ &
$\Phi_{0}^{(h_2),ABCD}$ &
${\scriptstyle 1-h_3}$ &
${\scriptstyle -\frac{1}{2}-h_3}$ &
${\cal O}_{+\frac{3}{2}}^{(h_3),E}$  &
${\scriptstyle \frac{3}{2}-h_2-h_3}$ &
${\scriptstyle -\frac{1}{2}-h_1-h_3}$ &
${\scriptstyle h_1+h_2}$
\\
\hline
${\scriptstyle  h_1+2}$ &
$\Phi_{+2}^{(h_1)}$ &
${\scriptstyle h_2}$ &
$\Phi_{0}^{(h_2),ABCD}$ &
${\scriptstyle 1-h_3}$ &
${\scriptstyle -h_3}$ &
${\cal O}_{+1}^{(h_3),EF}$  &
${\scriptstyle 2-h_2-h_3}$ &
${\scriptstyle -h_1-h_3}$ &
${\scriptstyle h_1+h_2}$
\\
\hline
${\scriptstyle  h_1+2}$ &
$\Phi_{+2}^{(h_1)}$ &
${\scriptstyle h_2}$ &
$\Phi_{0}^{(h_2),ABCD}$ &
${\scriptstyle 1-h_3}$ &
${\scriptstyle \frac{1}{2}-h_3}$ &
${\cal O}_{+\frac{1}{2}}^{(h_3),EFG}$  &
${\scriptstyle \frac{5}{2}-h_2-h_3}$ &
${\scriptstyle \frac{1}{2}-h_1-h_3}$ &
${\scriptstyle h_1+h_2}$
\\
\hline
${\scriptstyle  h_1+2}$ &
$\Phi_{+2}^{(h_1)}$ &
${\scriptstyle h_2}$ &
$\Phi_{0}^{(h_2),ABCD}$ &
${\scriptstyle 1-h_3}$ &
${\scriptstyle 1-h_3}$ &
${\cal O}_{0}^{(h_3),EFGH}$  &
${\scriptstyle 3-h_2-h_3}$ &
${\scriptstyle 1-h_1-h_3}$ &
${\scriptstyle h_1+h_2}$
\\
\hline
\hline
\end{tabular}
\caption{
This is a continuation of previous Table \ref{table3}
and we present the corresponding weights and the combinations
of the weights for the next  
eighteen relations of (\ref{wwO}) in the section \ref{twocurrents}. }
\label{table4}
\end{table}
\begin{table}[tbp]
\centering
\renewcommand{\arraystretch}{1.7}
\begin{tabular}{|c|c||c|c||c|c|c||c|c|c| }
\hline
$ p$ &
$ {\rm w^p}$ &
$ q$ &
$ {\rm w^q}$ &
$ h $ &
$ \bar{ h}$ &
$ {\cal O}_{{ h},\bar{ h}}$ &
${\scriptstyle \bar{h}+2-q}$ &
${\scriptstyle \bar{h}+2-p}$ &
${\scriptstyle p+q-2}$
\\
\hline
\hline
${\scriptstyle  h_1+2}$ &
$\Phi_{+2}^{(h_1)}$ &
${\scriptstyle h_2-\frac{1}{2}}$ &
$\Phi_{ABC,-\frac{1}{2}}^{(h_2)}$ &
${\scriptstyle 1-h_3}$ &
${\scriptstyle -1-h_3}$ &
${\cal O}_{+2}^{(h_3)}$ &
${\scriptstyle \frac{3}{2}-h_2-h_3}$ &
${\scriptstyle -1-h_1-h_3}$ &
${\scriptstyle h_1+h_2-\frac{1}{2}}$
\\
\hline
${\scriptstyle  h_1+2}$ &
$\Phi_{+2}^{(h_1)}$ &
${\scriptstyle h_2-\frac{1}{2}}$ &
$\Phi_{ABC,-\frac{1}{2}}^{(h_2)}$ &
${\scriptstyle 1-h_3}$ &
${\scriptstyle -\frac{1}{2}-h_3}$ &
${\cal O}_{+\frac{3}{2}}^{(h_3),D}$  &
${\scriptstyle 2-h_2-h_3}$ &
${\scriptstyle -\frac{1}{2}-h_1-h_3}$ &
${\scriptstyle h_1+h_2-\frac{1}{2}}$
\\
\hline
${\scriptstyle  h_1+2}$ &
$\Phi_{+2}^{(h_1)}$ &
${\scriptstyle h_2-\frac{1}{2}}$ &
$\Phi_{ABC,-\frac{1}{2}}^{(h_2)}$ &
${\scriptstyle 1-h_3}$ &
${\scriptstyle -h_3}$ &
${\cal O}_{+1}^{(h_3),DE}$  &
${\scriptstyle \frac{5}{2}-h_2-h_3}$ &
${\scriptstyle -h_1-h_3}$ &
${\scriptstyle h_1+h_2-\frac{1}{2}}$
\\
\hline
${\scriptstyle  h_1+2}$ &
$\Phi_{+2}^{(h_1)}$ &
${\scriptstyle h_2-\frac{1}{2}}$ &
$\Phi_{ABC,-\frac{1}{2}}^{(h_2)}$ &
${\scriptstyle 1-h_3}$ &
${\scriptstyle \frac{1}{2}-h_3}$ &
${\cal O}_{+\frac{1}{2}}^{(h_3),DEF}$  &
${\scriptstyle 3-h_2-h_3}$ &
${\scriptstyle \frac{1}{2}-h_1-h_3}$ &
${\scriptstyle h_1+h_2-\frac{1}{2}}$
\\
\hline
\hline
${\scriptstyle  h_1+2}$ &
$\Phi_{+2}^{(h_1)}$ &
${\scriptstyle h_2-1}$ &
$\Phi_{AB,-1}^{(h_2)}$ &
${\scriptstyle 1-h_3}$ &
${\scriptstyle -1-h_3}$ &
${\cal O}_{+2}^{(h_3)}$  &
${\scriptstyle 2-h_2-h_3}$ &
${\scriptstyle -1-h_1-h_3}$ &
${\scriptstyle h_1+h_2-1}$
\\
\hline
${\scriptstyle  h_1+2}$ &
$\Phi_{+2}^{(h_1)}$ &
${\scriptstyle h_2-1}$ &
$\Phi_{AB,-1}^{(h_2)}$ &
${\scriptstyle 1-h_3}$ &
${\scriptstyle -\frac{1}{2}-h_3}$ &
${\cal O}_{+\frac{3}{2}}^{(h_3),C}$  &
${\scriptstyle \frac{5}{2}-h_2-h_3}$ &
${\scriptstyle -\frac{1}{2}-h_1-h_3}$ &
${\scriptstyle h_1+h_2-1}$
\\
\hline
${\scriptstyle  h_1+2}$ &
$\Phi_{+2}^{(h_1)}$ &
${\scriptstyle h_2-1}$ &
$\Phi_{AB,-1}^{(h_2)}$ &
${\scriptstyle 1-h_3}$ &
${\scriptstyle -h_3}$ &
${\cal O}_{+1}^{(h_3),CD}$  &
${\scriptstyle 3-h_2-h_3}$ &
${\scriptstyle -h_1-h_3}$ &
${\scriptstyle h_1+h_2-1}$
\\
\hline
\hline
${\scriptstyle  h_1+2}$ &
$\Phi_{+2}^{(h_1)}$ &
${\scriptstyle h_2-\frac{3}{2}}$ &
$\Phi_{A,-\frac{3}{2}}^{(h_2)}$ &
${\scriptstyle 1-h_3}$ &
${\scriptstyle -1-h_3}$ &
${\cal O}_{+2}^{(h_3)}$  &
${\scriptstyle \frac{5}{2}-h_2-h_3}$ &
${\scriptstyle -1-h_1-h_3}$ &
${\scriptstyle h_1+h_2-\frac{3}{2}}$
\\
\hline
${\scriptstyle  h_1+2}$ &
$\Phi_{+2}^{(h_1)}$ &
${\scriptstyle h_2-\frac{3}{2}}$ &
$\Phi_{A,-\frac{3}{2}}^{(h_2)}$ &
${\scriptstyle 1-h_3}$ &
${\scriptstyle -\frac{1}{2}-h_3}$ &
${\cal O}_{+\frac{3}{2}}^{(h_3),B}$  &
${\scriptstyle 3-h_2-h_3}$ &
${\scriptstyle -\frac{1}{2}-h_1-h_3}$ &
${\scriptstyle h_1+h_2-\frac{3}{2}}$
\\
\hline
\hline
${\scriptstyle  h_1+2}$ &
$\Phi_{+2}^{(h_1)}$ &
${\scriptstyle h_2-2}$ &
$\Phi_{-2}^{(h_2)}$ &
${\scriptstyle 1-h_3}$ &
${\scriptstyle -1-h_3}$ &
${\cal O}_{+2}^{(h_3)}$  &
${\scriptstyle 3-h_2-h_3}$ &
${\scriptstyle -1-h_1-h_3}$ &
${\scriptstyle h_1+h_2-2}$
\\
\hline
\hline
\end{tabular}
\caption{
This is a continuation of previous Tables \ref{table3}
and \ref{table4} 
and we present the corresponding weights and the combinations
of the weights for the next  
ten relations of (\ref{wwO}) in the section \ref{twocurrents}.
We will see the remaining fifty cases in Tables \ref{table11},
\ref{table12}, \ref{table13}
and \ref{table14}.}
\label{table5}
\end{table}

We are ready to simplify (\ref{wwO}) by substituting
the previous (anti)commutator in (\ref{anticomm}) correctly
and obtain the following result
\bea
&&
\Bigg[
\la_{s_1,s_2+s_3-2,-s_1-s_2-s_3+4}\, \la_{s_2,s_3,-s_2-s_3+2} \,
{\mathscr D}_n^{q}(\bar{h}) \, {\mathscr D}_m^{p}(\bar{h}+2-q)
\nonu \\
&& \mp \la_{s_2,s_1+s_3-2,-s_1-s_2-s_3+4}\, \la_{s_1,s_3,-s_1-s_3+2}\,
{\mathscr D}_m^{p}(\bar{h}) \, {\mathscr D}_n^{q}(\bar{h}+2-p)
\nonu \\
&& -
\kappa_{s_1,s_2,-s_1-s_2+2} \, \la_{s_1+s_2-2,s_3,-s_1-s_2-s_3+4}
\, \bigg( (q-1)m-(p-1)n \bigg) \,
   {\mathscr D}_{m+n}^{p+q-2}(\bar{h})\, \Bigg]
\nonu \\
&& \,\times
{\cal O}_{h+s_2+s_1-p-q,\bar{h}+4-p-q}(z,\bar{z})=0.
\label{Iden}
\eea
In the first term of (\ref{Iden}), the current
$\hat{\rm w}^p_m$ commutes with ${\mathscr D}_n^{q}(\bar{h})$.
Similarly, $\hat{\rm w}^q_n$ commutes with ${\mathscr D}_m^{p}(\bar{h})$
in the second term.
The left- and right-weights of the matter field in (\ref{Iden})
can be obtained from the (anti)commutator on the right-hand side of
(\ref{wwO}) together with (\ref{anticomm}) by substituting
the helicity $(s_1+s_2-2)$ and the right-weight
$(p+q-2)$
appearing in the superscript
of the derivative.
This check can be also determined by the first term or the second term
in (\ref{Iden}).
Moreover, the two differential operators do not commute with each other
because the differential operator contains the complex coordinate
$\bar{z}$ in addition to the differentiation with
respect to
the complex coordinate $\bar{z}$.

Here we introduce the various couplings in front of
the right-hand side of (\ref{anticomm})
\footnote{There exist some normalization factors when we take
the (antiholomorphic) light transform for
the conformal celestial primary operator
in (\ref{OPEwO}).} and those
on the right-hand side of (\ref{wwcomm}) introduced in \cite{AK2509}.
We will find the constraints between the couplings. 
As described before, the combinations of
the right-weights appearing in the arguments of differential operators
and in the upper index of those operators are listed in the last
three columns of previous
Tables \ref{table3}, \ref{table4} and \ref{table5}. 
In order to classify the relation (\ref{Iden}) completely,
we should also specify the weights for the matter fields.
By identifying these weights properly, we
can write down all the matter fields explicitly.
In Tables \ref{table6}, \ref{table7} and \ref{table8},
we classify the left- and right-weights for the matter fields
in (\ref{Iden}) with
\bea
h_4 \equiv h+s_2+s_1-p-q, \qquad \bar{h}_4 \equiv
\bar{h}+4-p-q
\label{h4h4bar}
\eea
for the first forty-five cases
and 
in Tables \ref{table15}, \ref{table16}, \ref{table17} and \ref{table18}
(of Appendix C),
the remaining fifty cases are presented.
Note that the weights for the matter fields
appearing in (\ref{Iden}) or (\ref{h4h4bar}) depend on the weights of the
two ${\cal N}=8$ ${\rm w}_{1+\infty}$ currents and the matter fields
in (\ref{wwO}).
Therefore, in order to satisfy the equation (\ref{Iden})
for any matter fields, the differential operators
with coefficients which depend on the unknown couplings
$\la_{s_1,s_2,-s_3}$ and some known and unknown couplings
$\kappa_{s_1,s_2,-s_3}$ from \cite{AK2509}
inside the bracket should vanish.

\begin{table}[tbp]
\centering
\renewcommand{\arraystretch}{1.7}
\begin{tabular}{|c|c||c|c||c|c|||c|c|c|| }
\hline
$ p$ &
$s_1 $ &
$ q$ &
$s_2$ &
$ h $ &
$ \bar{h}$ &
$h_4$
&
$\bar{h}_4$
&
$ {\cal O}_{h_4,\bar{h}_4}$ 
\\
\hline
\hline
${\scriptstyle h_1+2}$ &
${\scriptstyle +2}$ &
${\scriptstyle h_2+2}$ &
${\scriptstyle +2}$ &
${\scriptstyle 1-h_3}$ &
${\scriptstyle -1-h_3}$ &
${\scriptstyle 1-(h_1+h_2+h_3)}$ &
${\scriptstyle -1-(h_1+h_2+h_3)}$ &
${\cal O}_{+2}^{(h_1+h_2+h_3)}$
\\
\hline
${\scriptstyle h_1+2}$ &
${\scriptstyle +2}$ &
${\scriptstyle h_2+2}$ &
${\scriptstyle +2}$ &
${\scriptstyle 1-h_3}$ &
${\scriptstyle -\frac{1}{2}-h_3}$ &
${\scriptstyle 1-(h_1+h_2+h_3)}$ &
${\scriptstyle -\frac{1}{2}-(h_1+h_2+h_3)}$ &
${\cal O}_{+\frac{3}{2}}^{(h_1+h_2+h_3),A}$
\\
\hline
${\scriptstyle h_1+2}$ &
${\scriptstyle +2}$ &
${\scriptstyle h_2+2}$ &
${\scriptstyle +2}$ &
${\scriptstyle 1-h_3}$ &
${\scriptstyle -h_3}$ &
${\scriptstyle 1-(h_1+h_2+h_3)}$ &
${\scriptstyle -(h_1+h_2+h_3)}$ &
${\cal O}_{+1}^{(h_1+h_2+h_3),AB}$
\\
\hline
${\scriptstyle h_1+2}$ &
${\scriptstyle +2}$ &
${\scriptstyle h_2+2}$ &
${\scriptstyle +2}$ &
${\scriptstyle 1-h_3}$ &
${\scriptstyle \frac{1}{2}-h_3}$ &
${\scriptstyle 1-(h_1+h_2+h_3)}$ &
${\scriptstyle \frac{1}{2}-(h_1+h_2+h_3)}$ &
${\cal O}_{+\frac{1}{2}}^{(h_1+h_2+h_3),ABC}$
\\
\hline
${\scriptstyle h_1+2}$ &
${\scriptstyle +2}$ &
${\scriptstyle h_2+2}$ &
${\scriptstyle +2}$ &
${\scriptstyle 1-h_3}$ &
${\scriptstyle 1-h_3}$ &
${\scriptstyle 1-(h_1+h_2+h_3)}$ &
${\scriptstyle 1-(h_1+h_2+h_3)}$ &
${\cal O}_{0}^{(h_1+h_2+h_3),ABCD}$
\\
\hline
${\scriptstyle h_1+2}$ &
${\scriptstyle +2}$ &
${\scriptstyle h_2+2}$ &
${\scriptstyle +2}$ &
${\scriptstyle 1-h_3}$ &
${\scriptstyle \frac{3}{2}-h_3}$ &
${\scriptstyle 1-(h_1+h_2+h_3)}$ &
${\scriptstyle \frac{3}{2}-(h_1+h_2+h_3)}$ &
${\cal O}_{ABC,-\frac{1}{2}}^{(h_1+h_2+h_3)}$
\\
\hline
${\scriptstyle h_1+2}$ &
${\scriptstyle +2}$ &
${\scriptstyle h_2+2}$ &
${\scriptstyle +2}$ &
${\scriptstyle 1-h_3}$ &
${\scriptstyle 2-h_3}$ &
${\scriptstyle 1-(h_1+h_2+h_3)}$ &
${\scriptstyle 2-(h_1+h_2+h_3)}$ &
${\cal O}_{AB,-1}^{(h_1+h_2+h_3)}$
\\
\hline
${\scriptstyle h_1+2}$ &
${\scriptstyle +2}$ &
${\scriptstyle h_2+2}$ &
${\scriptstyle +2}$ &
${\scriptstyle 1-h_3}$ &
${\scriptstyle \frac{5}{2}-h_3}$ &
${\scriptstyle 1-(h_1+h_2+h_3)}$ &
${\scriptstyle \frac{5}{2}-(h_1+h_2+h_3)}$ &
${\cal O}_{A,-\frac{3}{2}}^{(h_1+h_2+h_3)}$
\\
\hline
${\scriptstyle h_1+2}$ &
${\scriptstyle +2}$ &
${\scriptstyle h_2+2}$ &
${\scriptstyle +2}$ &
${\scriptstyle 1-h_3}$ &
${\scriptstyle 3-h_3}$ &
${\scriptstyle 1-(h_1+h_2+h_3)}$ &
${\scriptstyle 3-(h_1+h_2+h_3)}$ &
${\cal O}_{-2}^{(h_1+h_2+h_3)}$
\\
\hline
\hline
${\scriptstyle h_1+2}$ &
${\scriptstyle +2}$ &
${\scriptstyle h_2+\frac{3}{2}}$ &
${\scriptstyle +\frac{3}{2}}$ &
${\scriptstyle 1-h_3}$ &
${\scriptstyle -1-h_3}$ &
${\scriptstyle 1-(h_1+h_2+h_3)}$ &
${\scriptstyle -\frac{1}{2}-(h_1+h_2+h_3)}$ &
${\cal O}_{+\frac{3}{2}}^{(h_1+h_2+h_3),A}$
\\
\hline
${\scriptstyle h_1+2}$ &
${\scriptstyle +2}$ &
${\scriptstyle h_2+\frac{3}{2}}$ &
${\scriptstyle +\frac{3}{2}}$ &
${\scriptstyle 1-h_3}$ &
${\scriptstyle -\frac{1}{2}-h_3}$ &
${\scriptstyle 1-(h_1+h_2+h_3)}$ &
${\scriptstyle -(h_1+h_2+h_3)}$ &
${\cal O}_{+1}^{(h_1+h_2+h_3),AB}$
\\
\hline
${\scriptstyle h_1+2}$ &
${\scriptstyle +2}$ &
${\scriptstyle h_2+\frac{3}{2}}$ &
${\scriptstyle +\frac{3}{2}}$ &
${\scriptstyle 1-h_3}$ &
${\scriptstyle -h_3}$ &
${\scriptstyle 1-(h_1+h_2+h_3)}$ &
${\scriptstyle \frac{1}{2}-(h_1+h_2+h_3)}$ &
${\cal O}_{+\frac{1}{2}}^{(h_1+h_2+h_3),ABC}$
\\
\hline
${\scriptstyle h_1+2}$ &
${\scriptstyle +2}$ &
${\scriptstyle h_2+\frac{3}{2}}$ &
${\scriptstyle +\frac{3}{2}}$ &
${\scriptstyle 1-h_3}$ &
${\scriptstyle \frac{1}{2}-h_3}$ &
${\scriptstyle 1-(h_1+h_2+h_3)}$ &
${\scriptstyle 1-(h_1+h_2+h_3)}$ &
${\cal O}_{0}^{(h_1+h_2+h_3),ABCD}$
\\
\hline
${\scriptstyle h_1+2}$ &
${\scriptstyle +2}$ &
${\scriptstyle h_2+\frac{3}{2}}$ &
${\scriptstyle +\frac{3}{2}}$ &
${\scriptstyle 1-h_3}$ &
${\scriptstyle 1-h_3}$ &
${\scriptstyle 1-(h_1+h_2+h_3)}$ &
${\scriptstyle \frac{3}{2}-(h_1+h_2+h_3)}$ &
${\cal O}_{FGH,-\frac{1}{2}}^{(h_1+h_2+h_3)}$
\\
\hline
${\scriptstyle h_1+2}$ &
${\scriptstyle +2}$ &
${\scriptstyle h_2+\frac{3}{2}}$ &
${\scriptstyle +\frac{3}{2}}$ &
${\scriptstyle 1-h_3}$ &
${\scriptstyle \frac{3}{2}-h_3}$ &
${\scriptstyle 1-(h_1+h_2+h_3)}$ &
${\scriptstyle 2-(h_1+h_2+h_3)}$ &
${\cal O}_{CD,-1}^{(h_1+h_2+h_3)}+\cdots$
\\
\hline
${\scriptstyle h_1+2}$ &
${\scriptstyle +2}$ &
${\scriptstyle h_2+\frac{3}{2}}$ &
${\scriptstyle +\frac{3}{2}}$ &
${\scriptstyle 1-h_3}$ &
${\scriptstyle 2-h_3}$ &
${\scriptstyle 1-(h_1+h_2+h_3)}$ &
${\scriptstyle \frac{5}{2}-(h_1+h_2+h_3)}$ &
${\cal O}_{C,-\frac{3}{2}}^{(h_1+h_2+h_3)}+\cdots$
\\
\hline
${\scriptstyle h_1+2}$ &
${\scriptstyle +2}$ &
${\scriptstyle h_2+\frac{3}{2}}$ &
${\scriptstyle +\frac{3}{2}}$ &
${\scriptstyle 1-h_3}$ &
${\scriptstyle \frac{5}{2}-h_3}$ &
${\scriptstyle 1-(h_1+h_2+h_3)}$ &
${\scriptstyle 3-(h_1+h_2+h_3)}$ &
${\cal O}_{-2}^{(h_1+h_2+h_3)}$
\\
\hline
\hline
\end{tabular}
\caption{For given right-weights, helicities, and the left- and right-weights
appearing in Table \ref{table3},
the left- and right-weights for the seventeen matter fields of (\ref{Iden})
in the section \ref{twocurrents}
are presented. The relation (\ref{h4h4bar})
is used.}
\label{table6}
\end{table}
\begin{table}[tbp]
\centering
\renewcommand{\arraystretch}{1.7}
\begin{tabular}{|c|c||c|c||c|c|||c|c|c|| }
\hline
$ p$ &
$s_1 $ &
$ q$ &
$s_2$ &
$ h $ &
$ \bar{h}$ &
$h_4$
&
$\bar{h}_4$
&
$ {\cal O}_{h_4,\bar{h}_4}$ 
\\
\hline
\hline
${\scriptstyle h_1+2}$ &
${\scriptstyle +2}$ &
${\scriptstyle h_2+1}$ &
${\scriptstyle +1}$ &
${\scriptstyle 1-h_3}$ &
${\scriptstyle -1-h_3}$ &
${\scriptstyle 1-(h_1+h_2+h_3)}$ &
${\scriptstyle -(h_1+h_2+h_3)}$ &
${\cal O}_{+1}^{(h_1+h_2+h_3),AB}$
\\
\hline
${\scriptstyle h_1+2}$ &
${\scriptstyle +2}$ &
${\scriptstyle h_2+1}$ &
${\scriptstyle +1}$ &
${\scriptstyle 1-h_3}$ &
${\scriptstyle -\frac{1}{2}-h_3}$ &
${\scriptstyle 1-(h_1+h_2+h_3)}$ &
${\scriptstyle \frac{1}{2}-(h_1+h_2+h_3)}$ &
${\cal O}_{+\frac{1}{2}}^{(h_1+h_2+h_3),ABC}$
\\
\hline
${\scriptstyle h_1+2}$ &
${\scriptstyle +2}$ &
${\scriptstyle h_2+1}$ &
${\scriptstyle +1}$ &
${\scriptstyle 1-h_3}$ &
${\scriptstyle -h_3}$ &
${\scriptstyle 1-(h_1+h_2+h_3)}$ &
${\scriptstyle 1-(h_1+h_2+h_3)}$ &
${\cal O}_{0}^{(h_1+h_2+h_3),ABCD}$
\\
\hline
${\scriptstyle h_1+2}$ &
${\scriptstyle +2}$ &
${\scriptstyle h_2+1}$ &
${\scriptstyle +1}$ &
${\scriptstyle 1-h_3}$ &
${\scriptstyle \frac{1}{2}-h_3}$ &
${\scriptstyle 1-(h_1+h_2+h_3)}$ &
${\scriptstyle \frac{3}{2}-(h_1+h_2+h_3)}$ &
${\cal O}_{FGH,-\frac{1}{2}}^{(h_1+h_2+h_3)}$
\\
\hline
${\scriptstyle h_1+2}$ &
${\scriptstyle +2}$ &
${\scriptstyle h_2+1}$ &
${\scriptstyle +1}$ &
${\scriptstyle 1-h_3}$ &
${\scriptstyle 1-h_3}$ &
${\scriptstyle 1-(h_1+h_2+h_3)}$ &
${\scriptstyle 2-(h_1+h_2+h_3)}$ &
${\cal O}_{GH,-1}^{(h_1+h_2+h_3)}$
\\
\hline
${\scriptstyle h_1+2}$ &
${\scriptstyle +2}$ &
${\scriptstyle h_2+1}$ &
${\scriptstyle +1}$ &
${\scriptstyle 1-h_3}$ &
${\scriptstyle \frac{3}{2}-h_3}$ &
${\scriptstyle 1-(h_1+h_2+h_3)}$ &
${\scriptstyle \frac{5}{2}-(h_1+h_2+h_3)}$ &
${\cal O}_{C,-\frac{3}{2}}^{(h_1+h_2+h_3)}+\cdots$
\\
\hline
${\scriptstyle h_1+2}$ &
${\scriptstyle +2}$ &
${\scriptstyle h_2+1}$ &
${\scriptstyle +1}$ &
${\scriptstyle 1-h_3}$ &
${\scriptstyle 2-h_3}$ &
${\scriptstyle 1-(h_1+h_2+h_3)}$ &
${\scriptstyle 3-(h_1+h_2+h_3)}$ &
${\cal O}_{-2}^{(h_1+h_2+h_3)}$
\\
\hline
\hline
${\scriptstyle h_1+2}$ &
${\scriptstyle +2}$ &
${\scriptstyle h_2+\frac{1}{2}}$ &
${\scriptstyle +\frac{1}{2}}$ &
${\scriptstyle 1-h_3}$ &
${\scriptstyle -1-h_3}$ &
${\scriptstyle 1-(h_1+h_2+h_3)}$ &
${\scriptstyle \frac{1}{2}-(h_1+h_2+h_3)}$ &
${\cal O}_{+\frac{1}{2}}^{(h_1+h_2+h_3),ABC}$
\\
\hline
${\scriptstyle h_1+2}$ &
${\scriptstyle +2}$ &
${\scriptstyle h_2+\frac{1}{2}}$ &
${\scriptstyle +\frac{1}{2}}$ &
${\scriptstyle 1-h_3}$ &
${\scriptstyle -\frac{1}{2}-h_3}$ &
${\scriptstyle 1-(h_1+h_2+h_3)}$ &
${\scriptstyle 1-(h_1+h_2+h_3)}$ &
${\cal O}_{0}^{(h_1+h_2+h_3),ABCD}$
\\
\hline
${\scriptstyle h_1+2}$ &
${\scriptstyle +2}$ &
${\scriptstyle h_2+\frac{1}{2}}$ &
${\scriptstyle +\frac{1}{2}}$ &
${\scriptstyle 1-h_3}$ &
${\scriptstyle -h_3}$ &
${\scriptstyle 1-(h_1+h_2+h_3)}$ &
${\scriptstyle \frac{3}{2}-(h_1+h_2+h_3)}$ &
${\cal O}_{FGH,-\frac{1}{2}}^{(h_1+h_2+h_3)}$
\\
\hline
${\scriptstyle h_1+2}$ &
${\scriptstyle +2}$ &
${\scriptstyle h_2+\frac{1}{2}}$ &
${\scriptstyle +\frac{1}{2}}$ &
${\scriptstyle 1-h_3}$ &
${\scriptstyle \frac{1}{2}-h_3}$ &
${\scriptstyle 1-(h_1+h_2+h_3)}$ &
${\scriptstyle 2-(h_1+h_2+h_3)}$ &
${\cal O}_{GH,-1}^{(h_1+h_2+h_3)}$
\\
\hline
${\scriptstyle h_1+2}$ &
${\scriptstyle +2}$ &
${\scriptstyle h_2+\frac{1}{2}}$ &
${\scriptstyle +\frac{1}{2}}$ &
${\scriptstyle 1-h_3}$ &
${\scriptstyle 1-h_3}$ &
${\scriptstyle 1-(h_1+h_2+h_3)}$ &
${\scriptstyle \frac{5}{2}-(h_1+h_2+h_3)}$ &
${\cal O}_{H,-\frac{3}{2}}^{(h_1+h_2+h_3)}$
\\
\hline
${\scriptstyle h_1+2}$ &
${\scriptstyle +2}$ &
${\scriptstyle h_2+\frac{1}{2}}$ &
${\scriptstyle +\frac{1}{2}}$ &
${\scriptstyle 1-h_3}$ &
${\scriptstyle \frac{3}{2}-h_3}$ &
${\scriptstyle 1-(h_1+h_2+h_3)}$ &
${\scriptstyle 3-(h_1+h_2+h_3)}$ &
${\cal O}_{-2}^{(h_1+h_2+h_3)}$
\\
\hline
\hline
${\scriptstyle h_1+2}$ &
${\scriptstyle +2}$ &
${\scriptstyle h_2 }$ &
${\scriptstyle 0}$ &
${\scriptstyle 1-h_3}$ &
${\scriptstyle -1-h_3}$ &
${\scriptstyle 1-(h_1+h_2+h_3)}$ &
${\scriptstyle 1-(h_1+h_2+h_3)}$ &
${\cal O}_{0}^{(h_1+h_2+h_3),ABCD}$
\\
\hline
${\scriptstyle h_1+2}$ &
${\scriptstyle +2}$ &
${\scriptstyle h_2}$ &
${\scriptstyle 0}$ &
${\scriptstyle 1-h_3}$ &
${\scriptstyle -\frac{1}{2}-h_3}$ &
${\scriptstyle 1-(h_1+h_2+h_3)}$ &
${\scriptstyle \frac{3}{2}-(h_1+h_2+h_3)}$ &
${\cal O}_{FGH,-\frac{1}{2}}^{(h_1+h_2+h_3)}$
\\
\hline
${\scriptstyle h_1+2}$ &
${\scriptstyle +2}$ &
${\scriptstyle h_2}$ &
${\scriptstyle 0}$ &
${\scriptstyle 1-h_3}$ &
${\scriptstyle -h_3}$ &
${\scriptstyle 1-(h_1+h_2+h_3)}$ &
${\scriptstyle 2-(h_1+h_2+h_3)}$ &
${\cal O}_{GH,-1}^{(h_1+h_2+h_3)}$
\\
\hline
${\scriptstyle h_1+2}$ &
${\scriptstyle +2}$ &
${\scriptstyle h_2}$ &
${\scriptstyle 0}$ &
${\scriptstyle 1-h_3}$ &
${\scriptstyle \frac{1}{2}-h_3}$ &
${\scriptstyle 1-(h_1+h_2+h_3)}$ &
${\scriptstyle \frac{5}{2}-(h_1+h_2+h_3)}$ &
${\cal O}_{H,-\frac{3}{2}}^{(h_1+h_2+h_3)}$
\\
\hline
${\scriptstyle h_1+2}$ &
${\scriptstyle +2}$ &
${\scriptstyle h_2}$ &
${\scriptstyle 0}$ &
${\scriptstyle 1-h_3}$ &
${\scriptstyle 1-h_3}$ &
${\scriptstyle 1-(h_1+h_2+h_3)}$ &
${\scriptstyle 3-(h_1+h_2+h_3)}$ &
${\cal O}_{-2}^{(h_1+h_2+h_3)}$
\\
\hline
\hline
\end{tabular}
\caption{
For given right-weights, helicities, and the left- and right-weights
appearing in Table \ref{table4}, we present
the left- and right-weights for the eighteen matter fields of (\ref{Iden})
in the section \ref{twocurrents}.
This is a continuation of previous Table \ref{table6}. }
\label{table7}
\end{table}
\begin{table}[tbp]
\centering
\renewcommand{\arraystretch}{1.7}
\begin{tabular}{|c|c||c|c||c|c|||c|c|c|| }
\hline
$ p$ &
$s_1 $ &
$ q$ &
$s_2$ &
$ h $ &
$ \bar{h}$ &
$h_4$
&
$\bar{h}_4$
&
$ {\cal O}_{h_4,\bar{h}_4}$ 
\\
\hline
\hline
${\scriptstyle h_1+2}$ &
${\scriptstyle +2}$ &
${\scriptstyle h_2-\frac{1}{2}}$ &
${\scriptstyle -\frac{1}{2}}$ &
${\scriptstyle 1-h_3}$ &
${\scriptstyle -1-h_3}$ &
${\scriptstyle 1-(h_1+h_2+h_3)}$ &
${\scriptstyle \frac{3}{2}-(h_1+h_2+h_3)}$ &
${\cal O}_{ABC,-\frac{1}{2}}^{(h_1+h_2+h_3)}$
\\
\hline
${\scriptstyle h_1+2}$ &
${\scriptstyle +2}$ &
${\scriptstyle h_2-\frac{1}{2}}$ &
${\scriptstyle -\frac{1}{2}}$ &
${\scriptstyle 1-h_3}$ &
${\scriptstyle -\frac{1}{2}-h_3}$ &
${\scriptstyle 1-(h_1+h_2+h_3)}$ &
${\scriptstyle 2-(h_1+h_2+h_3)}$ &
${\cal O}_{AB,-1}^{(h_1+h_2+h_3)}+\cdots$
\\
\hline
${\scriptstyle h_1+2}$ &
${\scriptstyle +2}$ &
${\scriptstyle h_2-\frac{1}{2}}$ &
${\scriptstyle -\frac{1}{2}}$ &
${\scriptstyle 1-h_3}$ &
${\scriptstyle -h_3}$ &
${\scriptstyle 1-(h_1+h_2+h_3)}$ &
${\scriptstyle \frac{5}{2}-(h_1+h_2+h_3)}$ &
${\cal O}_{A,-\frac{3}{2}}^{(h_1+h_2+h_3)}+\cdots$
\\
\hline
${\scriptstyle h_1+2}$ &
${\scriptstyle +2}$ &
${\scriptstyle h_2-\frac{1}{2}}$ &
${\scriptstyle -\frac{1}{2}}$ &
${\scriptstyle 1-h_3}$ &
${\scriptstyle \frac{1}{2}-h_3}$ &
${\scriptstyle 1-(h_1+h_2+h_3)}$ &
${\scriptstyle 3-(h_1+h_2+h_3)}$ &
${\cal O}_{-2}^{(h_1+h_2+h_3)}$
\\
\hline
\hline
${\scriptstyle h_1+2}$ &
${\scriptstyle +2}$ &
${\scriptstyle h_2-1}$ &
${\scriptstyle -1}$ &
${\scriptstyle 1-h_3}$ &
${\scriptstyle -1-h_3}$ &
${\scriptstyle 1-(h_1+h_2+h_3)}$ &
${\scriptstyle 2-(h_1+h_2+h_3)}$ &
${\cal O}_{AB,-1}^{(h_1+h_2+h_3)}$
\\
\hline
${\scriptstyle h_1+2}$ &
${\scriptstyle +2}$ &
${\scriptstyle h_2-1}$ &
${\scriptstyle -1}$ &
${\scriptstyle 1-h_3}$ &
${\scriptstyle -\frac{1}{2}-h_3}$ &
${\scriptstyle 1-(h_1+h_2+h_3)}$ &
${\scriptstyle \frac{5}{2}-(h_1+h_2+h_3)}$ &
${\cal O}_{A,-\frac{3}{2}}^{(h_1+h_2+h_3)}+\cdots$
\\
\hline
${\scriptstyle h_1+2}$ &
${\scriptstyle +2}$ &
${\scriptstyle h_2-1}$ &
${\scriptstyle -1}$ &
${\scriptstyle 1-h_3}$ &
${\scriptstyle -h_3}$ &
${\scriptstyle 1-(h_1+h_2+h_3)}$ &
${\scriptstyle 3-(h_1+h_2+h_3)}$ &
${\cal O}_{-2}^{(h_1+h_2+h_3)}$
\\
\hline
\hline
${\scriptstyle h_1+2}$ &
${\scriptstyle +2}$ &
${\scriptstyle h_2-\frac{3}{2}}$ &
${\scriptstyle -\frac{3}{2}}$ &
${\scriptstyle 1-h_3}$ &
${\scriptstyle -1-h_3}$ &
${\scriptstyle 1-(h_1+h_2+h_3)}$ &
${\scriptstyle \frac{5}{2}-(h_1+h_2+h_3)}$ &
${\cal O}_{A,-\frac{3}{2}}^{(h_1+h_2+h_3)}$
\\
\hline
${\scriptstyle h_1+2}$ &
${\scriptstyle +2}$ &
${\scriptstyle h_2-\frac{3}{2}}$ &
${\scriptstyle -\frac{3}{2}}$ &
${\scriptstyle 1-h_3}$ &
${\scriptstyle -\frac{1}{2}-h_3}$ &
${\scriptstyle 1-(h_1+h_2+h_3)}$ &
${\scriptstyle 3-(h_1+h_2+h_3)}$ &
${\cal O}_{-2}^{(h_1+h_2+h_3)}$
\\
\hline
\hline
${\scriptstyle h_1+2}$ &
${\scriptstyle +2}$ &
${\scriptstyle h_2-2}$ &
${\scriptstyle -2}$ &
${\scriptstyle 1-h_3}$ &
${\scriptstyle -1-h_3}$ &
${\scriptstyle 1-(h_1+h_2+h_3)}$ &
${\scriptstyle 3-(h_1+h_2+h_3)}$ &
${\cal O}_{-2}^{(h_1+h_2+h_3)}$
\\
\hline
\hline
\end{tabular}
\caption{
For given right-weights, helicities, and the left- and right-weights
appearing in Table \ref{table5}, 
the left- and right-weights for the ten matter fields of (\ref{Iden})
in the section \ref{twocurrents} are given explicitly.
This is a continuation of previous Tables \ref{table6} and \ref{table7}.}
\label{table8}
\end{table}

Let us consider one of the nontrivial examples
for the two fermionic modes of ${\cal N}=8$ ${\rm w}_{1+\infty}$
currents $\hat{\Phi}_{+\frac{3}{2},m}^{(h_1),A}$
and $\hat{\Phi}_{+\frac{3}{2},n}^{(h_2),B}$ with the bosonic matter field
${\cal O}_0^{(h),CDEF}(z,\bar{z})$.
Then the first commutator inside the first term of (\ref{wwO})
is, from the thirteenth equation of (\ref{comm25OPEs}), given by
\bea
\bigg[ \hat{\Phi}_{+\frac{3}{2},n}^{(h_2),B}, {\cal O}_0^{(h),CDEF}(z,\bar{z}) \bigg]
= \la_{+\frac{3}{2},0,+\frac{1}{2}}\, \frac{1}{3!}\, \ep^{BCDEFGHI}\,
{\mathscr D}_n^{h_2+\frac{3}{2}}(1-h)\,
{\cal O}_{GHI,-\frac{1}{2}}^{(h_2+h)}(z,\bar{z}).
\label{firstexp}
\eea
Now the next anticommutator between the
first mode of the current and the matter field
on the right-hand side of (\ref{firstexp}), from the fourteenth
equation of (\ref{comm25OPEs}), can be written as
\bea
\bigg\{ \hat{\Phi}_{+\frac{3}{2},m}^{(h_1),A},  {\cal O}_{GHI,-\frac{1}{2}}^{(h_2+h)}(z,\bar{z})
\bigg\} = \la_{+\frac{3}{2},-\frac{1}{2},+1}\,
3 \, \de^{A}_{[G} \,  {\mathscr D}_m^{h_1+\frac{3}{2}}(\tfrac{3}{2}-h_2-h)\,
  {\cal O}_{HI],-1}^{(h_1+h_2+h)}(z,\bar{z}).
\label{secondexp}
\eea
Then the final contribution from the first term of (\ref{wwO}),
by combining (\ref{firstexp}) with (\ref{secondexp}),
can be described as
\bea
 \la_{+\frac{3}{2},-\frac{1}{2},+1}\,  \la_{+\frac{3}{2},0,+\frac{1}{2}}\,
\frac{1}{2!}\,  \ep^{BCDEFGHI} \, \de^{A}_{[G} \,
{\mathscr D}_n^{h_2+\frac{3}{2}}(1-h)\,  
 {\mathscr D}_m^{h_1+\frac{3}{2}}(\tfrac{3}{2}-h_2-h)\,
  {\cal O}_{HI],-1}^{(h_1+h_2+h)}(z,\bar{z}).
\label{firstresult}
\eea

On the other hand,
the first commutator
inside the second term of (\ref{wwO}) with plus sign
is,  from the thirteenth equation of (\ref{comm25OPEs}), given by
\bea
\bigg[ \hat{\Phi}_{+\frac{3}{2},m}^{(h_1),A},  {\cal O}_0^{(h),CDEF}(z,\bar{z})  \bigg]=
\la_{+\frac{3}{2},0,+\frac{1}{2}}\,
\frac{1}{3!}\, \ep^{ACDEFGHI}\,
 {\mathscr D}_m^{h_1+\frac{3}{2}}(1-h)\,
{\cal O}_{GHI,-\frac{1}{2}}^{(h_1+h)}(z,\bar{z}).
\label{res1}
\eea
The next anticommutator between
the
second mode of the current and the matter field
on the right-hand side of (\ref{res1})
is
\bea
\bigg\{ \hat{\Phi}_{+\frac{3}{2},n}^{(h_2),B},
{\cal O}_{GHI,-\frac{1}{2}}^{(h_1+h)}(z,\bar{z}) \bigg\}=
\la_{+\frac{3}{2},-\frac{1}{2},+1}\,
 3 \, \de^{B}_{[G} \,  {\mathscr D}_n^{h_2+\frac{3}{2}}(\tfrac{3}{2}-h_1-h)\,
  {\cal O}_{HI],-1}^{(h_1+h_2+h)}(z,\bar{z}).
\label{res2}
\eea
By combining (\ref{res1}) with (\ref{res2}),
the contribution from the second term of (\ref{wwO})
can be obtained as
\bea
\la_{+\frac{3}{2},-\frac{1}{2},+1}
\la_{+\frac{3}{2},0,+\frac{1}{2}}
\frac{1}{2!}  \ep^{ACDEFGHI}  \de^{B}_{[G} 
{\mathscr D}_m^{h_1+\frac{3}{2}}(1-h)  
{\mathscr D}_n^{h_2+\frac{3}{2}}(\tfrac{3}{2}-h_1-h)
{\cal O}_{HI],-1}^{(h_1+h_2+h)}(z,\bar{z}).
\label{secondresult}
\eea

The right-hand side of (\ref{wwO}) from the tenth equation of
(\ref{comm25OPEs})
is given by
\bea
&& \kappa_{+\frac{3}{2},+\frac{3}{2},-1}\,
\bigg( (h_2+\tfrac{1}{2})m-(h_1+\tfrac{1}{2})n \bigg) \,
\bigg[ \hat{\Phi}_{+1,m+n}^{(h_1+h_2),AB}, {\cal O}_{0}^{(h),CDEF}(z,\bar{z}) \bigg]
= \kappa_{+\frac{3}{2},+\frac{3}{2},-1}\,
\la_{+1,0,+1}\nonu \\
&& \times 
\bigg( (h_2+\tfrac{1}{2})m-(h_1+\tfrac{1}{2})n \bigg) \,
\frac{1}{2!} \, \ep^{ABCDEFGH}\,
{\mathscr D}_{m+n}^{h_1+h_2+1}(1-h)\,  
{\cal O}_{GH,-1}^{(h_1+h_2+h)}(z,\bar{z}),
\label{thirdresult}
\eea
where the nineteenth equation of (\ref{comm25OPEs}) is used.

Finally, we construct the relation (\ref{Iden})
by using (\ref{firstresult}), (\ref{secondresult}) and (\ref{thirdresult})
\bea
&& \Bigg[
(-1) \la_{+\fft32,-\fft12,+1}\,\la_{+\fft32,0,+\fft12} \,
{\mathscr D}_n^{h_2+\fft32}(1-h) \,
{\mathscr D}_m^{h_1+\fft32}(\tf32-h_2-h)
\nonu \\
&& - (-1) \la_{+\fft32,-\fft12,+1}\, \la_{+\fft32,0,+\fft12}\,
{\mathscr D}_m^{h_1+\fft32}(1-h)
{\mathscr D}_n^{h_2+\fft32}(\tf32-h_1-h)
\nonu \\
&& -
\la_{+1,0,+1} \, \kappa_{+\fft32,+\fft32,-1} \,\bigg(
(h_2+\tf12)m-(h_1+\tf12)n \bigg) \,
{\mathscr D}_{m+n}^{h_1+h_2+1}(1-h)\, \Bigg]\nonu \\
&& \times
\frac{1}{2!}\, \ep^{ABCDEFGH}\,
{\cal O}_{GH,-1}^{(h_1+h_2+h)}(z,\bar{z})=0.
\label{Iden1}
\eea
We can check the minus sign in the first term of (\ref{Iden1})
explicitly from (\ref{firstresult}).
This can be seen by focusing on the particular values for the $SU(8)$ indices
$ABCDEF=123456$.
The six terms contribute equally.
In the antisymmetric notation, there is a factor $\frac{1}{3!}$.
This leads to the fact that the number of terms
for fixed indices $H$ and $I$ is equal to $1$.
For the second term in (\ref{Iden1}), there exists
an overall plus sign which is written in terms of two minus signs.
In order to satisfy the relation (\ref{Iden1})
for any matter field, the three combinations inside the bracket
should vanish.
As shown in \cite{HPS},
the first coefficient should be equal to the negative
of
the third coefficient as follows 
\footnote{In \cite{HPS}, the way to show the identity (\ref{Iden})
for the gravitons is as follows. The simplest nontrivial example they
took is the case of $p=\frac{5}{2}$ and $m=-\frac{3}{2}$.
Then
the first two terms of (\ref{Iden}) with $+1$ coefficients
can be written as 
\bea
\bigg[{\mathscr D}_n^{q}(\bar{h}) \,
{\mathscr D}_{-\fft32}^{\fft52}(\bar{h}+2-q)
-
{\mathscr D}_{-\fft32}^{\fft52}(\bar{h})
{\mathscr D}_n^{q}(\bar{h}-\tfrac{1}{2})
\bigg] \, {\cal O}_{h+\frac{3}{2}-q,\bar{h}+\fft32-q}.
\nonu
\eea
Then by using the explicit form for the differential operators,
this becomes three terms with dummy variable $l$.
By using the change of variable, we can collect one single term
and this leads to
\bea
-\frac{3}{2}\, (q+n-1) \, {\mathscr D}_{n-\fft32}^{q+\fft12}(\bar{h})\,
{\cal O}_{h+\frac{3}{2}-q,\bar{h}+\fft32-q}.
\nonu
\eea
This implies the commutator relation (\ref{wwcomm})
for this particular case.
They further considered the more general case and
showed the identity (\ref{Iden}) or (\ref{wwcomm}).}
\bea
\la_{+\fft32,-\fft12,+1}\,\la_{+\fft32,0,+\fft12} = 
-\la_{+1,0,+1} \, \kappa_{+\fft32,+\fft32,-1}.
\label{oneexample}
\eea
In this way, we can obtain the relations between the couplings.
Then we are left with ninety-five
relations in (\ref{17}), (\ref{18}), (\ref{10}),
(\ref{other18}), where the relation (\ref{oneexample})
can be seen, (\ref{other10}), (\ref{15}), and  (\ref{7}),
where some of the relative minus signs arise from the
arrangements of $SU(8)$ indices of matter fields or the anticommutators
between the modes of ${\cal N}=8$ ${\rm w}_{1+\infty}$ currents.

In Appendix D,
by solving the constraints, (\ref{17-lambda}), (\ref{18-lambda}),
(\ref{10-lambda}),
(\ref{other18-lambda}), (\ref{other10-lambda}),
(\ref{15-lambda}), and  (\ref{7-lambda}),
where some of the relative minus signs appear
in the quadratic equations,
together with the constraints for the coupling $\kappa_{s_1,s_2,-s_3}$
(i.e., the seventeen couplings can be written in terms of eight
independent couplings) in \cite{AK2509},
we obtain the final solution
\bea
\lambda_{+2,-2,+2} & = & \kappa_{+2,+2,-2},
\qquad
\lambda_{+2,+2,-2}= \kappa_{+2,+2,-2},
\qquad
\lambda_{+2,-\frac{3}{2},+\frac{3}{2}}=
\kappa_{+2,+2,-2},
\nonu \\
\lambda_{+2,+\frac{3}{2},-\frac{3}{2}}  & = & 
\kappa_{+2,+2,-2},
\qquad
\lambda_{+2,-1,+1}  =  \kappa_{+2,+2,-2},
\qquad
\lambda_{+2,+1,-1}= \kappa_{+2,+2,-2},
\nonu \\
\lambda_{+2,-\frac{1}{2},+\frac{1}{2}}  & = & 
\kappa_{+2,+2,-2},
\qquad
\lambda_{+2,+\frac{1}{2},-\frac{1}{2}}=
\kappa_{+2,+2,-2},
\qquad
\lambda_{+2,0,0}  =  \kappa_{+2,+2,-2},
\nonu \\
\lambda_{+\frac{3}{2},-\frac{3}{2},+2} & = & 
\kappa_{+\frac{3}{2},-\frac{3}{2},+2},
\qquad
\lambda_{+\frac{3}{2},+\frac{3}{2},-1}=
\kappa_{+\frac{3}{2},+\frac{3}{2},-1},
\qquad
\lambda_{+\frac{3}{2},-1,+\frac{3}{2}}=
\kappa_{+\frac{3}{2},-1,+\frac{3}{2}},
\nonu \\
\lambda_{+\frac{3}{2},+1,-\frac{1}{2}} & = &
\kappa_{+\frac{3}{2},+1,-\frac{1}{2}},
\qquad
\lambda_{+\frac{3}{2},-\frac{1}{2},+1}=
\frac{\kappa_{0,0,+2}
\, \kappa_{+1,+\frac{1}{2},+\frac{1}{2}}
\, \kappa_{+\frac{3}{2},+1,-\frac{1}{2}}
\, \kappa_{+\frac{3}{2},+\frac{3}{2},-1}^2
}{\kappa_{+\frac{3}{2},-\frac{3}{2},+2}
\, \kappa_{+\frac{3}{2},-1,+\frac{3}{2}}
\, \kappa_{+\frac{3}{2},0,+\frac{1}{2}}^2},
\nonu \\
\lambda_{+\frac{3}{2},+\frac{1}{2},0}  & = & 
\frac{\kappa_{+1,+\frac{1}{2},+\frac{1}{2}}
\, \kappa_{+\frac{3}{2},+\frac{3}{2},-1}}
{\kappa_{+\frac{3}{2},0,+\frac{1}{2}}},
\qquad
\lambda_{+\frac{3}{2},0,+\frac{1}{2}}  = 
\kappa_{+\frac{3}{2},0,+\frac{1}{2}},
\nonu \\
\lambda_{+1,-1,+2}  & = & -\frac{\kappa_{+\frac{3}{2},-
\frac{3}{2},+2}\, \kappa_{+\frac{3}{2},-1,+\frac{3}{2}}}
{\kappa_{+\frac{3}{2},+\frac{3}{2},-1}},
\qquad
\lambda_{+1,+1,0}= \frac{\kappa_{+1,+\frac{1}{2},+\frac{1}{2}}
\, \kappa_{+\frac{3}{2},+1,-\frac{1}{2}}}
{\kappa_{+\frac{3}{2},0,+\frac{1}{2}}},
\nonu \\
\lambda_{+1,-\frac{1}{2},+\frac{3}{2}}
& = & \frac{\kappa_{0,0,+2}
\, \kappa_{+1,+\frac{1}{2},+\frac{1}{2}}
\, \kappa_{+\frac{3}{2},+1,-\frac{1}{2}}
\, \kappa_{+\frac{3}{2},+\frac{3}{2},-1}}
{\kappa_{+\frac{3}{2},-\frac{3}{2},+2}
\, \kappa_{+\frac{3}{2},0,+\frac{1}{2}}^2},
\qquad
\lambda_{+1,+\frac{1}{2},+\frac{1}{2}}=
\kappa_{+1,+\frac{1}{2},+\frac{1}{2}},
\nonu \\
\lambda_{+1,0,+1} & = & -\frac{\kappa_{0,0,+2}
\, \kappa_{+1,+\frac{1}{2},+\frac{1}{2}} 
\, \kappa_{+\frac{3}{2},+1,-\frac{1}{2}}
\, \kappa_{+\frac{3}{2},+\frac{3}{2},-1}}
{\kappa_{+\frac{3}{2},-\frac{3}{2},+2}
\, \kappa_{+\frac{3}{2},-1,+\frac{3}{2}}
\, \kappa_{+\frac{3}{2},0,+\frac{1}{2}}},
\nonu \\
\lambda_{+\frac{1}{2},-\frac{1}{2},+2}
 & = & -\frac{\kappa_{0,0,+2}\, \kappa_{+1,+\frac{1}{2},+\frac{1}{2}}
\kappa_{+\frac{3}{2},+\frac{3}{2},-1}}
{\kappa_{+\frac{3}{2},0,+\frac{1}{2}}^2},
\qquad
\lambda_{+\frac{1}{2},+\frac{1}{2},+1}  = 
-\frac{\kappa_{0,0,+2}\, \kappa_{+1,+\frac{1}{2},+\frac{1}{2}}^2
\kappa_{+\frac{3}{2},+\frac{3}{2},-1}^2}
{\kappa_{+\frac{3}{2},-\frac{3}{2},+2}
\, \kappa_{+\frac{3}{2},-1,+\frac{3}{2}}
\, \kappa_{+\frac{3}{2},0,+\frac{1}{2}}^2},
\nonu \\
\lambda_{+\frac{1}{2},0,+\frac{3}{2}} & = &
-\frac{\kappa_{0,0,+2}\, \kappa_{+1,+\frac{1}{2},+\frac{1}{2}}
\, \kappa_{+\frac{3}{2},+\frac{3}{2},-1}}
{\kappa_{+\frac{3}{2},-\frac{3}{2},+2}
\, \kappa_{+\frac{3}{2},0,+\frac{1}{2}}},
\qquad
\lambda_{0,0,+2}  =  \kappa_{0,0,+2},
\label{lambdacon}
\eea
where we assume that the coupling
$\la_{s_1,s_2,-s_3}$
satisfies $\la_{s_2,s_1,-s_3}= \pm \la_{s_1,s_2,-s_3}$
with the minus sign for half integers $s_1$ and $s_2$
and the plus sign for other cases of $s_1$ and $s_2$.  
In (\ref{lambdacon}), all the $\lambda_{s_1,s_2,-s_3}$
couplings can be written in terms of the
eight arbitrary
couplings in \cite{AK2509}.
Moreover, by comparing (\ref{lambdacon})
with the equations $(5.20)$ of \cite{AK2509},
we conclude that the coupling
$\la_{s_1,s_2,-s_3}$ is the same as  the coupling
$\kappa_{s_1,s_2,-s_3}$.

Or we can solve
ninety-five relations without imposing the constraints
between the couplings
in \cite{AK2509} directly by putting the constraint
$\la_{s_1,s_2,-s_3}=\kappa_{s_1,s_2,-s_3}$.
Then we arrive at the constraints between
the seventeen couplings (in terms of
eight independent couplings) found in \cite{AK2509}
together with the fact that the twelve (redundant) couplings
$\kappa_{s_1,s_2,-s_3}$ ($s_1 < s_2$)
can be written in terms of $\pm \kappa_{s_2,s_1,-s_3}$.
Here the minus signs appear only for half integers
$s_1$ and $s_2$ described before. Moreover,
the seven couplings can be written in terms of
another arbitrary coupling \footnote{
That is, $\kappa_{-\frac{3}{2},+2,+\frac{3}{2}}=\kappa_{+\frac{3}{2},
+2,-\frac{3}{2}}$, $\kappa_{-1,+2,+1}=\kappa_{+\frac{3}{2},
+2,-\frac{3}{2}}$, $\kappa_{-\frac{1}{2},+2,+\frac{1}{2}}=\kappa_{+\frac{3}{2},
+2,-\frac{3}{2}}$, $\kappa_{0,+2,+0}=\kappa_{+\frac{3}{2},
+2,-\frac{3}{2}}$, $\kappa_{+\frac{1}{2},+2,-\frac{1}{2}}=\kappa_{+\frac{3}{2},
+2,-\frac{3}{2}}$, $\kappa_{+1,+2,-1}=\kappa_{+\frac{3}{2},
+2,-\frac{3}{2}}$, and $\kappa_{-2,+2,+2}=\kappa_{+\frac{3}{2},
  +2,-\frac{3}{2}}$.}.
We expect that there should be further constraints
on these seven couplings once we use other identities (\ref{Iden})
having the conditions $s_1 < s_2$.

Therefore, as in the (bosonic) pure gravity theory \cite{HPS},
we prove that 
the charges (i.e., the modes of ${\cal N}=8$ ${\rm w}_{1+\infty}$ currents)
generate the action of the ${\cal N}=8$ ${\rm w}_{1+\infty}$ (super)symmetry. 

We can check
i) the commutator between the mode of Lorentz (or right Virasoro)
generators and the mode
of supertranslation generators, ii)
the commutator between the mode of Lorentz generators
and themselves and iii)
the commutator between the mode of Lorentz generators and the mode
of supercurrent generators
as follows:
\bea
\bigg[\hat{\Phi}_{+2,m}^{(0)},
{\Phi}_{+2,k,l}^{(-\frac{1}{2})}  \bigg] & = &
\bigg( \frac{1}{2}m-l\bigg)\, {\Phi}_{+2,k,m+l}^{(-\frac{1}{2})},
\nonu \\
\bigg[
\hat{\Phi}_{+2,m}^{(0)},
\hat{\Phi}_{+2,n}^{(0)}  \bigg] & = &
\bigg( m-n\bigg)\, \hat{\Phi}_{+2,m+n}^{(0)},
\nonu \\
\bigg[
\hat{\Phi}_{+2,m}^{(0)},
\hat{\Phi}_{+\frac{3}{2},n}^{(0),A}  \bigg] & = &
\bigg( \frac{1}{2}m-n\bigg)\, \hat{\Phi}_{+\frac{3}{2},m+n}^{(0),A}. 
\label{BMS}
\eea
These are a subalgebra of ${\cal N}=8$ supersymmetric BMS algebra
\footnote{
\label{BMScheck}
  In order to check the first equation of (\ref{BMS}),
$\bigg[\bar{L}_m, P_{k,l} \bigg]=\bigg( \frac{1}{2} m -l\bigg)
\, P_{k,m+l}$, in the notation of \cite{FSTZ}, we need to
use (\ref{wwO}). In particular, the second term of the left-hand side
contains the commutator $\bigg[{\Phi}_{+2,k,l}^{(-\frac{1}{2})},
\bar{\pa} \,
{\cal O}_{h,\bar{h}}(z, \bar{z}) \bigg]$. Then we should
compute the contour integral (\ref{wOdef}). For a
given OPE (\ref{OPEwO}) or (\ref{25OPEs}),
when we differentiate (\ref{translation}) 
with respect to the coordinate $\bar{z}_2$,
the contribution from
the right-hand side consists of two terms
because the $\bar{z}_2$ dependence appears in the $\bar{z}_{12}$ factor
and in the operator itself on the right-hand side.
Then the derivative term of
operator is canceled by the corresponding term from the contribution
from
the first term of the left-hand side of (\ref{wwO}).
The remaining term can combine with the contribution from the first
term of the left-hand side of (\ref{wwO}) and we obtain
the right-hand side of the first equation of (\ref{BMS}).
Note that the $k$ dependent
factor $z^{k+\frac{1}{2}}$ from the left-hand side
of the first equation of (\ref{BMS}) remains the same on the
right-hand side.

For the second equation, $\bigg[\bar{L}_m, \bar{L}_n \bigg]=
\bigg(m-n \bigg)\, \bar{L}_{m+n}$, 
we can use, after simplifying four terms from (\ref{PhiO}), 
\bea
 {\Phi}_{+2}^{(0)}(z_1,\bar{z}_1) \,
\bar{\pa}_{\bar{z}_2} {\cal O}_{+2}^{(h_2)}(z_2,\bar{z}_2)
=  \frac{1}{z_{12}}\bigg[ 
\frac{1}{\bar{z}_{12}^3}\, 2(-1-h_2)
+\frac{1}{\bar{z}_{12}^2}\,(-h_2) \, \bar{\pa}_{\bar{z}_2}
+ \frac{1}{\bar{z}_{12}}\, \bar{\pa}_{\bar{z}_2}^2 \bigg]
{\cal O}_{+2}^{(h_2)}(z_2,\bar{z}_2)
+  \cdots.
\nonu
\eea
With the help of (\ref{wOdef}), the antiholomorphic
second derivative terms of the operator are canceled
and the remaining first derivative terms with nonderivative terms
can be written as the right-hand side of the second equation.

For the last equation, $\bigg[\bar{L}_m, \bar{G}^A_n \bigg]=
\bigg(\frac{1}{2} m -n \bigg)\,  \bar{G}^A_{m+n}$,
we can use, from the first equation of (\ref{20OPEs})
(see also the right-hand side of (\ref{tenth})),
\bea
 {\Phi}_{+\frac{3}{2}}^{(0),A}(z_1,\bar{z}_1) \,
\bar{\pa}_{\bar{z}_2} {\cal O}_{+2}^{(h_2)}(z_2,\bar{z}_2)
=\frac{1}{z_{12}} \, \frac{1}{2} \, \bigg[ 
\frac{1}{\bar{z}_{12}^2}
+ \frac{1}{\bar{z}_{12}}\, \bar{\pa}_{\bar{z}_2} \bigg]
{\cal O}_{+\frac{3}{2}}^{(h_2),A}(z_2,\bar{z}_2)
+  \cdots.
\nonu
\eea
The derivative terms are canceled and the nonderivative terms
can be simplified as the right-hand side of the third equation in
(\ref{BMS}).
Although we take the graviton matter fields above,
the commutator relations (\ref{BMS}) should hold for
any matter fields.}.
In the next section \ref{holomorphic},
we will construct the stress energy tensor and the supercurrents
in the holomorphic sector. Then we will obtain the three nontrivial
commutator relations similar to (\ref{BMS}).
Moreover, according to the ${\cal N}=8$ BMS algebra, there exists
the nontrivial anticommutator between the supercurrent
found in this section and another
supercurrent which will be described
in section \ref{holomorphic}.

We can easily check that the second and third equations
of (\ref{BMS})
can be seen from the the first and the second equations
of $(4.1)$ in \cite{AK2509} respectively by putting $h_1=0=h_2$
with the couplings of one \footnote{
The first and the second equations
of $(4.1)$ in \cite{AK2509} are given by
\bea
\bigg[ \hat{{\Phi}}^{(h_1)}_{+2,m},
\hat{{\Phi}}^{(h_2)}_{+2,n} \bigg]&=&
{\kappa}_{+2,+2,-2}\,\Big((h_2+1)m-(h_1+1)n\Big)\,
\hat{{\Phi}}^{(h_1+h_2)}_{+2,m+n}
\, :\text{eq.1},
\nonu    \\
\bigg[
\hat{{\Phi}}^{(h_1)}_{+2,m},
\hat{{\Phi}}^{(h_2),A}_{+\frac{3}{2},n} \bigg]&=&
{\kappa}_{+2,+\frac{3}{2},-\frac{3}{2}}\,
\Big((h_2+\tfrac{1}{2})m-(h_1+1)n\Big)\,
\hat{{\Phi}}^{(h_1+h_2),A}_{+\frac{3}{2},m+n}
\, :\text{eq.2}.
\nonu
\eea}.
Finally,  the first equation
of (\ref{BMS}) with $k=-\frac{1}{2}$ (then the additional hatted notations
can be used on both sides and see also (\ref{anticomm}))
can be seen also from the the first equation
of $(4.1)$ in \cite{AK2509} by putting $h_1=0$, $h_2=-\frac{1}{2}$
with the couplings of one.
In order to cover the arbitrary $k (\neq -\frac{1}{2})$,
we should obtain the generalization of ${\cal N}=8$ ${\rm w}_{1+\infty}$
algebra described in \cite{AK2509} which is an open problem.
See also the similar analysis in (\ref{BMS1}).

\section{The ${\cal N}=8$ ${\rm \bar{w}}_{1+\infty}$
currents from holomorphic light transform 
\label{holomorphic}
}

\subsection{The OPEs of ${\cal N}=8$
${\rm \bar{w}}_{1+\infty}$ currents with matter fields}

So far, we have considered antiholomorphic light transform
which shifts the weights from $(h,\bar{h})$ to $(h,1-\bar{h})$.
We can apply the holomorphic light transform
\footnote{The holomorphic light transform of celestial
operators can be introduced
\bea
\int_{\mathbb{R}} \frac{d w}{2 \pi i}\,
\frac{{\cal O}_{h,\bar{h}}(w,\bar{z})}{(w-z)^{2-2h}}.
\label{hololight}
\eea
From (\ref{OOOPE}) or (\ref{otherOOOPE}),
the celestial operator in (\ref{hololight})
should be the first operator because we don't know
how to calculate the contour integral (in the holomorphic
light transform or in the antiholomorphic transform)
of the second operator appearing on the right-hand sides.
However, the $z_1$ and $\bar{z}_1$ dependence is known for
the OPE (\ref{OOOPE}) or the OPE (\ref{otherOOOPE}) and
the contour integral can be performed explicitly.

Furthermore, in the next section \ref{lastsection}, we will describe
another type of (two dimensional) integral transform, shadow transform
which is the composition of holomorphic and antiholomorphic light
transforms.}
to
the previous sections. This leads to the shift of the
weights from $(h,\bar{h})$ to $(1-h,\bar{h})$.
We can start with the following OPE 
between celestial primary operators 
\bea
{\cal O}_{h_1,\bar{h}_1}(z,\bar{z}) \, {\cal O}_{h_2,\bar{h}_2}(0,0)
& = &
\frac{z}{\bar{z}}\, \bar{\ga}_{s_3}^{s_1,s_2} \, \sum_{m=0}^{\infty}\,
\frac{1}{m!}\,
B(2h_1+1+m, 2h_2+1) \, {z}^{m}\nonu \\
& \times & {\pa}^m \, {\cal O}_{h_1+h_2+1,\bar{h}_1+\bar{h}_2-1}(0,0)
+ \cdots,
\label{otherOOOPE}
\eea
where the right-hand side of this OPE (\ref{otherOOOPE})
can be obtained by taking $z \leftrightarrow \bar{z}$
and $h \leftrightarrow \bar{h}$ in (\ref{OOOPE}).
The $O(\bar{z}^0)$ terms are ignored.
In \cite{CP},
the coupling has a relation
$\bar{\ga}_{s_3}^{s_1,s_2}={\ga}_{-s_3}^{-s_1,-s_2}$.
We consider the combination of
helicities $\bar{p}_{12I} \equiv -s_1-s_2+s_I-1$ which is equal to
$1$. That is, $s_1+s_2-s_3=-2$.
The left-weight of the right-hand side of (\ref{otherOOOPE}) is
given by $-1-m+m+(h_1+h_2+1)=h_1+h_2$ which is the left-weight of the
left-hand side while the right-weight of the right-hand side
is given by $1+(\bar{h}_1+\bar{h}_2-1)=\bar{h}_1+\bar{h}_2$
which is the right-weight of the left-hand side.

As done before in the previous sections or from the work of \cite{HPS},
the holomorphic light transform (\ref{hololight}) for the
first celestial operator in the OPE
(\ref{otherOOOPE}) can be taken.
Then the right-hand side of the OPE of ${\cal N}=8$
${\rm \bar{w}}_{1+\infty}$ currents
and matter fields contains 
i) the original Euler beta function in terms of dummy variable and
the left-weights of the operators and
ii) the ratio of the
Gamma functions in terms of dummy variable and
the left-weight of the first operator, which comes from the integration of
holomophic light transform.
By describing the left- and right-weights for the
two celestial operators
\bea
(h_1, \bar{h}_1) = ( 1-q+\varepsilon, -s_1+1-q), \qquad
(h_2, \bar{h}_2) = (h, \bar{h}),
\label{otherfourh}
\eea
where $q$ is the left-weight of the ${\rm \bar{w}}_{1+\infty}$
currents ($q=1, \frac{3}{2}, 2, \frac{5}{2}, \cdots $.),
the ratio of two Gamma functions with non-positive arguments 
can be further simplified with finite values as before.
The helicity $s_1$ can be read off by taking
the difference between left- and right-weights of the first operator
\footnote{The left-weight of the ${\cal N}=8$ ${\rm \bar{w}}_{1+\infty}$
currents
can be calculated and 
is the sum of $-1$ from $d {w}$, $(2-2{h}_1)$
from the integrand in (\ref{hololight})
and ${h}_1$ from the left-weight of
corresponding operator.  This leads to $(1-h_1)=q$.}.

Then
the OPE of the ${\cal N}=8$ ${\rm{\bar{w}}}_{1+\infty}$ currents with the
matter fields can be summarized by\footnote{
We are using the barred notation for the currents
associated with holomorphic transform.}
\bea
{\rm \bar{w}}^q(z_1,\bar{z}_1) \, {\cal O}_{h,\bar{h}}(z_2,\bar{z}_2)
& = & \frac{1}{\bar{z}_{12}}\, \frac{1}{2}\,
\sum_{m=0}^{2q-3}\, \frac{(m+1)\, \Gamma(2h+1)}{\Gamma(2h+4
-2q +m)}\, {z}_{12}^{2-2q+m}\, {\pa}^m \,
{\cal O}_{h+2-q,\bar{h}-s_1-q}(z_2,\bar{z}_2) \nonu \\
&+ & \cdots.
\label{otherOPEwO}
\eea
Note that $-1$ and  $\Gamma(2q)$
are absorbed inside the normaliztion of ${\rm{\bar{w}}}^q$ current.
We can check the left- and right-weights of
both sides of (\ref{otherOPEwO}).
The left-weight of the left-hand side is given by
$(q+h)$ which can be rewritten as $-(2-2q-m)+m+(h+2-q)$
and this is the left-weight of the right-hand side.
The right-weight of the left-hand side is given by
$(-s_1+1-q)+\bar{h}$ from (\ref{otherfourh}) which can be rewritten as
$1+(\bar{h}-s_1-q)$. This is equal to the right-weight of
the right-hand side.
Note that the left- and right-weights of the operator on the 
right-hand side are obtained from those in
(\ref{OPEwO}) by changing the two locations of two subscripts
and taking the previous relations $h \leftrightarrow \bar{h}$
and $s_1 \leftrightarrow -s_1$.

Now we can associate the above ${\cal N}=8$ ${\rm \bar{w}}_{1+\infty}$
currents with the graviton, the gravitinos, the graviphotons, the
graviphotinos and the scalars as follows:
\bea
{\rm \bar{w}}^q & \longleftrightarrow &
\bar{\Phi}_{-2}^{(h)}, \,
\bar{\Phi}_{A,-\frac{3}{2}}^{(h)}, \,
\bar{\Phi}_{AB,-1}^{(h)},
\,
\bar{\Phi}_{ABC,-\frac{1}{2}}^{(h)}, \,
\bar{\Phi}_{0}^{(h),ABCD},
\,
\bar{\Phi}_{+\frac{1}{2}}^{(h),ABC}, \,
\bar{\Phi}_{+1}^{(h),AB}, \,
\bar{\Phi}_{+\frac{3}{2}}^{(h),A},
\,
\bar{\Phi}_{+2}^{(h)},
\label{otherwO}
\\
{\cal O}_{h,\bar{h}}
&\longleftrightarrow &
{\cal O}_{+2}^{(h)}, \, {\cal O}_{+\frac{3}{2}}^{(h),A}, \, {\cal O}_{+1}^{(h),AB},
\, {\cal O}_{+\frac{1}{2}}^{(h),ABC}, \, {\cal O}_{0}^{(h),ABCD},
\, {\cal O}_{ABC,-\frac{1}{2}}^{(h)}, \, {\cal O}_{AB,-1}^{(h)},
\, {\cal O}_{A,-\frac{3}{2}}^{(h)},
\, {\cal O}_{-2}^{(h)}.
\nonu
\eea
We summarize them in Tables \ref{table1-1} and \ref{table2-1}
for the helicities $s_1 \leq s_2$ and Tables \ref{table9-1}
and \ref{table10-1} (of Appendix E) for the helicities $ s_1 > s_2$.
We can easily see that the matter fields appearing on the
left-hand side of (\ref{otherOPEwO}) in
Tables \ref{table1-1} and \ref{table2-1} (Tables \ref{table9-1}
and \ref{table10-1}) can be obtained as
those in Tables \ref{table1} and \ref{table2} (Tables \ref{table9}
and \ref{table10}) by taking
$h \leftrightarrow \bar{h}$ which implies the helicity
$s$ goes to $-s$ as before.
Similarly, the matter fields on the right-hand side
can be applied also as in the previous paragraph.
See also the footnote \ref{TWOmatters}.

\begin{table}[tbp]
\centering
\renewcommand{\arraystretch}{1.7}
\begin{tabular}{|c|c||c|c|c|||c|c|c|c|| }
\hline
$ q$ &
$ \bar{\rm w}^q$ &
$ h$ &
$ \bar{h}$ &
$ {\cal O}_{h,\bar{h}}$ &
$ s_1$ &
$ {\scriptstyle h+2-q }$ &
$ {\scriptstyle \bar{h}-s_1-q}$ &
${\cal O}_{h+2-q,\bar{h}-s_1-q}$  
\\
\hline
\hline
${\scriptstyle h_1+2}$ &
$  \bar{\Phi}_{-2}^{(h_1)}$ &
${\scriptstyle -1-h_2}$ &
${\scriptstyle 1-h_2}$ &
$ {\cal O}_{-2}^{(h_2)}$ &
${\scriptstyle -2}$ &
${\scriptstyle -1-(h_1+h_2)}$ &
${\scriptstyle 1-(h_1+h_2)} $ &
$ {\cal O}_{-2}^{(h_1+h_2)}$
\\
\hline
${\scriptstyle h_1+2}$ &
$\bar\Phi_{-2}^{(h_1)}$&
${\scriptstyle -\frac{1}{2}-h_2}$ &
${\scriptstyle 1-h_2}$ &
${\cal O}_{A,-\frac{3}{2}}^{(h_2)}$ &
${\scriptstyle -2}$ &
${\scriptstyle -\frac{1}{2}-(h_1+h_2)}$ &
${\scriptstyle 1-(h_1+h_2)} $ &
${\cal O}_{A,-\frac{3}{2}}^{(h_1+h_2)}$
\\
\hline
${\scriptstyle h_1+2}$ &
$\bar\Phi_{-2}^{(h_1)}$&
${\scriptstyle -h_2}$ &
${\scriptstyle 1-h_2}$ &
${\cal O}_{AB,-1}^{(h_2)}$ &
${\scriptstyle -2}$ &
${\scriptstyle -(h_1+h_2)}$ &
${\scriptstyle 1-(h_1+h_2)} $ &
${\cal O}_{AB,-1}^{(h_1+h_2)}$
\\
\hline
${\scriptstyle h_1+2}$ &
$\bar\Phi_{-2}^{(h_1)}$&
${\scriptstyle \frac{1}{2}-h_2}$ &
${\scriptstyle 1-h_2}$ &
${\cal O}_{ABC,-\frac{1}{2}}^{(h_2)}$ &
${\scriptstyle -2}$ &
${\scriptstyle \frac{1}{2}-(h_1+h_2)}$ &
${\scriptstyle 1-(h_1+h_2)} $ &
${\cal O}_{ABC,-\frac{1}{2}}^{(h_1+h_2)}$
\\
\hline
${\scriptstyle h_1+2}$ &
$\bar\Phi_{-2}^{(h_1)}$&
${\scriptstyle 1-h_2}$ &
${\scriptstyle 1-h_2}$ &
${\cal O}_{0}^{(h_2),ABCD}$ &
${\scriptstyle -2}$ &
${\scriptstyle 1-(h_1+h_2)}$ &
${\scriptstyle 1-(h_1+h_2)} $ &
${\cal O}_{0}^{(h_1+h_2),ABCD}$
\\
\hline
${\scriptstyle h_1+2}$ &
$\bar\Phi_{-2}^{(h_1)}$&
${\scriptstyle \frac{3}{2}-h_2}$ &
${\scriptstyle 1-h_2}$ &
${\cal O}_{+\frac{1}{2}}^{(h_2),ABC}$ &
${\scriptstyle -2}$ &
${\scriptstyle \frac{3}{2}-(h_1+h_2)}$ &
${\scriptstyle 1-(h_1+h_2)} $ &
${\cal O}_{+\frac{1}{2}}^{(h_1+h_2),ABC}$
\\
\hline
${\scriptstyle h_1+2}$ &
$\bar\Phi_{-2}^{(h_1)}$&
${\scriptstyle 2-h_2}$ &
${\scriptstyle 1-h_2}$ &
${\cal O}_{+1}^{(h_2),AB}$ &
${\scriptstyle -2}$ &
${\scriptstyle 2-(h_1+h_2)}$ &
${\scriptstyle 1-(h_1+h_2)} $ &
${\cal O}_{+1}^{(h_1+h_2),AB}$
\\
\hline
${\scriptstyle h_1+2}$ &
$\bar\Phi_{-2}^{(h_1)}$&
${\scriptstyle \frac{5}{2}-h_2}$ &
${\scriptstyle 1-h_2}$ &
${\cal O}_{+\frac{3}{2}}^{(h_2),A}$ &
${\scriptstyle -2}$ &
${\scriptstyle \frac{5}{2}-(h_1+h_2)}$ &
${\scriptstyle 1-(h_1+h_2)} $ &
${\cal O}_{+\frac{3}{2}}^{(h_1+h_2),A}$
\\
\hline
${\scriptstyle h_1+2}$ &
$\bar\Phi_{-2}^{(h_1)}$&
${\scriptstyle 3-h_2}$ &
${\scriptstyle 1-h_2}$ &
${\cal O}_{+2}^{(h_2)}$ &
${\scriptstyle -2}$ &
${\scriptstyle 3-(h_1+h_2)}$ &
${\scriptstyle 1-(h_1+h_2)} $ &
${\cal O}_{+2}^{(h_1+h_2)}$
\\
\hline
\hline
${\scriptstyle h_1+\frac{3}{2}}$ &
$\bar{\Phi}_{A,-\frac{3}{2}}^{(h_1)}$&
${\scriptstyle -\frac{1}{2}-h_2}$ &
${\scriptstyle 1-h_2}$ &
${\cal O}_{B,-\frac{3}{2}}^{(h_2)}$ &
${\scriptstyle -\frac{3}{2}}$ &
${\scriptstyle -(h_1+h_2)}$ &
${\scriptstyle 1-(h_1+h_2)} $ &
${\cal O}_{AB,-1}^{(h_1+h_2)}$
\\
\hline
${\scriptstyle h_1+\frac{3}{2}}$ &
$\bar{\Phi}_{A,-\frac{3}{2}}^{(h_1)}$&
${\scriptstyle -h_2}$ &
${\scriptstyle 1-h_2}$ &
${\cal O}_{BC,-1}^{(h_2)}$ &
${\scriptstyle -\frac{3}{2}}$ &
${\scriptstyle \frac{1}{2}-(h_1+h_2)}$ &
${\scriptstyle 1-(h_1+h_2)} $ &
${\cal O}_{ABC,-\frac{1}{2}}^{(h_1+h_2)}$
\\
\hline
${\scriptstyle h_1+\frac{3}{2}}$ &
$\bar{\Phi}_{A,-\frac{3}{2}}^{(h_1)}$&
${\scriptstyle \frac{1}{2}-h_2}$ &
${\scriptstyle 1-h_2}$ &
${\cal O}_{BCD,-\frac{1}{2}}^{(h_2)}$ &
${\scriptstyle -\frac{3}{2}}$ &
${\scriptstyle 1-(h_1+h_2)}$ &
${\scriptstyle 1-(h_1+h_2)} $ &
${\cal O}_{0}^{(h_1+h_2),ABCD}$
\\
\hline
${\scriptstyle h_1+\frac{3}{2}}$ &
$\bar{\Phi}_{A,-\frac{3}{2}}^{(h_1)}$&
${\scriptstyle 1-h_2}$ &
${\scriptstyle 1-h_2}$ &
${\cal O}_{0}^{(h_2),BCDE}$ &
${\scriptstyle -\frac{3}{2}}$ &
${\scriptstyle \frac{3}{2}-(h_1+h_2)}$ &
${\scriptstyle 1-(h_1+h_2)} $ &
${\cal O}_{+\frac{1}{2}}^{(h_1+h_2),FGH}$
\\
\hline
${\scriptstyle h_1+\frac{3}{2}}$ &
$\bar{\Phi}_{A,-\frac{3}{2}}^{(h_1)}$&
${\scriptstyle \frac{3}{2}-h_2}$ &
${\scriptstyle 1-h_2}$ &
${\cal O}_{+\frac{1}{2}}^{(h_2),BCD}$ &
${\scriptstyle -\frac{3}{2}}$ &
${\scriptstyle 2-(h_1+h_2)}$ &
${\scriptstyle 1-(h_1+h_2)} $ &
${\cal O}_{+1}^{(h_1+h_2),CD} +\cdots$
\\
\hline
${\scriptstyle h_1+\frac{3}{2}}$ &
$\bar{\Phi}_{A,-\frac{3}{2}}^{(h_1)}$&
${\scriptstyle 2-h_2}$ &
${\scriptstyle 1-h_2}$ &
${\cal O}_{+1}^{(h_2),BC}$ &
${\scriptstyle -\frac{3}{2}}$ &
${\scriptstyle \frac{5}{2}-(h_1+h_2)}$ &
${\scriptstyle 1-(h_1+h_2)} $ &
${\cal O}_{+\frac{3}{2}}^{(h_1+h_2),C}+\cdots$
\\
\hline
${\scriptstyle h_1+\frac{3}{2}}$ &
$\bar{\Phi}_{A,-\frac{3}{2}}^{(h_1)}$&
${\scriptstyle \frac{5}{2}-h_2}$ &
${\scriptstyle 1-h_2}$ &
${\cal O}_{+\frac{3}{2}}^{(h_2),B}$ &
${\scriptstyle -\frac{3}{2}}$ &
${\scriptstyle 3-(h_1+h_2)}$ &
${\scriptstyle 1-(h_1+h_2)} $ &
${\cal O}_{+2}^{(h_1+h_2)}$
\\
\hline
\hline
\end{tabular}
\caption{
  The left-weights $q$ (and helicities $s_1$) for the
${\cal N}=8$ ${\rm \bar{w}}_{1+\infty}$
currents,
the left- and right-weights
$(h, \bar{h})$ for the matter fields on the left-hand sides
and those $(h+2-q, \bar{h}-s_1-q)$
on the right-hand sides in (\ref{otherOPEwO}).
We list these weights for the 
first sixteen equations corresponding (\ref{otherOPEwO})
in the section \ref{holomorphic}.
As in Table \ref{table1}, we can identify the matter fields
appearing on the last columns from their weights
up to the $SU(8)$ indices completely.}
\label{table1-1}
\end{table}

\begin{table}[tbp]
\centering
\renewcommand{\arraystretch}{1.7}
\begin{tabular}{|c|c||c|c|c|||c|c|c|c|| }
\hline
$ q$ &
$ {\rm {\bar w}^q}$ &
$ h$ &
$ \bar{h}$ &
$ {\cal O}_{h,\bar{h}}$ &
$ s_1$ &
${\scriptstyle h+2-q }$ &
${\scriptstyle \bar{h}-s_1-q }$ &
${\cal O}_{h+2-q,\bar{h}-s_1-q}$  
\\
\hline
\hline
${\scriptstyle h_1+1}$ &
$\bar{\Phi}_{AB,-1}^{(h_1)}$&
${\scriptstyle -h_2}$ &
${\scriptstyle 1-h_2}$ &
${\cal O}_{CD,-1}^{(h_2)}$ &
${\scriptstyle -1}$ &
${\scriptstyle 1-(h_1+h_2)}$ &
${\scriptstyle 1-(h_1+h_2)} $ &
${\cal O}_{0}^{(h_1+h_2),ABCD}$
\\
\hline
${\scriptstyle h_1+1}$ &
$  \bar{\Phi}_{AB,-1}^{(h_1)}$ &
${\scriptstyle \frac{1}{2}-h_2}$ &
${\scriptstyle 1-h_2}$ &
$ {\cal O}_{CDE,-\frac{1}{2}}^{(h_2)}$ &
${\scriptstyle -1}$ &
${\scriptstyle \frac{3}{2}-(h_1+h_2)}$ &
${\scriptstyle 1-(h_1+h_2)} $ &
$ {\cal O}_{+\frac{1}{2}}^{(h_1+h_2),FGH}$
\\
\hline
${\scriptstyle h_1+1}$ &
$\bar{\Phi}_{AB,-1}^{(h_1)}$&
${\scriptstyle 1-h_2}$ &
${\scriptstyle 1-h_2}$ &
${\cal O}_{0}^{(h_2),CDEF}$ &
${\scriptstyle -1}$ &
${\scriptstyle 2-(h_1+h_2)}$ &
${\scriptstyle 1-(h_1+h_2)} $ &
${\cal O}_{+1}^{(h_1+h_2),GH}$
\\
\hline
${\scriptstyle h_1+1}$ &
$\bar{\Phi}_{AB,-1}^{(h_1)}$&
${\scriptstyle \frac{3}{2}-h_2}$ &
${\scriptstyle 1-h_2}$ &
${\cal O}_{+\frac{1}{2}}^{(h_2),CDE}$ &
${\scriptstyle -1}$ &
${\scriptstyle \frac{5}{2}-(h_1+h_2)}$ &
${\scriptstyle 1-(h_1+h_2)} $ &
${\cal O}_{+\frac{3}{2}}^{(h_1+h_2),D}+\cdots$
\\
\hline
${\scriptstyle h_1+1}$ &
$\bar{\Phi}_{AB,-1}^{(h_1)}$&
${\scriptstyle 2-h_2}$ &
${\scriptstyle 1-h_2}$ &
${\cal O}_{+1}^{(h_2),CD}$ &
${\scriptstyle -1}$ &
${\scriptstyle 3-(h_1+h_2)}$ &
${\scriptstyle 1-(h_1+h_2)} $ &
${\cal O}_{+2}^{(h_1+h_2)}$
\\
\hline
\hline
${\scriptstyle h_1+\frac{1}{2}}$ &
$\bar{\Phi}_{ABC,-\frac{1}{2}}^{(h_1)}$&
${\scriptstyle \frac{1}{2}-h_2}$ &
${\scriptstyle 1-h_2}$ &
${\cal O}_{DEF,-\frac{1}{2}}^{(h_2)}$ &
${\scriptstyle -\frac{1}{2}}$ &
${\scriptstyle 2-(h_1+h_2)}$ &
${\scriptstyle 1-(h_1+h_2)} $ &
${\cal O}_{+1}^{(h_1+h_2),GH}$
\\
\hline
${\scriptstyle h_1+\frac{1}{2}}$ &
$\bar{\Phi}_{ABC,-\frac{1}{2}}^{(h_1)}$&
${\scriptstyle 1-h_2}$ &
${\scriptstyle 1-h_2}$ &
${\cal O}_{0}^{(h_2),DEFG}$ &
${\scriptstyle -\frac{1}{2}}$ &
${\scriptstyle \frac{5}{2}-(h_1+h_2)}$ &
${\scriptstyle 1-(h_1+h_2)} $ &
${\cal O}_{+\frac{3}{2}}^{(h_1+h_2),H}$
\\
\hline
${\scriptstyle h_1+\frac{1}{2}}$ &
$\bar{\Phi}_{ABC,-\frac{1}{2}}^{(h_1)}$&
${\scriptstyle \frac{3}{2}-h_2}$ &
${\scriptstyle 1-h_2}$ &
${\cal O}_{+\frac{1}{2}}^{(h_2),DEF}$ &
${\scriptstyle -\frac{1}{2}}$ &
${\scriptstyle 3-(h_1+h_2)}$ &
${\scriptstyle 1-(h_1+h_2)} $ &
${\cal O}_{+2}^{(h_1+h_2)}$
\\
\hline
\hline
${\scriptstyle h_1}$ &
$\bar{\Phi}_{0}^{(h_1),ABCD}$&
${\scriptstyle 1-h_2}$ &
${\scriptstyle 1-h_2}$ &
${\cal O}_{0}^{(h_2),EFGH}$ &
${\scriptstyle 0}$ &
${\scriptstyle 3-(h_1+h_2)}$ &
${\scriptstyle 1-(h_1+h_2)} $ &
${\cal O}_{+2}^{(h_1+h_2)}$
\\
\hline
\hline
\end{tabular}
\caption{This is a continuation of previous Table \ref{table1-1}.
We list the corresponding weights (and helicities) for  
the remaining nine equations among twenty-five
equations (\ref{otherOPEwO}) in the section \ref{holomorphic}. }
\label{table2-1}
\end{table}

In Appendix E, the forty-five OPEs are given explicitly.
The twenty-five OPEs with $s_1 \leq s_2$ are given in (\ref{25OPEsholo})
while the twenty OPEs with $s_1 > s_2$
are given in (\ref{20OPEsholo}).

Now we are ready to reconsider
the analysis given in the footnote \ref{superBMS}. For $h_1=0$ in the current
$\bar{\Phi}_{-2}^{(h_1)}(z_1,\bar{z}_1)$, the corresponding current
$\bar{\Phi}_{-2}^{(0)}(z_1,\bar{z}_1)$ plays the role of stress energy tensor
in the holomorphic sector. The left- and right-weights
of this current are given by
$(q=2,\bar{h}=1)$.
The right-hand side of the ninth equation of (\ref{25OPEsholo}),
when we take consider the holomorphic light transform,
consists of
two terms
\bea
 \bar{\Phi}_{-2}^{(0)}(z,\bar{z}) \,
{\cal O}_{+2}^{(h_2)}(0,0)
=  \frac{1}{\bar{z}}\bigg[ 
\frac{1}{{z}^2}\, (3-h_2)
+ \frac{1}{{z}}\, {\pa} \bigg]
{\cal O}_{+2}^{(h_2)}(0,0)
+  \cdots,
\label{otherPhiO}
\eea
where $(3-h_2)$ at the second order pole in the holomorphic
coordinates is the left-weight of
the matter field ${\cal O}_{+2}^{(h_2)}(0,0)$
according to Table \ref{table1-1}
(or Table \ref{table1} with proper exchanges for the helicities
and the left- and right-weights).
Of course, we also understand
the above result of right-hand side in (\ref{otherPhiO}) from
the ninth equation of (\ref{25OPEs}) by taking
$s_i \rightarrow -s_i$
($i=1,2,3$) and $z \leftrightarrow \bar{z}$
with proper assignment of the weights.
See also (\ref{twovaluesofq}) for the structure of the complex
coordinates.

We can expand $ \bar{\Phi}_{-2}^{(0)}(z,\bar{z})$
in terms of the antiholomorphic coordinate $\bar{z}$.
Each coefficient operator denoted by $\bar{\Phi}_{-2}^{(0)}({z})$
depends on the holomorphic
coordinate ${z}$.
See also (\ref{otherwmode}) which contains $\frac{1}{\bar{z}}$
factor.
We take the contour integral over $\bar{z}$
on both sides of (\ref{otherPhiO}).
We obtain the following OPE of the (antiholomorphic) zero mode
of the current with the matter field
\bea
\bar{\Phi}_{-2}^{(0)}({z}) \,
{\cal O}_{+2}^{(h_2)}(0,0)
= \bigg[ 
\frac{1}{{z}^2}\, (3-h_2)
+ \frac{1}{{z}}\, {\pa} \bigg]
{\cal O}_{+2}^{(h_2)}(0,0)
+  \cdots.
\label{otherprimary}
\eea
We can check other eight matter fields
(corresponding to the negative graviton, the gravitinos,
the graviphotons, the graviphotinos and the
scalars) in (\ref{otherwO}) by using the corresponding equations
which are the first to the eighth relations in (\ref{25OPEsholo}).
They satisfy their primary conditions in (\ref{otherprimary})
with their own left-weights given in Table \ref{table1-1}
\footnote{
\label{Tz}
In the notation of \cite{FSTZ},
$\bar{\Phi}_{-2}^{(0)}({z})=T(z)$ which is the zero mode of $SL(2,\mathbb{R})_R$.}.

On the other hand, for $h_1=-\frac{1}{2}$, we have
the following OPE from the ninth equation of (\ref{25OPEsholo})
\bea
\bar{\Phi}_{-2}^{(-\frac{1}{2})}(z,\bar{z}) \,
{\cal O}_{+2}^{(h_2)}(0,0)
=\frac{1}{z\, \bar{z}} \, \frac{1}{2}\,
{\cal O}_{+2}^{(-\frac{1}{2}+h_2)}(0,0)
+  \cdots.
\label{othertranslation}
\eea
The left- and right-weights of the matter field
${\cal O}_{+2}^{(h_2)}(0,0)$ on the left-hand side,
$(3-h_2)$ and $(1-h_2)$
are increased by $(\frac{1}{2}$, $\frac{1}{2})$ respectively
in the matter field ${\cal O}_{+2}^{(-\frac{1}{2}+h_2)}(0,0)$
on the right-hand side, according to Table \ref{table1-1}
with proper shifts on the helicities and the weights.
Note that $\bar{\Phi}_{-2}^{(-\frac{1}{2})}(z,\bar{z})$
has the left- and right-weights $(q=\frac{3}{2},\bar{h}=\frac{3}{2})$
and is identified with the supertranslation generator \cite{HPS}.
As we did before in the previous paragraph,
other eight matter fields (the negative graviton, the gravitinos,
the graviphotons, the graviphotinos and the
scalars) in (\ref{otherwO}) can be analyzed similarly and obtain
the corresponding relations like as (\ref{othertranslation})
\footnote{
\label{sametranslation}
That is, there are similar equations to
(\ref{transla}) where the modes
${\Phi}_{+2,m,n}^{(-\frac{1}{2})}$ are replaced by
$\bar{\Phi}_{-2,m,n}^{(-\frac{1}{2})}$. Their actions to the matter fields
are exactly the same. See also (\ref{translation}).
}.

Let us focus on 
the sixteenth relation of (\ref{25OPEsholo}).
The ${\cal N}=8$ ${\rm \bar{w}}_{1+\infty}$
current $\bar\Phi_{A,-\frac{3}{2}}^{(h_1=0)}$ has the following
OPE
\bea
\bar{\Phi}_{A,-\frac{3}{2}}^{(0)}(z,\bar{z}) 
{\cal O}_{+\frac{3}{2}}^{(h_2),B}(0,0)= \frac{1}{z \,\bar{z} }
\,\frac{1}{2} \, \de^{B}_{A}\, {\cal O}_{+2}^{(h_2)}(0,0)
+  \cdots.
\label{sixteenth}
\eea
As described in the case of stress energy tensor,
the contour integral over $\bar{z}$ of both sides of (\ref{sixteenth})
can be applied together with the mode expansion of
$\bar{\Phi}_{A,-\frac{3}{2}}^{(0)}(z,\bar{z})$ in terms of
antiholomorphic coordinate.
This leads to the OPE of the (antiholomorphic) zero mode
of the current with the matter field
\bea
\bar{\Phi}_{A,-\frac{3}{2}}^{(0)}({z}) 
{\cal O}_{+\frac{3}{2}}^{(h_2),B}(0,0)= \frac{1}{{z} }
\, \frac{1}{2} \, \de^{B}_{A}\, {\cal O}_{+2}^{(h_2)}(0,0)
+  \cdots.
\label{simplesixteenth}
\eea
The left-weight of the left-hand side of
(\ref{simplesixteenth}) is given by $\frac{3}{2}+(\frac{5}{2}-h_2)$
which can be written as $1+(3-h_2)$ which is the left-weight
of the right-hand side of (\ref{simplesixteenth}).
The right-weight of the left-hand side
is $0+(1-h_2)$
which is equal to $(1-h_2)$ (Note that the right-weight of
$\bar{\Phi}_{A,-\frac{3}{2}}^{(0)}(z,\bar{z})$, $1$, becomes
zero for the right-weight of $\bar{\Phi}_{A,-\frac{3}{2}}^{(0)}(z)$
after the contour integration over $\bar{z}$).
This is the right-weight of the right-hand side.
That is, the sum of left- and right-weights
of the matter field on the left-hand side,
$(\frac{7}{2}-2h_2)$, is transformed into
$(4-2h_2)$ of the matter field on the right-hand side from Table
\ref{table1-1} (i.e., $\De \rightarrow \De +\frac{1}{2}$).
Or the left-weight of the matter field
is increased by $\frac{1}{2}$.
Therefore, 
the ${\cal N}=8$ ${\rm \bar{w}}_{1+\infty}$
current $\bar{\Phi}_{A,-\frac{3}{2}}^{(0)}({z})$
with weights $(\frac{3}{2},0)$
is other supercurrent of ${\cal N}=8$ BMS algebra
\cite{2212-2,Tropper}
by recognizing (\ref{simplesixteenth}).
The behavior of this
supercurrent in the context of shadow transform
for the currents will be described in section \ref{lastsection}
\footnote{
In the notation of \cite{FSTZ,2212-2},
this current corresponds to ${S}({z})$ or
${S}_A({z})$. Under the action of
the supercurrent, the matter field
${\cal O}^{(h_2)}_{-2}(z,\bar{z})$ is changed into
${\cal O}^{(h_2)}_{A,-\frac{3}{2}}(z,\bar{z})$ from
the first equation of (\ref{20OPEsholo}).
Similarly, there are
changes of the matter fields 
${\cal O}^{(h_2)}_{B,-\frac{3}{2}}(z,\bar{z}) \rightarrow 
{\cal O}^{(h_2)}_{AB,-1}(z,\bar{z}), \cdots,
{\cal O}^{(h_2),B}_{+\frac{3}{2}}(z,\bar{z}) \rightarrow
\de^B_{A}\,
{\cal O}^{(h_2)}_{+2}(z,\bar{z})$.
We observe that  each helicity of the matter field
is increased by $\frac{1}{2}$: $s_2 \rightarrow s_2+\frac{1}{2}$.
This implies that
under this supercurrent, the right-weights (or $\frac{(\De-s_2)}{2}$)
of the matter fields
remain the same while the left-weights (or $\frac{(\De+s_2)}{2}$)
are increased by $\frac{1}{2}$.}.

We can analyze from the eleventh equation to the sixteenth
equation of (\ref{25OPEsholo}) similarly.
By comparing the third column with the seventh coulumn
in Table \ref{table1-1} for fix $h_1=0$,
the left-weights are increased by $\frac{1}{2}$.
Equivalently, the sum of left- and right-weights
of the matter fields is increased by $\frac{1}{2}$
because the right-weights for the matter fields in
both sides are the same.
For the three negative helicities of the matter fields on the
left-hand sides, each $SU(8)$ index
on the left-hand sides appears as the product
of $SU(8)$ indices of the matter fields on the right-hand sides.
For the remaining four non-positive helicities of the matter fields on the
left-hand sides, the $SU(8)$ invariant tensors $\ep^{ABCDEFGH}$ or
$\de^A_B$ appear on the right-hand sides.


\subsection{The action of charges on matter fields}

In the mode expansion of ${\rm \bar{w}}^{q}(z,\bar{z})$
where the left- and right-weights are
given by $(h,\bar{h})=(q,-s_1+1-q)$, the following relation holds
\bea
{\rm \bar{w}}^q(z, \bar{z}) =\sum_{m,n} \,
\frac{{\rm \bar{w}}_{m,n}^q}{z^{q+m} \, \bar{z}^{-s_1+1-q+n}}.
\label{otherwmode}
\eea
Each mode ${\rm \bar{w}}^q_{m,n}$ can be extracted by performing 
the contour integrals over $z$ and $\bar{z}$ on
both sides of (\ref{otherwmode}).
Let us introduce the following (anti)commutators
\bea
\bigg[
{\rm \bar{w}}^q_{m,n},
{\cal O}_{h,\bar{h}}(z,\bar{z})   \bigg\}
=\oint_z \, \frac{d w}{2 \pi i} \, w^{q-1+m} \, \oint_{\bar{z}} \,
\frac{d \bar{w}}{2 \pi i}\, \bar{w}^{-s_1-q+n}\, {\rm \bar{w}}^q(w,\bar{w})\,
{\cal O}_{h,\bar{h}}(z,\bar{z}).
\label{otherwOdef}
\eea
The OPE of the ${\rm \bar{w}}_{1+\infty}$ currents with the matter fields
appearing on the right-hand side of (\ref{otherwOdef})
is described in (\ref{otherOPEwO}).
The left-weight of the right-hand side of (\ref{otherwOdef})
is given by $-1-(q-1+m)+q+h=-m+h$ which is the left-weight of left-hand
side. The right-weight of right-hand side is given by
$-1-(-s_1-q+n)+(-s_1+1-q)+\bar{h}=-n+\bar{h}$ which is the right-weight
of the left-hand side.

For $n=q+s_1$, we observe that there exists only $\frac{1}{\bar{z}}$ term
in the antiholomorphic coordinate (\ref{otherwmode}).
Moreover, the contour integral over $\bar{w}$ coordinate in (\ref{otherwOdef})
becomes unity.
After integrating over ${w}$ in (\ref{otherwOdef})
and using the change of variables, finally we obtain
\footnote{We are using the hatted notation $\hat{}$ for the modes
of the currents in the
(anti)commutators as in the footnote \ref{hat}.}
\bea
\bigg[
\hat{ \bar{{\rm w}}}^q_m,
{\cal O}_{h,\bar{h}}(z,\bar{z})   \bigg\}  & = &
\bar{\mathscr D}_m^{q}({h}) \, {\cal O}_{h+2-q,\bar{h}-s_1-q}(z,\bar{z}),
\qquad \hat{\bar{{\rm w}}}^q_m \equiv {\rm w}^q_{m,q+s_1},
\label{otheranticomm}
\eea
where the differential operator  appearing on the
right-hand side of (\ref{otheranticomm}) is given by
\bea
\bar{\mathscr D}_m^{q}({h}) \equiv \frac{1}{2}\, \sum_{l=0}^{2q-3}\,
\left(\begin{array}{c}
q+m-1 \\  l \\
\end{array}\right)    
\, \frac{(2q-2-l)\, \Gamma(2 {h}+1)}{\Gamma(2 {h}+1-l)} \,
{z}^{q+m-1-l} \, {\pa}_{{z}}^{2q-3-l}. 
\label{otherDiff}
\eea
Compared to the previous differential operator (\ref{Diff}),
this differential operator (\ref{otherDiff})
has the same form with the fact that the right-weight $ \bar{h}$ and
the complex coordinate $\bar{z}$
are replaced by the left-weight $h$ and the
complex coordinate $z$ respectively.
In this case, the left-weight of the right-hand side of
(\ref{otheranticomm}) is given by
$-(q+m-1-l)+(2q-3-l)+h+2-q=-m+h$ which is the left-weight of the
left-hand side while the right-weight of the right-hand side
is given by $\bar{h}-s_1-q=-(q+s_1)+\bar{h}$ which is the right-weight
of the left-hand side.
Note that the right-weight of $\hat{ \bar{{\rm w}}}^q_m$ is
$-(q+s_1)$ in (\ref{otheranticomm}).

Therefore, we obtain the twenty-five (anti)commutators in 
(\ref{25anticommsection5}) corresponding to (\ref{25OPEsholo})
for the case $s_1 \leq s_2$ 
by i) substituting the left-weights
$q$ and ${h}$ on the left-hand sides from Tables \ref{table9-1}
and \ref{table10-1} into the differential operators (\ref{otherDiff})
on the right-hand sides,
ii) identifying the ${\rm {\bar w}}_{1+\infty}$
currents with the hatted notation
and the two kinds of matter fields correctly and iii)
putting the $SU(8)$ indices together with
the matter fields on the right-hand sides.  
Similarly,
there are the remaining twenty (anti)commutators in 
(\ref{20anticommsection5}) corresponding to (\ref{20OPEsholo})
for the case $s_1 > s_2$. 

We can analyze the description given in the footnote \ref{q2and3half}.
For $q=2$, we can check the following
result corresponding to (\ref{otherPhiO})
and (\ref{otherprimary})
(from the ninth equation of (\ref{25OPEs})
with the replacements $s_i \rightarrow -s_i$, or
from the ninth equation of (\ref{25anticommsection5}) directly)
\bea
\bigg[
\hat{\bar{\Phi}}_{-2,m}^{(0)},
{\cal O}_{+2}^{(h_2)}(z,\bar{z}) \bigg]  & = &
{z}^m \bigg( (m+1)\, (3-h_2)+ {z}\, {\pa}_z
\bigg) \, {\cal O}_{+2}^{(h_2)}(z,\bar{z}).
\label{pricondition}
\eea
Of course, the expression $(3-h_2)$ in (\ref{pricondition})
is the left-weight
of the matter field on the left-hand side.
We can write down the similar expressions for other eight
kinds of matter fields in (\ref{otherwO}).
This implies that the mode $\hat{\bar{\Phi}}_{-2,m}^{(0)}$
plays the role of the 
Virasoro generators.
The other generators are given by
the mode $\hat{{\Phi}}_{+2,m}^{(0)}$
in the footnote \ref{q2and3half} \footnote{In the notation of
  \cite{FSTZ},  $\hat{\bar{\Phi}}_{-2,m}^{(0)} \equiv L_m$
  which is the mode of the left Virasoro generator
  $T(z)$ appearing in the footnote \ref{Tz}.}.

For $q=\frac{3}{2}$, the following result
corresponding to (\ref{othertranslation}) (or from the ninth equation of
(\ref{25anticommsection5}))
holds
\bea
\bigg[
\bar{\Phi}_{-2,m,n}^{(-\frac{1}{2})},
{\cal O}_{+2}^{(h_2)}(z,\bar{z}) \bigg]  & = &
\frac{1}{2}\, z^{m+\frac{1}{2}}\,
\bar{z}^{n+\frac{1}{2}}\,
{\cal O}_{+2}^{(-\frac{1}{2}+h_2)}(z,\bar{z}), \nonu \\
\bigg[
\hat{\bar{\Phi}}_{-2,m}^{(-\frac{1}{2})},
{\cal O}_{+2}^{(h_2)}(z,\bar{z}) \bigg]  & = & 
\frac{1}{2}\, {z}^{m+\frac{1}{2}}\,
{\cal O}_{+2}^{(-\frac{1}{2}+h_2)}(z,\bar{z}).
\label{other3half}
\eea
Compared to the two expressions appearing in the
footnote \ref{q2and3half},
the notation for the matter fields are the same but
the left- and right-weights are different from each other
\footnote{In other words, the left- and right-weights
of ${\cal O}_{+2}^{(h_2)}(z,\bar{z})$ appearing in Tables \ref{table1}
and \ref{table1-1} are different from each other.
We can identify $\hat{\bar{\Phi}}_{-2,m}^{(-\frac{1}{2})}$ with
$P_{m,-\frac{1}{2}}$ \cite{FSTZ}.  
See also the footnote \ref{TWOmatters}.}.
When we exchanges the helicities in 
those two equations (\ref{q3halftwocases})
in the footnote \ref{q2and3half},
then we obtain the different matter fields.
It is obvious to see that the right-hand side of second equation
in (\ref{other3half}) contains the holomorphic complex coordinate
$z$ dependence from the analysis of (\ref{otheranticomm}).
When we fix $n=-\frac{1}{2}$ in the first equation of (\ref{other3half}),
then this becomes the second equation of (\ref{other3half}).

We can write down the anticommutator corresponding to
(\ref{simplesixteenth}) as follows:
\bea
\bigg\{
\hat{\bar{\Phi}}_{A,-\frac{3}{2},m}^{(0)},
{\cal O}_{+\frac{3}{2}}^{(h_2),B}(z,\bar{z}) \bigg\}   = 
\frac{1}{2}\, {z}^{m+\frac{1}{2}}\,
\de^{B}_{A}\,
{\cal O}_{+2}^{(h_2)}(z,\bar{z}).
\label{two3half}
\eea
From Table \ref{table1-1}, the left- and right-weights
of the matter field with helicity $+\frac{3}{2}$
are given by $(\frac{5}{2}-h_2,1-h_2)$ while
those for the matter field with helicity $+2$
are given by $(3-h_2, 1-h_2)$.
The sum of these weights,
$(\frac{7}{2}-2h_2)$ on the left-hand side
is changed into $(4-2h_2)$ on the right-hand side. 
This behavior satisfies for the tenth to the fifteenth cases
of Table \ref{table1-1}
by summing over i) the third and the fourth columns and ii)
the seventh and the eighth and taking their differences
\footnote{This holds for the first row of Table \ref{table9-1}.
For the commutator $\bigg[
\hat{\bar{\Phi}}_{A,-\frac{3}{2},m}^{(0)},
{\cal O}_{+2}^{(h_2)}(z,\bar{z}) \bigg]$ where the matter field
with helicity $+2$ has the left- and right-weights
$(3-h_2,1-h_2)$, there is no matter field
due to the fact that $s_3 \neq s_1+s_2+2=+\frac{5}{2}$.
According to Table \ref{table1-1}, the above combination
$h+\bar{h}$ for the helicity $+2$ is the maximum value
and there is no room for the matter field having $(h+\bar{h})=
\frac{9}{2}-2h_2$. Here $\frac{9}{2}$ is the numerical value
of the sum of the left-weights of this commutator.

As in the footnote \ref{q2and3half}, the following relations
hold for the actions on the primary operators
\bea
\bigg[
\hat{\bar{\Phi}}_{-2,m}^{(0)},
{\cal O}_{+2}^{(h)}(z,\bar{z}) \bigg]  & = &
{z}^m \bigg( (m+1)\, (3-h)+ {z}\, {\pa}
\bigg) \, {\cal O}_{+2}^{(h)}(z,\bar{z}),
\nonu \\
\bigg[
\hat{\bar{\Phi}}_{-2,m}^{(0)},
{\cal O}_{+\frac{3}{2}}^{(h),A}(z,\bar{z}) \bigg]  & = &
{z}^m \bigg( (m+1)\, (\tfrac{5}{2}-h)+ {z}\, {\pa}
\bigg) \, {\cal O}_{+\frac{3}{2}}^{(h),A}(z,\bar{z}),
\nonu \\
\bigg[
\hat{\bar{\Phi}}_{-2,m}^{(0)},
{\cal O}_{+1}^{(h),AB}(z,\bar{z}) \bigg]  & = &
{z}^m \bigg( (m+1)\, (2-h)+ {z}\, {\pa}
\bigg) \, {\cal O}_{+1}^{(h),AB}(z,\bar{z}),
\nonu \\
\bigg[
\hat{\bar{\Phi}}_{-2,m}^{(0)},
{\cal O}_{+\frac{1}{2}}^{(h),ABC}(z,\bar{z}) \bigg]  & = &
{z}^m \bigg( (m+1)\, (\tfrac{3}{2}-h)+ {z}\, {\pa}
\bigg) \, {\cal O}_{+\frac{1}{2}}^{(h),ABC}(z,\bar{z}),
\nonu \\
\bigg[
\hat{\bar{\Phi}}_{-2,m}^{(0)},
{\cal O}_{0}^{(h),ABCD}(z,\bar{z}) \bigg]  & = &
{z}^m \bigg( (m+1)\, (1-h)+ {z}\, {\pa}
\bigg) \, {\cal O}_{0}^{(h),ABCD}(z,\bar{z}),
\nonu \\
\bigg[
\hat{\bar{\Phi}}_{-2,m}^{(0)},
{\cal O}_{ABC,-\frac{1}{2}}^{(h)}(z,\bar{z}) \bigg]  & = &
{z}^m \bigg( (m+1)\, (\tfrac{1}{2}-h)+ {z}\, {\pa}
\bigg) \, {\cal O}_{ABC,-\frac{1}{2}}^{(h)}(z,\bar{z}),
\nonu \\
\bigg[
\hat{\bar{\Phi}}_{-2,m}^{(0)},
{\cal O}_{AB,-1}^{(h)}(z,\bar{z}) \bigg]  & = &
{z}^m \bigg( (m+1)\, (-h)+ {z}\, {\pa}
\bigg) \, {\cal O}_{AB,-1}^{(h)}(z,\bar{z}),
\nonu \\
\bigg[
\hat{\bar{\Phi}}_{-2,m}^{(0)},
{\cal O}_{A,-\frac{3}{2}}^{(h)}(z,\bar{z}) \bigg]  & = &
{z}^m \bigg( (m+1)\, (-\tfrac{1}{2}-h)+ {z}\, {\pa}
\bigg) \, {\cal O}_{A,-\frac{3}{2}}^{(h)}(z,\bar{z}),
\nonu \\
\bigg[
\hat{\bar{\Phi}}_{-2,m}^{(0)},
{\cal O}_{-2}^{(h)}(z,\bar{z}) \bigg]  & = &
{z}^m \bigg( (m+1)\, (-1-h)+ {z}\, {\pa}
\bigg) \, {\cal O}_{-2}^{(h)}(z,\bar{z}).
\nonu
\eea
The left-weights from Table \ref{table1-1} are substituted properly.
We can check the left- and right-weights of both sides.
The weights of $\hat{\bar{\Phi}}_{-2,m}^{(0)}$
are given by $(-m,0)$.
}.

We also present the complete (anti)commutators
between the modes of
${\cal N}=8$ ${\rm \bar{w}}_{1+\infty}$ currents (i.e., supercurrents)
and the matter fields from (\ref{25anticommsection5})
(and (\ref{20anticommsection5}))
\bea
\bigg\{
\hat{\bar{\Phi}}_{A,-\frac{3}{2},m}^{(0)},
{\cal O}_{+\frac{3}{2}}^{(h),B}(z,\bar{z}) \bigg\} &  = & 
\frac{1}{2}\, {z}^{m+\frac{1}{2}}\,
\de^{B}_{A}\, {\cal O}_{+2}^{(h)}(z,\bar{z}),
\nonu \\
\bigg[
\hat{\bar{\Phi}}_{A,-\frac{3}{2},m}^{(0)},
{\cal O}_{+1}^{(h),BC}(z,\bar{z}) \bigg]  & =& 
\frac{1}{2}\, {z}^{m+\frac{1}{2}}\,
2! \, \de^{[B}_{A}\, {\cal O}_{+\frac{3}{2}}^{(h),C]}(z,\bar{z}),
\nonu \\
\bigg\{
\hat{\bar{\Phi}}_{A,-\frac{3}{2},m}^{(0)},
{\cal O}_{+\frac{1}{2}}^{(h),BCD}(z,\bar{z}) \bigg\}  & = &
\frac{1}{2}\, {z}^{m+\frac{1}{2}}\,
3 \, \de^{[B}_{A}\, 
{\cal O}_{+1}^{(h),CD]}(z,\bar{z}),
\nonu \\
\bigg[
\hat{\bar{\Phi}}_{A,-\frac{3}{2},m}^{(0)},
{\cal O}_{0}^{(h),BCDE}(z,\bar{z}) \bigg] &  = & 
\frac{1}{2}\, {z}^{m+\frac{1}{2}}\,
\frac{1}{4!} \, \frac{1}{3!} \, \ep^{BCDEIJKL}\, \ep_{AFGHIJKL}\,
{\cal O}_{+\frac{1}{2}}^{(h),FGH}(z,\bar{z}),
\nonu \\
\bigg\{
\hat{\bar{\Phi}}_{A,-\frac{3}{2},m}^{(0)},
{\cal O}_{BCD,-\frac{1}{2}}^{(h)}(z,\bar{z}) \bigg\} &  = &
\frac{1}{2}\, {z}^{m+\frac{1}{2}}\,
\frac{1}{4!} \, \ep_{ABCDEFGH} \, {\cal O}_{0}^{(h),EFGH}(z,\bar{z}),
\nonu \\
\bigg[
\hat{\bar{\Phi}}_{A,-\frac{3}{2},m}^{(0)},
{\cal O}_{BC,-1}^{(h)}(z,\bar{z}) \bigg]  & = & 
\frac{1}{2}\, {z}^{m+\frac{1}{2}}\,
{\cal O}_{ABC,-\frac{1}{2}}^{(h)}(z,\bar{z}),
\nonu \\
\bigg\{
\hat{\bar{\Phi}}_{A,-\frac{3}{2},m}^{(0)},
{\cal O}_{B,-\frac{3}{2}}^{(h)}(z,\bar{z}) \bigg\}  & = & 
\frac{1}{2}\, {z}^{m+\frac{1}{2}}\,
{\cal O}_{AB,-1}^{(h)}(z,\bar{z}),
\nonu \\
\bigg[
\hat{\bar{\Phi}}_{A,-\frac{3}{2},m}^{(0)},
{\cal O}_{-2}^{(h)}(z,\bar{z}) \bigg]  & =& 
\frac{1}{2}\, {z}^{m+\frac{1}{2}}\,
{\cal O}_{A,-\frac{3}{2}}^{(h)}(z,\bar{z}).
\label{supercur}
\eea
From Table \ref{table1-1},
the right-weights $\bar{h}$ in the fourth column
are the same as the right-weights
$(\bar{h}-s_1-q)$ in the eighth column
due to the fact $q=\frac{3}{2}=-s_1$.
On the other hand, the left-weights $h$
in the third column are increased by $\frac{1}{2}$
because the seventh column has $({h}+2-q)$
which is equal to $({h}+\frac{1}{2})$ for $q=\frac{3}{2}$.
Moreover, the antiholomorphic mode of
$\hat{\bar{\Phi}}_{A,-\frac{3}{2},m}^{(0)}$ is fixed by zero mode (from
$q+s_1=0$ in (\ref{otheranticomm})) in (\ref{supercur}).

\subsection{The action of
${\cal N}=8$ ${\rm \bar{w}}_{1+\infty}$ symmetry}

\subsubsection{ The ${\cal N}=8$ ${\rm \bar{w}}_{1+\infty}$ algebra}

For given (\ref{otheranticomm}),
we can perform the action of the modes of
${\cal N}=8$ ${\rm \bar{w}}_{1+\infty}$ currents
on both sides of  (\ref{otheranticomm}) and
we obtain 
\bea
\bigg[
{\rm \hat{\bar{w}}}^p_{m},
\bigg[{\rm \hat{\bar{w}}}^q_{n}, {\cal O}_{h,\bar{h}}(z,\bar{z}) \bigg\} \bigg\}
\mp \bigg[
{\rm \hat{\bar{w}}}^q_{n},
\bigg[{\rm \hat{\bar{w}}}^p_{m}, {\cal O}_{h,\bar{h}}(z,\bar{z}) \bigg\} \bigg\}
= \nonu \\ 
\bigg( (q-1)m-(p-1)n \bigg) \,
\bigg[{\rm \hat{\bar{w}}}^{p+q-2}_{m+n}, {\cal O}_{h,\bar{h}}(z,\bar{z}) \bigg\},
\label{wwOother}
\eea
where the following (anti)commutator relation between
the modes of the ${\cal N}=8$ $\bar{\rm w}_{1+\infty}$ currents
satisfies
\bea
\bigg[ {\rm \hat{\bar{w}}}_m^p, {\rm \hat{\bar{w}}}^q_n \bigg\}=
\bigg( (q-1)m-(p-1)n \bigg) \,
{\rm \hat{\bar{w}}}^{p+q-2}_{m+n}.
\label{wwcommother}
\eea
The right-weight of the right-hand side
of (\ref{wwcommother}) is given by
$-s_3-(p+q-2)=-s_1-s_2-2-(p+q-2)=-(s_1+p)-(s_2+q)$
which is the right-weight of the left-hand side.
The left-weight of the right-hand side
is given by $-(m+n)=-m-n$
which is the left-weight of the left-hand side.
The above relation (\ref{wwOother}) corresponds to the previous
relation in (\ref{wwO}) while
the above relation (\ref{wwcommother}) corresponds to the previous
relation in (\ref{wwcomm}).
In order to classify
all the forty-five cases with $s_1 \leq s_2$ for the left-hand side of
(\ref{wwOother}),
we can make some Tables corresponding to
Tables \ref{table3}, \ref{table4} and \ref{table5}
but we do not present them in this paper.

As in the previous section \ref{twocurrents},
we can further simplify (\ref{wwOother}) using the
relation (\ref{otheranticomm})
\bea
&&
\Bigg[
\bar{\la}_{s_1,s_2+s_3+2,-s_1-s_2-s_3-4}\, \bar{\la}_{s_2,s_3,-s_2-s_3-2} \,
\bar{\mathscr D}_n^{q}({h}) \, \bar{\mathscr D}_m^{p}({h}+2-q)
\nonu \\
&& \mp \bar{\la}_{s_2,s_1+s_3+2,-s_1-s_2-s_3-4}\, \bar{\la}_{s_1,s_3,-s_1-s_3-2}\,
\bar{\mathscr D}_m^{p}({h}) \, \bar{\mathscr D}_n^{q}({h}+2-p)
\nonu \\
&& -
\bar{\kappa}_{s_1,s_2,-s_1-s_2-2} \, \bar{\la}_{s_1+s_2+2,s_3,-s_1-s_2-s_3-4}
\, \bigg( (q-1)m-(p-1)n \bigg) \,
   \bar{\mathscr D}_{m+n}^{p+q-2}({h})\, \Bigg]
\nonu \\
&& \,\times
{\cal O}_{h+4-p-q,\bar{h}-s_1-s_2-p-q}(z,\bar{z})=0.
\label{Idenother}
\eea
Here the left- and right-weights for the matter field
in (\ref{Idenother}) after taking the two actions of the modes
of ${\cal N}=8$ ${\rm \bar{w}}_{1+\infty}$ currents
are given by
\bea
h_4 \equiv h+4-p-q, \qquad \bar{h}_4 \equiv
\bar{h}-s_1-s_2-p-q.
\label{h4h4bar1}
\eea

Let us consider one of the nontrivial examples
for the two fermionic modes of ${\cal N}=8$ ${\rm \bar{w}}_{1+\infty}$
currents $\hat{\bar{\Phi}}_{A,-\frac{3}{2},m}^{(h_1)}$
and $\hat{\bar{\Phi}}_{B,-\frac{3}{2},n}^{(h_2)}$ with the bosonic matter field
${\cal O}_0^{(h),CDEF}(z,\bar{z})$.
Then the first term of (\ref{wwOother})
is, from the thirteenth and fourteenth equations of (\ref{25anticommsection5}), 
can be described as
\bea
&& \bar{\la}_{-\frac{3}{2},+\frac{1}{2},-1}\,  \bar{\la}_{-\frac{3}{2},0,-\frac{1}{2}}\,
 \frac{1}{4!}\, \frac{1}{3!} \, 3 \, \ep^{CDEFIJKL}
 \, \ep_{BFGHIJKL} \, \nonu \\
&& \times \de^{[F}_{A} \,
\bar{\mathscr D}_n^{h_2+\frac{3}{2}}(1-h)\,  
 \bar{\mathscr D}_m^{h_1+\frac{3}{2}}(\tfrac{3}{2}-h_2-h)\,
  {\cal O}_{+1}^{(h_1+h_2+h),GH]}(z,\bar{z}).
\label{firstresultone}
\eea
On the other hand,
the second term of (\ref{wwOother}) with plus sign
is,  from the thirteenth and the fourteenth equations of
(\ref{25anticommsection5}),
can be obtained as
\bea
&& \bar{\la}_{-\frac{3}{2},+\frac{1}{2},-1}
\bar{\la}_{-\frac{3}{2},0,-\frac{1}{2}}
\frac{1}{4!}\, \frac{1}{3!} \, 3 \, \ep^{CDEFIJKL} \, \ep_{AFGHIJKL}
\nonu \\
&& \times \de^{F}_{[B} 
\bar{\mathscr D}_m^{h_1+\frac{3}{2}}(1-h)  
\bar{\mathscr D}_n^{h_2+\frac{3}{2}}(\tfrac{3}{2}-h_1-h)
{\cal O}_{+1}^{(h_1+h_2+h),GH]}(z,\bar{z}).
\label{secondresultone}
\eea
The right-hand side of (\ref{wwOother}) from the tenth equation of
(\ref{comm25OPEs}) (in addition to the equation $10$
of celestial soft symmetry algbra in \cite{AK2509} where all the
helicities are reversed and see also the tenth equation of (\ref{Eqs}))
is given by
\bea
&&
\bar{\kappa}_{-\frac{3}{2},-\frac{3}{2},+1}\,
\bar{\la}_{-1,0,-1}\, 
\bigg( (h_2+\tfrac{1}{2})m-(h_1+\tfrac{1}{2})n \bigg) \,
\frac{1}{4!}\, \frac{1}{2!} \, \ep^{CDEFIJKL}\,
\ep_{ABGHIJKL}
\nonu \\
&&
\times \bar{\mathscr D}_{m+n}^{h_1+h_2+1}(1-h)\,  
{\cal O}_{+1}^{(h_1+h_2+h),GH}(z,\bar{z}),
\label{thirdresultother}
\eea
where the nineteenth equation of (\ref{25anticommsection5}) is used.
Finally, we construct the relation (\ref{Idenother})
by using (\ref{firstresultone}), (\ref{secondresultone})
and (\ref{thirdresultother}) with (\ref{h4h4bar1})
\bea
&& \Bigg[
(-1) \bar{\la}_{-\fft32,+\fft12,-1}\,\bar{\la}_{-\fft32,0,-\fft12} \,
\bar{\mathscr D}_n^{h_2+\fft32}(1-h) \,
\bar{\mathscr D}_m^{h_1+\fft32}(\tf32-h_2-h)
\nonu \\
&& - (-1) \bar{\la}_{-\fft32,+\fft12,-1}\, \bar{\la}_{-\fft32,0,-\fft12}\,
\bar{\mathscr D}_m^{h_1+\fft32}(1-h)
\bar{\mathscr D}_n^{h_2+\fft32}(\tf32-h_1-h)
\nonu \\
&& -
\bar{\la}_{-1,0,-1} \, \bar{\kappa}_{-\fft32,-\fft32,+1} \,\bigg(
(h_2+\tf12)m-(h_1+\tf12)n \bigg) \,
\bar{\mathscr D}_{m+n}^{h_1+h_2+1}(1-h)\, \Bigg]\nonu \\
&& \times
\frac{1}{4!}\, \frac{1}{2!}\, \ep^{CDEFIJKL}\, \ep_{ABGHIJKL} \,
{\cal O}_{+1}^{(h_1+h_2+h),GH}(z,\bar{z})=0.
\label{Iden1other}
\eea
Note that the minus sign in the first term of (\ref{Iden1other})
appears.
For the second term in (\ref{Iden1other}), there exists
an overall plus sign which is written in terms of two minus signs.
In order to satisfy the relation (\ref{Iden1other})
for any matter field, the three combinations inside the bracket
should vanish as before.
The first coefficient should be equal to the minus of
the third coefficient as follows:
\bea
\bar{\la}_{-\fft32,+\fft12,-1}\,\bar{\la}_{-\fft32,0,-\fft12} = 
-\bar{\la}_{-1,0,-1} \, \bar{\kappa}_{-\fft32,-\fft32,+1}.
\label{oneexampleother}
\eea
Compared to (\ref{oneexample}),
this expression (\ref{oneexampleother}) has exact same behavior
where all the helicities are reversed with barred notations.

It is straightforward to continue to check (\ref{wwOother})
for other remaining cases.
We expect that the corresponding (anti)commutator relations of
(\ref{wwcommother}) can be obtained by constructing
the constraint equations and solving their solutions.

By following the procedure in \cite{AK2509},
the following twenty-five (anti)commutators with $s_1 \leq s_2$
can be obtained
in the holomorphic sector inside the Thielemans package
\cite{Thielemans}
\bea
\bigg[ \hat{\bar{\Phi}}^{(h_1)}_{-2,m},
\hat{\bar{\Phi}}^{(h_2)}_{-2,n} \bigg]&=&
\bar{\kappa}_{-2,-2,+2}\,\Big((h_2+1)m-(h_1+1)n\Big)\,
\hat{\bar{\Phi}}^{(h_1+h_2)}_{-2,m+n}
\, :\text{eq.1},
\nonu    \\
\bigg[
\hat{\bar{\Phi}}^{(h_1)}_{-2,m},
\hat{\bar{\Phi}}^{(h_2)}_{A,-\frac{3}{2},n} \bigg]&=&
\bar{\kappa}_{-2,-\frac{3}{2},+\frac{3}{2}}\,
\Big((h_2+\tfrac{1}{2})m-(h_1+1)n\Big)\,
\hat{\bar{\Phi}}^{(h_1+h_2)}_{A,-\frac{3}{2},m+n}
\, :\text{eq.2},
\nonu    \\
\bigg[\hat{\bar{\Phi}}^{(h_1)}_{-2,m},
\hat{\bar{\Phi}}^{(h_2)}_{AB,-1,n} \bigg]&=&
\bar{\kappa}_{-2,-1,+1}\,\Big(h_2\,m-(h_1+1)n \Big)\,
\hat{\bar{\Phi}}^{(h_1+h_2)}_{AB,-1,m+n}
\, :\text{eq.3},
\nonu    \\
\bigg[\hat{\bar{\Phi}}^{(h_1)}_{-2,m},
\hat{\bar{\Phi}}^{(h_2)}_{ABC,-\frac{1}{2},n}\bigg]
&=&
\bar{\kappa}_{-2,-\frac{1}{2},+\frac{1}{2}}\,
\Big((h_2-\tfrac{1}{2})m-(h_1+1)n\Big)\,
\hat{\bar{\Phi}}^{(h_1+h_2)}_{ABC,-\frac{1}{2},m+n}
\, :  \text{eq.4},
\nonu    \\
\bigg[\hat{\bar{\Phi}}^{(h_1)}_{-2,m},
\hat{\bar{\Phi}}^{(h_2),ABCD}_{0,n} \bigg]
&=&
\bar{\kappa}_{-2,0,0}\,\Big((h_2-1)m-(h_1+1)n\Big)\,
\hat{\bar{\Phi}}^{(h_1+h_2),ABCD}_{0,m+n}
\, :\text{eq.5},
\nonu    \\
\bigg[\hat{\bar{\Phi}}^{(h_1)}_{-2,m},
\hat{\bar{\Phi}}^{(h_2),ABC}_{+\frac{1}{2},n} \bigg]
&=&\bar{\kappa}_{-2,+\frac{1}{2},-\frac{1}{2}}
\Big((h_2-\tfrac{3}{2})m-(h_1+1)n\Big)
\hat{\bar{\Phi}}^{(h_1+h_2),ABC}_{+\frac{1}{2},m+n}
:\text{eq.6},
\nonu   \\
\bigg[\hat{\bar{\Phi}}^{(h_1)}_{-2,m},
\hat{\bar{\Phi}}^{(h_2),AB}_{+1,n} \bigg]
&=&\bar{\kappa}_{-2,+1,-1}\,
\Big((h_2-2)m-(h_1+1)n\Big)\,\hat{\bar{\Phi}}^{(h_1+h_2),AB}_{+1,m+n}
\, :\text{eq.7},
\nonu    \\
\bigg[\hat{\bar{\Phi}}^{(h_1)}_{-2,m},
\hat{\bar{\Phi}}^{(h_2),A}_{+\frac{3}{2},n} \bigg]
&=&
\bar{\kappa}_{-2,+\frac{3}{2},-\frac{3}{2}}\,
\Big((h_2-\tfrac{5}{2})m-(h_1+1)n\Big)\,
\hat{\bar{\Phi}}^{(h_1+h_2),A}_{+\frac{3}{2},m+n}
\, :\text{eq.8},
\nonu    \\
\bigg[\hat{\bar{\Phi}}^{(h_1)}_{-2,m},
\hat{\bar{\Phi}}^{(h_2)}_{+2,n} \bigg]&=&
\bar{\kappa}_{-2,+2,-2}\,
\Big((h_2-3)m-(h_1+1)n\Big)\,\hat{\bar{\Phi}}^{(h_1+h_2)}_{+2,m+n}
\, :\text{eq.9},
\nonu \\
\bigg\{ \hat{\bar{\Phi}}^{(h_1)}_{A,-\frac{3}{2},m},
\hat{\bar{\Phi}}^{(h_2)}_{B,-\frac{3}{2},n} \bigg\}
&=& \bar{\kappa}_{-\frac{3}{2},-\frac{3}{2},+1}\,
\Big((h_2+\tfrac{1}{2})m-(h_1+\tfrac{1}{2})n\Big)\,
\hat{\bar{\Phi}}^{(h_1+h_2)}_{AB,-1,m+n}
\, : \text{eq.10},
\nonu \\
\bigg[\hat{\bar{\Phi}}^{(h_1)}_{A,-\frac{3}{2},m},
\hat{\bar{\Phi}}^{(h_2)}_{BC,-1,n}\bigg]
&=&
\bar{\kappa}_{-\frac{3}{2},-1,+\frac{1}{2}}\,
\Big(h_2\,m-(h_1+\tfrac{1}{2})n\Big)\,
\hat{\bar{\Phi}}^{(h_1+h_2)}_{ABC,-\frac{1}{2},m+n}
\, :\text{eq.11},
\nonu \\
\bigg\{ \hat{\bar{\Phi}}^{(h_1)}_{A,-\frac{3}{2},m},
\hat{\bar{\Phi}}^{(h_2)}_{BCD,-\frac{1}{2},n} \bigg\}
&=&
\bar{\kappa}_{-\frac{3}{2},-\frac{1}{2},0}\,
\Big((h_2-\tfrac{1}{2})m-(h_1+\tfrac{1}{2})n\Big)
\frac{1}{4!}\, \ep_{ABCDEFGH} \nonu \\
& \times & \hat{\bar{\Phi}}^{(h_1+h_2),EFGH}_{0,m+n}
: \text{eq.12},
\nonu \\
\bigg[ \hat{\bar{\Phi}}^{(h_1)}_{A,-\frac{3}{2},m},
\hat{\bar{\Phi}}^{(h_2),BCDE}_{0,n} \bigg]&=&
\bar{\kappa}_{-\frac{3}{2},0,-\frac{1}{2}}\,
\Big((h_2-1)m-(h_1+\tfrac{1}{2})n\Big)\,
\frac{1}{3!}\, \frac{1}{4!}\, \ep^{BCDEIJKL}\nonu \\
& \times & \epsilon_{AFGHIJKL}
\, \hat{\bar{\Phi}}^{(h_1+h_2),FGH}_{+\frac{1}{2},m+n}
\, :  \text{eq.13},
\nonu \\
\bigg\{\hat{\bar{\Phi}}^{(h_1)}_{A,-\frac{3}{2},m},
\hat{\bar{\Phi}}^{(h_2),BCD}_{+\frac{1}{2},n} \bigg\}
&
=& \bar{\kappa}_{-\frac{3}{2},+\frac{1}{2},-1}\,
\Big((h_2-\tfrac{3}{2})m-(h_1+\tfrac{1}{2})n\Big)\,3
\delta^{[B}_{\,\,\,A} \hat{\bar{\Phi}}^{(h_1+h_2),CD]}_{+1,m+n}\,
: \text{eq.14},
\nonu \\
\bigg[\hat{\bar{\Phi}}^{(h_1)}_{A,-\frac{3}{2},m},
\hat{\bar{\Phi}}^{(h_2),BC}_{+1,n} \bigg]
&=&
\bar{\kappa}_{-\frac{3}{2},+1,-\frac{3}{2}}\,
\Big((h_2-2)m-(h_1+\tfrac{1}{2})n\Big)\,
2!\, \delta^{[B}_{\,\,\,A}\,
\hat{\bar{\Phi}}^{(h_1+h_2),C]}_{+\frac{3}{2},m+n}\, :\text{eq.15},
\nonu \\
\bigg\{\hat{\bar{\Phi}}^{(h_1)}_{A,-\frac{3}{2},m},
\hat{\bar{\Phi}}^{(h_2),B}_{+\frac{3}{2},n} \bigg\}
&=&
\bar{\kappa}_{-\frac{3}{2},+\frac{3}{2},-2}\,
\Big((h_2-\tfrac{5}{2})m-(h_1+\tfrac{1}{2})n\Big)\,
\delta^{B}_{\,\,\,A}\,\hat{\bar{\Phi}}^{(h_1+h_2)}_{+2,m+n}
\, : \text{eq.16},
\nonu \\
\bigg[\hat{\bar{\Phi}}^{(h_1)}_{AB,-1,m},
\hat{\bar{\Phi}}^{(h_2)}_{CD,-1,n} \bigg]&=&
\bar{\kappa}_{-1,-1,0}\,\Big(h_2\,m-h_1\,n\Big)\,
\frac{1}{4!}\, \ep_{ABCDEFGH}\nonu \\
& \times &
\hat{\bar{\Phi}}^{(h_1+h_2),EFGH}_{0,m+n}
\, :\text{eq.17},
\nonu \\
\bigg[\hat{\bar{\Phi}}^{(h_1)}_{AB,-1,m},
\hat{\bar{\Phi}}^{(h_2)}_{CDE,-\frac{1}{2},n} \bigg]&=&
\bar{\kappa}_{-1,-\frac{1}{2},-\frac{1}{2}}\,
\Big((h_2-\tfrac{1}{2})m-h_1\,n\Big)
\frac{1}{3!}\,\ep_{ABCDEFGH}
\nonu \\
& \times & \hat{\bar{\Phi}}^{(h_1+h_2),FGH}_{+\frac{1}{2},m+n}
\, :  \text{eq.18},
\nonu \\
\bigg[\hat{\bar{\Phi}}^{(h_1)}_{AB,-1,m},
\hat{\bar{\Phi}}^{(h_2),CDEF}_{0,n} \bigg]&=&
\bar{\kappa}_{-1,0,-1}\,\Big((h_2-1)m-h_1\,n\Big)\,
\frac{1}{4!}\,\frac{1}{2!}\, \epsilon^{CDEFIJKL}\nonu \\
& \times & \ep_{ABGHIJKL}\,
\hat{\bar{\Phi}}^{(h_1+h_2),GH}_{+1,m+n}
:  \text{eq.19},
\nonu \\
\bigg[\hat{\bar{\Phi}}^{(h_1)}_{AB,-1,m},
\hat{\bar{\Phi}}^{(h_2),CDE}_{+\frac{1}{2},n} \bigg]&=&
\bar{\kappa}_{-1,+\frac{1}{2},-\frac{3}{2}}\,
\Big((h_2-\tfrac{3}{2})m-h_1\,n\Big)\,
3! \delta^{[C}_{\,\,\,A}
\hat{\bar{\Phi}}^{(h_1+h_2),D}_{+\frac{3}{2},m+n} \delta_{B}^{\,\,\,E]}
\, :  \text{eq.20},
\nonu \\
\bigg[\hat{\bar{\Phi}}^{(h_1)}_{AB,-1,m},
\hat{\bar{\Phi}}^{(h_2),CD}_{+1,n} \bigg]&=&
\bar{\kappa}_{-1,+1,-2}\,\Big((h_2-2)m-h_1\,n\Big)\,
\delta^{CD}_{AB} \, \hat{\bar{\Phi}}^{(h_1+h_2)}_{+2,m+n}
\, : \text{eq.21},
\nonu \\
\bigg\{\hat{\bar{\Phi}}^{(h_1)}_{ABC,-\frac{1}{2},m},
\hat{\bar{\Phi}}^{(h_2)}_{DEF,-\frac{1}{2},n}\bigg\}&=&
\bar{\kappa}_{-\frac{1}{2},-\frac{1}{2},-1}\,
\Big((h_2-\tfrac{1}{2})m-(h_1-\tfrac{1}{2})n\Big)\,
\frac{1}{2!}\,\epsilon_{ABCDEFGH}
\nonu \\
& \times & \hat{\bar{\Phi}}^{(h_1+h_2),GH}_{+1,m+n}
\, :  \text{eq.22},
\nonu \\
\bigg[\hat{\bar{\Phi}}^{(h_1)}_{ABC,-\frac{1}{2},m},
\hat{\bar{\Phi}}^{(h_2),DEFG}_{0,n} \bigg]&=&
\bar{\kappa}_{-\frac{1}{2},0,-\frac{3}{2}}\,
\Big((h_2-1)m-(h_1-\tfrac{1}{2})n\Big)\,
\frac{1}{4!} \, \epsilon^{DEFGIJKL}
\nonu \\
& \times & \ep_{ABCHIJKL} \, \hat{\bar{\Phi}}^{(h_1+h_2),H}_{+\frac{3}{2},m+n}
\, :  \text{eq.23},
\label{Eqs}
\\
\bigg\{\hat{\bar{\Phi}}^{(h_1)}_{ABC,-\frac{1}{2},m},
\hat{\bar{\Phi}}^{(h_2),DEF}_{+\frac{1}{2},n} \bigg\}
&=&\bar{\kappa}_{-\frac{1}{2},+\frac{1}{2},-2}\,\Big((h_2-\tfrac{3}{2})m-(h_1-\tfrac{1}{2})n\Big)\,
\delta^{DEF}_{ABC}\,\hat{\bar{\Phi}}^{(h_1+h_2)}_{+2,m+n}
\, :  \text{eq.24},
\nonu \\
\bigg[\hat{\bar{\Phi}}^{(h_1),ABCD}_{0,m},
\hat{\bar{\Phi}}^{(h_2),EFGH}_{0,n} \bigg]&=&
\bar{\kappa}_{0,0,-2}\,\Big((h_2-1)m-(h_1-1)n\Big)\epsilon^{ABCDEFGH}\,
\hat{\bar{\Phi}}^{(h_1+h_2)}_{+2,m+n}
\, :  \text{eq.25}\,.
%
\nonu
\eea
Note that
the right-hand sides of the eq.12,
eq.13, eq.17, eq.19 and eq.23 of (\ref{Eqs})
where the scalars appear
have different $SU(8)$ indices compared to \cite{AK2509}.
After solving the Jacobi identities
for (\ref{Eqs}) where we need to change Appendix D of \cite{AK2509}
a little bit, 
we obtain the following relations between the couplings
\bea
\bar{\kappa}_{-\frac{1}{2}, +\frac{1}{2},-2} & = & 
-\frac{\bar{\kappa}_ {0, 0, -2}\,
\bar{\kappa}_{-1, -\frac{1}{2}, -\frac{1}{2}} \,
\bar{\kappa}_{-\frac{3}{2}, -\frac{3}{2}, +1}}
{ (\bar{\kappa}_{-\frac{3}{2}, 0, -\frac{1}{2}})^2}\, ,
\quad
\bar{\kappa}_{-\frac{1}{2}, 0, -\frac{3}{2}} = 
-\frac{\bar{\kappa}_ {0, 0, -2}\,
\bar{\kappa}_{-1, -\frac{1}{2}, -\frac{1}{2}} \,
\bar{\kappa}_{-\frac{3}{2}, -\frac{3}{2}, +1}}
{\bar{\kappa}_{-\frac{3}{2}, +\frac{3}{2}, -2}\,
\bar{\kappa}_{-\frac{3}{2}, 0, -\frac{1}{2}}}\, ,
\nonu \\
\bar{\kappa}_{-\frac{1}{2}, -\frac{1}{2}, -1} & = &
-\frac{ \bar{\kappa}_ {0, 0, -2} 
\,(\bar{\kappa}_{-1, -\frac{1}{2}, -\frac{1}{2}})^2 \,
(\bar{\kappa}_{-\frac{3}{2}, -\frac{3}{2}, +1})^2}
{\bar{\kappa}_{-\frac{3}{2}, +\frac{3}{2}, -2} \,
\bar{\kappa}_{-\frac{3}{2}, +1, -\frac{3}{2}}\,
(\bar{\kappa}_{-\frac{3}{2}, 0, -\frac{1}{2}})^2}\, ,
\quad 
\bar{\kappa}_{-1, +1,-2} = -
\frac{\bar{\kappa}_{-\frac{3}{2}, +\frac{3}{2}, -2}\,
\bar{\kappa}_{-\frac{3}{2}, +1, -\frac{3}{2}}}
{\bar{\kappa}_{-\frac{3}{2}, -\frac{3}{2}, +1}}\, ,
\nonu \\
\bar{\kappa}_{-1, +\frac{1}{2}, -\frac{3}{2}} & = &
\frac{\bar{\kappa}_{0, 0, -2}\,
\bar{\kappa}_{-1, -\frac{1}{2}, -\frac{1}{2}} \,
\bar{\kappa}_{-\frac{3}{2}, -1, +\frac{1}{2}}\,
\bar{\kappa}_{-\frac{3}{2}, -\frac{3}{2}, +1}} {
\bar{\kappa}_{-\frac{3}{2}, +\frac{3}{2},- 2}\,
(\bar{\kappa}_{-\frac{3}{2}, 0, -\frac{1}{2}})^2}\, ,
\nonu \\
\bar{\kappa}_{-1, 0,-1} & = &
-\frac{ \bar{\kappa}_ {0, 0, -2}\,
\bar{\kappa}_{-1, -\frac{1}{2}, -\frac{1}{2}} \,
\bar{\kappa}_{-\frac{3}{2}, -1, +\frac{1}{2}}\,
\bar{\kappa}_{-\frac{3}{2},-\frac{3}{2}, +1}} {
\bar{\kappa}_{-\frac{3}{2}, +\frac{3}{2}, -2}\,
\bar{\kappa}_{-\frac{3}{2}, +1, -\frac{3}{2}} \,
\bar{\kappa}_{-\frac{3}{2}, 0, -\frac{1}{2}}}\, ,
\quad
\bar{\kappa}_{-1, -1, 0} =
\frac{\bar{\kappa}_{-1, -\frac{1}{2}, -\frac{1}{2}}\,
\bar{\kappa}_{-\frac{3}{2}, 
-1, +\frac{1}{2}}}{\bar{\kappa}_{-\frac{3}{2}, 0, -\frac{1}{2}}}\, ,
\nonu \\
\bar{\kappa}_{-\frac{3}{2}, +\frac{1}{2}, -1} & = &
\frac{ \bar{\kappa}_ {0, 0, -2}\,
\bar{\kappa}_{-1, -\frac{1}{2}, -\frac{1}{2}}\,
\bar{\kappa}_{-\frac{3}{2}, -1, +\frac{1}{2}}\,
(\bar{\kappa}_{-\frac{3}{2}, -\frac{3}{2}, +1})^2}
{\bar{\kappa}_{-\frac{3}{2}, +\frac{3}{2}, -2}
\,\bar{\kappa}_{-\frac{3}{2}, +1, -\frac{3}{2}} \,
(\bar{\kappa}_{-\frac{3}{2}, 0, -\frac{1}{2}})^2}\, ,
\quad
\bar{\kappa}_{-\frac{3}{2}, -\frac{1}{2},0} =\frac{
\bar{\kappa}_{-1, -\frac{1}{2}, -\frac{1}{2}} \,
\bar{\kappa}_{-\frac{3}{2}, -\frac{3}{2}, +1}}
{\bar{\kappa}_{-\frac{3}{2}, 0, -\frac{1}{2}}}\, ,
\nonu \\
\bar{\kappa}_{-2, +2, -2} & = & \bar{\kappa}_{-2, -2, +2}\,,
\qquad
\bar{\kappa}_{-2, +\frac{3}{2}, -\frac{3}{2}} = \bar{\kappa}_{-2, -2, +2}\, ,
\qquad
\bar{\kappa}_{-2, +1, -1} = \bar{\kappa}_{-2, -2, +2}\, ,
\nonu \\
\bar{\kappa}_{-2, +\frac{1}{2}, -\frac{1}{2}} & = & \bar{\kappa}_{-2, -2, +2}\, ,
\qquad
\bar{\kappa}_{-2, 0, 0} = \bar{\kappa}_{-2, -2, +2}\, ,
\qquad
\bar{\kappa}_{-2, -\frac{1}{2}, +\frac{1}{2}} = \bar{\kappa}_{-2, -2, +2}\, ,
\nonu \\
\bar{\kappa}_{-2, -1, +1} & = & \bar{\kappa}_{-2, -2, +2}\, ,
\qquad 
\bar{\kappa}_{-2, -\frac{3}{2}, +\frac{3}{2}} = \bar{\kappa}_{-2, -2, +2}\, .
\label{kappabsol}
\eea
This implies that the twenty-five
couplings can be written in terms of
the following eight couplings
\bea
&& \bar{\kappa}_{0,0,-2}, \qquad
\bar{\kappa}_{-1,-\frac{1}{2},-\frac{1}{2}},
\qquad
\bar{\kappa}_{-\frac{3}{2},-\frac{3}{2},+1},
\qquad
\bar{\kappa}_{-\frac{3}{2},0,-\frac{1}{2}},
\nonu \\
&& \bar{\kappa}_{-\frac{3}{2},+\frac{3}{2},-2},
\qquad
\bar{\kappa}_{-\frac{3}{2},+1,-\frac{3}{2}},
\qquad
\bar{\kappa}_{-\frac{3}{2},-1,+\frac{1}{2}},
\qquad
\bar{\kappa}_{-2,-2,+2}.
\label{Eightkappab}
\eea
The behavior of these constraints for the couplings
with (\ref{kappabsol}) and (\ref{Eightkappab})
is exactly the same as the one in \cite{AK2509}.
As in \cite{AK2509}, one of the solutions satisfying
(\ref{kappabsol}) is given by
$\bar{\kappa}_{-\frac{3}{2},+\frac{3}{2},-2}=
\bar{\kappa}_{-1,-\frac{1}{2},-\frac{1}{2}}=
\bar{\kappa}_{-\frac{1}{2},0,-\frac{3}{2}}=
\bar{\kappa}_{-1,+\frac{1}{2},-\frac{3}{2}}=
\bar{\kappa}_{-\frac{3}{2},+\frac{1}{2},-1}=
\bar{\kappa}_{-\frac{3}{2},-\frac{1}{2},0}=
\bar{\kappa}_{-\frac{3}{2},-1,+\frac{1}{2}}=-1$ and
the remaining couplings
with $+1$.

The remaining twenty (anti)commutator relations
which can be obtained by interchanging of
two modes with different helicities appearing on the
left-hand sides of (\ref{Eqs})
are given by 
(\ref{20expression}), for convenience.
The ${\cal N}=8$ ${\rm \bar{w}}_{1+\infty}$ algebra
is given by (\ref{Eqs}) and (\ref{20expression}).

\subsubsection{Other subalgebra of ${\cal N}=8$ BMS algebra}

We have seen the subalgebra of
${\cal N}=8$ supersymmetric BMS algebra in (\ref{BMS}).
We can easily observe that
there exists another type of subalgebra of the ${\cal N}=8$ supersymmetric
BMS algebra as follows:
\bea
\bigg[\hat{\bar{\Phi}}_{-2,m}^{(0)},
{\bar{\Phi}}_{-2,k,l}^{(-\frac{1}{2})}  \bigg] & = &
\bigg( \frac{1}{2}m-k\bigg)\, {\bar{\Phi}}_{-2,m+k,l}^{(-\frac{1}{2})},
\nonu \\
\bigg[
\hat{\bar{\Phi}}_{-2,m}^{(0)},
\hat{\bar{\Phi}}_{-2,n}^{(0)}  \bigg] & = &
\bigg( m-n\bigg)\, \hat{\bar{\Phi}}_{-2,m+n}^{(0)},
\nonu \\
\bigg[
\hat{\bar{\Phi}}_{-2,m}^{(0)},
\hat{\bar{\Phi}}_{A,-\frac{3}{2},n}^{(0)}  \bigg] & = &
\bigg( \frac{1}{2}m-n\bigg)\, \hat{\bar{\Phi}}_{A,-\frac{3}{2},m+n}^{(0)}. 
\label{BMS1}
\eea
It is straightforward to show
the above commutator relations in the context of (\ref{wwOother}) by
following the descriptions given in the footnote \ref{BMScheck}
\footnote{In the notations of \cite{FSTZ},
the above three commutators become
\bea
\bigg[ L_m,
P_{k,l}  \bigg] & = &
\bigg( \frac{1}{2}m-k\bigg)\, P_{m+k,l},
\nonu \\
\bigg[
L_m, L_n  \bigg] & = &
\bigg( m-n\bigg)\, L_{m+n},
\nonu \\
\bigg[
L_m,
G_{A,n}  \bigg] & = &
\bigg( \frac{1}{2}m-n\bigg)\, G_{A,m+n}. 
\nonu
\eea
The corresponding commutators in the antiholomorphic sector
is given by (\ref{BMS}). Of course, it is obvious to observe that
the bosonic extended BMS algebra is described
by the first two commutators from (\ref{BMS})
and those from (\ref{BMS1}) where the two modes of
supertranslation generators are equal to each other
from the footnote \ref{sametranslation}.}.
We can easily check that the second and third equations
of (\ref{BMS1})
can be seen from the the first and the second equations
of (\ref{Eqs}) respectively by putting $h_1=0=h_2$
with the couplings of one.
Finally,  the first equation
of (\ref{BMS1}) with $l=-\frac{1}{2}$ (then the additional hatted notations
can be used on both sides and see also (\ref{otheranticomm}))
can be seen also from the the first equation
of (\ref{Eqs}) by putting $h_1=0$, $h_2=-\frac{1}{2}$
with the couplings of one.
The $l$ dependent factor $\bar{z}^{l+\frac{1}{2}}$ appears
on both sides of the first equation of (\ref{BMS1}).
In order to cover the arbitrary $l (\neq -\frac{1}{2})$,
we should obtain the generalization of ${\cal N}=8$ ${\rm \bar{w}}_{1+\infty}$
algebra described in (\ref{Eqs}).

\subsubsection{The nonvanishing anticommutator of ${\cal N}=8$
BMS algebra
\label{Non}}

There exists one additional nontrivial
anticommutator relation
between two kinds of supersymmetry currents
where the right-hand side becomes the supertranslation generator
in addition to (\ref{BMS}) and (\ref{BMS1}).

We would like to check this nontirvial
anticommutator relation for the particular (gravitino) matter field
with positive helicity which appears in Table \ref{table1}.
Let us consider  the modified relation of (\ref{wwO})
for the ($\hat{\Phi}_{+\frac{3}{2},m}^{(0),A}$,
$\hat{\bar{\Phi}}_{B,-\frac{3}{2},n}^{(0)}$,
${\cal O}_{+\frac{3}{2}}^{(h),C}(z,\bar{z})$) where the second current
comes from the one in (\ref{otherwO}).
Then we need to compute the following action
coming from (\ref{wwOother})
\bea
&&
\bigg[ \hat{\Phi}_{+\frac{3}{2},m}^{(0),A},
\bigg\{ \hat{\bar{\Phi}}_{B,-\frac{3}{2},n}^{(0)},
{\cal O}_{+\frac{3}{2}}^{(h),C}(z,\bar{z})\bigg\} \bigg]+
\bigg[ \hat{\bar{\Phi}}_{B,-\frac{3}{2},n}^{(0)},
\bigg\{  \hat{\Phi}_{+\frac{3}{2},m}^{(0),A},
{\cal O}_{+\frac{3}{2}}^{(h),C}(z,\bar{z})\bigg\} \bigg]
\nonu \\
&& =
\bigg[ \bigg\{ \hat{\Phi}_{+\frac{3}{2},m}^{(0),A},
\hat{\bar{\Phi}}_{B,-\frac{3}{2},n}^{(0)} \bigg\},
{\cal O}_{+\frac{3}{2}}^{(h),C}(z,\bar{z}) \bigg].  
\label{3halfcombi}
\eea
Then the anticommutator of inside the first term in (\ref{3halfcombi})
can be written as
\bea
\bigg\{ \hat{\bar{\Phi}}_{B,-\frac{3}{2},n}^{(0)},
{\cal O}_{+\frac{3}{2}}^{(h),C}(z,\bar{z})\bigg\} =
\frac{1}{2}\, {z}^{n+\frac{1}{2}}\,
\de^{C}_{B}\,
{\cal O}_{+2}^{(-\frac{1}{2}+h)}(z,\bar{z}),        
\label{two3halfs}
\eea
according to (\ref{two3half}).
Note that $s_1+s_2-s_3=-2$ and there exists the barred notation
for the mode of current on the left-hand side.
In other words,
for $h_2=h+\frac{3}{2}$,
the matter field on the left-hand side of
(\ref{two3half}) has the left- and right-weights
$(1-h,-\frac{1}{2}-h)$ for helicity $+\frac{3}{2}$
from Table \ref{table1-1} while
the matter field on the right-hand side of
(\ref{two3half}) has the left- and right-weights
$(\frac{3}{2}-h,-\frac{1}{2}-h)$ (from Table \ref{table1-1})
which can be rewritten as
$(1-(-\frac{1}{2}+h), -1-(-\frac{1}{2}+h))$
for helicity $+2$ operator
\footnote{
\label{TWOmatters}
We can associate the operators in Table \ref{table1} with
the ones in Table \ref{table1-1} as follows:
\bea
{\cal O}_{+2}^{(h)}(z, \bar{z}) & \leftrightarrow &
{\cal O}_{+2}^{(h+2)}(z, \bar{z}),
\qquad
{\cal O}_{+\frac{3}{2}}^{(h),A}(z, \bar{z}) \leftrightarrow
{\cal O}_{+\frac{3}{2}}^{(h+\frac{3}{2}),A}(z, \bar{z}),
\nonu \\
{\cal O}_{+1}^{(h),AB}(z, \bar{z}) & \leftrightarrow &
{\cal O}_{+1}^{(h+1),AB}(z, \bar{z}),
\qquad
{\cal O}_{+\frac{1}{2}}^{(h),ABC}(z, \bar{z}) \leftrightarrow
{\cal O}_{+\frac{1}{2}}^{(h+\frac{1}{2}),ABC}(z, \bar{z}),
\nonu \\
{\cal O}_{0}^{(h),ABCD}(z, \bar{z}) & \leftrightarrow &
{\cal O}_{0}^{(h),ABCD}(z, \bar{z}),
\qquad
{\cal O}_{ABC,-\frac{1}{2}}^{(h)}(z, \bar{z}) \leftrightarrow
{\cal O}_{ABC,-\frac{1}{2}}^{(h-\frac{1}{2})}(z, \bar{z}),
\nonu \\
{\cal O}_{AB,-1}^{(h)}(z, \bar{z}) & \leftrightarrow &
{\cal O}_{AB,-1}^{(h-1)}(z, \bar{z}),
\qquad
{\cal O}_{A,-\frac{3}{2}}^{(h)}(z, \bar{z}) \leftrightarrow
{\cal O}_{A,-\frac{3}{2}}^{(h-\frac{3}{2})}(z, \bar{z}),
\nonu \\
{\cal O}_{-2}^{(h)}(z, \bar{z}) & \leftrightarrow &
{\cal O}_{-2}^{(h-2)}(z, \bar{z}).
\label{tworelations}
\eea
The left- and right-weights of these operators
for each helicitiy
are the same. Note that there exist the changes of the superscripts
on the right-hand sides of (\ref{tworelations}).
A particular example for this is given by (\ref{iinter2}). }.
This implies that the two matter fields
in (\ref{two3half}) can be identified with
those in (\ref{two3halfs}).
We would like to express the matter field appearing on the
right-hand side of (\ref{two3halfs}) which lives in Table
\ref{table1}.

Now we can use the first equation of
Appendix (\ref{20OPEs}) because we already
changed the matter field in the basis of Table \ref{table1}
and obtain
\bea
\bigg[ \hat{\Phi}_{+\frac{3}{2},m}^{(0),A},
{\cal O}_{+2}^{(-\frac{1}{2}+h)}(z,\bar{z}) \bigg]=
\frac{1}{2}\, \bar{z}^{m+\frac{1}{2}} \,
{\cal O}_{+\frac{3}{2}}^{(-\frac{1}{2}+h),A}(z,\bar{z}).
\label{3halftwo}
\eea
Note that $s_1+s_2-s_3=+2$ (there is unbarred notation for the
mode of the current on the left-hand side).
Then by combining (\ref{two3halfs}) with (\ref{3halftwo}),
the contribution of the first term of 
(\ref{3halfcombi}) becomes
\bea
\frac{1}{4}\, \bar{z}^{m+\frac{1}{2}}\, z^{n+\frac{1}{2}}\,
\de^{C}_{B}\, {\cal O}_{+\frac{3}{2}}^{(-\frac{1}{2}+h),A}(z,\bar{z}).
\label{result1}
\eea

Let us consider the second term of (\ref{3halfcombi}).
The anticommutator inside of this term
can be obtained from (\ref{Two3half}) and is given by
\bea
\bigg\{  \hat{\Phi}_{+\frac{3}{2},m}^{(0),A},
{\cal O}_{+\frac{3}{2}}^{(h),C}(z,\bar{z})\bigg\} =
\frac{1}{2} \, \bar{z}^{m+\frac{1}{2}}\,
{\cal O}_{+1}^{(h),AC}(z,\bar{z}). 
\label{firstcon}
\eea
In this case, $s_1+s_2-s_3=+2$ with unbarred notation of the mode
on the left-hand side.
Next, we arrive at the commutator, from (\ref{25anticommsection5}),
\bea
\bigg[ \hat{\bar{\Phi}}_{B,-\frac{3}{2},n}^{(0)},
{\cal O}_{+1}^{(h),AC}(z,\bar{z}) \bigg] =\frac{1}{2}
\, z^{n+\frac{1}{2}} \,  2! \, \de^{[A}_{B}\,
{\cal O}_{+\frac{3}{2}}^{(-\frac{1}{2}+h),C]}(z,\bar{z}).
\label{secondcon}
\eea
Note that $s_1+s_2-s_3=-2$ with barred notation for the mode
on the left-hand side.
There are two terms rather than one on the right-hand side.
As described in (\ref{two3halfs}),
the left- and right-weights for helicity $+1$
in Table \ref{table1-1} are given by
$(2-h_2,1-h_2)$ which become $(1-h,-h)$ for $h_2=1+h$.
The left- and right-weights for the helicity
$+\frac{3}{2}$ in Table \ref{table1-1}
are given by $(\frac{5}{2}-h_2,1-h_2)$
which becomes $(1-(-\frac{1}{2}+h),-\frac{1}{2}-(-\frac{1}{2}+h))$.
The matter fields in (\ref{secondcon}) have these weights.
In other words, the matter field on the right-hand side
of (\ref{secondcon}) is in the basis of Table \ref{table1}.
Therefore, the contribution from the second term
is described by combining (\ref{firstcon}) and (\ref{secondcon})
\bea
\frac{1}{4}\, \bar{z}^{m+\frac{1}{2}}\, z^{n+\frac{1}{2}}\,
2!\,
\de^{[A}_{B}\, {\cal O}_{+\frac{3}{2}}^{(-\frac{1}{2}+h),C]}(z,\bar{z}).
\label{result2}
\eea
The final result of the left-hand side of (\ref{3halfcombi}),
by adding (\ref{result1}) and (\ref{result2}),
leads to
\bea
\frac{1}{4}\, \bar{z}^{m+\frac{1}{2}}\, z^{n+\frac{1}{2}}\,
\de^{A}_{B}\, {\cal O}_{+\frac{3}{2}}^{(-\frac{1}{2}+h),C}(z,\bar{z}).
\label{finalexp}
\eea

On the other hand, we have, from (\ref{transla}),
\bea
\bigg[ {{\Phi}}_{+2,n,m}^{(-\frac{1}{2})},
{\cal O}_{+\frac{3}{2}}^{(h),C}(z,\bar{z})
\bigg]= \frac{1}{2}\, z^{n+\frac{1}{2}}\, \bar{z}^{m+\frac{1}{2}}\,
     {\cal O}_{+\frac{3}{2}}^{(-\frac{1}{2}+h),C}(z,\bar{z}).
\label{finalexp1}
\eea
Then the left-hand side of (\ref{3halfcombi}), by noting
(\ref{finalexp}) and (\ref{finalexp1}),
can be written as
\bea
\frac{1}{2} \, \de^{A}_{B}\, \bigg[ {{\Phi}}_{+2,n,m}^{(-\frac{1}{2})},
{\cal O}_{+\frac{3}{2}}^{(h),C}(z,\bar{z})
\bigg],
\label{commcomm}
\eea
which should be equal to the right-hand side of (\ref{3halfcombi}).
This implies that the anticommutator, for any matter field,
can be written as
the supertranslation generator with (\ref{commcomm})
\bea
\bigg\{
\hat{\bar{\Phi}}_{A,-\frac{3}{2},m}^{(0)},
\hat{\Phi}_{+\frac{3}{2},n}^{(0),B} \bigg\}   =
\frac{1}{2}\,
\de^{B}_{A}\,
{{\Phi}}_{+2,m,n}^{(-\frac{1}{2})} \qquad \mbox{or} \qquad
\frac{1}{2}\, \de^{B}_{A}\,
{\bar{\Phi}}_{-2,m,n}^{(-\frac{1}{2})}   .
\label{BMS2}
\eea
Therefore, we obtain the nontrivial nonzero seven
(anti)commutators
(\ref{BMS}), (\ref{BMS1}) and (\ref{BMS2})
together with the trivial eight (anti)commutators below
for the ${\cal N}=8$ BMS algebra
\footnote{Note that from the weights $(1-h_1,2+h_1)$ of
the current $\Phi_{+2}^{(h_1)}$
the condition for the same left- and right-weights is given by
$h_1=-\frac{1}{2}$
and for the weights
$(2+h_2,1-h_2)$ of the current  $\bar{\Phi}_{-2}^{(h_2)}$,
this condition gives $h_2=-\frac{1}{2}$.
In both cases, the weights are given by
$(\frac{3}{2},\frac{3}{2})$ 
respectively.
If we take the gravitino matter field with positive helicity from Table
\ref{table1-1}, then we arrive at the second case of
(\ref{BMS2}). The two supertranslation generators in (\ref{BMS2})
are identical to each other. See also the footnote \ref{sametranslation}.
For the wedge modes, the global
${\cal N}=8$ super Poincare algebra, as a finite dimensional
subalgebra, is
reproduced.
}.
So far, the matter field is given by
the positive gravitino matter field but 
the above relation should satisfy for any matter fields
\footnote{We have checked that for other eight matter fields
in (\ref{wO}), the above anticommutator satisfies.
In particular, for the scalar matter field,
we need to use the (anti)commutators in the twelfth and the
thirteenth equations of (\ref{25anticommsection5}). We can write
down each contribution from the first term, the second term of
the left-hand side of (\ref{wwOother}) and the right-hand side
of (\ref{wwOother}). For fixed $SU(8)$ indices, we can check
both sides of (\ref{wwOother}). This implies the correct
normalizations of above two equations in (\ref{25anticommsection5}). }.
See also the equation $(8.5)$ of \cite{FSTZ}.
We can check the first equation of (\ref{BMS})
and those of (\ref{BMS1}) by considering the left-hand side
for the supertranslation generator (\ref{BMS2}) together with
the last equation of (\ref{BMS}), the last equation of
(\ref{BMS1}),
the second and sixth equations of (\ref{trivialeight}).

\subsubsection{The vanishing (anti)commutators of ${\cal N}=8$
BMS algebra}

Explicitly,
there exist the following eight trivial (anti)commutators
of ${\cal N}=8$ BMS algebra \footnote{In the notation of \cite{FSTZ},
  there exist
\bea
\bigg[{L}_m, \bar{L}_n \bigg] & = & 0,
\qquad \bigg[{L}_m, \bar{G}^A_n \bigg]=0,
\qquad
\bigg[{P}_{k,l}, {P}_{m,n} \bigg]  =  0,
\qquad
\bigg[{P}_{k,l}, {G}_{A,m} \bigg]=0, 
\nonu \\
\bigg[{P}_{k,l}, \bar{G}^A_m \bigg] & = & 0, \qquad
\bigg[\bar{L}_m, {G}_{A,n} \bigg]=0,
\qquad
\bigg\{ G_{A,m}, G_{B,n} \bigg\}  =  0,
\qquad \bigg\{ \bar{G}^A_{m}, \bar{G}^B_{n} \bigg\}=0.
\nonu
\eea
Note that although we present the modes of
supertranslation generators with barred notations in
(\ref{trivialeight}), they can be also written in terms of unbarred
ones.}
\bea
\bigg[\hat{\bar{\Phi}}_{-2,m}^{(0)},\hat{{\Phi}}_{+2,n}^{(0)} \bigg] & = &
0,
\qquad
\bigg[\hat{\bar{\Phi}}_{-2,m}^{(0)},
\hat{{\Phi}}_{+\frac{3}{2},n}^{(0),A}   \bigg]  = 
0,
\nonu \\
\bigg[ {\bar{\Phi}}_{-2,k,l}^{(-\frac{1}{2})},
{\bar{\Phi}}_{-2,m,n}^{(-\frac{1}{2})} \bigg] & = & 0,
\qquad
\bigg[ {\bar{\Phi}}_{-2,k,l}^{(-\frac{1}{2})},
\hat{\bar{\Phi}}_{A,-\frac{3}{2},m}^{(0)}   \bigg]  =  0,
\nonu \\
\bigg[ {\bar{\Phi}}_{-2,k,l}^{(-\frac{1}{2})},
\hat{{\Phi}}_{+\frac{3}{2},m}^{(0),A}   \bigg] & = & 0,
\qquad
\bigg[\hat{{\Phi}}_{+2,m}^{(0)},
\hat{\bar{\Phi}}_{A,-\frac{3}{2},n}^{(0)}   \bigg]  =  0,
\nonu \\
\bigg\{ \hat{\bar{\Phi}}_{A,-\frac{3}{2},m}^{(0)},
\hat{\bar{\Phi}}_{B,-\frac{3}{2},n}^{(0)}   \bigg\}  & = &  0,
\qquad
\bigg\{
\hat{{\Phi}}_{+\frac{3}{2},m}^{(0),A},
\hat{{\Phi}}_{+\frac{3}{2},n}^{(0),B} \bigg\} =0.
\label{trivialeight}
\eea
In the first equation of (\ref{trivialeight}),
we can use
the OPE with positive graviton matter field
as follows
\bea
 {\Phi}_{+2}^{(0)}(z_1,\bar{z}_1) \,
{\pa}_{{z}_2} {\cal O}_{+2}^{(h_2)}(z_2,\bar{z}_2)
& = &  \frac{1}{z_{12}^2}\bigg[ 
\frac{1}{\bar{z}_{12}^2}\,(-1-h_2) 
+ \frac{1}{\bar{z}_{12}}\, \bar{\pa}_{\bar{z}_2} \bigg]
{\cal O}_{+2}^{(h_2)}(z_2,\bar{z}_2)
\nonu \\
& + &
\frac{1}{z_{12}}\bigg[ 
\frac{1}{\bar{z}_{12}^2}\,(-1-h_2) 
+ \frac{1}{\bar{z}_{12}}\, \bar{\pa}_{\bar{z}_2} \bigg]\,
\pa_{z_2}\, 
{\cal O}_{+2}^{(h_2)}(z_2,\bar{z}_2)+
\cdots,
\label{twoterms}
\eea
which leads to
\bea
\bigg[  \hat{\Phi}_{+2,m}^{(0)}, 
{\pa}_{{z}_2} {\cal O}_{+2}^{(h_2)}(z_2,\bar{z}_2) \bigg]=
\bar{z}^m \bigg( (1+m)(-1-h_2)+ \bar{z} \, \bar{\pa}_{\bar{z}}
\bigg) \,  \pa_{z}\, {\cal O}_{+2}^{(h_2)}(z,\bar{z}).
\label{inter1}
\eea
Note that the above OPE (\ref{twoterms})
can be obtained from (\ref{PhiO}).
Similarly, the following OPE satisfies, from (\ref{otherPhiO}),
\bea
\bar{\Phi}_{-2}^{(0)}(z_1,\bar{z}_1) \,
\bar{\pa}_{\bar{z}_2} {\cal O}_{+2}^{(h_2)}(z_2,\bar{z}_2)
& = &  \frac{1}{\bar{z}_{12}^2}\bigg[ 
\frac{1}{{z}_{12}^2}\,(3-h_2) 
+ \frac{1}{{z}_{12}}\, {\pa}_{{z}_2} \bigg]
{\cal O}_{+2}^{(h_2)}(z_2,\bar{z}_2)
\nonu \\
& + &
\frac{1}{\bar{z}_{12}}\bigg[ 
\frac{1}{{z}_{12}^2}\,(3-h_2) 
+ \frac{1}{{z}_{12}}\, {\pa}_{{z}_2} \bigg]\,
\bar{\pa}_{\bar{z}_2}\, 
{\cal O}_{+2}^{(h_2)}(z_2,\bar{z}_2)+
\cdots.
\label{opederivative}
\eea
As before, the coefficient $(3-h_2)$ in (\ref{opederivative})
is the left-weight of the matter field on the left-hand side.
This implies
the following commutator
\bea
\bigg[\hat{\bar{\Phi}}_{-2,n}^{(0)}, 
\bar{\pa}_{\bar{z}_2} {\cal O}_{+2}^{(h_2)}(z_2,\bar{z}_2) \bigg]=
{z}^n \bigg( (1+n)(3-h_2)+ {z} \, {\pa}_{{z}}
\bigg) \,  \bar\pa_{\bar{z}}\, {\cal O}_{+2}^{(h_2)}(z,\bar{z}).
\label{inter2}
\eea
Then we can compute, by using (\ref{inter1}) and (\ref{inter2}),
\bea
\bigg[\hat{\Phi}_{+2,m}^{(0)}, \bigg[ \hat{\bar{\Phi}}_{-2,n}^{(0)},
{\cal O}_{+2}^{(h_2)}(z,\bar{z}) \bigg] \bigg]-
\bigg[\hat{\bar{\Phi}}_{-2,n}^{(0)}, \bigg[ \hat{\Phi}_{+2,m}^{(0)},
{\cal O}_{+2}^{(h_2)}(z,\bar{z}) \bigg] \bigg],
\label{Twocomm}
\eea
which becomes zero (the total eight terms
vanish) and the first equation of
(\ref{trivialeight}) is satisfied.
In (\ref{Twocomm}), the minus sign comes from the
commutator between the modes
\footnote{For different matter fields,
the first equation of (\ref{trivialeight}) should satisfy although
we don't consider in this paper.}.

For the last equation of (\ref{trivialeight}) with positive gravitino
matter field,
we consider the following action
\bea
&&
\bigg[ \hat{\Phi}_{+\frac{3}{2},m}^{(0),A},
\bigg\{ \hat{{\Phi}}_{+\frac{3}{2},n}^{(0),B},
{\cal O}_{+\frac{3}{2}}^{(h),C}(z,\bar{z})\bigg\} \bigg]+
\bigg[ \hat{{\Phi}}_{+\frac{3}{2},n}^{(0),B},
\bigg\{  \hat{\Phi}_{+\frac{3}{2},m}^{(0),A},
{\cal O}_{+\frac{3}{2}}^{(h),C}(z,\bar{z})\bigg\} \bigg].
\label{three3half}
\eea
Then the anticommutators appearing in (\ref{three3half}),
from (\ref{red25anticomm}), become
\bea
\frac{1}{2}\, \bar{z}^{n+\frac{1}{2}}\,
{\cal O}_{+1}^{(h),BC}(z,\bar{z}), \qquad
\frac{1}{2}\, \bar{z}^{m+\frac{1}{2}}\,
{\cal O}_{+1}^{(h),AC}(z,\bar{z}),     
\label{TWOanticomm}
\eea
respectively.
Then the first term of (\ref{three3half}) together with
(\ref{TWOanticomm}) becomes
\bea
\frac{1}{4}\, \bar{z}^{m+n+1}\, {\cal O}_{+\frac{1}{2}}^{(h),ABC}(z,\bar{z})
\label{Inter1}
\eea
while the second term leads to
\bea
\frac{1}{4}\, \bar{z}^{m+n+1}\, {\cal O}_{+\frac{1}{2}}^{(h),BAC}(z,\bar{z}).
\label{Inter2}
\eea
Here the relation (\ref{red25anticomm}) is used again.
Therefore, the sum of (\ref{Inter1}) and (\ref{Inter2})
is equal to zero due to the antisymmetric property for $SU(8)$
indices of
the graviphotino matter field
and this shows the last equation of
(\ref{trivialeight}). We can analyze the other remaining cases 
\footnote{
For the seventh equation of (\ref{trivialeight}),
we can follow the procedure done in the last equation
of (\ref{trivialeight}). 

For the fifth equation of (\ref{trivialeight}),
when we take the positive graviton
matter field ${\cal O}_{+2}^{(h_2)}(z,\bar{z})$,
the commutator inside the first commutator in (\ref{wwO})
is simply $\frac{1}{2}\, \bar{z}^{m+\frac{1}{2}}\,
{\cal O}_{+\frac{3}{2}}^{(h_2),A}(z,\bar{z})$. After acting
the mode of first current on this, the contribution from the
first term in (\ref{wwO}) becomes
$\frac{1}{4} \, z^{k+\frac{1}{2}}\, \bar{z}^{l+m+1}\,
{\cal O}_{+\frac{3}{2}}^{(-\frac{1}{2}+h_2),A}(z,\bar{z})$.
Similarly, the commutator inside the second commutator in (\ref{wwO})
is $\frac{1}{2}\, z^{k+\frac{1}{2}} \, \bar{z}^{l+\frac{1}{2}}\,
{\cal O}_{+2}^{(-\frac{1}{2}+h_2),A}(z,\bar{z})$.
After acting
the mode of second current on this, the contribution from the
second term in (\ref{wwO}) becomes the above one with minus sign.
Therefore, we observe the fifth equation of (\ref{trivialeight}).

We can check the fourth equation of (\ref{trivialeight}) with a little
change for above description from
the fifth equation of (\ref{trivialeight}).

For the sixth equation with positive gravitino matter field,
the relation of (\ref{two3halfs})
can be used in the anticommutator of
inside the first term in (\ref{wwO}). Then
after acting
the mode of first current on this, the contribution from the
first term in (\ref{wwO}) can be obtained. In the contribution of
second term, we need to calculate the anticommutator
$\bigg\{\hat{\bar{\Phi}}_{A,-\frac{3}{2},n}^{(0)}, \bar{\pa}_{\bar{z}}\,
{\cal O}_{+\frac{3}{2}}^{(h_2),B}(z,\bar{z})\bigg\}$
which becomes $\frac{1}{2} \, z^{n+\frac{1}{2}}\, \de^{B}_{A}\,
\bar{\pa}_{\bar{z}} \,
{\cal O}_{+2}^{(h_2)}(z,\bar{z})$ with the help of (\ref{sixteenth}) which
comes from the simple pole in the antiholomorphic coordinate.
There exists also nonderivative term.
By collecting the nonzero contributions from the second term
and realizing that the left-hand side of (\ref{wwO}) vanishes, we
arrive at the result of sixth equation.

The second equation of (\ref{trivialeight}) with a little
change for above description can be analyzed similarly.

It is straightforward to check the third equation of (\ref{trivialeight})
by using the property (\ref{q3halftwocases})
in the footnote \ref{q2and3half}.}. 

\subsection{Toward the extension of ${\cal N}=8$ ${\rm w}_{1+\infty}$
algebra (or the extension of ${\cal N}=8$ ${\rm \bar{w}}_{1+\infty}$
algebra)}

Due to the fact that there exists
nonzero anticommutator between the two supercurrents
in the ${\cal N}=8$ BMS algebra,
we expect that there are nontrivial (anti)commutators
between the ${\cal N}=8$ ${\rm w}_{1+\infty}$ currents
and the ${\rm \bar{w}}_{1+\infty}$ currents.

Let us consider the lower $q$ cases where
we can see the structures of the actions
of these modes on the matter fields.
For $q=\frac{3}{2}$,
there are
\bea
&& \hat{\bar{\Phi}}_{-2}^{(-\frac{1}{2})}, \qquad
\hat{\bar{\Phi}}_{A,-\frac{3}{2}}^{(0)}, \qquad
\hat{\bar{\Phi}}_{AB,-1}^{(\frac{1}{2})}, \qquad
\hat{\bar{\Phi}}_{ABC,-\frac{1}{2}}^{(1)}, \qquad
\hat{\bar{\Phi}}_{0}^{(\frac{3}{2}),ABCD},
\nonu \\
&& \hat{\bar{\Phi}}_{+\frac{1}{2}}^{(2),ABC}, \qquad
\hat{\bar{\Phi}}_{+1}^{(\frac{5}{2}),AB}, \qquad
\hat{\bar{\Phi}}_{+\frac{3}{2}}^{(3),A}, \qquad
\hat{\bar{\Phi}}_{+2}^{(\frac{7}{2})},
\label{threehalf1}
\eea
where the upper indices minus the lower indices are equal to
$\frac{3}{2}$,
in addition to
\bea
&&
\hat{\Phi}_{+2}^{(-\frac{1}{2})}, \qquad
\hat{\Phi}_{+\frac{3}{2}}^{(0),A}, \qquad
\hat{\Phi}_{+1}^{(\frac{1}{2}),AB}, \qquad
\hat{\Phi}_{+\frac{1}{2}}^{(1),ABC}, \qquad
\hat{\Phi}_{0}^{(\frac{3}{2}),ABCD},
\nonu \\
&&
\hat{\Phi}_{ABC,-\frac{1}{2}}^{(2)}, \qquad
\hat{\Phi}_{AB,-1}^{(\frac{5}{2})}, \qquad
\hat{\Phi}_{A,-\frac{3}{2}}^{(3)}, \qquad
\hat{\Phi}_{-2}^{(\frac{7}{2})},
\label{threehalf2}
\eea
where the sum of the upper indices and lower
indices is equal to $\frac{3}{2}$.
We would like to describe the action of the
two modes from the two kinds of the currents in (\ref{threehalf1})
and (\ref{threehalf2}) with
two opposite helicities on the matter field
for the helicity $+2$ graviton in Table \ref{table1}.
We summarize the results as follows
\footnote{
For this matter field, the previous relation (\ref{BMS2})
corresponds to
\bea
\bigg[ \bigg\{ \hat{\bar{\Phi}}_{A,-\frac{3}{2},m}^{(0)},
\hat{\Phi}_{+\frac{3}{2},n}^{(0),B} \bigg\}, {\cal O}_{+2}^{(h)}(z,\bar{z})
\bigg] & = &
\frac{1}{4}\, z^{m+\frac{1}{2}} \, \bar{z}^{n+\frac{1}{2}} \,
\de^{B}_{A} \,  {\cal O}_{+2}^{(-\frac{1}{2}+h)}(z,\bar{z}).
\nonu
\eea
The contribution from the second term
of the left-hand side 
in (\ref{wwO})
vanishes. The left- and right-weights of the matter field
are increased by $(\frac{1}{2},\frac{1}{2})$ under this action.
}:
\bea
\bigg[ \bigg[ \hat{\bar{\Phi}}_{AB,-1,m}^{(\frac{1}{2})},
\hat{\Phi}_{+1,n}^{(\frac{1}{2}),CD} \bigg], {\cal O}_{+2}^{(h)}(z,\bar{z}) \bigg]
& = &
\frac{1}{4}\, z^{m+\frac{1}{2}} \, \bar{z}^{n+\frac{1}{2}} \,
\de^{CD}_{AB} \,  {\cal O}_{+2}^{(h)}(z,\bar{z}),
\nonu \\
\bigg[ \bigg\{ \hat{\bar{\Phi}}_{ABC,-\frac{1}{2},m}^{(1)},
\hat{\Phi}_{+\frac{1}{2},n}^{(1),DEF} \bigg\}, {\cal O}_{+2}^{(h)}(z,\bar{z})
\bigg] & = &
\frac{1}{4}\, z^{m+\frac{1}{2}} \, \bar{z}^{n+\frac{1}{2}} \,
\de^{DEF}_{ABC} \,  {\cal O}_{+2}^{(\frac{1}{2}+h)}(z,\bar{z}),
\nonu \\
\bigg[ \bigg[ \hat{\bar{\Phi}}_{0,m}^{(\frac{3}{2}),ABCD},
\hat{\Phi}_{0,n}^{(\frac{3}{2}),EFGH} \bigg], {\cal O}_{+2}^{(h)}(z,\bar{z})
\bigg] & = &
\frac{1}{4}\, z^{m+\frac{1}{2}} \, \bar{z}^{n+\frac{1}{2}} \,
\ep^{ABCDEFGH} \,  {\cal O}_{+2}^{(1+h)}(z,\bar{z}),
\nonu \\
\bigg[ \bigg\{ \hat{\bar{\Phi}}_{+\frac{1}{2},m}^{(2),ABC},
\hat{\Phi}_{DEF,-\frac{1}{2},n}^{(2)} \bigg\}, {\cal O}_{+2}^{(h)}(z,\bar{z})
\bigg] & = &
\frac{1}{4}\, z^{m+\frac{1}{2}} \, \bar{z}^{n+\frac{1}{2}} \,
\de^{ABC}_{DEF} \,  {\cal O}_{+2}^{(\frac{3}{2}+h)}(z,\bar{z}),
\nonu \\
\bigg[ \bigg[ \hat{\bar{\Phi}}_{+1,m}^{(\frac{5}{2}),AB},
\hat{\Phi}_{CD,-1,n}^{(\frac{5}{2})} \bigg], {\cal O}_{+2}^{(h)}(z,\bar{z})
\bigg] & = &
\frac{1}{4}\, z^{m+\frac{1}{2}} \, \bar{z}^{n+\frac{1}{2}} \,
\de^{AB}_{CD} \,  {\cal O}_{+2}^{(2+h)}(z,\bar{z}),
\nonu \\
\bigg[ \bigg\{ \hat{\bar{\Phi}}_{+\frac{3}{2},m}^{(3),A},
\hat{\Phi}_{B,-\frac{3}{2},n}^{(3)} \bigg\}, {\cal O}_{+2}^{(h)}(z,\bar{z})
\bigg] & = &
\frac{1}{4}\, z^{m+\frac{1}{2}} \, \bar{z}^{n+\frac{1}{2}} \,
\de^{A}_{B} \,  {\cal O}_{+2}^{(\frac{5}{2}+h)}(z,\bar{z}),
\nonu \\
\bigg[ \bigg[ \hat{\bar{\Phi}}_{+2,m}^{(\frac{7}{2})},
\hat{\Phi}_{-2,n}^{(\frac{7}{2})} \bigg], {\cal O}_{+2}^{(h)}(z,\bar{z})
\bigg] & = &
\frac{1}{4}\, z^{m+\frac{1}{2}} \, \bar{z}^{n+\frac{1}{2}} \,
{\cal O}_{+2}^{(3+h)}(z,\bar{z}).
\label{q3halfexamples}
\eea
Due to the condition $-2 \leq s_3 =s_1+s_2+2 \leq 2$ ($-4 \leq s_1+s_2
\leq 0 $),
the contributions from the second term
of the left-hand side in (\ref{wwO})
vanish, and the only nonzero contributions come from the first term
of the left-hand side in (\ref{wwO}). For example,
for the first case of (\ref{q3halfexamples}),
the sum of helicities between the mode of first current
and the matter field is equal to one and this is not satisfied
because $s_3 = s_1+s_2+2=3$.
Then the first commutator from the first term of
left-hand side in (\ref{wwO})
is given by, from (\ref{comm20OPEs}),
\bea
\bigg[ \hat{\Phi}_{+1,n}^{(\frac{1}{2}),CD}, {\cal O}_{+2}^{(h)}(z,\bar{z})
\bigg]=
\frac{1}{2} \, \bar{z}^{n+\frac{1}{2}}\,
{\cal O}_{+1}^{(\frac{1}{2}+h),CD}(z,\bar{z}).
\label{iinter1}
\eea
Now we calculate the second commutator from the first term of
(\ref{wwO}) and we obtain
\bea
\bigg[ \hat{\bar{\Phi}}_{AB,-1,m}^{(\frac{1}{2})},
{\cal O}_{+1}^{(\frac{3}{2}+h),CD}(z,\bar{z}) 
\bigg] = \frac{1}{2} \, {z}^{m+\frac{1}{2}}\,
\de^{CD}_{AB}\, {\cal O}_{+2}^{(2+h)}(z,\bar{z}).
\label{iinter2}
\eea
Note that there exists a shift ($\frac{1}{2} \rightarrow
\frac{3}{2}$) in the upper index
for the matter field
of the left-hand side in (\ref{iinter2}),
according to (\ref{tworelations}).
Then from (\ref{iinter1}) and (\ref{iinter2}),
we arrive at the final expression in the first equation of
(\ref{q3halfexamples}) with proper shift in the upper index
$(2 \rightarrow 0)$.
The other cases in (\ref{q3halfexamples})
can be analyzed similarly.
By noticing
the upper indices
appearing on the right-hand sides of (\ref{q3halfexamples}),
the left- and right-weights for the matter fields
on the right-hand sides of the second to the seventh
equations are decreased by
$(-\frac{1}{2},-\frac{1}{2})$, $(-1,-1)$,
$(-\frac{3}{2},-\frac{3}{2})$, $(-2,-2)$,
$(-\frac{5}{2},-\frac{5}{2})$ and $(-3,-3)$ respectively
\footnote{In the first equation of (\ref{q3halfexamples}),
there is no change in the weights for the matter field.}.
We can also obtain the similar results for the matter fields
with the helicities $\pm \frac{3}{2}$, $ \pm 1$, $\pm \frac{1}{2}$,
$0$, and $-2$ in Table \ref{table1}.

It would be interesting to obtain the whole
extension of ${\cal N}=8$ ${\rm w}_{1+\infty}$ algebra
(or ${\rm \bar{w}}_{1+\infty}$ algebra) by examining the
actions of (\ref{wwO}) and (\ref{wwOother}) starting with the lower values
of $p$ and $q$ and extracting the algebraic structures from the
differential operators (\ref{Diff}) and (\ref{otherDiff}).
Of course, the whole algebra contains both 
the ${\cal N}=8$ ${\rm w}_{1+\infty}$ algebra and
the ${\cal N}=8$ ${\rm \bar{w}}_{1+\infty}$ algebra. 
The supertranslation generator is common to these algebras
and see also (\ref{BMS}) and (\ref{BMS1}).

\section{The OPEs and (anti)commutators of the ${\cal N}=8$ 
${\rm w}_{1+\infty}$ currents (and ${\rm \bar{w}}_{1+\infty}$
currents) with matter fields
in different bases
\label{lastsection}
}

In this section, we describe the two ${\cal N}=8$ algebras
from the  different viewpoints. 

\subsection{The ${\cal N}=8$ ${\rm w}_{1+\infty}$ algebra}

\subsubsection{The action of the charges on matter fields
from other approach}

Let us multiply both sides of (\ref{OOOPE})
by
\bea
\mbox{Lim}_{\De_1 \rightarrow k } \, (\De_1-k).
\label{Lim}
\eea
The left-hand side of (\ref{OOOPE})
becomes the currents $H^k(z_1,\bar{z}_1)$ associated with the
particles of ${\cal N}=8$ supergravity where the left- and right-weights
are given by $(\frac{k+s_1}{2},\frac{k-s_1}{2})$
with helicity $s_1$.
After multiplying the Gamma function by
(\ref{Lim}) and using the residue of the Gamma function
(from the Euler beta function), the following relation holds
\bea
\mbox{Lim}_{\De_1 \rightarrow k } \, (\De_1-k) \, \Gamma(\De_1-s_1+1+m) =
\frac{(-1)^{ (-k-m+s_1-1)}}{ (-k-m+s_1-1)!},
\label{Res}
\eea
and the OPE of the currents with the matter field can be written as
\bea
&& H^k(z_1,\bar{z}_1) \, {\cal O}_{h,\bar{h}}(z_2,\bar{z}_2)=-
\frac{\kappa_{s_1,s_2,-s_3}}{2}\, \frac{1}{z_{12}} \,
\sum_{m=0}^{-k+s_1-1} \, \frac{(-1)^{-k-m-1+s_1}}{m!\, (-k-m+s_1-1)!}
\nonu \\
&& \times \frac{\Gamma(2\bar{h}+1)}{\Gamma(2\bar{h}+k+m+2-s_1)}
\, \bar{z}_{12}^{m+1}\, \bar{\pa}_{\bar{z}_2}^m \, {\cal O}_{h+\frac{k}{2}+
\frac{1}{2}(s_1-2),\bar{h}+\frac{k}{2}-
\frac{1}{2}(s_1-2)}(z_2,\bar{z}_2)+\cdots,
\label{HOOPE}
\eea
where the left- and right-weights of the matter field
on the right-hand side of (\ref{HOOPE}) with
$(h_2,\bar{h}_2) \equiv (h, \bar{h})$ are used from the relations
\bea
h_1+h-1 = \frac{k+s_1}{2}+h -1, \qquad
\bar{h}_1 + \bar{h} +1= \frac{k-s_1}{2}+\bar{h}+1.
\label{Weights}
\eea
Note that the upper index in the dummy variable $m$
comes from the nonnegative argument of the
factorial function $(-k-m+s_1-1)!$ in the denominator
of (\ref{Res}) or (\ref{HOOPE})
\footnote{
\label{softgraviton}
As found in \cite{Jiang},
the leading, subleading and the sub-sub-leading soft theorems
\cite{AMS,Puhm,Guevara}  
for the gravitons in the context of Ward identities are checked
for $k=1,0,-1$ with $s_1=+2$ respectively.
It is obvious to see that from the upper limit of
the dummy variable in the summation of (\ref{HOOPE}),
each case has, single term (no derivative terms),
two terms (up to first derivative term) and three terms (up to two
derivative term).

That is, for graviton matter field, we obtain
from the first equation of (\ref{twentyfive})
\bea
{{\Phi}}_{+2}^{(-\frac{1}{2})}(z_1,\bar{z}_1) \,
{\cal O}_{+2}^{(h_2)}(z_2,\bar{z}_2) & = &
-\frac{{\kappa}_{+2,+2,-2}}{2}\,
\frac{\bar{z}_{12}}{{z}_{12}} \,
{\cal O}_{+2}^{(-\frac{1}{2}+h_2)}(z_2,\bar{z}_2)+\cdots,
\nonu \\
{{\Phi}}_{+2}^{(0)}(z_1,\bar{z}_1) \,
{\cal O}_{+2}^{(h_2)}(z_2,\bar{z}_2) & = &
\frac{{\kappa}_{+2,+2,-2}}{2}\,
\frac{\bar{z}_{12}^2}{z_{12}}\,
\bigg[ \frac{2(-1-h_2)}{\bar{z}_{12}}- \bar{\pa}_{\bar{z}_2} \bigg]
{\cal O}_{+2}^{(h_2)}(z_2,\bar{z}_2)+\cdots,
\nonu \\
{{\Phi}}_{+2}^{(\frac{1}{2})}(z_1,\bar{z}_1) \,
{\cal O}_{+2}^{(h_2)}(z_2,\bar{z}_2) & = &
-\frac{{\kappa}_{+2,+2,-2}}{4}\,
 \frac{\bar{z}_{12}^3}{{z}_{12}}
\bigg[ \frac{2(-1-h_2)(2(-1-h_2)-1)}{\bar{z}_{12}^2}
\nonu \\
& - &
\frac{4(-1-h_2)\, \bar{\pa}_{\bar{z}_2}}{\bar{z}_{12}}
+\bar{\pa}_{\bar{z}_2}^2 \bigg]
{\cal O}_{+2}^{(\frac{1}{2}+h_2)}(z_2,\bar{z}_2)+\cdots.
\label{Gravishadow}
\eea
The first equation of (\ref{Gravishadow})
corresponds to (\ref{translation}).
Although we use the same notations for the currents,
they are different from each other (\ref{wHrelation}).
The corresponding commutator has $\bar{z}^{n+\frac{1}{2}}$ from the
first equation of (\ref{anticomm25}). Also note
$\bar{\pa}_{\bar{z}}{{\Phi}}_{+2}^{(-\frac{1}{2})}(z,\bar{z})$
is identified with $P(z)$ in \cite{FSTZ1912,FSTZ}
because the corresponding first equation of (\ref{Gravishadow})
contains $\frac{1}{(z-z_2)}$ factor. One can check the weights
on both sides of (\ref{Gravishadow}).
For example, the weights of the left-hand side of
the first equation are given by $(\frac{3}{2},-\frac{1}{2})+
(1-h_2,-1-h_2)$ which can be written as $(1,0)+(0,-1)+
(1-(-\frac{1}{2}+h_2),-1-(-\frac{1}{2}+h_2))$ which are the
weights of the right-hand side.}.
Compared to the work of \cite{HPS},
after we substitute the helicity $s_1=+2$, then the above
equation (\ref{HOOPE}) becomes the $(B.5)$ of \cite{HPS}.
See also the work of \cite{Jiang}.


We present the twenty-five OPEs corresponding to (\ref{HOOPE})
\footnote{We are still using the same notations for the {\it soft} currents
as the ones in the previous sections \ref{OPE}, \ref{anticommutators} and
\ref{twocurrents}. The precise relations are given by
(\ref{wHrelation}).}
\bea
&& \Phi_{+2}^{(-\frac{k}{2})}(z_1,\bar{z}_1) \,
{\cal O}_{+2}^{(h_2)}(z_2,\bar{z}_2)=-
\frac{\kappa_{+2,+2,-2}}{2}\, \frac{1}{z_{12}} \,
\sum_{m=0}^{-k+1} \, \frac{(-1)^{-k-m+1}}{m!\, (-k-m+1)!}
\nonu \\
&& \times \frac{\Gamma(2(-1-h_2)+1)}{\Gamma(2(-1-h_2)+k+m)}
\, \bar{z}_{12}^{m+1}\, \bar{\pa}_{\bar{z}_2}^m \,
{\cal O}_{+2}^{(-\frac{k}{2}+h_2)}(z_2,\bar{z}_2)+\cdots,
\nonu \\
&& \Phi_{+2}^{(-\frac{k}{2})}(z_1,\bar{z}_1) \,
{\cal O}_{+\frac{3}{2}}^{(h_2),A}(z_2,\bar{z}_2)=-
\frac{\kappa_{+2,+\frac{3}{2},-\fft32}}{2}\, \frac{1}{z_{12}} \,
\sum_{m=0}^{-k+1} \, \frac{(-1)^{-k-m+1}}{m!\, (-k-m+1)!}
\nonu \\
&& \times \frac{\Gamma(2(-\frac{1}{2}-h_2)+1)}
{\Gamma(2(-\frac{1}{2}-h_2)+k+m)}
\, \bar{z}_{12}^{m+1}\, \bar{\pa}_{\bar{z}_2}^m \,
{\cal O}_{+\frac{3}{2}}^{(-\frac{k}{2}+h_2),A}(z_2,\bar{z}_2)+\cdots,
\nonu \\
&& \Phi_{+2}^{(-\frac{k}{2})}(z_1,\bar{z}_1) \,
{\cal O}_{+1}^{(h_2),AB}(z_2,\bar{z}_2)=-
\frac{\kappa_{+2,+1,-1}}{2}\, \frac{1}{z_{12}} \,
\sum_{m=0}^{-k+1} \, \frac{(-1)^{-k-m+1}}{m!\, (-k-m+1)!}
\nonu \\
&& \times \frac{\Gamma(2(-h_2)+1)}
{\Gamma(2(-h_2)+k+m)}
\, \bar{z}_{12}^{m+1}\, \bar{\pa}_{\bar{z}_2}^m \,
{\cal O}_{+1}^{(-\frac{k}{2}+h_2),AB}(z_2,\bar{z}_2)+\cdots,
\nonu \\
&& \Phi_{+2}^{(-\frac{k}{2})}(z_1,\bar{z}_1) \,
{\cal O}_{+\frac{1}{2}}^{(h_2),ABC}(z_2,\bar{z}_2)=-
\frac{\kappa_{+2,+\frac{1}{2},-\fft12}}{2}\, \frac{1}{z_{12}} \,
\sum_{m=0}^{-k+1} \, \frac{(-1)^{-k-m+1}}{m!\, (-k-m+1)!}
\nonu \\
&& \times \frac{\Gamma(2(\fft12-h_2)+1)}
{\Gamma(2(\fft12-h_2)+k+m)}
\, \bar{z}_{12}^{m+1}\, \bar{\pa}_{\bar{z}_2}^m \,
{\cal O}_{+\fft12}^{(-\frac{k}{2}+h_2),ABC}(z_2,\bar{z}_2)+\cdots,
\nonu \\
&& \Phi_{+2}^{(-\frac{k}{2})}(z_1,\bar{z}_1) \,
{\cal O}_{0}^{(h_2),ABCD}(z_2,\bar{z}_2)=-
\frac{\kappa_{+2,0,0}}{2}\, \frac{1}{z_{12}} \,
\sum_{m=0}^{-k+1} \, \frac{(-1)^{-k-m+1}}{m!\, (-k-m+1)!}
\nonu \\
&& \times \frac{\Gamma(2(1-h_2)+1)}
{\Gamma(2(1-h_2)+k+m)}
\, \bar{z}_{12}^{m+1}\, \bar{\pa}_{\bar{z}_2}^m \,
{\cal O}_{0}^{(-\frac{k}{2}+h_2),ABCD}(z_2,\bar{z}_2)+\cdots,
\nonu \\
&& \Phi_{+2}^{(-\frac{k}{2})}(z_1,\bar{z}_1) \,
{\cal O}_{ABC,-\frac{1}{2}}^{(h_2)}(z_2,\bar{z}_2)=-
\frac{\kappa_{+2,-\fft12,+\fft12}}{2}\, \frac{1}{z_{12}} \,
\sum_{m=0}^{-k+1} \, \frac{(-1)^{-k-m+1}}{m!\, (-k-m+1)!}
\nonu \\
&& \times \frac{\Gamma(2(\fft32-h_2)+1)}
{\Gamma(2(\fft32-h_2)+k+m)}
\, \bar{z}_{12}^{m+1}\, \bar{\pa}_{\bar{z}_2}^m \,
{\cal O}_{ABC,-\fft12}^{(-\frac{k}{2}+h_2)}(z_2,\bar{z}_2)+\cdots,
\nonu \\
&& \Phi_{+2}^{(-\frac{k}{2})}(z_1,\bar{z}_1) \,
{\cal O}_{AB,-1}^{(h_2)}(z_2,\bar{z}_2)=-
\frac{\kappa_{+2,-1,+1}}{2}\, \frac{1}{z_{12}} \,
\sum_{m=0}^{-k+1} \, \frac{(-1)^{-k-m+1}}{m!\, (-k-m+1)!}
\nonu \\
&& \times \frac{\Gamma(2(2-h_2)+1)}
{\Gamma(2(2-h_2)+k+m)}
\, \bar{z}_{12}^{m+1}\, \bar{\pa}_{\bar{z}_2}^m \,
{\cal O}_{AB,-1}^{(-\frac{k}{2}+h_2)}(z_2,\bar{z}_2)+\cdots,
\nonu \\
&& \Phi_{+2}^{(-\frac{k}{2})}(z_1,\bar{z}_1) \,
{\cal O}_{A,-\fft32}^{(h_2)}(z_2,\bar{z}_2)=-
\frac{\kappa_{+2,-\fft32,+\fft32}}{2}\, \frac{1}{z_{12}} \,
\sum_{m=0}^{-k+1} \, \frac{(-1)^{-k-m+1}}{m!\, (-k-m+1)!}
\nonu \\
&& \times \frac{\Gamma(2(\fft52-h_2)+1)}
{\Gamma(2(\fft52-h_2)+k+m)}
\, \bar{z}_{12}^{m+1}\, \bar{\pa}_{\bar{z}_2}^m \,
{\cal O}_{A,-\fft32}^{(-\frac{k}{2}+h_2)}(z_2,\bar{z}_2)+\cdots,
\nonu \\
&& \Phi_{+2}^{(-\frac{k}{2})}(z_1,\bar{z}_1) \,
{\cal O}_{-2}^{(h_2)}(z_2,\bar{z}_2)=-
\frac{\kappa_{+2,-2,+2}}{2}\, \frac{1}{z_{12}} \,
\sum_{m=0}^{-k+1} \, \frac{(-1)^{-k-m+1}}{m!\, (-k-m+1)!}
\nonu \\
&& \times \frac{\Gamma(2(3-h_2)+1)}
{\Gamma(2(3-h_2)+k+m)}
\, \bar{z}_{12}^{m+1}\, \bar{\pa}_{\bar{z}_2}^m \,
{\cal O}_{-2}^{(-\frac{k}{2}+h_2)}(z_2,\bar{z}_2)+\cdots,
\nonu \\
&& \Phi_{+\fft32}^{(-\frac{k}{2}+\fft14),A}(z_1,\bar{z}_1) \,
{\cal O}_{+\frac{3}{2}}^{(h_2),B}(z_2,\bar{z}_2)=-
\frac{\kappa_{+\frac{3}{2},+\fft32,-1}}{2}\, \frac{1}{z_{12}} \,
\sum_{m=0}^{-k+\fft12} \, \frac{(-1)^{-k-m+\fft12}}{m!\, (-k-m+\fft12)!}
\nonu \\
&& \times \frac{\Gamma(2(-\frac{1}{2}-h_2)+1)}
{\Gamma(2(-\frac{1}{2}-h_2)+k+m+\fft12)}
\, \bar{z}_{12}^{m+1}\, \bar{\pa}_{\bar{z}_2}^m \,
{\cal O}_{+1}^{(-\frac{k}{2}+\fft14+h_2),AB}(z_2,\bar{z}_2)+\cdots,
\nonu \\
&& \Phi_{+\fft32}^{(-\frac{k}{2}+\fft14),A}(z_1,\bar{z}_1) \,
{\cal O}_{+1}^{(h_2),BC}(z_2,\bar{z}_2)=-
\frac{\kappa_{+\frac{3}{2},+1,-\fft12}}{2}\, \frac{1}{z_{12}} \,
\sum_{m=0}^{-k+\fft12} \, \frac{(-1)^{-k-m+\fft12}}{m!\, (-k-m+\fft12)!}
\nonu \\
&& \times \frac{\Gamma(2(-h_2)+1)}
{\Gamma(2(-h_2)+k+m+\fft12)}
\, \bar{z}_{12}^{m+1}\, \bar{\pa}_{\bar{z}_2}^m \,
{\cal O}_{+\fft12}^{(-\frac{k}{2}+\fft14+h_2),ABC}(z_2,\bar{z}_2)+\cdots,
\nonu \\
&& \Phi_{+\fft32}^{(-\frac{k}{2}+\fft14),A}(z_1,\bar{z}_1) \,
{\cal O}_{+\fft12}^{(h_2),BCD}(z_2,\bar{z}_2)=-
\frac{\kappa_{+\frac{3}{2},+\fft12,0}}{2}\, \frac{1}{z_{12}} \,
\sum_{m=0}^{-k+\fft12} \, \frac{(-1)^{-k-m+\fft12}}{m!\, (-k-m+\fft12)!}
\nonu \\
&& \times \frac{\Gamma(2(\fft12-h_2)+1)}
{\Gamma(2(\fft12-h_2)+k+m+\fft12)}
\, \bar{z}_{12}^{m+1}\, \bar{\pa}_{\bar{z}_2}^m \,
{\cal O}_{0}^{(-\frac{k}{2}+\fft14+h_2),ABCD}(z_2,\bar{z}_2)+\cdots,
\nonu \\
&& \Phi_{+\fft32}^{(-\frac{k}{2}+\fft14),A}(z_1,\bar{z}_1) \,
{\cal O}_{0}^{(h_2),BCDE}(z_2,\bar{z}_2)=-
\frac{\kappa_{+\frac{3}{2},0,+\fft12}}{2}\, \frac{1}{z_{12}} \,
\sum_{m=0}^{-k+\fft12} \, \frac{(-1)^{-k-m+\fft12}}{m!\, (-k-m+\fft12)!}
\nonu \\
&& \times \frac{\Gamma(2(1-h_2)+1)}
{\Gamma(2(1-h_2)+k+m+\fft12)}
\, \bar{z}_{12}^{m+1}\,
\frac{1}{3!}\, \ep^{ABCDEFGH}\,
\bar{\pa}_{\bar{z}_2}^m \,
{\cal O}_{FGH,-\fft12}^{(-\frac{k}{2}+\fft14+h_2)}(z_2,\bar{z}_2)+\cdots,
\nonu \\
&& \Phi_{+\fft32}^{(-\frac{k}{2}+\fft14),A}(z_1,\bar{z}_1) \,
{\cal O}_{BCD,-\fft12}^{(h_2)}(z_2,\bar{z}_2)=-
\frac{\kappa_{+\frac{3}{2},-\fft12,+1}}{2}\, \frac{1}{z_{12}} \,
\sum_{m=0}^{-k+\fft12} \, \frac{(-1)^{-k-m+\fft12}}{m!\, (-k-m+\fft12)!}
\nonu \\
&& \times \frac{\Gamma(2(\fft32-h_2)+1)}
{\Gamma(2(\fft32-h_2)+k+m+\fft12)}
\, \bar{z}_{12}^{m+1}\,
3 \de^A_{[B} \,
\bar{\pa}_{\bar{z}_2}^m \,
{\cal O}_{CD],-1}^{(-\frac{k}{2}+\fft14+h_2)}(z_2,\bar{z}_2)+\cdots,
\nonu \\
&& \Phi_{+\fft32}^{(-\frac{k}{2}+\fft14),A}(z_1,\bar{z}_1) \,
{\cal O}_{BC,-1}^{(h_2)}(z_2,\bar{z}_2)=-
\frac{\kappa_{+\frac{3}{2},-1,+\fft32}}{2}\, \frac{1}{z_{12}} \,
\sum_{m=0}^{-k+\fft12} \, \frac{(-1)^{-k-m+\fft12}}{m!\, (-k-m+\fft12)!}
\nonu \\
&& \times \frac{\Gamma(2(2-h_2)+1)}
{\Gamma(2(2-h_2)+k+m+\fft12)}
\, \bar{z}_{12}^{m+1}\,
2! \de^A_{[B} \,
\bar{\pa}_{\bar{z}_2}^m \,
{\cal O}_{C],-\fft32}^{(-\frac{k}{2}+\fft14+h_2)}(z_2,\bar{z}_2)+\cdots,
\nonu \\
&& \Phi_{+\fft32}^{(-\frac{k}{2}+\fft14),A}(z_1,\bar{z}_1) \,
{\cal O}_{B,-\fft32}^{(h_2)}(z_2,\bar{z}_2)=-
\frac{\kappa_{+\frac{3}{2},-\fft32,+2}}{2}\, \frac{1}{z_{12}} \,
\sum_{m=0}^{-k+\fft12} \, \frac{(-1)^{-k-m+\fft12}}{m!\, (-k-m+\fft12)!}
\nonu \\
&& \times \frac{\Gamma(2(\fft52-h_2)+1)}
{\Gamma(2(\fft52-h_2)+k+m+\fft12)}
\, \bar{z}_{12}^{m+1}\,
\de^A_{B} \,
\bar{\pa}_{\bar{z}_2}^m \,
{\cal O}_{-2}^{(-\frac{k}{2}+\fft14+h_2)}(z_2,\bar{z}_2)+\cdots,
\nonu \\
&& \Phi_{+1}^{(-\frac{k}{2}+\fft12),AB}(z_1,\bar{z}_1) \,
{\cal O}_{+1}^{(h_2),CD}(z_2,\bar{z}_2)=-
\frac{\kappa_{+1,+1,0}}{2}\, \frac{1}{z_{12}} \,
\sum_{m=0}^{-k} \, \frac{(-1)^{-k-m}}{m!\, (-k-m)!}
\nonu \\
&& \times \frac{\Gamma(2(-h_2)+1)}
{\Gamma(2(-h_2)+k+m+1)}
\, \bar{z}_{12}^{m+1}\,
\bar{\pa}_{\bar{z}_2}^m \,
{\cal O}_{0}^{(-\frac{k}{2}+\fft12+h_2),ABCD}(z_2,\bar{z}_2)+\cdots,
\nonu \\
&& \Phi_{+1}^{(-\frac{k}{2}+\fft12),AB}(z_1,\bar{z}_1) \,
{\cal O}_{+\fft12}^{(h_2),CDE}(z_2,\bar{z}_2)=-
\frac{\kappa_{+1,+\fft12,+\fft12}}{2}\, \frac{1}{z_{12}} \,
\sum_{m=0}^{-k} \, \frac{(-1)^{-k-m}}{m!\, (-k-m)!}
\nonu \\
&& \times \frac{\Gamma(2(\fft12-h_2)+1)}
{\Gamma(2(\fft12-h_2)+k+m+1)}
\, \bar{z}_{12}^{m+1}\,
\frac{1}{3!}\, \ep^{ABCDEFGH}\,
\bar{\pa}_{\bar{z}_2}^m \,
{\cal O}_{FGH,-\fft12}^{(-\frac{k}{2}+\fft12+h_2)}(z_2,\bar{z}_2)+\cdots,
\nonu \\
&& \Phi_{+1}^{(-\frac{k}{2}+\fft12),AB}(z_1,\bar{z}_1) \,
{\cal O}_{0}^{(h_2),CDEF}(z_2,\bar{z}_2)=-
\frac{\kappa_{+1,0,+1}}{2}\, \frac{1}{z_{12}} \,
\sum_{m=0}^{-k} \, \frac{(-1)^{-k-m}}{m!\, (-k-m)!}
\nonu \\
&& \times \frac{\Gamma(2(1-h_2)+1)}
{\Gamma(2(1-h_2)+k+m+1)}
\, \bar{z}_{12}^{m+1}\,
\frac{1}{2!}\, \ep^{ABCDEFGH}\,
\bar{\pa}_{\bar{z}_2}^m \,
{\cal O}_{GH,-1}^{(-\frac{k}{2}+\fft12+h_2)}(z_2,\bar{z}_2)+\cdots,
\nonu \\
&& \Phi_{+1}^{(-\frac{k}{2}+\fft12),AB}(z_1,\bar{z}_1) \,
{\cal O}_{CDE,-\fft12}^{(h_2)}(z_2,\bar{z}_2)=-
\frac{\kappa_{+1,-\fft12,+\fft32}}{2}\, \frac{1}{z_{12}} \,
\sum_{m=0}^{-k} \, \frac{(-1)^{-k-m}}{m!\, (-k-m)!}
\nonu \\
&& \times \frac{\Gamma(2(\fft32-h_2)+1)}
{\Gamma(2(\fft32-h_2)+k+m+1)}
\, \bar{z}_{12}^{m+1}\,
3!\, \de^{A}_{[C} \,
\bar{\pa}_{\bar{z}_2}^m \,
{\cal O}_{D,-\fft32}^{(-\frac{k}{2}+\fft12+h_2)}(z_2,\bar{z}_2)
\, \de^{B}_{E]}+\cdots,
\nonu \\
&& \Phi_{+1}^{(-\frac{k}{2}+\fft12),AB}(z_1,\bar{z}_1) \,
{\cal O}_{CD,-1}^{(h_2)}(z_2,\bar{z}_2)=-
\frac{\kappa_{+1,-1,+2}}{2}\, \frac{1}{z_{12}} \,
\sum_{m=0}^{-k} \, \frac{(-1)^{-k-m}}{m!\, (-k-m)!}
\nonu \\
&& \times \frac{\Gamma(2(2-h_2)+1)}
{\Gamma(2(2-h_2)+k+m+1)}
\, \bar{z}_{12}^{m+1}\,
\de^{AB}_{CD} \,
\bar{\pa}_{\bar{z}_2}^m \,
{\cal O}_{-2}^{(-\frac{k}{2}+\fft12+h_2)}(z_2,\bar{z}_2)+\cdots,
\nonu \\
&& \Phi_{+\fft12}^{(-\frac{k}{2}+\fft34),ABC}(z_1,\bar{z}_1) \,
{\cal O}_{+\fft12}^{(h_2),DEF}(z_2,\bar{z}_2)=-
\frac{\kappa_{+\fft12,+\fft12,+1}}{2}\, \frac{1}{z_{12}} \,
\sum_{m=0}^{-k-\fft12} \, \frac{(-1)^{-k-m-\fft12}}{m!\,
(-k-m-\fft12)!}
\nonu \\
&& \times \frac{\Gamma(2(\fft12-h_2)+1)}
{\Gamma(2(\fft12-h_2)+k+m+\tf32)}
\, \bar{z}_{12}^{m+1}\,
\frac{1}{2!} \, \ep^{ABCDEFGH}
\bar{\pa}_{\bar{z}_2}^m \,
{\cal O}_{GH,-1}^{(-\frac{k}{2}+\fft34+h_2)}(z_2,\bar{z}_2)+\cdots,
\nonu \\
&& \Phi_{+\fft12}^{(-\frac{k}{2}+\fft34),ABC}(z_1,\bar{z}_1) \,
{\cal O}_{0}^{(h_2),DEFG}(z_2,\bar{z}_2)=-
\frac{\kappa_{+\fft12,0,+\fft32}}{2}\, \frac{1}{z_{12}} \,
\sum_{m=0}^{-k-\fft12} \, \frac{(-1)^{-k-m-\fft12}}{m!\,
(-k-m-\fft12)!}
\nonu \\
&& \times \frac{\Gamma(2(1-h_2)+1)}
{\Gamma(2(1-h_2)+k+m+\tf32)}
\, \bar{z}_{12}^{m+1}\,
\ep^{ABCDEFGH}
\bar{\pa}_{\bar{z}_2}^m \,
{\cal O}_{H,-\fft32}^{(-\frac{k}{2}+\fft34+h_2)}(z_2,\bar{z}_2)+\cdots,
\nonu \\
&& \Phi_{+\fft12}^{(-\frac{k}{2}+\fft34),ABC}(z_1,\bar{z}_1) \,
{\cal O}_{DEF,-\fft12}^{(h_2)}(z_2,\bar{z}_2)=-
\frac{\kappa_{+\fft12,-\fft12,+2}}{2}\, \frac{1}{z_{12}} \,
\sum_{m=0}^{-k-\fft12} \, \frac{(-1)^{-k-m-\fft12}}{m!\,
(-k-m-\fft12)!}
\nonu \\
&& \times \frac{\Gamma(2(\fft32-h_2)+1)}
{\Gamma(2(\fft32-h_2)+k+m+\tf32)}
\, \bar{z}_{12}^{m+1}\,
\de^{ABC}_{DEF}\,
\bar{\pa}_{\bar{z}_2}^m \,
{\cal O}_{-2}^{(-\frac{k}{2}+\fft34+h_2)}(z_2,\bar{z}_2)+\cdots,
\nonu \\
&& \Phi_{0}^{(-\frac{k}{2}+1),ABCD}(z_1,\bar{z}_1) \,
{\cal O}_{0}^{(h_2),EFGH}(z_2,\bar{z}_2)=-
\frac{\kappa_{0,0,+2}}{2}\, \frac{1}{z_{12}} \,
\sum_{m=0}^{-k-1} \, \frac{(-1)^{-k-m-1}}{m!\,
(-k-m-1)!}
\nonu \\
&& \times \frac{\Gamma(2(1-h_2)+1)}
{\Gamma(2(1-h_2)+k+m+2)}
\, \bar{z}_{12}^{m+1}\,
\ep^{ABCDEFGH}\,
\bar{\pa}_{\bar{z}_2}^m \,
{\cal O}_{-2}^{(-\frac{k}{2}+1+h_2)}(z_2,\bar{z}_2)+\cdots.
\label{twentyfive}
\eea
The remaining twenty OPEs are given by (\ref{twenty}) of Appendix F.

Because the $\bar{z}_1$ dependence
appears in the factor $\bar{z}_{12}^{m+1}$,
we can extract the $\bar{z}_1$ dependence of
the current from the right-hand side
of (\ref{HOOPE}) and we obtain the corresponding OPE of
holomorphic part of the current with the matter field as follows.
Now the binomial theorem implies
\bea
(\bar{z}_1-\bar{z}_2)^{m+1} =\sum_{k'=0}^{m+1} \,
\left(\begin{array}{c}
m+1 \\  k' \\
\end{array}\right) \, \bar{z}_1^{k'} \, (-\bar{z}_2)^{m+1-k'}.
\label{binomial}
\eea
Let us introduce the mode $n$ for the current from the dummy variable
$k'$ as follows
\bea
k' \equiv -n -\frac{k-s_1}{2}, \qquad H^k(z_1,\bar{z}_1) =
\sum_{n=\frac{k-s_1}{2}}^{-\frac{k-s_1}{2}} \, \frac{H_n^k(z_1)}{\bar{z}_1^{n+
\frac{k-s_1}{2}}}.
\label{modecurrent}
\eea
Then we arrive at the following OPE of the mode in (\ref{modecurrent})
with the matter field, by simplifying
the binomial coefficient in (\ref{binomial})
and collecting the power of $(-1)$ correctly,
\bea
H_n^k(z_1) \, {\cal O}_{h,\bar{h}}(z_2,\bar{z}_2) & = & -
\frac{\kappa_{s_1,s_2,-s_3}}{2}\, \frac{1}{z_{12}} \,
\sum_{m=0}^{-k+s_1-1} \, \frac{(-1)^{-\frac{k}{2}+n+\frac{s_1}{2}}
\, (m+1)}{ (-k-m+s_1-1)!}
\nonu \\
&\times& \frac{\Gamma(2\bar{h}+1)}{\Gamma(2\bar{h}+k+m+2-s_1)}
\, \frac{ \bar{z}_{2}^{m+\frac{k}{2}+n+
1-\frac{s_1}{2}}}{(\frac{s_1-k}{2}-n)! \, (m+\frac{k}{2}+n+1-
\frac{s_1}{2})!}
\nonu \\
& \times & \bar{\pa}_{\bar{z}_2}^{m}\, {\cal O}_{h+\frac{k}{2}+
\frac{1}{2}(s_1-2),\bar{h}+\frac{k}{2}-
\frac{1}{2}(s_1-2)}(z_2,\bar{z}_2)+\cdots.
\label{HOOPE1}
\eea
This is the $s_1$ dependent generalization of $(B.6)$ of \cite{HPS}.
The $s_1$ dependence appears at nine locations in (\ref{HOOPE1}).

By noting that $\oint_{\bar{z}}\, \frac{d \bar{w}}{2 \pi i} \, \bar{w}^{m+
\frac{k-s_1}{2}-1}\, H^k(w,\bar{w})=H_m^{k}(w)$ in (\ref{wOdef})
and the power of the holomorphic coordinate is equal to zero,
we obtain, after performing the integral over the $w$ in
(\ref{wOdef}) with (\ref{HOOPE1}),
\bea
\bigg[ \hat{H}_n^k, {\cal O}_{h,\bar{h}}(z,\bar{z}) \bigg\} & = & -
\frac{\kappa_{s_1,s_2,-s_3}}{2} \,
\sum_{m=0}^{-k+s_1-1} \, \frac{(-1)^{-\frac{k}{2}+n+\frac{s_1}{2}}\,
(m+1)}{ (-k-m+s_1-1)!}
\nonu \\
&\times& \frac{\Gamma(2\bar{h}+1)}{\Gamma(2\bar{h}+k+m+2-s_1)}
\, \frac{ \bar{z}^{m+\frac{k}{2}+n+
1-\frac{s_1}{2}}}{(\frac{s_1-k}{2}-n)! \, (m+\frac{k}{2}+n+1-
\frac{s_1}{2})!}
\nonu \\
& \times & \bar{\pa}_{\bar{z}}^{m}\, {\cal O}_{h+\frac{k}{2}+
\frac{1}{2}(s_1-2),\bar{h}+\frac{k}{2}-
\frac{1}{2}(s_1-2)}(z,\bar{z})+\cdots.
\label{HOcomm}
\eea
Again, this is the generalization of $(B.8)$ of \cite{HPS}.

We would like to compare (\ref{HOcomm}) with (\ref{anticomm})
and (\ref{Diff}).
Again, from the analysis of \cite{HPS},
we identify the mode in (\ref{anticomm}) with
the mode in (\ref{HOcomm}) as follows:
\bea
\hat{\rm w}_n^q \equiv \frac{1}{ (-1)^{n+q}\,\kappa_{s_1,s_2,-s_3}}\,
\Gamma(q+n) \, \Gamma(q-n)\, \hat{H}_n^{2+s_1-2q}.
\label{wHrelation}
\eea
Note that the upper index of $\hat{H}_n$ depends on the helicity $s_1$
explicitly. See also \cite{Ahn2111,MRSV}.
Then the following relation between
the previous $q$ and the boost weight $k$ holds
\bea
q \equiv 1+\frac{s_1}{2}-\frac{k}{2},
\label{qk}
\eea
by identifying the superscript of the mode of
current on the right-hand side
in (\ref{wHrelation}) with $k$.
Then it is easy to see that
the $\Gamma(q-n)$ in (\ref{wHrelation})
and the first factorial function in the second line of
(\ref{HOcomm}) are canceled out.
We introduce a change of dummy variable
\bea
l \equiv -k+s_1-1-m =2q-3-m,
\label{lm}
\eea
in order to match with the power of (antiholomorphic) derivative
in (\ref{anticomm}).
By using (\ref{qk}) and (\ref{lm}),
the following relation holds
\bea
\left(\begin{array}{c}
q+n-1 \\  l \\
\end{array}\right) =
\left(\begin{array}{c}
\frac{s_1}{2}-\frac{k}{2} +n \\
-k+s_1-1-m  \\
\end{array}\right)=
\frac{(\frac{s_1}{2}-\frac{k}{2} +n )!}{(m+\frac{k}{2}+n+1-
\frac{s_1}{2})! (-k+s_1-1-m)!}.
\nonu \\
\label{BIN}
\eea
This binomial coefficient (\ref{BIN}) is exactly the ratio of
$\Gamma(q+n)$ to the two factorials having
$m$ dependence in (\ref{HOcomm}).

Finally, we arrive at the equation (\ref{anticomm})
together with (\ref{Diff})
\bea
\bigg[
\hat{\rm w}^q_n,
{\cal O}_{h,\bar{h}}(z,\bar{z})   \bigg\}  & = &
{\mathscr D}_n^{q}(\bar{h}) \, {\cal O}_{h+s_1-q,\bar{h}+2-q}(z,\bar{z}),
\label{wOsection6}
\eea
after checking the power of $\bar{z}$ and
the derivative $\bar{\pa}_{\bar{z}}$.
The differential operator in (\ref{wOsection6})
is the same as the one in (\ref{Diff}).
We need to rewrite the left- and right-weights (\ref{Weights}) of the
matter field on the right-hand side in terms of
$q$ rather than $k$ via (\ref{qk}).
Then we can follow the descriptions of (\ref{wwO}), (\ref{wwcomm})
and (\ref{Iden}) and the implication of (\ref{lambdacon})
leads to the action of the ${\cal N}=8$ ${\rm w}_{1+\infty}$ symmetry.

Then the 
algebra from the whole (anti)commutator relations in (\ref{wwcomm}) 
is the complete ${\cal N}=8$ ${\rm w}_{1+\infty}$
algebra found in \cite{AK2509}.
In (\ref{twentyfive}) and (\ref{twenty}),
the whole forty-five OPEs corresponding to (\ref{HOOPE})
are given explicitly and
in (\ref{anticomm25}) and (\ref{anticomm20}),
the forty-five (anti)commutators corresponding to
(\ref{HOcomm}) are presented also.
It is obvious to check the equations from (\ref{wHrelation})
to (\ref{wOsection6}) starting with (\ref{anticomm25}) and
(\ref{anticomm20}).

\subsubsection{Other soft theorems for gravitinos and graviphotons
\label{softfirst}}

In Appendix \ref{lastapp}, the corresponding
equations (\ref{HOOPE}) and (\ref{HOcomm})
for the ${\cal N}=8$ supergravity are presented explicitly
as mentioned before.
In particular, the tenth equation  to the sixteenth equation in
(\ref{twentyfive}) and the first equation of
(\ref{twenty})
provide the leading soft theorems for the gravitinos
\cite{Liu,2212-1,Tropper}
when we consider the $n$-point celestial amplitude in the
${\cal N}=8$ supergravity. See also the footnote \ref{softgraviton}
for soft theorem for graviton case.

For $k=\frac{1}{2}$ in the
above eight cases, each summation has $m=0$ term
and this fact leads to the simple expression
$-\frac{\kappa_{+\frac{3}{2},s_2,-s_3}}{2}
\, \frac{\bar{z}_{12}}{z_{12}}$ acting on each matter field
of the right-hand sides.
For example, the OPE of the gravitino currents with the gravitino matter
fields is given by
\bea
\Phi_{+\fft32}^{(0),A}(z_1,\bar{z}_1) \,
{\cal O}_{+\frac{3}{2}}^{(h_2),B}(z_2,\bar{z}_2)=-
\frac{\kappa_{+\frac{3}{2},+\fft32,-1}}{2}\, \frac{\bar{z}_{12}}{z_{12}} \,
{\cal O}_{+1}^{(h_2),AB}(z_2,\bar{z}_2)+\cdots.
\label{softgravitino}
\eea
The left-weight of the right-hand side of
(\ref{softgravitino}) is given by $1+(1-h_2)$
which is the left-weight of the left-hand side
while the right-weight of the right-hand side is
given by $-1-h_2=-\frac{1}{2}+(-\frac{1}{2}-h_2)$
which is the right-weight of the left-hand side
from Table \ref{table1}. 
Then this OPE (\ref{softgravitino}) can be inserted into the
above $n$-point celestial amplitude where the remaining
$(n-1)$ operators can be any combinations of the nine
matter fields in (\ref{wO}).
The whole contribution in the
leading soft theorems for the gravitinos 
can be obtained by summing over all 
the other contributions from the OPEs of the gravitino currents
with other various matter fields located at $(z_3,\bar{z}_3), \cdots,
(z_{n+1}, \bar{z}_{n+1})$.
If the matter fields are given by the gravitinos, then
we can use the above OPE (\ref{softgravitino}) successively.
Otherwise, we should use other remaining seven OPEs
\footnote{That is, the following OPEs satisfy, at leading order, 
\bea
\Phi_{+\fft32}^{(0),A}(z_1,\bar{z}_1) \,
{\cal O}_{+1}^{(h_2),BC}(z_2,\bar{z}_2) & = & -
\frac{\kappa_{+\frac{3}{2},+1,-\frac{1}{2}}}{2}\, \frac{\bar{z}_{12}}{z_{12}} \,
{\cal O}_{+\frac{1}{2}}^{(h_2),ABC}(z_2,\bar{z}_2)+\cdots,
\nonu \\
\Phi_{+\fft32}^{(0),A}(z_1,\bar{z}_1) \,
{\cal O}_{+\frac{1}{2}}^{(h_2),BCD}(z_2,\bar{z}_2) & = & -
\frac{\kappa_{+\frac{3}{2},+\frac{1}{2},0}}{2}\, \frac{\bar{z}_{12}}{z_{12}} \,
{\cal O}_{0}^{(h_2),ABCD}(z_2,\bar{z}_2)+\cdots,
\nonu \\
\Phi_{+\fft32}^{(0),A}(z_1,\bar{z}_1) \,
{\cal O}_{0}^{(h_2),BCDE}(z_2,\bar{z}_2) & = & -
\frac{\kappa_{+\frac{3}{2},0,+\frac{1}{2}}}{2}\, \frac{\bar{z}_{12}}{z_{12}} \,
\frac{1}{3!} \, \ep^{ABCDEFGH}\,
{\cal O}_{FGH,-\frac{1}{2}}^{(h_2)}(z_2,\bar{z}_2)+\cdots,
\nonu \\
\Phi_{+\fft32}^{(0),A}(z_1,\bar{z}_1) \,
{\cal O}_{BCD,-\frac{1}{2}}^{(h_2)}(z_2,\bar{z}_2) & = & -
\frac{\kappa_{+\frac{3}{2},-\fft12,+1}}{2}\, \frac{\bar{z}_{12}}{z_{12}} \,
3 \, \de^{A}_{[B} \,
{\cal O}_{CD],-1}^{(h_2)}(z_2,\bar{z}_2)+\cdots,
\nonu \\
\Phi_{+\fft32}^{(0),A}(z_1,\bar{z}_1) \,
{\cal O}_{BC,-1}^{(h_2)}(z_2,\bar{z}_2) & = & -
\frac{\kappa_{+\frac{3}{2},-1,+\fft32}}{2}\, \frac{\bar{z}_{12}}{z_{12}} \,
2! \, \de^{A}_{[B}\,
{\cal O}_{C],-\frac{3}{2}}^{(h_2)}(z_2,\bar{z}_2)+\cdots,
\nonu \\
\Phi_{+\fft32}^{(0),A}(z_1,\bar{z}_1) \,
{\cal O}_{B,-\frac{3}{2}}^{(h_2)}(z_2,\bar{z}_2) & = & -
\frac{\kappa_{+\frac{3}{2},-\fft32,+2}}{2}\, \frac{\bar{z}_{12}}{z_{12}} \,
\de^{A}_{B}\,
{\cal O}_{-2}^{(h_2)}(z_2,\bar{z}_2)+\cdots,
\nonu \\
\Phi_{+\fft32}^{(0),A}(z_1,\bar{z}_1) \,
{\cal O}_{+2}^{(h_2)}(z_2,\bar{z}_2) & = & -
\frac{\kappa_{+\frac{3}{2},+2,-\fft32}}{2}\, \frac{\bar{z}_{12}}{z_{12}} \,
{\cal O}_{+\frac{3}{2}}^{(h_2),A}(z_2,\bar{z}_2)+\cdots.
\label{leadinggravitino}
\eea
We can check the consistent
weights of both sides of (\ref{leadinggravitino})
from Table \ref{table1}. Note that
the weights of the current are given by $(\frac{k+s_1}{2},
\frac{k-s_1}{2})$ with $k=\frac{1}{2}$ and $s_1=\frac{3}{2}$.}.

Let us consider the subleading order for the soft theorem for gravitinos.
For $k=-\frac{1}{2}$,
each summation in the above contains
$m=1$ term also.
From the general term in the dummy variable $m$,
the $m=0$ term leads to the coefficient $-2\bar{h}_2$
while the $m=1$ term leads to the coefficient $1$.
By collecting the other factors, we obtain the OPE
as follows
\bea
\Phi_{+\fft32}^{(\frac{1}{2}),A}(z_1,\bar{z}_1) \,
{\cal O}_{+\frac{3}{2}}^{(h_2),B}(z_2,\bar{z}_2) & = & -
\frac{\kappa_{+\frac{3}{2},+\fft32,-1}}{2}\, \frac{\bar{z}_{12}}{z_{12}} \,
\bigg[ -2 (-\tfrac{1}{2}-h_2) +\bar{z}_{12} \, \bar{\pa}_{\bar{z}_2} \bigg] \,
{\cal O}_{+1}^{(\frac{1}{2}+h_2),AB}(z_2,\bar{z}_2)
\nonu \\
& + & \cdots.
\label{softgravitino1}
\eea
Due to the nonzero value
for the upper index of the current on the left-hand side,
the counting of weights are different from the previous
case. That is, the left-weight of the right-hand side of
(\ref{softgravitino1}) is
$1+(1-(\frac{1}{2}+h_2))=\frac{1}{2}+(1-h_2)$
which is the left-weight of the left-hand side and
the right-weight of the right-hand side is
$-1-(\frac{1}{2}+h_2)$ which is the right-weight of the left-hand side.
Note that the right-weight of the matter field
on the left-hand side is given by $(-\tfrac{1}{2}-h_2)$
appearing on the right-hand side of (\ref{softgravitino1})
\footnote{
Then it is straightforword to write down
other seven cases at the subleading order like as
(\ref{leadinggravitino}) by writing down the correct matter fields
both sides, the couplings and the common factors appearing inside
the bracket in (\ref{softgravitino1}) with proper
right-weights of the matter fields.}. The complex coordinate
dependent terms
on the right-hand side in
(\ref{softgravitino1}) can be seen from the equation $(6.4)$ in \cite{FSTZ}. 

Similarly,
the seventeenth equation  to the twenty-first equation in
(\ref{twentyfive}) and the second equation and the ninth equation of
(\ref{twenty})
provide the leading soft theorems for the graviphotons
for the $n$-point celestial amplitude in the
${\cal N}=8$ supergravity.
For $k=0$ in the
above seven cases, each summation has $m=0$ term
and this gives the simple expression
$-\frac{\kappa_{+1,s_2,-s_3}}{2}
\, \frac{\bar{z}_{12}}{z_{12}}$ acting on each matter field
of the right-hand sides.
Then the OPE
of the graviphoton currents
with graviphoton matter fields, at leading-order, can be written as
\bea
\Phi_{+1}^{(\frac{1}{2}),AB}(z_1,\bar{z}_1) \,
{\cal O}_{+1}^{(h_2),CD}(z_2,\bar{z}_2)=-
\frac{\kappa_{+1,+1,0}}{2}\, \frac{\bar{z}_{12}}{z_{12}} \,
{\cal O}_{0}^{(\frac{1}{2}+h_2),ABCD}(z_2,\bar{z}_2)+\cdots.
\label{softphoton}
\eea
The left-weight of the right-hand side is given by
$1+(1-(\frac{1}{2}+h_2))=\frac{1}{2}+(1-h_2)$
which is the left-weight of the left-hand side and
the right-weight of the right-hand side is given by
$-1+(1-(\frac{1}{2}+h_2))=-\frac{1}{2}+(-h_2)$ which is the
right-weight of the left-hand side from Table \ref{table1} \footnote{
We can also express other six OPEs
similar to (\ref{softphoton}) by putting the $SU(8)$ indices
on the right-hand sides with proper matter fields and the couplings
as in the previous cases.}.

For the graviphotinos,
the relevant OPEs are given by the twenty-second,
the twenty-third and the twenty-fourth of
(\ref{twentyfive}) and 
the third equation, the tenth equation and the fifteenth
equation of
(\ref{twenty}).
The substitutions of $k=\frac{1}{2}$ don't produce
any singular behaviors due to the empty sum.
For the scalars,
the relevant OPEs are given by the last equation of
(\ref{twentyfive}) and 
the fourth equation, the eleventh equation, the sixteenth
equation and nineteenth equation of
(\ref{twenty}).
In this case,
the substitutions of $k=1$ don't produce
any singular behaviors for the same reason.

\subsubsection{A shadow transform in the antiholomorphic sector}

It is known that a shadow transform can be obtained
as the consecutive composition of holomorphic (described in section
\ref{holomorphic})
and antiholomorphic (described in sections \ref{OPE},
\ref{anticommutators} and \ref{twocurrents}) light transforms.
As explained in \cite{Jiang}, by changing
the currents $H^k$ in (\ref{HOOPE}) into
its shadow (integral) transform, 
the leading, subleading and the sub-sub-leading soft theorems
for the gravitons in the context of shadow Ward identities are checked
for $k=1,0,-1$ with the helicity $s_1=+2$ respectively.

Let us consider the shadow transform for the gravitino
(and graviphotons and so on) currents \footnote{We are using
the tilde $\tilde{}$ notation for {\it shadow} currents. At the moment,
it is not clear whether there exists an algbra between the shadow currents
for any weights 
as in (\ref{wwcomm}) or (\ref{wwcommother}) or not.
For lower weights, the ${\cal N}=8$ BMS algebra
from the shadow currents exists.}.
We introduce
\bea
\tilde{H}^k(z_1,\bar{z}_1) \equiv \int  d^2 z 
\frac{H^k(z,\bar{z})}{(z-z_1)^{2-2 h_1} 
(\bar{z}-\bar{z}_1)^{2-2\bar{h}_1}} =
\int  d^2 z 
\frac{H^k(z,\bar{z})}{(z-z_1)^{2-s_1-k} 
(\bar{z}-\bar{z}_1)^{2+s_1-k}},
\label{shadow}
\eea
where the weights are
$(h_1,\bar{h}_1)=(\frac{k+s_1}{2},\frac{k-s_1}{2})$.
When we apply the shadow transform (\ref{shadow}) to (\ref{HOOPE}),
we need to calculate the following double integrals
\bea
\int \, d^2 z \,\frac{1}{(z-z_1)^{2-s_1-k} \,
(\bar{z}-\bar{z}_1)^{2+s_1-k}} \, \frac{(\bar{z}-\bar{z}_2)^{m+1}}{(z-z_2)},
\label{twodimint}
\eea
which is given by, using the result of \cite{Jiang},
\bea
\frac{-\pi \, (-1)^{m+1}\, (m+1)! \,
(z_2-z_1)^{-2+s_1+k}\, (\bar{z}_2-\bar{z}_1)^{-2-s_1+k+m+2}}{
(k-2-s_1+1) \, (k-2-s_1+2) \, \cdots \, (k-2-s_1+m+2)}.
\label{integralvalues}
\eea
Note that the three powers appearing in
(\ref{twodimint}) are placed in (\ref{integralvalues})
properly.
The factorial function appearing in the first line of
(\ref{HOOPE}) can be combined with the denominator of
(\ref{integralvalues}) and be simplified further.
We can also simplify the power of $(-1)$ factor appropriately.

Then the OPE of {\it shadow} (gravitinos, graviphotons
and so on) current with various matter fields
can be obtained as 
\bea
&& \tilde{H}^k(z_1,\bar{z}_1) \, {\cal O}_{h,\bar{h}}(z_2,\bar{z}_2)=
\frac{\kappa_{s_1,s_2,-s_3}}{2}\,
\frac{\pi \, (-1)^{k+s_1}}{(1+s_1-k)!}\,
\frac{1}{z_{12}^{2-s_1-k}} \,
\sum_{m=0}^{-k+s_1-1} \bigg[ \, (m+1) \,
\nonu \\
&& \times \frac{\Gamma(2\bar{h}+1)}{\Gamma(2\bar{h}+k+m+2-s_1)}
\, \bar{z}_{12}^{k-s_1+m}\, \bar{\pa}_{\bar{z}_2}^{m} \, {\cal O}_{h+\frac{k}{2}+
\frac{1}{2}(s_1-2),\bar{h}+\frac{k}{2}-
\frac{1}{2}(s_1-2)}(z_2,\bar{z}_2) \bigg] +\cdots.
\label{HOOPEshadow}
\eea
This is a generalization of \cite{Jiang}.
The left-weight of the right-hand side in
(\ref{HOOPEshadow}) can be written as
\bea
(2-s_1-k)+h+\frac{k}{2}+
\frac{1}{2}(s_1-2)= 1-\frac{(k+s_1)}{2}+h.
\label{lw}
\eea
The right-weight of the right-hand side in
(\ref{HOOPEshadow}) can be given by
\bea
-(k-s_1+m)+m+\bar{h}+\frac{k}{2}-
\frac{1}{2}(s_1-2)= 1-\frac{(k-s_1)}{2}+\bar{h}.
\label{rw}
\eea
These (\ref{lw}) and (\ref{rw}) imply that
the left and right-weight  of 
{\it shadow} current is given by
\bea
\bigg(1-\frac{(k+s_1)}{2},1-\frac{(k-s_1)}{2}\bigg)=(1-h_1, 1-\bar{h}_1).
\label{LR}
\eea
If we add these weights, then we obtain $2-(h_1+\bar{h}_1)$.
This implies that the conformal dimension $(h_1+\bar{h}_1)$
before a shadow transform is changed into  the $2-(h_1+\bar{h}_1)$
after a shadow transform.
On the other hand, if we substract these weights, then
we obtain $-(h_1-\bar{h}_1)$ which is the opposite helicity
of the original helicity $(h_1-\bar{h}_1)$ before a shadow transform.


We present the twenty-five OPEs corresponding to (\ref{HOOPEshadow})
\bea
&& \tilde{\Phi}_{+2}^{(-\frac{k}{2})}(z_1,\bar{z}_1) \,
{\cal O}_{+2}^{(h_2)}(z_2,\bar{z}_2)=
\frac{\kappa_{+2,+2,-2}}{2}\, \frac{\pi (-1)^{k}}{(3-k)!}\,
\frac{1}{z_{12}^{-k}} \,
\sum_{m=0}^{-k+1} \, (m+1)
\nonu \\
&& \times \frac{\Gamma(2(-1-h_2)+1)}{\Gamma(2(-1-h_2)+k+m)}
\, \bar{z}_{12}^{k-2+m}\, \bar{\pa}_{\bar{z}_2}^m \,
{\cal O}_{+2}^{(-\frac{k}{2}+h_2)}(z_2,\bar{z}_2)+\cdots,
\nonu \\
&& \tilde{\Phi}_{+2}^{(-\frac{k}{2})}(z_1,\bar{z}_1) \,
{\cal O}_{+\frac{3}{2}}^{(h_2),A}(z_2,\bar{z}_2)=
\frac{\kappa_{+2,+\frac{3}{2},-\fft32}}{2}\,
\frac{\pi (-1)^{k}}{(3-k)!}\,
\frac{1}{z_{12}^{-k}}
\,
\sum_{m=0}^{-k+1} \, (m+1)
\nonu \\
&& \times \frac{\Gamma(2(-\frac{1}{2}-h_2)+1)}
{\Gamma(2(-\frac{1}{2}-h_2)+k+m)}
\, \bar{z}_{12}^{k-2+m}\, \bar{\pa}_{\bar{z}_2}^m \,
{\cal O}_{+\frac{3}{2}}^{(-\frac{k}{2}+h_2),A}(z_2,\bar{z}_2)+\cdots,
\nonu \\
&& \tilde{\Phi}_{+2}^{(-\frac{k}{2})}(z_1,\bar{z}_1) \,
{\cal O}_{+1}^{(h_2),AB}(z_2,\bar{z}_2)=
\frac{\kappa_{+2,+1,-1}}{2}\,
\frac{\pi (-1)^{k}}{(3-k)!}\,
\frac{1}{z_{12}^{-k}}\,
\sum_{m=0}^{-k+1} \, (m+1)
\nonu \\
&& \times \frac{\Gamma(2(-h_2)+1)}
{\Gamma(2(-h_2)+k+m)}
\, \bar{z}_{12}^{k-2+m}\, \bar{\pa}_{\bar{z}_2}^m \,
{\cal O}_{+1}^{(-\frac{k}{2}+h_2),AB}(z_2,\bar{z}_2)+\cdots,
\nonu \\
&& \tilde{\Phi}_{+2}^{(-\frac{k}{2})}(z_1,\bar{z}_1) \,
{\cal O}_{+\frac{1}{2}}^{(h_2),ABC}(z_2,\bar{z}_2)=
\frac{\kappa_{+2,+\frac{1}{2},-\fft12}}{2}\,
\frac{\pi (-1)^{k}}{(3-k)!}\,
\frac{1}{z_{12}^{-k}}
\,
\sum_{m=0}^{-k+1} \, (m+1)
\nonu \\
&& \times \frac{\Gamma(2(\fft12-h_2)+1)}
{\Gamma(2(\fft12-h_2)+k+m)}
\, \bar{z}_{12}^{k-2+m}\, \bar{\pa}_{\bar{z}_2}^m \,
{\cal O}_{+\fft12}^{(-\frac{k}{2}+h_2),ABC}(z_2,\bar{z}_2)+\cdots,
\nonu \\
&& \tilde{\Phi}_{+2}^{(-\frac{k}{2})}(z_1,\bar{z}_1) \,
{\cal O}_{0}^{(h_2),ABCD}(z_2,\bar{z}_2)=
\frac{\kappa_{+2,0,0}}{2}\,
\frac{\pi (-1)^{k}}{(3-k)!}\,
\frac{1}{z_{12}^{-k}}
\,
\sum_{m=0}^{-k+1} \, (m+1)
\nonu \\
&& \times \frac{\Gamma(2(1-h_2)+1)}
{\Gamma(2(1-h_2)+k+m)}
\, \bar{z}_{12}^{k-s1+m}\, \bar{\pa}_{\bar{z}_2}^m \,
{\cal O}_{0}^{(-\frac{k}{2}+h_2),ABCD}(z_2,\bar{z}_2)+\cdots,
\nonu \\
&& \tilde{\Phi}_{+2}^{(-\frac{k}{2})}(z_1,\bar{z}_1) \,
{\cal O}_{ABC,-\frac{1}{2}}^{(h_2)}(z_2,\bar{z}_2)=
\frac{\kappa_{+2,-\fft12,+\fft12}}{2}\,
\frac{\pi (-1)^{k}}{(3-k)!}\,
\frac{1}{z_{12}^{-k}}
\,
\sum_{m=0}^{-k+1} \, (m+1)
\nonu \\
&& \times \frac{\Gamma(2(\fft32-h_2)+1)}
{\Gamma(2(\fft32-h_2)+k+m)}
\, \bar{z}_{12}^{k-2+m}\, \bar{\pa}_{\bar{z}_2}^m \,
{\cal O}_{ABC,-\fft12}^{(-\frac{k}{2}+h_2)}(z_2,\bar{z}_2)+\cdots,
\nonu \\
&& \tilde{\Phi}_{+2}^{(-\frac{k}{2})}(z_1,\bar{z}_1) \,
{\cal O}_{AB,-1}^{(h_2)}(z_2,\bar{z}_2)=
\frac{\kappa_{+2,-1,+1}}{2}\,
\frac{\pi (-1)^{k}}{(3-k)!}\,
\frac{1}{z_{12}^{-k}}
\,
\sum_{m=0}^{-k+1} \, (m+1)
\nonu \\
&& \times \frac{\Gamma(2(2-h_2)+1)}
{\Gamma(2(2-h_2)+k+m)}
\, \bar{z}_{12}^{k-2+m}\, \bar{\pa}_{\bar{z}_2}^m \,
{\cal O}_{AB,-1}^{(-\frac{k}{2}+h_2)}(z_2,\bar{z}_2)+\cdots,
\nonu \\
&& \tilde{\Phi}_{+2}^{(-\frac{k}{2})}(z_1,\bar{z}_1) \,
{\cal O}_{A,-\fft32}^{(h_2)}(z_2,\bar{z}_2)=
\frac{\kappa_{+2,-\fft32,+\fft32}}{2}\,
\frac{\pi (-1)^{k}}{(3-k)!}\,
\frac{1}{z_{12}^{-k}}
\,
\sum_{m=0}^{-k+1} \, (m+1)
\nonu \\
&& \times \frac{\Gamma(2(\fft52-h_2)+1)}
{\Gamma(2(\fft52-h_2)+k+m)}
\, \bar{z}_{12}^{k-2+m}\, \bar{\pa}_{\bar{z}_2}^m \,
{\cal O}_{A,-\fft32}^{(-\frac{k}{2}+h_2)}(z_2,\bar{z}_2)+\cdots,
\nonu \\
&& \tilde{\Phi}_{+2}^{(-\frac{k}{2})}(z_1,\bar{z}_1) \,
{\cal O}_{-2}^{(h_2)}(z_2,\bar{z}_2)=
\frac{\kappa_{+2,-2,+2}}{2}\,
\frac{\pi (-1)^{k}}{(3-k)!}\,
\frac{1}{z_{12}^{-k}}
\,
\sum_{m=0}^{-k+1} \, (m+1)
\nonu \\
&& \times \frac{\Gamma(2(3-h_2)+1)}
{\Gamma(2(3-h_2)+k+m)}
\, \bar{z}_{12}^{k-2+m}\, \bar{\pa}_{\bar{z}_2}^m \,
{\cal O}_{-2}^{(-\frac{k}{2}+h_2)}(z_2,\bar{z}_2)+\cdots,
\nonu \\
&& \tilde{\Phi}_{+\fft32}^{(-\frac{k}{2}+\fft14),A}(z_1,\bar{z}_1) \,
{\cal O}_{+\frac{3}{2}}^{(h_2),B}(z_2,\bar{z}_2)=
\frac{\kappa_{+\frac{3}{2},+\fft32,-1}}{2}\,
\frac{\pi (-1)^{k+\fft32}}{(\fft52-k)!}\,
\frac{1}{z_{12}^{\fft12-k}}
\,
\sum_{m=0}^{-k+\fft12} \, (m+1)
\nonu \\
&& \times \frac{\Gamma(2(-\frac{1}{2}-h_2)+1)}
{\Gamma(2(-\frac{1}{2}-h_2)+k+m+\fft12)}
\, \bar{z}_{12}^{k-\fft32+m}\, \bar{\pa}_{\bar{z}_2}^m \,
{\cal O}_{+1}^{(-\frac{k}{2}+\fft14+h_2),AB}(z_2,\bar{z}_2)+\cdots,
\nonu \\
&& \tilde{\Phi}_{+\fft32}^{(-\frac{k}{2}+\fft14),A}(z_1,\bar{z}_1) \,
{\cal O}_{+1}^{(h_2),BC}(z_2,\bar{z}_2)=
\frac{\kappa_{+\frac{3}{2},+1,-\fft12}}{2}\,
\frac{\pi (-1)^{k+\fft32}}{(\fft52-k)!}\,
\frac{1}{z_{12}^{\fft12-k}}
\,
\sum_{m=0}^{-k+\fft12} \, (m+1)
\nonu \\
&& \times \frac{\Gamma(2(-h_2)+1)}
{\Gamma(2(-h_2)+k+m+\fft12)}
\, \bar{z}_{12}^{k-\fft32+m}\, \bar{\pa}_{\bar{z}_2}^m \,
{\cal O}_{+\fft12}^{(-\frac{k}{2}+\fft14+h_2),ABC}(z_2,\bar{z}_2)+\cdots,
\nonu \\
&& \tilde{\Phi}_{+\fft32}^{(-\frac{k}{2}+\fft14),A}(z_1,\bar{z}_1) \,
{\cal O}_{+\fft12}^{(h_2),BCD}(z_2,\bar{z}_2)=
\frac{\kappa_{+\frac{3}{2},+\fft12,0}}{2}\,
\frac{\pi (-1)^{k+\fft32}}{(\fft52-k)!}\,
\frac{1}{z_{12}^{\fft12-k}}
\,
\sum_{m=0}^{-k+\fft12} \, (m+1)
\nonu \\
&& \times \frac{\Gamma(2(\fft12-h_2)+1)}
{\Gamma(2(\fft12-h_2)+k+m+\fft12)}
\, \bar{z}_{12}^{k-\fft32+m}\, \bar{\pa}_{\bar{z}_2}^m \,
{\cal O}_{0}^{(-\frac{k}{2}+\fft14+h_2),ABCD}(z_2,\bar{z}_2)+\cdots,
\nonu \\
&& \tilde{\Phi}_{+\fft32}^{(-\frac{k}{2}+\fft14),A}(z_1,\bar{z}_1) \,
{\cal O}_{0}^{(h_2),BCDE}(z_2,\bar{z}_2)=
\frac{\kappa_{+\frac{3}{2},0,+\fft12}}{2}\,
\frac{\pi (-1)^{k+\fft32}}{(\fft52-k)!}\,
\frac{1}{z_{12}^{\fft12-k}}
\,
\sum_{m=0}^{-k+\fft12} \, (m+1)
\nonu \\
&& \times \frac{\Gamma(2(1-h_2)+1)}
{\Gamma(2(1-h_2)+k+m+\fft12)}
\, \bar{z}_{12}^{k-\fft32+m}\,
\frac{1}{3!}\, \ep^{ABCDEFGH}\,
\bar{\pa}_{\bar{z}_2}^m \,
{\cal O}_{FGH,-\fft12}^{(-\frac{k}{2}+\fft14+h_2)}(z_2,\bar{z}_2)+\cdots,
\nonu \\
&& \tilde{\Phi}_{+\fft32}^{(-\frac{k}{2}+\fft14),A}(z_1,\bar{z}_1) \,
{\cal O}_{BCD,-\fft12}^{(h_2)}(z_2,\bar{z}_2)=
\frac{\kappa_{+\frac{3}{2},-\fft12,+1}}{2}\,
\frac{\pi (-1)^{k+\fft32}}{(\fft52-k)!}\,
\frac{1}{z_{12}^{\fft12-k}}
\,
\sum_{m=0}^{-k+\fft12} \, (m+1)
\nonu \\
&& \times \frac{\Gamma(2(\fft32-h_2)+1)}{
\Gamma(2(\fft32-h_2)+k+m+\fft12)}
\, \bar{z}_{12}^{k-\fft32+m}\,
3 \de^A_{[B} \, \bar{\pa}_{\bar{z}_2}^m \,
{\cal O}_{CD],-1}^{(-\frac{k}{2}+\fft14+h_2)}(z_2,\bar{z}_2)+\cdots,
\nonu \\
&& \tilde{\Phi}_{+\fft32}^{(-\frac{k}{2}+\fft14),A}(z_1,\bar{z}_1) \,
{\cal O}_{BC,-1}^{(h_2)}(z_2,\bar{z}_2)=
\frac{\kappa_{+\frac{3}{2},-1,+\fft32}}{2}\,
\frac{\pi (-1)^{k+\fft32}}{(\fft52-k)!}\,
\frac{1}{z_{12}^{\fft12-k}}
\,
\sum_{m=0}^{-k+\fft12} \, (m+1)
\nonu \\
&& \times \frac{\Gamma(2(2-h_2)+1)}
{\Gamma(2(2-h_2)+k+m+\fft12)}
\, \bar{z}_{12}^{k-\fft32+m}\,
2! \de^A_{[B} \,
\bar{\pa}_{\bar{z}_2}^m \,
{\cal O}_{C],-\fft32}^{(-\frac{k}{2}+\fft14+h_2)}(z_2,\bar{z}_2)+\cdots,
\nonu \\
&& \tilde{\Phi}_{+\fft32}^{(-\frac{k}{2}+\fft14),A}(z_1,\bar{z}_1) \,
{\cal O}_{B,-\fft32}^{(h_2)}(z_2,\bar{z}_2)=
\frac{\kappa_{+\frac{3}{2},-\fft32,+2}}{2}\,
\frac{\pi (-1)^{k+\fft32}}{(\fft52-k)!}\,
\frac{1}{z_{12}^{\fft12-k}}
\,
\sum_{m=0}^{-k+\fft12} \, (m+1)
\nonu \\
&& \times \frac{\Gamma(2(\fft52-h_2)+1)}
{\Gamma(2(\fft52-h_2)+k+m+\fft12)}
\, \bar{z}_{12}^{k-\fft32+m}\,
\de^A_{B} \,
\bar{\pa}_{\bar{z}_2}^m \,
{\cal O}_{-2}^{(-\frac{k}{2}+\fft14+h_2)}(z_2,\bar{z}_2)+\cdots,
\nonu \\
&& \tilde{\Phi}_{+1}^{(-\frac{k}{2}+\fft12),AB}(z_1,\bar{z}_1) \,
{\cal O}_{+1}^{(h_2),CD}(z_2,\bar{z}_2)=
\frac{\kappa_{+1,+1,0}}{2}\,
\frac{\pi (-1)^{k+1}}{(2-k)!}\,
\frac{1}{z_{12}^{1-k}}
\,
\sum_{m=0}^{-k} \, (m+1)
\nonu \\
&& \times \frac{\Gamma(2(-h_2)+1)}
{\Gamma(2(-h_2)+k+m+1)}
\, \bar{z}_{12}^{k-1+m}\,
\bar{\pa}_{\bar{z}_2}^m \,
{\cal O}_{0}^{(-\frac{k}{2}+\fft12+h_2),ABCD}(z_2,\bar{z}_2)+\cdots,
\nonu \\
&& \tilde{\Phi}_{+1}^{(-\frac{k}{2}+\fft12),AB}(z_1,\bar{z}_1) \,
{\cal O}_{+\fft12}^{(h_2),CDE}(z_2,\bar{z}_2)=
\frac{\kappa_{+1,+\fft12,+\fft12}}{2}\,
\frac{\pi (-1)^{k+1}}{(2-k)!}\,
\frac{1}{z_{12}^{1-k}}
\,
\sum_{m=0}^{-k} \, (m+1)
\nonu \\
&& \times \frac{\Gamma(2(\fft12-h_2)+1)}
{\Gamma(2(\fft12-h_2)+k+m+1)}
\, \bar{z}_{12}^{k-1+m}\,
\frac{1}{3!}\, \ep^{ABCDEFGH}\,
\bar{\pa}_{\bar{z}_2}^m \,
{\cal O}_{FGH,-\fft12}^{(-\frac{k}{2}+\fft12+h_2)}(z_2,\bar{z}_2)+\cdots,
\nonu \\
&& \tilde{\Phi}_{+1}^{(-\frac{k}{2}+\fft12),AB}(z_1,\bar{z}_1) \,
{\cal O}_{0}^{(h_2),CDEF}(z_2,\bar{z}_2)=
\frac{\kappa_{+1,0,+1}}{2}\,
\frac{\pi (-1)^{k+1}}{(2-k)!}\,
\frac{1}{z_{12}^{1-k}}
\,
\sum_{m=0}^{-k} \, (m+1)
\nonu \\
&& \times \frac{\Gamma(2(1-h_2)+1)}
{\Gamma(2(1-h_2)+k+m+1)}
\, \bar{z}_{12}^{k-1+m}\,
\frac{1}{2!}\, \ep^{ABCDEFGH}\,
\bar{\pa}_{\bar{z}_2}^m \,
{\cal O}_{GH,-1}^{(-\frac{k}{2}+\fft12+h_2)}(z_2,\bar{z}_2)+\cdots,
\nonu \\
&& \tilde{\Phi}_{+1}^{(-\frac{k}{2}+\fft12),AB}(z_1,\bar{z}_1) \,
{\cal O}_{CDE,-\fft12}^{(h_2)}(z_2,\bar{z}_2)=
\frac{\kappa_{+1,-\fft12,+\fft32}}{2}\,
\frac{\pi (-1)^{k+1}}{(2-k)!}\,
\frac{1}{z_{12}^{1-k}}
\,
\sum_{m=0}^{-k} \, (m+1)
\nonu \\
&& \times \frac{\Gamma(2(\fft32-h_2)+1)}
{\Gamma(2(\fft32-h_2)+k+m+1)}
\, \bar{z}_{12}^{k-1+m}\,
3!\, \de^{A}_{[C} \,
\bar{\pa}_{\bar{z}_2}^m \,
{\cal O}_{D,-\fft32}^{(-\frac{k}{2}+\fft12+h_2)}(z_2,\bar{z}_2)
\, \de^{B}_{E]}+\cdots,
\nonu \\
&& \tilde{\Phi}_{+1}^{(-\frac{k}{2}+\fft12),AB}(z_1,\bar{z}_1) \,
{\cal O}_{CD,-1}^{(h_2)}(z_2,\bar{z}_2)=
\frac{\kappa_{+1,-1,+2}}{2}\,
\frac{\pi (-1)^{k+1}}{(2-k)!}\,
\frac{1}{z_{12}^{1-k}}
\,
\sum_{m=0}^{-k} \, (m+1)
\nonu \\
&& \times \frac{\Gamma(2(2-h_2)+1)}
{\Gamma(2(2-h_2)+k+m+1)}
\, \bar{z}_{12}^{k-1+m}\,
\de^{AB}_{CD} \,
\bar{\pa}_{\bar{z}_2}^m \,
{\cal O}_{-2}^{(-\frac{k}{2}+\fft12+h_2)}(z_2,\bar{z}_2)+\cdots,
\nonu \\
&& \tilde{\Phi}_{+\fft12}^{(-\frac{k}{2}+\fft34),ABC}(z_1,\bar{z}_1) \,
{\cal O}_{+\fft12}^{(h_2),DEF}(z_2,\bar{z}_2)=
\frac{\kappa_{+\fft12,+\fft12,+1}}{2}\,
\frac{\pi (-1)^{k+\fft12}}{(\fft32-k)!}\,
\frac{1}{z_{12}^{\fft32-k}}
\,
\sum_{m=0}^{-k-\fft12} \, (m+1)
\nonu \\
&& \times \frac{\Gamma(2(\fft12-h_2)+1)}
{\Gamma(2(\fft12-h_2)+k+m+\tf32)}
\, \bar{z}_{12}^{k-\fft12+m}\,
\frac{1}{2!} \, \ep^{ABCDEFGH}
\bar{\pa}_{\bar{z}_2}^m \,
{\cal O}_{GH,-1}^{(-\frac{k}{2}+\fft34+h_2)}(z_2,\bar{z}_2)+\cdots,
\nonu \\
&& \tilde{\Phi}_{+\fft12}^{(-\frac{k}{2}+\fft34),ABC}(z_1,\bar{z}_1) \,
{\cal O}_{0}^{(h_2),DEFG}(z_2,\bar{z}_2)=
\frac{\kappa_{+\fft12,0,+\fft32}}{2}\,
\frac{\pi (-1)^{k+\fft12}}{(\fft32-k)!}\,
\frac{1}{z_{12}^{\fft32-k}}
\,
\sum_{m=0}^{-k-\fft12} \, (m+1)
\nonu \\
&& \times \frac{\Gamma(2(1-h_2)+1)}
{\Gamma(2(1-h_2)+k+m+\tf32)}
\, \bar{z}_{12}^{k-\fft12+m}\,
\ep^{ABCDEFGH}
\bar{\pa}_{\bar{z}_2}^m \,
{\cal O}_{H,-\fft32}^{(-\frac{k}{2}+\fft34+h_2)}(z_2,\bar{z}_2)+\cdots,
\nonu \\
&& \tilde{\Phi}_{+\fft12}^{(-\frac{k}{2}+\fft34),ABC}(z_1,\bar{z}_1) \,
{\cal O}_{DEF,-\fft12}^{(h_2)}(z_2,\bar{z}_2)=
\frac{\kappa_{+\fft12,-\fft12,+2}}{2}\,
\frac{\pi (-1)^{k+\fft12}}{(\fft32-k)!}\,
\frac{1}{z_{12}^{\fft32-k}}
\,
\sum_{m=0}^{-k-\fft12} \, (m+1)
\nonu \\
&& \times \frac{\Gamma(2(\fft32-h_2)+1)}
{\Gamma(2(\fft32-h_2)+k+m+\tf32)}
\, \bar{z}_{12}^{k-\fft12+m}\,
\de^{ABC}_{DEF}\,
\bar{\pa}_{\bar{z}_2}^m \,
{\cal O}_{-2}^{(-\frac{k}{2}+\fft34+h_2)}(z_2,\bar{z}_2)+\cdots,
\nonu \\
&& \tilde{\Phi}_{0}^{(-\frac{k}{2}+1),ABCD}(z_1,\bar{z}_1) \,
{\cal O}_{0}^{(h_2),EFGH}(z_2,\bar{z}_2)=
\frac{\kappa_{0,0,+2}}{2}\,
\frac{\pi (-1)^{k}}{(1-k)!}\,
\frac{1}{z_{12}^{2-k}}
\,
\sum_{m=0}^{-k-1} \, (m+1)
\nonu \\
&& \times \frac{\Gamma(2(1-h_2)+1)}
{\Gamma(2(1-h_2)+k+m+2)}
\, \bar{z}_{12}^{k+m}\,
\ep^{ABCDEFGH}\,
\bar{\pa}_{\bar{z}_2}^m \,
{\cal O}_{-2}^{(-\frac{k}{2}+1+h_2)}(z_2,\bar{z}_2)+\cdots.
\label{twentyfivesection6}
\eea
The remaining twenty OPEs are given by (\ref{twentysection6}) of Appendix F.
In (\ref{twentyfivesection6}) and (\ref{twentysection6}),
the whole forty-five OPEs corresponding to (\ref{HOOPEshadow})
are described explicitly
\footnote{There are
the OPEs of {\it shadow} graviton currents with graviton matter field
at the leading, subleading and sub-sub-leading orders,
\bea
\tilde{{\Phi}}_{+2}^{(-\frac{1}{2})}(z_1,\bar{z}_1) \,
{\cal O}_{+2}^{(h_2)}(z_2,\bar{z}_2) & = &
-\frac{{\kappa}_{+2,+2,-2}}{2}\,
\frac{\pi}{2}\,
\frac{{z}_{12}}{\bar{z}_{12}} \,
{\cal O}_{+2}^{(-\frac{1}{2}+h_2)}(z_2,\bar{z}_2)+\cdots,
\nonu \\
\tilde{{\Phi}}_{+2}^{(0)}(z_1,\bar{z}_1) \,
{\cal O}_{+2}^{(h_2)}(z_2,\bar{z}_2) & = &
\frac{{\kappa}_{+2,+2,-2}}{3!}\,
\pi \,
\bigg[ \frac{(-1-h_2)}{\bar{z}_{12}^2}+ \frac{\bar{\pa}_{\bar{z}_2}}{
\bar{z}_{12}} \bigg]
{\cal O}_{+2}^{(h_2)}(z_2,\bar{z}_2)+\cdots,
\nonu \\
\tilde{{\Phi}}_{+2}^{(\frac{1}{2})}(z_1,\bar{z}_1) \,
{\cal O}_{+2}^{(h_2)}(z_2,\bar{z}_2) & = &
-\frac{{\kappa}_{+2,+2,-2}}{2}\,
\frac{\pi}{4!} \, \frac{1}{{z}_{12}}
\bigg[ \frac{2(-1-h_2)(2(-1-h_2)-1)}{\bar{z}_{12}^3}
\nonu \\
& + &
\frac{2(-1-h_2)\, \bar{\pa}_{\bar{z}_2}}{\bar{z}_{12}^2}
+3\, \frac{\bar{\pa}_{\bar{z}_2}^2}{\bar{z}_{12}} \bigg]
{\cal O}_{+2}^{(\frac{1}{2}+h_2)}(z_2,\bar{z}_2)+\cdots.
\label{Gravitonshadow}
\eea
The {\it shadow} current appearing in the first equation of (\ref{Gravitonshadow})
has the weights $(-\frac{1}{2},\frac{3}{2})$.
Then by taking the holomorphic derivative with respect to
$z_1$, there exists 
\bea
\pa_{z_1}\, \tilde{{\Phi}}_{+2}^{(-\frac{1}{2})}(z_1,\bar{z}_1) \,
{\cal O}_{+2}^{(h_2)}(z_2,\bar{z}_2) & = &
-\frac{{\kappa}_{+2,+2,-2}}{2}\,
\frac{\pi}{2}\,
\frac{1}{\bar{z}_{12}} \,
{\cal O}_{+2}^{(-\frac{1}{2}+h_2)}(z_2,\bar{z}_2)+\cdots.
\nonu
\eea
This descendant
$\pa_{z}\, \tilde{{\Phi}}_{+2}^{(-\frac{1}{2})}(z,\bar{z})$
with weights $(\frac{1}{2},\frac{3}{2})$ plays the role
of supertranslation current $\bar{P}(\bar{z})$
in the notation of \cite{FSTZ}.
See also \cite{2212-2,Tropper}.
The {\it shadow} current
$\tilde{{\Phi}}_{+2}^{(0)}(z,\bar{z})$ in the second equation of
(\ref{Gravitonshadow}) with the weights
$(0,2)$ doesn't depend on the complex coordinate
$z$ from (\ref{shadow}) for $k=0$ and $s_1=+2$
according to naive analysis
and can be identified with the stress energy tensor $\bar{T}(\bar{z})=
\tilde{{\Phi}}_{+2}^{(0)}(z,\bar{z})$ in the notation of \cite{FSTZ}.
See also \cite{KMRS}.
However, the precise analysis gives the nonholomorphic property of
$\tilde{{\Phi}}_{+2}^{(0)}(z,\bar{z})$ \cite{BP,2212-2}.
Then the modified stress-energy tensor provides the usual
commutator relation for the right Virasoro generator.

By following the descriptions in the footnote \ref{BMScheck},
we can write down the two relevant OPEs and  
obtain the corresponding third equation of (\ref{BMS})
where the {\it shadow} currents are substituted. In the standard OPE formalism
between the (modified)
stress energy tensor and the supercurrent, we have
\bea
\bar{T}(\bar{z}_1)\, \bar{S}^A(\bar{z}_2) =\frac{1}{(\bar{z}_1-\bar{z}_2)^2}
\, \frac{3}{2}\,  \bar{S}^A(\bar{z}_2) +
\frac{1}{(\bar{z}_1-\bar{z}_2)}\, \bar{\pa}_{\bar{z}_2} \,   \bar{S}^A(\bar{z}_2)
+ \cdots.
\label{TbarSbar}
\eea
The supercurrents are defined in (\ref{softgravitinoshadow}) and 
see also \cite{FSTZ,2212-2,Tropper} for the same results.
Note that the modified stress energy tensor has
the usual OPE for the primary hard operators.}.

Again,
the tenth equation  to the sixteenth equation in
(\ref{twentyfivesection6}) and the first equation of
(\ref{twentysection6})
lead to  the shadow Ward identities
to the leading soft theorems for the gravitinos
when we consider the $n$-point celestial amplitude in the
${\cal N}=8$ supergravity.
For $k=\frac{1}{2}$ in the
above eight cases, each summation in (\ref{HOOPEshadow})
has only $m=0$ term
and the results become the simple expression
acting on each matter field on the right-hand sides.
For example, the OPE of the {\it shadow}
gravitino currents with gravitino matter
fields is given by
\bea
\tilde{\Phi}_{+\fft32}^{(0),A}(z_1,\bar{z}_1) \,
{\cal O}_{+\frac{3}{2}}^{(h_2),B}(z_2,\bar{z}_2)=
\frac{\kappa_{+\frac{3}{2},+\fft32,-1}}{2}\,
\frac{\pi}{2}\,
\frac{1}{\bar{z}_{12}} \,
{\cal O}_{+1}^{(h_2),AB}(z_2,\bar{z}_2)+\cdots.
\label{softgravitinoshadow}
\eea
The left-weight of the right-hand side of (\ref{softgravitinoshadow})
is given by $(1-h_2)$ from Table \ref{table1}.
Then the left-weight of the {\it shadow} current
on the left thand side is
zero which can be also obtained from (\ref{LR})
with $k=\frac{1}{2}$ and $s_1=\frac{3}{2}$.
This implies that the left-weight of the left-hand side
is given by the above value $(1-h_2)$
which can be seen from Table \ref{table1}.
Similarly, the right-weight of the
right-hand side is $1+(-h_2)=\frac{3}{2}+(-\frac{1}{2}-h_2)$,
This implies that
the right-weight of the {\it shadow} current on the left-hand side is
$\frac{3}{2}$ which can be seen from (\ref{LR}).
The complex coordinate dependent factor
on the right-hand side in
(\ref{softgravitinoshadow})
can be seen from the equation $(6.14)$ in \cite{FSTZ}.

Therefore, $\tilde{\Phi}_{+\fft32}^{(0),A}(\bar{z})\equiv \bar{S}^A(\bar{z})$
with weights
$(0,\frac{3}{2})$ plays the role of one of
the supercurrents of the ${\cal N}=8$ BMS algebra
\footnote{
\label{vanishingLW}
Note that for the vanishing left-weight
of {\it shadow} current, the relation (\ref{LR}) implies
$h_1=1$. Moreover, the definition (\ref{shadow})
leads to the fact that the integrand does not depend on
$z_1$. Then naively speaking,
we can ignore the $z$ dependence in
$\tilde{\Phi}_{+\fft32}^{(0),A}(z,\bar{z})$ completely.
However, as noted in \cite{2212-2},
$\tilde{\Phi}_{+\fft32}^{(0),A}(z,\bar{z})$ is also nonholomorphic.

We emphasize that 
the helicity of  $\tilde{\Phi}_{+\fft32}^{(0),A}(\bar{z})$ is
given by $-\frac{3}{2}$ which is opposite to the subscript
of this current which is the original helicity before a shadow transform.}.

Let us consider the subleading order and
for $k=-\frac{1}{2}$,
each summation in the above contains
$m=1$ term also as before.
By collecting the two contributions, we obtain 
the following OPE
\bea
\tilde{\Phi}_{+\fft32}^{(\frac{1}{2}),A}(z_1,\bar{z}_1) \,
{\cal O}_{+\frac{3}{2}}^{(h_2),B}(z_2,\bar{z}_2) & = & -
\frac{\kappa_{+\frac{3}{2},+\fft32,-1}}{2}\,
\frac{2\,\pi}{3!}\,
\frac{1}{z_{12}} \,
\bigg[ \frac{1}{\bar{z}_{12}^2}(-\tfrac{1}{2}-h_2)
+\frac{1}{\bar{z}_{12}} \, \bar{\pa}_{\bar{z}_2} \bigg] \,
{\cal O}_{+1}^{(\frac{1}{2}+h_2),AB}(z_2,\bar{z}_2)
\nonu \\
& + & \cdots.
\label{softgravitino1shadow}
\eea
Again, the left-weight of the right-hand side is given by
$1+(1-(\frac{1}{2}+h_2))=\frac{1}{2}+(1-h_2)$
which is the left-weight of the left-hand side
and the right-weight of the right-hand side is
given by
$2-(\frac{1}{2}+h_2)=2+(-\frac{1}{2}-h_2)$
which is the right-weight of the left-hand side.
The coefficient appearing in the second-order pole
in the antiholomorphic coordinates (\ref{softgravitino1shadow}) is the
right-weight of the matter field on the left-hand side.

Moreover,
the seventeenth equation  to the twenty-first equation in
(\ref{twentyfive}) and the second equation and the ninth equation of
(\ref{twenty})
can provide the shadow Ward identities
to the leading soft theorems for the graviphotons
for the $n$-point celestial amplitude in the
${\cal N}=8$ supergravity.

For $k=0$ in the
above seven cases, each summation has $m=0$ term and
this gives the simple expression
acting on each matter field on the right-hand sides.
Then the OPE
of the graviphoton {\it shadow} currents
with graviphoton matter fields, at leading-order, can be written as
\bea
\tilde{\Phi}_{+1}^{(\frac{1}{2}),AB}(z_1,\bar{z}_1) \,
{\cal O}_{+1}^{(h_2),CD}(z_2,\bar{z}_2)=-
\frac{\kappa_{+1,+1,0}}{2}\, \frac{\pi}{2!} \,
\frac{1}{{z}_{12}\,  \bar{z}_{12}} \,
{\cal O}_{0}^{(\frac{1}{2}+h_2),ABCD}(z_2,\bar{z}_2)+\cdots.
\label{softphotonshadow}
\eea
The left-weight of the right-hand side of (\ref{softphotonshadow})
is $1+(1-(\frac{1}{2}+h_2))=\frac{1}{2}+(1-h_2)$.
The left-weight of the {\it shadow} current is given by
$\frac{1}{2}$ which can be seen from (\ref{LR}) with $k=0$
and $s_1=1$.
Then the above left-weight is the same as the one of the
left-hand side.
The right-weight of the right-hand side of (\ref{softphotonshadow})
is $1+(1-(\frac{1}{2}+h_2))=\frac{3}{2}-h_2$.
The right-weight of the {\it shadow} current is given by
$\frac{3}{2}$ which can be seen from (\ref{LR}) with $k=0$
and $s_1=1$. We have consistent right-weight on the left-hand side. 

For the graviphotinos,
the relevant OPEs are given by the twenty-second,
the twenty-third and the twenty-fourth of
(\ref{twentyfivesection6}) and 
the third equation, the tenth equation and the fifteenth
equation of
(\ref{twentysection6}).
The substitutions of $k=\frac{1}{2}$ don't provide
any singular behaviors due to the empty sum.
For the scalars,
the relevant OPEs are given by the last equation of
(\ref{twentyfivesection6}) and 
the fourth equation, the eleventh equation, the sixteenth
equation and nineteenth equation of
(\ref{twentysection6}) and 
the substitutions of $k=1$ don't give
any singular behaviors for the same reason.

\subsection{The ${\cal N}=8$ ${\rm \bar{w}}_{1+\infty}$ algebra} 

\subsubsection{The action of the charges on matter fields
from other approach}
Our starting point is
the OPE in (\ref{otherOOOPE}).
The OPE of the currents with the matter field can be written as
\bea
&& \bar{H}^k(z_1,\bar{z}_1) \, {\cal O}_{h,\bar{h}}(z_2,\bar{z}_2)=-
\frac{\bar{\kappa}_{s_1,s_2,-s_3}}{2}\, \frac{1}{\bar{z}_{12}} \,
\sum_{m=0}^{-k-s_1-1} \, \frac{(-1)^{-k-m-1-s_1}}{m!\, (-k-m-s_1-1)!}
\nonu \\
&& \times \frac{\Gamma(2{h}+1)}{\Gamma(2{h}+k+m+2+s_1)}
\, {z}_{12}^{m+1}\, {\pa}_{{z}_2}^m \, {\cal O}_{h+\frac{k}{2}+
\frac{1}{2}(s_1+2),\bar{h}+\frac{k}{2}-
 \frac{1}{2}(s_1+2)}(z_2,\bar{z}_2)+\cdots,
\label{HOOPEholo}
\eea
after multiplying (\ref{Lim}) by the Gamma function
$\Gamma(2h_1+1+m)$ appearing inside the Euler beta function of
(\ref{otherOOOPE}).
The helicity $s_1$ dependence appears in various places.
The left- and right-weights of the matter field
on the right-hand side of (\ref{HOOPEholo})
can be obtained from $(h_1+h+1)$ and $(\bar{h}_1+\bar{h}-1)$
similar to (\ref{Weights}).
 

Similarly,
we describe the twenty-five OPEs corresponding to (\ref{HOOPEholo})
\footnote{We use the same notations for the {\it soft} currents
as the ones in the previous section 
\ref{holomorphic}. The precise relations are given by
(\ref{wHrelation1}).}
\bea
&& \bar{\Phi}_{-2}^{(-\frac{k}{2})}(z_1,\bar{z}_1) \,
{\cal O}_{-2}^{(h_2)}(z_2,\bar{z}_2)=-
\frac{\bar{\kappa}_{-2,-2,+2}}{2}\, \frac{1}{\bar{z}_{12}} \,
\sum_{m=0}^{-k+1} \, \frac{(-1)^{-k-m+1}}{m!\, (-k-m+1)!}
\nonu \\
&& \times \frac{\Gamma(2(-1-h_2)+1)}{\Gamma(2(-1-h_2)+k+m)}
\, {z}_{12}^{m+1}\, {\pa}_{{z}_2}^m \,
{\cal O}_{-2}^{(-\frac{k}{2}+h_2)}(z_2,\bar{z}_2)+\cdots,
\nonu \\
&& \bar{\Phi}_{-2}^{(-\frac{k}{2})}(z_1,\bar{z}_1) \,
{\cal O}_{A,-\frac{3}{2}}^{(h_2)}(z_2,\bar{z}_2)=-
\frac{\bar{\kappa}_{-2,-\frac{3}{2},+\fft32}}{2}\, \frac{1}{\bar{z}_{12}} \,
\sum_{m=0}^{-k+1} \, \frac{(-1)^{-k-m+1}}{m!\, (-k-m+1)!}
\nonu \\
&& \times \frac{\Gamma(2(-\frac{1}{2}-h_2)+1)}
{\Gamma(2(-\frac{1}{2}-h_2)+k+m)}
\, {z}_{12}^{m+1}\, {\pa}_{{z}_2}^m \,
{\cal O}_{A,-\frac{3}{2}}^{(-\frac{k}{2}+h_2)}(z_2,\bar{z}_2)+\cdots,
\nonu \\
&& \bar{\Phi}_{-2}^{(-\frac{k}{2})}(z_1,\bar{z}_1) \,
{\cal O}_{AB,-1}^{(h_2)}(z_2,\bar{z}_2)=-
\frac{\bar{\kappa}_{-2,-1,+1}}{2}\, \frac{1}{\bar{z}_{12}} \,
\sum_{m=0}^{-k+1} \, \frac{(-1)^{-k-m+1}}{m!\, (-k-m+1)!}
\nonu \\
&& \times \frac{\Gamma(2(-h_2)+1)}
{\Gamma(2(-h_2)+k+m)}
\, {z}_{12}^{m+1}\, {\pa}_{{z}_2}^m \,
{\cal O}_{AB,-1}^{(-\frac{k}{2}+h_2)}(z_2,\bar{z}_2)+\cdots,
\nonu \\
&& \bar{\Phi}_{-2}^{(-\frac{k}{2})}(z_1,\bar{z}_1) \,
{\cal O}_{ABC,-\frac{1}{2}}^{(h_2)}(z_2,\bar{z}_2)=-
\frac{\bar{\kappa}_{-2,-\frac{1}{2},+\fft12}}{2}\, \frac{1}{\bar{z}_{12}} \,
\sum_{m=0}^{-k+1} \, \frac{(-1)^{-k-m+1}}{m!\, (-k-m+1)!}
\nonu \\
&& \times \frac{\Gamma(2(\fft12-h_2)+1)}
{\Gamma(2(\fft12-h_2)+k+m)}
\, {z}_{12}^{m+1}\, {\pa}_{{z}_2}^m \,
{\cal O}_{ABC,-\fft12}^{(-\frac{k}{2}+h_2)}(z_2,\bar{z}_2)+\cdots,
\nonu \\
&& \bar{\Phi}_{-2}^{(-\frac{k}{2})}(z_1,\bar{z}_1) \,
{\cal O}_{0}^{(h_2),ABCD}(z_2,\bar{z}_2)=-
\frac{\bar{\kappa}_{-2,0,0}}{2}\, \frac{1}{\bar{z}_{12}} \,
\sum_{m=0}^{-k+1} \, \frac{(-1)^{-k-m+1}}{m!\, (-k-m+1)!}
\nonu \\
&& \times \frac{\Gamma(2(1-h_2)+1)}
{\Gamma(2(1-h_2)+k+m)}
\, {z}_{12}^{m+1}\, {\pa}_{{z}_2}^m \,
{\cal O}_{0}^{(-\frac{k}{2}+h_2),ABCD}(z_2,\bar{z}_2)+\cdots,
\nonu \\
&& \bar{\Phi}_{-2}^{(-\frac{k}{2})}(z_1,\bar{z}_1) \,
{\cal O}_{+\frac{1}{2}}^{(h_2),ABC}(z_2,\bar{z}_2)=-
\frac{\bar{\kappa}_{-2,+\fft12,-\fft12}}{2}\, \frac{1}{\bar{z}_{12}} \,
\sum_{m=0}^{-k+1} \, \frac{(-1)^{-k-m+1}}{m!\, (-k-m+1)!}
\nonu \\
&& \times \frac{\Gamma(2(\fft32-h_2)+1)}
{\Gamma(2(\fft32-h_2)+k+m)}
\, {z}_{12}^{m+1}\, {\pa}_{{z}_2}^m \,
{\cal O}_{+\fft12}^{(-\frac{k}{2}+h_2),ABC}(z_2,\bar{z}_2)+\cdots,
\nonu \\
&& \bar{\Phi}_{-2}^{(-\frac{k}{2})}(z_1,\bar{z}_1) \,
{\cal O}_{+1}^{(h_2),AB}(z_2,\bar{z}_2)=-
\frac{\bar{\kappa}_{-2,+1,-1}}{2}\, \frac{1}{\bar{z}_{12}} \,
\sum_{m=0}^{-k+1} \, \frac{(-1)^{-k-m+1}}{m!\, (-k-m+1)!}
\nonu \\
&& \times \frac{\Gamma(2(2-h_2)+1)}
{\Gamma(2(2-h_2)+k+m)}
\, {z}_{12}^{m+1}\, {\pa}_{{z}_2}^m \,
{\cal O}_{+1}^{(-\frac{k}{2}+h_2),AB}(z_2,\bar{z}_2)+\cdots,
\nonu \\
&& \bar{\Phi}_{-2}^{(-\frac{k}{2})}(z_1,\bar{z}_1) \,
{\cal O}_{+\fft32}^{(h_2),A}(z_2,\bar{z}_2)=-
\frac{\bar{\kappa}_{-2,+\fft32,-\fft32}}{2}\, \frac{1}{\bar{z}_{12}} \,
\sum_{m=0}^{-k+1} \, \frac{(-1)^{-k-m+1}}{m!\, (-k-m+1)!}
\nonu \\
&& \times \frac{\Gamma(2(\fft52-h_2)+1)}
{\Gamma(2(\fft52-h_2)+k+m)}
\, {z}_{12}^{m+1}\, {\pa}_{{z}_2}^m \,
{\cal O}_{+\fft32}^{(-\frac{k}{2}+h_2),A}(z_2,\bar{z}_2)+\cdots,
\nonu \\
&& \bar{\Phi}_{-2}^{(-\frac{k}{2})}(z_1,\bar{z}_1) \,
{\cal O}_{+2}^{(h_2)}(z_2,\bar{z}_2)=-
\frac{\bar{\kappa}_{-2,+2,-2}}{2}\, \frac{1}{\bar{z}_{12}} \,
\sum_{m=0}^{-k+1} \, \frac{(-1)^{-k-m+1}}{m!\, (-k-m+1)!}
\nonu \\
&& \times \frac{\Gamma(2(3-h_2)+1)}
{\Gamma(2(3-h_2)+k+m)}
\, {z}_{12}^{m+1}\, {\pa}_{{z}_2}^m \,
{\cal O}_{+2}^{(-\frac{k}{2}+h_2)}(z_2,\bar{z}_2)+\cdots,
\nonu \\
&& \bar{\Phi}_{A,-\fft32}^{(-\frac{k}{2}+\fft14)}(z_1,\bar{z}_1) \,
{\cal O}_{B,-\frac{3}{2}}^{(h_2)}(z_2,\bar{z}_2)=-
\frac{\bar{\kappa}_{-\frac{3}{2},-\fft32,+1}}{2}\, \frac{1}{\bar{z}_{12}} \,
\sum_{m=0}^{-k+\fft12} \, \frac{(-1)^{-k-m+\fft12}}{m!\, (-k-m+\fft12)!}
\nonu \\
&& \times \frac{\Gamma(2(-\frac{1}{2}-h_2)+1)}
{\Gamma(2(-\frac{1}{2}-h_2)+k+m+\fft12)}
\, {z}_{12}^{m+1}\, {\pa}_{{z}_2}^m \,
{\cal O}_{AB,-1}^{(-\frac{k}{2}+\fft14+h_2)}(z_2,\bar{z}_2)+\cdots,
\nonu \\
&& \bar{\Phi}_{A,-\fft32}^{(-\frac{k}{2}+\fft14)}(z_1,\bar{z}_1) \,
{\cal O}_{BC,-1}^{(h_2)}(z_2,\bar{z}_2)=-
\frac{\bar{\kappa}_{-\frac{3}{2},-1,+\fft12}}{2}\, \frac{1}{\bar{z}_{12}} \,
\sum_{m=0}^{-k+\fft12} \, \frac{(-1)^{-k-m+\fft12}}{m!\, (-k-m+\fft12)!}
\nonu \\
&& \times \frac{\Gamma(2(-h_2)+1)}
{\Gamma(2(-h_2)+k+m+\fft12)}
\, {z}_{12}^{m+1}\, {\pa}_{{z}_2}^m \,
{\cal O}_{ABC,-\fft12}^{(-\frac{k}{2}+\fft14+h_2)}(z_2,\bar{z}_2)+\cdots,
\nonu \\
&& \bar{\Phi}_{A,-\fft32}^{(-\frac{k}{2}+\fft14)}(z_1,\bar{z}_1) \,
{\cal O}_{BCD,-\fft12}^{(h_2)}(z_2,\bar{z}_2)=-
\frac{\bar{\kappa}_{-\frac{3}{2},-\fft12,0}}{2}\, \frac{1}{\bar{z}_{12}} \,
\sum_{m=0}^{-k+\fft12} \, \frac{(-1)^{-k-m+\fft12}}{m!\, (-k-m+\fft12)!}
\nonu \\
&& \times \frac{\Gamma(2(\fft12-h_2)+1)}
{\Gamma(2(\fft12-h_2)+k+m+\fft12)}
\, {z}_{12}^{m+1}\, \frac{1}{4!} \, \ep_{ABCDEFGH}\, {\pa}_{{z}_2}^m \,
{\cal O}_{0}^{(-\frac{k}{2}+\fft14+h_2),EFGH}(z_2,\bar{z}_2)+\cdots,
\nonu \\
&& \bar{\Phi}_{A,-\fft32}^{(-\frac{k}{2}+\fft14)}(z_1,\bar{z}_1) \,
{\cal O}_{0}^{(h_2),BCDE}(z_2,\bar{z}_2)=-
\frac{\bar{\kappa}_{-\frac{3}{2},0,-\fft12}}{2}\, \frac{1}{\bar{z}_{12}} \,
\sum_{m=0}^{-k+\fft12} \, \frac{(-1)^{-k-m+\fft12}}{m!\, (-k-m+\fft12)!}
\nonu \\
&& \times \frac{\Gamma(2(1-h_2)+1)}
{\Gamma(2(1-h_2)+k+m+\fft12)}
\, {z}_{12}^{m+1}\,
\frac{1}{4!}\, \frac{1}{3!}\, \ep^{BCDEIJKL}\, \ep_{AFGHIJKL}
{\pa}_{{z}_2}^m \,
{\cal O}_{+\fft12}^{(-\frac{k}{2}+\fft14+h_2),FGH}(z_2,\bar{z}_2)
\nonu \\
&& +\cdots,
\nonu \\
&& \bar{\Phi}_{A,-\fft32}^{(-\frac{k}{2}+\fft14)}(z_1,\bar{z}_1) \,
{\cal O}_{+\fft12}^{(h_2),BCD}(z_2,\bar{z}_2)=-
\frac{\bar{\kappa}_{-\frac{3}{2},+\fft12,-1}}{2}\, \frac{1}{\bar{z}_{12}} \,
\sum_{m=0}^{-k+\fft12} \, \frac{(-1)^{-k-m+\fft12}}{m!\, (-k-m+\fft12)!}
\nonu \\
&& \times \frac{\Gamma(2(\fft32-h_2)+1)}
{\Gamma(2(\fft32-h_2)+k+m+\fft12)}
\, {z}_{12}^{m+1}\,
3 \de^{[B}_{A} \,
{\pa}_{{z}_2}^m \,
{\cal O}_{+1}^{(-\frac{k}{2}+\fft14+h_2),CD]}(z_2,\bar{z}_2)+\cdots,
\nonu \\
&& \bar{\Phi}_{A,-\fft32}^{(-\frac{k}{2}+\fft14)}(z_1,\bar{z}_1) \,
{\cal O}_{+1}^{(h_2),BC}(z_2,\bar{z}_2)=-
\frac{\bar{\kappa}_{-\frac{3}{2},+1,-\fft32}}{2}\, \frac{1}{\bar{z}_{12}} \,
\sum_{m=0}^{-k+\fft12} \, \frac{(-1)^{-k-m+\fft12}}{m!\, (-k-m+\fft12)!}
\nonu \\
&& \times \frac{\Gamma(2(2-h_2)+1)}
{\Gamma(2(2-h_2)+k+m+\fft12)}
\, {z}_{12}^{m+1}\,
2! \de^{[B}_{A} \,
{\pa}_{{z}_2}^m \,
{\cal O}_{+\fft32}^{(-\frac{k}{2}+\fft14+h_2),C]}(z_2,\bar{z}_2)+\cdots,
\nonu \\
&& \bar{\Phi}_{A,-\fft32}^{(-\frac{k}{2}+\fft14)}(z_1,\bar{z}_1) \,
{\cal O}_{+\fft32}^{(h_2),B}(z_2,\bar{z}_2)=-
\frac{\bar{\kappa}_{-\frac{3}{2},+\fft32,-2}}{2}\, \frac{1}{\bar{z}_{12}} \,
\sum_{m=0}^{-k+\fft12} \, \frac{(-1)^{-k-m+\fft12}}{m!\, (-k-m+\fft12)!}
\nonu \\
&& \times \frac{\Gamma(2(\fft52-h_2)+1)}
{\Gamma(2(\fft52-h_2)+k+m+\fft12)}
\, {z}_{12}^{m+1}\,
\de^B_{A} \,
{\pa}_{{z}_2}^m \,
{\cal O}_{+2}^{(-\frac{k}{2}+\fft14+h_2)}(z_2,\bar{z}_2)+\cdots,
\nonu \\
&& \bar{\Phi}_{AB,-1}^{(-\frac{k}{2}+\fft12)}(z_1,\bar{z}_1) \,
{\cal O}_{CD,-1}^{(h_2)}(z_2,\bar{z}_2)=-
\frac{\bar{\kappa}_{-1,-1,0}}{2}\, \frac{1}{\bar{z}_{12}} \,
\sum_{m=0}^{-k} \, \frac{(-1)^{-k-m}}{m!\, (-k-m)!}
\nonu \\
&& \times \frac{\Gamma(2(-h_2)+1)}
{\Gamma(2(-h_2)+k+m+1)}
\, {z}_{12}^{m+1}\, \frac{1}{4!}\, \ep_{ABCDEFGH}
{\pa}_{{z}_2}^m \,
{\cal O}_{0}^{(-\frac{k}{2}+\fft12+h_2),EFGH}(z_2,\bar{z}_2)+\cdots,
\nonu \\
&& \bar{\Phi}_{AB,-1}^{(-\frac{k}{2}+\fft12)}(z_1,\bar{z}_1) \,
{\cal O}_{CDE,-\fft12}^{(h_2)}(z_2,\bar{z}_2)=-
\frac{\bar{\kappa}_{-1,-\fft12,-\fft12}}{2}\, \frac{1}{\bar{z}_{12}} \,
\sum_{m=0}^{-k} \, \frac{(-1)^{-k-m}}{m!\, (-k-m)!}
\nonu \\
&& \times \frac{\Gamma(2(\fft12-h_2)+1)}
{\Gamma(2(\fft12-h_2)+k+m+1)}
\, {z}_{12}^{m+1}\,
\frac{1}{3!}\, \ep_{ABCDEFGH}\,
{\pa}_{{z}_2}^m \,
{\cal O}_{+\fft12}^{(-\frac{k}{2}+\fft12+h_2),FGH}(z_2,\bar{z}_2)+\cdots,
\nonu \\
&& \bar{\Phi}_{AB,-1}^{(-\frac{k}{2}+\fft12)}(z_1,\bar{z}_1) \,
{\cal O}_{0}^{(h_2),CDEF}(z_2,\bar{z}_2)=-
\frac{\bar{\kappa}_{-1,0,-1}}{2}\, \frac{1}{\bar{z}_{12}} \,
\sum_{m=0}^{-k} \, \frac{(-1)^{-k-m}}{m!\, (-k-m)!}
\nonu \\
&& \times \frac{\Gamma(2(1-h_2)+1)}
{\Gamma(2(1-h_2)+k+m+1)}
\, {z}_{12}^{m+1}\,\frac{1}{4!}\,
\frac{1}{2!}\, \ep^{CDEFIJKL}\, \ep_{ABGHIJKL}
{\pa}_{{z}_2}^m \,
{\cal O}_{+1}^{(-\frac{k}{2}+\fft12+h_2),GH}(z_2,\bar{z}_2)
\nonu \\
&& +\cdots,
\nonu \\
&& \bar{\Phi}_{AB,-1}^{(-\frac{k}{2}+\fft12)}(z_1,\bar{z}_1) \,
{\cal O}_{+\fft12}^{(h_2),CDE}(z_2,\bar{z}_2)=-
\frac{\bar{\kappa}_{-1,+\fft12,-\fft32}}{2}\, \frac{1}{\bar{z}_{12}} \,
\sum_{m=0}^{-k} \, \frac{(-1)^{-k-m}}{m!\, (-k-m)!}
\nonu \\
&& \times \frac{\Gamma(2(\fft32-h_2)+1)}
{\Gamma(2(\fft32-h_2)+k+m+1)}
\, {z}_{12}^{m+1}\,
3!\, \de^{[C}_{A} \,
{\pa}_{{z}_2}^m \,
{\cal O}_{+\fft32}^{(-\frac{k}{2}+\fft12+h_2),D}(z_2,\bar{z}_2)
\, \de^{E]}_{B}+\cdots,
\nonu \\
&& \bar{\Phi}_{AB,-1}^{(-\frac{k}{2}+\fft12)}(z_1,\bar{z}_1) \,
{\cal O}_{+1}^{(h_2),CD}(z_2,\bar{z}_2)=-
\frac{\bar{\kappa}_{-1,+1,-2}}{2}\, \frac{1}{\bar{z}_{12}} \,
\sum_{m=0}^{-k} \, \frac{(-1)^{-k-m}}{m!\, (-k-m)!}
\nonu \\
&& \times \frac{\Gamma(2(2-h_2)+1)}
{\Gamma(2(2-h_2)+k+m+1)}
\, {z}_{12}^{m+1}\,
\de^{CD}_{AB} \,
{\pa}_{{z}_2}^m \,
{\cal O}_{+2}^{(-\frac{k}{2}+\fft12+h_2)}(z_2,\bar{z}_2)+\cdots,
\nonu \\
&& \bar{\Phi}_{ABC,-\fft12}^{(-\frac{k}{2}+\fft34)}(z_1,\bar{z}_1) \,
{\cal O}_{DEF,-\fft12}^{(h_2)}(z_2,\bar{z}_2)=-
\frac{\bar{\kappa}_{-\fft12,-\fft12,-1}}{2}\, \frac{1}{\bar{z}_{12}} \,
\sum_{m=0}^{-k-\fft12} \, \frac{(-1)^{-k-m-\fft12}}{m!\,
(-k-m-\fft12)!}
\nonu \\
&& \times \frac{\Gamma(2(\fft12-h_2)+1)}
{\Gamma(2(\fft12-h_2)+k+m+\tf32)}
\, {z}_{12}^{m+1}\,
\frac{1}{2!} \, \ep_{ABCDEFGH}
{\pa}_{{z}_2}^m \,
{\cal O}_{+1}^{(-\frac{k}{2}+\fft34+h_2),GH}(z_2,\bar{z}_2)+\cdots,
\nonu \\
&& \bar{\Phi}_{ABC,-\fft12}^{(-\frac{k}{2}+\fft34)}(z_1,\bar{z}_1) \,
{\cal O}_{0}^{(h_2),DEFG}(z_2,\bar{z}_2)=-
\frac{\bar{\kappa}_{-\fft12,0,-\fft32}}{2}\, \frac{1}{\bar{z}_{12}} \,
\sum_{m=0}^{-k-\fft12} \, \frac{(-1)^{-k-m-\fft12}}{m!\,
(-k-m-\fft12)!}
\nonu \\
&& \times \frac{\Gamma(2(1-h_2)+1)}
{\Gamma(2(1-h_2)+k+m+\tf32)}
\, {z}_{12}^{m+1}\,\frac{1}{4!}
\ep^{DEFGIJKL}\, \ep_{ABCHIJKL}
{\pa}_{{z}_2}^m \,
{\cal O}_{+\fft32}^{(-\frac{k}{2}+\fft34+h_2),H}(z_2,\bar{z}_2)+\cdots,
\nonu \\
&& \bar{\Phi}_{ABC,-\fft12}^{(-\frac{k}{2}+\fft34)}(z_1,\bar{z}_1) \,
{\cal O}_{+\fft12}^{(h_2),DEF}(z_2,\bar{z}_2)=-
\frac{\bar{\kappa}_{-\fft12,+\fft12,-2}}{2}\, \frac{1}{\bar{z}_{12}} \,
\sum_{m=0}^{-k-\fft12} \, \frac{(-1)^{-k-m-\fft12}}{m!\,
(-k-m-\fft12)!}
\nonu \\
&& \times \frac{\Gamma(2(\fft32-h_2)+1)}
{\Gamma(2(\fft32-h_2)+k+m+\tf32)}
\, {z}_{12}^{m+1}\,
\de^{DEF}_{ABC}\,
{\pa}_{{z}_2}^m \,
{\cal O}_{+2}^{(-\frac{k}{2}+\fft34+h_2)}(z_2,\bar{z}_2)+\cdots,
\nonu \\
&& \bar{\Phi}_{0}^{(-\frac{k}{2}+1),ABCD}(z_1,\bar{z}_1) \,
{\cal O}_{0}^{(h_2),EFGH}(z_2,\bar{z}_2)=-
\frac{\bar{\kappa}_{0,0,-2}}{2}\, \frac{1}{\bar{z}_{12}} \,
\sum_{m=0}^{-k-1} \, \frac{(-1)^{-k-m-1}}{m!\,
(-k-m-1)!}
\nonu \\
&& \times \frac{\Gamma(2(1-h_2)+1)}
{\Gamma(2(1-h_2)+k+m+2)}
\, {z}_{12}^{m+1}\,
\ep^{ABCDEFGH}\,
{\pa}_{{z}_2}^m \,
{\cal O}_{+2}^{(-\frac{k}{2}+1+h_2)}(z_2,\bar{z}_2)+\cdots.
\label{othertwentyfive}
\eea
The remaining twenty OPEs are given by (\ref{othertwenty}) of Appendix F.

We also use the binomial expansion for $z_{12}^{m+1}$
as in (\ref{binomial}). 
As in (\ref{modecurrent}), let us introduce
the mode $n$ for the current
\bea
k' \equiv -n -\frac{k+s_1}{2}, \qquad \bar{H}^k(z_1,\bar{z}_1) =
\sum_{n=\frac{k+s_1}{2}}^{-\frac{k+s_1}{2}} \, \frac{\bar{H}_n^k(\bar{z}_1)}
{{z}_1^{n+
\frac{k+s_1}{2}}}.
\label{modecurrent1}
\eea

Then we obtain the OPE of the mode of the current (\ref{modecurrent1})
with
the matter field as follows
\bea
\bar{H}_n^k(\bar{z}_1) \, {\cal O}_{h,\bar{h}}(z_2,\bar{z}_2) & = & -
\frac{\bar{\kappa}_{s_1,s_2,-s_3}}{2}\, \frac{1}{\bar{z}_{12}} \,
\sum_{m=0}^{-k-s_1-1} \, \frac{(-1)^{-\frac{k}{2}+n-\frac{s_1}{2}}
\, (m+1)}{ (-k-m-s_1-1)!}
\nonu \\
&\times& \frac{\Gamma(2{h}+1)}{\Gamma(2{h}+k+m+2+s_1)}
\, \frac{ {z}_{2}^{m+\frac{k}{2}+n+
1+\frac{s_1}{2}}}{(\frac{-s_1-k}{2}-n)! \, (m+\frac{k}{2}+n+1+
\frac{s_1}{2})!}
\nonu \\
& \times & {\pa}_{{z}_2}^{m}\, {\cal O}_{h+\frac{k}{2}+
\frac{1}{2}(s_1+2),\bar{h}+\frac{k}{2}-
\frac{1}{2}(s_1+2)}(z_2,\bar{z}_2)+\cdots.
\label{HOOPE1holo}
\eea
Again, by introducing
\footnote{
The (anti)commutator, from (\ref{HOOPE1holo}), leads to
\bea
\bigg[ \hat{\bar{H}}_n^k, {\cal O}_{h,\bar{h}}(z,\bar{z}) \bigg\} & = & -
\frac{\bar{\kappa}_{s_1,s_2,-s_3}}{2} \,
\sum_{m=0}^{-k-s_1-1} \, \frac{(-1)^{-\frac{k}{2}+n-\frac{s_1}{2}}\,
(m+1)}{ (-k-m-s_1-1)!}
\nonu \\
&\times& \frac{\Gamma(2{h}+1)}{\Gamma(2{h}+k+m+2+s_1)}
\, \frac{ {z}^{m+\frac{k}{2}+n+
1+\frac{s_1}{2}}}{(\frac{-s_1-k}{2}-n)! \, (m+\frac{k}{2}+n+1+
\frac{s_1}{2})!}
\nonu \\
& \times & {\pa}_{{z}}^{m}\, {\cal O}_{h+\frac{k}{2}+
\frac{1}{2}(s_1+2),\bar{h}+\frac{k}{2}-
\frac{1}{2}(s_1+2)}(z,\bar{z})+\cdots,
\label{HOcomm1}
\eea
corresponding to the previous ones in (\ref{HOcomm}).}
\bea
\hat{ \bar{{\rm w}}}_n^q \equiv \frac{1}{ (-1)^{n+q}\,
\bar{\kappa}_{s_1,s_2,-s_3}}\,
\Gamma(q+n) \, \Gamma(q-n)\, \hat{\bar{H}}_n^{2-s_1-2q},
\label{wHrelation1}
\eea
and using the relation $q \equiv 1-\frac{s_1}{2}-\frac{k}{2}$
similar to (\ref{qk})
together with (\ref{lm}) and (\ref{BIN}) where the $s_1$
is replaced with $-s_1$,
we arrive at the following (anti)commutator (\ref{otheranticomm})
\bea
\bigg[
\hat{ \bar{{\rm w}}}^q_m,
{\cal O}_{h,\bar{h}}(z,\bar{z})   \bigg\}  & = &
\bar{\mathscr D}_m^{q}({h}) \, {\cal O}_{h+2-q,\bar{h}-s_1-q}(z,\bar{z}),
\label{otherwOsection6}
\eea
which can be obtained from (\ref{wHrelation1}) and (\ref{HOcomm1}).
The differential operator in (\ref{otherwOsection6})
is the same as the one in (\ref{otherDiff}).
We need to rewrite the left- and right-weights  of the
matter field on the right-hand side in terms of
$q$ rather than $k$ as before.
Then we can follow the descriptions of (\ref{wwOother}), (\ref{wwcommother})
and (\ref{Idenother}) and the implication of (\ref{oneexampleother})
leads to the action of the ${\cal N}=8$
${\rm \bar{w}}_{1+\infty}$ symmetry associated with
(\ref{Eqs}) and (\ref{20expression}).

That is, in (\ref{Eqs}) and (\ref{20expression}),
the complete ${\cal N}=8$ $\bar{\rm w}_{1+\infty}$
algebra is summarized.
In (\ref{othertwentyfive}) and (\ref{othertwenty}),
the whole forty-five OPEs corresponding to (\ref{HOOPEholo})
are given explicitly and
in (\ref{anticomm25other}) and (\ref{anticomm20other}),
the forty-five (anti)commutators corresponding to
(\ref{HOcomm1}) are presented also.

\subsubsection{Other soft theorems for gravitinos and graviphotons}

In particular, the tenth equation  to the sixteenth equation in
(\ref{othertwentyfive}) and the first equation of
(\ref{othertwenty})
provide the leading soft theorems for the gravitinos
when we consider the $n$-point celestial amplitude in the
${\cal N}=8$ supergravity. See also the footnote \ref{softgraviton}
for soft theorem for graviton case and they are given as follows
\footnote{
That is, for graviton matter field with the negative helicity graviton
current, we obtain
from the first equation of (\ref{othertwentyfive})
\bea
\bar{\Phi}_{-2}^{(-\frac{1}{2})}(z_1,\bar{z}_1) \,
{\cal O}_{-2}^{(h_2)}(z_2,\bar{z}_2) & = &
-\frac{\bar{\kappa}_{-2,-2,+2}}{2}\,
\frac{{z}_{12}}{\bar{z}_{12}} \,
{\cal O}_{-2}^{(-\frac{1}{2}+h_2)}(z_2,\bar{z}_2)+\cdots,
\nonu \\
\bar{\Phi}_{-2}^{(0)}(z_1,\bar{z}_1) \,
{\cal O}_{-2}^{(h_2)}(z_2,\bar{z}_2) & = &
\frac{\bar{\kappa}_{-2,-2,+2}}{2}\,
\frac{{z}_{12}^2}{\bar{z}_{12}}\,
\bigg[ \frac{2(-1-h_2)}{{z}_{12}}- {\pa}_{{z}_2} \bigg]
{\cal O}_{-2}^{(h_2)}(z_2,\bar{z}_2)+\cdots,
\nonu \\
\bar{\Phi}_{-2}^{(\frac{1}{2})}(z_1,\bar{z}_1) \,
{\cal O}_{-2}^{(h_2)}(z_2,\bar{z}_2) & = &
-\frac{\bar{\kappa}_{-2,-2,+2}}{4}\,
 \frac{{z}_{12}^3}{\bar{z}_{12}}
\bigg[ \frac{2(-1-h_2)(2(-1-h_2)-1)}{{z}_{12}^2}
\nonu \\
& - &
\frac{4(-1-h_2)\, {\pa}_{{z}_2}}{{z}_{12}}
+{\pa}_{{z}_2}^2 \bigg]
{\cal O}_{-2}^{(\frac{1}{2}+h_2)}(z_2,\bar{z}_2)+\cdots,
\label{Gravishadow1}
\eea
at the leading, subleading and subsubleading \cite{PRSY}
orders respectively.
The first equation of (\ref{Gravishadow1})
corresponds to (\ref{othertranslation}) with different helicity
of matter field.
Although we use the same notations for the currents,
they are different from each other (\ref{wHrelation1}).
The corresponding commutator has ${z}^{n+\frac{1}{2}}$
from the first equation of (\ref{anticomm25other}).
Note
${\pa}_{{z}}{\bar{\Phi}}_{+2}^{(-\frac{1}{2})}(z,\bar{z})$
is identified with $\bar{P}(\bar{z})$ in \cite{FSTZ1912,FSTZ}
because the corresponding first equation of (\ref{Gravishadow1})
contains $\frac{1}{(\bar{z}-\bar{z}_2)}$ factor.
Compared to (\ref{Gravishadow}), the dependence of the
complex coordinates of the right-hand sides in (\ref{Gravishadow1})
can be obtained also from the replacement $z_{12}
\leftrightarrow \bar{z}_{12}$ in (\ref{Gravishadow}).}.

For $k=\frac{1}{2}$ in the
above eight cases, each summation has $m=0$ term
and this fact leads to the simple expression
$-\frac{\bar{\kappa}_{-\frac{3}{2},s_2,-s_3}}{2}
\, \frac{{z}_{12}}{\bar{z}_{12}}$ acting on each matter field
of the right-hand sides.
For example, the OPE of the
gravitino currents (the helicity $-\frac{3}{2}$) with the gravitino matter
fields is given by
\bea
\bar{\Phi}_{A,-\fft32}^{(0)}(z_1,\bar{z}_1) \,
{\cal O}_{B,-\frac{3}{2}}^{(h_2)}(z_2,\bar{z}_2)=-
\frac{\bar{\kappa}_{-\frac{3}{2},-\fft32,+1}}{2}\,
\frac{{z}_{12}}{\bar{z}_{12}} \,
{\cal O}_{AB,-1}^{(h_2)}(z_2,\bar{z}_2)+\cdots.
\label{softgravitino1-1}
\eea
The left-weight of the right-hand side of
(\ref{softgravitino1-1}) is given by $-1-h_2=-\frac{1}{2}+(-\frac{1}{2}-h_2)$
which is the left-weight of the left-hand side
while the right-weight of the right-hand side is
given by $1+(1-h_2)$
which is the right-weight of the left-hand side
from Table \ref{table1-1}. 
Then this OPE (\ref{softgravitino1-1}) can be inserted into the
above $n$-point celestial amplitude where the remaining
$(n-1)$ operators can be any combinations of the nine
matter fields in (\ref{otherwO}) as before.
If the matter fields are given by the gravitinos, then
we can use the above OPE (\ref{softgravitino1-1}) successively.
Otherwise, we should use other remaining seven OPEs
\footnote{That is, the following OPEs satisfy, at leading order, 
\bea
\bar{\Phi}_{A,-\fft32}^{(0)}(z_1,\bar{z}_1) \,
{\cal O}_{BC,-1}^{(h_2)}(z_2,\bar{z}_2) & = & -
\frac{\bar{\kappa}_{-\frac{3}{2},-1,+\frac{1}{2}}}{2}\,
\frac{{z}_{12}}{\bar{z}_{12}} \,
{\cal O}_{ABC,-\frac{1}{2}}^{(h_2)}(z_2,\bar{z}_2)+\cdots,
\nonu \\
\bar{\Phi}_{A,-\fft32}^{(0)}(z_1,\bar{z}_1) \,
{\cal O}_{BCD,-\frac{1}{2}}^{(h_2)g}(z_2,\bar{z}_2) & = & -
\frac{\bar{\kappa}_{-\frac{3}{2},-\frac{1}{2},0}}{2}\,
\frac{{z}_{12}}{\bar{z}_{12}} \,
\frac{1}{4!}\, \ep_{ABCDEFGH}
{\cal O}_{0}^{(h_2),EFGH}(z_2,\bar{z}_2)+\cdots,
\nonu \\
\bar{\Phi}_{A,-\fft32}^{(0)}(z_1,\bar{z}_1) \,
{\cal O}_{0}^{(h_2),BCDE}(z_2,\bar{z}_2) & = & -
\frac{\bar{\kappa}_{-\frac{3}{2},0,-\frac{1}{2}}}{2}\,
\frac{{z}_{12}}{\bar{z}_{12}} \,
\frac{1}{4!}\,
\frac{1}{3!} \, \ep^{BCDEIJKL}\, \ep_{AFGHIJKL}\,
{\cal O}_{+\frac{1}{2}}^{(h_2),FGH}(z_2,\bar{z}_2)+\cdots,
\nonu \\
\bar{\Phi}_{A,-\fft32}^{(0)}(z_1,\bar{z}_1) \,
{\cal O}_{+\frac{1}{2}}^{(h_2),BCD}(z_2,\bar{z}_2) & = & -
\frac{\bar{\kappa}_{-\frac{3}{2},+\fft12,-1}}{2}\,
\frac{{z}_{12}}{\bar{z}_{12}} \,
3 \, \de^{[B}_{A} \,
{\cal O}_{+1}^{(h_2),CD]}(z_2,\bar{z}_2)+\cdots,
\nonu \\
\bar{\Phi}_{A,-\fft32}^{(0)}(z_1,\bar{z}_1) \,
{\cal O}_{+1}^{(h_2),BC}(z_2,\bar{z}_2) & = & -
\frac{\bar{\kappa}_{-\frac{3}{2},+1,-\fft32}}{2}\,
\frac{{z}_{12}}{\bar{z}_{12}} \,
2! \, \de^{[B}_{A}\,
{\cal O}_{+\frac{3}{2}}^{(h_2),C]}(z_2,\bar{z}_2)+\cdots,
\nonu \\
\bar{\Phi}_{A,-\fft32}^{(0)}(z_1,\bar{z}_1) \,
{\cal O}_{+\frac{3}{2}}^{(h_2),B}(z_2,\bar{z}_2) & = & -
\frac{\bar{\kappa}_{-\frac{3}{2},+\fft32,-2}}{2}\,
\frac{{z}_{12}}{\bar{z}_{12}} \,
\de^{B}_{A}\,
{\cal O}_{+2}^{(h_2)}(z_2,\bar{z}_2)+\cdots,
\nonu \\
\bar{\Phi}_{A,-\fft32}^{(0)}(z_1,\bar{z}_1) \,
{\cal O}_{-2}^{(h_2)}(z_2,\bar{z}_2) & = & -
\frac{\bar{\kappa}_{-\frac{3}{2},-2,+\fft32}}{2}\,
\frac{{z}_{12}}{\bar{z}_{12}} \,
{\cal O}_{A,-\frac{3}{2}}^{(h_2)}(z_2,\bar{z}_2)+\cdots.
\nonu
\eea
We can check the consistent
weights of both sides
from Table \ref{table1-1}.}.

Let us consider the subleading order for the soft theorem for gravitinos.
For $k=-\frac{1}{2}$,
each summation in the above contains
$m=1$ term also.
From the general term in the dummy variable $m$,
the $m=0$ term leads to the coefficient $-2{h}_2$
while the $m=1$ term leads to the coefficient $1$.
By collecting the other factors, we obtain OPE
\bea
\bar{\Phi}_{A,-\fft32}^{(\frac{1}{2})}(z_1,\bar{z}_1) \,
{\cal O}_{B,-\frac{3}{2}}^{(h_2)}(z_2,\bar{z}_2) & = & -
\frac{\bar{\kappa}_{-\frac{3}{2},-\fft32,+1}}{2}\,
\frac{{z}_{12}}{\bar{z}_{12}} \,
\bigg[ -2 (-\tfrac{1}{2}-h_2) +{z}_{12} \, {\pa}_{{z}_2} \bigg] \,
{\cal O}_{AB,-1}^{(\frac{1}{2}+h_2)}(z_2,\bar{z}_2)
\nonu \\
& + & \cdots.
\label{softgravitino2}
\eea
Due to the nonzero value
for the upper index of the current on the left-hand side,
the counting of weights are different from the previous
case. That is, the left-weight of the right-hand side of
(\ref{softgravitino2}) is
$-1-(\frac{1}{2}+h_2)$
which is the left-weight of the left-hand side and
the right-weight of the right-hand side is
$1+(1-(\frac{1}{2}+h_2))=\frac{1}{2}+(1-h_2)$
which is the right-weight of the left-hand side.
Note that the left-weight of the matter field
on the left-hand side is given by $(-\frac{1}{2}-h_2)$
appearing on the right-hand side of (\ref{softgravitino2}).
As before, there exists a simple relation $z_{12} \leftrightarrow
\bar{z}_{12}$ between the right-hand sides of (\ref{softgravitino}) and
(\ref{softgravitino1-1}). This can also apply to (\ref{softgravitino1})
and (\ref{softgravitino2}).

Similarly,
the seventeenth equation  to the twenty-first equation in
(\ref{othertwentyfive}) and the second equation and the ninth equation of
(\ref{othertwenty})
provide the leading soft theorems for the graviphotons
for the $n$-point celestial amplitude in the
${\cal N}=8$ supergravity.
For $k=0$ in the
above seven cases, each summation has $m=0$ term
and this gives the simple expression
$-\frac{\bar{\kappa}_{-1,s_2,-s_3}}{2}
\, \frac{{z}_{12}}{\bar{z}_{12}}$ acting on each matter field
of the right-hand sides.
Then the OPE
of the graviphoton currents (the helicity
$-1$)
with graviphoton matter fields, at leading-order, can be written as
\bea
\bar{\Phi}_{AB,-1}^{(\frac{1}{2})}(z_1,\bar{z}_1) \,
{\cal O}_{CD,-1}^{(h_2)}(z_2,\bar{z}_2) & = & -
\frac{\bar{\kappa}_{-1,-1,0}}{2}\,
\frac{{z}_{12}}{\bar{z}_{12}} \,
\frac{1}{4!}\, \ep_{ABCDEFGH}\,
     {\cal O}_{0}^{(\frac{1}{2}+h_2),EFGH}(z_2,\bar{z}_2)
\nonu \\
    & + & \cdots.
\label{softphoton-1}
\eea
The left-weight of the right-hand side in (\ref{softphoton-1})
is given by
$-1+(1-(\frac{1}{2}+h_2))=-\frac{1}{2}-h_2$
which is the left-weight of the left-hand side and
the right-weight of the right-hand side is given by
$1+(1-(\frac{1}{2}+h_2))=\frac{1}{2}+(1-h_2)$ which is the
right-weight of the left-hand side from Table \ref{table1-1}.

For the graviphotinos and the scalars,
the previous analysis at the end of
subsection \ref{softfirst} can be applied.

\subsubsection{A shadow transform in the holomorphic sector}

Let us consider the shadow transform for the graviton (and other)
currents.
Let us introduce
\bea
\tilde{\bar{H}}^k(z_1,\bar{z}_1) \equiv \int  d^2 z 
\frac{\bar{H}^k(z,\bar{z})}{(z-z_1)^{2-2 {h}_1} 
(\bar{z}-\bar{z}_1)^{2-2\bar{h}_1}} =
\int  d^2 z 
\frac{\bar{H}^k(z,\bar{z})}{(z-z_1)^{2-s_1-k} 
(\bar{z}-\bar{z}_1)^{2+s_1-k}},
\label{sha1}
\eea
where the weights are given by
$(h_1,\bar{h}_1)=(\frac{k+s_1}{2},\frac{k-s_1}{2})$.
When we apply the shadow transform (\ref{sha1}) to (\ref{HOOPEholo}),
the following double integrals are needed
\bea
\int \, d^2 z \,\frac{1}{(z-z_1)^{2-s_1-k} \,
(\bar{z}-\bar{z}_1)^{2+s_1-k}} \, \frac{({z}-{z}_2)^{m+1}}{(\bar{z}-\bar{z}_2)},
\label{doubleint}
\eea
which is given by, according to the formula in \cite{Jiang},
\bea
\frac{-\pi \, (-1)^{m+1}\, (m+1)! \,
(\bar{z}_2-\bar{z}_1)^{-2-s_1+k}\, ({z}_2-{z}_1)^{-2+s_1+k+m+2}}{
(k-2+s_1+1) \, (k-2+s_1+2) \, \cdots \, (k-2+s_1+m+2)}.
\label{integralvalues1}
\eea
As before,
the three powers of (\ref{doubleint})
appear in (\ref{integralvalues1}) appropriately.
The factorial function appearing in the first line of
(\ref{HOOPEholo}) can be combined with the denominator of
(\ref{integralvalues1}) and can be simplified further.
The power of $(-1)$ factor is simplified also.
The final OPE of the {\it shadow} currents with the matter fields is given by
\bea
&& \tilde{\bar{H}}^k(z_1,\bar{z}_1) \, {\cal O}_{h,\bar{h}}(z_2,\bar{z}_2)=
\frac{\bar{\kappa}_{s_1,s_2,-s_3}}{2}\, 
\frac{\pi\, (-1)^{k-s_1}}{(1-s_1-k)!}\,
\frac{1}
{\bar{z}_{12}^{2+s_1-k}} \,
\sum_{m=0}^{-k-s_1-1} \bigg[ (m+1) \,
\nonu \\
&& \times \frac{\Gamma(2{h}+1)}{\Gamma(2{h}+k+m+2+s_1)}
\, {z}_{12}^{k+s_1+m}\, {\pa}_{{z}_2}^{m} \, {\cal O}_{h+\frac{k}{2}+
\frac{1}{2}(s_1+2),\bar{h}+\frac{k}{2}-
\frac{1}{2}(s_1+2)}(z_2,\bar{z}_2) \bigg]+\cdots.
\label{HOOPEshadow1}
\eea
The left-weight of the right-hand side in
(\ref{HOOPEshadow1}) can be written as
\bea
-(k+s_1+m)+m+{h}+\frac{k}{2}+
\frac{1}{2}(s_1+2)= 1-\frac{(k+s_1)}{2}+{h}.
\label{lw1}
\eea
The right-weight of the right-hand side in
(\ref{HOOPEshadow1}) can be given by
\bea
(2+s_1-k)+\bar{h}+\frac{k}{2}-
\frac{1}{2}(s_1+2)= 1-\frac{(k-s_1)}{2}+\bar{h}.
\label{rw1}
\eea
These two relations (\ref{lw1}) and (\ref{rw1}) imply that
the left and right-weight  of 
{\it shadow} current is given by
\bea
\bigg(1-\frac{(k+s_1)}{2},1-\frac{(k-s_1)}{2}\bigg)=
(1-{h}_1, 1-\bar{h}_1).
\label{leftright}
\eea
When we add these weights, then we obtain $2-(h_1+\bar{h}_1)$.
The conformal dimension $(h_1+\bar{h}_1)$ 
before a shadow transform is changed into  the $2-(h_1+\bar{h}_1)$
after a shadow transform.
On the other hand, when we substract these weights, then
we obtain $-(h_1-\bar{h}_1)$ which is the opposite of original helicity
$(h_1-\bar{h}_1)$
before a shadow transform.


Similarly,
we describe the twenty-five OPEs corresponding to (\ref{HOOPEshadow1})
\bea
&& \tilde{\bar{\Phi}}_{-2}^{(-\frac{k}{2})}(z_1,\bar{z}_1) \,
{\cal O}_{-2}^{(h_2)}(z_2,\bar{z}_2)=
\frac{\bar{\kappa}_{-2,-2,+2}}{2}\,
\frac{\pi \, (-1)^{k}}{(3-k)!}
\frac{1}{\bar{z}_{12}^{-k}} \,
\sum_{m=0}^{-k+1} \, (m+1)
\nonu \\
&& \times \frac{\Gamma(2(-1-h_2)+1)}{\Gamma(2(-1-h_2)+k+m)}
\, z_{12}^{k-2+m}\, {\pa}_{{z}_2}^m \,
{\cal O}_{-2}^{(-\frac{k}{2}+h_2)}(z_2,\bar{z}_2)+\cdots,
\nonu \\
&& \tilde{\bar{\Phi}}_{-2}^{(-\frac{k}{2})}(z_1,\bar{z}_1) \,
{\cal O}_{A,-\frac{3}{2}}^{(h_2)}(z_2,\bar{z}_2)=
\frac{\bar{\kappa}_{-2,-\frac{3}{2},+\fft32}}{2}\,
\frac{\pi \, (-1)^{k}}{(3-k)!}
\frac{1}{\bar{z}_{12}^{-k}} \,
\sum_{m=0}^{-k+1} \,
(m+1)
\nonu \\
&& \times \frac{\Gamma(2(-\frac{1}{2}-h_2)+1)}
{\Gamma(2(-\frac{1}{2}-h_2)+k+m)}
\, z_{12}^{k-2+m}\, {\pa}_{{z}_2}^m \,
{\cal O}_{A,-\frac{3}{2}}^{(-\frac{k}{2}+h_2)}(z_2,\bar{z}_2)+\cdots,
\nonu \\
&& \tilde{\bar{\Phi}}_{-2}^{(-\frac{k}{2})}(z_1,\bar{z}_1) \,
{\cal O}_{AB,-1}^{(h_2)}(z_2,\bar{z}_2)=
\frac{\bar{\kappa}_{-2,-1,+1}}{2}\,
\frac{\pi \, (-1)^{k}}{(3-k)!}
\frac{1}{\bar{z}_{12}^{-k}} \,
\sum_{m=0}^{-k+1} \,
(m+1)
\nonu \\
&& \times \frac{\Gamma(2(-h_2)+1)}
{\Gamma(2(-h_2)+k+m)}
\, z_{12}^{k-2+m}\, {\pa}_{{z}_2}^m \,
{\cal O}_{AB,-1}^{(-\frac{k}{2}+h_2)}(z_2,\bar{z}_2)+\cdots,
\nonu \\
&& \tilde{\bar{\Phi}}_{-2}^{(-\frac{k}{2})}(z_1,\bar{z}_1) \,
{\cal O}_{ABC,-\frac{1}{2}}^{(h_2)}(z_2,\bar{z}_2)=
\frac{\bar{\kappa}_{-2,-\frac{1}{2},+\fft12}}{2}\,
\frac{\pi \, (-1)^{k}}{(3-k)!}
\frac{1}{\bar{z}_{12}^{-k}} \,
\sum_{m=0}^{-k+1} \,
(m+1)
\nonu \\
&& \times \frac{\Gamma(2(\fft12-h_2)+1)}
{\Gamma(2(\fft12-h_2)+k+m)}
\, z_{12}^{k-2+m}\, {\pa}_{{z}_2}^m \,
{\cal O}_{ABC,-\fft12}^{(-\frac{k}{2}+h_2)}(z_2,\bar{z}_2)+\cdots,
\nonu \\
&& \tilde{\bar{\Phi}}_{-2}^{(-\frac{k}{2})}(z_1,\bar{z}_1) \,
{\cal O}_{0}^{(h_2),ABCD}(z_2,\bar{z}_2)=
\frac{\bar{\kappa}_{-2,0,0}}{2}\,
\frac{\pi \, (-1)^{k}}{(3-k)!}
\frac{1}{\bar{z}_{12}^{-k}} \,
\sum_{m=0}^{-k+1} \,
(m+1)
\nonu \\
&& \times \frac{\Gamma(2(1-h_2)+1)}
{\Gamma(2(1-h_2)+k+m)}
\, z_{12}^{k-2+m}\, {\pa}_{{z}_2}^m \,
{\cal O}_{0}^{(-\frac{k}{2}+h_2),ABCD}(z_2,\bar{z}_2)+\cdots,
\nonu \\
&& \tilde{\bar{\Phi}}_{-2}^{(-\frac{k}{2})}(z_1,\bar{z}_1) \,
{\cal O}_{+\frac{1}{2}}^{(h_2),ABC}(z_2,\bar{z}_2)=
\frac{\bar{\kappa}_{-2,+\fft12,-\fft12}}{2}\,
\frac{\pi \, (-1)^{k}}{(3-k)!}
\frac{1}{\bar{z}_{12}^{-k}} \,
\sum_{m=0}^{-k+1} \,
(m+1)
\nonu \\
&& \times \frac{\Gamma(2(\fft32-h_2)+1)}
{\Gamma(2(\fft32-h_2)+k+m)}
\, z_{12}^{k-2+m}\, {\pa}_{{z}_2}^m \,
{\cal O}_{+\fft12}^{(-\frac{k}{2}+h_2),ABC}(z_2,\bar{z}_2)+\cdots,
\nonu \\
&& \tilde{\bar{\Phi}}_{-2}^{(-\frac{k}{2})}(z_1,\bar{z}_1) \,
{\cal O}_{+1}^{(h_2),AB}(z_2,\bar{z}_2)=
\frac{\bar{\kappa}_{-2,+1,-1}}{2}\,
\frac{\pi \, (-1)^{k}}{(3-k)!}
\frac{1}{\bar{z}_{12}^{-k}} \,
\sum_{m=0}^{-k+1} \,
(m+1)
\nonu \\
&& \times \frac{\Gamma(2(2-h_2)+1)}
{\Gamma(2(2-h_2)+k+m)}
\, z_{12}^{k-2+m}\, {\pa}_{{z}_2}^m \,
{\cal O}_{+1}^{(-\frac{k}{2}+h_2),AB}(z_2,\bar{z}_2)+\cdots,
\nonu \\
&& \tilde{\bar{\Phi}}_{-2}^{(-\frac{k}{2})}(z_1,\bar{z}_1) \,
{\cal O}_{+\fft32}^{(h_2),A}(z_2,\bar{z}_2)=
\frac{\bar{\kappa}_{-2,+\fft32,-\fft32}}{2}\,
\frac{\pi \, (-1)^{k}}{(3-k)!}
\frac{1}{\bar{z}_{12}^{-k}} \,
\sum_{m=0}^{-k+1} \,
(m+1)
\nonu \\
&& \times \frac{\Gamma(2(\fft52-h_2)+1)}
{\Gamma(2(\fft52-h_2)+k+m)}
\, z_{12}^{k-2+m}\, {\pa}_{{z}_2}^m \,
{\cal O}_{+\fft32}^{(-\frac{k}{2}+h_2),A}(z_2,\bar{z}_2)+\cdots,
\nonu \\
&& \tilde{\bar{\Phi}}_{-2}^{(-\frac{k}{2})}(z_1,\bar{z}_1) \,
      {\cal O}_{+2}^{(h_2)}(z_2,\bar{z}_2)=
      \frac{\bar{\kappa}_{-2,+2,-2}}{2}\,
\frac{\pi \, (-1)^{k}}{(3-k)!}
\frac{1}{\bar{z}_{12}^{-k}} \,
\sum_{m=0}^{-k+1} \,
(m+1)
\nonu \\
&& \times \frac{\Gamma(2(3-h_2)+1)}
{\Gamma(2(3-h_2)+k+m)}
\, z_{12}^{k-2+m}\, {\pa}_{{z}_2}^m \,
{\cal O}_{+2}^{(-\frac{k}{2}+h_2)}(z_2,\bar{z}_2)+\cdots,
\nonu \\
&& \tilde{\bar{\Phi}}_{A,-\fft32}^{(-\frac{k}{2}+\fft14)}(z_1,\bar{z}_1) \,
{\cal O}_{B,-\frac{3}{2}}^{(h_2)}(z_2,\bar{z}_2)=
\frac{\bar{\kappa}_{-\frac{3}{2},-\fft32,+1}}{2}\,
\frac{\pi \, (-1)^{k+\fft32}}{(\fft52-k)!}
\frac{1}{\bar{z}_{12}^{\fft12-k}} \,
\sum_{m=0}^{-k+\fft12} \,
(m+1)
\nonu \\
&& \times \frac{\Gamma(2(-\frac{1}{2}-h_2)+1)}
{\Gamma(2(-\frac{1}{2}-h_2)+k+m+\fft12)}
\, z_{12}^{k-\fft32+m}\, {\pa}_{{z}_2}^m \,
{\cal O}_{AB,-1}^{(-\frac{k}{2}+\fft14+h_2)}(z_2,\bar{z}_2)+\cdots,
\nonu \\
&& \tilde{\bar{\Phi}}_{A,-\fft32}^{(-\frac{k}{2}+\fft14)}(z_1,\bar{z}_1) \,
{\cal O}_{BC,-1}^{(h_2)}(z_2,\bar{z}_2)=
\frac{\bar{\kappa}_{-\frac{3}{2},-1,+\fft12}}{2}\,
\frac{\pi \, (-1)^{k+\fft32}}{(\fft52-k)!}
\frac{1}{\bar{z}_{12}^{\fft12-k}} \,
\sum_{m=0}^{-k+\fft12} \,
(m+1)
\nonu \\
&& \times \frac{\Gamma(2(-h_2)+1)}
{\Gamma(2(-h_2)+k+m+\fft12)}
\, z_{12}^{k-\fft32+m}\, {\pa}_{{z}_2}^m \,
{\cal O}_{ABC,-\fft12}^{(-\frac{k}{2}+\fft14+h_2)}(z_2,\bar{z}_2)+\cdots,
\nonu \\
&& \tilde{\bar{\Phi}}_{A,-\fft32}^{(-\frac{k}{2}+\fft14)}(z_1,\bar{z}_1) \,
      {\cal O}_{BCD,-\fft12}^{(h_2)}(z_2,\bar{z}_2)=
      \frac{\bar{\kappa}_{-\frac{3}{2},-\fft12,0}}{2}\,
\frac{\pi \, (-1)^{k+\fft32}}{(\fft52-k)!}
\frac{1}{\bar{z}_{12}^{\fft12-k}} \,
\sum_{m=0}^{-k+\fft12} \,
(m+1)
\nonu \\
&& \times \frac{\Gamma(2(\fft12-h_2)+1)}
{\Gamma(2(\fft12-h_2)+k+m+\fft12)}
\, z_{12}^{k-\fft32+m}\, \frac{1}{4!} \, \ep_{ABCDEFGH}\, {\pa}_{{z}_2}^m \,
{\cal O}_{0}^{(-\frac{k}{2}+\fft14+h_2),EFGH}(z_2,\bar{z}_2)+\cdots,
\nonu \\
&& \tilde{\bar{\Phi}}_{A,-\fft32}^{(-\frac{k}{2}+\fft14)}(z_1,\bar{z}_1) \,
{\cal O}_{0}^{(h_2),BCDE}(z_2,\bar{z}_2)=
\frac{\bar{\kappa}_{-\frac{3}{2},0,-\fft12}}{2}\,
\frac{\pi \, (-1)^{k+\fft32}}{(\fft52-k)!}
\frac{1}{\bar{z}_{12}^{\fft12-k}} \,
\sum_{m=0}^{-k+\fft12} \,
(m+1)
\nonu \\
&& \times \frac{\Gamma(2(1-h_2)+1)}
{\Gamma(2(1-h_2)+k+m+\fft12)}
\, z_{12}^{k-\fft32+m}\,
\frac{1}{4!}\, \frac{1}{3!}\, \ep^{BCDEIJKL}\, \ep_{AFGHIJKL}
{\pa}_{{z}_2}^m \,
{\cal O}_{+\fft12}^{(-\frac{k}{2}+\fft14+h_2),FGH}(z_2,\bar{z}_2)
\nonu \\
&& +\cdots,
\nonu \\
&& \tilde{\bar{\Phi}}_{A,-\fft32}^{(-\frac{k}{2}+\fft14)}(z_1,\bar{z}_1) \,
{\cal O}_{+\fft12}^{(h_2),BCD}(z_2,\bar{z}_2)=
\frac{\bar{\kappa}_{-\frac{3}{2},+\fft12,-1}}{2}\,
\frac{\pi \, (-1)^{k+\fft32}}{(\fft52-k)!}
\frac{1}{\bar{z}_{12}^{\fft12-k}} \,
\sum_{m=0}^{-k+\fft12} \,
(m+1)
\nonu \\
&& \times \frac{\Gamma(2(\fft32-h_2)+1)}
{\Gamma(2(\fft32-h_2)+k+m+\fft12)}
\, z_{12}^{k-\fft32+m}\,
3 \de^{[B}_{A} \,
{\pa}_{{z}_2}^m \,
{\cal O}_{+1}^{(-\frac{k}{2}+\fft14+h_2),CD]}(z_2,\bar{z}_2)+\cdots,
\nonu \\
&& \tilde{\bar{\Phi}}_{A,-\fft32}^{(-\frac{k}{2}+\fft14)}(z_1,\bar{z}_1) \,
{\cal O}_{+1}^{(h_2),BC}(z_2,\bar{z}_2)=
\frac{\bar{\kappa}_{-\frac{3}{2},+1,-\fft32}}{2}\,
\frac{\pi \, (-1)^{k+\fft32}}{(\fft52-k)!}
\frac{1}{\bar{z}_{12}^{\fft12-k}} \,
\sum_{m=0}^{-k+\fft12} \,
(m+1)
\nonu \\
&& \times \frac{\Gamma(2(2-h_2)+1)}
{\Gamma(2(2-h_2)+k+m+\fft12)}
\, z_{12}^{k-\fft32+m}\,
2! \de^{[B}_{A} \,
{\pa}_{{z}_2}^m \,
{\cal O}_{+\fft32}^{(-\frac{k}{2}+\fft14+h_2),C]}(z_2,\bar{z}_2)+\cdots,
\nonu \\
&& \tilde{\bar{\Phi}}_{A,-\fft32}^{(-\frac{k}{2}+\fft14)}(z_1,\bar{z}_1) \,
{\cal O}_{+\fft32}^{(h_2),B}(z_2,\bar{z}_2)=
\frac{\bar{\kappa}_{-\frac{3}{2},+\fft32,-2}}{2}\,
\frac{\pi \, (-1)^{k+\fft32}}{(\fft52-k)!}
\frac{1}{\bar{z}_{12}^{\fft12-k}} \,
\sum_{m=0}^{-k+\fft12} \,
(m+1)
\nonu \\
&& \times \frac{\Gamma(2(\fft52-h_2)+1)}
{\Gamma(2(\fft52-h_2)+k+m+\fft12)}
\, z_{12}^{k-\fft32+m}\,
\de^B_{A} \,
{\pa}_{{z}_2}^m \,
{\cal O}_{+2}^{(-\frac{k}{2}+\fft14+h_2)}(z_2,\bar{z}_2)+\cdots,
\nonu \\
&& \tilde{\bar{\Phi}}_{AB,-1}^{(-\frac{k}{2}+\fft12)}(z_1,\bar{z}_1) \,
{\cal O}_{CD,-1}^{(h_2)}(z_2,\bar{z}_2)=
\frac{\bar{\kappa}_{-1,-1,0}}{2}\,
\frac{\pi \, (-1)^{k+1}}{(2-k)!}
\frac{1}{\bar{z}_{12}^{1-k}} \,
\sum_{m=0}^{-k} \,
(m+1)
\nonu \\
&& \times \frac{\Gamma(2(-h_2)+1)}
{\Gamma(2(-h_2)+k+m+1)}
\, z_{12}^{k-1+m}\, \frac{1}{4!}\, \ep_{ABCDEFGH}
{\pa}_{{z}_2}^m \,
{\cal O}_{0}^{(-\frac{k}{2}+\fft12+h_2),EFGH}(z_2,\bar{z}_2)+\cdots,
\nonu \\
&& \tilde{\bar{\Phi}}_{AB,-1}^{(-\frac{k}{2}+\fft12)}(z_1,\bar{z}_1) \,
{\cal O}_{CDE,-\fft12}^{(h_2)}(z_2,\bar{z}_2)=
\frac{\bar{\kappa}_{-1,-\fft12,-\fft12}}{2}\,
\frac{\pi \, (-1)^{k+1}}{(2-k)!}
\frac{1}{\bar{z}_{12}^{1-k}} \,
\sum_{m=0}^{-k} \,
(m+1)
\nonu \\
&& \times \frac{\Gamma(2(\fft12-h_2)+1)}
{\Gamma(2(\fft12-h_2)+k+m+1)}
\, z_{12}^{k-1+m}\,
\frac{1}{3!}\, \ep_{ABCDEFGH}\,
{\pa}_{{z}_2}^m \,
{\cal O}_{+\fft12}^{(-\frac{k}{2}+\fft12+h_2),FGH}(z_2,\bar{z}_2)+\cdots,
\nonu \\
&& \tilde{\bar{\Phi}}_{AB,-1}^{(-\frac{k}{2}+\fft12)}(z_1,\bar{z}_1) \,
{\cal O}_{0}^{(h_2),CDEF}(z_2,\bar{z}_2)=
\frac{\bar{\kappa}_{-1,0,-1}}{2}\,
\frac{\pi \, (-1)^{k+1}}{(2-k)!}
\frac{1}{\bar{z}_{12}^{1-k}} \,
\sum_{m=0}^{-k} \,
(m+1)
\nonu \\
&& \times \frac{\Gamma(2(1-h_2)+1)}
{\Gamma(2(1-h_2)+k+m+1)}
\, z_{12}^{k-1+m}\,\frac{1}{4!}\,
\frac{1}{2!}\, \ep^{CDEFIJKL}\, \ep_{ABGHIJKL}
{\pa}_{{z}_2}^m \,
{\cal O}_{+1}^{(-\frac{k}{2}+\fft12+h_2),GH}(z_2,\bar{z}_2)
\nonu \\
&& +\cdots,
\nonu \\
&& \tilde{\bar{\Phi}}_{AB,-1}^{(-\frac{k}{2}+\fft12)}(z_1,\bar{z}_1) \,
{\cal O}_{+\fft12}^{(h_2),CDE}(z_2,\bar{z}_2)=
\frac{\bar{\kappa}_{-1,+\fft12,-\fft32}}{2}\,
\frac{\pi \, (-1)^{k+1}}{(2-k)!}
\frac{1}{\bar{z}_{12}^{1-k}} \,
\sum_{m=0}^{-k} \,
(m+1)
\nonu \\
&& \times \frac{\Gamma(2(\fft32-h_2)+1)}
{\Gamma(2(\fft32-h_2)+k+m+1)}
\, z_{12}^{k-1+m}\,
3!\, \de^{[C}_{A} \,
{\pa}_{{z}_2}^m \,
{\cal O}_{+\fft32}^{(-\frac{k}{2}+\fft12+h_2),D}(z_2,\bar{z}_2)
\, \de^{E]}_{B}+\cdots,
\nonu \\
&& \tilde{\bar{\Phi}}_{AB,-1}^{(-\frac{k}{2}+\fft12)}(z_1,\bar{z}_1) \,
{\cal O}_{+1}^{(h_2),CD}(z_2,\bar{z}_2)=
\frac{\bar{\kappa}_{-1,+1,-2}}{2}\,
\frac{\pi \, (-1)^{k+1}}{(2-k)!}
\frac{1}{\bar{z}_{12}^{1-k}} \,
\sum_{m=0}^{-k} \,
(m+1)
\nonu \\
&& \times \frac{\Gamma(2(2-h_2)+1)}
{\Gamma(2(2-h_2)+k+m+1)}
\, z_{12}^{k-1+m}\,
\de^{CD}_{AB} \,
{\pa}_{{z}_2}^m \,
{\cal O}_{+2}^{(-\frac{k}{2}+\fft12+h_2)}(z_2,\bar{z}_2)+\cdots,
\nonu \\
&& \tilde{\bar{\Phi}}_{ABC,-\fft12}^{(-\frac{k}{2}+\fft34)}(z_1,\bar{z}_1) \,
{\cal O}_{DEF,-\fft12}^{(h_2)}(z_2,\bar{z}_2)=
\frac{\bar{\kappa}_{-\fft12,-\fft12,-1}}{2}\,
\frac{\pi \, (-1)^{k+\fft12}}{(\fft32-k)!}
\frac{1}{\bar{z}_{12}^{\fft32-k}} \,
\sum_{m=0}^{-k-\fft12} \,
(m+1)
\nonu \\
&& \times \frac{\Gamma(2(\fft12-h_2)+1)}
{\Gamma(2(\fft12-h_2)+k+m+\tf32)}
\, z_{12}^{k-\fft12+m}\,
\frac{1}{2!} \, \ep_{ABCDEFGH}
{\pa}_{{z}_2}^m \,
{\cal O}_{+1}^{(-\frac{k}{2}+\fft34+h_2),GH}(z_2,\bar{z}_2)+\cdots,
\nonu \\
&& \tilde{\bar{\Phi}}_{ABC,-\fft12}^{(-\frac{k}{2}+\fft34)}(z_1,\bar{z}_1) \,
{\cal O}_{0}^{(h_2),DEFG}(z_2,\bar{z}_2)=
\frac{\bar{\kappa}_{-\fft12,0,-\fft32}}{2}\,
\frac{\pi \, (-1)^{k+\fft12}}{(\fft12-k)!}
\frac{1}{\bar{z}_{12}^{\fft32-k}} \,
\sum_{m=0}^{-k-\fft12} \,
(m+1)
\nonu \\
&& \times \frac{\Gamma(2(1-h_2)+1)}
{\Gamma(2(1-h_2)+k+m+\tf32)}
\, z_{12}^{k-\fft12+m}\,\frac{1}{4!}
\ep^{DEFGIJKL}\, \ep_{ABCHIJKL}
{\pa}_{{z}_2}^m \,
{\cal O}_{+\fft32}^{(-\frac{k}{2}+\fft34+h_2),H}(z_2,\bar{z}_2)+\cdots,
\nonu \\
&& \tilde{\bar{\Phi}}_{ABC,-\fft12}^{(-\frac{k}{2}+\fft34)}(z_1,\bar{z}_1) \,
{\cal O}_{+\fft12}^{(h_2),DEF}(z_2,\bar{z}_2)=
\frac{\bar{\kappa}_{-\fft12,+\fft12,-2}}{2}\,
\frac{\pi \, (-1)^{k+\fft12}}{(\fft12-k)!}
\frac{1}{\bar{z}_{12}^{\fft32-k}} \,
\sum_{m=0}^{-k-\fft12} \,
(m+1)
\nonu \\
&& \times \frac{\Gamma(2(\fft32-h_2)+1)}
{\Gamma(2(\fft32-h_2)+k+m+\tf32)}
\, z_{12}^{k-\fft12+m}\,
\de^{DEF}_{ABC}\,
{\pa}_{{z}_2}^m \,
{\cal O}_{+2}^{(-\frac{k}{2}+\fft34+h_2)}(z_2,\bar{z}_2)+\cdots,
\nonu \\
&& \tilde{\bar{\Phi}}_{0}^{(-\frac{k}{2}+1),ABCD}(z_1,\bar{z}_1) \,
{\cal O}_{0}^{(h_2),EFGH}(z_2,\bar{z}_2)=
\frac{\bar{\kappa}_{0,0,-2}}{2}\,
\frac{\pi \, (-1)^{k}}{(1-k)!}
\frac{1}{\bar{z}_{12}^{2-k}} \,
\sum_{m=0}^{-k-1} \,
(m+1)
\nonu \\
&& \times \frac{\Gamma(2(1-h_2)+1)}
{\Gamma(2(1-h_2)+k+m+2)}
\, z_{12}^{k+m}\,
\ep^{ABCDEFGH}\,
{\pa}_{{z}_2}^m \,
{\cal O}_{+2}^{(-\frac{k}{2}+1+h_2)}(z_2,\bar{z}_2)+\cdots.
\label{othertwentyfivesection6}
\eea
The remaining twenty OPEs are given by (\ref{othertwentysection6})
of Appendix F.
In (\ref{othertwentyfivesection6}) and (\ref{othertwentysection6}),
the whole forty-five OPEs corresponding to (\ref{HOOPEshadow1})
are described explicitly.

Again,
the first nine equations in
(\ref{othertwentyfivesection6})
lead to  the shadow Ward identities
to the leading soft theorems for the gravitons
when we consider the $n$-point celestial amplitude in the
${\cal N}=8$ supergravity.
See also (\ref{othertwentysection6}).
For $k=1$ in the
above nine cases, each summation in (\ref{HOOPEshadow1})
has only $m=0$ term
and the results become the simple expression
acting on each matter field on the right-hand sides.
For example, the OPE of the {\it shadow}
graviton currents with graviton matter
fields is given by
\bea
\tilde{\bar{\Phi}}_{-2}^{(-\frac{1}{2})}(z_1,\bar{z}_1) \,
{\cal O}_{-2}^{(h_2)}(z_2,\bar{z}_2)=
-\frac{\bar{\kappa}_{-2,-2,+2}}{2}\,
\frac{\pi}{2}\,
\frac{\bar{z}_{12}}{{z}_{12}} \,
{\cal O}_{-2}^{(-\frac{1}{2}+h_2)}(z_2,\bar{z}_2)+\cdots.
\label{softgravitonshadow}
\eea
The left-weight on the right-hand side of (\ref{softgravitonshadow})
is given by $1+(-1-(-\frac{1}{2}+h_2))$
which can be written as
$\frac{3}{2}+(-1-h_2)$. This implies that
the left-weight of the {\it shadow} current is given by $\frac{3}{2}$.
This behavior also can be seen from (\ref{leftright})
with $k=1$ and $s_1=-2$.
On the other hand, the right-weight
on the right-hand side of 
(\ref{softgravitonshadow}) is
$-1+(1-(-\frac{1}{2}+h_2))$ which becomes
$-\frac{1}{2} + (1-h_2)$.
This implies that
the right-weight of the {\it shadow} current is given by $-\frac{1}{2}$.
This behavior also can be seen from (\ref{leftright})
with $k=1$ and $s_1=-2$.

We can easily obtain the following OPE, from (\ref{softgravitonshadow})
by taking the antiholomorphic derivative,
\bea
\bar{\pa}_{\bar{z}_1} \, \tilde{\bar{\Phi}}_{-2}^{(-\frac{1}{2})}(z_1,\bar{z}_1) \,
{\cal O}_{-2}^{(h_2)}(z_2,\bar{z}_2)=
-\frac{\bar{\kappa}_{-2,-2,+2}}{2}\,
\frac{\pi}{2}\,
\frac{1}{{z}_{12}} \,
{\cal O}_{-2}^{(-\frac{1}{2}+h_2)}(z_2,\bar{z}_2)+\cdots.
\label{derivative}
\eea
Then the original weights $(\frac{3}{2},-\frac{1}{2})$ of
the {\it shadow} current becomes $(\frac{3}{2},\frac{1}{2})$
and this descendant in (\ref{derivative})
plays the role of supertranslation current
$P(z) \equiv \bar{\pa}_{\bar{z}} \,
\tilde{\bar{\Phi}}_{-2}^{(-\frac{1}{2})}(z,\bar{z})$
with weights $(\frac{3}{2},\frac{1}{2})$
in the notation of \cite{FSTZ}
\footnote{
The OPE of the {\it shadow}
graviton currents with graviton matter
fields, at the subleading order, is given by
\bea
\tilde{\bar{\Phi}}_{-2}^{(0)}(z_1,\bar{z}_1) \,
{\cal O}_{-2}^{(h_2)}(z_2,\bar{z}_2)=
\frac{\bar{\kappa}_{-2,-2,+2}}{2}\,
\pi \,
\bigg[ \frac{(-1-h_2)}{{z}_{12}^2}+ \frac{\pa_{z_2}}{z_{12}} \bigg]
{\cal O}_{-2}^{(h_2)}(z_2,\bar{z}_2)+\cdots.
\label{softgravitonshadow1}
\eea
We can easily check that the left-weight of the
right-hand side of (\ref{softgravitonshadow1}) is given by
$2+(-1-h_2)$. This is consistent with the fact that
the left-weight of the {\it shadow} current is equal to $2$.
Moreover,  the right-weight of the
right-hand side of (\ref{softgravitonshadow1}) is given by
$(1-h_2)$. This implies that the right-weight of {\it shadow} current is
equal to zero which can be also seen from (\ref{leftright}).
Therefore,  $\tilde{\bar{\Phi}}_{-2}^{(0)}(z)$ with weights
$(2,0)$ (plus the modified term \cite{BP}) plays the role of
the left Virasoro generator $T(z)$ (in the notation of
\cite{FSTZ}) of the ${\cal N}=8$ BMS algebra.

At the sub-sub-leading order,
the corresponding OPE for $k=-1$ is given by 
\bea
\tilde{\bar{\Phi}}_{-2}^{(\frac{1}{2})}(z_1,\bar{z}_1) \,
{\cal O}_{-2}^{(h_2)}(z_2,\bar{z}_2) & = &
-\frac{\bar{\kappa}_{-2,-2,+2}}{2}\,
\frac{\pi}{4!} \, \frac{1}{\bar{z}_{12}}
\bigg[ \frac{2(-1-h_2)(2(-1-h_2)-1)}{{z}_{12}^3}
\nonu \\
& + &
\frac{2(-1-h_2)\, \pa_{z_2}}{z_{12}^2}
+3\, \frac{\pa_{z_2}^2}{z_{12}} \bigg]
{\cal O}_{-2}^{(\frac{1}{2}+h_2)}(z_2,\bar{z}_2)+\cdots.
\label{softgravitonshadow2}
\eea
It is obvious to see that the left-weight of the
right-hand side of (\ref{softgravitonshadow2}) is given by
$3+(-1-(\frac{1}{2}+h_2))$. This is consistent with the fact that
the left-weight of the {\it shadow} current is equal to $\frac{5}{2}$.
Moreover,  the right-weight of the
right-hand side of (\ref{softgravitonshadow1}) is given by
$1+(1-(\frac{1}{2}+h_2))$. This implies that the right-weight
of {\it shadow} current is
equal to $\frac{1}{2}$ which can be also seen from (\ref{leftright}).
}.

For $k=\frac{1}{2}$ in the
previous eight cases, each summation in (\ref{HOOPEshadow1})
has only $m=0$ term.
For example, the OPE of the {\it shadow}
gravitino currents with gravitino matter
fields can be expressed as
\bea
\tilde{\bar{\Phi}}_{A,-\fft32}^{(0)}(z_1,\bar{z}_1) \,
{\cal O}_{B,-\frac{3}{2}}^{(h_2)}(z_2,\bar{z}_2)=
\frac{\bar{\kappa}_{-\frac{3}{2},-\fft32,+1}}{2}\,
\frac{\pi}{2}\,
\frac{1}{{z}_{12}} \,
{\cal O}_{AB,-1}^{(h_2)}(z_2,\bar{z}_2)+\cdots.
\label{softgravitinoshadow1}
\eea
The left-weight of the right-hand side of (\ref{softgravitinoshadow1})
is given by $1-h_2=\frac{3}{2}+(-\frac{1}{2}-h_2)$.
Then the left-weight of the {\it shadow} current on the left-hand side is
$\frac{3}{2}$ which can be also obtained from (\ref{leftright})
with $k=\frac{1}{2}$ and $s_1=-\frac{3}{2}$.
Similarly, the right-weight of the
right-hand side is $(1-h_2)$. This implies that
the right-weight of the {\it shadow} current on the left-hand side is
zero. See also (\ref{leftright}).
The complex coordinate dependent factor on the right-hand side in
(\ref{softgravitinoshadow1})
can also be seen from the equation $(6.13)$ in \cite{FSTZ}.
Therefore, $\tilde{\bar{\Phi}}_{A,-\fft32}^{(0)}(z) \equiv
S_{A}(z)$ with weights
$(\frac{3}{2},0)$ (having the non-antiholomorphic property \cite{2212-2})
plays the role of
other supercurrent of the ${\cal N}=8$ BMS algebra.
See also the page 16 of \cite{FSTZ}
\footnote{We can analyze the $z$ dependence
in this {\it shadow} current by following the description of the
footnote \ref{vanishingLW}.
By following the descriptions in the footnote \ref{BMScheck},
we can write down the two relevant OPEs and  
obtain the corresponding third equation of (\ref{BMS1})
where the {\it shadow} currents are substituted. The OPE of
(modified) stress energy tensor
with the supercurrent is given by
\bea
T({z}_1)\, {S}_A({z}_2) =\frac{1}{({z}_1-{z}_2)^2}
\, \frac{3}{2}\,  {S}_A({z}_2) +
\frac{1}{({z}_1-{z}_2)}\, {\pa}_{{z}_2} \,   {S}_A({z}_2)
+ \cdots.
\label{TS}
\eea
We saw the similar OPE in (\ref{TbarSbar}) in the antiholomorphic
sector.}.

Let us consider the subleading order and
for $k=-\frac{1}{2}$,
each summation in the above contains
$m=1$ term also as before.
By collecting the two contributions, we obtain 
\bea
\tilde{\bar{\Phi}}_{A,-\fft32}^{(\frac{1}{2})}(z_1,\bar{z}_1) \,
{\cal O}_{B,-\frac{3}{2}}^{(h_2)}(z_2,\bar{z}_2) & = & -
\frac{\bar{\kappa}_{-\frac{3}{2},-\fft32,+1}}{2}\,
\frac{2\,\pi}{3!}\,
\frac{1}{\bar{z}_{12}} \,
\bigg[ \frac{1}{{z}_{12}^2}(-\tfrac{1}{2}-h_2)
+\frac{1}{{z}_{12}} \, {\pa}_{{z}_2} \bigg] \,
{\cal O}_{AB,-1}^{(\frac{1}{2}+h_2)}(z_2,\bar{z}_2)
\nonu \\
& + & \cdots.
\label{softgravitino1shadow2}
\eea
The coefficient appearing in the second-order pole
in the holomorphic coordinates is the
left-weight of the matter field on the left-hand side.
The left-weight of the right-hand side of (\ref{softgravitino1shadow2})
is given by $2-(\frac{1}{2}+h_2)$.
Then the left-weight of the {\it shadow} current on the left-hand side is
$2$ which can be also obtained from (\ref{leftright})
with $k=-\frac{1}{2}$ and $s_1=-\frac{3}{2}$.
Similarly, the right-weight of the
right-hand side is $1+(1-(\frac{1}{2}+h_2))=
\frac{1}{2}+(1-h_2)$. This implies that
the right-weight of the {\it shadow} current on the left-hand side is
$\frac{1}{2}$ which can be seen from (\ref{leftright}).

For $k=0$ in the
previous seven cases, each summation gives the simple expression
acting on each matter field on the right-hand side.
Then the OPE
of the graviphoton currents
with graviphoton matter fields,
at leading-order, can be written as
\bea
\tilde{\bar{\Phi}}_{AB,-1}^{(\frac{1}{2})}(z_1,\bar{z}_1) \,
{\cal O}_{CD,-1}^{(h_2)}(z_2,\bar{z}_2) & = & -
\frac{\bar{\kappa}_{-1,-1,0}}{2}\,
\frac{\pi}{2!} \, \frac{1}{{z}_{12}\,  \bar{z}_{12}} \,
\frac{1}{4!}\, \ep_{ABCDEFGH}
{\cal O}_{0}^{(\frac{1}{2}+h_2),EFGH}(z_2,\bar{z}_2)
\nonu \\
&&+\cdots.
\label{softphotonshadow2}
\eea
The left-weight of the right-hand side of
(\ref{softphotonshadow2})
is given by $1+(1-(\frac{1}{2}+h_2))=\frac{3}{2}-h_2$.
Then the left-weight of the {\it shadow} current on the left-hand side is
$\frac{3}{2}$ which can be obtained from (\ref{leftright})
with $k=0$ and $s_1=-1$.
Similarly, the right-weight of the
right-hand side is $1+(1-(\frac{1}{2}+h_2))=
\frac{1}{2}+(1-h_2)$, This implies that
the right-weight of the {\it shadow} current on the left-hand side is
$\frac{1}{2}$ which can be seen from (\ref{leftright}).

For
the graviphotons with $k=1$
and $s_1=-1$,
the graviphotinos with $k=\frac{1}{2}$
and $s_1=-\frac{1}{2}$
and the scalars with $k=1$ and $s_1=0$, there are no 
any singular behaviors due to the empty sum
\footnote{
Let us emphasize that
the Virasoro generators, the supercurrents
and the supertranslation generator
of ${\cal N}=8$ BMS algebra
are given by
\bea
\tilde{\Phi}_{+2}^{(0)}(\bar{z}) & \equiv & \bar{T}(\bar{z}), \qquad
\tilde{\bar{\Phi}}_{-2}^{(0)}(z)\equiv T(z), \qquad
\tilde{\Phi}_{+\frac{3}{2}}^{(0),A}(\bar{z}) \equiv \bar{S}^A(\bar{z}),
\nonu \\
\tilde{\bar{\Phi}}_{A,-\frac{3}{2}}^{(0)}(z) & \equiv & S_A(z),
\qquad
\bigg( \bigg\{ 
\tilde{\bar{\Phi}}_{A,-\frac{3}{2}}^{(0)}(z),
\tilde{\Phi}_{+\frac{3}{2}}^{(0),B}(\bar{z})  \bigg\} \bigg) \equiv
\bigg( \bigg\{ 
S_{A}(z),
\bar{S}^{B}(\bar{z}) \bigg\} \bigg) 
=  \de^{B}_{A}\, {\cal P}(z,\bar{z}),
\label{BMSgenerators}
\eea
having the weights $(0,2)$, $(2,0)$, $(0,\frac{3}{2})$,
$(\frac{3}{2},0)$ and $(\frac{3}{2},\frac{3}{2})$ respectively.
Here the round bracket in the
last element is the notation for normal ordering.
We can construct the equation (\ref{two3halfs}) where the
mode of
current on the left-hand side is replaced by the mode of
{\it shadow} current
$\tilde{\bar{\Phi}}^{(0)}_{B,-\frac{3}{2},n}$ with proper normalization.
This comes from the sixteenth equation of
(\ref{othertwentyfivesection6}) together with the analysis of
(\ref{softgravitinoshadow1}) after the integration over
the holomorphic coordinate for the {\it shadow} current.
Similarly, the equation (\ref{3halftwo})
where the
mode of
current on the left-hand side is replaced by the mode of
{\it shadow} current
$\tilde{{\Phi}}^{(0),A}_{+\frac{3}{2},m}$ with corresponding
normalization
can be obtained from the first equation of
(\ref{twentysection6}) and the description of
(\ref{softgravitino}). Then we arrive at the equation
(\ref{result1}) exactly. Moreover, the equations (\ref{firstcon})
and (\ref{secondcon})
where the corresponding {\it shadow} currents appear on the left-hand sides 
can be obtained also.
Finally we end up with the equation (\ref{finalexp}).
This implies that the action of the composite mode
of the {\it shadow} current on the matter fields
$\bigg[ \bigg\{ \tilde{\Phi}_{+\frac{3}{2},m}^{(0),A},
\tilde{\bar{\Phi}}_{B,-\frac{3}{2},n}^{(0)} \bigg\},
{\cal O}_{+\frac{3}{2}}^{(h),C}(z,\bar{z}) \bigg]=
\frac{1}{4}\, \bar{z}^{m+\frac{1}{2}}\, z^{n+\frac{1}{2}}\,
\de^{A}_{B}\, {\cal O}_{+\frac{3}{2}}^{(-\frac{1}{2}+h),C}(z,\bar{z})$.
Therefore, we can identify the last element of
(\ref{BMSgenerators}) with the
supertranslation generator based on (\ref{finalexp1})
after interchanging the two currents and using the different
$SU(8)$ indices.

Moreover, by using (\ref{TbarSbar}) and (\ref{TS}), the following OPEs
satisfy
\bea
T(z_1) \, \de^B_A \, {\cal P}(z_2,\bar{z}_2) &=& \frac{1}{(z_1-z_2)^2}
\, \frac{3}{2}\,  \de^B_A \,{\cal P}(z_2,\bar{z}_2) +
\frac{1}{(z_1-z_2)} \, \de^B_A \, \pa_{z_2} \,  {\cal P}(z_2,\bar{z}_2)+ \cdots,
\nonu \\
\bar{T}(\bar{z}_1) \, \de^B_A \, {\cal P}(z_2,\bar{z}_2) & = &
\frac{1}{(\bar{z}_1-\bar{z}_2)^2}
\, \frac{3}{2}\,  \de^B_A \, {\cal P}(z_2,\bar{z}_2) +
\frac{1}{(\bar{z}_1-\bar{z}_2)} \, \de^B_A \, \bar{\pa}_{\bar{z}_2} \,
{\cal P}(z_2,\bar{z}_2)+ \cdots,
\nonu
\eea
where the relations $T(z_1) \, \bar{S}^{A}(\bar{z}_2)= 0 + \cdots$
and $\bar{T}(\bar{z}_1) \, {S}_{A}({z}_2)= 0 + \cdots$
are used.
If we expand the supercurrents and the supertranslation in terms of modes,
then the last relation of (\ref{BMSgenerators}) becomes the equation
(\ref{BMS2}). There are the OPEs
$S_A(z_1)\, \bar{S}^{B}(\bar{z}_2)= 0 + (S_A\, \bar{S}^B)
(z_2,\bar{z}_2)+ \cdots $ and
$\bar{S}^B(\bar{z}_1)\, {S}_A({z}_2)= 0 + (\bar{S}^B\, S_A)
(z_2,\bar{z}_2)+ \cdots $ with the normal orderings $ (S_A\, \bar{S}^B)
(z_2,\bar{z}_2) =  S_A(z_2) \, \bar{S}^B(\bar{z}_2)$ and
$ (\bar{S}^B\, S_A)
(z_2,\bar{z}_2)=  \bar{S}^B(\bar{z}_2)\, S_A(z_2)$.}.

\section{ Conclusions and outlook}

The summary of this paper is already given in the abstract.
In particular, the main equations found newly are
given by (\ref{25OPEs}), (\ref{comm25OPEs}),
(\ref{lambdacon}), (\ref{Eqs}), (\ref{BMS2}), (\ref{twentyfive}),
(\ref{twentyfivesection6}), (\ref{othertwentyfive}),
(\ref{othertwentyfivesection6}) and the equations of Appendices.

In the full antiholomorphic and holomorphic sectors,
the two independent ${\cal N}=8$ celestial {\it soft} symmetry algebras
(found in \cite{AK2509} and given by (\ref{Eqs}) respectively)
containing the ${\cal N}=8$ extended BMS algebra
with some modes of supertranslation generator
exist. Moreover, for the complete modes of the supertranslation
generator leading to the full ${\cal N}=8$ extended BMS algebra, 
the previous two commutator relations (each commutator for each soft
algebra given by the first equation of (\ref{BMS})
and the first equation of (\ref{BMS1})) remain the same.
Then the above two soft algebras are no longer independent of each other
because the anticommutator between the supercurrent
in the antiholomorphic sector and another supercurrent
in the holomorphic sector
is nonzero
\footnote{For two given
independent ${\cal N}=8$ celestial
{\it soft} symmetry algebras, if we add the equation
(\ref{BMS2}), then  the first equation of (\ref{BMS})
and the first equation of (\ref{BMS1}) are satisfied
automatically as at the end of the subsection \ref{Non}.
This implies that the nontrivial result
of (\ref{BMS2}) together with
these two soft algebras leads to the ${\cal N}=8$ extended
BMS algebra as a subalgebra.}.

The immediate question is
how to construct the extension of
${\cal N}=8$ celestial {\it soft} symmetry algebra
by ${\rm w}_{1+\infty}$ (or ${\rm \bar{w}}_{1+\infty}$) currents.
We already saw some examples
for the actions of these two kinds of currents
on the matter fields in the triple (anti)commutators
in the last subsection of section \ref{holomorphic}.
It is an open problem to obtain
the whole extension of
${\cal N}=8$ 
${\rm w}_{1+\infty}$ (or  ${\rm \bar{w}}_{1+\infty}$) algebra.

Recently, in \cite{CP}, the multi-particle OPEs of single-particle
celestial operators with two-particle celestial operators in
pure Einstein gravity were studied.
Can we define the antiholomorphic or holomorphic
light transforms for the composite operator?
If we find the corresponding currents, then the next step is
to obtain the OPEs of these currents (from the composite
operators) with the matter fields. After that
we will consider the commutator relations between the
currents (from the single-particle operators) and 
the currents (from the two-particle operators) for the single-particle
contributions.

In arriving at
the relations of (\ref{lambdacon}), we have
considered only the cases of the helicity condition $s_1 \geq s_2$.
What occurs if one considers the remaining cases
where $s_1 < s_2$? Are there any further restrictions on the
couplings $\la_{s_1,s_2,-s_3}$?

For given OPEs in (\ref{HOOPEshadow}) and
(\ref{othertwentyfivesection6}), in principle,
the (anti)commutators can be obtained. The singular factors
in the holomorphic (or antiholomorphic) coordinates
outside the summations are complicated compared to
other OPEs in this paper. It is an open problem to understand
the relations of these (anti)commutators obtained from the shadow
transforms to other (anti)commutators obtained from different
approaches.

It is an open problem to
check the equations (\ref{Eqs}) from the triple (anti)commutator
relations by following the descriptions in Appendix D.
Moreover, as a consistency check, we can verify each equation
of (\ref{Eqs}) from the structure of the first and the second
${\cal N}=4$ super Yang-Mills theories with index structures and
factorization of states given in \cite{BEF,2212-1,KLT}.

So far we considered
the case of $s_1+s_2-s_3 =\pm 2$.
For the different combinations of the helicities,
we expect that there are other operators on the right-hand sides of
OPEs or (anti)commutators. In \cite{EJN},
the possible terms in the effective field theory
corresponding to the corrections for the subsubleading {\it soft} graviton
theorem
are presented
for $s_1+s_2-s_3=4$. See also \cite{BSYV}.
It would be interesting to understand the appearance of these
modifications in the context of this paper.

%
%

\vspace{.7cm}

\centerline{\bf Acknowledgments}

CA thanks M. Pate, A. Tropper, H. Jiang and M.H. Kim for  discussions.
This work by CA was supported by the National
Research Foundation of Korea(NRF) grant funded by the
Korea government(MSIT) 
(No. 2023R1A2C1003750).
CA
acknowledges warm hospitality from
Institute for Convergent Fundamental Studies,
Seoul National University of Science and Technology.

CA
acknowledges the use of Google Gemini to
assist in drafting portions of the introduction.


\newpage

\appendix

\renewcommand{\theequation}{\Alph{section}\mbox{.}\arabic{equation}}

\section{The remaining
twenty OPEs of the ${\cal N}=8$ ${\rm w}_{1+\infty}$
currents with matter fields in section \ref{OPE}
\label{appA}}

We present the twenty OPEs of the currents with the matter fields
where the helicities for the currents and the matter fields
are satisfied with
$s_1 < s_2$ as follows:
\bea
&&
{\Phi}_{+\frac{3}{2}}^{(h_1),A}(z_1,\bar{z}_1) \,
{\cal O}_{+2}^{(h_2)}(z_2,\bar{z}_2)
=  \frac{1}{z_{12}}\, \frac{1}{2}\,
\sum_{m=0}^{2(h_1+\frac{3}{2})-3}\,
\frac{(m+1)\, \Gamma(2(-1-h_2)+1)}{
\Gamma(-1-2(h_1+h_2) +m)}\, \bar{z}_{12}^{2-2(h_1+\frac{3}{2})+m}
\nonu \\
&& \times   
\bar{\pa}^m \,
{\cal O}_{+\frac{3}{2}}^{(h_1+h_2),A}(z_2,\bar{z}_2)
+  \cdots,
\nonu \\
&&
{\Phi}_{+1}^{(h_1),AB}(z_1,\bar{z}_1) \,
{\cal O}_{+2}^{(h_2)}(z_2,\bar{z}_2)
=  \frac{1}{z_{12}}\, \frac{1}{2}\,
\sum_{m=0}^{2(h_1+1)-3}\,
\frac{(m+1)\, \Gamma(2(-1-h_2)+1)}{
  \Gamma(-2(h_1+h_2) +m)}\, \bar{z}_{12}^{2-2(h_1+1)+m}
\nonu \\
&& \times   
\bar{\pa}^m \,
{\cal O}_{+1}^{(h_1+h_2),AB}(z_2,\bar{z}_2)
+  \cdots,
\nonu \\
&&
{\Phi}_{+\frac{1}{2}}^{(h_1),ABC}(z_1,\bar{z}_1) \,
{\cal O}_{+2}^{(h_2)}(z_2,\bar{z}_2)
=  \frac{1}{z_{12}}\, \frac{1}{2}\,
\sum_{m=0}^{2(h_1+\frac{1}{2})-3}\,
\frac{(m+1)\, \Gamma(2(-1-h_2)+1)}{
  \Gamma(1-2(h_1+h_2) +m)}\, \bar{z}_{12}^{2-2(h_1+\frac{1}{2})+m}
\nonu \\
&& \times   
\bar{\pa}^m \,
{\cal O}_{+\frac{1}{2}}^{(h_1+h_2),ABC}(z_2,\bar{z}_2)
+  \cdots,
\nonu \\
&&
{\Phi}_{0}^{(h_1),ABCD}(z_1,\bar{z}_1) \,
{\cal O}_{+2}^{(h_2)}(z_2,\bar{z}_2)
=  \frac{1}{z_{12}}\, \frac{1}{2}\,
\sum_{m=0}^{2(h_1)-3}\,
\frac{(m+1)\, \Gamma(2(-1-h_2)+1)}{
  \Gamma(2-2(h_1+h_2) +m)}\, \bar{z}_{12}^{2-2(h_1)+m}
\nonu \\
&& \times   
\bar{\pa}^m \,
{\cal O}_{0}^{(h_1+h_2),ABCD}(z_2,\bar{z}_2)
+  \cdots,
\nonu \\
&&
{\Phi}_{ABC,-\frac{1}{2}}^{(h_1)}(z_1,\bar{z}_1) \,
{\cal O}_{+2}^{(h_2)}(z_2,\bar{z}_2)
=  \frac{1}{z_{12}}\, \frac{1}{2}\,
\sum_{m=0}^{2(h_1-\frac{1}{2})-3}\,
\frac{(m+1)\, \Gamma(2(-1-h_2)+1)}{
  \Gamma(3-2(h_1+h_2) +m)}\, \bar{z}_{12}^{2-2(h_1-\frac{1}{2})+m}
\nonu \\
&& \times   
\bar{\pa}^m \,
{\cal O}_{ABC,-\frac{1}{2}}^{(h_1+h_2)}(z_2,\bar{z}_2)
+  \cdots,
\nonu \\
&&
{\Phi}_{AB,-1}^{(h_1)}(z_1,\bar{z}_1) \,
{\cal O}_{+2}^{(h_2)}(z_2,\bar{z}_2)
=  \frac{1}{z_{12}}\, \frac{1}{2}\,
\sum_{m=0}^{2(h_1-1)-3}\,
\frac{(m+1)\, \Gamma(2(-1-h_2)+1)}{
  \Gamma(4-2(h_1+h_2) +m)}\, \bar{z}_{12}^{2-2(h_1-1)+m}
\nonu \\
&& \times   
\bar{\pa}^m \,
{\cal O}_{AB,-1}^{(h_1+h_2)}(z_2,\bar{z}_2)
+  \cdots,
\nonu \\
&&
{\Phi}_{A,-\frac{3}{2}}^{(h_1)}(z_1,\bar{z}_1) \,
{\cal O}_{+2}^{(h_2)}(z_2,\bar{z}_2)
=  \frac{1}{z_{12}}\, \frac{1}{2}\,
\sum_{m=0}^{2(h_1-\frac{3}{2})-3}\,
\frac{(m+1)\, \Gamma(2(-1-h_2)+1)}{
  \Gamma(5-2(h_1+h_2) +m)}\, \bar{z}_{12}^{2-2(h_1-\frac{3}{2})+m}
\nonu \\
&& \times   
\bar{\pa}^m \,
{\cal O}_{A,-\frac{3}{2}}^{(h_1+h_2)}(z_2,\bar{z}_2)
+  \cdots,
\nonu \\
&&
{\Phi}_{-2}^{(h_1)}(z_1,\bar{z}_1) \,
{\cal O}_{+2}^{(h_2)}(z_2,\bar{z}_2)
=  \frac{1}{z_{12}}\, \frac{1}{2}\,
\sum_{m=0}^{2(h_1-2)-3}\,
\frac{(m+1)\, \Gamma(2(-1-h_2)+1)}{
  \Gamma(6-2(h_1+h_2) +m)}\, \bar{z}_{12}^{2-2(h_1-2)+m}
\nonu \\
&& \times   
\bar{\pa}^m \,
{\cal O}_{-2}^{(h_1+h_2)}(z_2,\bar{z}_2)
+  \cdots,
\nonu \\
&&
{\Phi}_{+1}^{(h_1),AB}(z_1,\bar{z}_1) \,
{\cal O}_{+\frac{3}{2}}^{(h_2),C}(z_2,\bar{z}_2)
=  \frac{1}{z_{12}}\, \frac{1}{2}\,
\sum_{m=0}^{2(h_1+1)-3}\,
\frac{(m+1)\, \Gamma(2(-\frac{1}{2}-h_2)+1)}{
  \Gamma(1-2(h_1+h_2) +m)}\, \bar{z}_{12}^{2-2(h_1+1)+m}
\nonu \\
&& \times   
\bar{\pa}^m \,
{\cal O}_{\frac{1}{2}}^{(h_1+h_2),ABC}(z_2,\bar{z}_2)
+  \cdots,
\nonu \\
&&
{\Phi}_{+\frac{1}{2}}^{(h_1),ABC}(z_1,\bar{z}_1) \,
{\cal O}_{+\frac{3}{2}}^{(h_2),D}(z_2,\bar{z}_2)
= \frac{1}{z_{12}}\, \frac{1}{2}\,
\sum_{m=0}^{2(h_1+\frac{1}{2})-3}\,
\frac{(m+1)\, \Gamma(2(-\frac{1}{2}-h_2)+1)}{
  \Gamma(2-2(h_1+h_2) +m)}\, \bar{z}_{12}^{2-2(h_1+\frac{1}{2})+m}
\nonu \\
&& \times   
\bar{\pa}^m \,
{\cal O}_{0}^{(h_1+h_2),ABCD}(z_2,\bar{z}_2)
+  \cdots,
\nonu \\
&&
{\Phi}_{0}^{(h_1),ABCD}(z_1,\bar{z}_1) \,
{\cal O}_{+\frac{3}{2}}^{(h_2),E}(z_2,\bar{z}_2)
=  \frac{1}{z_{12}}\, \frac{1}{2}\,
\sum_{m=0}^{2(h_1)-3}\,
\frac{(m+1)\, \Gamma(2(-\frac{1}{2}-h_2)+1)}{
  \Gamma(3-2(h_1+h_2) +m)}\, \bar{z}_{12}^{2-2(h_1)+m}
\nonu \\
&& \times 
\frac{1}{3!}\, \ep^{ABCDEFGH}\,
\bar{\pa}^m \,
{\cal O}_{FGH,-\frac{1}{2}}^{(h_1+h_2)}(z_2,\bar{z}_2)
+  \cdots,
\nonu \\
&&
{\Phi}_{ABC,-\frac{1}{2}}^{(h_1)}(z_1,\bar{z}_1) \,
{\cal O}_{+\frac{3}{2}}^{(h_2),D}(z_2,\bar{z}_2)
=  -\frac{1}{z_{12}}\, \frac{1}{2}\,
\sum_{m=0}^{2(h_1-\frac{1}{2})-3}\,
\frac{(m+1)\, \Gamma(2(-\frac{1}{2}-h_2)+1)}{
  \Gamma(4-2(h_1+h_2) +m)}\nonu \\
&& \times  \bar{z}_{12}^{2-2(h_1-\frac{1}{2})+m}
\,
3\,
\bar{\pa}^m \,
{\cal O}_{[AB,-1}^{(h_1+h_2)}(z_2,\bar{z}_2) \, \de^{D}_{C]}
+  \cdots,
\nonu \\
&&
{\Phi}_{AB,-1}^{(h_1)}(z_1,\bar{z}_1) \,
{\cal O}_{+\frac{3}{2}}^{(h_2),C}(z_2,\bar{z}_2)
=  \frac{1}{z_{12}}\, \frac{1}{2}\,
\sum_{m=0}^{2(h_1-1)-3}\,
\frac{(m+1)\, \Gamma(2(-\frac{1}{2}-h_2)+1)}{
  \Gamma(5-2(h_1+h_2) +m)}\, \bar{z}_{12}^{2-2(h_1-1)+m}
\nonu \\
&& \times 
2\,
\bar{\pa}^m \, \de^{C}_{[A}
{\cal O}_{B],-\frac{3}{2}}^{(h_1+h_2)}(z_2,\bar{z}_2)
+  \cdots,
\nonu \\
&&
{\Phi}_{A,-\frac{3}{2}}^{(h_1)}(z_1,\bar{z}_1) \,
{\cal O}_{+\frac{3}{2}}^{(h_2),B}(z_2,\bar{z}_2)
=  -\frac{1}{z_{12}}\, \frac{1}{2}\,
\sum_{m=0}^{2(h_1-\frac{3}{2})-3}\,
\frac{(m+1)\, \Gamma(2(-\frac{1}{2}-h_2)+1)}{
  \Gamma(6-2(h_1+h_2) +m)}\nonu \\
&& \times \bar{z}_{12}^{2-2(h_1-\frac{3}{2})+m}
\,
\de^B_{A} \,
\bar{\pa}^m \, 
{\cal O}_{-2}^{(h_1+h_2)}(z_2,\bar{z}_2)
+  \cdots,
\nonu \\
&&
{\Phi}_{+\frac{1}{2}}^{(h_1),ABC}(z_1,\bar{z}_1) \,
{\cal O}_{+1}^{(h_2),DE}(z_2,\bar{z}_2)
=  \frac{1}{z_{12}}\, \frac{1}{2}\,
\sum_{m=0}^{2(h_1+\frac{1}{2})-3}\,
\frac{(m+1)\, \Gamma(2(-h_2)+1)}{
  \Gamma(3-2(h_1+h_2) +m)}\, \bar{z}_{12}^{2-2(h_1+\frac{1}{2})+m}
\nonu \\
&& \times 
\frac{1}{3!} \,
\ep^{ABCDEFGH} \,
\bar{\pa}^m \, 
{\cal O}_{FGH,-\frac{1}{2}}^{(h_1+h_2)}(z_2,\bar{z}_2)
+  \cdots,
\nonu \\
&&
{\Phi}_{0}^{(h_1),ABCD}(z_1,\bar{z}_1) \,
{\cal O}_{+1}^{(h_2),EF}(z_2,\bar{z}_2)
=  \frac{1}{z_{12}}\, \frac{1}{2}\,
\sum_{m=0}^{2(h_1)-3}\,
\frac{(m+1)\, \Gamma(2(-h_2)+1)}{
  \Gamma(4-2(h_1+h_2) +m)}\, \bar{z}_{12}^{2-2(h_1)+m}
\nonu \\
&& \times 
\frac{1}{2!} \,
\ep^{ABCDEFGH} \,
\bar{\pa}^m \, 
{\cal O}_{GH,-1}^{(h_1+h_2)}(z_2,\bar{z}_2)
+  \cdots,
\nonu \\
&&
{\Phi}_{ABC,-\frac{1}{2}}^{(h_1)}(z_1,\bar{z}_1) \,
{\cal O}_{+1}^{(h_2),DE}(z_2,\bar{z}_2)
=  \frac{1}{z_{12}}\, \frac{1}{2}\,
\sum_{m=0}^{2(h_1-\frac{1}{2})-3}\,
\frac{(m+1)\, \Gamma(2(-h_2)+1)}{
  \Gamma(5-2(h_1+h_2) +m)}\, \bar{z}_{12}^{2-2(h_1-\frac{1}{2})+m}
\nonu \\
&& \times 
3! \,
\de^{D}_{[A} \,
\bar{\pa}^m \, 
{\cal O}_{B,-\frac{3}{2}}^{(h_1+h_2)}(z_2,\bar{z}_2) \, \de^{E}_{C]}
+  \cdots,
\nonu \\
&&
{\Phi}_{AB,-1}^{(h_1)}(z_1,\bar{z}_1) \,
{\cal O}_{+1}^{(h_2),CD}(z_2,\bar{z}_2)
=  \frac{1}{z_{12}}\, \frac{1}{2}\,
\sum_{m=0}^{2(h_1-1)-3}\,
\frac{(m+1)\, \Gamma(2(-h_2)+1)}{
  \Gamma(6-2(h_1+h_2) +m)}\, \bar{z}_{12}^{2-2(h_1-1)+m}
\nonu \\
&& \times 
\de^{CD}_{AB} \,
\bar{\pa}^m \, 
{\cal O}_{-2}^{(h_1+h_2)}(z_2,\bar{z}_2)
+  \cdots,
\nonu \\
&&
{\Phi}_{0}^{(h_1),ABCD}(z_1,\bar{z}_1) \,
{\cal O}_{+\frac{1}{2}}^{(h_2),EFG}(z_2,\bar{z}_2)
=  \frac{1}{z_{12}}\, \frac{1}{2}\,
\sum_{m=0}^{2(h_1)-3}\,
\frac{(m+1)\, \Gamma(2(\frac{1}{2}-h_2)+1)}{
  \Gamma(5-2(h_1+h_2) +m)}\, \bar{z}_{12}^{2-2(h_1)+m}
\nonu \\
&& \times 
\ep^{ABCDEFGH} \,
\bar{\pa}^m \, 
{\cal O}_{H,-\frac{3}{2}}^{(h_1+h_2)}(z_2,\bar{z}_2)
+  \cdots,
\nonu \\
&&
{\Phi}_{ABC,-\frac{1}{2}}^{(h_1)}(z_1,\bar{z}_1) \,
{\cal O}_{+\frac{1}{2}}^{(h_2),DEF}(z_2,\bar{z}_2)
=  -\frac{1}{z_{12}}\, \frac{1}{2}\,
\sum_{m=0}^{2(h_1-\frac{1}{2})-3}\,
\frac{(m+1)\, \Gamma(2(\frac{1}{2}-h_2)+1)}{
  \Gamma(6-2(h_1+h_2) +m)}\, \bar{z}_{12}^{2-2(h_1-\frac{1}{2})+m}
\nonu \\
&& 
\de^{DEF}_{ABC} \,
\bar{\pa}^m \, 
{\cal O}_{-2}^{(h_1+h_2)}(z_2,\bar{z}_2)
+  \cdots.
\label{20OPEs}
\eea
There are three cases (the twelfth, the fourteenth and the
twentieth) with an overall minus signs
on the right-hand sides of (\ref{20OPEs}).
In Tables \ref{table9} and \ref{table10},
we summarize the right-weights and the helicities for the currents and
the left- and right-weights for the matter fields.
Then the complete forty-five OPEs of the currents with matter fields
with the helicity condition $s_1+s_2-s_3=2$
are given by (\ref{25OPEs}) and (\ref{20OPEs}).

\begin{table}[tbp]
\centering
\renewcommand{\arraystretch}{1.7}
\begin{tabular}{|c|c||c|c|c|||c|c|c|c|| }
\hline
$ q$ &
$ {\rm w^q}$ &
$ h$ &
$ \bar{h}$ &
$ {\cal O}_{h,\bar{h}}$ &
$ s_1$ &
$ h+s_1-q$ &
$ \bar{h}+2-q$ &
${\cal O}_{h+s_1-q,\bar{h}+2-q}$  
\\
\hline
\hline
${\scriptstyle h_1+\frac{3}{2}}$ &
$  \Phi_{+\frac{3}{2}}^{(h_1),A}$ &
${\scriptstyle 1-h_2}$ &
${\scriptstyle -1-h_2}$ &
$ {\cal O}_{+2}^{(h_2)}$ &
${\scriptstyle +\frac{3}{2}}$ &
${\scriptstyle 1-(h_1+h_2)}$ &
${\scriptstyle -\frac{1}{2}-(h_1+h_2)} $ &
$ {\cal O}_{+\frac{3}{2}}^{(h_1+h_2),A}$
\\
\hline
${\scriptstyle h_1+1}$ &
$\Phi_{+1}^{(h_1),AB}$&
${\scriptstyle 1-h_2}$ &
${\scriptstyle -1-h_2}$ &
${\cal O}_{+2}^{(h_2)}$ &
${\scriptstyle +1}$ &
${\scriptstyle 1-(h_1+h_2)}$ &
${\scriptstyle -(h_1+h_2)} $ &
${\cal O}_{+1}^{(h_1+h_2),AB}$
\\
\hline
${\scriptstyle h_1+\frac{1}{2}}$ &
$\Phi_{+\frac{1}{2}}^{(h_1),ABC}$&
${\scriptstyle 1-h_2}$ &
${\scriptstyle -1-h_2}$ &
${\cal O}_{+2}^{(h_2)}$ &
${\scriptstyle +\frac{1}{2}}$ &
${\scriptstyle 1-(h_1+h_2)}$ &
${\scriptstyle \frac{1}{2}-(h_1+h_2)} $ &
${\cal O}_{+\frac{1}{2}}^{(h_1+h_2),ABC}$
\\
\hline
${\scriptstyle h_1}$ &
$\Phi_{0}^{(h_1),ABCD}$&
${\scriptstyle 1-h_2}$ &
${\scriptstyle -1-h_2}$ &
${\cal O}_{+2}^{(h_2)}$ &
${\scriptstyle 0}$ &
${\scriptstyle 1-(h_1+h_2)}$ &
${\scriptstyle 1-(h_1+h_2)} $ &
${\cal O}_{0}^{(h_1+h_2),ABCD}$
\\
\hline
${\scriptstyle h_1-\frac{1}{2}}$ &
$\Phi_{ABC,-\frac{1}{2}}^{(h_1)}$&
${\scriptstyle 1-h_2}$ &
${\scriptstyle -1-h_2}$ &
${\cal O}_{+2}^{(h_2)}$ &
${\scriptstyle -\frac{1}{2}}$ &
${\scriptstyle 1-(h_1+h_2)}$ &
${\scriptstyle \frac{3}{2}-(h_1+h_2)} $ &
${\cal O}_{ABC,-\frac{1}{2}}^{(h_1+h_2)}$
\\
\hline
${\scriptstyle h_1-1}$ &
$\Phi_{AB,-1}^{(h_1)}$&
${\scriptstyle 1-h_2}$ &
${\scriptstyle -1-h_2}$ &
${\cal O}_{+2}^{(h_2)}$ &
${\scriptstyle -1}$ &
${\scriptstyle 1-(h_1+h_2)}$ &
${\scriptstyle 2-(h_1+h_2)} $ &
${\cal O}_{AB,-1}^{(h_1+h_2)}$
\\
\hline
${\scriptstyle h_1-\frac{3}{2}}$ &
$\Phi_{A,-\frac{3}{2}}^{(h_1)}$&
${\scriptstyle 1-h_2}$ &
${\scriptstyle -1-h_2}$ &
${\cal O}_{+2}^{(h_2)}$ &
${\scriptstyle -\frac{3}{2}}$ &
${\scriptstyle 1-(h_1+h_2)}$ &
${\scriptstyle \frac{5}{2}-(h_1+h_2)} $ &
${\cal O}_{A,-\frac{3}{2}}^{(h_1+h_2)}$
\\
\hline
${\scriptstyle h_1-2}$ &
$\Phi_{-2}^{(h_1)}$&
${\scriptstyle 1-h_2}$ &
${\scriptstyle -1-h_2}$ &
${\cal O}_{+2}^{(h_2)}$ &
${\scriptstyle -2}$ &
${\scriptstyle 1-(h_1+h_2)}$ &
${\scriptstyle 3-(h_1+h_2)} $ &
${\cal O}_{-2}^{(h_1+h_2)}$
\\
\hline
\hline
${\scriptstyle h_1+1}$ &
$\Phi_{+1}^{(h_1),AB}$&
${\scriptstyle 1-h_2}$ &
${\scriptstyle -\frac{1}{2}-h_2}$ &
${\cal O}_{+\frac{3}{2}}^{(h_2),C}$ &
${\scriptstyle +1}$ &
${\scriptstyle 1-(h_1+h_2)}$ &
${\scriptstyle \frac{1}{2}-(h_1+h_2)} $ &
${\cal O}_{+\frac{1}{2}}^{(h_1+h_2),ABC}$
\\
\hline
${\scriptstyle h_1+\frac{1}{2}}$ &
$\Phi_{+\frac{1}{2}}^{(h_1),ABC}$&
${\scriptstyle 1-h_2}$ &
${\scriptstyle -\frac{1}{2}-h_2}$ &
${\cal O}_{+\frac{3}{2}}^{(h_2),D}$ &
${\scriptstyle +\frac{1}{2}}$ &
${\scriptstyle 1-(h_1+h_2)}$ &
${\scriptstyle 1-(h_1+h_2)} $ &
${\cal O}_{0}^{(h_1+h_2),ABCD}$
\\
\hline
${\scriptstyle h_1}$ &
$\Phi_{0}^{(h_1),ABCD}$&
${\scriptstyle 1-h_2}$ &
${\scriptstyle -\frac{1}{2}-h_2}$ &
${\cal O}_{+\frac{3}{2}}^{(h_2),E}$ &
${\scriptstyle 0}$ &
${\scriptstyle 1-(h_1+h_2)}$ &
${\scriptstyle \frac{3}{2}-(h_1+h_2)} $ &
${\cal O}_{FGH,-\frac{1}{2}}^{(h_1+h_2)}$
\\
\hline
${\scriptstyle h_1-\frac{1}{2}}$ &
$\Phi_{ABC,-\frac{1}{2}}^{(h_1)}$&
${\scriptstyle 1-h_2}$ &
${\scriptstyle -\frac{1}{2}-h_2}$ &
${\cal O}_{+\frac{3}{2}}^{(h_2),D}$ &
${\scriptstyle -\frac{1}{2}}$ &
${\scriptstyle 1-(h_1+h_2)}$ &
${\scriptstyle 2-(h_1+h_2)} $ &
${\cal O}_{AB,-1}^{(h_1+h_2)}+\cdots$
\\
\hline
${\scriptstyle h_1-1}$ &
$\Phi_{AB,-1}^{(h_1)}$&
${\scriptstyle 1-h_2}$ &
${\scriptstyle -\frac{1}{2}-h_2}$ &
${\cal O}_{+\frac{3}{2}}^{(h_2),C}$ &
${\scriptstyle -1}$ &
${\scriptstyle 1-(h_1+h_2)}$ &
${\scriptstyle \frac{5}{2}-(h_1+h_2)} $ &
${\cal O}_{A,-\frac{3}{2}}^{(h_1+h_2)}+\cdots$
\\
\hline
${\scriptstyle h_1-\frac{3}{2}}$ &
$\Phi_{A,-\frac{3}{2}}^{(h_1)}$&
${\scriptstyle 1-h_2}$ &
${\scriptstyle -\frac{1}{2}-h_2}$ &
${\cal O}_{+\frac{3}{2}}^{(h_2),B}$ &
${\scriptstyle -\frac{3}{2}}$ &
${\scriptstyle 1-(h_1+h_2)}$ &
${\scriptstyle 3-(h_1+h_2)} $ &
${\cal O}_{-2}^{(h_1+h_2)}$
\\
\hline
\hline
\end{tabular}
\caption{
The right-weights (and helicities) for the currents,
the left- and right-weights for the matter fields on the left-hand sides
and those on the right-hand sides in (\ref{20OPEs})
are given for the 
first fourteen equations of (\ref{20OPEs}) in Appendix \ref{appA}.
See also Tables \ref{table1} and \ref{table2}.}
\label{table9}
\end{table}

\begin{table}[tbp]
\centering
\renewcommand{\arraystretch}{1.7}
\begin{tabular}{|c|c||c|c|c|||c|c|c|c|| }
\hline
$ q$ &
$ {\rm w^q}$ &
$ h$ &
$ \bar{h}$ &
$ {\cal O}_{h,\bar{h}}$ &
$ s_1$ &
$ h+s_1-q$ &
$ \bar{h}+2-q$ &
${\cal O}_{h+s_1-q,\bar{h}+2-q}$  
\\
\hline
\hline
${\scriptstyle h_1+\frac{1}{2}}$ &
$  \Phi_{+\frac{1}{2}}^{(h_1),ABC}$ &
${\scriptstyle 1-h_2}$ &
${\scriptstyle -h_2}$ &
$ {\cal O}_{+1}^{(h_2),DE}$ &
${\scriptstyle +\frac{1}{2}}$ &
${\scriptstyle 1-(h_1+h_2)}$ &
${\scriptstyle \frac{3}{2}-(h_1+h_2)} $ &
$ {\cal O}_{FGH,-\frac{1}{2}}^{(h_1+h_2)}$
\\
\hline
${\scriptstyle h_1}$ &
$\Phi_{0}^{(h_1),ABCD}$&
${\scriptstyle 1-h_2}$ &
${\scriptstyle -h_2}$ &
${\cal O}_{+1}^{(h_2),EF}$ &
${\scriptstyle 0}$ &
${\scriptstyle 1-(h_1+h_2)}$ &
${\scriptstyle 2-(h_1+h_2)} $ &
${\cal O}_{GH,-1}^{(h_1+h_2)}$
\\
\hline
${\scriptstyle h_1-\frac{1}{2}}$ &
$\Phi_{ABC,-\frac{1}{2}}^{(h_1)}$&
${\scriptstyle 1-h_2}$ &
${\scriptstyle -h_2}$ &
${\cal O}_{+1}^{(h_2),DE}$ &
${\scriptstyle -\frac{1}{2}}$ &
${\scriptstyle 1-(h_1+h_2)}$ &
${\scriptstyle \frac{5}{2}-(h_1+h_2)} $ &
${\cal O}_{B,-\frac{3}{2}}^{(h_1+h_2)}+\cdots$
\\
\hline
${\scriptstyle h_1-1}$ &
$\Phi_{AB,-1}^{(h_1)}$&
${\scriptstyle 1-h_2}$ &
${\scriptstyle -h_2}$ &
${\cal O}_{+1}^{(h_2),CD}$ &
${\scriptstyle -1}$ &
${\scriptstyle 1-(h_1+h_2)}$ &
${\scriptstyle 3-(h_1+h_2)} $ &
${\cal O}_{-2}^{(h_1+h_2)}$
\\
\hline
\hline
${\scriptstyle h_1}$ &
$\Phi_{0}^{(h_1),ABCD}$&
${\scriptstyle 1-h_2}$ &
${\scriptstyle \frac{1}{2}-h_2}$ &
${\cal O}_{+\frac{1}{2}}^{(h_2),EFG}$ &
${\scriptstyle 0}$ &
${\scriptstyle 1-(h_1+h_2)}$ &
${\scriptstyle \frac{5}{2}-(h_1+h_2)} $ &
${\cal O}_{H,-\frac{3}{2}}^{(h_1+h_2)}$
\\
\hline
${\scriptstyle h_1-\frac{1}{2}}$ &
$\Phi_{ABC,-\frac{1}{2}}^{(h_1)}$&
${\scriptstyle 1-h_2}$ &
${\scriptstyle \frac{1}{2}-h_2}$ &
${\cal O}_{+\frac{1}{2}}^{(h_2),DEF}$ &
${\scriptstyle -\frac{1}{2}}$ &
${\scriptstyle 1-(h_1+h_2)}$ &
${\scriptstyle 3-(h_1+h_2)} $ &
${\cal O}_{-2}^{(h_1+h_2)}$
\\
\hline
\hline
\end{tabular}
\caption{This is a continuation of previous Table \ref{table9}.
We list the corresponding weights (and helicities) for  
the remaining six equations of (\ref{20OPEs}) in Appendix \ref{appA}.
See also Tables \ref{table1} and \ref{table2}.}
\label{table10}
\end{table}

\section{The remaining twenty (anti)commutators
between the ${\cal N}=8$ ${\rm w}_{1+\infty}$
currents and  matter fields  in section \ref{anticommutators}}

In this Appendix, we describe the twenty (anti)commutators
corresponding to 
(\ref{20OPEs})
\bea
\bigg[
\hat{\Phi}_{+\frac{3}{2},m}^{(h_1),A},
{\cal O}_{+2}^{(h_2)}(z,\bar{z}) \bigg]  & = &
{\mathscr D}_m^{h_1+\frac{3}{2}}(-1-h_2) \,
{\cal O}_{+\frac{3}{2}}^{(h_1+h_2),A}(z,\bar{z}),
\nonu \\
\bigg[
\hat{\Phi}_{+1,m}^{(h_1),AB},
{\cal O}_{+2}^{(h_2)}(z,\bar{z}) \bigg]  & = &
{\mathscr D}_m^{h_1+1}(-1-h_2) \,
{\cal O}_{+1}^{(h_1+h_2),AB}(z,\bar{z}),
\nonu \\
\bigg[
\hat{\Phi}_{+\frac{1}{2},m}^{(h_1),ABC},
{\cal O}_{+2}^{(h_2)}(z,\bar{z}) \bigg]  & = &
{\mathscr D}_m^{h_1+\frac{1}{2}}(-1-h_2) \,
{\cal O}_{+\frac{1}{2}}^{(h_1+h_2),ABC}(z,\bar{z}),
\nonu \\
\bigg[
\hat{\Phi}_{0,m}^{(h_1),ABCD},
{\cal O}_{+2}^{(h_2)}(z,\bar{z}) \bigg]  & = &
{\mathscr D}_m^{h_1}(-1-h_2) \,
{\cal O}_{0}^{(h_1+h_2),ABCD}(z,\bar{z}),
\nonu \\
\bigg[
\hat{\Phi}_{ABC,-\frac{1}{2},m}^{(h_1)},
{\cal O}_{+2}^{(h_2)}(z,\bar{z}) \bigg]  & = &
{\mathscr D}_m^{h_1-\frac{1}{2}}(-1-h_2) \,
{\cal O}_{ABC,-\frac{1}{2}}^{(h_1+h_2)}(z,\bar{z}),
\nonu \\
\bigg[
\hat{\Phi}_{AB,-1,m}^{(h_1)},
{\cal O}_{+2}^{(h_2)}(z,\bar{z}) \bigg]  & = &
{\mathscr D}_m^{h_1-1}(-1-h_2) \,
{\cal O}_{AB,-1}^{(h_1+h_2)}(z,\bar{z}),
\nonu \\
\bigg[
\hat{\Phi}_{A,-\frac{3}{2},m}^{(h_1)},
{\cal O}_{+2}^{(h_2)}(z,\bar{z}) \bigg]  & = &
{\mathscr D}_m^{h_1-\frac{3}{2}}(-1-h_2) \,
{\cal O}_{A,-\frac{3}{2}}^{(h_1+h_2)}(z,\bar{z}),
\nonu \\
\bigg[
\hat{\Phi}_{-2,m}^{(h_1)},
{\cal O}_{+2}^{(h_2)}(z,\bar{z}) \bigg]  & = &
{\mathscr D}_m^{h_1-2}(-1-h_2) \,
{\cal O}_{-2}^{(h_1+h_2)}(z,\bar{z}),
\nonu \\
\bigg[
\hat{\Phi}_{+1,m}^{(h_1),AB},
{\cal O}_{+\frac{3}{2}}^{(h_2),C}(z,\bar{z}) \bigg]  & = &
{\mathscr D}_m^{h_1+1}(-\tfrac{1}{2}-h_2) \,
{\cal O}_{+\frac{1}{2}}^{(h_1+h_2),ABC}(z,\bar{z}),
\nonu \\
\bigg\{
\hat{\Phi}_{+\frac{1}{2},m}^{(h_1),ABC},
{\cal O}_{+\frac{3}{2}}^{(h_2),D}(z,\bar{z}) \bigg\}  & = &
{\mathscr D}_m^{h_1+\frac{1}{2}}(-\tfrac{1}{2}-h_2) \,
{\cal O}_{0}^{(h_1+h_2),ABCD}(z,\bar{z}),
\nonu \\
\bigg[
\hat{\Phi}_{0,m}^{(h_1),ABCD},
{\cal O}_{+\frac{3}{2}}^{(h_2),E}(z,\bar{z}) \bigg]  & = &
\frac{1}{3!} \, \ep^{ABCDEFGH} \,
{\mathscr D}_m^{h_1}(-\tfrac{1}{2}-h_2) \,
{\cal O}_{FGH,-\frac{1}{2}}^{(h_1+h_2)}(z,\bar{z}),
\nonu \\
\bigg\{
\hat{\Phi}_{ABC,-\frac{1}{2},m}^{(h_1)},
{\cal O}_{+\frac{3}{2}}^{(h_2),D}(z,\bar{z}) \bigg\}  & = &
-3 \,
{\mathscr D}_m^{h_1-\frac{1}{2}}(-\tfrac{1}{2}-h_2) \, 
{\cal O}_{[AB,-1}^{(h_1+h_2)}(z,\bar{z}) \, \de^{D}_{C]},
\nonu \\
\bigg[
\hat{\Phi}_{AB,-1,m}^{(h_1)},
{\cal O}_{+\frac{3}{2}}^{(h_2),C}(z,\bar{z}) \bigg]  & = &
2 \,
{\mathscr D}_m^{h_1-1}(-\tfrac{1}{2}-h_2) \,
\de^{C}_{[A}
{\cal O}_{B],-\frac{3}{2}}^{(h_1+h_2)}(z,\bar{z}),
\nonu \\
\bigg\{
\hat{\Phi}_{A,-\frac{3}{2},m}^{(h_1)},
{\cal O}_{+\frac{3}{2}}^{(h_2),B}(z,\bar{z}) \bigg \}  & = &
-{\mathscr D}_m^{h_1-\frac{3}{2}}(-\tfrac{1}{2}-h_2) \,
\de^{B}_{A}
{\cal O}_{-2}^{(h_1+h_2)}(z,\bar{z}),
\nonu \\
\bigg[
\hat{\Phi}_{+\frac{1}{2},m}^{(h_1),ABC},
{\cal O}_{+1}^{(h_2),DE}(z,\bar{z}) \bigg]  & = &
\frac{1}{3!} \, \ep^{ABCDEFGH} \,
{\mathscr D}_m^{h_1+\frac{1}{2}}(-h_2) \,
{\cal O}_{FGH,-\frac{1}{2}}^{(h_1+h_2)}(z,\bar{z}),
\nonu \\
\bigg[
\hat{\Phi}_{0,m}^{(h_1),ABCD},
{\cal O}_{+1}^{(h_2),EF}(z,\bar{z}) \bigg]  & = &
\frac{1}{2!} \, \ep^{ABCDEFGH}
\, {\mathscr D}_m^{h_1}(-h_2) \,
{\cal O}_{GH,-1}^{(h_1+h_2)}(z,\bar{z}),
\nonu \\
\bigg[
\hat{\Phi}_{ABC,-\frac{1}{2},m}^{(h_1)},
{\cal O}_{+1}^{(h_2),DE}(z,\bar{z}) \bigg]  & = &
3!\,
\de^{D}_{[A} \,
{\mathscr D}_m^{h_1-\frac{1}{2}}(-h_2) \,
{\cal O}_{B,-\frac{3}{2}}^{(h_1+h_2)}(z,\bar{z}) \, \de^{E}_{C]},
\nonu \\
\bigg[
\hat{\Phi}_{AB,-1,m}^{(h_1)},
{\cal O}_{+1}^{(h_2),CD}(z,\bar{z}) \bigg]  & = &
\de^{CD}_{AB} \, {\mathscr D}_m^{h_1-1}(-h_2) \,
{\cal O}_{-2}^{(h_1+h_2)}(z,\bar{z}),
\nonu \\
\bigg[
\hat{\Phi}_{0,m}^{(h_1),ABCD},
{\cal O}_{+\frac{1}{2}}^{(h_2),EFG}(z,\bar{z}) \bigg]  & = &
\ep^{ABCDEFGH} \,
{\mathscr D}_m^{h_1}(\tfrac{1}{2}-h_2) \,
{\cal O}_{H,-\frac{3}{2}}^{(h_1+h_2)}(z,\bar{z}),
\nonu \\
\bigg\{
\hat{\Phi}_{ABC,-\frac{1}{2},m}^{(h_1)},
{\cal O}_{+\frac{1}{2}}^{(h_2),DEF}(z,\bar{z}) \bigg\}  & = &
-\de^{DEF}_{ABC} \,
{\mathscr D}_m^{h_1-\frac{1}{2}}(\tfrac{1}{2}-h_2) \,
{\cal O}_{-2}^{(h_1+h_2)}(z,\bar{z}).
\label{comm20OPEs}
\eea
As mentioned in Appendix A,
the minus signs on the right-hand sides
appear in three cases among four anticommutators.
As explained in \cite{Ahn2607},
the remaining anticommutator (the tenth case)
can also have minus sign when we define
this anticommutator with the matter field
${\cal O}_0^{(h_1+h_2),DABC}(z,\bar{z})$
rather than ${\cal O}_0^{(h_1+h_2),ABCD}(z,\bar{z})$.
The full forty-five (anti)commutators
of the currents with matter fields
with the helicity condition $s_1+s_2-s_3=2$
are summarized by (\ref{comm25OPEs}) and (\ref{comm20OPEs}).

\section{The complete list of
the weights, the helicities,
the ${\cal N}=8$ ${\rm w}_{1+\infty}$ currents, and the matter fields
for the fifty relations in section \ref{twocurrents}}

In this Appendix, we classify
the right-weights for the two currents, the left- and right-weights
for the matter fields in (\ref{wwO}) and
the combinations of the weights in (\ref{Iden})
for the remaining fifty cases
in Tables \ref{table11}, \ref{table12}, \ref{table13}
and \ref{table14}.

\begin{table}[tbp]
\centering
\renewcommand{\arraystretch}{1.7}
\begin{tabular}{|c|c||c|c||c|c|c||c|c|c| }
\hline
$ p$ &
$ {\rm w^p}$ &
$ q$ &
$ {\rm w^q}$ &
$ h $ &
$ \bar{ h}$ &
${\scriptstyle {\cal O}_{{ h},\bar{ h}}} $  &
${\scriptstyle \bar{h}+2-q}$ &
${\scriptstyle \bar{h}+2-p}$ &
${\scriptstyle p+q-2}$
\\
\hline
\hline
${\scriptstyle  h_1+\frac{3}{2}}$ &
$\Phi_{+\frac{3}{2}}^{(h_1),A}$ &
${\scriptstyle h_2+\frac{3}{2}}$ &
$\Phi_{+\frac{3}{2}}^{(h_2),B}$ &
${\scriptstyle 1-h_3}$ &
${\scriptstyle -1-h_3}$ &
${\scriptstyle {\cal O}_{+2}^{(h_3)}}$ &
${\scriptstyle -\frac{1}{2}-h_2-h_3}$ &
${\scriptstyle -\frac{1}{2}-h_1-h_3}$ &
${\scriptstyle h_1+h_2+1}$
\\
\hline
${\scriptstyle  h_1+\frac{3}{2}}$ &
$\Phi_{+\frac{3}{2}}^{(h_1),A}$ &
${\scriptstyle h_2+\frac{3}{2}}$ &
$\Phi_{+\frac{3}{2}}^{(h_2),B}$ &
${\scriptstyle 1-h_3}$ &
${\scriptstyle -\frac{1}{2}-h_3}$ &
${\scriptstyle {\cal O}_{+\frac{3}{2}}^{(h_3),C}}$  &
${\scriptstyle -h_2-h_3}$ &
${\scriptstyle -h_1-h_3}$ &
${\scriptstyle h_1+h_2+1}$
\\
\hline
${\scriptstyle  h_1+\frac{3}{2}}$ &
$\Phi_{+\frac{3}{2}}^{(h_1),A}$ &
${\scriptstyle h_2+\frac{3}{2}}$ &
$\Phi_{+\frac{3}{2}}^{(h_2),B}$ &
${\scriptstyle 1-h_3}$ &
${\scriptstyle -h_3}$ &
${\scriptstyle {\cal O}_{+1}^{(h_3),CD}}$  &
${\scriptstyle \frac{1}{2}-h_2-h_3}$ &
${\scriptstyle \frac{1}{2}-h_1-h_3}$ &
${\scriptstyle h_1+h_2+1}$
\\
\hline
${\scriptstyle  h_1+\frac{3}{2}}$ &
$\Phi_{+\frac{3}{2}}^{(h_1),A}$ &
${\scriptstyle h_2+\frac{3}{2}}$ &
$\Phi_{+\frac{3}{2}}^{(h_2),B}$ &
${\scriptstyle 1-h_3}$ &
${\scriptstyle \frac{1}{2}-h_3}$ &
${\scriptstyle {\cal O}_{+\frac{1}{2}}^{(h_3),CDE}}$  &
${\scriptstyle 1-h_2-h_3}$ &
${\scriptstyle 1-h_1-h_3}$ &
${\scriptstyle h_1+h_2+1}$
\\
\hline
${\scriptstyle  h_1+\frac{3}{2}}$ &
$\Phi_{+\frac{3}{2}}^{(h_1),A}$ &
${\scriptstyle h_2+\frac{3}{2}}$ &
$\Phi_{+\frac{3}{2}}^{(h_2),B}$ &
${\scriptstyle 1-h_3}$ &
${\scriptstyle 1-h_3}$ &
${\scriptstyle {\cal O}_{0}^{(h_3),CDEF}}$  &
${\scriptstyle \frac{3}{2}-h_2-h_3}$ &
${\scriptstyle \frac{3}{2}-h_1-h_3}$ &
${\scriptstyle h_1+h_2+1}$
\\
\hline
${\scriptstyle  h_1+\frac{3}{2}}$ &
$\Phi_{+\frac{3}{2}}^{(h_1),A}$ &
${\scriptstyle h_2+\frac{3}{2}}$ &
$\Phi_{+\frac{3}{2}}^{(h_2),B}$ &
${\scriptstyle 1-h_3}$ &
${\scriptstyle \frac{3}{2}-h_3}$ &
${\scriptstyle {\cal O}_{CDE,-\frac{1}{2}}^{(h_3)}}$  &
${\scriptstyle 2-h_2-h_3}$ &
${\scriptstyle 2-h_1-h_3}$ &
${\scriptstyle h_1+h_2+1}$
\\
\hline
${\scriptstyle  h_1+\frac{3}{2}}$ &
$\Phi_{+\frac{3}{2}}^{(h_1),A}$ &
${\scriptstyle h_2+\frac{3}{2}}$ &
$\Phi_{+\frac{3}{2}}^{(h_2),B}$ &
${\scriptstyle 1-h_3}$ &
${\scriptstyle 2-h_3}$ &
${\scriptstyle {\cal O}_{CD,-1}^{(h_3)}}$  &
${\scriptstyle \frac{5}{2}-h_2-h_3}$ &
${\scriptstyle \frac{5}{2}-h_1-h_3}$ &
${\scriptstyle h_1+h_2+1}$
\\
\hline
\hline
${\scriptstyle  h_1+\frac{3}{2}}$ &
$\Phi_{+\frac{3}{2}}^{(h_1),A}$ &
${\scriptstyle h_2+1}$ &
$\Phi_{+1}^{(h_2),BC}$ &
${\scriptstyle 1-h_3}$ &
${\scriptstyle -1-h_3}$ &
${\scriptstyle {\cal O}_{+2}^{(h_3)}}$  &
${\scriptstyle -h_2-h_3}$ &
${\scriptstyle -\frac{1}{2}-h_1-h_3}$ &
${\scriptstyle h_1+h_2+\frac{1}{2}}$
\\
\hline
${\scriptstyle  h_1+\frac{3}{2}}$ &
$\Phi_{+\frac{3}{2}}^{(h_1),A}$ &
${\scriptstyle h_2+1}$ &
$\Phi_{+1}^{(h_2),BC}$ &
${\scriptstyle 1-h_3}$ &
${\scriptstyle -\frac{1}{2}-h_3}$ &
${\scriptstyle {\cal O}_{+\frac{3}{2}}^{(h_3),D}}$  &
${\scriptstyle \frac{1}{2}-h_2-h_3}$ &
${\scriptstyle -h_1-h_3}$ &
${\scriptstyle h_1+h_2+\frac{1}{2}}$
\\
\hline
${\scriptstyle  h_1+\frac{3}{2}}$ &
$\Phi_{+\frac{3}{2}}^{(h_1),A}$ &
${\scriptstyle h_2+1}$ &
$\Phi_{+1}^{(h_2),BC}$ &
${\scriptstyle 1-h_3}$ &
${\scriptstyle -h_3}$ &
${\scriptstyle {\cal O}_{+1}^{(h_3),DE}}$  &
${\scriptstyle 1-h_2-h_3}$ &
${\scriptstyle \frac{1}{2}-h_1-h_3}$ &
${\scriptstyle h_1+h_2+\frac{1}{2}}$
\\
\hline
${\scriptstyle  h_1+\frac{3}{2}}$ &
$\Phi_{+\frac{3}{2}}^{(h_1),A}$ &
${\scriptstyle h_2+1}$ &
$\Phi_{+1}^{(h_2),BC}$ &
${\scriptstyle 1-h_3}$ &
${\scriptstyle \frac{1}{2}-h_3}$ &
${\scriptstyle {\cal O}_{+\frac{1}{2}}^{(h_3),DEF}}$  &
${\scriptstyle \frac{3}{2}-h_2-h_3}$ &
${\scriptstyle 1-h_1-h_3}$ &
${\scriptstyle h_1+h_2+\frac{1}{2}}$
\\
\hline
${\scriptstyle  h_1+\frac{3}{2}}$ &
$\Phi_{+\frac{3}{2}}^{(h_1),A}$ &
${\scriptstyle h_2+1}$ &
$\Phi_{+1}^{(h_2),BC}$ &
${\scriptstyle 1-h_3}$ &
${\scriptstyle 1-h_3}$ &
${\scriptstyle {\cal O}_{0}^{(h_3),DEFG}}$  &
${\scriptstyle 2-h_2-h_3}$ &
${\scriptstyle \frac{3}{2}-h_1-h_3}$ &
${\scriptstyle h_1+h_2+\frac{1}{2}}$
\\
\hline
${\scriptstyle  h_1+\frac{3}{2}}$ &
$\Phi_{+\frac{3}{2}}^{(h_1),A}$ &
${\scriptstyle h_2+1}$ &
$\Phi_{+1}^{(h_2),BC}$ &
${\scriptstyle 1-h_3}$ &
${\scriptstyle \frac{3}{2}-h_3}$ &
${\scriptstyle {\cal O}_{DEF,-\frac{1}{2}}^{(h_3)}}$  &
${\scriptstyle \frac{5}{2}-h_2-h_3}$ &
${\scriptstyle 2-h_1-h_3}$ &
${\scriptstyle h_1+h_2+\frac{1}{2}}$
\\
\hline
\hline
${\scriptstyle  h_1+\frac{3}{2}}$ &
$\Phi_{+\frac{3}{2}}^{(h_1),A}$ &
${\scriptstyle h_2+\frac{1}{2}}$ &
$\Phi_{+\frac{1}{2}}^{(h_2),BCD}$ &
${\scriptstyle 1-h_3}$ &
${\scriptstyle -1-h_3}$ &
${\scriptstyle {\cal O}_{+2}^{(h_3)}}$  &
${\scriptstyle \frac{1}{2}-h_2-h_3}$ &
${\scriptstyle -\frac{1}{2}-h_1-h_3}$ &
${\scriptstyle h_1+h_2}$
\\
\hline
${\scriptstyle  h_1+\frac{3}{2}}$ &
$\Phi_{+\frac{3}{2}}^{(h_1),A}$ &
${\scriptstyle h_2+\frac{1}{2}}$ &
$\Phi_{+\frac{1}{2}}^{(h_2),BCD}$ &
${\scriptstyle 1-h_3}$ &
${\scriptstyle -\frac{1}{2}-h_3}$ &
${\scriptstyle {\cal O}_{+\frac{3}{2}}^{(h_3),E}}$  &
${\scriptstyle 1-h_2-h_3}$ &
${\scriptstyle -h_1-h_3}$ &
${\scriptstyle h_1+h_2}$
\\
\hline
${\scriptstyle  h_1+\frac{3}{2}}$ &
$\Phi_{+\frac{3}{2}}^{(h_1),A}$ &
${\scriptstyle h_2+\frac{1}{2}}$ &
$\Phi_{+\frac{1}{2}}^{(h_2),BCD}$ &
${\scriptstyle 1-h_3}$ &
${\scriptstyle -h_3}$ &
${\scriptstyle {\cal O}_{+1}^{(h_3),EF}}$  &
${\scriptstyle \frac{3}{2}-h_2-h_3}$ &
${\scriptstyle \frac{1}{2}-h_1-h_3}$ &
${\scriptstyle h_1+h_2}$
\\
\hline
${\scriptstyle  h_1+\frac{3}{2}}$ &
$\Phi_{+\frac{3}{2}}^{(h_1),A}$ &
${\scriptstyle h_2+\frac{1}{2}}$ &
$\Phi_{+\frac{1}{2}}^{(h_2),BCD}$ &
${\scriptstyle 1-h_3}$ &
${\scriptstyle \frac{1}{2}-h_3}$ &
${\scriptstyle {\cal O}_{+\frac{1}{2}}^{(h_3),EFG}}$  &
${\scriptstyle 2-h_2-h_3}$ &
${\scriptstyle 1-h_1-h_3}$ &
${\scriptstyle h_1+h_2}$
\\
\hline
${\scriptstyle  h_1+\frac{3}{2}}$ &
$\Phi_{+\frac{3}{2}}^{(h_1),A}$ &
${\scriptstyle h_2+\frac{1}{2}}$ &
$\Phi_{+\frac{1}{2}}^{(h_2),BCD}$ &
${\scriptstyle 1-h_3}$ &
${\scriptstyle 1-h_3}$ &
${\scriptstyle {\cal O}_{0}^{(h_3),EFGH}}$  &
${\scriptstyle \frac{5}{2}-h_2-h_3}$ &
${\scriptstyle \frac{3}{2}-h_1-h_3}$ &
${\scriptstyle h_1+h_2}$
\\
\hline
\hline
\end{tabular}
\caption{
This is a continuation of previous Tables \ref{table3}, \ref{table4}
and \ref{table5}
and the corresponding weights and the combinations
of the weights for the next  
eighteen relations of (\ref{wwO}) in the section \ref{twocurrents}
are given explicitly.   }
\label{table11}
\end{table}

\begin{table}[tbp]
\centering
\renewcommand{\arraystretch}{1.7}
\begin{tabular}{|c|c||c|c||c|c|c||c|c|c| }
\hline
$ p$ &
$ {\rm w^p}$ &
$ q$ &
$ {\rm w^q}$ &
$ h $ &
$ \bar{ h}$ &
$ {\cal O}_{{ h},\bar{ h}}$  &
${\scriptstyle \bar{h}+2-q}$ &
${\scriptstyle \bar{h}+2-p}$ &
${\scriptstyle p+q-2}$
\\
\hline
\hline
${\scriptstyle  h_1+\frac{3}{2}}$ &
$\Phi_{+\frac{3}{2}}^{(h_1),A}$ &
${\scriptstyle h_2}$ &
$\Phi_{0}^{(h_2),BCDE}$ &
${\scriptstyle 1-h_3}$ &
${\scriptstyle -1-h_3}$ &
${\cal O}_{+2}^{(h_3)}$ &
${\scriptstyle 1-h_2-h_3}$ &
${\scriptstyle -\frac{1}{2}-h_1-h_3}$ &
${\scriptstyle h_1+h_2-\frac{1}{2}}$
\\
\hline
${\scriptstyle  h_1+\frac{3}{2}}$ &
$\Phi_{+\frac{3}{2}}^{(h_1),A}$ &
${\scriptstyle h_2}$ &
$\Phi_{0}^{(h_2),BCDE}$ &
${\scriptstyle 1-h_3}$ &
${\scriptstyle -\frac{1}{2}-h_3}$ &
${\cal O}_{+\frac{3}{2}}^{(h_3),F}$  &
${\scriptstyle \frac{3}{2}-h_2-h_3}$ &
${\scriptstyle -h_1-h_3}$ &
${\scriptstyle h_1+h_2-\frac{1}{2}}$
\\
\hline
${\scriptstyle  h_1+\frac{3}{2}}$ &
$\Phi_{+\frac{3}{2}}^{(h_1),A}$ &
${\scriptstyle h_2}$ &
$\Phi_{0}^{(h_2),BCDE}$ &
${\scriptstyle 1-h_3}$ &
${\scriptstyle -h_3}$ &
${\cal O}_{+1}^{(h_3),FG}$  &
${\scriptstyle 2-h_2-h_3}$ &
${\scriptstyle \frac{1}{2}-h_1-h_3}$ &
${\scriptstyle h_1+h_2-\frac{1}{2}}$
\\
\hline
${\scriptstyle  h_1+\frac{3}{2}}$ &
$\Phi_{+\frac{3}{2}}^{(h_1),A}$ &
${\scriptstyle h_2}$ &
$\Phi_{0}^{(h_2),BCDE}$ &
${\scriptstyle 1-h_3}$ &
${\scriptstyle \frac{1}{2}-h_3}$ &
${\cal O}_{+\frac{1}{2}}^{(h_3),FGH}$  &
${\scriptstyle \frac{5}{2}-h_2-h_3}$ &
${\scriptstyle 1-h_1-h_3}$ &
${\scriptstyle h_1+h_2-\frac{1}{2}}$
\\
\hline
\hline
${\scriptstyle  h_1+\frac{3}{2}}$ &
$\Phi_{+\frac{3}{2}}^{(h_1),A}$ &
${\scriptstyle h_2-\frac{1}{2}}$ &
$\Phi_{BCD,-\frac{1}{2}}^{(h_2)}$ &
${\scriptstyle 1-h_3}$ &
${\scriptstyle -1-h_3}$ &
${\cal O}_{+2}^{(h_3)}$  &
${\scriptstyle \frac{3}{2}-h_2-h_3}$ &
${\scriptstyle -\frac{1}{2}-h_1-h_3}$ &
${\scriptstyle h_1+h_2-1}$
\\
\hline
${\scriptstyle  h_1+\frac{3}{2}}$ &
$\Phi_{+\frac{3}{2}}^{(h_1),A}$ &
${\scriptstyle h_2-\frac{1}{2}}$ &
$\Phi_{BCD,-\frac{1}{2}}^{(h_2)}$ &
${\scriptstyle 1-h_3}$ &
${\scriptstyle -\frac{1}{2}-h_3}$ &
${\cal O}_{+\frac{3}{2}}^{(h_3),E}$  &
${\scriptstyle 2-h_2-h_3}$ &
${\scriptstyle -h_1-h_3}$ &
${\scriptstyle h_1+h_2-1}$
\\
\hline
${\scriptstyle  h_1+\frac{3}{2}}$ &
$\Phi_{+\frac{3}{2}}^{(h_1),A}$ &
${\scriptstyle h_2-\frac{1}{2}}$ &
$\Phi_{BCD,-\frac{1}{2}}^{(h_2)}$ &
${\scriptstyle 1-h_3}$ &
${\scriptstyle -h_3}$ &
${\cal O}_{+1}^{(h_3),EF}$  &
${\scriptstyle \frac{5}{2}-h_2-h_3}$ &
${\scriptstyle \frac{1}{2}-h_1-h_3}$ &
${\scriptstyle h_1+h_2-1}$
\\
\hline
\hline
${\scriptstyle  h_1+\frac{3}{2}}$ &
$\Phi_{+\frac{3}{2}}^{(h_1),A}$ &
${\scriptstyle h_2-1}$ &
$\Phi_{BC,-1}^{(h_2)}$ &
${\scriptstyle 1-h_3}$ &
${\scriptstyle -1-h_3}$ &
${\cal O}_{+2}^{(h_3)}$  &
${\scriptstyle 2-h_2-h_3}$ &
${\scriptstyle -\frac{1}{2}-h_1-h_3}$ &
${\scriptstyle h_1+h_2-\frac{3}{2}}$
\\
\hline
${\scriptstyle  h_1+\frac{3}{2}}$ &
$\Phi_{+\frac{3}{2}}^{(h_1),A}$ &
${\scriptstyle h_2-1}$ &
$\Phi_{BC,-1}^{(h_2)}$ &
${\scriptstyle 1-h_3}$ &
${\scriptstyle -\frac{1}{2}-h_3}$ &
${\cal O}_{+\frac{3}{2}}^{(h_3),D}$  &
${\scriptstyle \frac{5}{2}-h_2-h_3}$ &
${\scriptstyle -h_1-h_3}$ &
${\scriptstyle h_1+h_2-\frac{3}{2}}$
\\
\hline
\hline
${\scriptstyle  h_1+\frac{3}{2}}$ &
$\Phi_{+\frac{3}{2}}^{(h_1),A}$ &
${\scriptstyle h_2-\frac{3}{2}}$ &
$\Phi_{B,-\frac{3}{2}}^{(h_2)}$ &
${\scriptstyle 1-h_3}$ &
${\scriptstyle -1-h_3}$ &
${\cal O}_{+2}^{(h_3)}$  &
${\scriptstyle \frac{5}{2}-h_2-h_3}$ &
${\scriptstyle -\frac{1}{2}-h_1-h_3}$ &
${\scriptstyle h_1+h_2-2}$
\\
\hline
\hline
\end{tabular}
\caption{
This is a continuation of previous Table \ref{table11}
and we list the corresponding weights and the combinations
of the weights for the next  
ten relations of (\ref{wwO}) in the section \ref{twocurrents}
explicitly. }
\label{table12}
\end{table}

\begin{table}[tbp]
\centering
\renewcommand{\arraystretch}{1.7}
\begin{tabular}{|c|c||c|c||c|c|c||c|c|c| }
\hline
$ p$ &
$ {\rm w^p}$ &
$ q$ &
$ {\rm w^q}$ &
$ h $ &
$ \bar{ h}$ &
${\scriptstyle {\cal O}_{{ h},\bar{ h}}}$ &
${\scriptstyle \bar{h}+2-q}$ &
${\scriptstyle \bar{h}+2-p}$ &
${\scriptstyle p+q-2}$
\\
\hline
\hline
${\scriptstyle  h_1+1}$ &
$\Phi_{+1}^{(h_1),AB}$ &
${\scriptstyle h_2+1}$ &
${\scriptstyle \Phi_{+1}^{(h_2),CD}}$ &
${\scriptstyle 1-h_3}$ &
${\scriptstyle -1-h_3}$ &
${\scriptstyle {\cal O}_{+2}^{(h_3)}}$ &
${\scriptstyle -h_2-h_3}$ &
${\scriptstyle -h_1-h_3}$ &
${\scriptstyle h_1+h_2}$
\\
\hline
${\scriptstyle  h_1+1}$ &
$\Phi_{+1}^{(h_1),AB}$ &
${\scriptstyle h_2+1}$ &
${\scriptstyle \Phi_{+1}^{(h_2),CD}}$ &
${\scriptstyle 1-h_3}$ &
${\scriptstyle -\frac{1}{2}-h_3}$ &
${\scriptstyle {\cal O}_{+\frac{3}{2}}^{(h_3),E}}$  &
${\scriptstyle \frac{1}{2}-h_2-h_3}$ &
${\scriptstyle \frac{1}{2}-h_1-h_3}$ &
${\scriptstyle h_1+h_2}$
\\
\hline
${\scriptstyle  h_1+1}$ &
$\Phi_{+1}^{(h_1),AB}$ &
${\scriptstyle h_2+1}$ &
${\scriptstyle \Phi_{+1}^{(h_2),CD}}$ &
${\scriptstyle 1-h_3}$ &
${\scriptstyle -h_3}$ &
${\scriptstyle {\cal O}_{+1}^{(h_3),EF}}$  &
${\scriptstyle 1-h_2-h_3}$ &
${\scriptstyle 1-h_1-h_3}$ &
${\scriptstyle h_1+h_2}$
\\
\hline
${\scriptstyle  h_1+1}$ &
$\Phi_{+1}^{(h_1),AB}$ &
${\scriptstyle h_2+1}$ &
${\scriptstyle \Phi_{+1}^{(h_2),CD}}$ &
${\scriptstyle 1-h_3}$ &
${\scriptstyle \frac{1}{2}-h_3}$ &
${\scriptstyle {\cal O}_{+\frac{1}{2}}^{(h_3),EFG}}$  &
${\scriptstyle \frac{3}{2}-h_2-h_3}$ &
${\scriptstyle \frac{3}{2}-h_1-h_3}$ &
${\scriptstyle h_1+h_2}$
\\
\hline
${\scriptstyle  h_1+1}$ &
$\Phi_{+1}^{(h_1),AB}$ &
${\scriptstyle h_2+1}$ &
${\scriptstyle \Phi_{+1}^{(h_2),CD}}$ &
${\scriptstyle 1-h_3}$ &
${\scriptstyle 1-h_3}$ &
${\scriptstyle {\cal O}_{0}^{(h_3),EFGH}}$  &
${\scriptstyle 2-h_2-h_3}$ &
${\scriptstyle 2-h_1-h_3}$ &
${\scriptstyle h_1+h_2}$
\\
\hline
\hline
${\scriptstyle  h_1+1}$ &
$\Phi_{+1}^{(h_1),AB}$ &
${\scriptstyle h_2+\frac{1}{2}}$ &
${\scriptstyle \Phi_{+\frac{1}{2}}^{(h_2),CDE}}$ &
${\scriptstyle 1-h_3}$ &
${\scriptstyle -1-h_3}$ &
${\scriptstyle {\cal O}_{+2}^{(h_3)}}$  &
${\scriptstyle \frac{1}{2}-h_2-h_3}$ &
${\scriptstyle -h_1-h_3}$ &
${\scriptstyle h_1+h_2-\frac{1}{2}}$
\\
\hline
${\scriptstyle  h_1+1}$ &
$\Phi_{+1}^{(h_1),AB}$ &
${\scriptstyle h_2+\frac{1}{2}}$ &
${\scriptstyle \Phi_{+\frac{1}{2}}^{(h_2),CDE}}$ &
${\scriptstyle 1-h_3}$ &
${\scriptstyle -\frac{1}{2}-h_3}$ &
${\scriptstyle {\cal O}_{+\frac{3}{2}}^{(h_3),F}}$  &
${\scriptstyle 1-h_2-h_3}$ &
${\scriptstyle \fft12-h_1-h_3}$ &
${\scriptstyle h_1+h_2-\frac{1}{2}}$
\\
\hline
${\scriptstyle  h_1+1}$ &
$\Phi_{+1}^{(h_1),AB}$ &
${\scriptstyle h_2+\frac{1}{2}}$ &
${\scriptstyle \Phi_{+\frac{1}{2}}^{(h_2),CDE}}$ &
${\scriptstyle 1-h_3}$ &
${\scriptstyle -h_3}$ &
${\scriptstyle {\cal O}_{+1}^{(h_3),FG}}$  &
${\scriptstyle \frac{3}{2}-h_2-h_3}$ &
${\scriptstyle 1-h_1-h_3}$ &
${\scriptstyle h_1+h_2-\frac{1}{2}}$
\\
\hline
${\scriptstyle  h_1+1}$ &
$\Phi_{+1}^{(h_1),AB}$ &
${\scriptstyle h_2+\frac{1}{2}}$ &
${\scriptstyle \Phi_{+\frac{1}{2}}^{(h_2),CDE}}$ &
${\scriptstyle 1-h_3}$ &
${\scriptstyle \frac{1}{2}-h_3}$ &
${\scriptstyle {\cal O}_{+\frac{1}{2}}^{(h_3),FGH}}$  &
${\scriptstyle 2-h_2-h_3}$ &
${\scriptstyle \fft32-h_1-h_3}$ &
${\scriptstyle h_1+h_2-\frac{1}{2}}$
\\
\hline
\hline
${\scriptstyle  h_1+1}$ &
$\Phi_{+1}^{(h_1),AB}$ &
${\scriptstyle h_2}$ &
${\scriptstyle \Phi_{0}^{(h_2),CDEF}}$ &
${\scriptstyle 1-h_3}$ &
${\scriptstyle -1-h_3}$ &
${\scriptstyle {\cal O}_{+2}^{(h_3)}}$  &
${\scriptstyle 1-h_2-h_3}$ &
${\scriptstyle -h_1-h_3}$ &
${\scriptstyle h_1+h_2-1}$
\\
\hline
${\scriptstyle  h_1+1}$ &
$\Phi_{+1}^{(h_1),AB}$ &
${\scriptstyle h_2}$ &
${\scriptstyle \Phi_{0}^{(h_2),CDEF}}$ &
${\scriptstyle 1-h_3}$ &
${\scriptstyle -\frac{1}{2}-h_3}$ &
${\scriptstyle {\cal O}_{+\frac{3}{2}}^{(h_3),G}}$  &
${\scriptstyle \frac{3}{2}-h_2-h_3}$ &
${\scriptstyle \frac{1}{2}-h_1-h_3}$ &
${\scriptstyle h_1+h_2-1}$
\\
\hline
${\scriptstyle  h_1+1}$ &
$\Phi_{+1}^{(h_1),AB}$ &
${\scriptstyle h_2}$ &
${\scriptstyle \Phi_{0}^{(h_2),CDEF}}$ &
${\scriptstyle 1-h_3}$ &
${\scriptstyle -h_3}$ &
${\scriptstyle {\cal O}_{+1}^{(h_3),GH}}$  &
${\scriptstyle 2-h_2-h_3}$ &
${\scriptstyle 1-h_1-h_3}$ &
${\scriptstyle h_1+h_2-1}$
\\
\hline
\hline
${\scriptstyle  h_1+1}$ &
$\Phi_{+1}^{(h_1),AB}$ &
${\scriptstyle h_2-\frac{1}{2}}$ &
${\scriptstyle \Phi_{CDE,-\frac{1}{2}}^{(h_2)}}$ &
${\scriptstyle 1-h_3}$ &
${\scriptstyle -1-h_3}$ &
${\scriptstyle {\cal O}_{+2}^{(h_3)}}$  &
${\scriptstyle \frac{3}{2}-h_2-h_3}$ &
${\scriptstyle -h_1-h_3}$ &
${\scriptstyle h_1+h_2-\frac{3}{2}}$
\\
\hline
${\scriptstyle  h_1+1}$ &
$\Phi_{+1}^{(h_1),AB}$ &
${\scriptstyle h_2-\frac{1}{2}}$ &
${\scriptstyle \Phi_{CDE,-\frac{1}{2}}^{(h_2)}}$ &
${\scriptstyle 1-h_3}$ &
${\scriptstyle -\frac{1}{2}-h_3}$ &
${\scriptstyle {\cal O}_{+\frac{3}{2}}^{(h_3),F}}$  &
${\scriptstyle 2-h_2-h_3}$ &
${\scriptstyle \frac{1}{2}-h_1-h_3}$ &
${\scriptstyle h_1+h_2-\frac{3}{2}}$
\\
\hline
\hline
${\scriptstyle  h_1+1}$ &
$\Phi_{+1}^{(h_1),AB}$ &
${\scriptstyle h_2-1}$ &
${\scriptstyle \Phi_{CD,-1}^{(h_2)}}$ &
${\scriptstyle 1-h_3}$ &
${\scriptstyle -1-h_3}$ &
${\scriptstyle {\cal O}_{+2}^{(h_3)}}$  &
${\scriptstyle 2-h_2-h_3}$ &
${\scriptstyle -h_1-h_3}$ &
${\scriptstyle h_1+h_2-2}$
\\
\hline
\hline
\end{tabular}
\caption{
This is a continuation of previous Tables \ref{table11} and \ref{table12}
and the corresponding weights and the combinations
of the weights for the next  
fifteen relations of (\ref{wwO}) in the section \ref{twocurrents}
are presented.  }
\label{table13}
\end{table}

\begin{table}[tbp]
\centering
\renewcommand{\arraystretch}{1.7}
\begin{tabular}{|c|c||c|c||c|c|c||c|c|c| }
\hline
$ p$ &
$ {\rm w^p}$ &
$ q$ &
$ {\rm w^q}$ &
$ h $ &
$ \bar{ h}$ &
${\scriptstyle {\scriptstyle {\cal O}_{{ h},\bar{ h}}}}$ &
${\scriptstyle \bar{h}+2-q}$ &
${\scriptstyle \bar{h}+2-p}$ &
${\scriptstyle p+q-2}$
\\
\hline
\hline
${\scriptstyle  h_1+\frac{1}{2}}$ &
$\Phi_{+\frac{1}{2}}^{(h_1),ABC}$ &
${\scriptstyle h_2+\frac{1}{2}}$ &
${\scriptstyle \Phi_{+\frac{1}{2}}^{(h_2),DEF}}$ &
${\scriptstyle 1-h_3}$ &
${\scriptstyle -1-h_3}$ &
${\scriptstyle {\cal O}_{+2}^{(h_3)}}$ &
${\scriptstyle \fft12-h_2-h_3}$ &
${\scriptstyle \fft12-h_1-h_3}$ &
${\scriptstyle h_1+h_2-1}$
\\
\hline
${\scriptstyle  h_1+\frac{1}{2}}$ &
$\Phi_{+\frac{1}{2}}^{(h_1),ABC}$ &
${\scriptstyle h_2+\frac{1}{2}}$ &
${\scriptstyle \Phi_{+\frac{1}{2}}^{(h_2),DEF}}$ &
${\scriptstyle 1-h_3}$ &
${\scriptstyle -\frac{1}{2}-h_3}$ &
${\scriptstyle {\cal O}_{+\frac{3}{2}}^{(h_3),G}}$  &
${\scriptstyle 1-h_2-h_3}$ &
${\scriptstyle 1-h_1-h_3}$ &
${\scriptstyle h_1+h_2-1}$
\\
\hline
${\scriptstyle  h_1+\frac{1}{2}}$ &
$\Phi_{+\frac{1}{2}}^{(h_1),ABC}$ &
${\scriptstyle h_2+\frac{1}{2}}$ &
${\scriptstyle \Phi_{+\frac{1}{2}}^{(h_2),DEF}}$ &
${\scriptstyle 1-h_3}$ &
${\scriptstyle -h_3}$ &
${\scriptstyle {\cal O}_{+1}^{(h_3),GH}}$  &
${\scriptstyle \fft32-h_2-h_3}$ &
${\scriptstyle \fft32-h_1-h_3}$ &
${\scriptstyle h_1+h_2-1}$
\\
\hline
\hline
${\scriptstyle  h_1+\frac{1}{2}}$ &
$\Phi_{+\frac{1}{2}}^{(h_1),ABC}$ &
${\scriptstyle h_2}$ &
${\scriptstyle \Phi_{0}^{(h_2),DEFG}}$ &
${\scriptstyle 1-h_3}$ &
${\scriptstyle -1-h_3}$ &
${\scriptstyle {\cal O}_{+2}^{(h_3)}}$  &
${\scriptstyle 1-h_2-h_3}$ &
${\scriptstyle \fft12-h_1-h_3}$ &
${\scriptstyle h_1+h_2-\fft32}$
\\
\hline
${\scriptstyle  h_1+\frac{1}{2}}$ &
$\Phi_{+\frac{1}{2}}^{(h_1),ABC}$ &
${\scriptstyle h_2}$ &
${\scriptstyle \Phi_{0}^{(h_2),DEFG}}$ &
${\scriptstyle 1-h_3}$ &
${\scriptstyle -\frac{1}{2}-h_3}$ &
${\scriptstyle {\cal O}_{+\frac{3}{2}}^{(h_3),H}}$  &
${\scriptstyle \fft32-h_2-h_3}$ &
${\scriptstyle 1-h_1-h_3}$ &
${\scriptstyle h_1+h_2-\fft32}$
\\
\hline
\hline
${\scriptstyle  h_1+\frac{1}{2}}$ &
$\Phi_{+\frac{1}{2}}^{(h_1),ABC}$ &
${\scriptstyle h_2-\frac{1}{2}}$ &
${\scriptstyle \Phi_{DEF,-\frac{1}{2}}^{(h_2)}}$ &
${\scriptstyle 1-h_3}$ &
${\scriptstyle -1-h_3}$ &
${\scriptstyle {\cal O}_{+2}^{(h_3)}}$  &
${\scriptstyle \fft32-h_2-h_3}$ &
${\scriptstyle \fft12-h_1-h_3}$ &
${\scriptstyle h_1+h_2-2}$
\\
\hline
\hline
${\scriptstyle  h_1}$ &
$\Phi_{0}^{(h_1),ABCD}$ &
${\scriptstyle h_2}$ &
${\scriptstyle \Phi_{0}^{(h_2),EFGH}}$ &
${\scriptstyle 1-h_3}$ &
${\scriptstyle -1-h_3}$ &
${\scriptstyle {\cal O}_{+2}^{(h_3)}}$  &
${\scriptstyle 1-h_2-h_3}$ &
${\scriptstyle 1-h_1-h_3}$ &
${\scriptstyle h_1+h_2-2}$
\\
\hline
\hline
\end{tabular}
\caption{
This is a continuation of previous Tables \ref{table11},
\ref{table12} and \ref{table13}
and the corresponding weights and the combinations
of the weights for the final  
seven relations of (\ref{wwO}) in the section \ref{twocurrents}
are given. }
\label{table14}
\end{table}

Moreover,
the left- and right-weights for the matter fields
appearing in (\ref{Iden})
for the remaining fifty cases
with weights and helicities in (\ref{h4h4bar})
are given explicitly
in Tables \ref{table15}, \ref{table16}, \ref{table17}
and \ref{table18}.

\begin{table}[tbp]
\centering
\renewcommand{\arraystretch}{1.7}
\begin{tabular}{|c|c||c|c||c|c|||c|c|c|| }
\hline
$ p$ &
$s_1 $ &
$ q$ &
$s_2$ &
$ h $ &
$ \bar{h}$ &
$h_4$
&
$\bar{h}_4$
&
$ {\cal O}_{h_4,\bar{h}_4}$ 
\\
\hline
\hline
${\scriptstyle h_1+\frac{3}{2}}$ &
${\scriptstyle +\frac{3}{2}}$ &
${\scriptstyle h_2+\frac{3}{2}}$ &
${\scriptstyle +\frac{3}{2}}$ &
${\scriptstyle 1-h_3}$ &
${\scriptstyle -1-h_3}$ &
${\scriptstyle 1-(h_1+h_2+h_3)}$ &
${\scriptstyle -(h_1+h_2+h_3)}$ &
${\cal O}_{+1}^{(h_1+h_2+h_3),AB}$
\\
\hline
${\scriptstyle h_1+\frac{3}{2}}$ &
${\scriptstyle +\frac{3}{2}}$ &
${\scriptstyle h_2+\frac{3}{2}}$ &
${\scriptstyle +\frac{3}{2}}$ &
${\scriptstyle 1-h_3}$ &
${\scriptstyle -\frac{1}{2}-h_3}$ &
${\scriptstyle 1-(h_1+h_2+h_3)}$ &
${\scriptstyle \frac{1}{2}-(h_1+h_2+h_3)}$ &
${\cal O}_{+\frac{1}{2}}^{(h_1+h_2+h_3),ABC}$
\\
\hline
${\scriptstyle h_1+\frac{3}{2}}$ &
${\scriptstyle +\frac{3}{2}}$ &
${\scriptstyle h_2+\frac{3}{2}}$ &
${\scriptstyle +\frac{3}{2}}$ &
${\scriptstyle 1-h_3}$ &
${\scriptstyle -h_3}$ &
${\scriptstyle 1-(h_1+h_2+h_3)}$ &
${\scriptstyle 1-(h_1+h_2+h_3)}$ &
${\cal O}_{0}^{(h_1+h_2+h_3),ABCD}$
\\
\hline
${\scriptstyle h_1+\frac{3}{2}}$ &
${\scriptstyle +\frac{3}{2}}$ &
${\scriptstyle h_2+\frac{3}{2}}$ &
${\scriptstyle +\frac{3}{2}}$ &
${\scriptstyle 1-h_3}$ &
${\scriptstyle \frac{1}{2}-h_3}$ &
${\scriptstyle 1-(h_1+h_2+h_3)}$ &
${\scriptstyle \frac{3}{2}-(h_1+h_2+h_3)}$ &
${\cal O}_{FGH,-\frac{1}{2}}^{(h_1+h_2+h_3)}$
\\
\hline
${\scriptstyle h_1+\frac{3}{2}}$ &
${\scriptstyle +\frac{3}{2}}$ &
${\scriptstyle h_2+\frac{3}{2}}$ &
${\scriptstyle +\frac{3}{2}}$ &
${\scriptstyle 1-h_3}$ &
${\scriptstyle 1-h_3}$ &
${\scriptstyle 1-(h_1+h_2+h_3)}$ &
${\scriptstyle 2-(h_1+h_2+h_3)}$ &
${\cal O}_{GH,-1}^{(h_1+h_2+h_3)}$
\\
\hline
${\scriptstyle h_1+\frac{3}{2}}$ &
${\scriptstyle +\frac{3}{2}}$ &
${\scriptstyle h_2+\frac{3}{2}}$ &
${\scriptstyle +\frac{3}{2}}$ &
${\scriptstyle 1-h_3}$ &
${\scriptstyle \frac{3}{2}-h_3}$ &
${\scriptstyle 1-(h_1+h_2+h_3)}$ &
${\scriptstyle \frac{5}{2}-(h_1+h_2+h_3)}$ &
${\cal O}_{C,-\frac{3}{2}}^{(h_1+h_2+h_3)}+\cdots$
\\
\hline
${\scriptstyle h_1+\frac{3}{2}}$ &
${\scriptstyle +\frac{3}{2}}$ &
${\scriptstyle h_2+\frac{3}{2}}$ &
${\scriptstyle +\frac{3}{2}}$ &
${\scriptstyle 1-h_3}$ &
${\scriptstyle 2-h_3}$ &
${\scriptstyle 1-(h_1+h_2+h_3)}$ &
${\scriptstyle 3-(h_1+h_2+h_3)}$ &
${\cal O}_{-2}^{(h_1+h_2+h_3)}$
\\
\hline
\hline
${\scriptstyle h_1+\frac{3}{2}}$ &
${\scriptstyle +\frac{3}{2}}$ &
${\scriptstyle h_2+1}$ &
${\scriptstyle +1}$ &
${\scriptstyle 1-h_3}$ &
${\scriptstyle -1-h_3}$ &
${\scriptstyle 1-(h_1+h_2+h_3)}$ &
${\scriptstyle \frac{1}{2}-(h_1+h_2+h_3)}$ &
${\cal O}_{+\frac{1}{2}}^{(h_1+h_2+h_3),ABC}$
\\
\hline
${\scriptstyle h_1+\frac{3}{2}}$ &
${\scriptstyle +\frac{3}{2}}$ &
${\scriptstyle h_2+1}$ &
${\scriptstyle +1}$ &
${\scriptstyle 1-h_3}$ &
${\scriptstyle -\frac{1}{2}-h_3}$ &
${\scriptstyle 1-(h_1+h_2+h_3)}$ &
${\scriptstyle 1-(h_1+h_2+h_3)}$ &
${\cal O}_{0}^{(h_1+h_2+h_3),ABCD}$
\\
\hline
${\scriptstyle h_1+\frac{3}{2}}$ &
${\scriptstyle +\frac{3}{2}}$ &
${\scriptstyle h_2+1}$ &
${\scriptstyle +1}$ &
${\scriptstyle 1-h_3}$ &
${\scriptstyle -h_3}$ &
${\scriptstyle 1-(h_1+h_2+h_3)}$ &
${\scriptstyle \frac{3}{2}-(h_1+h_2+h_3)}$ &
${\cal O}_{FGH,-\frac{1}{2}}^{(h_1+h_2+h_3)}$
\\
\hline
${\scriptstyle h_1+\frac{3}{2}}$ &
${\scriptstyle +\frac{3}{2}}$ &
${\scriptstyle h_2+1}$ &
${\scriptstyle +1}$ &
${\scriptstyle 1-h_3}$ &
${\scriptstyle \frac{1}{2}-h_3}$ &
${\scriptstyle 1-(h_1+h_2+h_3)}$ &
${\scriptstyle 2-(h_1+h_2+h_3)}$ &
${\cal O}_{GH,-1}^{(h_1+h_2+h_3)}$
\\
\hline
${\scriptstyle h_1+\frac{3}{2}}$ &
${\scriptstyle +\frac{3}{2}}$ &
${\scriptstyle h_2+1}$ &
${\scriptstyle +1}$ &
${\scriptstyle 1-h_3}$ &
${\scriptstyle 1-h_3}$ &
${\scriptstyle 1-(h_1+h_2+h_3)}$ &
${\scriptstyle \frac{5}{2}-(h_1+h_2+h_3)}$ &
${\cal O}_{H,-\frac{3}{2}}^{(h_1+h_2+h_3)}$
\\
\hline
${\scriptstyle h_1+\frac{3}{2}}$ &
${\scriptstyle +\frac{3}{2}}$ &
${\scriptstyle h_2+1}$ &
${\scriptstyle +1}$ &
${\scriptstyle 1-h_3}$ &
${\scriptstyle \frac{3}{2}-h_3}$ &
${\scriptstyle 1-(h_1+h_2+h_3)}$ &
${\scriptstyle 3-(h_1+h_2+h_3)}$ &
${\cal O}_{-2}^{(h_1+h_2+h_3)}$
\\
\hline
\hline
${\scriptstyle h_1+\frac{3}{2}}$ &
${\scriptstyle +\frac{3}{2}}$ &
${\scriptstyle h_2+\frac{1}{2}}$ &
${\scriptstyle +\frac{1}{2}}$ &
${\scriptstyle 1-h_3}$ &
${\scriptstyle -1-h_3}$ &
${\scriptstyle 1-(h_1+h_2+h_3)}$ &
${\scriptstyle 1-(h_1+h_2+h_3)}$ &
${\cal O}_{0}^{(h_1+h_2+h_3),ABCD}$
\\
\hline
${\scriptstyle h_1+\frac{3}{2}}$ &
${\scriptstyle +\frac{3}{2}}$ &
${\scriptstyle h_2+\frac{1}{2}}$ &
${\scriptstyle +\frac{1}{2}}$ &
${\scriptstyle 1-h_3}$ &
${\scriptstyle -\frac{1}{2}-h_3}$ &
${\scriptstyle 1-(h_1+h_2+h_3)}$ &
${\scriptstyle \frac{3}{2}-(h_1+h_2+h_3)}$ &
${\cal O}_{FGH,-\frac{1}{2}}^{(h_1+h_2+h_3)}$
\\
\hline
${\scriptstyle h_1+\frac{3}{2}}$ &
${\scriptstyle +\frac{3}{2}}$ &
${\scriptstyle h_2+\frac{1}{2}}$ &
${\scriptstyle +\frac{1}{2}}$ &
${\scriptstyle 1-h_3}$ &
${\scriptstyle -h_3}$ &
${\scriptstyle 1-(h_1+h_2+h_3)}$ &
${\scriptstyle 2-(h_1+h_2+h_3)}$ &
${\cal O}_{GH,-1}^{(h_1+h_2+h_3)}$
\\
\hline
${\scriptstyle h_1+\frac{3}{2}}$ &
${\scriptstyle +\frac{3}{2}}$ &
${\scriptstyle h_2+\frac{1}{2}}$ &
${\scriptstyle +\frac{1}{2}}$ &
${\scriptstyle 1-h_3}$ &
${\scriptstyle \frac{1}{2}-h_3}$ &
${\scriptstyle 1-(h_1+h_2+h_3)}$ &
${\scriptstyle \frac{5}{2}-(h_1+h_2+h_3)}$ &
${\cal O}_{H,-\frac{3}{2}}^{(h_1+h_2+h_3)}$
\\
\hline
${\scriptstyle h_1+\frac{3}{2}}$ &
${\scriptstyle +\frac{3}{2}}$ &
${\scriptstyle h_2+\frac{1}{2}}$ &
${\scriptstyle +\frac{1}{2}}$ &
${\scriptstyle 1-h_3}$ &
${\scriptstyle 1-h_3}$ &
${\scriptstyle 1-(h_1+h_2+h_3)}$ &
${\scriptstyle 3-(h_1+h_2+h_3)}$ &
${\cal O}_{-2}^{(h_1+h_2+h_3)}$
\\
\hline
\hline
\end{tabular}
\caption{For given right-weights, helicities, and the left- and right-weights
appearing in Table \ref{table11}, 
the left- and right-weights for the eighteen matter fields of (\ref{Iden})
in the section \ref{twocurrents} are given.
The relation (\ref{h4h4bar}) is used.
This is a continuation of previous Tables \ref{table6},
\ref{table7} and \ref{table8}.}
\label{table15}
\end{table}

\begin{table}[tbp]
\centering
\renewcommand{\arraystretch}{1.7}
\begin{tabular}{|c|c||c|c||c|c|||c|c|c|| }
\hline
$ p$ &
$s_1 $ &
$ q$ &
$s_2$ &
$ h $ &
$ \bar{h}$ &
$h_4$
&
$\bar{h}_4$
&
$ {\cal O}_{h_4,\bar{h}_4}$ 
\\
\hline
\hline
${\scriptstyle h_1+\frac{3}{2}}$ &
${\scriptstyle +\frac{3}{2}}$ &
${\scriptstyle h_2}$ &
${\scriptstyle 0}$ &
${\scriptstyle 1-h_3}$ &
${\scriptstyle -1-h_3}$ &
${\scriptstyle 1-(h_1+h_2+h_3)}$ &
${\scriptstyle \frac{3}{2}-(h_1+h_2+h_3)}$ &
${\cal O}_{FGH,-\frac{1}{2}}^{(h_1+h_2+h_3)}$
\\
\hline
${\scriptstyle h_1+\frac{3}{2}}$ &
${\scriptstyle +\frac{3}{2}}$ &
${\scriptstyle h_2}$ &
${\scriptstyle 0}$ &
${\scriptstyle 1-h_3}$ &
${\scriptstyle -\frac{1}{2}-h_3}$ &
${\scriptstyle 1-(h_1+h_2+h_3)}$ &
${\scriptstyle 2-(h_1+h_2+h_3)}$ &
${\cal O}_{GH,-1}^{(h_1+h_2+h_3)}+\cdots$
\\
\hline
${\scriptstyle h_1+\frac{3}{2}}$ &
${\scriptstyle +\frac{3}{2}}$ &
${\scriptstyle h_2}$ &
${\scriptstyle 0}$ &
${\scriptstyle 1-h_3}$ &
${\scriptstyle -h_3}$ &
${\scriptstyle 1-(h_1+h_2+h_3)}$ &
${\scriptstyle \frac{5}{2}-(h_1+h_2+h_3)}$ &
${\cal O}_{H,-\frac{3}{2}}^{(h_1+h_2+h_3)}+\cdots$
\\
\hline
${\scriptstyle h_1+\frac{3}{2}}$ &
${\scriptstyle +\frac{3}{2}}$ &
${\scriptstyle h_2}$ &
${\scriptstyle 0}$ &
${\scriptstyle 1-h_3}$ &
${\scriptstyle \frac{1}{2}-h_3}$ &
${\scriptstyle 1-(h_1+h_2+h_3)}$ &
${\scriptstyle 3-(h_1+h_2+h_3)}$ &
${\cal O}_{-2}^{(h_1+h_2+h_3)}$
\\
\hline
\hline
${\scriptstyle h_1+\frac{3}{2}}$ &
${\scriptstyle +\frac{3}{2}}$ &
${\scriptstyle h_2-\frac{1}{2}}$ &
${\scriptstyle -\frac{1}{2}}$ &
${\scriptstyle 1-h_3}$ &
${\scriptstyle -1-h_3}$ &
${\scriptstyle 1-(h_1+h_2+h_3)}$ &
${\scriptstyle 2-(h_1+h_2+h_3)}$ &
${\cal O}_{CD,-1}^{(h_1+h_2+h_3)}+\cdots$
\\
\hline
${\scriptstyle h_1+\frac{3}{2}}$ &
${\scriptstyle +\frac{3}{2}}$ &
${\scriptstyle h_2-\frac{1}{2}}$ &
${\scriptstyle -\frac{1}{2}}$ &
${\scriptstyle 1-h_3}$ &
${\scriptstyle -\frac{1}{2}-h_3}$ &
${\scriptstyle 1-(h_1+h_2+h_3)}$ &
${\scriptstyle \frac{5}{2}-(h_1+h_2+h_3)}$ &
${\cal O}_{D,-\frac{3}{2}}^{(h_1+h_2+h_3)}+\cdots$
\\
\hline
${\scriptstyle h_1+\frac{3}{2}}$ &
${\scriptstyle +\frac{3}{2}}$ &
${\scriptstyle h_2-\frac{1}{2}}$ &
${\scriptstyle -\frac{1}{2}}$ &
${\scriptstyle 1-h_3}$ &
${\scriptstyle -h_3}$ &
${\scriptstyle 1-(h_1+h_2+h_3)}$ &
${\scriptstyle 3-(h_1+h_2+h_3)}$ &
${\cal O}_{-2}^{(h_1+h_2+h_3)}$
\\
\hline
\hline
${\scriptstyle h_1+\frac{3}{2}}$ &
${\scriptstyle +\frac{3}{2}}$ &
${\scriptstyle h_2-1}$ &
${\scriptstyle -1}$ &
${\scriptstyle 1-h_3}$ &
${\scriptstyle -1-h_3}$ &
${\scriptstyle 1-(h_1+h_2+h_3)}$ &
${\scriptstyle \frac{5}{2}-(h_1+h_2+h_3)}$ &
${\cal O}_{C,-\frac{3}{2}}^{(h_1+h_2+h_3)}+\cdots$
\\
\hline
${\scriptstyle h_1+\frac{3}{2}}$ &
${\scriptstyle +\frac{3}{2}}$ &
${\scriptstyle h_2-1}$ &
${\scriptstyle -1}$ &
${\scriptstyle 1-h_3}$ &
${\scriptstyle -\frac{1}{2}-h_3}$ &
${\scriptstyle 1-(h_1+h_2+h_3)}$ &
${\scriptstyle 3-(h_1+h_2+h_3)}$ &
${\cal O}_{-2}^{(h_1+h_2+h_3)}$
\\
\hline
\hline
${\scriptstyle h_1+\frac{3}{2}}$ &
${\scriptstyle +\frac{3}{2}}$ &
${\scriptstyle h_2-\frac{3}{2}}$ &
${\scriptstyle -\frac{3}{2}}$ &
${\scriptstyle 1-h_3}$ &
${\scriptstyle -1-h_3}$ &
${\scriptstyle 1-(h_1+h_2+h_3)}$ &
${\scriptstyle 3-(h_1+h_2+h_3)}$ &
${\cal O}_{-2}^{(h_1+h_2+h_3)}$
\\
\hline
\hline
\end{tabular}
\caption{For given right-weights, helicities, and the left- and right-weights
appearing in Table \ref{table12}, 
the left- and right-weights for the ten matter fields of (\ref{Iden})
in the section \ref{twocurrents} are written. }
\label{table16}
\end{table}

\begin{table}[tbp]
\centering
\renewcommand{\arraystretch}{1.7}
\begin{tabular}{|c|c||c|c||c|c|||c|c|c|| }
\hline
$ p$ &
$s_1 $ &
$ q$ &
$s_2$ &
$ h $ &
$ \bar{h}$ &
$h_4$
&
$\bar{h}_4$
&
$ {\cal O}_{h_4,\bar{h}_4}$ 
\\
\hline
\hline
${\scriptstyle h_1+1}$ &
${\scriptstyle +1}$ &
${\scriptstyle h_2+1}$ &
${\scriptstyle +1}$ &
${\scriptstyle 1-h_3}$ &
${\scriptstyle -1-h_3}$ &
${\scriptstyle 1-(h_1+h_2+h_3)}$ &
${\scriptstyle 1-(h_1+h_2+h_3)}$ &
${\cal O}_{0}^{(h_1+h_2+h_3),ABCD}$
\\
\hline
${\scriptstyle h_1+1}$ &
${\scriptstyle +1}$ &
${\scriptstyle h_2+1}$ &
${\scriptstyle +1}$ &
${\scriptstyle 1-h_3}$ &
${\scriptstyle -\frac{1}{2}-h_3}$ &
${\scriptstyle 1-(h_1+h_2+h_3)}$ &
${\scriptstyle \frac{3}{2}-(h_1+h_2+h_3)}$ &
${\cal O}_{FGH,-\frac{1}{2}}^{(h_1+h_2+h_3)}$
\\
\hline
${\scriptstyle h_1+1}$ &
${\scriptstyle +1}$ &
${\scriptstyle h_2+1}$ &
${\scriptstyle +1}$ &
${\scriptstyle 1-h_3}$ &
${\scriptstyle -h_3}$ &
${\scriptstyle 1-(h_1+h_2+h_3)}$ &
${\scriptstyle 2-(h_1+h_2+h_3)}$ &
${\cal O}_{GH,-1}^{(h_1+h_2+h_3)}$
\\
\hline
${\scriptstyle h_1+1}$ &
${\scriptstyle +1}$ &
${\scriptstyle h_2+1}$ &
${\scriptstyle +1}$ &
${\scriptstyle 1-h_3}$ &
${\scriptstyle \frac{1}{2}-h_3}$ &
${\scriptstyle 1-(h_1+h_2+h_3)}$ &
${\scriptstyle \frac{5}{2}-(h_1+h_2+h_3)}$ &
${\cal O}_{H,-\frac{3}{2}}^{(h_1+h_2+h_3)}$
\\
\hline
${\scriptstyle h_1+1}$ &
${\scriptstyle +1}$ &
${\scriptstyle h_2+1}$ &
${\scriptstyle +1}$ &
${\scriptstyle 1-h_3}$ &
${\scriptstyle 1-h_3}$ &
${\scriptstyle 1-(h_1+h_2+h_3)}$ &
${\scriptstyle 3-(h_1+h_2+h_3)}$ &
${\cal O}_{-2}^{(h_1+h_2+h_3)}$
\\
\hline
\hline
${\scriptstyle h_1+1}$ &
${\scriptstyle +1}$ &
${\scriptstyle h_2+\frac{1}{2}}$ &
${\scriptstyle +\frac{1}{2}}$ &
${\scriptstyle 1-h_3}$ &
${\scriptstyle -1-h_3}$ &
${\scriptstyle 1-(h_1+h_2+h_3)}$ &
${\scriptstyle \frac{3}{2}-(h_1+h_2+h_3)}$ &
${\cal O}_{FGH,-\frac{1}{2}}^{(h_1+h_2+h_3)}$
\\
\hline
${\scriptstyle h_1+1}$ &
${\scriptstyle +1}$ &
${\scriptstyle h_2+\frac{1}{2}}$ &
${\scriptstyle +\frac{1}{2}}$ &
${\scriptstyle 1-h_3}$ &
${\scriptstyle -\frac{1}{2}-h_3}$ &
${\scriptstyle 1-(h_1+h_2+h_3)}$ &
${\scriptstyle 2-(h_1+h_2+h_3)}$ &
${\cal O}_{GH,-1}^{(h_1+h_2+h_3)}+\cdots$
\\
\hline
${\scriptstyle h_1+1}$ &
${\scriptstyle +1}$ &
${\scriptstyle h_2+\frac{1}{2}}$ &
${\scriptstyle +\frac{1}{2}}$ &
${\scriptstyle 1-h_3}$ &
${\scriptstyle -h_3}$ &
${\scriptstyle 1-(h_1+h_2+h_3)}$ &
${\scriptstyle \frac{5}{2}-(h_1+h_2+h_3)}$ &
${\cal O}_{H,-\frac{3}{2}}^{(h_1+h_2+h_3)}+\cdots$
\\
\hline
${\scriptstyle h_1+1}$ &
${\scriptstyle +1}$ &
${\scriptstyle h_2+\frac{1}{2}}$ &
${\scriptstyle +\frac{1}{2}}$ &
${\scriptstyle 1-h_3}$ &
${\scriptstyle \frac{1}{2}-h_3}$ &
${\scriptstyle 1-(h_1+h_2+h_3)}$ &
${\scriptstyle 3-(h_1+h_2+h_3)}$ &
${\cal O}_{-2}^{(h_1+h_2+h_3)}$
\\
\hline
\hline
${\scriptstyle h_1+1}$ &
${\scriptstyle +1}$ &
${\scriptstyle h_2}$ &
${\scriptstyle 0}$ &
${\scriptstyle 1-h_3}$ &
${\scriptstyle -1-h_3}$ &
${\scriptstyle 1-(h_1+h_2+h_3)}$ &
${\scriptstyle 2-(h_1+h_2+h_3)}$ &
${\cal O}_{GH,-1}^{(h_1+h_2+h_3)}$
\\
\hline
${\scriptstyle h_1+1}$ &
${\scriptstyle +1}$ &
${\scriptstyle h_2}$ &
${\scriptstyle 0}$ &
${\scriptstyle 1-h_3}$ &
${\scriptstyle -\frac{1}{2}-h_3}$ &
${\scriptstyle 1-(h_1+h_2+h_3)}$ &
${\scriptstyle \frac{5}{2}-(h_1+h_2+h_3)}$ &
${\cal O}_{H,-\frac{3}{2}}^{(h_1+h_2+h_3)}+\cdots$
\\
\hline
${\scriptstyle h_1+1}$ &
${\scriptstyle +1}$ &
${\scriptstyle h_2}$ &
${\scriptstyle 0}$ &
${\scriptstyle 1-h_3}$ &
${\scriptstyle -h_3}$ &
${\scriptstyle 1-(h_1+h_2+h_3)}$ &
${\scriptstyle 3-(h_1+h_2+h_3)}$ &
${\cal O}_{-2}^{(h_1+h_2+h_3)}$
\\
\hline
\hline
${\scriptstyle h_1+1}$ &
${\scriptstyle +1}$ &
${\scriptstyle h_2-\frac{1}{2}}$ &
${\scriptstyle -\frac{1}{2}}$ &
${\scriptstyle 1-h_3}$ &
${\scriptstyle -1-h_3}$ &
${\scriptstyle 1-(h_1+h_2+h_3)}$ &
${\scriptstyle \frac{5}{2}-(h_1+h_2+h_3)}$ &
${\cal O}_{C,-\frac{3}{2}}^{(h_1+h_2+h_3)}+\cdots$
\\
\hline
${\scriptstyle h_1+1}$ &
${\scriptstyle +1}$ &
${\scriptstyle h_2-\frac{1}{2}}$ &
${\scriptstyle -\frac{1}{2}}$ &
${\scriptstyle 1-h_3}$ &
${\scriptstyle -\frac{1}{2}-h_3}$ &
${\scriptstyle 1-(h_1+h_2+h_3)}$ &
${\scriptstyle 3-(h_1+h_2+h_3)}$ &
${\cal O}_{-2}^{(h_1+h_2+h_3)}$
\\
\hline
\hline
${\scriptstyle h_1+1}$ &
${\scriptstyle +1}$ &
${\scriptstyle h_2-1} $ &
${\scriptstyle -1}$ &
${\scriptstyle 1-h_3}$ &
${\scriptstyle -1-h_3}$ &
${\scriptstyle 1-(h_1+h_2+h_3)}$ &
${\scriptstyle 3-(h_1+h_2+h_3)}$ &
${\cal O}_{-2}^{(h_1+h_2+h_3)}$
\\
\hline
\hline
\end{tabular}
\caption{
For given right-weights, helicities, and the left- and right-weights
appearing in Table \ref{table13}, we list 
the left- and right-weights for the fifteen matter fields of (\ref{Iden})
in the section \ref{twocurrents}.   }
\label{table17}
\end{table}

\begin{table}[tbp]
\centering
\renewcommand{\arraystretch}{1.7}
\begin{tabular}{|c|c||c|c||c|c|||c|c|c|| }
\hline
$ p$ &
$s_1 $ &
$ q$ &
$s_2$ &
$ h $ &
$ \bar{h}$ &
$h_4$
&
$\bar{h}_4$
&
$ {\cal O}_{h_4,\bar{h}_4}$ 
\\
\hline
\hline
${\scriptstyle h_1+\frac{1}{2}}$ &
${\scriptstyle +\frac{1}{2}}$ &
${\scriptstyle h_2+\frac{1}{2}}$ &
${\scriptstyle +\frac{1}{2}}$ &
${\scriptstyle 1-h_3}$ &
${\scriptstyle -1-h_3}$ &
${\scriptstyle 1-(h_1+h_2+h_3)}$ &
${\scriptstyle 2-(h_1+h_2+h_3)}$ &
${\cal O}_{GH,-1}^{(h_1+h_2+h_3)}$
\\
\hline
${\scriptstyle h_1+\frac{1}{2}}$ &
${\scriptstyle +\frac{1}{2}}$ &
${\scriptstyle h_2+\frac{1}{2}}$ &
${\scriptstyle +\frac{1}{2}}$ &
${\scriptstyle 1-h_3}$ &
${\scriptstyle -\frac{1}{2}-h_3}$ &
${\scriptstyle 1-(h_1+h_2+h_3)}$ &
${\scriptstyle \frac{5}{2}-(h_1+h_2+h_3)}$ &
${\cal O}_{H,-\frac{3}{2}}^{(h_1+h_2+h_3)}+\cdots$
\\
\hline
${\scriptstyle h_1+\frac{1}{2}}$ &
${\scriptstyle +\frac{1}{2}}$ &
${\scriptstyle h_2+\frac{1}{2}}$ &
${\scriptstyle +\frac{1}{2}}$ &
${\scriptstyle 1-h_3}$ &
${\scriptstyle -h_3}$ &
${\scriptstyle 1-(h_1+h_2+h_3)}$ &
${\scriptstyle 3-(h_1+h_2+h_3)}$ &
${\cal O}_{-2}^{(h_1+h_2+h_3)}$
\\
\hline
\hline
${\scriptstyle h_1+\frac{1}{2}}$ &
${\scriptstyle +\frac{1}{2}}$ &
${\scriptstyle h_2}$ &
${\scriptstyle 0}$ &
${\scriptstyle 1-h_3}$ &
${\scriptstyle -1-h_3}$ &
${\scriptstyle 1-(h_1+h_2+h_3)}$ &
${\scriptstyle \frac{5}{2}-(h_1+h_2+h_3)}$ &
${\cal O}_{H,-\frac{3}{2}}^{(h_1+h_2+h_3)}$
\\
\hline
${\scriptstyle h_1+\frac{1}{2}}$ &
${\scriptstyle +\frac{1}{2}}$ &
${\scriptstyle h_2}$ &
${\scriptstyle 0}$ &
${\scriptstyle 1-h_3}$ &
${\scriptstyle -\frac{1}{2}-h_3}$ &
${\scriptstyle 1-(h_1+h_2+h_3)}$ &
${\scriptstyle 3-(h_1+h_2+h_3)}$ &
${\cal O}_{-2}^{(h_1+h_2+h_3)}$
\\
\hline
\hline
${\scriptstyle h_1+\frac{1}{2}}$ &
${\scriptstyle +\frac{1}{2}}$ &
${\scriptstyle h_2-\frac{1}{2}}$ &
${\scriptstyle -\frac{1}{2}}$ &
${\scriptstyle 1-h_3}$ &
${\scriptstyle -1-h_3}$ &
${\scriptstyle 1-(h_1+h_2+h_3)}$ &
${\scriptstyle 3-(h_1+h_2+h_3)}$ &
${\cal O}_{-2}^{(h_1+h_2+h_3)}$
\\
\hline
\hline
${\scriptstyle h_1}$ &
${\scriptstyle 0}$ &
${\scriptstyle h_2}$ &
${\scriptstyle 0}$ &
${\scriptstyle 1-h_3}$ &
${\scriptstyle -1-h_3}$ &
${\scriptstyle 1-(h_1+h_2+h_3)}$ &
${\scriptstyle 3-(h_1+h_2+h_3)}$ &
${\cal O}_{-2}^{(h_1+h_2+h_3)}$
\\
\hline
\hline
\end{tabular}
\caption{
For given right-weights, helicities, and the left- and right-weights
appearing in Table \ref{table14}, 
the left- and right-weights for the seven matter fields of (\ref{Iden})
in the section \ref{twocurrents} are described. }
\label{table18}
\end{table}


\section{The complete ninety-five
relations of two successive actions of the
${\cal N}=8$ ${\rm w}_{1+\infty}$ currents on the matter fields
in section \ref{twocurrents}}

\subsection{The forty-five relations of (\ref{Iden})
\label{45rel}}

We present the relations in (\ref{Iden})
for the forty-five cases.

\subsubsection{The seventeen relations }

The seventeen relations for (\ref{Iden}) corresponding to
Table \ref{table6} 
are given by
\bea
&& \Bigg[
\la_{+2,+2,-2} \,\la_{+2,+2,-2} \,
{\mathscr D}_n^{h_2+2}(-1-h_3) \, {\mathscr D}_m^{h_1+2}(-1-h_2-h_3)
\nonu \\
&& - \la_{+2,+2,-2} \,\la_{+2,+2,-2}\,
{\mathscr D}_m^{h_1+2}(-1-h_3) \, {\mathscr D}_n^{h_2+2}(-1-h_1-h_3)
\nonu \\
&& -
 \la_{+2,+2,-2} \,\kappa_{+2,+2,-2} \, \bigg( (h_2+1)m-(h_1+1)n \bigg) \,
{\mathscr D}_{m+n}^{h_1+h_2+2}(-1-h_3)\, \Bigg]\,
{\cal O}_{+2}^{(h_1+h_2+h_3)}(z,\bar{z})=0,
\nonu \\
&& \Bigg[
\la_{+2,+\frac{3}{2},-\frac{3}{2}}\,\la_{+2,+\frac{3}{2},-\frac{3}{2}} \,
{\mathscr D}_n^{h_2+2}(-\tfrac{1}{2}-h_3) \,
{\mathscr D}_m^{h_1+2}(-\tfrac{1}{2}-h_2-h_3)
\nonu \\
&& -\la_{+2,+\frac{3}{2},-\frac{3}{2}}\,\la_{+2,+\frac{3}{2},-\frac{3}{2}} \,
{\mathscr D}_m^{h_1+2}(-\tfrac{1}{2}-h_3) \,
{\mathscr D}_n^{h_2+2}(-\tfrac{1}{2}-h_1-h_3)
\nonu \\
&& -
\la_{+2,+\frac{3}{2},-\frac{3}{2}} \,\kappa_{+2,+2,-2} \,
\bigg( (h_2+1)m-(h_1+1)n \bigg) \,
{\mathscr D}_{m+n}^{h_1+h_2+2}(-\tfrac{1}{2}-h_3)\, \Bigg]
\nonu \\
&& \times
{\cal O}_{+\frac{3}{2}}^{(h_1+h_2+h_3),A}(z,\bar{z})=0,
\nonu \\
&& \Bigg[
\la_{+2,+1,-1}\,\la_{+2,+1,-1}\,
{\mathscr D}_n^{h_2+2}(-h_3) \,
{\mathscr D}_m^{h_1+2}(-h_2-h_3)
\nonu \\
&& -\la_{+2,+1,-1}\,\la_{+2,+1,-1}\,
{\mathscr D}_m^{h_1+2}(-h_3) \,
{\mathscr D}_n^{h_2+2}(-h_1-h_3)
\nonu \\
&& -
\la_{+2,+1,-1} \, \kappa_{+2,+2,-2} \,
\bigg( (h_2+1)m-(h_1+1)n \bigg) \,
{\mathscr D}_{m+n}^{h_1+h_2+2}(-h_3)\, \Bigg]\,
{\cal O}_{+1}^{(h_1+h_2+h_3),AB}(z,\bar{z})=0,
\nonu \\
&& \Bigg[
\la_{+2,+\frac{1}{2},-\frac{1}{2}} \,\la_{+2,+\frac{1}{2},-\frac{1}{2}} \,
{\mathscr D}_n^{h_2+2}(\tfrac{1}{2}-h_3) \,
{\mathscr D}_m^{h_1+2}(\tfrac{1}{2}-h_2-h_3)
\nonu \\
&& -\la_{+2,+\frac{1}{2},-\frac{1}{2}} \,\la_{+2,+\frac{1}{2},-\frac{1}{2}} \,
{\mathscr D}_m^{h_1+2}(\tfrac{1}{2}-h_3) \,
{\mathscr D}_n^{h_2+2}(\tfrac{1}{2}-h_1-h_3)
\nonu \\
&& -
\la_{+2,+\frac{1}{2},-\frac{1}{2}} \, \kappa_{+2,+2,-2} \,
\bigg( (h_2+1)m-(h_1+1)n \bigg) \,
{\mathscr D}_{m+n}^{h_1+h_2+2}(\tfrac{1}{2}-h_3)\, \Bigg]
\nonu \\
&& \times
{\cal O}_{+\frac{1}{2}}^{(h_1+h_2+h_3),ABC}(z,\bar{z})=0,
\nonu \\
&& \Bigg[
 \la_{+2,0,0}\, \la_{+2,0,0} \,
{\mathscr D}_n^{h_2+2}(1-h_3) \,
{\mathscr D}_m^{h_1+2}(1-h_2-h_3)
\nonu \\
&& - \la_{+2,0,0}\, \la_{+2,0,0} \,
{\mathscr D}_m^{h_1+2}(1-h_3) \,
{\mathscr D}_n^{h_2+2}(1-h_1-h_3)
\nonu \\
&& -
 \la_{+2,0,0} \, \kappa_{+2,+2,-2} \,\bigg( (h_2+1)m-(h_1+1)n \bigg) \,
    {\mathscr D}_{m+n}^{h_1+h_2+2}(1-h_3)\, \Bigg]\nonu \\
&& \times
{\cal O}_{0}^{(h_1+h_2+h_3),ABCD}(z,\bar{z})=0,
\nonu \\
&& \Bigg[
\la_{+2,-\frac{1}{2},+\frac{1}{2}} \,\la_{+2,-\frac{1}{2},+\frac{1}{2}} \,
{\mathscr D}_n^{h_2+2}(\tfrac{3}{2}-h_3) \,
{\mathscr D}_m^{h_1+2}(\tfrac{3}{2}-h_2-h_3)
\nonu \\
&& -\la_{+2,-\frac{1}{2},+\frac{1}{2}} \,\la_{+2,-\frac{1}{2},+\frac{1}{2}} \,
{\mathscr D}_m^{h_1+2}(\tfrac{3}{2}-h_3) \,
{\mathscr D}_n^{h_2+2}(\tfrac{3}{2}-h_1-h_3)
\nonu \\
&& -
\la_{+2,-\frac{1}{2},+\frac{1}{2}} \, \kappa_{+2,+2,-2} \, \bigg( (h_2+1)m-(h_1+1)n \bigg) \,
{\mathscr D}_{m+n}^{h_1+h_2+2}(\tfrac{3}{2}-h_3)\, \Bigg]\,
{\cal O}_{ABC,-\frac{1}{2}}^{(h_1+h_2+h_3)}(z,\bar{z})=0,
\nonu \\
&& \Bigg[
\la_{+2,-1,+1}\,\la_{+2,-1,+1} \,
{\mathscr D}_n^{h_2+2}(2-h_3) \,
{\mathscr D}_m^{h_1+2}(2-h_2-h_3)
\nonu \\
&& -\la_{+2,-1,+1}\,\la_{+2,-1,+1} \,
{\mathscr D}_m^{h_1+2}(2-h_3) \,
{\mathscr D}_n^{h_2+2}(2-h_1-h_3)
\nonu \\
&& -
 \la_{+2,-1,+1} \, \kappa_{+2,+2,-2} \,\bigg( (h_2+1)m-(h_1+1)n \bigg) \,
{\mathscr D}_{m+n}^{h_1+h_2+2}(2-h_3)\, \Bigg]\,
{\cal O}_{AB,-1}^{(h_1+h_2+h_3)}(z,\bar{z})=0,
\nonu \\
&& \Bigg[
\la_{+2,-\frac{3}{2},+\frac{3}{2}} \,\la_{+2,-\frac{3}{2},+\frac{3}{2}} \,
{\mathscr D}_n^{h_2+2}(\tfrac{5}{2}-h_3) \,
{\mathscr D}_m^{h_1+2}(\tfrac{5}{2}-h_2-h_3)
\nonu \\
&& -\la_{+2,-\frac{3}{2},+\frac{3}{2}} \,\la_{+2,-\frac{3}{2},+\frac{3}{2}} \,
{\mathscr D}_m^{h_1+2}(\tfrac{5}{2}-h_3) \,
{\mathscr D}_n^{h_2+2}(\tfrac{5}{2}-h_1-h_3)
\nonu \\
&& -
\la_{+2,-\frac{3}{2},+\frac{3}{2}} \,\kappa_{+2,+2,-2} \,
\bigg( (h_2+1)m-(h_1+1)n \bigg) \,
{\mathscr D}_{m+n}^{h_1+h_2+2}(\tfrac{5}{2}-h_3)\, \Bigg]\,
{\cal O}_{A,-\frac{3}{2}}^{(h_1+h_2+h_3)}(z,\bar{z})=0,
\nonu \\
&& \Bigg[
 \la_{+2,-2,+2} \, \la_{+2,-2,+2} \,
{\mathscr D}_n^{h_2+2}(3-h_3) \,
{\mathscr D}_m^{h_1+2}(3-h_2-h_3)
\nonu \\
&& - \la_{+2,-2,+2} \, \la_{+2,-2,+2} \,
{\mathscr D}_m^{h_1+2}(3-h_3) \,
{\mathscr D}_n^{h_2+2}(3-h_1-h_3)
\nonu \\
&& -
 \la_{+2,-2,+2} \, \kappa_{+2,+2,-2} \,\bigg( (h_2+1)m-(h_1+1)n \bigg) \,
{\mathscr D}_{m+n}^{h_1+h_2+2}(3-h_3)\, \Bigg]\,
{\cal O}_{-2}^{(h_1+h_2+h_3)}(z,\bar{z})=0,
\nonu \\
&& \Bigg[
\la_{+2,+\frac{3}{2},-\frac{3}{2}} \,\la_{+\frac{3}{2},+2,-\frac{3}{2}} \,
{\mathscr D}_n^{h_2+\frac{3}{2}}(-1-h_3) \,
{\mathscr D}_m^{h_1+2}(-\tfrac{1}{2}-h_2-h_3)
\nonu \\
&& -\la_{+\frac{3}{2},+2,-\frac{3}{2}} \,\la_{+2,+2,-2}\,
{\mathscr D}_m^{h_1+2}(-1-h_3) \,
{\mathscr D}_n^{h_2+\frac{3}{2}}(-1-h_1-h_3)
\nonu \\
&& -
\la_{+\frac{3}{2},+2,-\frac{3}{2}} \,
\kappa_{+2,+\frac{3}{2},-\frac{3}{2}} \,
\bigg(
(h_2+\tfrac{1}{2})m-(h_1+1)n \bigg) \,
{\mathscr D}_{m+n}^{h_1+h_2+\frac{3}{2}}(-1-h_3)\, \Bigg]
\nonu \\
&& \times
{\cal O}_{+\frac{3}{2}}^{(h_1+h_2+h_3),A}(z,\bar{z})=0,
\nonu \\
&& \Bigg[
 \la_{+2,+1,-1}\,\la_{+\frac{3}{2},+\frac{3}{2},-1} \,
{\mathscr D}_n^{h_2+\frac{3}{2}}(-\tfrac{1}{2}-h_3) \,
{\mathscr D}_m^{h_1+2}(-h_2-h_3)
\nonu \\
&& - \la_{+\frac{3}{2},+\frac{3}{2},-1} \,\la_{+2,+\frac{3}{2},-\frac{3}{2}}\,
{\mathscr D}_m^{h_1+2}(-\tfrac{1}{2}-h_3) \,
{\mathscr D}_n^{h_2+\frac{3}{2}}(-\tfrac{1}{2}-h_1-h_3)
\nonu \\
&& -
\la_{+\frac{3}{2},+\frac{3}{2},-1} \,\kappa_{+2,+\frac{3}{2},-\frac{3}{2}} \,
\bigg(
(h_2+\tfrac{1}{2})m-(h_1+1)n \bigg) \,
{\mathscr D}_{m+n}^{h_1+h_2+\frac{3}{2}}(-\tfrac{1}{2}-h_3)\, \Bigg]
\nonu \\
&& \times
{\cal O}_{+1}^{(h_1+h_2+h_3),AB}(z,\bar{z})=0,
\nonu \\
&& \Bigg[
 \la_{+2,+\frac{1}{2},-\frac{1}{2}} \,\la_{+\frac{3}{2},+1,-\frac{1}{2}} \,
{\mathscr D}_n^{h_2+\frac{3}{2}}(-h_3) \,
{\mathscr D}_m^{h_1+2}(\tfrac{1}{2}-h_2-h_3)
\nonu \\
&& - \la_{+\frac{3}{2},+1,-\frac{1}{2}} \,\la_{+2,+1,-1}\,
{\mathscr D}_m^{h_1+2}(-h_3) \,
{\mathscr D}_n^{h_2+\frac{3}{2}}(-h_1-h_3)
\nonu \\
&& -
\la_{+\frac{3}{2},+1,-\frac{1}{2}} \,\kappa_{+2,+\frac{3}{2},-\frac{3}{2}} \,
\bigg(
(h_2+\tfrac{1}{2})m-(h_1+1)n \bigg) \,
{\mathscr D}_{m+n}^{h_1+h_2+\frac{3}{2}}(-h_3)
\, \Bigg]
\nonu \\
&& \times
{\cal O}_{+\frac{1}{2}}^{(h_1+h_2+h_3),ABC}(z,\bar{z})=0,
\nonu \\
&& \Bigg[
 \la_{+2,0,0}\,\la_{+\frac{3}{2},+\frac{1}{2},0} \,
{\mathscr D}_n^{h_2+\frac{3}{2}}(\tfrac{1}{2}-h_3) \,
{\mathscr D}_m^{h_1+2}(1-h_2-h_3)
\nonu \\
&& -\la_{+2,+\frac{1}{2},-\frac{1}{2}}\,\la_{+\fft32,+\fft12,0}\,
{\mathscr D}_m^{h_1+2}(\tfrac{1}{2}-h_3) \,
{\mathscr D}_n^{h_2+\frac{3}{2}}(\tfrac{1}{2}-h_1-h_3)
\nonu \\
&& -
\la_{+\frac{3}{2},+\frac{1}{2},0} \,
\kappa_{+2,+\frac{3}{2},-\frac{3}{2}} \,
\bigg(
(h_2+\tfrac{1}{2})m-(h_1+1)n \bigg) \,
{\mathscr D}_{m+n}^{h_1+h_2+\frac{3}{2}}(\tfrac{1}{2}-h_3)\, \Bigg]
\nonu \\
&\times & 
{\cal O}_{0}^{(h_1+h_2+h_3),ABCD}(z,\bar{z})=0,
\nonu \\
&& \Bigg[
 \la_{+2,-\frac{1}{2},+\frac{1}{2}}\,\la_{+\frac{3}{2},0,+\frac{1}{2}}\,
{\mathscr D}_n^{h_2+\frac{3}{2}}(1-h_3) \,
{\mathscr D}_m^{h_1+2}(\tfrac{3}{2}-h_2-h_3)
\nonu \\
&& -\la_{+\frac{3}{2},0,+\frac{1}{2}}\,\la_{+2,0,0}\, 
{\mathscr D}_m^{h_1+2}(1-h_3) \,
{\mathscr D}_n^{h_2+\frac{3}{2}}(1-h_1-h_3)
\nonu \\
&& -
\la_{+\frac{3}{2},0,+\frac{1}{2}} \,
\kappa_{+2,+\frac{3}{2},-\frac{3}{2}} \,
\bigg(
(h_2+\tfrac{1}{2})m-(h_1+1)n \bigg) \,
{\mathscr D}_{m+n}^{h_1+h_2+\frac{3}{2}}(1-h_3)\, \Bigg]\nonu \\
&& \times 
\frac{1}{3!}\, \ep^{ABCDEFGH} \,
{\cal O}_{FGH,-\frac{1}{2}}^{(h_1+h_2+h_3)}(z,\bar{z})=0,
\nonu \\
&& \Bigg[
\la_{+2,-1,+1} \,\la_{+\frac{3}{2},-\frac{1}{2},+1} \, 
{\mathscr D}_n^{h_2+\frac{3}{2}}(\tfrac{3}{2}-h_3) \,
{\mathscr D}_m^{h_1+2}(2-h_2-h_3)
\nonu \\
&& - \la_{+\frac{3}{2},-\frac{1}{2},+1} \,\la_{+2,-\frac{1}{2},+\frac{1}{2}}\,
{\mathscr D}_m^{h_1+2}(\tfrac{3}{2}-h_3) \,
{\mathscr D}_n^{h_2+\frac{3}{2}}(\tfrac{3}{2}-h_1-h_3)
\nonu \\
&& -
\la_{+\frac{3}{2},-\frac{1}{2},+1} \, \kappa_{+2,+\frac{3}{2},-\frac{3}{2}} \,
\bigg(
(h_2+\tfrac{1}{2})m-(h_1+1)n \bigg) \,
{\mathscr D}_{m+n}^{h_1+h_2+\frac{3}{2}}(\tfrac{3}{2}-h_3)\, \Bigg]\nonu \\
&& \times 
3 \, \de^{A}_{[B} \,
{\cal O}_{CD],-1}^{(h_1+h_2+h_3)}(z,\bar{z})=0,
\nonu \\
&& \Bigg[
 \la_{+2,-\frac{3}{2},+\frac{3}{2}}\,\la_{+\frac{3}{2},-1,+\frac{3}{2}} \,
{\mathscr D}_n^{h_2+\frac{3}{2}}(2-h_3) \,
{\mathscr D}_m^{h_1+2}(\tfrac{5}{2}-h_2-h_3)
\nonu \\
&& -\la_{+\frac{3}{2},-1,+\frac{3}{2}} \,\la_{+2,-1,+1}\, 
{\mathscr D}_m^{h_1+2}(2-h_3) \,
{\mathscr D}_n^{h_2+\frac{3}{2}}(2-h_1-h_3)
\nonu \\
&& -
\la_{+\frac{3}{2},-1,+\frac{3}{2}} \,
\kappa_{+2,+\frac{3}{2},-\frac{3}{2}} \,
\bigg(
(h_2+\tfrac{1}{2})m-(h_1+1)n \bigg) \,
{\mathscr D}_{m+n}^{h_1+h_2+\frac{3}{2}}(2-h_3)\, \Bigg]\nonu \\
&& \times 
2! \, \de^{A}_{[B} \,
{\cal O}_{C],-\frac{3}{2}}^{(h_1+h_2+h_3)}(z,\bar{z})=0,
\nonu \\
&& \Bigg[
 \la_{+2,-2,+2}\,\la_{+\frac{3}{2},-\fft32,+2} \,
{\mathscr D}_n^{h_2+\frac{3}{2}}(\tfrac{5}{2}-h_3) \,
{\mathscr D}_m^{h_1+2}(3-h_2-h_3)
\nonu \\
&& - \la_{+\fft32,-\fft32,+2}\,\la_{+2,-\fft32,+\fft32}\,
{\mathscr D}_m^{h_1+2}(\tfrac{5}{2}-h_3) \,
{\mathscr D}_n^{h_2+\frac{3}{2}}(\tfrac{5}{2}-h_1-h_3)
\nonu \\
&& -
\la_{+\fft32,-\fft32,+2} \,\kappa_{+2,+\fft32,-\fft32} \,  \bigg(
(h_2+\tfrac{1}{2})m-(h_1+1)n \bigg) \,
{\mathscr D}_{m+n}^{h_1+h_2+\frac{3}{2}}(\tfrac{5}{2}-h_3)\, \Bigg]\nonu \\
&& \times 
\de^{A}_{B} \,
{\cal O}_{-2}^{(h_1+h_2+h_3)}(z,\bar{z})=0.
\label{17}
\eea

\subsubsection{The eighteen relations }

The eighteen relations for (\ref{Iden}) corresponding to
Table \ref{table7} 
are given by
\bea
&& \Bigg[
\la_{+2,+1,-1} \,\la_{+1,+2,-1} \, 
{\mathscr D}_n^{h_2+1}(-1-h_3) \,
{\mathscr D}_m^{h_1+2}(-h_2-h_3)
\nonu \\
&& -\la_{+1,+2,-1}\,\la_{+2,+2,-2}\,
{\mathscr D}_m^{h_1+2}(-1-h_3) \,
{\mathscr D}_n^{h_2+1}(-1-h_1-h_3)
\nonu \\
&& -
 \la_{+1,+2,-1} \, \kappa_{+2,+1,-1} \,\bigg(
(h_2)m-(h_1+1)n \bigg) \,
{\mathscr D}_{m+n}^{h_1+h_2+1}(-1-h_3)\, \Bigg]\,
{\cal O}_{+1}^{(h_1+h_2+h_3),AB}(z,\bar{z})=0,
\nonu \\
&& \Bigg[
 \la_{+2,+\fft12,-\fft12}\,\la_{+1,+\fft32,-\fft12} \,
{\mathscr D}_n^{h_2+1}(-\tfrac{1}{2}-h_3) \,
{\mathscr D}_m^{h_1+2}(\tfrac{1}{2}-h_2-h_3)
\nonu \\
&& - \la_{+1,+\fft32,-\fft12}\, \la_{+2,+\fft32,-\fft32}\,
{\mathscr D}_m^{h_1+2}(-\tfrac{1}{2}-h_3) \,
{\mathscr D}_n^{h_2+1}(-\tfrac{1}{2}-h_1-h_3)
\nonu \\
&& -
 \la_{+1,+\fft32,-\fft12} \,\kappa_{+2,+1,-1} \, \bigg(
(h_2)m-(h_1+1)n \bigg) \,
{\mathscr D}_{m+n}^{h_1+h_2+1}(-\tfrac{1}{2}-h_3)\, \Bigg]\,
{\cal O}_{+\frac{1}{2}}^{(h_1+h_2+h_3),ABC}(z,\bar{z})=0,
\nonu \\
&& \Bigg[
 \la_{+2,0,0}\,\la_{+1,+1,0} \,
{\mathscr D}_n^{h_2+1}(-h_3) \,
{\mathscr D}_m^{h_1+2}(1-h_2-h_3)
\nonu \\
&& - \la_{+1,+1,0}\,\la_{+2,+1,-1}\,
{\mathscr D}_m^{h_1+2}(-h_3) \,
{\mathscr D}_n^{h_2+1}(-h_1-h_3)
\nonu \\
&& -
 \la_{+1,+1,0} \, \kappa_{+2,+1,-1} \,\bigg(
(h_2)m-(h_1+1)n \bigg) \,
{\mathscr D}_{m+n}^{h_1+h_2+1}(-h_3)\, \Bigg]\nonu \\
&& \times 
{\cal O}_{0}^{(h_1+h_2+h_3),ABCD}(z,\bar{z})=0,
\nonu \\
&& \Bigg[
 \la_{+2,-\fft12,+\fft12}\,\la_{+1,+\fft12,+\fft12} \,
{\mathscr D}_n^{h_2+1}(\tf12-h_3) \,
{\mathscr D}_m^{h_1+2}(\tf32-h_2-h_3)
\nonu \\
&& - \la_{+1,+\fft12,+\fft12}\, \la_{+2,+\fft12,-\fft12}\,
{\mathscr D}_m^{h_1+2}(\tf12-h_3) \,
{\mathscr D}_n^{h_2+1}(\tf12-h_1-h_3)
\nonu \\
&& -
 \la_{+1,+\fft12,+\fft12} \, \kappa_{+2,+1,-1} \,\bigg(
(h_2)m-(h_1+1)n \bigg) \,
{\mathscr D}_{m+n}^{h_1+h_2+1}(\tf12-h_3)\, \Bigg]\nonu \\
&& \times
\frac{1}{3!} \, \ep^{ABCDEFGH}\,
{\cal O}_{FGH,-\fft12}^{(h_1+h_2+h_3)}(z,\bar{z})=0,
\nonu \\
&& \Bigg[
 \la_{+2,-1,+1}\,\la_{+1,0,+1} \,
{\mathscr D}_n^{h_2+1}(1-h_3) \,
{\mathscr D}_m^{h_1+2}(2-h_2-h_3)
\nonu \\'
&& - \la_{+1,0,+1}\, \la_{+2,0,0}\,
{\mathscr D}_m^{h_1+2}(1-h_3) \,
{\mathscr D}_n^{h_2+1}(1-h_1-h_3)
\nonu \\
&& -
 \la_{+1,0,+1} \,\kappa_{+2,+1,-1} \, \bigg(
(h_2)m-(h_1+1)n \bigg) \,
{\mathscr D}_{m+n}^{h_1+h_2+1}(1-h_3)\, \Bigg]\nonu \\
&& \times
\frac{1}{2!} \, \ep^{ABCDEFGH}\,
{\cal O}_{GH,-1}^{(h_1+h_2+h_3)}(z,\bar{z})=0,
\nonu \\
&& \Bigg[
 \la_{+2,-\fft32,+\fft32}\, \la_{+1,-\fft12,+\fft32} \,
{\mathscr D}_n^{h_2+1}(\tf32-h_3) \,
{\mathscr D}_m^{h_1+2}(\tf52-h_2-h_3)
\nonu \\
&& - \la_{+1,-\fft12,+\fft32} \, \la_{+2,-\fft12,+\fft12}\,
{\mathscr D}_m^{h_1+2}(\tf32-h_3) \,
{\mathscr D}_n^{h_2+1}(\tf32-h_1-h_3)
\nonu \\
&& -
 \la_{+1,-\fft12,+\fft32} \, \kappa_{+2,+1,-1} \,\bigg(
(h_2)m-(h_1+1)n \bigg) \,
{\mathscr D}_{m+n}^{h_1+h_2+1}(\tf32-h_3)\, \Bigg]\nonu \\
&& \times
3! \, \de^{A}_{[C}\,
{\cal O}_{D,-\frac{3}{2}}^{(h_1+h_2+h_3)}(z,\bar{z})\, \de^{B}_{E]}=0,
\nonu \\
&& \Bigg[
 \la_{+2,-2,+2}\,\la_{+1,-1,+2} \,
{\mathscr D}_n^{h_2+1}(2-h_3) \,
{\mathscr D}_m^{h_1+2}(3-h_2-h_3)
\nonu \\
&& - \la_{+1,-1,+2}\, \la_{+2,-1,+1}\,
{\mathscr D}_m^{h_1+2}(2-h_3) \,
{\mathscr D}_n^{h_2+1}(2-h_1-h_3)
\nonu \\
&& -
 \la_{+1,-1,+2} \, \kappa_{+2,+1,-1} \,\bigg(
(h_2)m-(h_1+1)n \bigg) \,
{\mathscr D}_{m+n}^{h_1+h_2+1}(2-h_3)\, \Bigg]\,
\de^{AB}_{CD}\,
{\cal O}_{-2}^{(h_1+h_2+h_3)}(z,\bar{z})=0,
\nonu \\
&& \Bigg[
 \la_{+2,+\fft12,-\fft12}\,\la_{+\fft12,+2,-\fft12} \,
{\mathscr D}_n^{h_2+\fft12}(-1-h_3) \,
{\mathscr D}_m^{h_1+2}(\tf12-h_2-h_3)
\nonu \\
&& - \la_{+\fft12,+2,-\fft12}\, \la_{+2,+2,-2}\,
{\mathscr D}_m^{h_1+2}(-1-h_3) \,
{\mathscr D}_n^{h_2+\fft12}(-1-h_1-h_3)
\nonu \\
&& -
 \la_{+\fft12,+2,-\fft12} \, \kappa_{+2,+\fft12,-\fft12} \, \bigg(
(h_2-\tf12)m-(h_1+1)n \bigg) \,
{\mathscr D}_{m+n}^{h_1+h_2+\fft12}(-1-h_3)\, \Bigg]\nonu \\
&& \times
{\cal O}_{+\fft12}^{(h_1+h_2+h_3),ABC}(z,\bar{z})=0,
\nonu \\
&& \Bigg[
 \la_{+2,+0,0}\,\la_{+\fft12,+\fft32,0} \,
{\mathscr D}_n^{h_2+\fft12}(-\tf12-h_3) \,
{\mathscr D}_m^{h_1+2}(1-h_2-h_3)
\nonu \\
&& - \la_{+\fft12,+\fft32,0}\, \la_{+2,+\fft32,-\fft32}\,
{\mathscr D}_m^{h_1+2}(-\tf12-h_3) \,
{\mathscr D}_n^{h_2+\fft12}(-\tf12-h_1-h_3)
\nonu \\
&& -
 \la_{+\fft12,+\fft32,0} \, \kappa_{+2,+\fft12,-\fft12} \,\bigg(
(h_2-\tf12)m-(h_1+1)n \bigg) \,
{\mathscr D}_{m+n}^{h_1+h_2+\fft12}(-\tf12-h_3)\, \Bigg]\nonu \\
&& \times
{\cal O}_{0}^{(h_1+h_2+h_3),ABCD}(z,\bar{z})=0,
\nonu \\
&& \Bigg[
\la_{+2,-\fft12,+\fft12}\,\la_{+\fft12,+1,+\fft12} \,
{\mathscr D}_n^{h_2+\fft12}(-h_3) \,
{\mathscr D}_m^{h_1+2}(\tf32-h_2-h_3)
\nonu \\
&& - \la_{+\fft12,+1,+\fft12}\, \la_{+2,+1,-1}\,
{\mathscr D}_m^{h_1+2}(-h_3) \,
{\mathscr D}_n^{h_2+\fft12}(-h_1-h_3)
\nonu \\
&& -
 \la_{+\fft12,+1,\fft12} \, \kappa_{+2,+\fft12,-\fft12} \,\bigg(
(h_2-\tf12)m-(h_1+1)n \bigg) \,
{\mathscr D}_{m+n}^{h_1+h_2+\fft12}(-h_3)\, \Bigg]\nonu \\
&& \times
\frac{1}{3!}\, \ep^{ABCDEFGH}\,
{\cal O}_{FGH,-\fft12}^{(h_1+h_2+h_3)}(z,\bar{z})=0,
\nonu \\
&& \Bigg[
 \la_{+2,+1,-1}\,\la_{+\fft12,+\fft12,+1} \,
{\mathscr D}_n^{h_2+\fft12}(\tf12-h_3) \,
{\mathscr D}_m^{h_1+2}(2-h_2-h_3)
\nonu \\
&& - \la_{+\fft12,+\fft12,+1}\, \la_{+2,+\fft12,-\fft12}\,
{\mathscr D}_m^{h_1+2}(\tf12-h_3) \,
{\mathscr D}_n^{h_2+\fft12}(\tf12-h_1-h_3)
\nonu \\
&& -
\la_{+\fft12,+\fft12,+1} \, \kappa_{+2,+\fft12,-\fft12} \, \bigg(
(h_2-\tf12)m-(h_1+1)n \bigg) \,
{\mathscr D}_{m+n}^{h_1+h_2+\fft12}(\tf12-h_3)\, \Bigg]\nonu \\
&& \times
\frac{1}{2!}\, \ep^{ABCDEFGH}\,
{\cal O}_{GH,-1}^{(h_1+h_2+h_3)}(z,\bar{z})=0,
\nonu \\
&& \Bigg[
 \la_{+2,-\fft32,+\fft32}\,\la_{+\fft12,0,+\fft32} \,
{\mathscr D}_n^{h_2+\fft12}(1-h_3) \,
{\mathscr D}_m^{h_1+2}(\tf52-h_2-h_3)
\nonu \\
&& - \la_{+\fft12,0,+\fft32}\, \la_{+2,0,0}\,
{\mathscr D}_m^{h_1+2}(1-h_3) \,
{\mathscr D}_n^{h_2+\fft12}(1-h_1-h_3)
\nonu \\
&& -
 \la_{+\fft12,0,+\fft32} \, \kappa_{+2,+\fft12,-\fft12} \,\bigg(
(h_2-\tf12)m-(h_1+1)n \bigg) \,
{\mathscr D}_{m+n}^{h_1+h_2+\fft12}(1-h_3)\, \Bigg]\nonu \\
&& \times
\ep^{ABCDEFGH}\,
{\cal O}_{H,-\frac{3}{2}}^{(h_1+h_2+h_3)}(z,\bar{z})=0,
\nonu \\
&& \Bigg[
 \la_{+2,-2,+2}\,\la_{+\fft12,-\fft12,+2} \,
{\mathscr D}_n^{h_2+\fft12}(\tf32-h_3) \,
{\mathscr D}_m^{h_1+2}(3-h_2-h_3)
\nonu \\
&& - \la_{+\fft12,-\fft12,+2}\, \la_{+2,-\fft12,+\fft12}\,
{\mathscr D}_m^{h_1+2}(\tf32-h_3) \,
{\mathscr D}_n^{h_2+\fft12}(\ft32-h_1-h_3)
\nonu \\
&& -
 \la_{+\fft12,-\fft12,+2} \, \kappa_{+2,+\fft12,-\fft12} \,\bigg(
(h_2-\tf12)m-(h_1+1)n \bigg) \,
{\mathscr D}_{m+n}^{h_1+h_2+\fft12}(\tf32-h_3)\, \Bigg]\nonu \\
&&\times \de^{ABC}_{DEF}\,
{\cal O}_{-2}^{(h_1+h_2+h_3)}(z,\bar{z})=0,
\nonu \\
&& \Bigg[
 \la_{+2,0,0}\,\la_{0,+2,0} \,
{\mathscr D}_n^{h_2}(-1-h_3) \,
{\mathscr D}_m^{h_1+2}(1-h_2-h_3)
\nonu \\
&& - \la_{0,+2,0}\, \la_{+2,+2,-2}\,
{\mathscr D}_m^{h_1+2}(-1-h_3) \,
{\mathscr D}_n^{h_2}(-1-h_1-h_3)
\nonu \\
&& -
 \la_{0,+2,0} \, \kappa_{+2,0,0} \,\bigg(
(h_2-1)m-(h_1+1)n \bigg) \,
{\mathscr D}_{m+n}^{h_1+h_2}(-1-h_3)\, \Bigg]\,
{\cal O}_{0}^{(h_1+h_2+h_3),ABCD}(z,\bar{z})=0,
\nonu \\
&& \Bigg[
 \la_{+2,-\fft12,+\fft12}\,\la_{0,+\fft32,+\fft12} \,
{\mathscr D}_n^{h_2}(-\tf12-h_3) \,
{\mathscr D}_m^{h_1+2}(\tf32-h_2-h_3)
\nonu \\
&& - \la_{0,+\fft32,+\fft12}\, \la_{+2,+\fft32,-\fft32}\,
{\mathscr D}_m^{h_1+2}(-\tf12-h_3) \,
{\mathscr D}_n^{h_2}(-\tf12-h_1-h_3)
\nonu \\
&& -
 \la_{0,+\fft32,+\fft12} \, \kappa_{+2,0,0} \,\bigg(
(h_2-1)m-(h_1+1)n \bigg) \,
{\mathscr D}_{m+n}^{h_1+h_2}(-\tf12-h_3)\, \Bigg]\nonu \\
&& \times
\frac{1}{3!} \, \ep^{ABCDEFGH}\,
{\cal O}_{FGH,-\fft12}^{(h_1+h_2+h_3)}(z,\bar{z})=0,
\nonu \\
&& \Bigg[
 \la_{+2,-1,+1}\,\la_{0,+1,+1} \,
{\mathscr D}_n^{h_2}(-h_3) \,
{\mathscr D}_m^{h_1+2}(2-h_2-h_3)
\nonu \\
&& - \la_{0,+1,+1}\, \la_{+2,+1,-1}\,
{\mathscr D}_m^{h_1+2}(-h_3)
{\mathscr D}_n^{h_2}(-h_1-h_3)
\nonu \\
&& -
 \la_{0,+1,+1} \, \kappa_{+2,0,0} \,\bigg(
(h_2-1)m-(h_1+1)n \bigg) \,
{\mathscr D}_{m+n}^{h_1+h_2}(-h_3)\, \Bigg]\nonu \\
&& \times
\frac{1}{2!} \, \ep^{ABCDEFGH}\,
{\cal O}_{GH,-1}^{(h_1+h_2+h_3)}(z,\bar{z})=0,
\nonu \\
&& \Bigg[
 \la_{+2,-\fft32,+\fft32}\,\la_{0,+\fft12,+\fft32} \,
{\mathscr D}_n^{h_2}(\tf12-h_3) \,
{\mathscr D}_m^{h_1+2}(\tf52-h_2-]h_3)
\nonu \\
&& - \la_{0,+\fft12,+\fft32}\, \la_{+2,+\fft12,-\fft12}\,
{\mathscr D}_m^{h_1+2}(\tf12-h_3)
{\mathscr D}_n^{h_2}(\tf12-h_1-h_3)
\nonu \\
&& -
 \la_{0,+\fft12,+\fft32} \, \kappa_{+2,0,0} \,\bigg(
(h_2-1)m-(h_1+1)n \bigg) \,
{\mathscr D}_{m+n}^{h_1+h_2}(\tf12-h_3)\, \Bigg]\nonu \\
&& \times
\ep^{ABCDEFGH}\,
{\cal O}_{H,-\frac{3}{2}}^{(h_1+h_2+h_3)}(z,\bar{z})=0,
\nonu \\
&& \Bigg[
 \la_{+2,-2,+2}\,\la_{0,0,+2} \,
{\mathscr D}_n^{h_2}(1-h_3) \,
{\mathscr D}_m^{h_1+2}(3-h_2-]h_3)
\nonu \\
&& - \la_{0,0,+2}\, \la_{+2,0,0}\,
{\mathscr D}_m^{h_1+2}(1-h_3)
{\mathscr D}_n^{h_2}(1-h_1-h_3)
\nonu \\
&& -
 \la_{0,0,+2} \, \kappa_{+2,0,0} \,\bigg(
(h_2-1)m-(h_1+1)n \bigg) \,
{\mathscr D}_{m+n}^{h_1+h_2}(1-h_3)\, \Bigg]\nonu \\
&& \times
\ep^{ABCDEFGH}\,
{\cal O}_{-2}^{(h_1+h_2+h_3)}(z,\bar{z})=0.
\label{18}
\eea

\subsubsection{The ten relations }

Finally, the ten relations for (\ref{Iden}) corresponding to
Table \ref{table8} 
are given by
\bea
&& \Bigg[
 \la_{+2,-\fft12,+\fft12}\,\la_{-\fft12,+2+\fft12} \,
{\mathscr D}_n^{h_2-\fft12}(-1-h_3) \,
{\mathscr D}_m^{h_1+2}(\tf32-h_2-h_3)
\nonu \\
&& - \la_{-\fft12,+2,+\fft12}\, \la_{+2,+2,-2}\,
{\mathscr D}_m^{h_1+2}(-1-h_3)
{\mathscr D}_n^{h_2-\fft12}(-1-h_1-h_3)
\nonu \\
&& -
 \la_{-\fft12,+2,+\fft12} \, \kappa_{+2,-\fft12,+\fft12} \,\bigg(
(h_2-\tf32)m-(h_1+1)n \bigg) \,
{\mathscr D}_{m+n}^{h_1+h_2-\fft12}(-1-h_3)\, \Bigg]\,
{\cal O}_{ABC,-\fft12}^{(h_1+h_2+h_3)}(z,\bar{z})=0,
\nonu \\
&& \Bigg[
 \la_{+2,-1,+1}\,\la_{-\fft12,+\fft32,+1} \,
{\mathscr D}_n^{h_2-\fft12}(-\tf12-h_3) \,
{\mathscr D}_m^{h_1+2}(2-h_2-h_3)
\nonu \\
&& - \la_{-\fft12,+\fft32,+1}\, \la_{+2,+\fft32,-\fft32}\,
{\mathscr D}_m^{h_1+2}(-\tf12-h_3)
{\mathscr D}_n^{h_2-\fft12}(-\tf12-h_1-h_3)
\nonu \\
&& -
 \la_{-\fft12,+\fft32,+1} \, \kappa_{+2,-\fft12,+\fft12} \,\bigg(
(h_2-\tf32)m-(h_1+1)n \bigg) \,
{\mathscr D}_{m+n}^{h_1+h_2-\fft12}(-\tf12-h_3)\, \Bigg]\nonu \\
&& \times
3 \, \de^{D}_{[A} \,
{\cal O}_{BC],-1}^{(h_1+h_2+h_3)}(z,\bar{z})=0,
\nonu \\
&& \Bigg[
 \la_{+2,-\fft32,+\fft32}\,\la_{-\fft12,+1,+\fft32} \,
{\mathscr D}_n^{h_2-\fft12}(-h_3) \,
{\mathscr D}_m^{h_1+2}(\tf52-h_2-h_3)
\nonu \\
&& - \la_{-\fft12,+1,+\fft32}\, \la_{+2,+1,-1}\,
{\mathscr D}_m^{h_1+2}(-h_3)
{\mathscr D}_n^{h_2-\fft12}(-h_1-h_3)
\nonu \\
&& -
 \la_{-\fft12,+1,+\fft32} \, \kappa_{+2,-\fft12,+\fft12} \,\bigg(
(h_2-\tf32)m-(h_1+1)n \bigg) \,
{\mathscr D}_{m+n}^{h_1+h_2-\fft12}(-h_3)\, \Bigg]\nonu \\
&& \times
3! \, \de^{D}_{[A} \,
{\cal O}_{B,-\fft32}^{(h_1+h_2+h_3)}(z,\bar{z}) \, \de^{E}_{C]}=0,
\nonu \\
&& \Bigg[
 \la_{+2,-2,+2}\,\la_{-\fft12,+\fft12,+2} \,
{\mathscr D}_n^{h_2-\fft12}(\tf12-h_3) \,
{\mathscr D}_m^{h_1+2}(3-h_2-h_3)
\nonu \\
&& - \la_{-\fft12,+\fft12,+2}\, \la_{+2,+\fft12,-\fft12}\,
{\mathscr D}_m^{h_1+2}(\tf12-h_3)
{\mathscr D}_n^{h_2-\fft12}(\tf12-h_1-h_3)
\nonu \\
&& -
 \la_{-\fft12,+\fft12,+2} \, \kappa_{+2,-\fft12,+\fft12} \,\bigg(
(h_2-\tf32)m-(h_1+1)n \bigg) \,
{\mathscr D}_{m+n}^{h_1+h_2-\fft12}(\tf12-h_3)\, \Bigg]\nonu \\
&& \times
\de^{DEF}_{ABC} \,
{\cal O}_{-2}^{(h_1+h_2+h_3)}(z,\bar{z})=0,
\nonu \\
&& \Bigg[
 \la_{+2,-1,+1}\,\la_{-1,+2,+1} \,
{\mathscr D}_n^{h_2-1}(-1-h_3) \,
{\mathscr D}_m^{h_1+2}(2-h_2-h_3)
\nonu \\
&& - \la_{-1,+2,+1}\, \la_{+2,+2,-2}\,
{\mathscr D}_m^{h_1+2}(-1-h_3)
{\mathscr D}_n^{h_2-1}(-1-h_1-h_3)
\nonu \\
&& -
\la_{-1,+2,+1} \, \kappa_{+2,-1,+1} \, \bigg(
(h_2-2)m-(h_1+1)n \bigg) \,
{\mathscr D}_{m+n}^{h_1+h_2-1}(-1-h_3)\, \Bigg]\,
{\cal O}_{AB,-1}^{(h_1+h_2+h_3)}(z,\bar{z})=0,
\nonu \\
&& \Bigg[
 \la_{+2,-\fft32,+\fft32}\,\la_{-1,+\fft32,+\fft32} \,
{\mathscr D}_n^{h_2-1}(-\tf12-h_3) \,
{\mathscr D}_m^{h_1+2}(\tf52-h_2-h_3)
\nonu \\
&& - \la_{-1,+\fft32,+\fft32}\, \la_{+2,+\fft32,-\fft32}\,
{\mathscr D}_m^{h_1+2}(-\tf12-h_3)
{\mathscr D}_n^{h_2-1}(-\tf12-h_1-h_3)
\nonu \\
&& -
 \la_{-1,+\fft32,+\fft32} \, \kappa_{+2,-1,+1} \,\bigg(
(h_2-2)m-(h_1+1)n \bigg) \,
{\mathscr D}_{m+n}^{h_1+h_2-1}(-\tf12-h_3)\, \Bigg]\nonu \\
&& \times
2! \, \de^{C}_{[A}\,
{\cal O}_{B],-\fft32}^{(h_1+h_2+h_3)}(z,\bar{z})=0,
\nonu \\
&& \Bigg[
 \la_{+2,-2,+2}\,\la_{-1,+1,+2} \,
{\mathscr D}_n^{h_2-1}(-h_3) \,
{\mathscr D}_m^{h_1+2}(3-h_2-h_3)
\nonu \\
&& - \la_{-1,+1,+2}\, \la_{+2,+1,-1}\,
{\mathscr D}_m^{h_1+2}(-h_3)
{\mathscr D}_n^{h_2-1}(-h_1-h_3)
\nonu \\
&& -
 \la_{-1,+1,+2} \,\kappa_{+2,-1,+1} \, \bigg(
(h_2-2)m-(h_1+1)n \bigg) \,
{\mathscr D}_{m+n}^{h_1+h_2-1}(-h_3)\, \Bigg]\nonu \\
&& \times
\de^{CD}_{AB}\,
{\cal O}_{-2}^{(h_1+h_2+h_3)}(z,\bar{z})=0,
\nonu \\
&& \Bigg[
 \la_{+2,-\fft32,+\fft32}\,\la_{-\fft32,+2,+\fft32} \,
{\mathscr D}_n^{h_2-\fft32}(-1-h_3) \,
{\mathscr D}_m^{h_1+2}(\tf52-h_2-h_3)
\nonu \\
&& - \la_{-\fft32,+2,+\fft32}\, \la_{+2,+2,-2}\,
{\mathscr D}_m^{h_1+2}(-1-h_3)
{\mathscr D}_n^{h_2-\fft32}(-1-h_1-h_3)
\nonu \\
&& -
 \la_{-\fft32,+2,+\fft32} \, \kappa_{+2,-\fft32,+\fft32} \,\bigg(
(h_2-\tf52)m-(h_1+1)n \bigg) \,
{\mathscr D}_{m+n}^{h_1+h_2-\fft32}(-1-h_3)\, \Bigg]\,
{\cal O}_{A,-\fft32}^{(h_1+h_2+h_3)}(z,\bar{z})=0,
\nonu \\
&& \Bigg[
 \la_{+2,-2,+2}\,\la_{-\fft32,+\fft32,+2} \,
{\mathscr D}_n^{h_2-\fft32}(-\tf12-h_3) \,
{\mathscr D}_m^{h_1+2}(3-h_2-h_3)
\nonu \\
&& - \la_{-\fft32,+\fft32,+2}\, \la_{+2,+\fft32,-\fft32}\,
{\mathscr D}_m^{h_1+2}(-\tf12-h_3)
{\mathscr D}_n^{h_2-\fft32}(-\tf12-h_1-h_3)
\nonu \\
&& -
 \la_{-\fft32,+\fft32,+2} \, \kappa_{+2,-\fft32,+\fft32} \,\bigg(
(h_2-\tf52)m-(h_1+1)n \bigg) \,
{\mathscr D}_{m+n}^{h_1+h_2-\fft32}(-\tf12-h_3)\, \Bigg]\nonu \\
&& \times
\de^B_A\,
{\cal O}_{-2}^{(h_1+h_2+h_3)}(z,\bar{z})=0,
\nonu \\
&& \Bigg[
 \la_{+2,-2,+2}\,\la_{-2,+2,+2} \,
{\mathscr D}_n^{h_2-2}(-1-h_3) \,
{\mathscr D}_m^{h_1+2}(3-h_2-h_3)
\nonu \\
&& - \la_{-2,+2,+2}\, \la_{+2,+2,-2}\,
{\mathscr D}_m^{h_1+2}(-1-h_3)
{\mathscr D}_n^{h_2-2}(-1-h_1-h_3)
\label{10}
\\
&& -
 \la_{-2,+2,+2} \, \kappa_{+2,-2,+2} \,\bigg(
(h_2-3)m-(h_1+1)n \bigg) \,
{\mathscr D}_{m+n}^{h_1+h_2-2}(-1-h_3)\, \Bigg]\,
{\cal O}_{-2}^{(h_1+h_2+h_3)}(z,\bar{z})=0.
\nonu
\eea

\subsection{The  relations for the couplings in the subsection \ref{45rel}}

We solve the previous forty-five relations and
obtain the relations between
the couplings.

\subsubsection{The constraints for the seventeen relations }

The constraints for the seventeen relations
(\ref{17}) can be summarized by
\bea
\lambda_{+2,+2,-2} \, \lambda_{+2,+2,-2} & = &
\lambda_{+2,+2,-2}\, \kappa_{+2,+2,-2},
\qquad
\lambda_{+2,+\frac{3}{2},-\frac{3}{2}} \,
\lambda_{+2,+\frac{3}{2},-\frac{3}{2}}  = 
\lambda_{+2,+\frac{3}{2},-\frac{3}{2}}\,
\kappa_{+2,+2,-2},
\nonu \\
\lambda_{+2,+1,-1}\, \lambda_{+2,+1,-1} & = &
\lambda_{+2,+1,-1} \, \kappa_{+2,+2,-2},
\qquad
\lambda_{+2,+\frac{1}{2},-\frac{1}{2}}
\,\lambda_{+2,+\frac{1}{2},-\frac{1}{2}}  = 
\lambda_{+2,+\frac{1}{2},-\frac{1}{2}}
\, \kappa_{+2,+2,-2},
\nonu \\
\lambda_{+2,0,0}\, \lambda_{+2,0,0} & = &
\lambda_{+2,0,0} \, \kappa_{+2,+2,-2},
\qquad
\lambda_{+2,-\frac{1}{2},+\frac{1}{2}}\,
\lambda_{+2,-\frac{1}{2},+\frac{1}{2}}  = 
\lambda_{+2,-\frac{1}{2},+\frac{1}{2}}\,
\kappa_{+2,+2,-2},
\nonu \\
\lambda_{+2,-1,+1}\, \lambda_{+2,-1,+1} & = &
\lambda_{+2,-1,+1} \, \kappa_{+2,+2,-2},
\qquad
\lambda_{+2,-\frac{3}{2},+\frac{3}{2}}\,
\lambda_{+2,-\frac{3}{2},+\frac{3}{2}}  = 
\lambda_{+2,-\frac{3}{2},+\frac{3}{2}} \,
\kappa_{+2,+2,-2},
\nonu \\
\lambda_{+2,-2,+2}\, \lambda_{+2,-2,+2} & = &
\lambda_{+2,-2,+2} \, \kappa_{+2,+2,-2},
\qquad
\lambda_{+2,+\frac{3}{2},-\frac{3}{2}} \,
\lambda_{+\frac{3}{2},+2,-\frac{3}{2}}  = 
\lambda_{+\frac{3}{2},+2,-\frac{3}{2}}\,
\kappa_{+2,+\frac{3}{2},-\frac{3}{2}},
\nonu \\
\lambda_{+\frac{3}{2},+2,-\frac{3}{2}}\, \lambda_{+2,+2,-2}
& = & \lambda_{+\frac{3}{2},+2,-\frac{3}{2}}
\, \kappa_{+2,+\frac{3}{2},-\frac{3}{2}},
\qquad
\lambda_{+2,+1,-1}\, \lambda_{+\frac{3}{2},+\frac{3}{2},-1}
=  \lambda_{+\frac{3}{2},+\frac{3}{2},-1}
\kappa_{+2,+\frac{3}{2},-\frac{3}{2}},
\nonu \\
\lambda_{+\frac{3}{2},+\frac{3}{2},-1} \,
\lambda_{+2,+\frac{3}{2},-\frac{3}{2}} & = &
\lambda_{+\frac{3}{2},+\frac{3}{2},-1}\, \kappa_{+2,+\frac{3}{2},-\frac{3}{2}},
\qquad
\lambda_{+2,+\frac{1}{2},-\frac{1}{2}}\,
\lambda_{+\frac{3}{2},+1,-\frac{1}{2}}  = 
\lambda_{+\frac{3}{2},+1,-\frac{1}{2}}\, \kappa_{+2,+\frac{3}{2},-\frac{3}{2}},
\nonu \\
\lambda_{+\frac{3}{2},+1,-\frac{1}{2}}\, \lambda_{+2,+1,-1}
& = & \lambda_{+\frac{3}{2},+1,-\frac{1}{2}}\, \kappa_{+2,+\frac{3}{2},-\frac{3}{2}},
\qquad
\lambda_{+2,0,0} \, \lambda_{+\frac{3}{2},+\frac{1}{2},0}
=  \lambda_{+\frac{3}{2},+\frac{1}{2},0}\, \kappa_{+2,+\frac{3}{2},-\frac{3}{2}},
\nonu \\
\lambda_{+\frac{3}{2},+\frac{1}{2},0}\, \lambda_{+2,+\frac{1}{2},-\frac{1}{2}} & = &
\lambda_{+\frac{3}{2},+\frac{1}{2},0}\, \kappa_{+2,+\frac{3}{2},-\frac{3}{2}},
\qquad
\lambda_{+2,-\frac{1}{2},+\frac{1}{2}}\, \lambda_{+\frac{3}{2},0,+\frac{1}{2}}
= 
\lambda_{+\frac{3}{2},0,+\frac{1}{2}} \, \kappa_{+2,+\frac{3}{2},-\frac{3}{2}},
\nonu \\
\lambda_{+\frac{3}{2},0,+\frac{1}{2}} \, \lambda_{+2,0,0}
& = &
\lambda_{+\frac{3}{2},0,+\frac{1}{2}}\, \kappa_{+2,+\frac{3}{2},-\frac{3}{2}},
\qquad
\lambda_{+2,-1,+1} \,\lambda_{+\frac{3}{2},-\frac{1}{2},+1}
=  \lambda_{+\frac{3}{2},-\frac{1}{2},+1} \,
\kappa_{+2,+\frac{3}{2},-\frac{3}{2}},
\nonu \\
\lambda_{+\frac{3}{2},-\frac{1}{2},+1}\, \lambda_{+2,-\frac{1}{2},+\frac{1}{2}} & = &
\lambda_{+\frac{3}{2},-\frac{1}{2},+1}\, \kappa_{+2,+\frac{3}{2},-\frac{3}{2}},
\qquad
\lambda_{+2,-\frac{3}{2},+\frac{3}{2}} \,
\lambda_{+\frac{3}{2},-1,+\frac{3}{2}}  = 
\lambda_{+\frac{3}{2},-1,+\frac{3}{2}}\, \kappa_{+2,+\frac{3}{2},-\frac{3}{2}},
\nonu \\
\lambda_{+\frac{3}{2},-1,+\frac{3}{2}} \, \lambda_{+2,-1,+1}
& = &
\lambda_{+\frac{3}{2},-1,+\frac{3}{2}}\, \kappa_{+2,+\frac{3}{2},-\frac{3}{2}},
\qquad
\lambda_{+2,-2,+2}\,  \lambda_{+\frac{3}{2},-\frac{3}{2},+2}
= 
\lambda_{+\frac{3}{2},-\frac{3}{2},+2} \, \kappa_{+2,+\frac{3}{2},-\frac{3}{2}},
\nonu \\
\lambda_{+\frac{3}{2},-\frac{3}{2},+2} \, \lambda_{+2,-\frac{3}{2},+\frac{3}{2}} & = &
\lambda_{+\frac{3}{2},-\frac{3}{2},+2} \, \kappa_{+2,+\frac{3}{2},-\frac{3}{2}}.
\label{17-lambda}
\eea

\subsubsection{The constraints for the eighteen relations }

The constraints for the eighteen relations
(\ref{18}) can be written as
\bea
\lambda_{+2,+1,-1} \, \lambda_{+1,+2,-1}  & = & 
\lambda_{+1,+2,-1} \, \kappa_{+2,+1,-1},
\nonu \\
\lambda_{+1,+2,-1} \, \lambda_{+2,+2,-2}& = &
\lambda_{+1,+2,-1} \kappa_{+2,+1,-1},
\qquad
\lambda_{+2,+\frac{1}{2},-\frac{1}{2}} \,
\lambda_{+1,+\frac{3}{2},-\frac{1}{2}}  = 
\lambda_{+1,+\frac{3}{2},-\frac{1}{2}}
\, \kappa_{+2,+1,-1},
\nonu \\
\lambda_{+1,+\frac{3}{2},-\frac{1}{2}}\,
\lambda_{+2,+\frac{3}{2},-\frac{3}{2}} & = &
\lambda_{+1,+\frac{3}{2},-\frac{1}{2}}
\, \kappa_{+2,+1,-1},
\qquad
\lambda_{+2,0,0} \, \lambda_{+1,+1,0}  = 
\lambda_{+1,+1,0}\, \kappa_{+2,+1,-1},
\nonu \\
\lambda_{+1,+1,0} \, \lambda_{+2,+1,-1} & = &
\lambda_{+1,+1,0} \, \kappa_{+2,+1,-1},
\qquad
\lambda_{+2,-\frac{1}{2},+\frac{1}{2}}
\, \lambda_{+1,+\frac{1}{2},+\frac{1}{2}}  = 
\lambda_{+1,+\frac{1}{2},+\frac{1}{2}}
\, \kappa_{+2,+1,-1},
\nonu \\
\lambda_{+1,+\frac{1}{2},+\frac{1}{2}}
\, \lambda_{+2,+\frac{1}{2},-\frac{1}{2}} & = &
\lambda_{+1,+\frac{1}{2},+\frac{1}{2}}
\, \kappa_{+2,+1,-1},
\qquad
\lambda_{+2,-1,+1}\, \lambda_{+1,0,+1}  = 
\lambda_{+1,0,+1} \, \kappa_{+2,+1,-1},
\nonu \\
\lambda_{+1,0,+1} \, \lambda_{+2,0,0} & = &
\lambda_{+1,0,+1} \, \kappa_{+2,+1,-1},
\qquad
\lambda_{+2,-\frac{3}{2},+\frac{3}{2}}
\, \lambda_{+1,-\frac{1}{2},+\frac{3}{2}}  = 
\lambda_{+1,-\frac{1}{2},+\frac{3}{2}}
\, \kappa_{+2,+1,-1},
\nonu \\
\lambda_{+1,-\frac{1}{2},+\frac{3}{2}}
\, \lambda_{+2,-\frac{1}{2},+\frac{1}{2}}
& = & \lambda_{+1,-\frac{1}{2},+\frac{3}{2}}
\, \kappa_{+2,+1,-1},
\qquad
\lambda_{+2,-2,+2} \, \lambda_{+1,-1,+2}  = 
\lambda_{+1,-1,+2} \, \kappa_{+2,+1,-1},
\nonu \\
\lambda_{+1,-1,+2} \, \lambda_{+2,-1,+1} & = &
\lambda_{+1,-1,+2} \, \kappa_{+2,+1,-1},
\qquad
\lambda_{+2,+\frac{1}{2},-\frac{1}{2}}
\, \lambda_{+\frac{1}{2},+2,-\frac{1}{2}}  =
\lambda_{+\frac{1}{2},+2,-\frac{1}{2}}
\, \kappa_{+2,+\frac{1}{2},-\frac{1}{2}},
\nonu \\
\lambda_{+\frac{1}{2},+2,-\frac{1}{2}}
\, \lambda_{+2,+2,-2} & = & \lambda_{+\frac{1}{2},+2,-\frac{1}{2}}
\, \kappa_{+2,+\frac{1}{2},-\frac{1}{2}},
\qquad
\lambda_{+2,0,0}\, \lambda_{+\frac{1}{2},+\frac{3}{2},0}
= 
\lambda_{+\frac{1}{2},+\frac{3}{2},0} \,
\kappa_{+2,+\frac{1}{2},-\frac{1}{2}},
\nonu \\
\lambda_{+\frac{1}{2},+\frac{3}{2},0}
\, \lambda_{+2,+\frac{3}{2},-\frac{3}{2}} & = &
\lambda_{+\frac{1}{2},+\frac{3}{2},0}\, \kappa_{+2,+\frac{1}{2},-\frac{1}{2}},
\qquad
\lambda_{+2,-\frac{1}{2},+\frac{1}{2}} \,
\lambda_{+\frac{1}{2},+1,+\frac{1}{2}}  = 
\lambda_{+\frac{1}{2},+1,+\frac{1}{2}}\, \kappa_{+2,+\frac{1}{2},-\frac{1}{2}},
\nonu \\
\lambda_{+\frac{1}{2},+1,+\frac{1}{2}}
\, \lambda_{+2,+1,-1} & = &
\lambda_{+\frac{1}{2},+1,+\frac{1}{2}} \, \kappa_{+2,+\frac{1}{2},-\frac{1}{2}},
\qquad
\lambda_{+2,-1,+1} \, \lambda_{+\frac{1}{2},+\frac{1}{2},+1}
=  \lambda_{+\frac{1}{2},+\frac{1}{2},+1} \,
\kappa_{+2,+\frac{1}{2},-\frac{1}{2}},
\nonu \\
\lambda_{+\frac{1}{2},+\frac{1}{2},+1}\,
\lambda_{+2,+\frac{1}{2},-\frac{1}{2}} & = &
\lambda_{+\frac{1}{2},+\frac{1}{2},+1}\, \kappa_{+2,+\frac{1}{2},-\frac{1}{2}},
\qquad
\lambda_{+2,-\frac{3}{2},+\frac{3}{2}} \,
\lambda_{+\frac{1}{2},0,+\frac{3}{2}}  = 
\lambda_{+\frac{1}{2},0,+\frac{3}{2}}\, \kappa_{+2,+\frac{1}{2},-\frac{1}{2}},
\nonu \\
\lambda_{+\frac{1}{2},0,+\frac{3}{2}} \, \lambda_{+2,0,0}
& = &
\lambda_{+\frac{1}{2},0,+\frac{3}{2}}\, \kappa_{+2,+\frac{1}{2},-\frac{1}{2}},
\qquad
\lambda_{+2,-2,+2}\, \lambda_{+\frac{1}{2},-\frac{1}{2},+2}
=  \lambda_{+\frac{1}{2},-\frac{1}{2},+2}\, \kappa_{+2,+\frac{1}{2},-\frac{1}{2}},
\nonu \\
\lambda_{+\frac{1}{2},-\frac{1}{2},+2} \,
\lambda_{+2,-\frac{1}{2},+\frac{1}{2}} & = &
\lambda_{+\frac{1}{2},-\frac{1}{2},+2}\, \kappa_{+2,+\frac{1}{2},-\frac{1}{2}},
\qquad
\lambda_{+2,0,0} \, \lambda_{0,+2,0}  = 
\lambda_{0,+2,0} \, \kappa_{+2,0,0},
\nonu \\
\lambda_{0,+2,0} \, \lambda _{+2,+2,-2} & = &
\lambda_{0,+2,0} \, \kappa_{+2,0,0},
\qquad
\lambda_{+2,-\frac{1}{2},+\frac{1}{2}}
\, \lambda_{0,+\frac{3}{2},+\frac{1}{2}}
=  \lambda_{0,+\frac{3}{2},+\frac{1}{2}}
\, \kappa_{+2,0,0},
\nonu \\
\lambda_{0,+\frac{3}{2},+\frac{1}{2}}
\, \lambda_{+2,+\frac{3}{2},-\frac{3}{2}} & = &
\lambda_{0,+\frac{3}{2},+\frac{1}{2}}
\, \kappa_{+2,0,0},
\qquad
\lambda_{+2,-1,+1}\, \lambda_{0,+1,+1}  = 
\lambda_{0,+1,+1} \, \kappa_{+2,0,0},
\nonu \\
\lambda_{0,+1,+1} \,\lambda _{+2,+1,-1}& = & \lambda_{0,+1,+1}\, \kappa_{+2,0,0},
\qquad
\lambda_{+2,-\frac{3}{2},+\frac{3}{2}} \,
\lambda_{0,+\frac{1}{2},+\frac{3}{2}}  = 
\lambda_{0,+\frac{1}{2},+\frac{3}{2}} \kappa_{+2,0,0},
\nonu \\
\lambda_{0,+\frac{1}{2},+\frac{3}{2}} \, \lambda_{+2,+\frac{1}{2},-\frac{1}{2}} & = &
\lambda_{0,+\frac{1}{2},+\frac{3}{2}} \,
\kappa_{+2,0,0},
\qquad
\lambda_{+2,-2,+2}\, \lambda_{0,0,+2}  = 
\lambda_{0,0,+2} \, \kappa_{+2,0,0},
\nonu \\
\lambda_{0,0,+2} \, \lambda_{+2,0,0} & = &
\lambda_{0,0,+2} \, \kappa_{+2,0,0}.
\label{18-lambda}
\eea

\subsubsection{The constraints for the ten relations  }

Finally, the constraints for the ten relations
(\ref{10}) can be obtained as 
\bea
\lambda_{+2,-\frac{1}{2},+\frac{1}{2}} \,
\lambda_{-\frac{1}{2},+2,+\frac{1}{2}}  & = & 
\lambda_{-\frac{1}{2},+2,+\frac{1}{2}}\, \kappa_{+2,-\frac{1}{2},\frac{1}{2}},
\nonu \\
\lambda_{-\frac{1}{2},+2,+\frac{1}{2}}\, \lambda_{+2,+2,-2}
& = & \lambda_{-\frac{1}{2},+2,+\frac{1}{2}}\,
\kappa_{+2,-\frac{1}{2},+\frac{1}{2}},
\qquad
\lambda_{+2,-1,+1}\, \lambda_{-\frac{1}{2},+\frac{3}{2},+1}
= 
\lambda_{-\frac{1}{2},+\frac{3}{2},+1}\, \kappa_{+2,-\frac{1}{2},+\frac{1}{2}},
\nonu \\
\lambda_{-\frac{1}{2},+\frac{3}{2},+1} \,
\lambda_{+2,+\frac{3}{2},-\frac{3}{2}} & = &
\lambda_{-\frac{1}{2},+\frac{3}{2},+1} \, \kappa_{+2,-\frac{1}{2},+\frac{1}{2}},
\qquad
\lambda_{+2,-\frac{3}{2},+\frac{3}{2}}\,
\lambda_{-\frac{1}{2},+1,+\frac{3}{2}}  = 
\lambda_{-\frac{1}{2},+1,+\frac{3}{2}}\, \kappa_{+2,-\frac{1}{2},+\frac{1}{2}},
\nonu \\
\lambda_{-\frac{1}{2},+1,+\frac{3}{2}} \, \lambda_{+2,+1,-1}
& = & \lambda_{-\frac{1}{2},+1,+\frac{3}{2}}
\, \kappa_{+2,-\frac{1}{2},+\frac{1}{2}},
\qquad
\lambda_{+2,-2,+2}\, \lambda_{-\frac{1}{2},+\frac{1}{2},+2}
= 
\lambda_{-\frac{1}{2},+\frac{1}{2},+2} \, \kappa_{+2,-\frac{1}{2},+\frac{1}{2}},
\nonu \\
\lambda_{-\frac{1}{2},+\frac{1}{2},+2} \,
\lambda_{+2,\frac{1}{2},-\frac{1}{2}} & = &
\lambda_{-\frac{1}{2},+\frac{1}{2},+2} \, \kappa_{+2,-\frac{1}{2},+\frac{1}{2}},
\qquad
\lambda_{+2,-1,+1} \, \lambda_{-1,+2,+1}  = 
\lambda_{-1,+2,+1} \, \kappa_{+2,-1,+1},
\nonu \\
\lambda_{-1,+2,+1} \,
\lambda_{+2,+2,-2} & = &
\lambda_{-1,+2,+1} \, \kappa_{+2,-1,+1},
\qquad
\lambda_{+2,-\frac{3}{2},+\frac{3}{2}}\,
\lambda_{-1,+\frac{3}{2},+\frac{3}{2}}
=  \lambda_{-1,+\frac{3}{2},+\frac{3}{2}}\, \kappa_{+2,-1,+1},
\nonu \\
\lambda_{-1,+\frac{3}{2},+\frac{3}{2}} \,
\lambda_{+2,+\frac{3}{2},-\frac{3}{2}} & = &
\lambda_{-1,+\frac{3}{2},+\frac{3}{2}} \,
\kappa_{+2,-1,+1},
\qquad
\lambda_{+2,-2,+2} \, \lambda_{-1,+1,+2}  = 
\lambda_{-1,+1,+2} \, \kappa_{+2,-1,+1},
\nonu \\
\lambda_{-1,+1,+2} \, \lambda_{+2,+1,-1} & = &
\lambda_{-1,+1,+2} \, \kappa_{+2,-1,+1},
\qquad
\lambda_{+2,-\frac{3}{2},+\frac{3}{2}} \,
\lambda_{-\frac{3}{2},+2,+\frac{3}{2}}  = 
\lambda_{-\frac{3}{2},+2,+\frac{3}{2}}
\, \kappa_{+2,-\frac{3}{2},+\frac{3}{2}},
\nonu \\
\lambda_{-\frac{3}{2},+2,+\frac{3}{2}}\, \lambda_{+2,+2,-2}
& = & \lambda_{-\frac{3}{2},+2,+\frac{3}{2}} \,
\kappa_{+2,-\frac{3}{2},+\frac{3}{2}},
\qquad
\lambda_{+2,-2,+2}\, \lambda_{-\frac{3}{2},+\frac{3}{2},+2}
=  \lambda_{-\frac{3}{2},+\frac{3}{2},+2} \,
\kappa_{+2,-\frac{3}{2},+\frac{3}{2}},
\nonu \\
\lambda_{-\frac{3}{2},+\frac{3}{2},+2} \,
\lambda_{+2,+\frac{3}{2},-\frac{3}{2}} & = &
\lambda_{-\frac{3}{2},+\frac{3}{2},+2}
\, \kappa_{+2,-\frac{3}{2},+\frac{3}{2}},
\qquad
\lambda_{+2,-2,+2}\, \lambda_{-2,+2,+2}  = 
\lambda_{-2,+2,+2} \,\kappa_{+2,-2,+2},
\nonu \\
\lambda_{-2,+2,+2} \,\lambda_{+2,+2,-2} & = &
\lambda_{-2,+2,+2} \, \kappa_{+2,-2,+2}.
\label{10-lambda}
\eea

\subsection{The twenty-eight relations of (\ref{Iden})
\label{28rel}}

The next twenty-eight relations (\ref{Iden})
are described.

\subsubsection{The eighteen relations }

The eighteen relations for (\ref{Iden}) corresponding to
Table \ref{table15} can be described as 
\bea
&& \Bigg[
 \la_{+\fft32,+\fft32,-1}\,\la_{+\fft32,+2,-\fft32} \,
{\mathscr D}_n^{h_2+\fft32}(-1-h_3) \,
{\mathscr D}_m^{h_1+\fft32}(-\tf12-h_2-h_3)
\nonu \\
&& - \la_{+\fft32,+\fft32,-1}\, \la_{+\fft32,+2,-\fft32}\,
{\mathscr D}_m^{h_1+\fft32}(-1-h_3)
{\mathscr D}_n^{h_2+\fft32}(-\tf12-h_1-h_3)
\nonu \\
&& -
 \la_{+1,+2,-1} \, \kappa_{+\fft32,+\fft32,-1} \,\bigg(
(h_2+\tf12)m-(h_1+\tf12)n \bigg) \,
{\mathscr D}_{m+n}^{h_1+h_2+1}(-1-h_3)\, \Bigg]\nonu \\
&& \times {\cal O}_{+1}^{(h_1+h_2+h_3),AB}(z,\bar{z})=0,
\nonu \\
&& \Bigg[
 \la_{+\fft32,+1,-\fft12}\,\la_{+\fft32,+\fft32,-1} \,
{\mathscr D}_n^{h_2+\fft32}(-\tf12-h_3) \,
{\mathscr D}_m^{h_1+\fft32}(-h_2-h_3)
\nonu \\
&& - \la_{+\fft32,+1,-\fft12}\, \la_{+\fft32,+\fft32,-1}\,
{\mathscr D}_m^{h_1+\fft32}(-\tf12-h_3)
{\mathscr D}_n^{h_2+\fft32}(-h_1-h_3)
\nonu \\
&& -
 \la_{+1,+\fft32,-\fft12} \, \kappa_{+\fft32,+\fft32,-1} \,\bigg(
(h_2+\tf12)m-(h_1+\tf12)n \bigg) \,
{\mathscr D}_{m+n}^{h_1+h_2+1}(-\tf12-h_3)\, \Bigg]\nonu \\
&& \times
{\cal O}_{+\fft12}^{(h_1+h_2+h_3),ABC}(z,\bar{z})=0,
\nonu \\
&& \Bigg[
 \la_{+\fft32,+\fft12,0}\,\la_{+\fft32,+1,-\fft12} \,
{\mathscr D}_n^{h_2+\fft32}(-h_3) \,
{\mathscr D}_m^{h_1+\fft32}(\tf12-h_2-h_3)
\nonu \\
&& - \la_{+\fft32,+\fft12,0}\, \la_{+\fft32,+1,-\fft12}\,
{\mathscr D}_m^{h_1+\fft32}(-h_3)
{\mathscr D}_n^{h_2+\fft32}(\tf12-h_1-h_3)
\nonu \\
&& -
 \la_{+1,+1,0} \, \kappa_{+\fft32,+\fft32,-1} \,\bigg(
(h_2+\tf12)m-(h_1+\tf12)n \bigg) \,
{\mathscr D}_{m+n}^{h_1+h_2+1}(-h_3)\, \Bigg]\nonu \\
&& \times
{\cal O}_{0}^{(h_1+h_2+h_3),ABCD}(z,\bar{z})=0,
\nonu \\
&& \Bigg[
 \la_{+\fft32,0,+\fft12}\,\la_{+\fft32,+\fft12,0} \,
{\mathscr D}_n^{h_2+\fft32}(\tf12-h_3) \,
{\mathscr D}_m^{h_1+\fft32}(1-h_2-h_3)
\nonu \\
&& - \la_{+\fft32,0,+\fft12}\, \la_{+\fft32,+\fft12,0}\,
{\mathscr D}_m^{h_1+\fft32}(\tf12-h_3)
{\mathscr D}_n^{h_2+\fft32}(1-h_1-h_3)
\nonu \\
&& -
 \la_{+1,+\fft12,+\fft12} \, \kappa_{+\fft32,+\fft32,-1} \,\bigg(
(h_2+\tf12)m-(h_1+\tf12)n \bigg) \,
{\mathscr D}_{m+n}^{h_1+h_2+1}(\tf12-h_3)\, \Bigg]\nonu \\
&& \times
\frac{1}{3!}\, \ep^{ABCDEFGH}\,
{\cal O}_{FGH,-\fft12}^{(h_1+h_2+h_3)}(z,\bar{z})=0,
\nonu \\
&& \Bigg[
 (-1) \la_{+\fft32,-\fft12,+1}\,\la_{+\fft32,0,+\fft12} \,
{\mathscr D}_n^{h_2+\fft32}(1-h_3) \,
{\mathscr D}_m^{h_1+\fft32}(\tf32-h_2-h_3)
\nonu \\
&& - (-1) \la_{+\fft32,-\fft12,+1}\, \la_{+\fft32,0,+\fft12}\,
{\mathscr D}_m^{h_1+\fft32}(1-h_3)
{\mathscr D}_n^{h_2+\fft32}(\tf32-h_1-h_3)
\nonu \\
&& -
 \la_{+1,0,+1} \, \kappa_{+\fft32,+\fft32,-1} \,\bigg(
(h_2+\tf12)m-(h_1+\tf12)n \bigg) \,
{\mathscr D}_{m+n}^{h_1+h_2+1}(1-h_3)\, \Bigg]\nonu \\
&& \times
\frac{1}{2!}\, \ep^{ABCDEFGH}\,
{\cal O}_{GH,-1}^{(h_1+h_2+h_3)}(z,\bar{z})=0,
\nonu \\
&& \Bigg[
  \la_{+\fft32,-1,+\fft32}\,\la_{+\fft32,-\fft12,+1} \,
{\mathscr D}_n^{h_2+\fft32}(\tf32-h_3) \,
{\mathscr D}_m^{h_1+\fft32}(2-h_2-h_3)
\nonu \\
&& -  \la_{+\fft32,-1,+\fft32}\, \la_{+\fft32,-\fft12,+1}\,
{\mathscr D}_m^{h_1+\fft32}(\tf32-h_3)
{\mathscr D}_n^{h_2+\fft32}(2-h_1-h_3)
\nonu \\
&& -
 \la_{+1,-\fft12,+\fft32} \, \kappa_{+\fft32,+\fft32,-1} \,\bigg(
(h_2+\tf12)m-(h_1+\tf12)n \bigg) \,
{\mathscr D}_{m+n}^{h_1+h_2+1}(\tf32-h_3)\, \Bigg]\nonu \\
&& \times
3!\, \de^{A}_{[C} \,
{\cal O}_{D,-\fft32}^{(h_1+h_2+h_3)}(z,\bar{z}) \, \de^{B}_{E]}=0,
\nonu \\
&& \Bigg[
  (-1) \la_{+\fft32,-\fft32,+2}\,\la_{+\fft32,-1,+\fft32} \,
{\mathscr D}_n^{h_2+\fft32}(2-h_3) \,
{\mathscr D}_m^{h_1+\fft32}(\tf52-h_2-h_3)
\nonu \\
&& -  (-1) \la_{+\fft32,-\fft32,+2} \la_{+\fft32,-1,+\fft32}
{\mathscr D}_m^{h_1+\fft32}(2-h_3)
{\mathscr D}_n^{h_2+\fft32}(\tf52-h_1-h_3)
\nonu \\
&& -
 \la_{+1,-1,+2}  \kappa_{+\fft32,+\fft32,-1} \bigg(
(h_2+\tf12)m-(h_1+\tf12)n \bigg) 
    {\mathscr D}_{m+n}^{h_1+h_2+1}(2-h_3) \Bigg]
\de^{AB}_{CD} 
{\cal O}_{-2}^{(h_1+h_2+h_3)}(z,\bar{z})=0,
\nonu \\
&& \Bigg[
   \la_{+\fft32,+1,-\fft12}\,\la_{+1,+2,-1} \,
{\mathscr D}_n^{h_2+1}(-1-h_3) \,
{\mathscr D}_m^{h_1+\fft32}(-h_2-h_3)
\nonu \\
&& -   \la_{+1,+\fft32,-\fft12}\, \la_{+\fft32,+2,-\fft32}\,
{\mathscr D}_m^{h_1+\fft32}(-1-h_3)
{\mathscr D}_n^{h_2+1}(-\tf12-h_1-h_3)
\nonu \\
&& -
 \la_{+\fft12,+2,-\fft12} \, \kappa_{+\fft32,+1,-\fft12} \,\bigg(
(h_2)m-(h_1+\tf12)n \bigg) \,
{\mathscr D}_{m+n}^{h_1+h_2+\fft12}(-1-h_3)\, \Bigg]\nonu \\
&& \times
{\cal O}_{+\fft12}^{(h_1+h_2+h_3),ABC}(z,\bar{z})=0,
\nonu \\
&& \Bigg[
   (-1) \la_{+\fft32,+\fft12,0}\,\la_{+1,+\fft32,-\fft12} \,
{\mathscr D}_n^{h_2+1}(-\tf12-h_3) \,
{\mathscr D}_m^{h_1+\fft32}(\tf12-h_2-h_3)
\nonu \\
&& - (-1)  \la_{+1,+1,0}\, \la_{+\fft32,+\fft32,-1}\,
{\mathscr D}_m^{h_1+\fft32}(-\tf12-h_3)
{\mathscr D}_n^{h_2+1}(-h_1-h_3)
\nonu \\
&& -
 \la_{+\fft12,+\fft32,0} \, \kappa_{+\fft32,+1,-\fft12} \,\bigg(
(h_2)m-(h_1+\tf12)n \bigg) \,
{\mathscr D}_{m+n}^{h_1+h_2+\fft12}(-\tf12-h_3)\, \Bigg]\nonu \\
&& \times
{\cal O}_{0}^{(h_1+h_2+h_3),ABCD}(z,\bar{z})=0,
\nonu \\
&& \Bigg[
    \la_{+\fft32,0,+\fft12}\,\la_{+1,+1,0} \,
{\mathscr D}_n^{h_2+1}(-h_3) \,
{\mathscr D}_m^{h_1+\fft32}(1-h_2-h_3)
\nonu \\
&& -   \la_{+1,+\fft12,+\fft12}\, \la_{+\fft32,+1,-\fft12}\,
{\mathscr D}_m^{h_1+\fft32}(-h_3)
{\mathscr D}_n^{h_2+1}(\tf12-h_1-h_3)
\nonu \\
&& -
 \la_{+\fft12,+1,+\fft12} \, \kappa_{+\fft32,+1,-\fft12} \,\bigg(
(h_2)m-(h_1+\tf12)n \bigg) \,
{\mathscr D}_{m+n}^{h_1+h_2+\fft12}(-h_3)\, \Bigg]\nonu \\
&& \times
\frac{1}{3!}\, \ep^{ABCDEFGH}\,
{\cal O}_{FGH,-\fft12}^{(h_1+h_2+h_3)}(z,\bar{z})=0,
\nonu \\
&& \Bigg[
   (-1) \la_{+\fft32,-\fft12,+1}\,\la_{+1,+\fft12,+\fft12} \,
{\mathscr D}_n^{h_2+1}(\tf12-h_3) \,
{\mathscr D}_m^{h_1+\fft32}(\tf32-h_2-h_3)
\nonu \\
&& -   \la_{+1,0,+1}\, \la_{+\fft32,+\fft12,0}\,
{\mathscr D}_m^{h_1+\fft32}(\tf12-h_3)
{\mathscr D}_n^{h_2+1}(1-h_1-h_3)
\nonu \\
&& -
 \la_{+\fft12,+\fft12,+1} \, \kappa_{+\fft32,+1,-\fft12} \,\bigg(
(h_2)m-(h_1+\tf12)n \bigg) \,
{\mathscr D}_{m+n}^{h_1+h_2+\fft12}(\tf12-h_3)\, \Bigg]\nonu \\
&& \times
\frac{1}{2!}\, \ep^{ABCDEFGH}\,
{\cal O}_{GH,-1}^{(h_1+h_2+h_3)}(z,\bar{z})=0,
\nonu \\
&& \Bigg[
   \la_{+\fft32,-1,+\fft32}\,\la_{+1,0,+1} \,
{\mathscr D}_n^{h_2+1}(1-h_3) \,
{\mathscr D}_m^{h_1+\fft32}(2-h_2-h_3)
\nonu \\
&& - (-1)  \la_{+1,-\fft12,+\fft32}\, \la_{+\fft32,0,+\fft12}\,
{\mathscr D}_m^{h_1+\fft32}(1-h_3)
{\mathscr D}_n^{h_2+1}(\tf32-h_1-h_3)
\nonu \\
&& -
 \la_{+\fft12,0,+\fft32} \, \kappa_{+\fft32,+1,-\fft12} \,\bigg(
(h_2)m-(h_1+\tf12)n \bigg) \,
{\mathscr D}_{m+n}^{h_1+h_2+\fft12}(1-h_3)\, \Bigg]\nonu \\
&& \times
\ep^{ABCDEFGH}\,
{\cal O}_{H,-\fft32}^{(h_1+h_2+h_3)}(z,\bar{z})=0,
\nonu \\
&& \Bigg[
  (-1) \la_{+\fft32,-\fft32,+2}\,\la_{+1-\fft12,+\fft32} \,
{\mathscr D}_n^{h_2+1}(\tf32-h_3) \,
{\mathscr D}_m^{h_1+\fft32}(\tf52-h_2-h_3)
\nonu \\
&& -   \la_{+1,-1,+2}\, \la_{+\fft32,-\fft12,+1}\,
{\mathscr D}_m^{h_1+\fft32}(\tf32-h_3)
{\mathscr D}_n^{h_2+1}(2-h_1-h_3)
\nonu \\
&& -
 \la_{+\fft12,-\fft12,+2} \, \kappa_{+\fft32,+1,-\fft12} \,\bigg(
(h_2)m-(h_1+\tf12)n \bigg) \,
{\mathscr D}_{m+n}^{h_1+h_2+\fft12}(\tf32-h_3)\, \Bigg]\nonu \\
&& \times
\de^{ABC}_{DEF}\,
{\cal O}_{-2}^{(h_1+h_2+h_3)}(z,\bar{z})=0,
\nonu \\
&& \Bigg[
  \la_{+\fft32,+\fft12,0}\,\la_{+\fft12,+2,-\fft12} \,
{\mathscr D}_n^{h_2+\fft12}(-1-h_3) \,
{\mathscr D}_m^{h_1+\fft32}(\tf12-h_2-h_3)
\nonu \\
&& - (-1)  \la_{+\fft12,+\fft32,0}\, \la_{+\fft32,+2,-\fft32}\,
{\mathscr D}_m^{h_1+\fft32}(-1-h_3)
{\mathscr D}_n^{h_2+\fft12}(-\tf12-h_1-h_3)
\nonu \\
&& -
 \la_{0,+2,0} \, \kappa_{+\fft32,+\fft12,0} \,\bigg(
(h_2-\tf12)m-(h_1+\tf12)n \bigg) \,
{\mathscr D}_{m+n}^{h_1+h_2}(-1-h_3)\, \Bigg]\nonu \\
&& \times
{\cal O}_{0}^{(h_1+h_2+h_3),ABCD}(z,\bar{z})=0,
\nonu \\
&& \Bigg[
 (-1) \la_{+\fft32,0,+\fft12}\,\la_{+\fft12,+\fft32,0} \,
{\mathscr D}_n^{h_2+\fft12}(-\tf12-h_3) \,
{\mathscr D}_m^{h_1+\fft32}(1-h_2-h_3)
\nonu \\
&& -   \la_{+\fft12,+1,+\fft12}\, \la_{+\fft32,+\fft32,-1}\,
{\mathscr D}_m^{h_1+\fft32}(-\tf12-h_3)
{\mathscr D}_n^{h_2+\fft12}(-h_1-h_3)
\nonu \\
&& -
 \la_{0,+\fft32,+\fft12} \, \kappa_{+\fft32,+\fft12,0} \,\bigg(
(h_2-\tf12)m-(h_1+\tf12)n \bigg) \,
{\mathscr D}_{m+n}^{h_1+h_2}(-\tf12-h_3)\, \Bigg]\nonu \\
&& \times
\frac{1}{3!}\, \ep^{ABCDEFGH}\,
{\cal O}_{FGH,-\fft12}^{(h_1+h_2+h_3)}(z,\bar{z})=0,
\nonu \\
&& \Bigg[
 (-1) \la_{+\fft32,-\fft12,+1}\,\la_{+\fft12,+1,+\fft12} \,
{\mathscr D}_n^{h_2+\fft12}(-h_3) \,
{\mathscr D}_m^{h_1+\fft32}(\tf32-h_2-h_3)
\nonu \\
&& -   \la_{+\fft12,+\fft12,+1}\, \la_{+\fft32,+1,-\fft12}\,
{\mathscr D}_m^{h_1+\fft32}(-h_3)
{\mathscr D}_n^{h_2+\fft12}(\tf12-h_1-h_3)
\nonu \\
&& -
 \la_{0,+1,+1} \, \kappa_{+\fft32,+\fft12,0} \,\bigg(
(h_2-\tf12)m-(h_1+\tf12)n \bigg) \,
{\mathscr D}_{m+n}^{h_1+h_2}(-h_3)\, \Bigg]\nonu \\
&& \times
\frac{1}{2!}\, \ep^{ABCDEFGH}\,
{\cal O}_{GH,-1}^{(h_1+h_2+h_3)}(z,\bar{z})=0,
\nonu \\
&& \Bigg[
  \la_{+\fft32,-1,+\fft32}\,\la_{+\fft12,+\fft12,+1} \,
{\mathscr D}_n^{h_2+\fft12}(\tf12-h_3) \,
{\mathscr D}_m^{h_1+\fft32}(2-h_2-h_3)
\nonu \\
&& -   \la_{+\fft12,0,+\fft32}\, \la_{+\fft32,+\fft12,0}\,
{\mathscr D}_m^{h_1+\fft32}(\tf12-h_3)
{\mathscr D}_n^{h_2+\fft12}(1-h_1-h_3)
\nonu \\
&& -
 \la_{0,+\fft12,+\fft32} \, \kappa_{+\fft32,+\fft12,0} \,\bigg(
(h_2-\tf12)m-(h_1+\tf12)n \bigg) \,
{\mathscr D}_{m+n}^{h_1+h_2}(\tf12-h_3)\, \Bigg]\nonu \\
&& \times
\ep^{ABCDEFGH}\,
{\cal O}_{H,-\fft32}^{(h_1+h_2+h_3)}(z,\bar{z})=0,
\nonu \\
&& \Bigg[
 (-1) \la_{+\fft32,-\fft32,+2}\,\la_{+\fft12,0,+\fft32} \,
{\mathscr D}_n^{h_2+\fft12}(1-h_3) \,
{\mathscr D}_m^{h_1+\fft32}(\tf52-h_2-h_3)
\nonu \\
&& - (-1)  \la_{+\fft12,-\fft12,+2}\, \la_{+\fft32,0,+\fft12}\,
{\mathscr D}_m^{h_1+\fft32}(1-h_3)
{\mathscr D}_n^{h_2+\fft12}(\tf32-h_1-h_3)
\nonu \\
&& -
 \la_{0,0,+2} \, \kappa_{+\fft32,+\fft12,0} \,\bigg(
(h_2-\tf12)m-(h_1+\tf12)n \bigg) \,
{\mathscr D}_{m+n}^{h_1+h_2}(1-h_3)\, \Bigg]\nonu \\
&& \times
\ep^{ABCDEFGH}\,
{\cal O}_{-2}^{(h_1+h_2+h_3)}(z,\bar{z})=0.
\label{other18}
\eea
In fifth relation of (\ref{other18}), 
due to the anticommutator beween the modes of
two positive gravitino currents,
the second term of (\ref{Iden}) has the plus sign.
Moreover, the extra minus sign from the
first term in (\ref{Iden}) comes from the
rearrangements of $SU(8)$ indices in order to factorize the
common factor which is the matter field appearing in (\ref{Iden}).

\subsubsection{The ten relations }

Similarly,
the ten relations for (\ref{Iden}) corresponding to
Table \ref{table16} can be summarized 
\bea
&& \Bigg[
  \la_{+\fft32,0,+\fft12}\,\la_{0,+2,0} \,
{\mathscr D}_n^{h_2}(-1-h_3) \,
{\mathscr D}_m^{h_1+\fft32}(1-h_2-h_3)
\nonu \\
&& -   \la_{0,+\fft32,+\fft12}\, \la_{+\fft32,+2,-\fft32}\,
{\mathscr D}_m^{h_1+\fft32}(-1-h_3)
{\mathscr D}_n^{h_2}(-\tf12-h_1-h_3)
\nonu \\
&& -
 \la_{-\fft12,+2,+\fft12} \, \kappa_{+\fft32,0,+\fft12} \,\bigg(
(h_2-1)m-(h_1+\tf12)n \bigg) \,
{\mathscr D}_{m+n}^{h_1+h_2-\fft12}(-1-h_3)\, \Bigg]\nonu \\
&& \times
\frac{1}{3!} \, \ep^{ABCDEFGH}\,
{\cal O}_{FGH,-\fft12}^{(h_1+h_2+h_3)}(z,\bar{z})=0,
\nonu \\
&& \Bigg[
 (-1) \la_{+\fft32,-\fft12,+1}\,\la_{0,+\fft32,+\fft12} \,
{\mathscr D}_n^{h_2}(-\tf12-h_3) \,
{\mathscr D}_m^{h_1+\fft32}(\tf32-h_2-h_3)
\nonu \\
&& -   \la_{0,+1,+1}\, \la_{+\fft32,+\fft32,-1}\,
{\mathscr D}_m^{h_1+\fft32}(-\tf12-h_3)
{\mathscr D}_n^{h_2}(-h_1-h_3)
\nonu \\
&& -
 \la_{-\fft12,+\fft32,+1} \, \kappa_{+\fft32,0,+\fft12} \,\bigg(
(h_2-1)m-(h_1+\tf12)n \bigg) \,
{\mathscr D}_{m+n}^{h_1+h_2-\fft12}(-\tf12-h_3)\, \Bigg]\nonu \\
&& \times
\frac{1}{3!} \, \ep^{ABCDEIGH}\, 3 \, \de^{F}_{[I} \,
{\cal O}_{GH],-1}^{(h_1+h_2+h_3)}(z,\bar{z})=0,
\nonu \\
&& \Bigg[
 (-1) \la_{+\fft32,-1,+\fft32}\,\la_{0,+1,+1} \,
{\mathscr D}_n^{h_2}(-h_3) \,
{\mathscr D}_m^{h_1+\fft32}(2-h_2-h_3)
\nonu \\
&& - (-1)  \la_{0,+\fft12,+\fft32}\, \la_{+\fft32,+1,-\fft12}\,
{\mathscr D}_m^{h_1+\fft32}(-h_3)
{\mathscr D}_n^{h_2}(\tf12-h_1-h_3)
\nonu \\
&& -
 \la_{-\fft12,+1,+\fft32} \, \kappa_{+\fft32,0,+\fft12} \,\bigg(
(h_2-1)m-(h_1+\tf12)n \bigg) \,
{\mathscr D}_{m+n}^{h_1+h_2-\fft12}(-h_3)\, \Bigg]\nonu \\
&& \times
\frac{1}{3!} \, \ep^{ABCDEIJH}\, 3! \, \de^{F}_{[I} \,
{\cal O}_{J,-\fft32}^{(h_1+h_2+h_3)}(z,\bar{z}) \, \de^{G}_{H]}=0,
\nonu \\
&& \Bigg[
 (-1) \la_{+\fft32,-\fft32,+2}\,\la_{0,+\fft12,+\fft32} \,
{\mathscr D}_n^{h_2}(\tf12-h_3) \,
{\mathscr D}_m^{h_1+\fft32}(\tf52-h_2-h_3)
\nonu \\
&& -   \la_{0,0,+2}\, \la_{+\fft32,+\fft12,0}\,
{\mathscr D}_m^{h_1+\fft32}(\tf12-h_3)
{\mathscr D}_n^{h_2}(1-h_1-h_3)
\nonu \\
&& -
 \la_{-\fft12,+\fft12,+2} \, \kappa_{+\fft32,0,+\fft12} \,\bigg(
(h_2-1)m-(h_1+\tf12)n \bigg) \,
{\mathscr D}_{m+n}^{h_1+h_2-\fft12}(\tf12-h_3)\, \Bigg]\nonu \\
&& \times
\frac{1}{3!} \, \ep^{ABCDEIJK} \, \de^{FGH}_{IJK} \,
{\cal O}_{-2}^{(h_1+h_2+h_3)}(z,\bar{z})=0,
\nonu \\
&& \Bigg[
 \la_{+\fft32,-\fft12,+1}\,\la_{-\fft12,+2,+\fft12} \,
{\mathscr D}_n^{h_2-\fft12}(-1-h_3) \,
{\mathscr D}_m^{h_1+\fft32}(\tf32-h_2-h_3)
\nonu \\
&& - (-1)  \la_{-\fft12,+\fft32,+1}\, \la_{+\fft32,+2,-\fft32}\,
{\mathscr D}_m^{h_1+\fft32}(-1-h_3)
{\mathscr D}_n^{h_2-\fft12}(-\tf12-h_1-h_3)
\nonu \\
&& -
\la_{-1,+2,+1} \, \kappa_{+\fft32,-\fft12,+1} \,\bigg(
(h_2-\tf32)m-(h_1+\tf12)n \bigg) \,
{\mathscr D}_{m+n}^{h_1+h_2-1}(-1-h_3)\, \Bigg]\nonu \\
&& \times
3 \,  \de^{A}_{[B} \,
{\cal O}_{CD],-1}^{(h_1+h_2+h_3)}(z,\bar{z})=0,
\nonu \\
&& \Bigg[
(-1)  \la_{+\fft32,-1,+\fft32}\,\la_{-\fft12,+\fft32,+1} \,
{\mathscr D}_n^{h_2-\fft12}(-\tf12-h_3) \,
{\mathscr D}_m^{h_1+\fft32}(2-h_2-h_3)
\nonu \\
&& -   \la_{-\fft12,+1,+\fft32}\, \la_{+\fft32,+\fft32,-1}\,
{\mathscr D}_m^{h_1+\fft32}(-\tf12-h_3)
{\mathscr D}_n^{h_2-\fft12}(-h_1-h_3)
\nonu \\
&& -
\la_{-1,+\fft32,+\fft32} \, \kappa_{+\fft32,-\fft12,+1} \,\bigg(
(h_2-\tf32)m-(h_1+\tf12)n \bigg) \,
{\mathscr D}_{m+n}^{h_1+h_2-1}(-\tf12-h_3)\, \Bigg]\nonu \\
&& \times
3 \times  2! \,  \de^{A}_{[B} \,\de^{E}_{C}
{\cal O}_{D],-\fft32}^{(h_1+h_2+h_3)}(z,\bar{z})=0,
\nonu \\
&& \Bigg[
 (-1) \la_{+\fft32,-\fft32,+2}\,\la_{-\fft12,+1,+\fft32} \,
{\mathscr D}_n^{h_2-\fft12}(-h_3) \,
{\mathscr D}_m^{h_1+\fft32}(\tf52-h_2-h_3)
\nonu \\
&& - (-1)  \la_{-\fft12,+\fft12,+2}\, \la_{+\fft32,+1,-\fft12}\,
{\mathscr D}_m^{h_1+\fft32}(-h_3)
{\mathscr D}_n^{h_2-\fft12}(\tf12-h_1-h_3)
\nonu \\
&& -
\la_{-1,+1,+2} \, \kappa_{+\fft32,-\fft12,+1} \,\bigg(
(h_2-\tf32)m-(h_1+\tf12)n \bigg) \,
{\mathscr D}_{m+n}^{h_1+h_2-1}(-h_3)\, \Bigg]\nonu \\
&& \times
3 \,  \de^{A}_{[B}  \, \de^{EF}_{CD]}
{\cal O}_{-2}^{(h_1+h_2+h_3)}(z,\bar{z})=0,
\nonu \\
&& \Bigg[
  \la_{+\fft32,-1,+\fft32}\,\la_{-1,+2,+1} \,
{\mathscr D}_n^{h_2-1}(-1-h_3) \,
{\mathscr D}_m^{h_1+\fft32}(2-h_2-h_3)
\nonu \\
&& -   \la_{-1,+\fft32,+\fft32}\, \la_{+\fft32,+2,-\fft32}\,
{\mathscr D}_m^{h_1+\fft32}(-1-h_3)
{\mathscr D}_n^{h_2-1}(-\tf12-h_1-h_3)
\nonu \\
&& -
\la_{-\fft32,+2,+\fft32} \, \kappa_{+\fft32,-1,+\fft32} \,\bigg(
(h_2-2)m-(h_1+\tf12)n \bigg) \,
{\mathscr D}_{m+n}^{h_1+h_2-\fft32}(-1-h_3)\, \Bigg]\nonu \\
&& \times
2! \,  \de^{A}_{[B}  \, 
{\cal O}_{C],-\fft32}^{(h_1+h_2+h_3)}(z,\bar{z})=0,
\nonu \\
&& \Bigg[
(-1)  \la_{+\fft32,-\fft32,+2}\,\la_{-1,+\fft32,+\fft32} \,
{\mathscr D}_n^{h_2-1}(-\tf12-h_3) \,
{\mathscr D}_m^{h_1+\fft32}(\tf52-h_2-h_3)
\nonu \\
&& -   \la_{-1,+1,+2}\, \la_{+\fft32,+\fft32,-1}\,
{\mathscr D}_m^{h_1+\fft32}(-\tf12-h_3)
{\mathscr D}_n^{h_2-1}(-h_1-h_3)
\nonu \\
&& -
\la_{-\fft32,+\fft32,+2} \, \kappa_{+\fft32,-1,+\fft32} \,\bigg(
(h_2-2)m-(h_1+\tf12)n \bigg) \,
{\mathscr D}_{m+n}^{h_1+h_2-\fft32}(-\tf12-h_3)\, \Bigg]\nonu \\
&& \times
2! \,  \de^{A}_{[B}  \, 
{\cal O}_{-2}^{(h_1+h_2+h_3)}(z,\bar{z}) \, \de^{D}_{C]}=0,
\nonu \\
&& \Bigg[
  \la_{+\fft32,-\fft32,+2}\,\la_{-\fft32,+2,+\fft32} \,
{\mathscr D}_n^{h_2-\fft32}(-1-h_3) \,
{\mathscr D}_m^{h_1+\fft32}(\tf52-h_2-h_3)
\nonu \\
&& - (-1)  \la_{-\fft32,+\fft32,+2} \la_{+\fft32,+2,-\fft32}
{\mathscr D}_m^{h_1+\fft32}(-1-h_3)
{\mathscr D}_n^{h_2-\fft32}(-\tf12-h_1-h_3)
\label{other10}
\\
&& -
\la_{-2,+2,+2}  \kappa_{+\fft32,-\fft32,+2} \bigg(
(h_2-\tf52)m-(h_1+\tf12)n \bigg) 
   {\mathscr D}_{m+n}^{h_1+h_2-2}(-1-h_3) \Bigg]
\de^{A}_{B}  \, 
{\cal O}_{-2}^{(h_1+h_2+h_3)}(z,\bar{z})=0.
\nonu
\eea
In the second relation of (\ref{other10}),
the two kinds of $SU(8)$ indices coming from two successive
commutator or anticommutator appear in the matter field. 

\subsection{The relations for the couplings in the subsection \ref{28rel}}

The previous twenty-eight relations can be solved
and we present the relations of the couplings. 

\subsubsection{The constraints for the eighteen relations  }

The constraints for the eighteen relations (\ref{other18}) 
are given by
\bea
\la_{+\fft32,+\fft32,-1}\,\la_{+\fft32,+2,-\fft32} & = &
\kappa_{+\fft32,+\fft32,-1} \,\la_{+1,+2,-1},
\qquad
\la_{+\fft32,+1,-\fft12}\,\la_{+\fft32,+\fft32,-1}  = 
\kappa_{+\fft32,+\fft32,-1} \,\la_{+1,+\fft32,-\fft12}, 
\nonu \\
\la_{+\fft32,+\fft12,0}\,\la_{+\fft32,+1,-\fft12} & = &
\kappa_{+\fft32,+\fft32,-1} \,\la_{+1,+1,0},
\qquad
\la_{+\fft32,0,+\fft12}\,\la_{+\fft32,+\fft12,0}  = 
\kappa_{+\fft32,+\fft32,-1} \,\la_{+1,+\fft12,+\fft12},
\nonu \\
\la_{+\fft32,-\fft12,+1}\,\la_{+\fft32,0,+\fft12} & = &
-\kappa_{+\fft32,+\fft32,-1} \,\la_{+1,0,+1},
\qquad
\la_{+\fft32,-1,+\fft32}\,\la_{+\fft32,-\fft12,+1}  = 
\kappa_{+\fft32,+\fft32,-1} \,\la_{+1,-\fft12,+\fft32},
\nonu \\
\la_{+\fft32,-\fft32,+2}\,\la_{+\fft32,-1,+\fft32} & = &
-\kappa_{+\fft32,+\fft32,-1} \,\la_{+1,-1,+2},
\qquad
\la_{+\fft32,+1,-\fft12}\,\la_{+1,+2,-1}  = 
\kappa_{+\fft32,+1,-\fft12} \,\la_{+\fft12,+2,-\fft12},
\nonu \\
\la_{+1,+\fft32,-\fft12}\, \la_{+\fft32,+2,-\fft32} & = &
\kappa_{+\fft32,+1,-\fft12} \,\la_{+\fft12,+2,-\fft12},
\qquad
\la_{+\fft32,+\fft12,0}\,\la_{+1,+\fft32,-\fft12}  = 
-\kappa_{+\fft32,+1,-\fft12} \,\la_{+\fft12,+\fft32,0},
\nonu \\
\la_{+1,+1,0}\, \la_{+\fft32,+\fft32,-1} & = &
-\kappa_{+\fft32,+1,-\fft12} \,\la_{+\fft12,+\fft32,0},
\qquad
\la_{+\fft32,0,+\fft12}\,\la_{+1,+1,0}  = 
\kappa_{+\fft32,+1,-\fft12} \,\la_{+\fft12,+1,+\fft12},
\nonu \\
\la_{+1,+\fft12,+\fft12}\, \la_{+\fft32,+1,-\fft12} & = &
\kappa_{+\fft32,+1,-\fft12} \,\la_{+\fft12,+1,+\fft12},
\qquad
\la_{+\fft32,-\fft12,+1}\,\la_{+1,+\fft12,+\fft12}  = 
-\kappa_{+\fft32,+1,-\fft12} \,\la_{+\fft12,+\fft12,+1},
\nonu \\
\la_{+1,0,+1}\, \la_{+\fft32,+\fft12,0} & = &
\kappa_{+\fft32,+1,-\fft12} \,\la_{+\fft12,+\fft12,+1},
\qquad
\la_{+\fft32,-1,+\fft32}\,\la_{+1,0,+1}  = 
\kappa_{+\fft32,+1,-\fft12} \,\la_{+\fft12,0,+\fft32},
\nonu \\
\la_{+1,-\fft12,+\fft32}\, \la_{+\fft32,0,+\fft12} & = &
-\kappa_{+\fft32,+1,-\fft12} \,\la_{+\fft12,0,+\fft32},
\qquad
\la_{+\fft32,-\fft32,+2}\,\la_{+1-\fft12,+\fft32}  = 
-\kappa_{+\fft32,+1,-\fft12} \,\la_{+\fft12,-\fft12,+2},
\nonu \\
\la_{+1,-1,+2}\, \la_{+\fft32,-\fft12,+1} & = &
\kappa_{+\fft32,+1,-\fft12} \,\la_{+\fft12,-\fft12,+2},
\qquad
\la_{+\fft32,+\fft12,0}\,\la_{+\fft12,+2,-\fft12}  = 
\kappa_{+\fft32,+\fft12,0} \,\la_{0,+2,0},
\nonu \\
\la_{+\fft12,+\fft12,+1}\, \la_{+\fft32,+2,-\fft12} & = &
-\kappa_{+\fft32,+\fft12,0} \,\la_{0,+2,0},
\qquad
\la_{+\fft32,0,+\fft12}\,\la_{+\fft12,+\fft32,0}  = 
-\kappa_{+\fft32,+\fft12,0} \,\la_{0,+\fft32,+\fft12},
\nonu \\
\la_{+\fft12,+1,+\fft12}\, \la_{+\fft32,+\fft32,-1} & = &
\kappa_{+\fft32,+\fft12,0} \,\la_{0,+\fft32,+\fft12},
\qquad
\la_{+\fft32,-\fft12,+1}\,\la_{+\fft12,+1,+\fft12}  = 
-\kappa_{+\fft32,+\fft12,0} \,\la_{0,+1,+1},
\nonu \\
\la_{+\fft12,+\fft12,+1}\, \la_{+\fft32,+1,-\fft12} & = &
\kappa_{+\fft32,+\fft12,0} \,\la_{0,+1,+1},
\qquad
\la_{+\fft32,-1,+\fft32}\,\la_{+\fft12,+\fft12,+1}  = 
\kappa_{+\fft32,+\fft12,0} \,\la_{0,+\fft12,+\fft32},
\nonu \\
\la_{+\fft12,0,+\fft32}\, \la_{+\fft32,+\fft12,0} & = &
\kappa_{+\fft32,+\fft12,0} \,\la_{0,+\fft12,+\fft32},
\qquad
\la_{+\fft32,-\fft32,+2}\,\la_{+\fft12,0,+\fft32}  = 
-\kappa_{+\fft32,+\fft12,0} \, \la_{0,0,+2},
\nonu \\
\la_{+\fft12,-\fft12,+2}\, \la_{+\fft32,0,+\fft12}] & = &
-\kappa_{+\fft32,+\fft12,0} \, \la_{0,0,+2}.
\label{other18-lambda}
\eea

\subsubsection{The constraints for the ten relations
}

The constraints for the ten relations (\ref{other10}) 
are obtained
\bea
\la_{+\fft32,0,+\fft12}\,\la_{0,+2,0}  &=& 
\kappa_{+\fft32,0,+\fft12} \,\la_{-\fft12,+2,+\fft12},
\nonu \\
\la_{0,+\fft32,+\fft12}\, \la_{+\fft32,+2,-\fft32} & = &
\kappa_{+\fft32,0,+\fft12} \,\la_{-\fft12,+2,+\fft12},
\qquad
\la_{+\fft32,-\fft12,+1}\,\la_{0,+\fft32,+\fft12}  = 
-\kappa_{+\fft32,0,+\fft12} \,\la_{-\fft12,+\fft32,+1}, 
\nonu \\
\la_{0,+1,+1}\, \la_{+\fft32,+\fft32,-1} & = &
\kappa_{+\fft32,0,+\fft12} \,\la_{-\fft12,+\fft32,+1}, 
\qquad
\la_{+\fft32,-1,+\fft32}\,\la_{0,+1,+1}  = 
-\kappa_{+\fft32,0,+\fft12} \, \la_{-\fft12,+1,+\fft32},
\nonu \\
\la_{0,+\fft12,+\fft32}\, \la_{+\fft32,+1,-\fft12} & = &
-\kappa_{+\fft32,0,+\fft12} \, \la_{-\fft12,+1,+\fft32},
\qquad
\la_{+\fft32,-\fft32,+2}\,\la_{0,+\fft12,+\fft32}  = 
-\kappa_{+\fft32,0,+\fft12} \,\la_{-\fft12,+\fft12,+2},
\nonu \\
\la_{0,0,+2}\, \la_{+\fft32,+\fft12,0} & = &
\kappa_{+\fft32,0,+\fft12} \,\la_{-\fft12,+\fft12,+2},
\qquad
\la_{+\fft32,-\fft12,+1}\,\la_{-\fft12,+2,+\fft12}  = 
\kappa_{+\fft32,-\fft12,+1} \,\la_{-1,+2,+1},
\nonu \\
\la_{-\fft12,+\fft32,+1}\, \la_{+\fft32,+2,-\fft32} & = &
-\kappa_{+\fft32,-\fft12,+1} \,\la_{-1,+2,+1},
\qquad
\la_{+\fft32,-1,+\fft32}\,\la_{-\fft12,+\fft32,+1}  = 
-\kappa_{+\fft32,-\fft12,+1} \,\la_{-1,+\fft32,+\fft32},
\nonu \\
\la_{-\fft12,+1,+\fft32}\, \la_{+\fft32,+\fft32,-1} & = &
\kappa_{+\fft32,-\fft12,+1} \,\la_{-1,+1,+2},
\qquad
\la_{+\fft32,-\fft32,+2}\,\la_{-\fft12,+1,+\fft32}  = 
-\kappa_{+\fft32,-\fft12,+1} \,\la_{-1,+1,+2},
\nonu \\
\la_{-\fft12,+\fft12,+2}\, \la_{+\fft32,+1,-\fft12} & = &
-\kappa_{+\fft32,-\fft12,+1} \,\la_{-1,+1,+2},
\qquad
\la_{+\fft32,-1,+\fft32}\,\la_{-1,+2,+1}  = 
\kappa_{+\fft32,-1,+\fft32} \,\la_{-\fft32,+2,+\fft32},
\nonu \\
\la_{-1,+\fft32,+\fft32}\, \la_{+\fft32,+2,-\fft32} & = &
\kappa_{+\fft32,-1,+\fft32} \,\la_{-\fft32,+2,+\fft32},
\qquad
\la_{+\fft32,-\fft32,+2}\,\la_{-1,+\fft32,+\fft32}  = 
-\kappa_{+\fft32,-1,+\fft32} \,\la_{-\fft32,+\fft32,+2},
\nonu \\
\la_{-1,+1,+2}\, \la_{+\fft32,+\fft32,-1} & = &
\kappa_{+\fft32,-1,+\fft32} \,\la_{-\fft32,+\fft32,+2},
\qquad
\la_{+\fft32,-\fft32,+2}\,\la_{-\fft32,+2,+\fft32}  = 
\kappa_{+\fft32,-\fft32,+2} \,\la_{-2,+2,+2},
\nonu \\
\la_{-\fft32,+\fft32,+2}\, \la_{+\fft32,+2,-\fft32} & = &
-\kappa_{+\fft32,-\fft32,+2} \,\la_{-2,+2,+2}.
\label{other10-lambda}
\eea

\subsection{The twenty-two relations of (\ref{Iden})
\label{22rel}}

The final twenty-two relations (\ref{Iden})
are given explicitly.

\subsubsection{The  fifteen relations}

The  fifteen relations
for (\ref{Iden}) corresponding to Table \ref{table17}
can be written as
\bea
&& \Bigg[
  \la_{+1,+1,0}\,\la_{+1,+2,-1} \,
{\mathscr D}_n^{h_2+1}(-1-h_3) \,
{\mathscr D}_m^{h_1+1}(-h_2-h_3)
\nonu \\
&& -   \la_{+1,+1,0}\, \la_{+1,+2,-1}\,
{\mathscr D}_m^{h_1+1}(-1-h_3)
{\mathscr D}_n^{h_2+1}(-h_1-h_3)
\nonu \\
&& -
\la_{0,+2,0} \, \kappa_{+1,+1,0} \,\bigg(
(h_2)m-(h_1)n \bigg) \,
{\mathscr D}_{m+n}^{h_1+h_2}(-1-h_3)\, \Bigg]\,
{\cal O}_{0}^{(h_1+h_2+h_3),ABCD}(z,\bar{z})=0,
\nonu \\
&& \Bigg[
  \la_{+1,+\fft12,\fft12}\,\la_{+1,+\fft32,-\fft12} \,
{\mathscr D}_n^{h_2+1}(-\tf12-h_3) \,
{\mathscr D}_m^{h_1+1}(\tf12-h_2-h_3)
\nonu \\
&& -   \la_{+1,+\fft12,+\fft12}\, \la_{+1,+\fft32,-\fft12}\,
{\mathscr D}_m^{h_1+1}(-\tf12-h_3)
{\mathscr D}_n^{h_2+1}(\tf12-h_1-h_3)
\nonu \\
&& -
\la_{0,+\fft32,+\fft12} \, \kappa_{+1,+1,0} \,\bigg(
(h_2)m-(h_1)n \bigg) \,
{\mathscr D}_{m+n}^{h_1+h_2}(-\tf12-h_3)\, \Bigg]\nonu \\
&& \times
\frac{1}{3!}\, \ep^{ABCDEFGH}\,
{\cal O}_{FGH,-\fft12}^{(h_1+h_2+h_3)}(z,\bar{z})=0,
\nonu \\
&& \Bigg[
  \la_{+1,0,+1}\,\la_{+1,+1,0} \,
{\mathscr D}_n^{h_2+1}(-h_3) \,
{\mathscr D}_m^{h_1+1}(1-h_2-h_3)
\nonu \\
&& -   \la_{+1,0,+1}\, \la_{+1,+1,0}\,
{\mathscr D}_m^{h_1+1}(-h_3)
{\mathscr D}_n^{h_2+1}(1-h_1-h_3)
\nonu \\
&& -
\la_{0,+1,+1} \, \kappa_{+1,+1,0} \,\bigg(
(h_2)m-(h_1)n \bigg) \,
{\mathscr D}_{m+n}^{h_1+h_2}(-h_3)\, \Bigg]\nonu \\
&& \times
\frac{1}{2!}\, \ep^{ABCDEFGH}\,
{\cal O}_{GH,-1}^{(h_1+h_2+h_3)}(z,\bar{z})=0,
\nonu \\
&& \Bigg[
 (-1) \la_{+1,-\fft12,+\fft32}\,\la_{+1,+\fft12,+\fft12} \,
{\mathscr D}_n^{h_2+1}(\tf12-h_3) \,
{\mathscr D}_m^{h_1+1}(\tf32-h_2-h_3)
\nonu \\
&& - (-1)  \la_{+1,-\fft12,+\fft32}\, \la_{+1,+\fft12,+\fft12}\,
{\mathscr D}_m^{h_1+1}(\tf12-h_3)
{\mathscr D}_n^{h_2+1}(\tf32-h_1-h_3)
\nonu \\
&& -
\la_{0,+\fft12,+\fft32} \, \kappa_{+1,+1,0} \,\bigg(
(h_2)m-(h_1)n \bigg) \,
{\mathscr D}_{m+n}^{h_1+h_2}(\tf12-h_3)\, \Bigg]\nonu \\
&& \times
\ep^{ABCDEFGH}\,
{\cal O}_{H,-\fft32}^{(h_1+h_2+h_3)}(z,\bar{z})=0,
\nonu \\
&& \Bigg[
  \la_{+1,-1,+2}\,\la_{+1,0,+1} \,
{\mathscr D}_n^{h_2+1}(1-h_3) \,
{\mathscr D}_m^{h_1+1}(2-h_2-h_3)
\nonu \\
&& -   \la_{+1,-1,+2}\, \la_{+1,0,+1}\,
{\mathscr D}_m^{h_1+1}(1-h_3)
{\mathscr D}_n^{h_2+1}(2-h_1-h_3)
\nonu \\
&& -
\la_{0,0,+2} \, \kappa_{+1,+1,0} \,\bigg(
(h_2)m-(h_1)n \bigg) \,
{\mathscr D}_{m+n}^{h_1+h_2}(1-h_3)\, \Bigg]\nonu \\
&& \times
\ep^{ABCDEFGH}\,
{\cal O}_{-2}^{(h_1+h_2+h_3)}(z,\bar{z})=0,
\nonu \\
&& \Bigg[
  \la_{+1,+\fft12,+\fft12}\,\la_{+\fft12,+2,-\fft12} \,
{\mathscr D}_n^{h_2+\fft12}(-1-h_3) \,
{\mathscr D}_m^{h_1+1}(\tf12-h_2-h_3)
\nonu \\
&& -   \la_{+\fft12,+1,+\fft12}\, \la_{+1,+2,-1}\,
{\mathscr D}_m^{h_1+1}(-1-h_3)
{\mathscr D}_n^{h_2+\fft12}(-h_1-h_3)
\nonu \\
&& -
\la_{-\fft12,+2,+\fft12} \, \kappa_{+1,+\fft12,+\fft12} \,\bigg(
(h_2-\tf12)m-(h_1)n \bigg) \,
{\mathscr D}_{m+n}^{h_1+h_2-\fft12}(-1-h_3)\, \Bigg]\nonu \\
&& \times
\frac{1}{3!}\,
\ep^{ABCDEFGH}\,
{\cal O}_{FGH,-\fft12}^{(h_1+h_2+h_3)}(z,\bar{z})=0,
\nonu \\
&& \Bigg[
  (-1) \la_{+1,0,+1}\,\la_{+\fft12,+\fft32,0} \,
{\mathscr D}_n^{h_2+\fft12}(-\tf12-h_3) \,
{\mathscr D}_m^{h_1+1}(1-h_2-h_3)
\nonu \\
&& -   \la_{+\fft12,+\fft12,+1}\, \la_{+1,+\fft32,-\fft12}\,
{\mathscr D}_m^{h_1+1}(-\tf12-h_3)
{\mathscr D}_n^{h_2+\fft12}(\tf12-h_1-h_3)
\nonu \\
&& -
\la_{-\fft12,+\fft32,+1} \, \kappa_{+1,+\fft12,+\fft12} \,\bigg(
(h_2-\tf12)m-(h_1)n \bigg) \,
{\mathscr D}_{m+n}^{h_1+h_2-\fft12}(-\tf12-h_3)\, \Bigg]\nonu \\
&& \times
\frac{1}{3!}\,
\ep^{ABCDEIGH}\, 3 \, \de^{F}_{[I}
{\cal O}_{GH],-1}^{(h_1+h_2+h_3)}(z,\bar{z})=0,
\nonu \\
&& \Bigg[
   \la_{+1,-\fft12,+\fft32}\,\la_{+\fft12,+1,+\fft12} \,
{\mathscr D}_n^{h_2+\fft12}(-h_3) \,
{\mathscr D}_m^{h_1+1}(\tf32-h_2-h_3)
\nonu \\
&& - (-1)  \la_{+\fft12,0,+\fft32}\, \la_{+1,+1,0}\,
{\mathscr D}_m^{h_1+1}(-h_3)
{\mathscr D}_n^{h_2+\fft12}(1-h_1-h_3)
\nonu \\
&& -
\la_{-\fft12,+1,+\fft32} \, \kappa_{+1,+\fft12,+\fft12} \,\bigg(
(h_2-\tf12)m-(h_1)n \bigg) \,
{\mathscr D}_{m+n}^{h_1+h_2-\fft12}(-h_3)\, \Bigg]\nonu \\
&& \times
\frac{1}{3!}\,
\ep^{ABCDEIJH}\, 3! \, \de^{F}_{[I}
{\cal O}_{J,-\fft32}^{(h_1+h_2+h_3)}(z,\bar{z}) \, \de^{G}_{H]}=0,
\nonu \\
&& \Bigg[
   \la_{+1,-1,+2}\,\la_{+\fft12,\fft12,+1} \,
{\mathscr D}_n^{h_2+\fft12}(\tf12-h_3) \,
{\mathscr D}_m^{h_1+1}(2-h_2-h_3)
\nonu \\
&& - (-1)  \la_{+\fft12,-\fft12,+2}\, \la_{+1,+\fft12,+\fft12}\,
{\mathscr D}_m^{h_1+1}(\tf12-h_3)
{\mathscr D}_n^{h_2+\fft12}(\tf32-h_1-h_3)
\nonu \\
&& -
\la_{-\fft12,+\fft12,+2} \, \kappa_{+1,+\fft12,+\fft12} \,\bigg(
(h_2-\tf12)m-(h_1)n \bigg) \,
{\mathscr D}_{m+n}^{h_1+h_2-\fft12}(\tf12-h_3)\, \Bigg]\nonu \\
&& \times
\frac{1}{3!}\,
\ep^{ABCDEIJK}\, \de^{FGH}_{IJK}
{\cal O}_{-2}^{(h_1+h_2+h_3)}(z,\bar{z})=0,
\nonu \\
&& \Bigg[
   \la_{+1,0,+1}\,\la_{0,+2,0} \,
{\mathscr D}_n^{h_2}(-1-h_3) \,
{\mathscr D}_m^{h_1+1}(1-h_2-h_3)
\nonu \\
&& -   \la_{0,+1,+1}\, \la_{+1,+2,-1}\,
{\mathscr D}_m^{h_1+1}(-1-h_3)
{\mathscr D}_n^{h_2}(-h_1-h_3)
\nonu \\
&& -
\la_{-1,+2,+1} \, \kappa_{+1,0,+1} \,\bigg(
(h_2-1)m-(h_1)n \bigg) \,
{\mathscr D}_{m+n}^{h_1+h_2-1}(-1-h_3)\, \Bigg]\nonu \\
&& \times
\frac{1}{2!}\,
\ep^{ABCDEFGH}\, 
{\cal O}_{GH,-1}^{(h_1+h_2+h_3)}(z,\bar{z})=0,
\nonu \\
&& \Bigg[
  (-1) \la_{+1,-\fft12,+\fft32}\,\la_{0,+\fft32,+\fft12} \,
{\mathscr D}_n^{h_2}(-\tf12-h_3) \,
{\mathscr D}_m^{h_1+1}(\tf32-h_2-h_3)
\nonu \\
&& -   \la_{0,+\fft12,+\fft32}\, \la_{+1,+\fft32,-\fft12}\,
{\mathscr D}_m^{h_1+1}(-\tf12-h_3)
{\mathscr D}_n^{h_2}(\tf12-h_1-h_3)
\nonu \\
&& -
\la_{-1,+\fft32,+\fft32} \, \kappa_{+1,0,+1} \,\bigg(
(h_2-1)m-(h_1)n \bigg) \,
{\mathscr D}_{m+n}^{h_1+h_2-1}(-\tf12-h_3)\, \Bigg]\nonu \\
&& \times
\frac{1}{2!}\,
\ep^{ABCDEFIH}\, 2!\, \de^{G}_{[I}
{\cal O}_{H],-\fft32}^{(h_1+h_2+h_3)}(z,\bar{z})=0,
\nonu \\
&& \Bigg[
 \la_{+1,-1,+2}\,\la_{0,+1,+1} \,
{\mathscr D}_n^{h_2}(-h_3) \,
{\mathscr D}_m^{h_1+1}(2-h_2-h_3)
\nonu \\
&& -   \la_{0,0,+2}\, \la_{+1,+1,0}\,
{\mathscr D}_m^{h_1+1}(-h_3)
{\mathscr D}_n^{h_2}(1-h_1-h_3)
\nonu \\
&& -
\la_{-1,+1,+2} \, \kappa_{+1,0,+1} \,\bigg(
(h_2-1)m-(h_1)n \bigg) \,
{\mathscr D}_{m+n}^{h_1+h_2-1}(-h_3)\, \Bigg]\nonu \\
&& \times
\frac{1}{2!}\,
\ep^{ABCDEFIJ}\, \de^{GH}_{IJ}
{\cal O}_{-2}^{(h_1+h_2+h_3)}(z,\bar{z})=0,
\nonu \\
&& \Bigg[
   \la_{+1,-\fft12,+\fft32}\,\la_{-\fft12,+2,+\fft12} \,
{\mathscr D}_n^{h_2-\fft12}(-1-h_3) \,
{\mathscr D}_m^{h_1+1}(\tf32-h_2-h_3)
\nonu \\
&& -   \la_{-\fft12,+1,+\fft32}\, \la_{+1,+2,-1}\,
{\mathscr D}_m^{h_1+1}(-1-h_3)
{\mathscr D}_n^{h_2-\fft12}(-h_1-h_3)
\nonu \\
&& -
\la_{-\fft32,+2,+\fft32} \, \kappa_{+1,-\fft12,+\fft32} \,\bigg(
(h_2-\tf32)m-(h_1)n \bigg) \,
{\mathscr D}_{m+n}^{h_1+h_2-\fft32}(-1-h_3)\, \Bigg]\nonu \\
&& \times
3!\,
\de^{A}_{[C}
{\cal O}_{D,-\fft32}^{(h_1+h_2+h_3)}(z,\bar{z}) \, \de^{B}_{E]}=0,
\nonu \\
&& \Bigg[
 (-1)  \la_{+1,-1,+2}\,\la_{-\fft12,+\fft32,+1} \,
{\mathscr D}_n^{h_2-\fft12}(-\tf12-h_3) \,
{\mathscr D}_m^{h_1+1}(2-h_2-h_3)
\nonu \\
&& - (-1)  \la_{-\fft12,+\fft12,+2}\, \la_{+1,+\fft32,-\fft12}\,
{\mathscr D}_m^{h_1+1}(-\tf12-h_3)
{\mathscr D}_n^{h_2-\fft12}(\tf12-h_1-h_3)
\nonu \\
&& -
\la_{-\fft32,+\fft32,+2} \, \kappa_{+1,-\fft12,+\fft32} \,\bigg(
(h_2-\tf32)m-(h_1)n \bigg) \,
{\mathscr D}_{m+n}^{h_1+h_2-\fft32}(-\tf12-h_3)\, \Bigg]\nonu \\
&& \times
3!\,
\de^{A}_{[C} \, \de^{F}_{D}
{\cal O}_{-2}^{(h_1+h_2+h_3)}(z,\bar{z}) \, \de^{B}_{E]}=0,
\nonu \\
&& \Bigg[
   \la_{+1,-1,+2}\,\la_{-1,+2,+1} \,
{\mathscr D}_n^{h_2-1}(-1-h_3) \,
{\mathscr D}_m^{h_1+1}(2-h_2-h_3)
\nonu \\
&& -   \la_{-1,+1,+2}\, \la_{+1,+2,-1}\,
{\mathscr D}_m^{h_1+1}(-1-h_3)
{\mathscr D}_n^{h_2-1}(-h_1-h_3)
\label{15}
\\
&& -
\la_{-2,+2,+2} \, \kappa_{+1,-1,+2} \,\bigg(
(h_2-2)m-(h_1)n \bigg) \,
{\mathscr D}_{m+n}^{h_1+h_2-2}(-1-h_3)\, \Bigg]\,
\de^{AB}_{CD} \, 
{\cal O}_{-2}^{(h_1+h_2+h_3)}(z,\bar{z})=0.
\nonu
\eea

\subsubsection{The seven relations }

Moreover, the final seven relations (\ref{Iden}) corresponding to Table
\ref{table18}
can be described as 
\bea
&& \Bigg[
   \la_{+\fft12,+\fft12,+1}\,\la_{+\fft12,+2,-\fft12} \,
{\mathscr D}_n^{h_2+\fft12}(-1-h_3) \,
{\mathscr D}_m^{h_1+\fft12}(\tf12-h_2-h_3)
\nonu \\
&& -   \la_{+\fft12,+\fft12,+1}\, \la_{+\fft12,+2,-\fft12}\,
{\mathscr D}_m^{h_1+\fft12}(-1-h_3)
{\mathscr D}_n^{h_2+\fft12}(\tf12-h_1-h_3)
\nonu \\
&& -
\la_{-1,+2,+1} \, \kappa_{+\fft12,+\fft12,+1} \,\bigg(
(h_2-\tf12)m-(h_1-\tf12)n \bigg) \,
{\mathscr D}_{m+n}^{h_1+h_2-1}(-1-h_3)\, \Bigg]\nonu \\
&& \times
\frac{1}{2!}\, \ep^{ABCDEFGH} \, 
{\cal O}_{GH,-1}^{(h_1+h_2+h_3)}(z,\bar{z})=0,
\nonu \\
&& \Bigg[
 (-1)  \la_{+\fft12,0,+\fft32}\,\la_{+\fft12,+\fft32,0} \,
{\mathscr D}_n^{h_2+\fft12}(-\tf12-h_3) \,
{\mathscr D}_m^{h_1+\fft12}(1-h_2-h_3)
\nonu \\
&& - (-1)  \la_{+\fft12,0,+\fft32}\, \la_{+\fft12,+\fft32,0}\,
{\mathscr D}_m^{h_1+\fft12}(-\tf12-h_3)
{\mathscr D}_n^{h_2+\fft12}(1-h_1-h_3)
\nonu \\
&& -
\la_{-1,+\fft32,+\fft32} \, \kappa_{+\fft12,+\fft12,+1} \,\bigg(
(h_2-\tf12)m-(h_1-\tf12)n \bigg) \,
{\mathscr D}_{m+n}^{h_1+h_2-1}(-\tf12-h_3)\, \Bigg]\nonu \\
&& \times
\frac{1}{2!}\, \ep^{ABCDEFIH} \, \de^{G}_{[I}\,
{\cal O}_{H],-\fft32}^{(h_1+h_2+h_3)}(z,\bar{z})=0,
\nonu \\
&& \Bigg[
 (-1)  \la_{+\fft12,-\fft12,+2}\,\la_{+\fft12,+1,\fft12} \,
{\mathscr D}_n^{h_2+\fft12}(-h_3) \,
{\mathscr D}_m^{h_1+\fft12}(\tf32-h_2-h_3)
\nonu \\
&& - (-1)  \la_{+\fft12,-\fft12,+2}\, \la_{+\fft12,+1,\fft12}\,
{\mathscr D}_m^{h_1+\fft12}(-h_3)
{\mathscr D}_n^{h_2+\fft12}(\tf32-h_1-h_3)
\nonu \\
&& -
\la_{-1,+1,+2} \, \kappa_{+\fft12,+\fft12,+1} \,\bigg(
(h_2-\tf12)m-(h_1-\tf12)n \bigg) \,
{\mathscr D}_{m+n}^{h_1+h_2-1}(-h_3)\, \Bigg]\nonu \\
&& \times
\frac{1}{2!}\, \ep^{ABCDEFIJ} \, \de^{GH}_{IJ}\,
{\cal O}_{-2}^{(h_1+h_2+h_3)}(z,\bar{z})=0,
\nonu \\
&& \Bigg[
   \la_{+\fft12,0,+\fft32}\,\la_{0,+2,0} \,
{\mathscr D}_n^{h_2}(-1-h_3) \,
{\mathscr D}_m^{h_1+\fft12}(1-h_2-h_3)
\nonu \\
&& -  \la_{0,+\fft12,+\fft32}\, \la_{+\fft12,+2,-\fft12}\,
{\mathscr D}_m^{h_1+\fft12}(-1-h_3)
{\mathscr D}_n^{h_2}(\tf12-h_1-h_3)
\nonu \\
&& -
\la_{-\fft32,+2,+\fft32} \, \kappa_{+\fft12,0,+\fft32} \,\bigg(
(h_2-1)m-(h_1-\tf12)n \bigg) \,
{\mathscr D}_{m+n}^{h_1+h_2-\fft32}(-1-h_3)\, \Bigg]\nonu \\
&& \times
 \ep^{ABCDEFGH} \,
{\cal O}_{H,-\fft32}^{(h_1+h_2+h_3)}(z,\bar{z})=0,
\nonu \\
&& \Bigg[
 (-1)  \la_{+\fft12,-\fft12,+2}\,\la_{0,+\fft32,+\fft12} \,
{\mathscr D}_n^{h_2}(-\tf12-h_3) \,
{\mathscr D}_m^{h_1+\fft12}(\tf32-h_2-h_3)
\nonu \\
&& - (-1) \la_{0,0,+2}\, \la_{+\fft12,+\fft32,0}\,
{\mathscr D}_m^{h_1+\fft12}(-\tf12-h_3)
{\mathscr D}_n^{h_2}(1-h_1-h_3)
\nonu \\
&& -
\la_{-\fft32,+\fft32,+2} \, \kappa_{+\fft12,0,+\fft32} \,\bigg(
(h_2-1)m-(h_1-\tf12)n \bigg) \,
{\mathscr D}_{m+n}^{h_1+h_2-\fft32}(-\tf12-h_3)\, \Bigg]\nonu \\
&& \times
 \ep^{ABCDEFGI} \,\de^{H}_{I}\,
{\cal O}_{-2}^{(h_1+h_2+h_3)}(z,\bar{z})=0,
\nonu \\
&& \Bigg[
   \la_{+\fft12,-\fft12,+2}\la_{-\fft12,+2,+\fft12} 
{\mathscr D}_n^{h_2-\fft12}(-1-h_3) 
{\mathscr D}_m^{h_1+\fft12}(\tf32-h_2-h_3)
\nonu \\
&& - (-1) \la_{-\fft12,+\fft12,+2} \la_{+\fft12,+2,-\fft12}
{\mathscr D}_m^{h_1+\fft12}(-1-h_3)
{\mathscr D}_n^{h_2-\fft12}(\tf12-h_1-h_3)
\nonu \\
&& -
\la_{-2,+2,+2}  \kappa_{+\fft12,-\fft12,+2} \big(
(h_2-\tf32)m-(h_1-\tf12)n \big) 
   {\mathscr D}_{m+n}^{h_1+h_2-2}(-1-h_3) \Bigg]
\de^{ABC}_{DEF}
{\cal O}_{-2}^{(h_1+h_2+h_3)}(z,\bar{z})\nonu \\
&&=0,
\nonu \\
&& \Bigg[
   \la_{0,0,+2}\,\la_{0,+2,0} \,
{\mathscr D}_n^{h_2}(-1-h_3) \,
{\mathscr D}_m^{h_1}(1-h_2-h_3)
\nonu \\
&& -  \la_{0,0,+2}\, \la_{0,+2,0}\,
{\mathscr D}_m^{h_1}(-1-h_3)
{\mathscr D}_n^{h_2}(1-h_1-h_3)
\nonu \\
&& -
\la_{-2,+2,+2} \, \kappa_{0,0,+2} \,\bigg(
(h_2-1)m-(h_1-1)n \bigg) \,
{\mathscr D}_{m+n}^{h_1+h_2-2}(-1-h_3)\, \Bigg]\nonu \\
&& \times
\ep^{ABCDEFGH}\,
{\cal O}_{-2}^{(h_1+h_2+h_3)}(z,\bar{z})=0.
\label{7}
\eea

\subsection{The relations for the couplings in the subsection \ref{22rel} }

Finally, in this subsection,
the previous twenty-two relations can be solved
explicitly and we obtain the relations
between the couplings.

\subsubsection{The  constraints for the fifteen relations}

The  constraints for the fifteen relations (\ref{15})
are given by
\bea
\la_{+1,+1,0}\,\la_{+1,+2,-1} & = &  \kappa_{+1,+1,0} \,\la_{0,+2,0},
\qquad
\la_{+1,+\fft12,+\fft12}\,\la_{+1,+\fft32,-\fft12}  = 
\kappa_{+1,+1,0} \, \la_{0,+\fft32,+\fft12},
\nonu \\
\la_{+1,0,+1}\,\la_{+1,+1,0} & = & \kappa_{+1,+1,0} \,\la_{0,+1,+1},
\qquad
\la_{+1,-\fft12,+\fft32}\,\la_{+1,+\fft12,+\fft12}  = 
-\kappa_{+1,+1,0} \,\la_{0,+\fft12,+\fft32},
\nonu \\
\la_{+1,-1,+2}\,\la_{+1,0,+1} & = & \kappa_{+1,+1,0} \,\la_{0,0,+2},
\qquad
\la_{+1,+\fft12,+\fft12}\,\la_{+\fft12,+2,-\fft12}  = 
\kappa_{+1,+\fft12,+\fft12} \, \la_{-\fft12,+2,+\fft12},
\nonu \\
\la_{+\fft12,+1,+\fft12}\, \la_{+1,+2,-1} & = &
\kappa_{+1,+\fft12,+\fft12} \, \la_{-\fft12,+2,+\fft12},
\qquad
\la_{+1,0,+1}\,\la_{+\fft12,+\fft32,0}  = 
-\kappa_{+1,+\fft12,+\fft12} \,\la_{-\fft12,+\fft32,+1},
\nonu \\
\la_{+\fft12,+\fft12,+1}\, \la_{+1,+\fft32,-\fft12} & = &
\kappa_{+1,+\fft12,+\fft12} \,\la_{-\fft12,+\fft32,+1},
\qquad
\la_{+1,-\fft12,+\fft32}\,\la_{+\fft12,+1,+\fft12}  =
\kappa_{+1,+\fft12,+\fft12} \,\la_{-\fft12,+1,+\fft32},
\nonu \\
\la_{+\fft12,0,+\fft32}\, \la_{+1,+1,0} & = &
-\kappa_{+1,+\fft12,+\fft12} \,\la_{-\fft12,+1,+\fft32},
\qquad
\la_{+1,-1,+2}\,\la_{+\fft12,+\fft12,+1}  = 
\kappa_{+1,+\fft12,+\fft12} \,\la_{-\fft12,+\fft12,+2},
\nonu \\
\la_{+\fft12,-\fft12,+2}\, \la_{+1,+\fft12,+\fft12}
& = & -\kappa_{+1,+\fft12,+\fft12} \,\la_{-\fft12,+\fft12,+2},
\qquad
\la_{+1,0,+1}\,\la_{0,+2,0}  =  \kappa_{+1,0,+1} \,\la_{-1,+2,+1},
\nonu \\
\la_{0,+1,+1}\, \la_{+1,+2,-1} & = &
\kappa_{+1,0,+1} \,\la_{-1,+2,+1},
\qquad
\la_{+1,-\fft12,+\fft32}\,\la_{0,+\fft32,+\fft12}  = 
-\kappa_{+1,0,+1} \,\la_{-1,+\fft32,+\fft32},
\nonu \\
\la_{0,+\fft12,+\fft32}\, \la_{+1,+\fft32,-\fft12} & = &
\kappa_{+1,0,+1} \,\la_{-1,+\fft32,+\fft32},
\qquad
\la_{+1,-1,+2}\,\la_{0,+1,+1}  = 
\kappa_{+1,0,+1} \,\la_{-1,+1,+2},
\nonu \\
\la_{0,0,+2}\, \la_{+1,+1,0} &= &
\kappa_{+1,0,+1} \,\la_{-1,+1,+2},
\qquad
\la_{+1,-\fft12,+\fft32}\,\la_{-\fft12,+2,+\fft12}  = 
\kappa_{+1,-\fft12,+\fft32} \, \la_{-\fft32,+2,+\fft32},
\nonu \\
\la_{-\fft12,+1,+\fft32}\, \la_{+1,+2,-1} & = &
\kappa_{+1,-\fft12,+\fft32} \, \la_{-\fft32,+2,+\fft32},
\qquad
\la_{+1,-1,+2}\,\la_{-\fft12,+\fft32,+1}  = 
-\kappa_{+1,-\fft12,+\fft32} \, \la_{-\fft32,+\fft32,+2},
\nonu \\
\la_{-\fft12,+\fft12,+2}\, \la_{+1,+\fft32,-\fft12} & = &
-\kappa_{+1,-\fft12,+\fft32} \, \la_{-\fft32,+\fft32,+2},
\qquad
\la_{+1,-1,+2}\,\la_{-1,+2,+1}  = 
\kappa_{+1,-1,+2} \,\la_{-2,+2,+2},
\nonu \\
\la_{-1,+1,+2}\, \la_{+1,+2,-1} & = &
\kappa_{+1,-1,+2} \,\la_{-2,+2,+2}.
\label{15-lambda}
\eea

\subsubsection{The  constraints for the seven relations}

Finally, the  constraints for the seven relations (\ref{7})
can be reduced to
\bea
\la_{+\fft12,+\fft12,+1}\,\la_{+\fft12,+2,-\fft12}  & = & 
\kappa_{+\fft12,+\fft12,+1} \,\la_{-1,+2,+1},
\nonu \\
\la_{+\fft12,0,+\fft32}\,\la_{+\fft12,+\fft32,0} & = &
-\kappa_{+\fft12,+\fft12,+1} \,\la_{-1,+\fft32,+\fft32},
\qquad
\la_{+\fft12,-\fft12,+2}\,\la_{+\fft12,+1,\fft12}  = 
-\kappa_{+\fft12,+\fft12,+1} \,\la_{-1,+1,+2},
\nonu \\
\la_{+\fft12,0,+\fft32}\,\la_{0,+2,0} & = &
\kappa_{+\fft12,0,+\fft32} \,\la_{-\fft32,+2,+\fft32},
\qquad
\la_{0,+\fft12,+\fft32}\, \la_{+\fft12,+2,-\fft12}  = 
\kappa_{+\fft12,0,+\fft32} \,\la_{-\fft32,+2,+\fft32},
\nonu \\
\la_{+\fft12,-\fft12,+2}\,\la_{0,+\fft32,+\fft12} & = &
-\kappa_{+\fft12,0,+\fft32} \,\la_{-\fft32,+\fft32,+2},
\qquad
\la_{0,0,+2}\, \la_{+\fft12,+\fft32,0}  = 
-\kappa_{+\fft12,0,+\fft32} \,\la_{-\fft32,+\fft32,+2},
\nonu \\
\la_{+\fft12,-\fft12,+2}\,\la_{-\fft12,+2,+\fft12} & = &
\kappa_{+\fft12,-\fft12,+2} \,\la_{-2,+2,+2},
\qquad
\la_{-\fft12,+\fft12,+2}\, \la_{+\fft12,+2,-\fft12}  = 
-\kappa_{+\fft12,-\fft12,+2} \,\la_{-2,+2,+2},
\nonu \\
\la_{0,0,+2}\,\la_{0,+2,0} & = &
\kappa_{0,0,+2} \,\la_{-2,+2,+2}.
\label{7-lambda}
\eea
By solving the constraint equations, (\ref{17-lambda}),
(\ref{18-lambda}),  (\ref{10-lambda}),
(\ref{other18-lambda}), (\ref{other10-lambda}),
(\ref{15-lambda}) and (\ref{7-lambda}),
one of the solutions is presented in (\ref{lambdacon})
\footnote{Note that these quadratic constraint relations
(the total number is $165$)  
between the couplings are nontrivial in the sense that
if there exists a single typo (in the signs) inside a
Mathematica \cite{mathematica} by mistake, there exists no solution.   
}.

\section{The OPEs
and (anti)commutators
of the ${\cal N}=8$ ${\rm \bar{w}_{1+\infty}}$ 
currents with the matter fields and the remaining twenty
(anti)commutators
between the celestial soft symmetry currents
in section \ref{holomorphic}}

\subsection{The twenty-five OPEs}

The twenty-five OPEs corresponding to (\ref{otherOPEwO})
with condition $s_1 \leq s_2$ are given by
\bea
&& \bar{\Phi}_{-2}^{(h_1)}(z_1,\bar{z}_1) \,
{\cal O}_{-2}^{(h_2)}(z_2,\bar{z}_2)
=  \frac{1}{\bar{z}_{12}}\, \frac{1}{2}\,
\sum_{m=0}^{2(h_1+2)-3}\,
\frac{(m+1)\, \Gamma(2(-1-h_2)+1)}{
  \Gamma(-2-2(h_1+h_2) +m)}
\nonu \\
&& \times  {z}_{12}^{2-2(h_1+2)+m}
\, {\pa}^m \,
{\cal O}_{-2}^{(h_1+h_2)}(z_2,\bar{z}_2)
+  \cdots,
\nonu \\
&& \bar{\Phi}_{-2}^{(h_1)}(z_1,\bar{z}_1) \,
{\cal O}_{A,-\frac{3}{2}}^{(h_2)}(z_2,\bar{z}_2)
=  \frac{1}{\bar{z}_{12}}\, \frac{1}{2}\,
\sum_{m=0}^{2(h_1+2)-3}\,
\frac{(m+1)\, \Gamma(2(-\frac{1}{2}-h_2)+1)}{
  \Gamma(-1-2(h_1+h_2) +m)}\nonu \\
&& \times  {z}_{12}^{2-2(h_1+2)+m}
\, {\pa}^m \,
{\cal O}_{A,-\frac{3}{2}}^{(h_1+h_2)}(z_2,\bar{z}_2)
+  \cdots,
\nonu \\
&& \bar{\Phi}_{-2}^{(h_1)}(z_1,\bar{z}_1) \,
{\cal O}_{AB,-1}^{(h_2)}(z_2,\bar{z}_2)
=  \frac{1}{\bar{z}_{12}}\, \frac{1}{2}\,
\sum_{m=0}^{2(h_1+2)-3}\,
\frac{(m+1)\, \Gamma(2(-h_2)+1)}{
  \Gamma(-2(h_1+h_2) +m)}\, {z}_{12}^{2-2(h_1+2)+m}
\nonu \\
&& \times  {\pa}^m \,
{\cal O}_{AB,-1}^{(h_1+h_2)}(z_2,\bar{z}_2)
+  \cdots,
\nonu \\
&&
\bar{\Phi}_{-2}^{(h_1)}(z_1,\bar{z}_1) \,
{\cal O}_{ABC,-\frac{1}{2}}^{(h_2)}(z_2,\bar{z}_2)
=  \frac{1}{\bar{z}_{12}}\, \frac{1}{2}\,
\sum_{m=0}^{2(h_1+2)-3}\,
\frac{(m+1)\, \Gamma(2(\frac{1}{2}-h_2)+1)}{
  \Gamma(1-2(h_1+h_2) +m)}\, {z}_{12}^{2-2(h_1+2)+m}
\nonu \\
&& \times  {\pa}^m \,
{\cal O}_{ABC,-\frac{1}{2}}^{(h_1+h_2)}(z_2,\bar{z}_2)
+  \cdots,
\nonu \\
&&
\bar{\Phi}_{-2}^{(h_1)}(z_1,\bar{z}_1) \,
{\cal O}_{0}^{(h_2),ABCD}(z_2,\bar{z}_2)
=  \frac{1}{\bar{z}_{12}}\, \frac{1}{2}\,
\sum_{m=0}^{2(h_1+2)-3}\,
\frac{(m+1)\, \Gamma(2(1-h_2)+1)}{
  \Gamma(2-2(h_1+h_2) +m)}\, {z}_{12}^{2-2(h_1+2)+m}
\nonu \\
&& \times  {\pa}^m \,
{\cal O}_{0}^{(h_1+h_2),ABCD}(z_2,\bar{z}_2)
+  \cdots,
\nonu \\
&&
\bar{\Phi}_{-2}^{(h_1)}(z_1,\bar{z}_1) \,
{\cal O}_{+\frac{1}{2}}^{(h_2),ABC}(z_2,\bar{z}_2)
=  \frac{1}{\bar{z}_{12}}\, \frac{1}{2}\,
\sum_{m=0}^{2(h_1+2)-3}\,
\frac{(m+1)\, \Gamma(2(\frac{3}{2}-h_2)+1)}{
  \Gamma(3-2(h_1+h_2) +m)}\, {z}_{12}^{2-2(h_1+2)+m} 
\nonu \\
&& \times  {\pa}^m \,
{\cal O}_{+\frac{1}{2}}^{(h_1+h_2),ABC}(z_2,\bar{z}_2)
+  \cdots,
\nonu \\
&& \bar{\Phi}_{-2}^{(h_1)}(z_1,\bar{z}_1) \,
{\cal O}_{+1}^{(h_2),AB}(z_2,\bar{z}_2)
=  \frac{1}{\bar{z}_{12}}\, \frac{1}{2}\,
\sum_{m=0}^{2(h_1+2)-3}\,
\frac{(m+1)\, \Gamma(2(2-h_2)+1)}{
  \Gamma(4-2(h_1+h_2) +m)}\, {z}_{12}^{2-2(h_1+2)+m}
\nonu \\
&& \times   {\pa}^m \,
{\cal O}_{+1}^{(h_1+h_2),AB}(z_2,\bar{z}_2)
+  \cdots,
\nonu \\
&&
\bar{\Phi}_{-2}^{(h_1)}(z_1,\bar{z}_1) \,
{\cal O}_{+\frac{3}{2}}^{(h_2),A}(z_2,\bar{z}_2)
=  \frac{1}{\bar{z}_{12}}\, \frac{1}{2}\,
\sum_{m=0}^{2(h_1+2)-3}\,
\frac{(m+1)\, \Gamma(2(\frac{5}{2}-h_2)+1)}{
  \Gamma(5-2(h_1+h_2) +m)}\, {z}_{12}^{2-2(h_1+2)+m}
\nonu \\
&& \times   {\pa}^m \,
{\cal O}_{+\frac{3}{2}}^{(h_1+h_2),A}(z_2,\bar{z}_2)
+  \cdots,
\nonu \\
&&
\bar{\Phi}_{-2}^{(h_1)}(z_1,\bar{z}_1) \,
{\cal O}_{+2}^{(h_2)}(z_2,\bar{z}_2)
=  \frac{1}{\bar{z}_{12}}\, \frac{1}{2}\,
\sum_{m=0}^{2(h_1+2)-3}\,
\frac{(m+1)\, \Gamma(2(3-h_2)+1)}{
  \Gamma(6-2(h_1+h_2) +m)}\, {z}_{12}^{2-2(h_1+2)+m}
\nonu \\
&& \times   {\pa}^m \,
{\cal O}_{+2}^{(h_1+h_2)}(z_2,\bar{z}_2)
+  \cdots,
\nonu \\
&&
\bar{\Phi}_{A,-\frac{3}{2}}^{(h_1)}(z_1,\bar{z}_1) 
{\cal O}_{B,-\frac{3}{2}}^{(h_2)}(z_2,\bar{z}_2)
=  \frac{1}{\bar{z}_{12}} \frac{1}{2}
\sum_{m=0}^{2(h_1+\frac{3}{2})-3}
\frac{(m+1) \Gamma(2(-\frac{1}{2}-h_2)+1)}{
\Gamma(-2(h_1+h_2) +m)}  {z}_{12}^{2-2(h_1+\frac{3}{2})+m}
\nonu \\
&& \times  {\pa}^m \,
{\cal O}_{AB,-1}^{(h_1+h_2)}(z_2,\bar{z}_2)
+  \cdots,
\nonu \\
&&
\bar{\Phi}_{A,-\frac{3}{2}}^{(h_1)}(z_1,\bar{z}_1) \,
{\cal O}_{BC,-1}^{(h_2)}(z_2,\bar{z}_2)
=  \frac{1}{\bar{z}_{12}}\, \frac{1}{2}\,
\sum_{m=0}^{2(h_1+\frac{3}{2})-3}\,
\frac{(m+1)\, \Gamma(2(-h_2)+1)}{
  \Gamma(1-2(h_1+h_2) +m)}\, {z}_{12}^{2-2(h_1+\frac{3}{2})+m}
\nonu \\
&& \times   {\pa}^m \,
{\cal O}_{ABC,-\frac{1}{2}}^{(h_1+h_2)}(z_2,\bar{z}_2)
+  \cdots,
\nonu \\
&&
\bar{\Phi}_{A,-\frac{3}{2}}^{(h_1)}(z_1,\bar{z}_1) 
{\cal O}_{BCD,-\frac{1}{2}}^{(h_2)}(z_2,\bar{z}_2)
=  \frac{1}{\bar{z}_{12}} \frac{1}{2}
\sum_{m=0}^{2(h_1+\frac{3}{2})-3}
\frac{(m+1) \Gamma(2(\frac{1}{2}-h_2)+1)}{
\Gamma(2-2(h_1+h_2) +m)} {z}_{12}^{2-2(h_1+\frac{3}{2})+m}
\nonu \\
&& \times  \frac{1}{4!} \ep_{ABCDEFGH}\, {\pa}^m \,
 {\cal O}_{0}^{(h_1+h_2),EFGH}(z_2,\bar{z}_2)
+  \cdots,
\nonu \\
&&
\bar{\Phi}_{A,-\frac{3}{2}}^{(h_1)}(z_1,\bar{z}_1) \,
{\cal O}_{0}^{(h_2),BCDE}(z_2,\bar{z}_2)
=  \frac{1}{\bar{z}_{12}}\, \frac{1}{2}\,
\sum_{m=0}^{2(h_1+\frac{3}{2})-3}\,
\frac{(m+1)\, \Gamma(2(1-h_2)+1)}{
  \Gamma(3-2(h_1+h_2) +m)}\, {z}_{12}^{2-2(h_1+\frac{3}{2})+m}
\nonu \\
&& \times  
\frac{1}{4!} \, \frac{1}{3!}\, \ep^{BCDEIJKL}\, \ep_{AFGHIJKL}
{\pa}^m \, {\cal O}_{+\frac{1}{2}}^{(h_1+h_2),FGH}(z_2,\bar{z}_2)
+  \cdots,
\nonu \\
&&
\bar{\Phi}_{A,-\frac{3}{2}}^{(h_1)}(z_1,\bar{z}_1) \,
{\cal O}_{+\frac{1}{2}}^{(h_2),BCD}(z_2,\bar{z}_2)
=  \frac{1}{\bar{z}_{12}}\, \frac{1}{2}\,
\sum_{m=0}^{2(h_1+\frac{3}{2})-3}\,
\frac{(m+1)\, \Gamma(2(\frac{3}{2}-h_2)+1)}{
  \Gamma(4-2(h_1+h_2) +m)}\, {z}_{12}^{2-2(h_1+\frac{3}{2})+m}
\nonu \\
&& \times   
3 \, \de^{[B}_{A} \, {\pa}^m \,
{\cal O}_{+1}^{(h_1+h_2),CD]}(z_2,\bar{z}_2)
+  \cdots,
\nonu \\
&&
\bar{\Phi}_{A,-\frac{3}{2}}^{(h_1)}(z_1,\bar{z}_1) \,
{\cal O}_{+1}^{(h_2),BC}(z_2,\bar{z}_2)
=  \frac{1}{\bar{z}_{12}}\, \frac{1}{2}\,
\sum_{m=0}^{2(h_1+\frac{3}{2})-3}\,
\frac{(m+1)\, \Gamma(2(2-h_2)+1)}{
  \Gamma(5-2(h_1+h_2) +m)}\, {z}_{12}^{2-2(h_1+\frac{3}{2})+m}
\nonu \\
&& \times  
2! \, \de^{[B}_{A} \, {\pa}^m \,
{\cal O}_{+\frac{3}{2}}^{(h_1+h_2),C]}(z_2,\bar{z}_2)
+  \cdots,
\nonu \\
&&
\bar{\Phi}_{A,-\frac{3}{2}}^{(h_1)}(z_1,\bar{z}_1) \,
{\cal O}_{+\frac{3}{2}}^{(h_2),B}(z_2,\bar{z}_2)
=  \frac{1}{\bar{z}_{12}}\, \frac{1}{2}\,
\sum_{m=0}^{2(h_1+\frac{3}{2})-3}\,
\frac{(m+1)\, \Gamma(2(\frac{5}{2}-h_2)+1)}{
  \Gamma(6-2(h_1+h_2) +m)}\, {z}_{12}^{2-2(h_1+\frac{3}{2})+m}
\nonu \\
&& \times 
\de^{B}_{A} \, {\pa}^m \,
{\cal O}_{+2}^{(h_1+h_2)}(z_2,\bar{z}_2)
+  \cdots,
\nonu \\
&&
\bar{\Phi}_{AB,-1}^{(h_1)}(z_1,\bar{z}_1) \,
{\cal O}_{CD,-1}^{(h_2)}(z_2,\bar{z}_2)
=  \frac{1}{\bar{z}_{12}}\, \frac{1}{2}\,
\sum_{m=0}^{2(h_1+1)-3}\,
\frac{(m+1)\, \Gamma(2(-h_2)+1)}{
  \Gamma(2-2(h_1+h_2) +m)}\, {z}_{12}^{2-2(h_1+1)+m}
\nonu \\
&& \times   \frac{1}{4!} \, \ep_{ABCDEFGH}\, {\pa}^m \,
{\cal O}_{0}^{(h_1+h_2),EFGH}(z_2,\bar{z}_2)
+  \cdots,
\nonu \\
&&
\bar{\Phi}_{AB,-1}^{(h_1)}(z_1,\bar{z}_1) \,
{\cal O}_{CDE,-\frac{1}{2}}^{(h_2)}(z_2,\bar{z}_2)
=  \frac{1}{\bar{z}_{12}}\, \frac{1}{2}\,
\sum_{m=0}^{2(h_1+1)-3}\,
\frac{(m+1)\, \Gamma(2(\frac{1}{2}-h_2)+1)}{
  \Gamma(3-2(h_1+h_2) +m)}\, {z}_{12}^{2-2(h_1+1)+m}\, 
\nonu \\
&& \times 
\frac{1}{3!}\, \ep_{ABCDEFGH}\, {\pa}^m \,
{\cal O}_{+\frac{1}{2}}^{(h_1+h_2),FGH}(z_2,\bar{z}_2)
+  \cdots,
\nonu \\
&&
\bar{\Phi}_{AB,-1}^{(h_1)}(z_1,\bar{z}_1) \,
{\cal O}_{0}^{(h_2),CDEF}(z_2,\bar{z}_2)
=  \frac{1}{\bar{z}_{12}}\, \frac{1}{2}\,
\sum_{m=0}^{2(h_1+1)-3}\,
\frac{(m+1)\, \Gamma(2(1-h_2)+1)}{
  \Gamma(4-2(h_1+h_2) +m)}\, {z}_{12}^{2-2(h_1+1)+m}
\nonu \\
&& \times
\frac{1}{4!}
\frac{1}{2!}\, \ep^{CDEFIJKL}\,\ep_{ABGHIJKL}\,  {\pa}^m \,
{\cal O}_{+1}^{(h_1+h_2),GH}(z_2,\bar{z}_2)
+  \cdots,
\nonu \\
&&
\bar{\Phi}_{AB,-1}^{(h_1)}(z_1,\bar{z}_1) \,
{\cal O}_{+\frac{1}{2}}^{(h_2),CDE}(z_2,\bar{z}_2)
=  \frac{1}{\bar{z}_{12}}\, \frac{1}{2}\,
\sum_{m=0}^{2(h_1+1)-3}\,
\frac{(m+1)\, \Gamma(2(\frac{3}{2}-h_2)+1)}{
  \Gamma(5-2(h_1+h_2) +m)}\, {z}_{12}^{2-2(h_1+1)+m}
\nonu \\
&& \times  3!\,
\de^{[C}_{A} \, {\pa}^m \,
{\cal O}_{+\frac{3}{2}}^{(h_1+h_2),D}(z_2,\bar{z}_2) \, \de^{E]}_{B}
+  \cdots,
\nonu \\
&&
\bar{\Phi}_{AB,-1}^{(h_1)}(z_1,\bar{z}_1) \,
{\cal O}_{+1}^{(h_2),CD}(z_2,\bar{z}_2)
=  \frac{1}{\bar{z}_{12}}\, \frac{1}{2}\,
\sum_{m=0}^{2(h_1+1)-3}\,
\frac{(m+1)\, \Gamma(2(2-h_2)+1)}{
  \Gamma(6-2(h_1+h_2) +m)}\, {z}_{12}^{2-2(h_1+1)+m}
\nonu \\
&& \times  
\de^{CD}_{AB} \, {\pa}^m \,
{\cal O}_{+2}^{(h_1+h_2)}(z_2,\bar{z}_2)
+  \cdots,
\nonu \\
&&
\bar{\Phi}_{ABC,-\frac{1}{2}}^{(h_1)}(z_1,\bar{z}_1) \,
{\cal O}_{DEF,-\frac{1}{2}}^{(h_2)}(z_2,\bar{z}_2)
=  \frac{1}{\bar{z}_{12}}\, \frac{1}{2}\,
\sum_{m=0}^{2(h_1+\frac{1}{2})-3}\,
\frac{(m+1)\, \Gamma(2(\frac{1}{2}-h_2)+1)}{
  \Gamma(4-2(h_1+h_2) +m)}\, {z}_{12}^{2-2(h_1+\frac{1}{2})+m}
\nonu \\
&& \times  
\frac{1}{2!} \, \ep_{ABCDEFGH} \, {\pa}^m \,
{\cal O}_{+1}^{(h_1+h_2),GH}(z_2,\bar{z}_2)
+  \cdots,
\nonu \\
&&
\bar{\Phi}_{ABC,-\frac{1}{2}}^{(h_1)}(z_1,\bar{z}_1) \,
{\cal O}_{0}^{(h_2),DEFG}(z_2,\bar{z}_2)
=  \frac{1}{\bar{z}_{12}}\, \frac{1}{2}\,
\sum_{m=0}^{2(h_1+\frac{1}{2})-3}\,
\frac{(m+1)\, \Gamma(2(1-h_2)+1)}{
  \Gamma(5-2(h_1+h_2) +m)}\, {z}_{12}^{2-2(h_1+\frac{1}{2})+m}
\nonu \\
&& \times
\frac{1}{4!}\,
\ep^{DEFGIJKL} \,\ep_{ABCHIJKL}\,  {\pa}^m \,
{\cal O}_{+\frac{3}{2}}^{(h_1+h_2),H}(z_2,\bar{z}_2)
+  \cdots,
\nonu \\
&&
\bar{\Phi}_{ABC,-\frac{1}{2}}^{(h_1)}(z_1,\bar{z}_1) \,
{\cal O}_{+\frac{1}{2}}^{(h_2),DEF}(z_2,\bar{z}_2)
=  \frac{1}{\bar{z}_{12}}\, \frac{1}{2}\,
\sum_{m=0}^{2(h_1+\frac{1}{2})-3}\,
\frac{(m+1)\, \Gamma(2(\frac{3}{2}-h_2)+1)}{
  \Gamma(6-2(h_1+h_2) +m)}\, {z}_{12}^{2-2(h_1+\frac{1}{2})+m} 
\nonu \\
&& \times 
\de^{DEF}_{ABC} \, {\pa}^m \,
{\cal O}_{+2}^{(h_1+h_2)}(z_2,\bar{z}_2)
+  \cdots,
\nonu \\
&&
\bar{\Phi}_{0}^{(h_1),ABCD}(z_1,\bar{z}_1) \,
{\cal O}_{0}^{(h_2),EFGH}(z_2,\bar{z}_2)
=  \frac{1}{\bar{z}_{12}}\, \frac{1}{2}\,
\sum_{m=0}^{2(h_1)-3}\,
\frac{(m+1)\, \Gamma(2(1-h_2)+1)}{
  \Gamma(6-2(h_1+h_2) +m)}\, {z}_{12}^{2-2(h_1)+m}
\nonu \\
&& \times   
\ep^{ABCDEFGH} \,{\pa}^m \,
{\cal O}_{+2}^{(h_1+h_2)}(z_2,\bar{z}_2)
+  \cdots.
\label{25OPEsholo}
\eea
Note that
the right-hand sides of the
twelfth, the thirteenth, the seventeenth, the
nineteenth, and the twenty-third
of (\ref{25OPEsholo}) are different from those
in (\ref{25OPEs}) due to the structure of $SU(8)$ indices on the
right-hand sides.
In Tables \ref{table9-1} and \ref{table10-1},
the left-weights and the helicities for the currents and
the left- and right-weights for the matter fields corresponding to
(\ref{20OPEsholo}) are described.

\begin{table}[tbp]
\centering
\renewcommand{\arraystretch}{1.7}
\begin{tabular}{|c|c||c|c|c|||c|c|c|c|| }
\hline
$ q$ &
$ {\rm {\bar w}^q}$ &
$ h$ &
$ \bar{h}$ &
$ {\cal O}_{h,\bar{h}}$ &
$ s_1$ &
$ h+2-q$ &
$ \bar{h}-s_1-q$ &
${\cal O}_{h+2-q,\bar{h}-s_1-q}$  
\\
\hline
\hline
${\scriptstyle h_1+\frac{3}{2}}$ &
$  \bar{\Phi}_{A,-\frac{3}{2}}^{(h_1)}$ &
${\scriptstyle -1-h_2}$ &
${\scriptstyle 1-h_2}$ &
$ {\cal O}_{-2}^{(h_2)}$ &
${\scriptstyle -\frac{3}{2}}$ &
${\scriptstyle -\frac{1}{2}-(h_1+h_2)}$ &
${\scriptstyle 1-(h_1+h_2)} $ &
$ {\cal O}_{A,-\frac{3}{2}}^{(h_1+h_2)}$
\\
\hline
${\scriptstyle h_1+1}$ &
$\bar{\Phi}_{AB,-1}^{(h_1)}$&
${\scriptstyle -1-h_2}$ &
${\scriptstyle 1-h_2}$ &
${\cal O}_{-2}^{(h_2)}$ &
${\scriptstyle -1}$ &
${\scriptstyle -(h_1+h_2)}$ &
${\scriptstyle 1-(h_1+h_2)} $ &
${\cal O}_{AB,-1}^{(h_1+h_2)}$
\\
\hline
${\scriptstyle h_1+\frac{1}{2}}$ &
$\bar{\Phi}_{ABC,-\frac{1}{2}}^{(h_1)}$&
${\scriptstyle -1-h_2}$ &
${\scriptstyle 1-h_2}$ &
${\cal O}_{-2}^{(h_2)}$ &
${\scriptstyle -\frac{1}{2}}$ &
${\scriptstyle \frac{1}{2}-(h_1+h_2)}$ &
${\scriptstyle 1-(h_1+h_2)} $ &
${\cal O}_{ABC,-\frac{1}{2}}^{(h_1+h_2)}$
\\
\hline
${\scriptstyle h_1}$ &
$\bar{\Phi}_{0}^{(h_1),ABCD}$&
${\scriptstyle -1-h_2}$ &
${\scriptstyle 1-h_2}$ &
${\cal O}_{-2}^{(h_2)}$ &
${\scriptstyle 0}$ &
${\scriptstyle 1-(h_1+h_2)}$ &
${\scriptstyle 1-(h_1+h_2)} $ &
${\cal O}_{0}^{(h_1+h_2),ABCD}$
\\
\hline
${\scriptstyle h_1-\frac{1}{2}}$ &
$\bar{\Phi}_{+\frac{1}{2}}^{(h_1),ABC}$&
${\scriptstyle -1-h_2}$ &
${\scriptstyle 1-h_2}$ &
${\cal O}_{-2}^{(h_2)}$ &
${\scriptstyle +\frac{1}{2}}$ &
${\scriptstyle \frac{3}{2}-(h_1+h_2)}$ &
${\scriptstyle 1-(h_1+h_2)} $ &
${\cal O}_{+\frac{1}{2}}^{(h_1+h_2),ABC}$
\\
\hline
${\scriptstyle h_1-1}$ &
$\bar{\Phi}_{+1}^{(h_1),AB}$&
${\scriptstyle -1-h_2}$ &
${\scriptstyle 1-h_2}$ &
${\cal O}_{-2}^{(h_2)}$ &
${\scriptstyle +1}$ &
${\scriptstyle 2-(h_1+h_2)}$ &
${\scriptstyle 1-(h_1+h_2)} $ &
${\cal O}_{+1}^{(h_1+h_2),AB}$
\\
\hline
${\scriptstyle h_1-\frac{3}{2}}$ &
$\bar{\Phi}_{+\frac{3}{2}}^{(h_1),A}$&
${\scriptstyle -1-h_2}$ &
${\scriptstyle 1-h_2}$ &
${\cal O}_{+2}^{(h_2)}$ &
${\scriptstyle +\frac{3}{2}}$ &
${\scriptstyle \frac{5}{2}-(h_1+h_2)}$ &
${\scriptstyle 1-(h_1+h_2)} $ &
${\cal O}_{+\frac{3}{2}}^{(h_1+h_2),A}$
\\
\hline
${\scriptstyle h_1-2}$ &
$\bar{\Phi}_{+2}^{(h_1)}$&
${\scriptstyle -1-h_2}$ &
${\scriptstyle 1-h_2}$ &
${\cal O}_{-2}^{(h_2)}$ &
${\scriptstyle +2}$ &
${\scriptstyle 3-(h_1+h_2)}$ &
${\scriptstyle 1-(h_1+h_2)} $ &
${\cal O}_{+2}^{(h_1+h_2)}$
\\
\hline
\hline
${\scriptstyle h_1+1}$ &
$\bar{\Phi}_{AB,-1}^{(h_1)}$&
${\scriptstyle -\frac{1}{2}-h_2}$ &
${\scriptstyle 1-h_2}$ &
${\cal O}_{C,-\frac{3}{2}}^{(h_2)}$ &
${\scriptstyle -1}$ &
${\scriptstyle \frac{1}{2}-(h_1+h_2)}$ &
${\scriptstyle 1-(h_1+h_2)} $ &
${\cal O}_{ABC,-\frac{1}{2}}^{(h_1+h_2)}$
\\
\hline
${\scriptstyle h_1+\frac{1}{2}}$ &
$\bar{\Phi}_{ABC,-\frac{1}{2}}^{(h_1)}$&
${\scriptstyle -\frac{1}{2}-h_2}$ &
${\scriptstyle 1-h_2}$ &
${\cal O}_{D,-\frac{3}{2}}^{(h_2)}$ &
${\scriptstyle -\frac{1}{2}}$ &
${\scriptstyle 1-(h_1+h_2)}$ &
${\scriptstyle 1-(h_1+h_2)} $ &
${\cal O}_{0}^{(h_1+h_2),ABCD}$
\\
\hline
${\scriptstyle h_1}$ &
$\bar{\Phi}_{0}^{(h_1),ABCD}$&
${\scriptstyle -\frac{1}{2}-h_2}$ &
${\scriptstyle 1-h_2}$ &
${\cal O}_{E,-\frac{3}{2}}^{(h_2)}$ &
${\scriptstyle 0}$ &
${\scriptstyle \frac{3}{2}-(h_1+h_2)}$ &
${\scriptstyle 1-(h_1+h_2)} $ &
${\cal O}_{+\frac{1}{2}}^{(h_1+h_2),FGH}$
\\
\hline
${\scriptstyle h_1-\frac{1}{2}}$ &
$\bar{\Phi}_{+\frac{1}{2}}^{(h_1),ABC}$&
${\scriptstyle -\frac{1}{2}-h_2}$ &
${\scriptstyle 1-h_2}$ &
${\cal O}_{D,-\frac{3}{2}}^{(h_2)}$ &
${\scriptstyle +\frac{1}{2}}$ &
${\scriptstyle 2-(h_1+h_2)}$ &
${\scriptstyle 1-(h_1+h_2)} $ &
${\cal O}_{+1}^{(h_1+h_2),AB}+\cdots$
\\
\hline
${\scriptstyle h_1-1}$ &
$\bar{\Phi}_{+1}^{(h_1),AB}$&
${\scriptstyle -\frac{1}{2}-h_2}$ &
${\scriptstyle 1-h_2}$ &
${\cal O}_{C,-\frac{3}{2}}^{(h_2)}$ &
${\scriptstyle +1}$ &
${\scriptstyle \frac{5}{2}-(h_1+h_2)}$ &
${\scriptstyle 1-(h_1+h_2)} $ &
${\cal O}_{+\frac{3}{2}}^{(h_1+h_2),A}+\cdots$
\\
\hline
${\scriptstyle h_1-\frac{3}{2}}$ &
$\bar{\Phi}_{+\frac{3}{2}}^{(h_1),A}$&
${\scriptstyle -\frac{1}{2}-h_2}$ &
${\scriptstyle 1-h_2}$ &
${\cal O}_{B,-\frac{3}{2}}^{(h_2)}$ &
${\scriptstyle +\frac{3}{2}}$ &
${\scriptstyle 3-(h_1+h_2)}$ &
${\scriptstyle 1-(h_1+h_2)} $ &
${\cal O}_{+2}^{(h_1+h_2)}$
\\
\hline
\hline
\end{tabular}
\caption{
The left-weights (and helicities) for the currents,
the left- and right-weights for the matter fields on the left-hand sides
and those on the right-hand sides in (\ref{20OPEs})
with exchanges for the helicities and left- and right-weights
are given for the 
first fourteen equations in Appendix \ref{appA}
(or (\ref{20OPEsholo}) directly)).
See also Tables \ref{table1-1} and \ref{table2-1}.}
\label{table9-1}
\end{table}

\begin{table}[tbp]
\centering
\renewcommand{\arraystretch}{1.7}
\begin{tabular}{|c|c||c|c|c|||c|c|c|c|| }
\hline
$ q$ &
$ {\rm {\bar w}^q}$ &
$ h$ &
$ \bar{h}$ &
$ {\cal O}_{h,\bar{h}}$ &
$ s_1$ &
$ h+2-q$ &
$ \bar{h}-s_1-q$ &
${\cal O}_{h+2-q,\bar{h}-s_1-q}$  
\\
\hline
\hline
${\scriptstyle h_1+\frac{1}{2}}$ &
$ \bar{ \Phi}_{ABC,-\frac{1}{2}}^{(h_1)}$ &
${\scriptstyle -h_2}$ &
${\scriptstyle 1-h_2}$ &
$ {\cal O}_{DE,-1}^{(h_2)}$ &
${\scriptstyle -\frac{1}{2}}$ &
${\scriptstyle \frac{3}{2}-(h_1+h_2)}$ &
${\scriptstyle 1-(h_1+h_2)} $ &
$ {\cal O}_{+\frac{1}{2}}^{(h_1+h_2),FGH}$
\\
\hline
${\scriptstyle h_1}$ &
$\bar{\Phi}_{0}^{(h_1),ABCD}$&
${\scriptstyle -h_2}$ &
${\scriptstyle 1-h_2}$ &
${\cal O}_{EF,-1}^{(h_2)}$ &
${\scriptstyle 0}$ &
${\scriptstyle 2-(h_1+h_2)}$ &
${\scriptstyle 1-(h_1+h_2)} $ &
${\cal O}_{+1}^{(h_1+h_2),GH}$
\\
\hline
${\scriptstyle h_1-\frac{1}{2}}$ &
$\bar{\Phi}_{+\frac{1}{2}}^{(h_1),ABC}$&
${\scriptstyle -h_2}$ &
${\scriptstyle 1-h_2}$ &
${\cal O}_{DE,-1}^{(h_2)}$ &
${\scriptstyle +\frac{1}{2}}$ &
${\scriptstyle \frac{5}{2}-(h_1+h_2)}$ &
${\scriptstyle 1-(h_1+h_2)} $ &
${\cal O}_{+\frac{3}{2}}^{(h_1+h_2),B}+\cdots$
\\
\hline
${\scriptstyle h_1-1}$ &
$\bar{\Phi}_{+1}^{(h_1),AB}$&
${\scriptstyle -h_2}$ &
${\scriptstyle 1-h_2}$ &
${\cal O}_{CD,-1}^{(h_2)}$ &
${\scriptstyle +1}$ &
${\scriptstyle 3-(h_1+h_2)}$ &
${\scriptstyle 1-(h_1+h_2)} $ &
${\cal O}_{+2}^{(h_1+h_2)}$
\\
\hline
\hline
${\scriptstyle h_1}$ &
$\bar{\Phi}_{0}^{(h_1),ABCD}$&
${\scriptstyle \frac{1}{2}-h_2}$ &
${\scriptstyle 1-h_2}$ &
${\cal O}_{EFG,-\frac{1}{2}}^{(h_2)}$ &
${\scriptstyle 0}$ &
${\scriptstyle \frac{5}{2}-(h_1+h_2)}$ &
${\scriptstyle 1-(h_1+h_2)} $ &
${\cal O}_{+\frac{3}{2}}^{(h_1+h_2),H}$
\\
\hline
${\scriptstyle h_1-\frac{1}{2}}$ &
$\bar{\Phi}_{+\frac{1}{2}}^{(h_1),ABC}$&
${\scriptstyle \frac{1}{2}-h_2}$ &
${\scriptstyle 1-h_2}$ &
${\cal O}_{DEF,-\frac{1}{2}}^{(h_2)}$ &
${\scriptstyle +\frac{1}{2}}$ &
${\scriptstyle 3-(h_1+h_2)}$ &
${\scriptstyle 1-(h_1+h_2)} $ &
${\cal O}_{+2}^{(h_1+h_2)}$
\\
\hline
\hline
\end{tabular}
\caption{This is a continuation of previous Table \ref{table9-1}.
We list the corresponding weights (and helicities) for  
the remaining six equations in Appendix \ref{appA} (or (\ref{20OPEsholo})).
See also Tables \ref{table1-1} and \ref{table2-1}.}
\label{table10-1}
\end{table}

\subsection{The twenty-five (anti)commutators}

The  twenty-five (anti)commutators corresponding to
(\ref{25OPEsholo})
are given by
\bea
\bigg[
\hat{\bar{\Phi}}_{-2,m}^{(h_1)},
{\cal O}_{-2}^{(h_2)}(z,\bar{z}) \bigg]  & = &
\bar{\mathscr D}_m^{h_1+2}(-1-h_2) \, {\cal O}_{-2}^{(h_1+h_2)}(z,\bar{z}),
\nonu \\
\bigg[
\hat{\bar{\Phi}}_{-2,m}^{(h_1)},
{\cal O}_{A,-\frac{3}{2}}^{(h_2)}(z,\bar{z}) \bigg]  & = &
\bar{\mathscr D}_m^{h_1+2}(-\tfrac{1}{2}-h_2) \,
{\cal O}_{A,-\frac{3}{2}}^{(h_1+h_2)}(z,\bar{z}),
\nonu \\
\bigg[
\hat{\bar{\Phi}}_{-2,m}^{(h_1)},
{\cal O}_{AB,-1}^{(h_2)}(z,\bar{z}) \bigg]  & = &
\bar{\mathscr D}_m^{h_1+2}(-h_2) \,
{\cal O}_{AB,-1}^{(h_1+h_2)}(z,\bar{z}),
\nonu \\
\bigg[
\hat{\bar{\Phi}}_{-2,m}^{(h_1)},
{\cal O}_{ABC,-\frac{1}{2}}^{(h_2)}(z,\bar{z}) \bigg]  & = &
\bar{\mathscr D}_m^{h_1+2}(\tfrac{1}{2}-h_2) \,
{\cal O}_{ABC,-\frac{1}{2}}^{(h_1+h_2)}(z,\bar{z}),
\nonu \\
\bigg[
\hat{\bar{\Phi}}_{-2,m}^{(h_1)},
{\cal O}_{0}^{(h_2),ABCD}(z,\bar{z}) \bigg]  & = &
\bar{\mathscr D}_m^{h_1+2}(1-h_2) \,
{\cal O}_{0}^{(h_1+h_2),ABCD}(z,\bar{z}),
\nonu \\
\bigg[
\hat{\bar{\Phi}}_{-2,m}^{(h_1)},
{\cal O}_{+\frac{1}{2}}^{(h_2),ABC}(z,\bar{z}) \bigg]  & = &
\bar{\mathscr D}_m^{h_1+2}(\tfrac{3}{2}-h_2) \,
{\cal O}_{+\frac{1}{2}}^{(h_1+h_2),ABC}(z,\bar{z}),
\nonu \\
\bigg[
\hat{\bar{\Phi}}_{-2,m}^{(h_1)},
{\cal O}_{+1}^{(h_2),AB}(z,\bar{z}) \bigg]  & = &
\bar{\mathscr D}_m^{h_1+2}(2-h_2) \,
{\cal O}_{+1}^{(h_1+h_2),AB}(z,\bar{z}),
\nonu \\
\bigg[
\hat{\bar{\Phi}}_{-2,m}^{(h_1)},
{\cal O}_{+\frac{3}{2}}^{(h_2),A}(z,\bar{z}) \bigg]  & = &
\bar{\mathscr D}_m^{h_1+2}(\tfrac{5}{2}-h_2) \,
{\cal O}_{+\frac{3}{2}}^{(h_1+h_2),A}(z,\bar{z}),
\nonu \\
\bigg[
\hat{\bar{\Phi}}_{-2,m}^{(h_1)},
{\cal O}_{+2}^{(h_2)}(z,\bar{z}) \bigg]  & = &
\bar{\mathscr D}_m^{h_1+2}(3-h_2) \,
{\cal O}_{+2}^{(h_1+h_2)}(z,\bar{z}),
\nonu \\
\bigg\{
\hat{\bar{\Phi}}_{A,-\frac{3}{2},m}^{(h_1)},
{\cal O}_{B,-\frac{3}{2}}^{(h_2)}(z,\bar{z}) \bigg\}  & = &
\bar{\mathscr D}_m^{h_1+\frac{3}{2}}(-\tfrac{1}{2}-h_2) \,
{\cal O}_{AB,-1}^{(h_1+h_2)}(z,\bar{z}),
\nonu \\
\bigg[
\hat{\bar{\Phi}}_{A,-\frac{3}{2},m}^{(h_1)},
{\cal O}_{BC,-1    }^{(h_2)}(z,\bar{z}) \bigg]  & = &
\bar{\mathscr D}_m^{h_1+\frac{3}{2}}(-h_2) \,
{\cal O}_{ABC,-\frac{1}{2}}^{(h_1+h_2)}(z,\bar{z}),
\nonu \\
\bigg\{
\hat{\bar{\Phi}}_{A,-\frac{3}{2},m}^{(h_1)},
{\cal O}_{BCD,-\frac{1}{2}}^{(h_2)}(z,\bar{z}) \bigg\}  & = &
\frac{1}{4!} \, \ep_{ABCDEFGH}\,
\bar{\mathscr D}_m^{h_1+\frac{3}{2}}(\tfrac{1}{2}-h_2) \,
{\cal O}_{0}^{(h_1+h_2),EFGH}(z,\bar{z}),
\nonu \\
\bigg[
\hat{\bar{\Phi}}_{A,-\frac{3}{2},m}^{(h_1)},
{\cal O}_{0}^{(h_2),BCDE}(z,\bar{z}) \bigg]  & = &
\frac{1}{4!} \, \frac{1}{3!} \, \ep^{BCDEIJKL} \, \ep_{AFGHIJKL}\nonu \\
&\times &
\bar{\mathscr D}_m^{h_1+\frac{3}{2}}(1-h_2) \,
{\cal O}_{+\frac{1}{2}}^{(h_1+h_2),FGH}(z,\bar{z}),
\nonu \\
\bigg\{
\hat{\bar{\Phi}}_{A,-\frac{3}{2},m}^{(h_1)},
{\cal O}_{+\frac{1}{2}}^{(h_2),BCD}(z,\bar{z}) \bigg\}  & = &
3 \, \de^{[B}_{A} \,
\bar {\mathscr D}_m^{h_1+\frac{3}{2}}(\tfrac{3}{2}-h_2) \,
{\cal O}_{+1}^{(h_1+h_2),CD]}(z,\bar{z}),
\nonu \\
\bigg[
\hat{\bar{\Phi}}_{A,-\frac{3}{2},m}^{(h_1)},
{\cal O}_{+1}^{(h_2),BC}(z,\bar{z}) \bigg]  & = &
2! \, \de^{[B}_{A} \,
\bar{\mathscr D}_m^{h_1+\frac{3}{2}}(2-h_2) \,
{\cal O}_{+\frac{3}{2}}^{(h_1+h_2),C]}(z,\bar{z}),
\nonu \\
\bigg\{
\hat{\bar{\Phi}}_{A,-\frac{3}{2},m}^{(h_1)},
{\cal O}_{+\frac{3}{2}}^{(h_2),B}(z,\bar{z}) \bigg\}  & = &
\de^B_{A} \, \bar{\mathscr D}_m^{h_1+\frac{3}{2}}(\tfrac{5}{2}-h_2) \,
{\cal O}_{+2}^{(h_1+h_2)}(z,\bar{z}),
\nonu \\
\bigg[
\hat{\bar{\Phi}}_{AB,-1,m}^{(h_1)},
{\cal O}_{CD,-1}^{(h_2)}(z,\bar{z}) \bigg]  & = &
\frac{1}{4!}\, \ep_{ABCDEFGH}\, 
\bar{\mathscr D}_m^{h_1+1}(-h_2) \,
{\cal O}_{0}^{(h_1+h_2),EFGH}(z,\bar{z}),
\nonu \\
\bigg[
\hat{\bar{\Phi}}_{AB,-1,m}^{(h_1)},
{\cal O}_{CDE,-\frac{1}{2}}^{(h_2)}(z,\bar{z}) \bigg]  & = &
\frac{1}{3!} \, \ep_{ABCDEFGH}\,
\bar{\mathscr D}_m^{h_1+1}(\tfrac{1}{2}-h_2) \,
{\cal O}_{+\frac{1}{2}}^{(h_1+h_2),FGH}(z,\bar{z}),
\nonu \\
\bigg[
\hat{\bar{\Phi}}_{AB,-1,m}^{(h_1)},
{\cal O}_{0}^{(h_2),CDEF}(z,\bar{z}) \bigg]  & = &
\frac{1}{4!} \, \frac{1}{2!} \, \ep^{CDEFIJKL}\, \ep_{ABGHIJKL}
\nonu \\
& \times & 
\bar{\mathscr D}_m^{h_1+1}(1-h_2) \,
{\cal O}_{+1}^{(h_1+h_2),GH}(z,\bar{z}),
\nonu \\
\bigg[
\hat{\bar{\Phi}}_{AB,-1,m}^{(h_1)},
{\cal O}_{+\frac{1}{2}}^{(h_2),CDE}(z,\bar{z}) \bigg]  & = &
3!\, \de^{[C}_{A} \,
\bar{\mathscr D}_m^{h_1+1}(\tf32-h_2) \,
{\cal O}_{+\frac{3}{2}}^{(h_1+h_2),D}(z,\bar{z}) \, \de^{E]}_{B},
\nonu \\
\bigg[
\hat{\bar{\Phi}}_{AB,-1,m}^{(h_1)},
{\cal O}_{+1}^{(h_2),CD}(z,\bar{z}) \bigg]  & = &
\de^{CD}_{AB} \,
\bar{\mathscr D}_m^{h_1+1}(2-h_2) \,
{\cal O}_{+2}^{(h_1+h_2)}(z,\bar{z}),
\nonu \\
\bigg\{
\hat{\bar{\Phi}}_{ABC,-\frac{1}{2},m}^{(h_1)},
{\cal O}_{DEF,-\frac{1}{2}}^{(h_2)}(z,\bar{z}) \bigg\}  & = &
\frac{1}{2!} \, \ep_{ABCDEFGH} \,
\bar{\mathscr D}_m^{h_1+\frac{1}{2}}(\tfrac{1}{2}-h_2) \,
{\cal O}_{+1}^{(h_1+h_2),GH}(z,\bar{z}),
\nonu \\
\bigg[
\hat{\bar{\Phi}}_{ABC,-\frac{1}{2},m}^{(h_1)},
{\cal O}_{0}^{(h_2),DEFG}(z,\bar{z}) \bigg]  & = &
\frac{1}{4!} \, \ep^{DEFGIJKL} \, \ep_{ABCHIJKL}\,
\bar{\mathscr D}_m^{h_1+\frac{1}{2}}(1-h_2) \,
{\cal O}_{+\frac{3}{2}}^{(h_1+h_2),H}(z,\bar{z}),
\nonu \\
\bigg\{
\hat{\bar{\Phi}}_{ABC,-\frac{1}{2},m}^{(h_1)},
{\cal O}_{+\frac{1}{2}}^{(h_2),DEF}(z,\bar{z}) \bigg\}  & = &
\de^{DEF}_{ABC} \,
\bar{\mathscr D}_m^{h_1+\frac{1}{2}}(\tf32-h_2) \,
{\cal O}_{+2}^{(h_1+h_2)}(z,\bar{z}),
\nonu \\
\bigg[
\hat{\bar{\Phi}}_{0,m}^{(h_1),ABCD},
{\cal O}_{0}^{(h_2),EFGH}(z,\bar{z}) \bigg]  & = &
\ep^{ABCDEFGH} \,
\bar{\mathscr D}_m^{h_1}(1-h_2) \,
{\cal O}_{+2}^{(h_1+h_2)}(z,\bar{z}).
\label{25anticommsection5}
\eea

\subsection{The remaining twenty OPEs}

The twenty OPEs corresponding to (\ref{otherOPEwO}) with $s_1 > s_2$
can be written as 
\bea
&&
\bar{\Phi}_{A,-\frac{3}{2}}^{(h_1)}(z_1,\bar{z}_1) \,
{\cal O}_{-2}^{(h_2)}(z_2,\bar{z}_2)
=  \frac{1}{\bar{z}_{12}}\, \frac{1}{2}\,
\sum_{m=0}^{2(h_1+\frac{3}{2})-3}\,
\frac{(m+1)\, \Gamma(2(-1-h_2)+1)}{
\Gamma(-1-2(h_1+h_2) +m)}\, {z}_{12}^{2-2(h_1+\frac{3}{2})+m}
\nonu \\
&& \times   
{\pa}^m \,
{\cal O}_{A,-\frac{3}{2}}^{(h_1+h_2)}(z_2,\bar{z}_2)
+  \cdots,
\nonu \\
&&
\bar{\Phi}_{AB,-1}^{(h_1)}(z_1,\bar{z}_1) \,
{\cal O}_{-2}^{(h_2)}(z_2,\bar{z}_2)
=  \frac{1}{\bar{z}_{12}}\, \frac{1}{2}\,
\sum_{m=0}^{2(h_1+1)-3}\,
\frac{(m+1)\, \Gamma(2(-1-h_2)+1)}{
  \Gamma(-2(h_1+h_2) +m)}\, {z}_{12}^{2-2(h_1+1)+m}
\nonu \\
&& \times   
{\pa}^m \,
{\cal O}_{AB,-1}^{(h_1+h_2)}(z_2,\bar{z}_2)
+  \cdots,
\nonu \\
&&
\bar{\Phi}_{ABC,-\frac{1}{2}}^{(h_1)}(z_1,\bar{z}_1) \,
{\cal O}_{-2}^{(h_2)}(z_2,\bar{z}_2)
=  \frac{1}{\bar{z}_{12}}\, \frac{1}{2}\,
\sum_{m=0}^{2(h_1+\frac{1}{2})-3}\,
\frac{(m+1)\, \Gamma(2(-1-h_2)+1)}{
  \Gamma(1-2(h_1+h_2) +m)}\, {z}_{12}^{2-2(h_1+\frac{1}{2})+m}
\nonu \\
&& \times   
{\pa}^m \,
{\cal O}_{ABC,-\frac{1}{2}}^{(h_1+h_2)}(z_2,\bar{z}_2)
+  \cdots,
\nonu \\
&&
\bar{\Phi}_{0}^{(h_1),ABCD}(z_1,\bar{z}_1) \,
{\cal O}_{-2}^{(h_2)}(z_2,\bar{z}_2)
=  \frac{1}{\bar{z}_{12}}\, \frac{1}{2}\,
\sum_{m=0}^{2(h_1)-3}\,
\frac{(m+1)\, \Gamma(2(-1-h_2)+1)}{
  \Gamma(2-2(h_1+h_2) +m)}\, {z}_{12}^{2-2(h_1)+m}
\nonu \\
&& \times   
{\pa}^m \,
{\cal O}_{0}^{(h_1+h_2),ABCD}(z_2,\bar{z}_2)
+  \cdots,
\nonu \\
&&
\bar{\Phi}_{+\frac{1}{2}}^{(h_1),ABC}(z_1,\bar{z}_1) \,
{\cal O}_{-2}^{(h_2)}(z_2,\bar{z}_2)
=  \frac{1}{\bar{z}_{12}}\, \frac{1}{2}\,
\sum_{m=0}^{2(h_1-\frac{1}{2})-3}\,
\frac{(m+1)\, \Gamma(2(-1-h_2)+1)}{
  \Gamma(3-2(h_1+h_2) +m)}\, {z}_{12}^{2-2(h_1-\frac{1}{2})+m}
\nonu \\
&& \times   
{\pa}^m \,
{\cal O}_{+\frac{1}{2}}^{(h_1+h_2),ABC}(z_2,\bar{z}_2)
+  \cdots,
\nonu \\
&&
\bar{\Phi}_{+1}^{(h_1),AB}(z_1,\bar{z}_1) \,
{\cal O}_{-2}^{(h_2)}(z_2,\bar{z}_2)
=  \frac{1}{\bar{z}_{12}}\, \frac{1}{2}\,
\sum_{m=0}^{2(h_1-1)-3}\,
\frac{(m+1)\, \Gamma(2(-1-h_2)+1)}{
  \Gamma(4-2(h_1+h_2) +m)}\, {z}_{12}^{2-2(h_1-1)+m}
\nonu \\
&& \times   
{\pa}^m \,
{\cal O}_{+1}^{(h_1+h_2),AB}(z_2,\bar{z}_2)
+  \cdots,
\nonu \\
&&
\bar{\Phi}_{+\frac{3}{2}}^{(h_1),A}(z_1,\bar{z}_1) \,
{\cal O}_{-2}^{(h_2)}(z_2,\bar{z}_2)
=  \frac{1}{\bar{z}_{12}}\, \frac{1}{2}\,
\sum_{m=0}^{2(h_1-\frac{3}{2})-3}\,
\frac{(m+1)\, \Gamma(2(-1-h_2)+1)}{
  \Gamma(5-2(h_1+h_2) +m)}\, {z}_{12}^{2-2(h_1-\frac{3}{2})+m}
\nonu \\
&& \times   
{\pa}^m \,
{\cal O}_{+\frac{3}{2}}^{(h_1+h_2),A}(z_2,\bar{z}_2)
+  \cdots,
\nonu \\
&&
\bar{\Phi}_{+2}^{(h_1)}(z_1,\bar{z}_1) \,
{\cal O}_{-2}^{(h_2)}(z_2,\bar{z}_2)
=  \frac{1}{\bar{z}_{12}}\, \frac{1}{2}\,
\sum_{m=0}^{2(h_1-2)-3}\,
\frac{(m+1)\, \Gamma(2(-1-h_2)+1)}{
  \Gamma(6-2(h_1+h_2) +m)}\, {z}_{12}^{2-2(h_1-2)+m}
\nonu \\
&& \times   
{\pa}^m \,
{\cal O}_{+2}^{(h_1+h_2)}(z_2,\bar{z}_2)
+  \cdots,
\nonu \\
&&
\bar{\Phi}_{AB,-1}^{(h_1)}(z_1,\bar{z}_1) \,
{\cal O}_{C,-\frac{3}{2}}^{(h_2)}(z_2,\bar{z}_2)
=  \frac{1}{\bar{z}_{12}}\, \frac{1}{2}\,
\sum_{m=0}^{2(h_1+1)-3}\,
\frac{(m+1)\, \Gamma(2(-\frac{1}{2}-h_2)+1)}{
  \Gamma(1-2(h_1+h_2) +m)}\, {z}_{12}^{2-2(h_1+1)+m}
\nonu \\
&& \times   
{\pa}^m \,
{\cal O}_{ABC,-\frac{1}{2}}^{(h_1+h_2)}(z_2,\bar{z}_2)
+  \cdots,
\nonu \\
&&
\bar{\Phi}_{ABC,-\frac{1}{2}}^{(h_1)}(z_1,\bar{z}_1) \,
{\cal O}_{D,-\frac{3}{2}}^{(h_2)}(z_2,\bar{z}_2)
= \frac{1}{\bar{z}_{12}}\, \frac{1}{2}\,
\sum_{m=0}^{2(h_1+\frac{1}{2})-3}\,
\frac{(m+1)\, \Gamma(2(-\frac{1}{2}-h_2)+1)}{
  \Gamma(2-2(h_1+h_2) +m)}\, {z}_{12}^{2-2(h_1+\frac{1}{2})+m}
\nonu \\
&& \times   \frac{1}{4!}\, \ep_{ABCDEFGH} \, 
{\pa}^m \,
{\cal O}_{0}^{(h_1+h_2),EFGH}(z_2,\bar{z}_2)
+  \cdots,
\nonu \\
&&
\bar{\Phi}_{0}^{(h_1),ABCD}(z_1,\bar{z}_1) \,
{\cal O}_{E,-\frac{3}{2}}^{(h_2)}(z_2,\bar{z}_2)
=  \frac{1}{\bar{z}_{12}}\, \frac{1}{2}\,
\sum_{m=0}^{2(h_1)-3}\,
\frac{(m+1)\, \Gamma(2(-\frac{1}{2}-h_2)+1)}{
  \Gamma(3-2(h_1+h_2) +m)}\, {z}_{12}^{2-2(h_1)+m}
\nonu \\
&& \times
\frac{1}{4!}\,
\frac{1}{3!}\, \ep^{ABCDIJKL}\, \ep_{EFGHIJKL}\,
{\pa}^m \,
{\cal O}_{+\frac{1}{2}}^{(h_1+h_2),FGH}(z_2,\bar{z}_2)
+  \cdots,
\nonu \\
&&
\bar{\Phi}_{+\frac{1}{2}}^{(h_1),ABC}(z_1,\bar{z}_1) \,
{\cal O}_{D,-\frac{3}{2}}^{(h_2)}(z_2,\bar{z}_2)
=  -\frac{1}{\bar{z}_{12}}\, \frac{1}{2}\,
\sum_{m=0}^{2(h_1-\frac{1}{2})-3}\,
\frac{(m+1)\, \Gamma(2(-\frac{1}{2}-h_2)+1)}{
  \Gamma(4-2(h_1+h_2) +m)}\nonu \\
&& \times  {z}_{12}^{2-2(h_1-\frac{1}{2})+m}
\,
3\,
{\pa}^m \,
{\cal O}_{+1}^{(h_1+h_2),[AB}(z_2,\bar{z}_2) \, \de^{C]}_{D}
+  \cdots,
\nonu \\
&&
\bar{\Phi}_{+1}^{(h_1),AB}(z_1,\bar{z}_1) \,
{\cal O}_{C,-\frac{3}{2}}^{(h_2)}(z_2,\bar{z}_2)
=  \frac{1}{\bar{z}_{12}}\, \frac{1}{2}\,
\sum_{m=0}^{2(h_1-1)-3}\,
\frac{(m+1)\, \Gamma(2(-\frac{1}{2}-h_2)+1)}{
  \Gamma(5-2(h_1+h_2) +m)}\, {z}_{12}^{2-2(h_1-1)+m}
\nonu \\
&& \times 
2\,
{\pa}^m \, \de^{[A}_{C}
{\cal O}_{+\frac{3}{2}}^{(h_1+h_2),B]}(z_2,\bar{z}_2)
+  \cdots,
\nonu \\
&&
\bar{\Phi}_{+\frac{3}{2}}^{(h_1),A}(z_1,\bar{z}_1) \,
{\cal O}_{B,-\frac{3}{2}}^{(h_2)}(z_2,\bar{z}_2)
=  -\frac{1}{\bar{z}_{12}}\, \frac{1}{2}\,
\sum_{m=0}^{2(h_1-\frac{3}{2})-3}\,
\frac{(m+1)\, \Gamma(2(-\frac{1}{2}-h_2)+1)}{
  \Gamma(6-2(h_1+h_2) +m)}\nonu \\
&& \times {z}_{12}^{2-2(h_1-\frac{3}{2})+m}
\,
\de^A_{B} \,
{\pa}^m \, 
{\cal O}_{+2}^{(h_1+h_2)}(z_2,\bar{z}_2)
+  \cdots,
\nonu \\
&&
\bar{\Phi}_{ABC,-\frac{1}{2}}^{(h_1)}(z_1,\bar{z}_1) \,
{\cal O}_{DE,-1}^{(h_2)}(z_2,\bar{z}_2)
=  \frac{1}{\bar{z}_{12}}\, \frac{1}{2}\,
\sum_{m=0}^{2(h_1+\frac{1}{2})-3}\,
\frac{(m+1)\, \Gamma(2(-h_2)+1)}{
  \Gamma(3-2(h_1+h_2) +m)}\, {z}_{12}^{2-2(h_1+\frac{1}{2})+m}
\nonu \\
&& \times 
\frac{1}{3!} \,
\ep_{ABCDEFGH} \,
{\pa}^m \, 
{\cal O}_{+\frac{1}{2}}^{(h_1+h_2),FGH}(z_2,\bar{z}_2)
+  \cdots,
\nonu \\
&&
\bar{\Phi}_{0}^{(h_1),ABCD}(z_1,\bar{z}_1) \,
{\cal O}_{EF,-1}^{(h_2)}(z_2,\bar{z}_2)
=  \frac{1}{\bar{z}_{12}}\, \frac{1}{2}\,
\sum_{m=0}^{2(h_1)-3}\,
\frac{(m+1)\, \Gamma(2(-h_2)+1)}{
  \Gamma(4-2(h_1+h_2) +m)}\, {z}_{12}^{2-2(h_1)+m}
\nonu \\
&& \times
\frac{1}{4!}\,
\frac{1}{2!} \,
\ep^{ABCDIJKL} \, \ep_{EFGHIJKL}\,
{\pa}^m \, 
{\cal O}_{+1}^{(h_1+h_2),GH}(z_2,\bar{z}_2)
+  \cdots,
\nonu \\
&&
\bar{\Phi}_{+\frac{1}{2}}^{(h_1),ABC}(z_1,\bar{z}_1) \,
{\cal O}_{DE,-1}^{(h_2)}(z_2,\bar{z}_2)
=  \frac{1}{\bar{z}_{12}}\, \frac{1}{2}\,
\sum_{m=0}^{2(h_1-\frac{1}{2})-3}\,
\frac{(m+1)\, \Gamma(2(-h_2)+1)}{
  \Gamma(5-2(h_1+h_2) +m)}\, {z}_{12}^{2-2(h_1-\frac{1}{2})+m}
\nonu \\
&& \times 
3! \,
\de^{[A}_{D} \,
{\pa}^m \, 
{\cal O}_{+\frac{3}{2}}^{(h_1+h_2),B}(z_2,\bar{z}_2) \, \de^{C]}_{E}
+  \cdots,
\nonu \\
&&
\bar{\Phi}_{+1}^{(h_1),AB}(z_1,\bar{z}_1) \,
{\cal O}_{CD,-1}^{(h_2)}(z_2,\bar{z}_2)
=  \frac{1}{\bar{z}_{12}}\, \frac{1}{2}\,
\sum_{m=0}^{2(h_1-1)-3}\,
\frac{(m+1)\, \Gamma(2(-h_2)+1)}{
  \Gamma(6-2(h_1+h_2) +m)}\, {z}_{12}^{2-2(h_1-1)+m}
\nonu \\
&& \times 
\de^{AB}_{CD} \,
{\pa}^m \, 
{\cal O}_{+2}^{(h_1+h_2)}(z_2,\bar{z}_2)
+  \cdots,
\nonu \\
&&
\bar{\Phi}_{0}^{(h_1),ABCD}(z_1,\bar{z}_1) \,
{\cal O}_{EFG,-\frac{1}{2}}^{(h_2)}(z_2,\bar{z}_2)
=  \frac{1}{\bar{z}_{12}}\, \frac{1}{2}\,
\sum_{m=0}^{2(h_1)-3}\,
\frac{(m+1)\, \Gamma(2(\frac{1}{2}-h_2)+1)}{
  \Gamma(5-2(h_1+h_2) +m)}\, {z}_{12}^{2-2(h_1)+m}
\nonu \\
&& \times
\frac{1}{4!}\,
\ep^{ABCDIJKL} \, \ep_{EFGHIJKL}\,
{\pa}^m \, 
{\cal O}_{+\frac{3}{2}}^{(h_1+h_2),H}(z_2,\bar{z}_2)
+  \cdots,
\nonu \\
&&
\bar{\Phi}_{+\frac{1}{2}}^{(h_1),ABC}(z_1,\bar{z}_1) \,
{\cal O}_{DEF,-\frac{1}{2}}^{(h_2)}(z_2,\bar{z}_2)
=  -\frac{1}{\bar{z}_{12}}\, \frac{1}{2}\,
\sum_{m=0}^{2(h_1-\frac{1}{2})-3}\,
\frac{(m+1)\, \Gamma(2(\frac{1}{2}-h_2)+1)}{
  \Gamma(6-2(h_1+h_2) +m)}\, {z}_{12}^{2-2(h_1-\frac{1}{2})+m}
\nonu \\
&& 
\de^{ABC}_{DEF} \,
{\pa}^m \, 
{\cal O}_{+2}^{(h_1+h_2)}(z_2,\bar{z}_2)
+  \cdots.
\label{20OPEsholo}
\eea
The right-hand sides of the
tenth, the eleventh, the sixteenth, and the
nineteenth relations
of (\ref{20OPEsholo}) are different from those
in (\ref{20OPEs}).

\subsection{The remaining twenty (anti)commutators}

The twenty (anti)commutators corresponding to
(\ref{20OPEsholo})
are given by
\bea
\bigg[
\hat{\bar{\Phi}}_{A,-\frac{3}{2},m}^{(h_1)},
{\cal O}_{-2}^{(h_2)}(z,\bar{z}) \bigg]  & = &
\bar{\mathscr D}_m^{h_1+\frac{3}{2}}(-1-h_2) \,
{\cal O}_{A,-\frac{3}{2}}^{(h_1+h_2)}(z,\bar{z}),
\nonu \\
\bigg[
\hat{\bar{\Phi}}_{AB,-1,m}^{(h_1)},
{\cal O}_{-2}^{(h_2)}(z,\bar{z}) \bigg]  & = &
\bar{\mathscr D}_m^{h_1+1}(-1-h_2) \,
{\cal O}_{AB,-1}^{(h_1+h_2)}(z,\bar{z}),
\nonu \\
\bigg[
\hat{\bar{\Phi}}_{ABC,-\frac{1}{2},m}^{(h_1)},
{\cal O}_{-2}^{(h_2)}(z,\bar{z}) \bigg]  & = &
\bar{\mathscr D}_m^{h_1+\frac{1}{2}}(-1-h_2) \,
{\cal O}_{ABC,-\frac{1}{2}}^{(h_1+h_2)}(z,\bar{z}),
\nonu \\
\bigg[
\hat{\bar{\Phi}}_{0,m}^{(h_1),ABCD},
{\cal O}_{-2}^{(h_2)}(z,\bar{z}) \bigg]  & = &
\bar{\mathscr D}_m^{h_1}(-1-h_2) \,
{\cal O}_{0}^{(h_1+h_2),ABCD}(z,\bar{z}),
\nonu \\
\bigg[
\hat{\bar{\Phi}}_{+\frac{1}{2},m}^{(h_1),ABC},
{\cal O}_{-2}^{(h_2)}(z,\bar{z}) \bigg]  & = &
\bar{\mathscr D}_m^{h_1-\frac{1}{2}}(-1-h_2) \,
{\cal O}_{+\frac{1}{2}}^{(h_1+h_2),ABC}(z,\bar{z}),
\nonu \\
\bigg[
\hat{\bar{\Phi}}_{+1,m}^{(h_1),AB},
{\cal O}_{-2}^{(h_2)}(z,\bar{z}) \bigg]  & = &
\bar{\mathscr D}_m^{h_1-1}(-1-h_2) \,
{\cal O}_{+1}^{(h_1+h_2),AB}(z,\bar{z}),
\nonu \\
\bigg[
\hat{\bar{\Phi}}_{+\frac{3}{2},m}^{(h_1),A},
{\cal O}_{-2}^{(h_2)}(z,\bar{z}) \bigg]  & = &
\bar{\mathscr D}_m^{h_1-\frac{3}{2}}(-1-h_2) \,
{\cal O}_{+\frac{3}{2}}^{(h_1+h_2),A}(z,\bar{z}),
\nonu \\
\bigg[
\hat{\bar{\Phi}}_{+2,m}^{(h_1)},
{\cal O}_{-2}^{(h_2)}(z,\bar{z}) \bigg]  & = &
\bar{\mathscr D}_m^{h_1-2}(-1-h_2) \,
{\cal O}_{+2}^{(h_1+h_2)}(z,\bar{z}),
\nonu \\
\bigg[
\hat{\bar{\Phi}}_{AB,-1,m}^{(h_1)},
{\cal O}_{C,-\frac{3}{2}}^{(h_2)}(z,\bar{z}) \bigg]  & = &
\bar{\mathscr D}_m^{h_1+1}(-\tfrac{1}{2}-h_2) \,
{\cal O}_{ABC,-\frac{1}{2}}^{(h_1+h_2)}(z,\bar{z}),
\nonu \\
\bigg\{
\hat{\bar{\Phi}}_{ABC,-\frac{1}{2},m}^{(h_1)},
{\cal O}_{D,-\frac{3}{2}}^{(h_2)}(z,\bar{z}) \bigg\}  & = &
\frac{1}{4!}\, \ep_{ABCDEFGH}\,
\bar{\mathscr D}_m^{h_1+\frac{1}{2}}(-\tfrac{1}{2}-h_2) \,
{\cal O}_{0}^{(h_1+h_2),EFGH}(z,\bar{z}),
\nonu \\
\bigg[
\hat{\bar{\Phi}}_{0,m}^{(h_1),ABCD},
{\cal O}_{E,-\frac{3}{2}}^{(h_2)}(z,\bar{z}) \bigg]  & = &
\frac{1}{4!}\, \frac{1}{3!} \, \ep^{ABCDIJKL} \, \ep_{EFGHIJKL}
\nonu \\
& \times & 
\bar{\mathscr D}_m^{h_1}(-\tfrac{1}{2}-h_2) \,
{\cal O}_{+\frac{1}{2}}^{(h_1+h_2),FGH}(z,\bar{z}),
\nonu \\
\bigg\{
\hat{\bar{\Phi}}_{+\frac{1}{2},m}^{(h_1),ABC},
{\cal O}_{D,-\frac{3}{2}}^{(h_2)}(z,\bar{z}) \bigg\}  & = &
-3 \,
\bar{\mathscr D}_m^{h_1-\frac{1}{2}}(-\tfrac{1}{2}-h_2) \, 
{\cal O}_{+1}^{(h_1+h_2),[AB}(z,\bar{z}) \, \de^{C]}_{D},
\nonu \\
\bigg[
\hat{\bar{\Phi}}_{+1,m}^{(h_1),AB},
{\cal O}_{C,-\frac{3}{2}}^{(h_2)}(z,\bar{z}) \bigg]  & = &
2 \,
\bar{\mathscr D}_m^{h_1-1}(-\tfrac{1}{2}-h_2) \,
\de^{[A}_{C}
{\cal O}_{+\frac{3}{2}}^{(h_1+h_2),B]}(z,\bar{z}),
\nonu \\
\bigg\{
\hat{\bar{\Phi}}_{+\frac{3}{2},m}^{(h_1),A},
{\cal O}_{B,-\frac{3}{2}}^{(h_2)}(z,\bar{z}) \bigg \}  & = &
-\bar{\mathscr D}_m^{h_1-\frac{3}{2}}(-\tfrac{1}{2}-h_2) \,
\de^{A}_{B}
{\cal O}_{+2}^{(h_1+h_2)}(z,\bar{z}),
\nonu \\
\bigg[
\hat{\bar{\Phi}}_{ABC,-\frac{1}{2},m}^{(h_1)},
{\cal O}_{DE,-1}^{(h_2)}(z,\bar{z}) \bigg]  & = &
\frac{1}{3!} \, \ep_{ABCDEFGH} \,
\bar{\mathscr D}_m^{h_1+\frac{1}{2}}(-h_2) \,
{\cal O}_{+\frac{1}{2}}^{(h_1+h_2),FGH}(z,\bar{z}),
\nonu \\
\bigg[
\hat{\bar{\Phi}}_{0,m}^{(h_1),ABCD},
{\cal O}_{EF,-1}^{(h_2)}(z,\bar{z}) \bigg]  & = &
\frac{1}{4!}\,
\frac{1}{2!} \, \ep^{ABCDIJKL} \, \ep_{EFGHIJKL}\,
\, \bar{\mathscr D}_m^{h_1}(-h_2) \,
{\cal O}_{+1}^{(h_1+h_2),GH}(z,\bar{z}),
\nonu \\
\bigg[
\hat{\bar{\Phi}}_{+\frac{1}{2},m}^{(h_1),ABC},
{\cal O}_{DE,-1}^{(h_2)}(z,\bar{z}) \bigg]  & = &
3!\,
\de^{[A}_{D} \,
\bar{\mathscr D}_m^{h_1-\frac{1}{2}}(-h_2) \,
{\cal O}_{+\frac{3}{2}}^{(h_1+h_2),B}(z,\bar{z}) \, \de^{C]}_{E},
\nonu \\
\bigg[
\hat{\bar{\Phi}}_{+1,m}^{(h_1),AB},
{\cal O}_{CD,-1}^{(h_2)}(z,\bar{z}) \bigg]  & = &
\de^{AB}_{CD} \, \bar{\mathscr D}_m^{h_1-1}(-h_2) \,
{\cal O}_{+2}^{(h_1+h_2)}(z,\bar{z}),
\nonu \\
\bigg[
\hat{\bar{\Phi}}_{0,m}^{(h_1),ABCD},
{\cal O}_{EFG,-\frac{1}{2}}^{(h_2)}(z,\bar{z}) \bigg]  & = &
\frac{1}{4!} \, \ep^{ABCDIJKL} \, \ep_{EFGHIJKL}\,
\bar{\mathscr D}_m^{h_1}(\tfrac{1}{2}-h_2) \,
{\cal O}_{+\frac{3}{2}}^{(h_1+h_2),H}(z,\bar{z}),
\nonu \\
\bigg\{
\hat{\bar{\Phi}}_{+\frac{1}{2},m}^{(h_1),ABC},
{\cal O}_{DEF,-\frac{1}{2}}^{(h_2)}(z,\bar{z}) \bigg\}  & = &
-\de^{ABC}_{DEF} \,
\bar{\mathscr D}_m^{h_1-\frac{1}{2}}(\tfrac{1}{2}-h_2) \,
{\cal O}_{+2}^{(h_1+h_2)}(z,\bar{z}).
\label{20anticommsection5}
\eea


\subsection{The remaining twenty (anti)commutators in the
holomorphic sector}

For the cases $s_1 > s_2$, there are following twenty (anti)commutators 
by changing the ordering of the two modes for two different currents
in (\ref{Eqs})
\bea
\bigg[\hat{\bar{\Phi}}^{(h_1)}_{A,-\frac{3}{2},m},
\hat{\bar{\Phi}}^{(h_2)}_{-2,n} \bigg]&=&
\bar{\kappa}_{-\frac{3}{2},-2,+\frac{3}{2}}\,\Big((h_2+1)m-
(h_1+\tfrac{1}{2})n\Big)\,\hat{\bar{\Phi}}^{(h_1+h_2)}_{A,-\frac{3}{2},m+n}\,\,\, :\text{eq.2},
\nonu \\
\bigg[\hat{\bar{\Phi}}^{(h_1)}_{AB,-1,m},
\hat{\bar{\Phi}}^{(h_2)}_{-2,n} \bigg]&=&
\bar{\kappa}_{-1,-2,+1}\,\Big((h_2+1)m-h_1\,n\Big)\,
\hat{\bar{\Phi}}^{(h_1+h_2)}_{AB,-1,m+n}
\,\,\, :\text{eq.3},
\nonu \\
\bigg[\hat{\bar{\Phi}}^{(h_1)}_{ABC,-\frac{1}{2},m},
\hat{\bar{\Phi}}^{(h_2)}_{-2,n} \bigg]&=&
\bar{\kappa}_{-\frac{1}{2},-2,+\frac{1}{2}}\,\Big((h_2+1)m-
(h_1-\tfrac{1}{2})n\Big)\, \hat{\bar{\Phi}}^{(h_1+h_2)}_{ABC,-\frac{1}{2},m+n}
\, :  \text{eq.4},
\nonu \\
\bigg[\hat{\bar{\Phi}}^{(h_1),ABCD}_{0,m},
\hat{\bar{\Phi}}^{(h_2)}_{-2,n} \bigg]&=&
\bar{\kappa}_{0,-2,0}\,\Big((h_2+1)m-(h_1-1)n\Big)\,
\hat{\bar{\Phi}}^{(h_1+h_2),ABCD}_{0,m+n}
\,\,\, :\text{eq.5},
\nonu \\
\bigg[\hat{\bar{\Phi}}^{(h_1),ABC}_{+\frac{1}{2},m},
\hat{\bar{\Phi}}^{(h_2)}_{-2,n} \bigg]&=&
\bar{\kappa}_{+\frac{1}{2},-2,-\frac{1}{2}}
\Big((h_2+1)m-(h_1-\tfrac{3}{2})n\Big)
\hat{\bar{\Phi}}^{(h_1+h_2),ABC}_{+\frac{1}{2},m+n} :\text{eq.6},
\nonu \\
\bigg[\hat{\bar{\Phi}}^{(h_1),AB}_{+1,m},
\hat{\bar{\Phi}}^{(h_2)}_{-2,n} \bigg]&=&
\bar{\kappa}_{+1,-2,-1}\,\Big((h_2+1)m-(h_1-2)n\Big)\,
\hat{\bar{\Phi}}^{(h_1+h_2),AB}_{+1,m+n}
\,\,\, :\text{eq.7},
\nonu \\
\bigg[\hat{\bar{\Phi}}^{(h_1),A}_{+\frac{3}{2},m},
\hat{\bar{\Phi}}^{(h_2)}_{-2,n} \bigg]&=&
\bar{\kappa}_{+\frac{3}{2},-2,-\frac{3}{2}}\,\Big((h_2+1)m-
(h_1-\tfrac{5}{2})n\Big)\,\hat{\bar{\Phi}}^{(h_1+h_2),A}_{+\frac{3}{2},m+n}
\,\,\, :\text{eq.8},
\nonu \\
\bigg[\hat{\bar{\Phi}}^{(h_1)}_{+2,m},
\hat{\bar{\Phi}}^{(h_2)}_{-2,n} \bigg]&=&
\bar{\kappa}_{+2,-2,-2}\,\Big((h_2+1)m-(h_1-3)n\Big)\,
\hat{\bar{\Phi}}^{(h_1+h_2)}_{+2,m+n}
\,\,\, :\text{eq.9},
\nonu \\
\bigg[\hat{\bar{\Phi}}^{(h_1)}_{AB,-1,m},
\hat{\bar{\Phi}}^{(h_2)}_{C,-\frac{3}{2},n} \bigg]
&=& \bar{\kappa}_{-1,-\frac{3}{2},+\frac{1}{2}}\,
\Big((h_2+\tfrac{1}{2})m-h_1\,n\Big)\,
\hat{\bar{\Phi}}^{(h_1+h_2)}_{ABC,-\frac{1}{2},m+n}
\,\,\, :\text{eq.11},
\nonu \\
\bigg\{\hat{\bar{\Phi}}^{(h_1)}_{ABC,-\frac{1}{2},m},
\hat{\bar{\Phi}}^{(h_2)}_{D,-\frac{3}{2},n}\bigg\}
&=& \bar{\kappa}_{-\frac{1}{2},-\frac{3}{2},0}\,\Big((h_2+\tfrac{1}{2})m
-(h_1-\tfrac{1}{2})n\Big) \nonu \\
& \times & \frac{1}{4!} \, \ep_{ABCDEFGH}\,
\hat{\bar{\Phi}}^{(h_1+h_2),EFGH}_{0,m+n}
\,\,\, :\text{eq.12},
\nonu \\
\bigg[\hat{\bar{\Phi}}^{(h_1),ABCD}_{0,m},
\hat{\bar{\Phi}}^{(h_2)}_{E,-\frac{3}{2},n}\bigg]&=&
\bar{\kappa}_{0,-\frac{3}{2},-\frac{1}{2}}\,\Big((h_2+\tfrac{1}{2})m-
(h_1-1)n\Big)\,\frac{1}{4!}\, \frac{1}{3!}\epsilon^{ABCDIJKL}
\nonu \\
& \times & \ep_{EFGHIJKL}\, \hat{\bar{\Phi}}^{(h_1+h_2),FGH}_{+\frac{1}{2},m+n}
\,\,\, : 
\text{eq.13},
\nonu \\
\bigg\{\hat{\bar{\Phi}}^{(h_1),ABC}_{+\frac{1}{2},m},
\hat{\bar{\Phi}}^{(h_2)}_{D,-\frac{3}{2},n}\bigg\}
&
=& \bar{\kappa}_{+\frac{1}{2},-\frac{3}{2},-1}\,
\Big((h_2+\tfrac{1}{2})m-(h_1-\tfrac{3}{2})n\Big)\,3\,
\hat{\bar{\Phi}}^{(h_1+h_2),[AB}_{+1,m+n} \delta^{\,\,\,C]}_{D}
\, :  \text{eq.14},
\nonu \\
\bigg[\hat{\bar{\Phi}}^{(h_1),AB}_{+1,m},
\hat{\bar{\Phi}}^{(h_2)}_{C,-\frac{3}{2},n} \bigg]
&=&
\bar{\kappa}_{+1,-\frac{3}{2},-\frac{3}{2}}\,
\Big((h_2+\tfrac{1}{2})m-(h_1-2)n\Big)\,2\,\delta^{\,\,\,[A}_{C}
\hat{\bar{\Phi}}^{(h_1+h_2),B]}_{+\frac{3}{2},m+n}
\, :  \text{eq.15},
\nonu \\
\bigg\{\hat{\bar{\Phi}}^{(h_1),A}_{+\frac{3}{2},m},
\hat{\bar{\Phi}}^{(h_2)}_{B,-\frac{3}{2},n}\bigg\}
&=&
\bar{\kappa}_{+\frac{3}{2},-\frac{3}{2},-2}\,\Big((h_2+\tfrac{1}{2})m
-(h_1-\tfrac{5}{2})n\Big)\,\delta^{\,\,\,A}_{B}\,\hat{\bar{\Phi}}^{(h_1+h_2)}_{+2,m+n}
\, :  \text{eq.16},
\nonu \\
\bigg[\hat{\bar{\Phi}}^{(h_1)}_{ABC,-\frac{1}{2},m},
\hat{\bar{\Phi}}^{(h_2)}_{DE,-1,n} \bigg]&=&
\bar{\kappa}_{-\frac{1}{2},-1,-\frac{1}{2}}\,\Big(h_2\,m
-(h_1-\tfrac{1}{2})n\Big)\frac{1}{3!}\,\epsilon_{ABCDEFGH}
\nonu \\
& \times & \hat{\bar{\Phi}}^{(h_1+h_2),FGH}_{+\frac{1}{2},m+n}
:  \text{eq.18},
\nonu \\
\bigg[\hat{\bar{\Phi}}^{(h_1),ABCD}_{0,m},
\hat{\bar{\Phi}}^{(h_2)}_{EF,-1,n} \bigg]&=&
\bar{\kappa}_{0,-1,-1}\,\Big(h_2\,m-(h_1-1)n\Big)\,
\frac{1}{4!} \, \frac{1}{2!} \, \epsilon^{ABCDIJKL}
\nonu \\
& \times & \ep_{EFGHIJKL}\, 
\hat{\bar{\Phi}}^{(h_1+h_2),GH}_{+1,m+n}
: \text{eq.19},
\nonu \\
\bigg[\hat{\bar{\Phi}}^{(h_1),ABC}_{+\frac{1}{2},m},
\hat{\bar{\Phi}}^{(h_2)}_{DE,-1,n} \bigg]&=&
\bar{\kappa}_{+\frac{1}{2},-1,-\frac{3}{2}}\,\Big(h_2\,m
-(h_1-\tfrac{3}{2})n\Big)\,3!\,\delta^{[A}_{\,\,\,D} \hat{\bar{\Phi}}^{(h_1+h_2),B}_{+\frac{3}{2},m+n}
\delta_{E}^{\,\,\,C]}
\, :  \text{eq.20},
\nonu \\
\bigg[\hat{\bar{\Phi}}^{(h_1),AB}_{+1,m},
\hat{\bar{\Phi}}^{(h_2)}_{CD,-1,n} \bigg]&=&
\bar{\kappa}_{+1,-1,-2}\,\Big(h_2\,m-(h_1-2)n\Big)\,\delta^{AB}_{CD}
\,\hat{\bar{\Phi}}^{(h_1+h_2)}_{+2,m+n}
\,\,\, :\text{eq.21},
\nonu \\
\bigg[\hat{\bar{\Phi}}^{(h_1),ABCD}_{0,m},
\hat{\bar{\Phi}}^{(h_2)}_{EFG,-\frac{1}{2},n} \bigg]&=&
\bar{\kappa}_{0,-\frac{1}{2},-\frac{3}{2}}\,\Big((h_2-\tfrac{1}{2})m
-(h_1-1)n\Big)\,
\frac{1}{4!} \, \epsilon^{ABCDIJKL}\,\nonu \\
& \times & \ep_{EFGHIJKL}\, \hat{\bar{\Phi}}^{(h_1+h_2),H}_{+\frac{3}{2},m+n}
\,\,\, :  \text{eq.23},
\nonu \\
\bigg\{\hat{\bar{\Phi}}^{(h_1),ABC}_{+\frac{1}{2},m},
\hat{\bar{\Phi}}^{(h_2)}_{DEF,-\frac{1}{2},n}\bigg\}
&=& \bar{\kappa}_{+\frac{1}{2},-\frac{1}{2},-2}\,\Big((h_2-\tfrac{1}{2})m
-(h_1-\tfrac{3}{2})n\Big)\,
\delta^{ABC}_{DEF}\,\hat{\bar{\Phi}}^{(h_1+h_2)}_{+2,m+n}
\, :  \text{eq.24}.
\label{20expression}
\eea
We also put the number of equations on the right-hand sides
of (\ref{20expression}) corresponding to the previous ones
in (\ref{Eqs}).

\section{The 
OPEs and (anti)commutators of ${\cal N}=8$ soft (and shadow)
${\rm w}_{1+\infty}$
currents (and
${\rm \bar{w}}_{1+\infty}$
currents)
with matter fields in section \ref{lastsection}
\label{lastapp}
}

In this Appendix,
the OPEs of the ${\cal N}=8$ currents with matter fields
in section \ref{lastsection} are given explicitly.

\subsection{
The last twenty OPE relations of (\ref{HOOPE}) 
}

The next twenty OPEs corresponding to (\ref{HOOPE})
are described as
\bea
&& \Phi_{+\fft32}^{(-\frac{k}{2}+\fft14),A}(z_1,\bar{z}_1) \,
{\cal O}_{+2}^{(h_2)}(z_2,\bar{z}_2)=-
\frac{\kappa_{+\frac{3}{2},+2,-\fft32}}{2}\, \frac{1}{z_{12}} \,
\sum_{m=0}^{-k+\fft12} \, \frac{(-1)^{-k-m+\fft12}}{m!\, (-k-m+\fft12)!}
\nonu \\
&& \times \frac{\Gamma(2(-1-h_2)+1)}
{\Gamma(2(-1-h_2)+k+m+\fft12)}
\, \bar{z}_{12}^{m+1}\, \bar{\pa}_{\bar{z}_2}^m \,
{\cal O}_{+\fft32}^{(-\frac{k}{2}+\fft14+h_2),A}(z_2,\bar{z}_2)+\cdots,
\nonu \\
&& \Phi_{+1}^{(-\frac{k}{2}+\fft12),AB}(z_1,\bar{z}_1) \,
{\cal O}_{+2}^{(h_2)}(z_2,\bar{z}_2)=-
\frac{\kappa_{+1,+2,-1}}{2}\, \frac{1}{z_{12}} \,
\sum_{m=0}^{-k} \, \frac{(-1)^{-k-m}}{m!\, (-k-m)!}
\nonu \\
&& \times \frac{\Gamma(2(-1-h_2)+1)}
{\Gamma(2(-1-h_2)+k+m+1)}
\, \bar{z}_{12}^{m+1}\,
\bar{\pa}_{\bar{z}_2}^m \,
{\cal O}_{+1}^{(-\frac{k}{2}+\fft12+h_2),AB}(z_2,\bar{z}_2)+\cdots,
\nonu \\
&& \Phi_{+\fft12}^{(-\frac{k}{2}+\fft34),ABC}(z_1,\bar{z}_1) \,
{\cal O}_{+2}^{(h_2)}(z_2,\bar{z}_2)=-
\frac{\kappa_{+\fft12,+2,-\fft12}}{2}\, \frac{1}{z_{12}} \,
\sum_{m=0}^{-k-\fft12} \, \frac{(-1)^{-k-m-\fft12}}{m!\,
(-k-m-\fft12)!}
\nonu \\
&& \times \frac{\Gamma(2(-1-h_2)+1)}
{\Gamma(2(-1-h_2)+k+m+\tf32)}
\, \bar{z}_{12}^{m+1}\,
\bar{\pa}_{\bar{z}_2}^m \,
{\cal O}_{\fft12}^{(-\frac{k}{2}+\fft34+h_2),ABC}(z_2,\bar{z}_2)+\cdots,
\nonu \\
&& \Phi_{0}^{(-\frac{k}{2}+1),ABCD}(z_1,\bar{z}_1) \,
{\cal O}_{+2}^{(h_2)}(z_2,\bar{z}_2)=-
\frac{\kappa_{0,+2,0}}{2}\, \frac{1}{z_{12}} \,
\sum_{m=0}^{-k-1} \, \frac{(-1)^{-k-m-1}}{m!\,
(-k-m-1)!}
\nonu \\
&& \times \frac{\Gamma(2(-1-h_2)+1)}
{\Gamma(2(-1-h_2)+k+m+2)}
\, \bar{z}_{12}^{m+1}\,
\bar{\pa}_{\bar{z}_2}^m \,
{\cal O}_{0}^{(-\frac{k}{2}+1+h_2),ABCD}(z_2,\bar{z}_2)+\cdots,
\nonu \\
&& \Phi_{ABC,-\fft12}^{(-\frac{k}{2}+\fft54)}(z_1,\bar{z}_1) \,
{\cal O}_{+2}^{(h_2)}(z_2,\bar{z}_2)=-
\frac{\kappa_{-\fft12,+2,+\fft12}}{2}\, \frac{1}{z_{12}} \,
\sum_{m=0}^{-k-\fft32} \, \frac{(-1)^{-k-m-\fft32}}{m!\,
(-k-m-\fft32)!}
\nonu \\
&& \times \frac{\Gamma(2(-1-h_2)+1)}
{\Gamma(2(-1-h_2)+k+m+\tf52)}
\, \bar{z}_{12}^{m+1}\,
\bar{\pa}_{\bar{z}_2}^m \,
{\cal O}_{ABC,-\fft12}^{(-\frac{k}{2}+\fft54+h_2)}(z_2,\bar{z}_2)+\cdots,
\nonu \\
&& \Phi_{AB,-1}^{(-\frac{k}{2}+\fft32)}(z_1,\bar{z}_1) \,
{\cal O}_{+2}^{(h_2)}(z_2,\bar{z}_2)=-
\frac{\kappa_{-1,+2,+1}}{2}\, \frac{1}{z_{12}} \,
\sum_{m=0}^{-k-2} \, \frac{(-1)^{-k-m-2}}{m!\,
(-k-m-2)!}
\nonu \\
&& \times \frac{\Gamma(2(-1-h_2)+1)}
{\Gamma(2(-1-h_2)+k+m+3)}
\, \bar{z}_{12}^{m+1}\,
\bar{\pa}_{\bar{z}_2}^m \,
{\cal O}_{AB,-1}^{(-\frac{k}{2}+\fft32+h_2)}(z_2,\bar{z}_2)+\cdots,
\nonu \\
&& \Phi_{A,-\fft32}^{(-\frac{k}{2}+\fft74)}(z_1,\bar{z}_1) \,
{\cal O}_{+2}^{(h_2)}(z_2,\bar{z}_2)=-
\frac{\kappa_{-\fft32,+2,+\fft32}}{2}\, \frac{1}{z_{12}} \,
\sum_{m=0}^{-k-\fft52} \, \frac{(-1)^{-k-m-\fft52}}{m!\,
(-k-m-\fft52)!}
\nonu \\
&& \times \frac{\Gamma(2(-1-h_2)+1)}
{\Gamma(2(-1-h_2)+k+m+\tf72)}
\, \bar{z}_{12}^{m+1}\,
\bar{\pa}_{\bar{z}_2}^m \,
{\cal O}_{A,-\fft32}^{(-\frac{k}{2}+\fft74+h_2)}(z_2,\bar{z}_2)+\cdots,
\nonu \\
&& \Phi_{-2}^{(-\frac{k}{2}+2)}(z_1,\bar{z}_1) \,
{\cal O}_{+2}^{(h_2)}(z_2,\bar{z}_2)=-
\frac{\kappa_{-2,+2,+2}}{2}\, \frac{1}{z_{12}} \,
\sum_{m=0}^{-k-3} \, \frac{(-1)^{-k-m-3}}{m!\,
(-k-m-3)!}
\nonu \\
&& \times \frac{\Gamma(2(-1-h_2)+1)}
{\Gamma(2(-1-h_2)+k+m+4)}
\, \bar{z}_{12}^{m+1}\,
\bar{\pa}_{\bar{z}_2}^m \,
{\cal O}_{-2}^{(-\frac{k}{2}+2+h_2)}(z_2,\bar{z}_2)+\cdots,
\nonu \\
&& \Phi_{+1}^{(-\frac{k}{2}+\fft12),AB}(z_1,\bar{z}_1) \,
{\cal O}_{+\fft32}^{(h_2),C}(z_2,\bar{z}_2)=-
\frac{\kappa_{+1,+\fft32,-\fft12}}{2}\, \frac{1}{z_{12}} \,
\sum_{m=0}^{-k} \, \frac{(-1)^{-k-m}}{m!\,
(-k-m)!}
\nonu \\
&& \times \frac{\Gamma(2(-\fft12-h_2)+1)}
{\Gamma(2(-\fft12-h_2)+k+m+1)}
\, \bar{z}_{12}^{m+1}\,
\bar{\pa}_{\bar{z}_2}^m \,
{\cal O}_{+\fft12}^{(-\frac{k}{2}+\fft12+h_2),ABC}(z_2,\bar{z}_2)+\cdots,
\nonu \\
&& \Phi_{+\fft12}^{(-\frac{k}{2}+\fft34),ABC}(z_1,\bar{z}_1) \,
{\cal O}_{+\fft32}^{(h_2),D}(z_2,\bar{z}_2)=-
\frac{\kappa_{+\fft12,+\fft32,0}}{2}\, \frac{1}{z_{12}} \,
\sum_{m=0}^{-k-\fft12} \, \frac{(-1)^{-k-m-\fft12}}{m!\,
(-k-m-\fft12)!}
\nonu \\
&& \times \frac{\Gamma(2(-\fft12-h_2)+1)}
{\Gamma(2(-\fft12-h_2)+k+m+\tf32)}
\, \bar{z}_{12}^{m+1}\,
\bar{\pa}_{\bar{z}_2}^m \,
{\cal O}_{0}^{(-\frac{k}{2}+\fft34+h_2),ABCD}(z_2,\bar{z}_2)+\cdots,
\nonu \\
&& \Phi_{0}^{(-\frac{k}{2}+1),ABCD}(z_1,\bar{z}_1) \,
{\cal O}_{+\fft32}^{(h_2),E}(z_2,\bar{z}_2)=-
\frac{\kappa_{0,+\fft32,\fft12}}{2}\, \frac{1}{z_{12}} \,
\sum_{m=0}^{-k-1} \, \frac{(-1)^{-k-m-1}}{m!\,
(-k-m-1)!}
\nonu \\
&& \times \frac{\Gamma(2(-\fft12-h_2)+1)}
{\Gamma(2(-\fft12-h_2)+k+m+2)}
\, \bar{z}_{12}^{m+1}\,
\frac{1}{3!}\, \ep^{ABCDEFGH}\,
\bar{\pa}_{\bar{z}_2}^m \,
{\cal O}_{FGH,-\fft12}^{(-\frac{k}{2}+1+h_2)}(z_2,\bar{z}_2)+\cdots,
\nonu \\
&& \Phi_{ABC,-\fft12}^{(-\frac{k}{2}+\fft54)}(z_1,\bar{z}_1) \,
{\cal O}_{+\fft32}^{(h_2),D}(z_2,\bar{z}_2)=-
\frac{\kappa_{-\fft12,+\fft32,+1}}{2}\, \frac{1}{z_{12}} \,
\sum_{m=0}^{-k-\fft32} \, \frac{(-1)^{-k-m-\fft32}}{m!\,
(-k-m-\fft32)!}
\nonu \\
&& \times \frac{\Gamma(2(-\fft12-h_2)+1)}
{\Gamma(2(-\fft12-h_2)+k+m+\tf52)}
\, \bar{z}_{12}^{m+1}\,
3\, 
\bar{\pa}_{\bar{z}_2}^m \,
{\cal O}_{[AB,-1}^{(-\frac{k}{2}+\fft54+h_2)}(z_2,\bar{z}_2)\,
\de^{D}_{C]}+\cdots,
\nonu \\
&& \Phi_{AB,-1}^{(-\frac{k}{2}+\fft32)}(z_1,\bar{z}_1) \,
{\cal O}_{+\fft32}^{(h_2),C}(z_2,\bar{z}_2)=-
\frac{\kappa_{-1,+\fft32,+\fft32}}{2}\, \frac{1}{z_{12}} \,
\sum_{m=0}^{-k-2} \, \frac{(-1)^{-k-m-2}}{m!\,
(-k-m-2)!}
\nonu \\
&& \times \frac{\Gamma(2(-\fft12-h_2)+1)}
{\Gamma(2(-\fft12-h_2)+k+m+3)}
\, \bar{z}_{12}^{m+1}\,
2\, \de^{C}_{[A} 
\bar{\pa}_{\bar{z}_2}^m \,
{\cal O}_{B],-\fft32}^{(-\frac{k}{2}+\fft32+h_2)}(z_2,\bar{z}_2)+\cdots,
\nonu \\
&& \Phi_{A,-\fft32}^{(-\frac{k}{2}+\fft74)}(z_1,\bar{z}_1) \,
{\cal O}_{+\fft32}^{(h_2),B}(z_2,\bar{z}_2)=-
\frac{\kappa_{-\fft32,+\fft32,+2}}{2}\, \frac{1}{z_{12}} \,
\sum_{m=0}^{-k-\fft52} \, \frac{(-1)^{-k-m-\fft52}}{m!\,
(-k-m-\fft52)!}
\nonu \\
&& \times \frac{\Gamma(2(-\fft12-h_2)+1)}
{\Gamma(2(-\fft12-h_2)+k+m+\tf72)}
\, \bar{z}_{12}^{m+1}\,
\de^{B}_{A} 
\bar{\pa}_{\bar{z}_2}^m \,
{\cal O}_{-2}^{(-\frac{k}{2}+\fft74+h_2)}(z_2,\bar{z}_2)+\cdots,
\nonu \\
&& \Phi_{+\fft12}^{(-\frac{k}{2}+\fft34),ABC}(z_1,\bar{z}_1) \,
{\cal O}_{+1}^{(h_2),DE}(z_2,\bar{z}_2)=-
\frac{\kappa_{+\fft12,+1,+\fft12}}{2}\, \frac{1}{z_{12}} \,
\sum_{m=0}^{-k-\fft12} \, \frac{(-1)^{-k-m-\fft12}}{m!\,
(-k-m-\fft12)!}
\nonu \\
&& \times \frac{\Gamma(2(-h_2)+1)}
{\Gamma(2(-h_2)+k+m+\tf32)}
\, \bar{z}_{12}^{m+1}\,
\frac{1}{3!}\, \ep^{ABCDEFGH}\,
\bar{\pa}_{\bar{z}_2}^m \,
{\cal O}_{FGH,-\fft12}^{(-\frac{k}{2}+\fft34+h_2)}(z_2,\bar{z}_2)+\cdots,
\nonu \\
&& \Phi_{0}^{(-\frac{k}{2}+1),ABCD}(z_1,\bar{z}_1) \,
{\cal O}_{+1}^{(h_2),EF}(z_2,\bar{z}_2)=-
\frac{\kappa_{0,+1,+1}}{2}\, \frac{1}{z_{12}} \,
\sum_{m=0}^{-k-1} \, \frac{(-1)^{-k-m-1}}{m!\,
(-k-m-1)!}
\nonu \\
&& \times \frac{\Gamma(2(-h_2)+1)}
{\Gamma(2(-h_2)+k+m+2)}
\, \bar{z}_{12}^{m+1}\,
\frac{1}{2!}\, \ep^{ABCDEFGH}\,
\bar{\pa}_{\bar{z}_2}^m \,
{\cal O}_{GH,-1}^{(-\frac{k}{2}+1+h_2)}(z_2,\bar{z}_2)+\cdots,
\nonu \\
&& \Phi_{ABC,-\fft12}^{(-\frac{k}{2}+\fft54)}(z_1,\bar{z}_1) \,
{\cal O}_{+1}^{(h_2),DE}(z_2,\bar{z}_2)=-
\frac{\kappa_{-\fft12,+1,+\fft32}}{2}\, \frac{1}{z_{12}} \,
\sum_{m=0}^{-k-\fft32} \, \frac{(-1)^{-k-m-\fft32}}{m!\,
(-k-m-\fft32)!}
\nonu \\
&& \times \frac{\Gamma(2(-h_2)+1)}
{\Gamma(2(-h_2)+k+m+\tf52)}
 \, \bar{z}_{12}^{m+1}\, 3!\,
\de^{D}_{[A}\,
\bar{\pa}_{\bar{z}_2}^m \,
{\cal O}_{B,-\fft32}^{(-\frac{k}{2}+\fft54+h_2)}(z_2,\bar{z}_2) \,
\de^{E}_{C]}+\cdots,
\nonu \\
&& \Phi_{AB,-1}^{(-\frac{k}{2}+\fft32)}(z_1,\bar{z}_1) \,
{\cal O}_{+1}^{(h_2),CD}(z_2,\bar{z}_2)=-
\frac{\kappa_{-1,+1,+2}}{2}\, \frac{1}{z_{12}} \,
\sum_{m=0}^{-k-2} \, \frac{(-1)^{-k-m-2}}{m!\,
(-k-m-2)!}
\nonu \\
&& \times \frac{\Gamma(2(-h_2)+1)}
{\Gamma(2(-h_2)+k+m+3)}
\, \bar{z}_{12}^{m+1}\,
\de^{CD}_{AB}\,
\bar{\pa}_{\bar{z}_2}^m \,
{\cal O}_{-2}^{(-\frac{k}{2}+\fft32+h_2)}(z_2,\bar{z}_2)+\cdots,
\nonu \\
&& \Phi_{0}^{(-\frac{k}{2}+1),ABCD}(z_1,\bar{z}_1) \,
{\cal O}_{+\fft12}^{(h_2),EFG}(z_2,\bar{z}_2)=-
\frac{\kappa_{0,+\fft12,+\fft32}}{2}\, \frac{1}{z_{12}} \,
\sum_{m=0}^{-k-1} \, \frac{(-1)^{-k-m-1}}{m!\,
(-k-m-1)!}
\nonu \\
&& \times \frac{\Gamma(2(\fft12-h_2)+1)}
{\Gamma(2(\fft12-h_2)+k+m+2)}
\, \bar{z}_{12}^{m+1}\,
\ep^{ABCDEFGH}\,
\bar{\pa}_{\bar{z}_2}^m \,
{\cal O}_{H,-\fft32}^{(-\frac{k}{2}+1+h_2)}(z_2,\bar{z}_2)+\cdots,
\nonu \\
&& \Phi_{ABC,-\fft12}^{(-\frac{k}{2}+\fft54)}(z_1,\bar{z}_1) \,
{\cal O}_{+\fft12}^{(h_2),DEF}(z_2,\bar{z}_2)=-
\frac{\kappa_{-\fft12,+\fft12,+2}}{2}\, \frac{1}{z_{12}} \,
\sum_{m=0}^{-k-\fft32} \, \frac{(-1)^{-k-m-\fft32}}{m!\,
(-k-m-\fft32)!}
\nonu \\
&& \times \frac{\Gamma(2(\fft12-h_2)+1)}
{\Gamma(2(\fft12-h_2)+k+m+\tf52)}
\, \bar{z}_{12}^{m+1}\,
\de^{DEF}_{ABC}\,
\bar{\pa}_{\bar{z}_2}^m \,
{\cal O}_{-2}^{(-\frac{k}{2}+\fft54+h_2)}(z_2,\bar{z}_2)+\cdots.
\label{twenty}
\eea


\subsection{
The first twenty-five (anti)commutator relations of (\ref{HOcomm})
}

The (anti)commutators corresponding to
(\ref{HOcomm}) are obtained
\bea
&& \bigg[ \hat{\Phi}_{+2,n}^{(-\frac{k}{2})},
{\cal O}_{+2}^{(h_2)}(z,\bar{z}) \bigg]  =  -
\frac{\kappa_{+2,+2,-2}}{2} \,
\sum_{m=0}^{-k+1} \, \frac{(-1)^{-\frac{k}{2}+n+1}\,
(m+1)}{ (-k-m+1)!}
\, \frac{\Gamma(2(-1-h_2)+1)}{\Gamma(2(-1-h_2)+k+m)}
\nonu \\
&& \times
\frac{ \bar{z}^{m+\frac{k}{2}+n}}{(\frac{2-k}{2}-n)! \,
(m+\frac{k}{2}+n)!}
\, \bar{\pa}_{\bar{z}_2}^{m}\,
{\cal O}^{(-\frac{k}{2}+h_2)}_{+2}(z,\bar{z}),
\nonu \\
&& \bigg[ \hat{\Phi}_{+2,n}^{(-\frac{k}{2})},
{\cal O}_{+\fft32}^{(h_2),A}(z,\bar{z}) \bigg]  =  -
\frac{\kappa_{+2,+\fft32,-\fft32}}{2} \,
\sum_{m=0}^{-k+1} \, \frac{(-1)^{-\frac{k}{2}+n+1}\,
(m+1)}{ (-k-m+1)!}
\, \frac{\Gamma(2(-\fft12-h_2)+1)}{\Gamma(2(-\fft12-h_2)+k+m)}
\nonu \\
&& \times
\frac{ \bar{z}^{m+\frac{k}{2}+n}}{(\frac{2-k}{2}-n)! \,
(m+\frac{k}{2}+n)!}
\, \bar{\pa}_{\bar{z}_2}^{m}\,
{\cal O}^{(-\frac{k}{2}+h_2),A}_{+\fft32}(z,\bar{z}),
\nonu \\
&& \bigg[ \hat{\Phi}_{+2,n}^{(-\frac{k}{2})},
{\cal O}_{+1}^{(h_2),AB}(z,\bar{z}) \bigg]  =  -
\frac{\kappa_{+2,+1,-1}}{2} \,
\sum_{m=0}^{-k+1} \, \frac{(-1)^{-\frac{k}{2}+n+1}\,
(m+1)}{ (-k-m+1)!}
\, \frac{\Gamma(2(-h_2)+1)}{\Gamma(2(-h_2)+k+m)}
\nonu \\
&& \times
\frac{ \bar{z}^{m+\frac{k}{2}+n}}{(\frac{2-k}{2}-n)! \,
(m+\frac{k}{2}+n)!}
\, \bar{\pa}_{\bar{z}_2}^{m}\,
{\cal O}^{(-\frac{k}{2}+h_2),AB}_{+1}(z,\bar{z}),
\nonu \\
&& \bigg[ \hat{\Phi}_{+2,n}^{(-\frac{k}{2})},
{\cal O}_{+\fft12}^{(h_2),ABC}(z,\bar{z}) \bigg]  =  -
\frac{\kappa_{+2,+\fft12,-\fft12}}{2} \,
\sum_{m=0}^{-k+1} \, \frac{(-1)^{-\frac{k}{2}+n+1}\,
(m+1)}{ (-k-m+1)!}
\, \frac{\Gamma(2(\fft12-h_2)+1)}{\Gamma(2(\fft12-h_2)+k+m)}
\nonu \\
&& \times
\frac{ \bar{z}^{m+\frac{k}{2}+n}}{(\frac{2-k}{2}-n)! \,
(m+\frac{k}{2}+n)!}
\, \bar{\pa}_{\bar{z}_2}^{m}\,
{\cal O}^{(-\frac{k}{2}+h_2),ABC}_{+\fft12}(z,\bar{z}),
\nonu \\
&& \bigg[ \hat{\Phi}_{+2,n}^{(-\frac{k}{2})},
{\cal O}_{0}^{(h_2),ABCD}(z,\bar{z}) \bigg]  =  -
\frac{\kappa_{+2,0,0}}{2} \,
\sum_{m=0}^{-k+1} \, \frac{(-1)^{-\frac{k}{2}+n+1}\,
(m+1)}{ (-k-m+1)!}
\, \frac{\Gamma(2(1-h_2)+1)}{\Gamma(2(1-h_2)+k+m)}
\nonu \\
&& \times
\frac{ \bar{z}^{m+\frac{k}{2}+n}}{(\frac{2-k}{2}-n)! \,
(m+\frac{k}{2}+n)!}
\, \bar{\pa}_{\bar{z}_2}^{m}\,
{\cal O}^{(-\frac{k}{2}+h_2),ABCD}_{0}(z,\bar{z}),
\nonu \\
&& \bigg[ \hat{\Phi}_{+2,n}^{(-\frac{k}{2})},
{\cal O}_{ABC,-\fft12}^{(h_2)}(z,\bar{z}) \bigg]  =  -
\frac{\kappa_{+2,-\fft12,+\fft12}}{2} \,
\sum_{m=0}^{-k+1} \, \frac{(-1)^{-\frac{k}{2}+n+1}\,
(m+1)}{ (-k-m+1)!}
\, \frac{\Gamma(2(\fft32-h_2)+1)}{\Gamma(2(\fft32-h_2)+k+m)}
\nonu \\
&& \times
\frac{ \bar{z}^{m+\frac{k}{2}+n}}{(\frac{2-k}{2}-n)! \,
(m+\frac{k}{2}+n)!}
\, \bar{\pa}_{\bar{z}_2}^{m}\,
{\cal O}^{(-\frac{k}{2}+h_2)}_{ABC,-\fft12}(z,\bar{z}),
\nonu \\
&& \bigg[ \hat{\Phi}_{+2,n}^{(-\frac{k}{2})},
{\cal O}_{AB,-1}^{(h_2)}(z,\bar{z}) \bigg]  =  -
\frac{\kappa_{+2,-1,+1}}{2} \,
\sum_{m=0}^{-k+1} \, \frac{(-1)^{-\frac{k}{2}+n+1}\,
(m+1)}{ (-k-m+1)!}
\, \frac{\Gamma(2(2-h_2)+1)}{\Gamma(2(2-h_2)+k+m)}
\nonu \\
&& \times
\frac{ \bar{z}^{m+\frac{k}{2}+n}}{(\frac{2-k}{2}-n)! \,
(m+\frac{k}{2}+n)!}
\, \bar{\pa}_{\bar{z}_2}^{m}\,
{\cal O}^{(-\frac{k}{2}+h_2)}_{AB,-1}(z,\bar{z}),
\nonu \\
&& \bigg[ \hat{\Phi}_{+2,n}^{(-\frac{k}{2})},
{\cal O}_{A,-\fft32}^{(h_2)}(z,\bar{z}) \bigg]  =  -
\frac{\kappa_{+2,-\fft32,+\fft32}}{2} \,
\sum_{m=0}^{-k+1} \, \frac{(-1)^{-\frac{k}{2}+n+1}\,
(m+1)}{ (-k-m+1)!}
\, \frac{\Gamma(2(\fft52-h_2)+1)}{\Gamma(2(\fft52-h_2)+k+m)}
\nonu \\
&& \times
\frac{ \bar{z}^{m+\frac{k}{2}+n}}{(\frac{2-k}{2}-n)! \,
(m+\frac{k}{2}+n)!}
\, \bar{\pa}_{\bar{z}_2}^{m}\,
{\cal O}^{(-\frac{k}{2}+h_2)}_{A,-\fft32}(z,\bar{z}),
\nonu \\
&& \bigg[ \hat{\Phi}_{+2,n}^{(-\frac{k}{2})},
{\cal O}_{-2}^{(h_2)}(z,\bar{z}) \bigg]  =  -
\frac{\kappa_{+2,-2,+2}}{2} \,
\sum_{m=0}^{-k+1} \, \frac{(-1)^{-\frac{k}{2}+n+1}\,
(m+1)}{ (-k-m+1)!}
\, \frac{\Gamma(2(3-h_2)+1)}{\Gamma(2(3-h_2)+k+m)}
\nonu \\
&& \times
\frac{ \bar{z}^{m+\frac{k}{2}+n}}{(\frac{2-k}{2}-n)! \,
(m+\frac{k}{2}+n)!}
\, \bar{\pa}_{\bar{z}_2}^{m}\,
{\cal O}^{(-\frac{k}{2}+h_2)}_{-2}(z,\bar{z}),
\nonu \\
&& \bigg\{ \hat{\Phi}_{+\fft32,n}^{(-\frac{k}{2}+\fft14),A},
{\cal O}_{+\fft32}^{(h_2),B}(z,\bar{z}) \bigg\}  =  -
\frac{\kappa_{+\fft32,+\fft32,-1}}{2} \,
\sum_{m=0}^{-k+\fft12} \, \frac{(-1)^{-\frac{k}{2}+n+\fft34}\,
(m+1)}{ (-k-m+\fft12)!}
\nonu \\
&& \times
\frac{\Gamma(2(-\fft12-h_2)+1)}{\Gamma(2(-\fft12-h_2)+k+m+
\fft12)}
\,
\frac{ \bar{z}^{m+\frac{k}{2}+n+\fft14}}{(\frac{\fft32-k}{2}-n)! \,
(m+\frac{k}{2}+n+\fft14)!}
\, \bar{\pa}_{\bar{z}_2}^{m}\,
{\cal O}^{(-\frac{k}{2}+\fft14+h_2),AB}_{+1}(z,\bar{z}),
\nonu \\
&& \bigg[ \hat{\Phi}_{+\fft32,n}^{(-\frac{k}{2}+\fft14),A},
{\cal O}_{+1}^{(h_2),BC}(z,\bar{z}) \bigg]  =  -
\frac{\kappa_{+\fft32,+1,-\fft12}}{2} \,
\sum_{m=0}^{-k+\fft12} \, \frac{(-1)^{-\frac{k}{2}+n+\fft34}\,
(m+1)}{ (-k-m+\fft12)!}
\nonu \\
&& \times
\frac{\Gamma(2(-h_2)+1)}{\Gamma(2(-h_2)+k+m+
\fft12)}
\,
\frac{ \bar{z}^{m+\frac{k}{2}+n+\fft14}}{(\frac{\fft32-k}{2}-n)! \,
(m+\frac{k}{2}+n+\fft14)!}
\, \bar{\pa}_{\bar{z}_2}^{m}\,
{\cal O}^{(-\frac{k}{2}+\fft14+h_2),ABC}_{+\fft12}(z,\bar{z}),
\nonu \\
&& \bigg\{ \hat{\Phi}_{+\fft32,n}^{(-\frac{k}{2}+\fft14),A},
{\cal O}_{+\fft12}^{(h_2),BCD}(z,\bar{z}) \bigg\}  =  -
\frac{\kappa_{+\fft32,+\fft12,0}}{2} \,
\sum_{m=0}^{-k+\fft12} \, \frac{(-1)^{-\frac{k}{2}+n+\fft34}\,
(m+1)}{ (-k-m+\fft12)!}
\nonu \\
&& \times
\frac{\Gamma(2(\fft12-h_2)+1)}{\Gamma(2(\fft12-h_2)+k+m+
\fft12)}
\,
\frac{ \bar{z}^{m+\frac{k}{2}+n+\fft14}}{(\frac{\fft32-k}{2}-n)! \,
(m+\frac{k}{2}+n+\fft14)!}
\, \bar{\pa}_{\bar{z}_2}^{m}\,
{\cal O}^{(-\frac{k}{2}+\fft14+h_2),ABCD}_{0}(z,\bar{z}),
\nonu \\
&& \bigg[ \hat{\Phi}_{+\fft32,n}^{(-\frac{k}{2}+\fft14),A},
{\cal O}_{0}^{(h_2),BCDE}(z,\bar{z}) \bigg]  =  -
\frac{\kappa_{+\fft32,0,+\fft12}}{2} \,
\sum_{m=0}^{-k+\fft12} \, \frac{(-1)^{-\frac{k}{2}+n+\fft34}\,
(m+1)}{ (-k-m+\fft12)!}
\nonu \\
&& \times
\frac{\Gamma(2(1-h_2)+1)}{\Gamma(2(1-h_2)+k+m+
\fft12)}
\,
\frac{ \bar{z}^{m+\frac{k}{2}+n+\fft14}}{(\frac{\fft32-k}{2}-n)! \,
(m+\frac{k}{2}+n+\fft14)!}
\frac{1}{3!}\, \ep^{ABCDEFGH}\,
\nonu \\
&& \times \bar{\pa}_{\bar{z}_2}^{m}\,
{\cal O}^{(-\frac{k}{2}+\fft14+h_2)}_{FGH,-\ft12}(z,\bar{z}),
\nonu \\
&& \bigg\{ \hat{\Phi}_{+\fft32,n}^{(-\frac{k}{2}+\fft14),A},
{\cal O}_{BCD,-\fft12}^{(h_2)}(z,\bar{z}) \bigg\}  =  -
\frac{\kappa_{+\fft32,-\fft12,+1}}{2} \,
\sum_{m=0}^{-k+\fft12} \, \frac{(-1)^{-\frac{k}{2}+n+\fft34}\,
(m+1)}{ (-k-m+\fft12)!}
\nonu \\
&& \times
\frac{\Gamma(2(\fft32-h_2)+1)}{\Gamma(2(\fft32-h_2)+k+m+
\fft12)}
\,
\frac{ \bar{z}^{m+\frac{k}{2}+n+\fft14}}{(\frac{\fft32-k}{2}-n)! \,
(m+\frac{k}{2}+n+\fft14)!}
3\, \de^{A}_{[B}     \,
\bar{\pa}_{\bar{z}_2}^{m}\,
{\cal O}^{(-\frac{k}{2}+\fft14+h_2)}_{CD],-1}(z,\bar{z}),
\nonu \\
&& \bigg[ \hat{\Phi}_{+\fft32,n}^{(-\frac{k}{2}+\fft14),A},
{\cal O}_{BC,-1}^{(h_2)}(z,\bar{z}) \bigg]  =  -
\frac{\kappa_{+\fft32,-1,+\fft32}}{2} \,
\sum_{m=0}^{-k+\fft12} \, \frac{(-1)^{-\frac{k}{2}+n+\fft34}\,
(m+1)}{ (-k-m+\fft12)!}
\nonu \\
&& \times
\frac{\Gamma(2(2-h_2)+1)}{\Gamma(2(2-h_2)+k+m+
\fft12)}
\,
\frac{ \bar{z}^{m+\frac{k}{2}+n+\fft14}}{(\frac{\fft32-k}{2}-n)! \,
(m+\frac{k}{2}+n+\fft14)!}
2!\, \de^{A}_{[B}     \,
\bar{\pa}_{\bar{z}_2}^{m}\,
{\cal O}^{(-\frac{k}{2}+\fft14+h_2)}_{C],-\fft32}(z,\bar{z}),
\nonu \\
&& \bigg\{ \hat{\Phi}_{+\fft32,n}^{(-\frac{k}{2}+\fft14),A},
{\cal O}_{B,-\fft32}^{(h_2)}(z,\bar{z}) \bigg\}  =  -
\frac{\kappa_{+\fft32,-\fft32,+2}}{2} \,
\sum_{m=0}^{-k+\fft12} \, \frac{(-1)^{-\frac{k}{2}+n+\fft34}\,
(m+1)}{ (-k-m+\fft12)!}
\nonu \\
&& \times
\frac{\Gamma(2(\fft52-h_2)+1)}{\Gamma(2(\fft52-h_2)+k+m+
\fft12)}
\,
\frac{ \bar{z}^{m+\frac{k}{2}+n+\fft14}}{(\frac{\fft32-k}{2}-n)! \,
(m+\frac{k}{2}+n+\fft14)!}
\, \de^{A}_{B}     \,
\bar{\pa}_{\bar{z}_2}^{m}\,
{\cal O}^{(-\frac{k}{2}+\fft14+h_2)}_{-2}(z,\bar{z}),
\nonu \\
&& \bigg[ \hat{\Phi}_{+1,n}^{(-\frac{k}{2}+\fft12),AB},
{\cal O}_{+1}^{(h_2),CD}(z,\bar{z}) \bigg]  =  -
\frac{\kappa_{+1,+1,0}}{2} \,
\sum_{m=0}^{-k} \, \frac{(-1)^{-\frac{k}{2}+n+\fft12}\,
(m+1)}{ (-k-m)!}
\nonu \\
&& \times
\frac{\Gamma(2(-h_2)+1)}{\Gamma(2(-h_2)+k+m+1)}
\,
\frac{ \bar{z}^{m+\frac{k}{2}+n+\fft12}}{(\frac{1-k}{2}-n)! \,
(m+\frac{k}{2}+n+\fft12)!}
\,
\bar{\pa}_{\bar{z}_2}^{m}\,
{\cal O}^{(-\frac{k}{2}+\fft12+h_2),ABCD}_{0}(z,\bar{z}),
\nonu \\
&& \bigg[ \hat{\Phi}_{+1,n}^{(-\frac{k}{2}+\fft12),AB},
{\cal O}_{+\fft12}^{(h_2),CDE}(z,\bar{z}) \bigg]  =  -
\frac{\kappa_{+1,+\fft12,+\fft12}}{2} \,
\sum_{m=0}^{-k} \, \frac{(-1)^{-\frac{k}{2}+n+\fft12}\,
(m+1)}{ (-k-m)!}
\nonu \\
&& \times
\frac{\Gamma(2(\fft12-h_2)+1)}{\Gamma(2(\fft12-h_2)+k+m+1)}
\,
\frac{ \bar{z}^{m+\frac{k}{2}+n+\fft12}}{(\frac{1-k}{2}-n)! \,
(m+\frac{k}{2}+n+\fft12)!}
\,
\frac{1}{3!}\, \ep^{ABCDEFGH}\nonu \\
&& \times
\bar{\pa}_{\bar{z}_2}^{m}\,
{\cal O}^{(-\frac{k}{2}+\fft12+h_2)}_{FGH,-\fft12}(z,\bar{z}),
\nonu \\
&& \bigg[ \hat{\Phi}_{+1,n}^{(-\frac{k}{2}+\fft12),AB},
{\cal O}_{0}^{(h_2),CDEF}(z,\bar{z}) \bigg]  =  -
\frac{\kappa_{+1,0,+1}}{2} \,
\sum_{m=0}^{-k} \, \frac{(-1)^{-\frac{k}{2}+n+\fft12}\,
(m+1)}{ (-k-m)!}
\nonu \\
&& \times
\frac{\Gamma(2(1-h_2)+1)}{\Gamma(2(1-h_2)+k+m+1)}
\,
\frac{ \bar{z}^{m+\frac{k}{2}+n+\fft12}}{(\frac{1-k}{2}-n)! \,
(m+\frac{k}{2}+n+\fft12)!}
\,
\frac{1}{2!}\, \ep^{ABCDEFGH}\nonu \\
&& \times
\bar{\pa}_{\bar{z}_2}^{m}\,
{\cal O}^{(-\frac{k}{2}+\fft12+h_2)}_{GH,-1}(z,\bar{z}),
\nonu \\
&& \bigg[ \hat{\Phi}_{+1,n}^{(-\frac{k}{2}+\fft12),AB},
{\cal O}_{CDE,-\fft12}^{(h_2)}(z,\bar{z}) \bigg]  =  -
\frac{\kappa_{+1,-\fft12,+\fft32}}{2} \,
\sum_{m=0}^{-k} \, \frac{(-1)^{-\frac{k}{2}+n+\fft12}\,
(m+1)}{ (-k-m)!}
\nonu \\
&& \times
\frac{\Gamma(2(\fft32-h_2)+1)}{\Gamma(2(\fft32-h_2)+k+m+1)}
\,
\frac{ \bar{z}^{m+\frac{k}{2}+n+\fft12}}{(\frac{1-k}{2}-n)! \,
(m+\frac{k}{2}+n+\fft12)!}
\,
3!\, \de^{A}_{[C} \, 
\bar{\pa}_{\bar{z}_2}^{m}\,
{\cal O}^{(-\frac{k}{2}+\fft12+h_2)}_{D,-\fft32}(z,\bar{z})
\, \de^{B}_{E]},
\nonu \\
&& \bigg[ \hat{\Phi}_{+1,n}^{(-\frac{k}{2}+\fft12),AB},
{\cal O}_{CD,-1}^{(h_2)}(z,\bar{z}) \bigg]  =  -
\frac{\kappa_{+1,-1,+2}}{2} \,
\sum_{m=0}^{-k} \, \frac{(-1)^{-\frac{k}{2}+n+\fft12}\,
(m+1)}{ (-k-m)!}
\nonu \\
&& \times
\frac{\Gamma(2(2-h_2)+1)}{\Gamma(2(2-h_2)+k+m+1)}
\,
\frac{ \bar{z}^{m+\frac{k}{2}+n+\fft12}}{(\frac{1-k}{2}-n)! \,
(m+\frac{k}{2}+n+\fft12)!}
\,
\de^{AB}_{CD} \, 
\bar{\pa}_{\bar{z}_2}^{m}\,
{\cal O}^{(-\frac{k}{2}+\fft12+h_2)}_{-2}(z,\bar{z}),
\nonu \\
&& \bigg\{ \hat{\Phi}_{+\fft12,n}^{(-\frac{k}{2}+\fft34),ABC},
{\cal O}_{+\fft12}^{(h_2),DEF}(z,\bar{z}) \bigg\}  =  -
\frac{\kappa_{+\fft12,+\fft12,+1}}{2} \,
\sum_{m=0}^{-k-\fft12} \, \frac{(-1)^{-\frac{k}{2}+n+\fft14}\,
(m+1)}{ (-k-m-\fft12)!}
\nonu \\
&& \times
\frac{\Gamma(2(\fft12-h_2)+1)}{\Gamma(2(\fft12-h_2)+k+m+\fft32)}
\,
\frac{ \bar{z}^{m+\frac{k}{2}+n+\fft34}}{(\frac{\fft12-k}{2}-n)! \,
(m+\frac{k}{2}+n+\fft34)!}
\, \frac{1}{2!}\, \ep^{ABCDEFGH}
\nonu \\
&& \times
\bar{\pa}_{\bar{z}_2}^{m}\,
{\cal O}^{(-\frac{k}{2}+\fft34+h_2)}_{GH,-1}(z,\bar{z}),
\nonu \\
&& \bigg[ \hat{\Phi}_{+\fft12,n}^{(-\frac{k}{2}+\fft34),ABC},
{\cal O}_{0}^{(h_2),DEFG}(z,\bar{z}) \bigg]  =  -
\frac{\kappa_{+\fft12,0,+\fft32}}{2} \,
\sum_{m=0}^{-k-\fft12} \, \frac{(-1)^{-\frac{k}{2}+n+\fft14}\,
(m+1)}{ (-k-m-\fft12)!}
\nonu \\
&& \times
\frac{\Gamma(2(1-h_2)+1)}{\Gamma(2(1-h_2)+k+m+\fft32)}
\,
\frac{ \bar{z}^{m+\frac{k}{2}+n+\fft34}}{(\frac{\fft12-k}{2}-n)! \,
(m+\frac{k}{2}+n+\fft34)!}
\, \ep^{ABCDEFGH}
\nonu \\
&& \times
\bar{\pa}_{\bar{z}_2}^{m}\,
{\cal O}^{(-\frac{k}{2}+\fft34+h_2)}_{H,-\fft32}(z,\bar{z}),
\nonu \\
&& \bigg\{ \hat{\Phi}_{+\fft12,n}^{(-\frac{k}{2}+\fft34),ABC},
{\cal O}_{DEF,-\fft12}^{(h_2)}(z,\bar{z}) \bigg\}  =  -
\frac{\kappa_{+\fft12,-\fft12,+2}}{2} \,
\sum_{m=0}^{-k-\fft12} \, \frac{(-1)^{-\frac{k}{2}+n+\fft14}\,
(m+1)}{ (-k-m-\fft12)!}
\nonu \\
&& \times
\frac{\Gamma(2(\fft32-h_2)+1)}{\Gamma(2(\fft32-h_2)+k+m+\fft32)}
\,
\frac{ \bar{z}^{m+\frac{k}{2}+n+\fft34}}{(\frac{\fft12-k}{2}-n)! \,
(m+\frac{k}{2}+n+\fft34)!}
\, \de^{ABC}_{DEF}
\,
\bar{\pa}_{\bar{z}_2}^{m}\,
{\cal O}^{(-\frac{k}{2}+\fft34+h_2)}_{-2}(z,\bar{z}),
\nonu \\
&& \bigg[ \hat{\Phi}_{0,n}^{(-\frac{k}{2}+1),ABCD},
{\cal O}_{0}^{(h_2),EFGH}(z,\bar{z}) \bigg]  =  -
\frac{\kappa_{0,0,+2}}{2} \,
\sum_{m=0}^{-k-1} \, \frac{(-1)^{-\frac{k}{2}+n}\,
(m+1)}{ (-k-m-1)!}
\label{anticomm25}
\\
&& \times
\frac{\Gamma(2(1-h_2)+1)}{\Gamma(2(1-h_2)+k+m+2)}
\,
\frac{ \bar{z}^{m+\frac{k}{2}+n+1}}{(\frac{-k}{2}-n)! \,
(m+\frac{k}{2}+n+1)!}
\, \ep^{ABCDEFGH}
\,
\bar{\pa}_{\bar{z}_2}^{m}\,
{\cal O}^{(-\frac{k}{2}+1+h_2)}_{-2}(z,\bar{z}).
\nonu
\eea


\subsection{
 The last twenty (anti)commutator relations of (\ref{HOcomm})
}

The (anti)commutators corresponding to
(\ref{twenty}) are obtained
\bea
&& \bigg[ \hat{\Phi}_{+\fft32,n}^{(-\frac{k}{2}+\fft14),A},
{\cal O}_{+2}^{(h_2)}(z,\bar{z}) \bigg]  =  -
\frac{\kappa_{+\fft32,+2,-\fft32}}{2} \,
\sum_{m=0}^{-k+\fft12} \, \frac{(-1)^{-\frac{k}{2}+n+\fft34}\,
(m+1)}{ (-k-m+\fft12)!}
\nonu \\
&& \times
\frac{\Gamma(2(-1-h_2)+1)}{\Gamma(2(-1-h_2)+k+m+
\fft12)}
\,
\frac{ \bar{z}^{m+\frac{k}{2}+n+\fft14}}{(\frac{\fft32-k}{2}-n)! \,
(m+\frac{k}{2}+n+\fft14)!}   \,
\bar{\pa}_{\bar{z}_2}^{m}\,
{\cal O}^{(-\frac{k}{2}+\fft14+h_2),A}_{+\fft32}(z,\bar{z}),
\nonu \\
&& \bigg[ \hat{\Phi}_{+1,n}^{(-\frac{k}{2}+\fft12),AB},
{\cal O}_{+2}^{(h_2)}(z,\bar{z}) \bigg]  =  -
\frac{\kappa_{+1,+2,-1}}{2} \,
\sum_{m=0}^{-k} \, \frac{(-1)^{-\frac{k}{2}+n+\fft12}\,
(m+1)}{ (-k-m)!}
\nonu \\
&& \times
\frac{\Gamma(2(-1-h_2)+1)}{\Gamma(2(-1-h_2)+k+m+1)}
\,
\frac{ \bar{z}^{m+\frac{k}{2}+n+\fft12}}{(\frac{1-k}{2}-n)! \,
(m+\frac{k}{2}+n+\fft12)!}   \,
\bar{\pa}_{\bar{z}_2}^{m}\,
{\cal O}^{(-\frac{k}{2}+\fft12+h_2),AB}_{+1}(z,\bar{z}),
\nonu \\
&& \bigg[ \hat{\Phi}_{+\fft12,n}^{(-\frac{k}{2}+\fft34),ABC},
{\cal O}_{+2}^{(h_2)}(z,\bar{z}) \bigg]  =  -
\frac{\kappa_{+\fft12,+2,-\fft12}}{2} \,
\sum_{m=0}^{-k-\fft12} \, \frac{(-1)^{-\frac{k}{2}+n+\fft14}\,
(m+1)}{ (-k-m-\fft12)!}
\nonu \\
&& \times
\frac{\Gamma(2(-1-h_2)+1)}{\Gamma(2(-1-h_2)+k+m+\fft32)}
\,
\frac{ \bar{z}^{m+\frac{k}{2}+n+\fft34}}{(\frac{\fft12-k}{2}-n)! \,
(m+\frac{k}{2}+n+\fft34)!}   \,
\bar{\pa}_{\bar{z}_2}^{m}\,
{\cal O}^{(-\frac{k}{2}+\fft34+h_2),ABC}_{+\fft12}(z,\bar{z}),
\nonu \\
&& \bigg[ \hat{\Phi}_{0,n}^{(-\frac{k}{2}+1),ABCD},
{\cal O}_{+2}^{(h_2)}(z,\bar{z}) \bigg]  =  -
\frac{\kappa_{0,+2,0}}{2} \,
\sum_{m=0}^{-k-1} \, \frac{(-1)^{-\frac{k}{2}+n}\,
(m+1)}{ (-k-m-1)!}
\nonu \\
&& \times
\frac{\Gamma(2(-1-h_2)+1)}{\Gamma(2(-1-h_2)+k+m+2)}
\,
\frac{ \bar{z}^{m+\frac{k}{2}+n+1}}{(\frac{-k}{2}-n)! \,
(m+\frac{k}{2}+n+1)!}   \,
\bar{\pa}_{\bar{z}_2}^{m}\,
{\cal O}^{(-\frac{k}{2}+1+h_2),ABCD}_{0}(z,\bar{z}),
\nonu \\
&& \bigg[ \hat{\Phi}_{ABC,-\fft12,n}^{(-\frac{k}{2}+\fft54)},
{\cal O}_{+2}^{(h_2)}(z,\bar{z}) \bigg]  =  -
\frac{\kappa_{-\fft12,+2,+\fft12}}{2} \,
\sum_{m=0}^{-k-\fft32} \, \frac{(-1)^{-\frac{k}{2}+n-\fft14}\,
(m+1)}{ (-k-m-\fft32)!}
\nonu \\
&& \times
\frac{\Gamma(2(-1-h_2)+1)}{\Gamma(2(-1-h_2)+k+m+\fft52)}
\,
\frac{ \bar{z}^{m+\frac{k}{2}+n+\fft54}}{(\frac{-\fft12-k}{2}-n)! \,
(m+\frac{k}{2}+n+\fft54)!}   \,
\bar{\pa}_{\bar{z}_2}^{m}\,
{\cal O}^{(-\frac{k}{2}+\fft54+h_2)}_{ABC,-\fft12}(z,\bar{z}),
\nonu \\
&& \bigg[ \hat{\Phi}_{AB,-1,n}^{(-\frac{k}{2}+\fft32)},
{\cal O}_{+2}^{(h_2)}(z,\bar{z}) \bigg]  =  -
\frac{\kappa_{-1,+2,+1}}{2} \,
\sum_{m=0}^{-k-2} \, \frac{(-1)^{-\frac{k}{2}+n-\fft12}\,
(m+1)}{ (-k-m-2)!}
\nonu \\
&& \times
\frac{\Gamma(2(-1-h_2)+1)}{\Gamma(2(-1-h_2)+k+m+3)}
\,
\frac{ \bar{z}^{m+\frac{k}{2}+n+\fft32}}{(\frac{-1-k}{2}-n)! \,
(m+\frac{k}{2}+n+\fft32)!}   \,
\bar{\pa}_{\bar{z}_2}^{m}\,
{\cal O}^{(-\frac{k}{2}+\fft32+h_2)}_{AB,-1}(z,\bar{z}),
\nonu \\
&& \bigg[ \hat{\Phi}_{A,-\fft32,n}^{(-\frac{k}{2}+\fft74)},
{\cal O}_{+2}^{(h_2)}(z,\bar{z}) \bigg]  =  -
\frac{\kappa_{-\fft32,+2,+\fft32}}{2} \,
\sum_{m=0}^{-k-\fft52} \, \frac{(-1)^{-\frac{k}{2}+n-\fft34}\,
(m+1)}{ (-k-m-\fft52)!}
\nonu \\
&& \times
\frac{\Gamma(2(-1-h_2)+1)}{\Gamma(2(-1-h_2)+k+m+\fft72)}
\,
\frac{ \bar{z}^{m+\frac{k}{2}+n+\fft74}}{(\frac{-\fft32-k}{2}-n)! \,
(m+\frac{k}{2}+n+\fft74)!}   \,
\bar{\pa}_{\bar{z}_2}^{m}\,
{\cal O}^{(-\frac{k}{2}+\fft74+h_2)}_{A,-\fft32}(z,\bar{z}),
\nonu \\
&& \bigg[ \hat{\Phi}_{-2,n}^{(-\frac{k}{2}+2)},
{\cal O}_{+2}^{(h_2)}(z,\bar{z}) \bigg]  =  -
\frac{\kappa_{-2,+2,+2}}{2} \,
\sum_{m=0}^{-k-3} \, \frac{(-1)^{-\frac{k}{2}+n-1}\,
(m+1)}{ (-k-m-3)!}
\nonu \\
&& \times
\frac{\Gamma(2(-1-h_2)+1)}{\Gamma(2(-1-h_2)+k+m+4)}
\,
\frac{ \bar{z}^{m+\frac{k}{2}+n+2}}{(\frac{-2-k}{2}-n)! \,
(m+\frac{k}{2}+n+2)!}   \,
\bar{\pa}_{\bar{z}_2}^{m}\,
{\cal O}^{(-\frac{k}{2}+2+h_2)}_{-2}(z,\bar{z}),
\nonu \\
&& \bigg[ \hat{\Phi}_{+1,n}^{(-\frac{k}{2}+\fft12),AB},
{\cal O}_{+\fft32}^{(h_2),C}(z,\bar{z}) \bigg]  =  -
\frac{\kappa_{+1,+\fft32,-\fft12}}{2} \,
\sum_{m=0}^{-k} \, \frac{(-1)^{-\frac{k}{2}+n+\fft12}\,
(m+1)}{ (-k-m)!}
\nonu \\
&& \times
\frac{\Gamma(2(-\fft12-h_2)+1)}{\Gamma(2(-\fft12-h_2)+k+m+1)}
\,
\frac{ \bar{z}^{m+\frac{k}{2}+n+\fft12}}{(\frac{1-k}{2}-n)! \,
(m+\frac{k}{2}+n+\fft12)!}
\, \bar{\pa}_{\bar{z}_2}^{m}\,
{\cal O}^{(-\frac{k}{2}+\fft12+h_2),ABC}_{+\fft12}(z,\bar{z}),
\nonu \\
&& \bigg\{ \hat{\Phi}_{+\fft12,n}^{(-\frac{k}{2}+\fft34),ABC},
{\cal O}_{+\fft32}^{(h_2),D}(z,\bar{z}) \bigg\}  =  -
\frac{\kappa_{+\fft12,+\fft32,0}}{2} \,
\sum_{m=0}^{-k-\fft12} \, \frac{(-1)^{-\frac{k}{2}+n+\fft14}\,
(m+1)}{ (-k-m-\fft12)!}
\nonu \\
&& \times
\frac{\Gamma(2(-\fft12-h_2)+1)}{\Gamma(2(-\fft12-h_2)+k+m+\tf32)}
\,
\frac{ \bar{z}^{m+\frac{k}{2}+n+\fft34}}{(\frac{\fft12-k}{2}-n)! \,
(m+\frac{k}{2}+n+\fft34)!}
\, \bar{\pa}_{\bar{z}_2}^{m}\,
{\cal O}^{(-\frac{k}{2}+\fft34+h_2),ABCD}_{0}(z,\bar{z}),
\nonu \\
&& \bigg[ \hat{\Phi}_{0,n}^{(-\frac{k}{2}+1),ABCD},
{\cal O}_{+\fft32}^{(h_2),E}(z,\bar{z}) \bigg]  =  -
\frac{\kappa_{0,+\fft32,\fft12}}{2} \,
\sum_{m=0}^{-k-1} \, \frac{(-1)^{-\frac{k}{2}+n}\,
(m+1)}{ (-k-m-1)!}
\nonu \\
&& \times
\frac{\Gamma(2(-\fft12-h_2)+1)}{\Gamma(2(-\fft12-h_2)+k+m+2)}
\,
\frac{ \bar{z}^{m+\frac{k}{2}+n+1}}{(\frac{-k}{2}-n)! \,
(m+\frac{k}{2}+n+1)!}
\, \frac{1}{3!}\, \ep^{ABCDEFGH} \nonu \\
&& \times \bar{\pa}_{\bar{z}_2}^{m}\,
{\cal O}^{(-\frac{k}{2}+1+h_2)}_{FGH,-\fft12}(z,\bar{z}),
\nonu \\
&& \bigg\{ \hat{\Phi}_{ABC,-\fft12,n}^{(-\frac{k}{2}+\fft54)},
{\cal O}_{+\fft32}^{(h_2),D}(z,\bar{z}) \bigg\}  =  -
\frac{\kappa_{-\fft12,+\fft32,+1}}{2} \,
\sum_{m=0}^{-k-\fft32} \, \frac{(-1)^{-\frac{k}{2}+n-\fft14}\,
(m+1)}{ (-k-m-\fft32)!}
\nonu \\
&& \times
\frac{\Gamma(2(-\fft12-h_2)+1)}{\Gamma(2(-\fft12-h_2)+k+m+\fft52)}
\,
\frac{ \bar{z}^{m+\frac{k}{2}+n+\fft54}}{(\frac{-\fft12-k}{2}-n)! \,
(m+\frac{k}{2}+n+\fft54)!}
\, 3 \,   \bar{\pa}_{\bar{z}_2}^{m}\,
{\cal O}^{(-\frac{k}{2}+\fft54+h_2)}_{[AB,-1}(z,\bar{z}) \,
\de^{D}_{C]},
\nonu \\
&& \bigg[ \hat{\Phi}_{AB,-1,n}^{(-\frac{k}{2}+\fft32)},
{\cal O}_{+\fft32}^{(h_2),C}(z,\bar{z}) \bigg]  =  -
\frac{\kappa_{-1,+\fft32,+\fft32}}{2} \,
\sum_{m=0}^{-k-2} \, \frac{(-1)^{-\frac{k}{2}+n-\fft12}\,
(m+1)}{ (-k-m-2)!}
\nonu \\
&& \times
\frac{\Gamma(2(-\fft12-h_2)+1)}{\Gamma(2(-\fft12-h_2)+k+m+3)}
\,
\frac{ \bar{z}^{m+\frac{k}{2}+n+\fft32}}{(\frac{-1-k}{2}-n)! \,
(m+\frac{k}{2}+n+\fft32)!}
\, 2 \,  \de^{C}_{[A}\,  \bar{\pa}_{\bar{z}_2}^{m}\,
{\cal O}^{(-\frac{k}{2}+\fft32+h_2)}_{B],-\fft32}(z,\bar{z}),
\nonu \\
&& \bigg\{ \hat{\Phi}_{A,-\fft32,n}^{(-\frac{k}{2}+\fft74)},
{\cal O}_{+\fft32}^{(h_2),B}(z,\bar{z}) \bigg\}  =  -
\frac{\kappa_{-\fft32,+\fft32,+2}}{2} \,
\sum_{m=0}^{-k-\fft52} \, \frac{(-1)^{-\frac{k}{2}+n-\fft34}\,
(m+1)}{ (-k-m-\fft52)!}
\nonu \\
&& \times
\frac{\Gamma(2(-\fft12-h_2)+1)}{\Gamma(2(-\fft12-h_2)+k+m+\fft72)}
\,
\frac{ \bar{z}^{m+\frac{k}{2}+n+\fft74}}{(\frac{-\fft32-k}{2}-n)! \,
(m+\frac{k}{2}+n+\fft74)!}
\,  \de^{B}_{A}\,  \bar{\pa}_{\bar{z}_2}^{m}\,
{\cal O}^{(-\frac{k}{2}+\fft74+h_2)}_{-2}(z,\bar{z}),
\nonu \\
&& \bigg[ \hat{\Phi}_{+\fft12,n}^{(-\frac{k}{2}+\fft34),ABC},
{\cal O}_{+1}^{(h_2),CD}(z,\bar{z}) \bigg]  =  -
\frac{\kappa_{+\fft12,+1,+\fft12}}{2} \,
\sum_{m=0}^{-k-\fft12} \, \frac{(-1)^{-\frac{k}{2}+n+\fft14}\,
(m+1)}{ (-k-m-\fft12)!}
\nonu \\
&& \times
\frac{\Gamma(2(-h_2)+1)}{\Gamma(2(-h_2)+k+m+\fft32)}
\,
\frac{ \bar{z}^{m+\frac{k}{2}+n+\fft34}}{(\frac{\fft12-k}{2}-n)! \,
(m+\frac{k}{2}+n+\fft34)!}
\, \frac{1}{3!}\, \ep^{ABCDEFGH} \nonu \\
&& \times 
\bar{\pa}_{\bar{z}_2}^{m}\,
{\cal O}^{(-\frac{k}{2}+\fft34+h_2)}_{FGH,-\fft12}(z,\bar{z}),
\nonu \\
&& \bigg[ \hat{\Phi}_{0,n}^{(-\frac{k}{2}+1),ABCD},
{\cal O}_{+1}^{(h_2),EF}(z,\bar{z}) \bigg]  =  -
\frac{\kappa_{0,+1,+1}}{2} \,
\sum_{m=0}^{-k-1} \, \frac{(-1)^{-\frac{k}{2}+n}\,
(m+1)}{ (-k-m-1)!}
\nonu \\
&& \times
\frac{\Gamma(2(-h_2)+1)}{\Gamma(2(-h_2)+k+m+2)}
\,
\frac{ \bar{z}^{m+\frac{k}{2}+n+1}}{(\frac{-k}{2}-n)! \,
(m+\frac{k}{2}+n+1)!}
\, \frac{1}{2!}\, \ep^{ABCDEFGH} \nonu \\
&& \times 
\bar{\pa}_{\bar{z}_2}^{m}\,
{\cal O}^{(-\frac{k}{2}+1+h_2)}_{GH,-1}(z,\bar{z}),
\nonu \\
&& \bigg[ \hat{\Phi}_{ABC,-\fft12,n}^{(-\frac{k}{2}+\fft54)},
{\cal O}_{+1}^{(h_2),DE}(z,\bar{z}) \bigg]  =  -
\frac{\kappa_{-\fft12,+1,+\fft32}}{2} \,
\sum_{m=0}^{-k-\fft32} \, \frac{(-1)^{-\frac{k}{2}+n-\fft14}\,
(m+1)}{ (-k-m-\fft32)!}
\nonu \\
&& \times
\frac{\Gamma(2(-h_2)+1)}{\Gamma(2(-h_2)+k+m+\fft52)}
\,
\frac{ \bar{z}^{m+\frac{k}{2}+n+\fft54}}{(\frac{-\fft12-k}{2}-n)! \,
(m+\frac{k}{2}+n+\fft54)!}
3!\, \de^{D}_{[A}
\,
\bar{\pa}_{\bar{z}_2}^{m}\,
{\cal O}^{(-\frac{k}{2}+\fft54+h_2)}_{B,-\fft32}(z,\bar{z}) \, \de^{E}_{C]},
\nonu \\
&& \bigg[ \hat{\Phi}_{AB,-1,n}^{(-\frac{k}{2}+\fft32)},
{\cal O}_{+1}^{(h_2),CD}(z,\bar{z}) \bigg]  =  -
\frac{\kappa_{-1,+1,+2}}{2} \,
\sum_{m=0}^{-k-2} \, \frac{(-1)^{-\frac{k}{2}+n-\fft12}\,
(m+1)}{ (-k-m-2)!}
\nonu \\
&& \times
\frac{\Gamma(2(-h_2)+1)}{\Gamma(2(-h_2)+k+m+3)}
\,
\frac{ \bar{z}^{m+\frac{k}{2}+n+\fft32}}{(\frac{-1-k}{2}-n)! \,
(m+\frac{k}{2}+n+\fft32)!}
\,  \de^{CD}_{AB}
\,
\bar{\pa}_{\bar{z}_2}^{m}\,
{\cal O}^{(-\frac{k}{2}+\fft32+h_2)}_{-2}(z,\bar{z}) \, \de^{E}_{C]},
\nonu \\
&& \bigg[ \hat{\Phi}_{0,n}^{(-\frac{k}{2}+1),ABCD},
{\cal O}_{+\fft12}^{(h_2),EFG}(z,\bar{z}) \bigg]  =  -
\frac{\kappa_{0,+\fft12,+\fft32}}{2} \,
\sum_{m=0}^{-k-1} \, \frac{(-1)^{-\frac{k}{2}+n}\,
(m+1)}{ (-k-m-1)!}
\nonu \\
&& \times
\frac{\Gamma(2(\tf12-h_2)+1)}{\Gamma(2(\tf12-h_2)+k+m+2)}
\,
\frac{ \bar{z}^{m+\frac{k}{2}+n+1}}{(\frac{-k}{2}-n)! \,
(m+\frac{k}{2}+n+1)!}
\,  \ep^{ABCDEFGH} \nonu \\
&& \times 
\bar{\pa}_{\bar{z}_2}^{m}\,
{\cal O}^{(-\frac{k}{2}+1+h_2)}_{H,-\fft32}(z,\bar{z}),
\nonu \\
&& \bigg\{ \hat{\Phi}_{ABC,-\fft12,n}^{(-\frac{k}{2}+\fft54)},
{\cal O}_{+\fft12}^{(h_2),DEF}(z,\bar{z}) \bigg\}  =  -
\frac{\kappa_{-\fft12,+\fft12,+2}}{2} \,
\sum_{m=0}^{-k-\fft32} \, \frac{(-1)^{-\frac{k}{2}+n-\fft14}\,
(m+1)}{ (-k-m-\fft32)!}
\label{anticomm20}
\\
&& \times
\frac{\Gamma(2(\fft12-h_2)+1)}{\Gamma(2(\fft12-h_2)+k+m+\fft52)}
\,
\frac{ \bar{z}^{m+\frac{k}{2}+n+\fft54}}{(\frac{-\fft12-k}{2}-n)! \,
(m+\frac{k}{2}+n+\fft54)!}
\, \de^{DEF}_{ABC}
\,
\bar{\pa}_{\bar{z}_2}^{m}\,
{\cal O}^{(-\frac{k}{2}+\fft54+h_2)}_{-2}(z,\bar{z}).
\nonu
\eea

\subsection{
The last twenty OPE relations of (\ref{HOOPEshadow})
from shadow transform
}

The next twenty OPEs corresponding to (\ref{HOOPEshadow})
are described as
\bea
&& \tilde{\Phi}_{+\fft32}^{(-\frac{k}{2}+\fft14),A}(z_1,\bar{z}_1) \,
{\cal O}_{+2}^{(h_2)}(z_2,\bar{z}_2)=-
\frac{\kappa_{+\frac{3}{2},+2,-\fft32}}{2}\,
\frac{\pi (-1)^{k+\fft32}}{(\fft52-k)!}\,
\frac{1}{z_{12}^{\fft12-k}}
\,
\sum_{m=0}^{-k+\fft12} \, (m+1)
\nonu \\
&& \times \frac{\Gamma(2(-1-h_2)+1)}
{\Gamma(2(-1-h_2)+k+m+\fft12)}
\, \bar{z}_{12}^{k-\fft32+m}\, \bar{\pa}_{\bar{z}_2}^m \,
{\cal O}_{+\fft32}^{(-\frac{k}{2}+\fft14+h_2),A}(z_2,\bar{z}_2)+\cdots,
\nonu \\
&& \tilde{\Phi}_{+1}^{(-\frac{k}{2}+\fft12),AB}(z_1,\bar{z}_1) \,
{\cal O}_{+2}^{(h_2)}(z_2,\bar{z}_2)=-
\frac{\kappa_{+1,+2,-1}}{2}\,
\frac{\pi (-1)^{k+1}}{(2-k)!}\,
\frac{1}{z_{12}^{1-k}}
\,
\sum_{m=0}^{-k} \, (m+1)
\nonu \\
&& \times \frac{\Gamma(2(-1-h_2)+1)}
{\Gamma(2(-1-h_2)+k+m+1)}
\, \bar{z}_{12}^{k-1+m}\,
\bar{\pa}_{\bar{z}_2}^m \,
{\cal O}_{+1}^{(-\frac{k}{2}+\fft12+h_2),AB}(z_2,\bar{z}_2)+\cdots,
\nonu \\
&& \tilde{\Phi}_{+\fft12}^{(-\frac{k}{2}+\fft34),ABC}(z_1,\bar{z}_1) \,
{\cal O}_{+2}^{(h_2)}(z_2,\bar{z}_2)=-
\frac{\kappa_{+\fft12,+2,-\fft12}}{2}\,
\frac{\pi (-1)^{k+\fft12}}{(\fft32-k)!}\,
\frac{1}{z_{12}^{\fft32-k}}
\,
\sum_{m=0}^{-k-\fft12} \, (m+1)
\nonu \\
&& \times \frac{\Gamma(2(-1-h_2)+1)}
{\Gamma(2(-1-h_2)+k+m+\tf32)}
\, \bar{z}_{12}^{k-\fft12+m}\,
\bar{\pa}_{\bar{z}_2}^m \,
{\cal O}_{\fft12}^{(-\frac{k}{2}+\fft34+h_2),ABC}(z_2,\bar{z}_2)+\cdots,
\nonu \\
&& \tilde{\Phi}_{0}^{(-\frac{k}{2}+1),ABCD}(z_1,\bar{z}_1) \,
{\cal O}_{+2}^{(h_2)}(z_2,\bar{z}_2)=-
\frac{\kappa_{0,+2,0}}{2}\,
\frac{\pi (-1)^{k}}{(1-k)!}\,
\frac{1}{z_{12}^{2-k}}
\,
\sum_{m=0}^{-k-1} \, (m+1)
\nonu \\
&& \times \frac{\Gamma(2(-1-h_2)+1)}
{\Gamma(2(-1-h_2)+k+m+2)}
\, \bar{z}_{12}^{k+m}\,
\bar{\pa}_{\bar{z}_2}^m \,
{\cal O}_{0}^{(-\frac{k}{2}+1+h_2),ABCD}(z_2,\bar{z}_2)+\cdots,
\nonu \\
&& \tilde{\Phi}_{ABC,-\fft12}^{(-\frac{k}{2}+\fft54)}(z_1,\bar{z}_1) \,
{\cal O}_{+2}^{(h_2)}(z_2,\bar{z}_2)=-
\frac{\kappa_{-\fft12,+2,+\fft12}}{2}\,
\frac{\pi (-1)^{k-\fft12}}{(\fft12-k)!}\,
\frac{1}{z_{12}^{\fft52-k}}
\,
\sum_{m=0}^{-k-\fft32} \, (m+1)
\nonu \\
&& \times \frac{\Gamma(2(-1-h_2)+1)}
{\Gamma(2(-1-h_2)+k+m+\tf52)}
\, \bar{z}_{12}^{k+\fft12+m}\,
\bar{\pa}_{\bar{z}_2}^m \,
{\cal O}_{ABC,-\fft12}^{(-\frac{k}{2}+\fft54+h_2)}(z_2,\bar{z}_2)+\cdots,
\nonu \\
&& \tilde{\Phi}_{AB,-1}^{(-\frac{k}{2}+\fft32)}(z_1,\bar{z}_1) \,
{\cal O}_{+2}^{(h_2)}(z_2,\bar{z}_2)=-
\frac{\kappa_{-1,+2,+1}}{2}\,
\frac{\pi (-1)^{k-1}}{(-k)!}\,
\frac{1}{z_{12}^{3-k}}
\,
\sum_{m=0}^{-k-2} \, (m+1)
\nonu \\
&& \times \frac{\Gamma(2(-1-h_2)+1)}
{\Gamma(2(-1-h_2)+k+m+3)}
\, \bar{z}_{12}^{k+1+m}\,
\bar{\pa}_{\bar{z}_2}^m \,
{\cal O}_{AB,-1}^{(-\frac{k}{2}+\fft32+h_2)}(z_2,\bar{z}_2)+\cdots,
\nonu \\
&& \tilde{\Phi}_{A,-\fft32}^{(-\frac{k}{2}+\fft74)}(z_1,\bar{z}_1) \,
{\cal O}_{+2}^{(h_2)}(z_2,\bar{z}_2)=-
\frac{\kappa_{-\fft32,+2,+\fft32}}{2}\,
\frac{\pi (-1)^{k-\fft32}}{(-\fft12-k)!}\,
\frac{1}{z_{12}^{\fft72-k}}
\,
\sum_{m=0}^{-k-\fft52} \, (m+1)
\nonu \\
&& \times \frac{\Gamma(2(-1-h_2)+1)}
{\Gamma(2(-1-h_2)+k+m+\tf72)}
\, \bar{z}_{12}^{k+\fft32+m}\,
\bar{\pa}_{\bar{z}_2}^m \,
{\cal O}_{A,-\fft32}^{(-\frac{k}{2}+\fft74+h_2)}(z_2,\bar{z}_2)+\cdots,
\nonu \\
&& \tilde{\Phi}_{-2}^{(-\frac{k}{2}+2)}(z_1,\bar{z}_1) \,
{\cal O}_{+2}^{(h_2)}(z_2,\bar{z}_2)=-
\frac{\kappa_{-2,+2,+2}}{2}\,
\frac{\pi (-1)^{k}}{(-1-k)!}\,
\frac{1}{z_{12}^{4-k}}
\,
\sum_{m=0}^{-k-3} \, (m+1)
\nonu \\
&& \times \frac{\Gamma(2(-1-h_2)+1)}
{\Gamma(2(-1-h_2)+k+m+4)}
\, \bar{z}_{12}^{k+2+m}\,
\bar{\pa}_{\bar{z}_2}^m \,
{\cal O}_{-2}^{(-\frac{k}{2}+2+h_2)}(z_2,\bar{z}_2)+\cdots,
\nonu \\
&& \tilde{\Phi}_{+1}^{(-\frac{k}{2}+\fft12),AB}(z_1,\bar{z}_1) \,
{\cal O}_{+\fft32}^{(h_2),C}(z_2,\bar{z}_2)=-
\frac{\kappa_{+1,+\fft32,-\fft12}}{2}\,
\frac{\pi (-1)^{k+1}}{(2-k)!}\,
\frac{1}{z_{12}^{1-k}}
\,
\sum_{m=0}^{-k} \, (m+1)
\nonu \\
&& \times \frac{\Gamma(2(-\fft12-h_2)+1)}
{\Gamma(2(-\fft12-h_2)+k+m+1)}
\, \bar{z}_{12}^{k-1+m}\,
\bar{\pa}_{\bar{z}_2}^m \,
{\cal O}_{+\fft12}^{(-\frac{k}{2}+\fft12+h_2),ABC}(z_2,\bar{z}_2)+\cdots,
\nonu \\
&& \tilde{\Phi}_{+\fft12}^{(-\frac{k}{2}+\fft34),ABC}(z_1,\bar{z}_1) \,
{\cal O}_{+\fft32}^{(h_2),D}(z_2,\bar{z}_2)=-
\frac{\kappa_{+\fft12,+\fft32,0}}{2}\,
\frac{\pi (-1)^{k+\fft12}}{(\fft32-k)!}\,
\frac{1}{z_{12}^{\fft32-k}}
\,
\sum_{m=0}^{-k-\fft12} \, (m+1)
\nonu \\
&& \times \frac{\Gamma(2(-\fft12-h_2)+1)}
{\Gamma(2(-\fft12-h_2)+k+m+\tf32)}
\, \bar{z}_{12}^{k-\fft12+m}\,
\bar{\pa}_{\bar{z}_2}^m \,
{\cal O}_{0}^{(-\frac{k}{2}+\fft34+h_2),ABCD}(z_2,\bar{z}_2)+\cdots,
\nonu \\
&& \tilde{\Phi}_{0}^{(-\frac{k}{2}+1),ABCD}(z_1,\bar{z}_1) \,
{\cal O}_{+\fft32}^{(h_2),E}(z_2,\bar{z}_2)=-
\frac{\kappa_{0,+\fft32,\fft12}}{2}\,
\frac{\pi (-1)^{k}}{(1-k)!}\,
\frac{1}{z_{12}^{2-k}}
\,
\sum_{m=0}^{-k-1} \, (m+1)
\nonu \\
&& \times \frac{\Gamma(2(-\fft12-h_2)+1)}
{\Gamma(2(-\fft12-h_2)+k+m+2)}
\, \bar{z}_{12}^{k+m}\,
\frac{1}{3!}\, \ep^{ABCDEFGH}\,
\bar{\pa}_{\bar{z}_2}^m \,
{\cal O}_{FGH,-\fft12}^{(-\frac{k}{2}+1+h_2)}(z_2,\bar{z}_2)+\cdots,
\nonu \\
&& \tilde{\Phi}_{ABC,-\fft12}^{(-\frac{k}{2}+\fft54)}(z_1,\bar{z}_1) \,
{\cal O}_{+\fft32}^{(h_2),D}(z_2,\bar{z}_2)=-
\frac{\kappa_{-\fft12,+\fft32,+1}}{2}\,
\frac{\pi (-1)^{k-\fft12}}{(\fft12-k)!}\,
\frac{1}{z_{12}^{\fft52-k}}
\,
\sum_{m=0}^{-k-\fft32} \, (m+1)
\nonu \\
&& \times \frac{\Gamma(2(-\fft12-h_2)+1)}
{\Gamma(2(-\fft12-h_2)+k+m+\tf52)}
\, \bar{z}_{12}^{k+\fft12+m}\,
3\, 
\bar{\pa}_{\bar{z}_2}^m \,
{\cal O}_{[AB,-1}^{(-\frac{k}{2}+\fft54+h_2)}(z_2,\bar{z}_2)\,
\de^{D}_{C]}+\cdots,
\nonu \\
&& \tilde{\Phi}_{AB,-1}^{(-\frac{k}{2}+\fft32)}(z_1,\bar{z}_1) \,
{\cal O}_{+\fft32}^{(h_2),C}(z_2,\bar{z}_2)=-
\frac{\kappa_{-1,+\fft32,+\fft32}}{2}\,
\frac{\pi (-1)^{k-1}}{(-k)!}\,
\frac{1}{z_{12}^{3-k}}
\,
\sum_{m=0}^{-k-2} \, (m+1)
\nonu \\
&& \times \frac{\Gamma(2(-\fft12-h_2)+1)}
{\Gamma(2(-\fft12-h_2)+k+m+3)}
\, \bar{z}_{12}^{k+1+m}\,
2\, \de^{C}_{[A} 
\bar{\pa}_{\bar{z}_2}^m \,
{\cal O}_{B],-\fft32}^{(-\frac{k}{2}+\fft32+h_2)}(z_2,\bar{z}_2)+\cdots,
\nonu \\
&& \tilde{\Phi}_{A,-\fft32}^{(-\frac{k}{2}+\fft74)}(z_1,\bar{z}_1) \,
{\cal O}_{+\fft32}^{(h_2),B}(z_2,\bar{z}_2)=-
\frac{\kappa_{-\fft32,+\fft32,+2}}{2}\,
\frac{\pi (-1)^{k-\fft32}}{(-\fft12-k)!}\,
\frac{1}{z_{12}^{\fft72-k}}
\,
\sum_{m=0}^{-k-\fft52} \, (m+1)
\nonu \\
&& \times \frac{\Gamma(2(-\fft12-h_2)+1)}
{\Gamma(2(-\fft12-h_2)+k+m+\tf72)}
\, \bar{z}_{12}^{k+\fft32+m}\,
\de^{B}_{A} 
\bar{\pa}_{\bar{z}_2}^m \,
{\cal O}_{-2}^{(-\frac{k}{2}+\fft74+h_2)}(z_2,\bar{z}_2)+\cdots,
\nonu \\
&& \tilde{\Phi}_{+\fft12}^{(-\frac{k}{2}+\fft34),ABC}(z_1,\bar{z}_1) \,
{\cal O}_{+1}^{(h_2),DE}(z_2,\bar{z}_2)=-
\frac{\kappa_{+\fft12,+1,+\fft12}}{2}\,
\frac{\pi (-1)^{k+\fft12}}{(\fft32-k)!}\,
\frac{1}{z_{12}^{\fft32-k}}
\,
\sum_{m=0}^{-k-\fft12} \, (m+1)
\nonu \\
&& \times \frac{\Gamma(2(-h_2)+1)}
{\Gamma(2(-h_2)+k+m+\tf32)}
\, \bar{z}_{12}^{k-\fft12+m}\,
\frac{1}{3!}\, \ep^{ABCDEFGH}\,
\bar{\pa}_{\bar{z}_2}^m \,
{\cal O}_{FGH,-\fft12}^{(-\frac{k}{2}+\fft34+h_2)}(z_2,\bar{z}_2)+\cdots,
\nonu \\
&& \tilde{\Phi}_{0}^{(-\frac{k}{2}+1),ABCD}(z_1,\bar{z}_1) \,
{\cal O}_{+1}^{(h_2),EF}(z_2,\bar{z}_2)=-
\frac{\kappa_{0,+1,+1}}{2}\,
\frac{\pi (-1)^{k}}{(1-k)!}\,
\frac{1}{z_{12}^{2-k}}
\,
\sum_{m=0}^{-k-1} \, (m+1)
\nonu \\
&& \times \frac{\Gamma(2(-h_2)+1)}
{\Gamma(2(-h_2)+k+m+2)}
\, \bar{z}_{12}^{k+m}\,
\frac{1}{2!}\, \ep^{ABCDEFGH}\,
\bar{\pa}_{\bar{z}_2}^m \,
{\cal O}_{GH,-1}^{(-\frac{k}{2}+1+h_2)}(z_2,\bar{z}_2)+\cdots,
\nonu \\
&& \tilde{\Phi}_{ABC,-\fft12}^{(-\frac{k}{2}+\fft54)}(z_1,\bar{z}_1) \,
{\cal O}_{+1}^{(h_2),DE}(z_2,\bar{z}_2)=-
\frac{\kappa_{-\fft12,+1,+\fft32}}{2}\,
\frac{\pi (-1)^{k-\fft12}}{(\fft12-k)!}\,
\frac{1}{z_{12}^{\fft52-k}}
\,
\sum_{m=0}^{-k-\fft32} \, (m+1)
\nonu \\
&& \times \frac{\Gamma(2(-h_2)+1)}
{\Gamma(2(-h_2)+k+m+\tf52)}
\, \bar{z}_{12}^{k+\fft12+m}\, 3!\, 
\de^{D}_{[A}\,
\bar{\pa}_{\bar{z}_2}^m \,
{\cal O}_{B,-\fft32}^{(-\frac{k}{2}+\fft54+h_2)}(z_2,\bar{z}_2) \,
\de^{E}_{C]}+\cdots,
\nonu \\
&& \tilde{\Phi}_{AB,-1}^{(-\frac{k}{2}+\fft32)}(z_1,\bar{z}_1) \,
{\cal O}_{+1}^{(h_2),CD}(z_2,\bar{z}_2)=-
\frac{\kappa_{-1,+1,+2}}{2}\,
\frac{\pi (-1)^{k-1}}{(-k)!}\,
\frac{1}{z_{12}^{3-k}}
\,
\sum_{m=0}^{-k-2} \, (m+1)
\nonu \\
&& \times \frac{\Gamma(2(-h_2)+1)}
{\Gamma(2(-h_2)+k+m+3)}
\, \bar{z}_{12}^{k+1+m}\,
\de^{CD}_{AB}\,
\bar{\pa}_{\bar{z}_2}^m \,
{\cal O}_{-2}^{(-\frac{k}{2}+\fft32+h_2)}(z_2,\bar{z}_2)+\cdots,
\nonu \\
&& \tilde{\Phi}_{0}^{(-\frac{k}{2}+1),ABCD}(z_1,\bar{z}_1) \,
{\cal O}_{+\fft12}^{(h_2),EFG}(z_2,\bar{z}_2)=-
\frac{\kappa_{0,+\fft12,+\fft32}}{2}\,
\frac{\pi (-1)^{k}}{(1-k)!}\,
\frac{1}{z_{12}^{2-k}}
\,
\sum_{m=0}^{-k-1} \, (m+1)
\nonu \\
&& \times \frac{\Gamma(2(\fft12-h_2)+1)}
{\Gamma(2(\fft12-h_2)+k+m+2)}
\, \bar{z}_{12}^{k+m}\,
\ep^{ABCDEFGH}\,
\bar{\pa}_{\bar{z}_2}^m \,
{\cal O}_{H,-\fft32}^{(-\frac{k}{2}+1+h_2)}(z_2,\bar{z}_2)+\cdots,
\nonu \\
&& \tilde{\Phi}_{ABC,-\fft12}^{(-\frac{k}{2}+\fft54)}(z_1,\bar{z}_1) \,
{\cal O}_{+\fft12}^{(h_2),DEF}(z_2,\bar{z}_2)=-
\frac{\kappa_{-\fft12,+\fft12,+2}}{2}\,
\frac{\pi (-1)^{k-\fft12}}{(\fft12-k)!}\,
\frac{1}{z_{12}^{\fft52-k}}
\,
\sum_{m=0}^{-k-\fft32} \, (m+1)
\nonu \\
&& \times \frac{\Gamma(2(\fft12-h_2)+1)}
{\Gamma(2(\fft12-h_2)+k+m+\tf52)}
\, \bar{z}_{12}^{k+\fft12+m}\,
\de^{DEF}_{ABC}\,
\bar{\pa}_{\bar{z}_2}^m \,
{\cal O}_{-2}^{(-\frac{k}{2}+\fft54+h_2)}(z_2,\bar{z}_2)+\cdots.
\label{twentysection6}
\eea

\subsection{
The last twenty OPE relations of (\ref{HOOPEholo}) 
}

The remaining twenty OPEs corresponding to (\ref{HOOPEholo})
can be described as
\bea
&& \bar{\Phi}_{A,-\fft32}^{(-\frac{k}{2}+\fft14)}(z_1,\bar{z}_1) \,
{\cal O}_{-2}^{(h_2)}(z_2,\bar{z}_2)=-
\frac{\bar{\kappa}_{-\frac{3}{2},-2,+\fft32}}{2}\, \frac{1}{\bar{z}_{12}} \,
\sum_{m=0}^{-k+\fft12} \, \frac{(-1)^{-k-m+\fft12}}{m!\, (-k-m+\fft12)!}
\nonu \\
&& \times \frac{\Gamma(2(-1-h_2)+1)}
{\Gamma(2(-1-h_2)+k+m+\fft12)}
\, {z}_{12}^{m+1}\, {\pa}_{{z}_2}^m \,
{\cal O}_{A,-\fft32}^{(-\frac{k}{2}+\fft14+h_2)}(z_2,\bar{z}_2)+\cdots,
\nonu \\
&& \bar{\Phi}_{AB,-1}^{(-\frac{k}{2}+\fft12)}(z_1,\bar{z}_1) \,
{\cal O}_{-2}^{(h_2)}(z_2,\bar{z}_2)=-
\frac{\bar{\kappa}_{-1,-2,+1}}{2}\, \frac{1}{\bar{z}_{12}} \,
\sum_{m=0}^{-k} \, \frac{(-1)^{-k-m}}{m!\, (-k-m)!}
\nonu \\
&& \times \frac{\Gamma(2(-1-h_2)+1)}
{\Gamma(2(-1-h_2)+k+m+1)}
\, {z}_{12}^{m+1}\,
{\pa}_{{z}_2}^m \,
{\cal O}_{AB,-1}^{(-\frac{k}{2}+\fft12+h_2)}(z_2,\bar{z}_2)+\cdots,
\nonu \\
&& \bar{\Phi}_{ABC,-\fft12}^{(-\frac{k}{2}+\fft34)}(z_1,\bar{z}_1) \,
{\cal O}_{-2}^{(h_2)}(z_2,\bar{z}_2)=-
\frac{\bar{\kappa}_{-\fft12,-2,+\fft12}}{2}\, \frac{1}{\bar{z}_{12}} \,
\sum_{m=0}^{-k-\fft12} \, \frac{(-1)^{-k-m-\fft12}}{m!\,
(-k-m-\fft12)!}
\nonu \\
&& \times \frac{\Gamma(2(-1-h_2)+1)}
{\Gamma(2(-1-h_2)+k+m+\tf32)}
\, {z}_{12}^{m+1}\,
{\pa}_{{z}_2}^m \,
{\cal O}_{ABC,-\fft12}^{(-\frac{k}{2}+\fft34+h_2)}(z_2,\bar{z}_2)+\cdots,
\nonu \\
&& \bar{\Phi}_{0}^{(-\frac{k}{2}+1),ABCD}(z_1,\bar{z}_1) \,
{\cal O}_{-2}^{(h_2)}(z_2,\bar{z}_2)=-
\frac{\bar{\kappa}_{0,-2,0}}{2}\, \frac{1}{\bar{z}_{12}} \,
\sum_{m=0}^{-k-1} \, \frac{(-1)^{-k-m-1}}{m!\,
(-k-m-1)!}
\nonu \\
&& \times \frac{\Gamma(2(-1-h_2)+1)}
{\Gamma(2(-1-h_2)+k+m+2)}
\, {z}_{12}^{m+1}\,
{\pa}_{{z}_2}^m \,
{\cal O}_{0}^{(-\frac{k}{2}+1+h_2),ABCD}(z_2,\bar{z}_2)+\cdots,
\nonu \\
&& \bar{\Phi}_{+\fft12}^{(-\frac{k}{2}+\fft54),ABC}(z_1,\bar{z}_1) \,
{\cal O}_{-2}^{(h_2)}(z_2,\bar{z}_2)=-
\frac{\bar{\kappa}_{+\fft12,-2,-\fft12}}{2}\, \frac{1}{\bar{z}_{12}} \,
\sum_{m=0}^{-k-\fft32} \, \frac{(-1)^{-k-m-\fft32}}{m!\,
(-k-m-\fft32)!}
\nonu \\
&& \times \frac{\Gamma(2(-1-h_2)+1)}
{\Gamma(2(-1-h_2)+k+m+\tf52)}
\, {z}_{12}^{m+1}\,
{\pa}_{{z}_2}^m \,
{\cal O}_{+\fft12}^{(-\frac{k}{2}+\fft54+h_2),ABC}(z_2,\bar{z}_2)+\cdots,
\nonu \\
&& \bar{\Phi}_{+1}^{(-\frac{k}{2}+\fft32),AB}(z_1,\bar{z}_1) \,
{\cal O}_{-2}^{(h_2)}(z_2,\bar{z}_2)=-
\frac{\bar{\kappa}_{+1,-2,-1}}{2}\, \frac{1}{\bar{z}_{12}} \,
\sum_{m=0}^{-k-2} \, \frac{(-1)^{-k-m-2}}{m!\,
(-k-m-2)!}
\nonu \\
&& \times \frac{\Gamma(2(-1-h_2)+1)}
{\Gamma(2(-1-h_2)+k+m+3)}
\, {z}_{12}^{m+1}\,
{\pa}_{{z}_2}^m \,
{\cal O}_{+1}^{(-\frac{k}{2}+\fft32+h_2),AB}(z_2,\bar{z}_2)+\cdots,
\nonu \\
&& \bar{\Phi}_{+\fft32}^{(-\frac{k}{2}+\fft74),A}(z_1,\bar{z}_1) \,
{\cal O}_{-2}^{(h_2)}(z_2,\bar{z}_2)=-
\frac{\bar{\kappa}_{+\fft32,-2,-\fft32}}{2}\, \frac{1}{\bar{z}_{12}} \,
\sum_{m=0}^{-k-\fft52} \, \frac{(-1)^{-k-m-\fft52}}{m!\,
(-k-m-\fft52)!}
\nonu \\
&& \times \frac{\Gamma(2(-1-h_2)+1)}
{\Gamma(2(-1-h_2)+k+m+\tf72)}
\, {z}_{12}^{m+1}\,
{\pa}_{{z}_2}^m \,
{\cal O}_{+\fft32}^{(-\frac{k}{2}+\fft74+h_2),A}(z_2,\bar{z}_2)+\cdots,
\nonu \\
&& \bar{\Phi}_{+2}^{(-\frac{k}{2}+2)}(z_1,\bar{z}_1) \,
{\cal O}_{-2}^{(h_2)}(z_2,\bar{z}_2)=-
\frac{\bar{\kappa}_{+2,-2,-2}}{2}\, \frac{1}{\bar{z}_{12}} \,
\sum_{m=0}^{-k-3} \, \frac{(-1)^{-k-m-3}}{m!\,
(-k-m-3)!}
\nonu \\
&& \times \frac{\Gamma(2(-1-h_2)+1)}
{\Gamma(2(-1-h_2)+k+m+4)}
\, {z}_{12}^{m+1}\,
{\pa}_{{z}_2}^m \,
{\cal O}_{+2}^{(-\frac{k}{2}+2+h_2)}(z_2,\bar{z}_2)+\cdots,
\nonu \\
&& \bar{\Phi}_{AB,-1}^{(-\frac{k}{2}+\fft12)}(z_1,\bar{z}_1) \,
{\cal O}_{C,-\fft32}^{(h_2)}(z_2,\bar{z}_2)=-
\frac{\bar{\kappa}_{-1,-\fft32,+\fft12}}{2}\, \frac{1}{\bar{z}_{12}} \,
\sum_{m=0}^{-k} \, \frac{(-1)^{-k-m}}{m!\,
(-k-m)!}
\nonu \\
&& \times \frac{\Gamma(2(-\fft12-h_2)+1)}
{\Gamma(2(-\fft12-h_2)+k+m+1)}
\, {z}_{12}^{m+1}\,
{\pa}_{{z}_2}^m \,
{\cal O}_{ABC,-\fft12}^{(-\frac{k}{2}+\fft12+h_2)}(z_2,\bar{z}_2)+\cdots,
\nonu \\
&& \bar{\Phi}_{ABC,-\fft12}^{(-\frac{k}{2}+\fft34)}(z_1,\bar{z}_1) \,
{\cal O}_{D,-\fft32}^{(h_2)}(z_2,\bar{z}_2)=-
\frac{\bar{\kappa}_{-\fft12,-\fft32,0}}{2}\, \frac{1}{\bar{z}_{12}} \,
\sum_{m=0}^{-k-\fft12} \, \frac{(-1)^{-k-m-\fft12}}{m!\,
(-k-m-\fft12)!}
\nonu \\
&& \times \frac{\Gamma(2(-\fft12-h_2)+1)}
{\Gamma(2(-\fft12-h_2)+k+m+\tf32)}
\, {z}_{12}^{m+1}\, \frac{1}{4!}\, \ep_{ABCDEFGH}\,
{\pa}_{{z}_2}^m \,
{\cal O}_{0}^{(-\frac{k}{2}+\fft34+h_2),EFGH}(z_2,\bar{z}_2)+\cdots,
\nonu \\
&& \bar{\Phi}_{0}^{(-\frac{k}{2}+1),ABCD}(z_1,\bar{z}_1) \,
{\cal O}_{E,-\fft32}^{(h_2)}(z_2,\bar{z}_2)=-
\frac{\bar{\kappa}_{0,-\fft32,-\fft12}}{2}\, \frac{1}{\bar{z}_{12}} \,
\sum_{m=0}^{-k-1} \, \frac{(-1)^{-k-m-1}}{m!\,
(-k-m-1)!}
\nonu \\
&& \times \frac{\Gamma(2(-\fft12-h_2)+1)}
{\Gamma(2(-\fft12-h_2)+k+m+2)}
\, {z}_{12}^{m+1}\,\frac{1}{4!}\,
\frac{1}{3!}\, \ep^{ABCDIJKL}\, \ep_{EFGHIJKL}
{\pa}_{{z}_2}^m \,
{\cal O}_{+\fft12}^{(-\frac{k}{2}+1+h_2),FGH}(z_2,\bar{z}_2)
\nonu \\
&& +\cdots,
\nonu \\
&& \bar{\Phi}_{+\fft12}^{(-\frac{k}{2}+\fft54),ABC}(z_1,\bar{z}_1) \,
{\cal O}_{D,-\fft32}^{(h_2)}(z_2,\bar{z}_2)=-
\frac{\bar{\kappa}_{+\fft12,-\fft32,-1}}{2}\, \frac{1}{\bar{z}_{12}} \,
\sum_{m=0}^{-k-\fft32} \, \frac{(-1)^{-k-m-\fft32}}{m!\,
(-k-m-\fft32)!}
\nonu \\
&& \times \frac{\Gamma(2(-\fft12-h_2)+1)}
{\Gamma(2(-\fft12-h_2)+k+m+\tf52)}
\, {z}_{12}^{m+1}\,
3\, 
{\pa}_{{z}_2}^m \,
{\cal O}_{+1}^{(-\frac{k}{2}+\fft54+h_2),[AB}(z_2,\bar{z}_2)\,
\de^{C]}_{D}+\cdots,
\nonu \\
&& \bar{\Phi}_{+1}^{(-\frac{k}{2}+\fft32),AB}(z_1,\bar{z}_1) \,
{\cal O}_{C,-\fft32}^{(h_2)}(z_2,\bar{z}_2)=-
\frac{\bar{\kappa}_{+1,-\fft32,-\fft32}}{2}\, \frac{1}{\bar{z}_{12}} \,
\sum_{m=0}^{-k-2} \, \frac{(-1)^{-k-m-2}}{m!\,
(-k-m-2)!}
\nonu \\
&& \times \frac{\Gamma(2(-\fft12-h_2)+1)}
{\Gamma(2(-\fft12-h_2)+k+m+3)}
\, {z}_{12}^{m+1}\,
2\, \de^{[A}_{C} 
{\pa}_{{z}_2}^m \,
{\cal O}_{+\fft32}^{(-\frac{k}{2}+\fft32+h_2),B]}(z_2,\bar{z}_2)+\cdots,
\nonu \\
&& \bar{\Phi}_{+\fft32}^{(-\frac{k}{2}+\fft74),A}(z_1,\bar{z}_1) \,
{\cal O}_{B,-\fft32}^{(h_2)}(z_2,\bar{z}_2)=-
\frac{\bar{\kappa}_{+\fft32,-\fft32,-2}}{2}\, \frac{1}{\bar{z}_{12}} \,
\sum_{m=0}^{-k-\fft52} \, \frac{(-1)^{-k-m-\fft52}}{m!\,
(-k-m-\fft52)!}
\nonu \\
&& \times \frac{\Gamma(2(-\fft12-h_2)+1)}
{\Gamma(2(-\fft12-h_2)+k+m+\tf72)}
\, {z}_{12}^{m+1}\,
\de^{A}_{B} 
{\pa}_{{z}_2}^m \,
{\cal O}_{+2}^{(-\frac{k}{2}+\fft74+h_2)}(z_2,\bar{z}_2)+\cdots,
\nonu \\
&& \bar{\Phi}_{ABC,-\fft12}^{(-\frac{k}{2}+\fft34)}(z_1,\bar{z}_1) \,
{\cal O}_{DE,-1}^{(h_2)}(z_2,\bar{z}_2)=-
\frac{\bar{\kappa}_{-\fft12,-1,-\fft12}}{2}\, \frac{1}{\bar{z}_{12}} \,
\sum_{m=0}^{-k-\fft12} \, \frac{(-1)^{-k-m-\fft12}}{m!\,
(-k-m-\fft12)!}
\nonu \\
&& \times \frac{\Gamma(2(-h_2)+1)}
{\Gamma(2(-h_2)+k+m+\tf32)}
\, {z}_{12}^{m+1}\,
\frac{1}{3!}\, \ep_{ABCDEFGH}\,
{\pa}_{{z}_2}^m \,
{\cal O}_{+\fft12}^{(-\frac{k}{2}+\fft34+h_2),FGH}(z_2,\bar{z}_2)+\cdots,
\nonu \\
&& \bar{\Phi}_{0}^{(-\frac{k}{2}+1),ABCD}(z_1,\bar{z}_1) \,
{\cal O}_{EF,-1}^{(h_2)}(z_2,\bar{z}_2)=-
\frac{\bar{\kappa}_{0,-1,-1}}{2}\, \frac{1}{\bar{z}_{12}} \,
\sum_{m=0}^{-k-1} \, \frac{(-1)^{-k-m-1}}{m!\,
(-k-m-1)!}
\nonu \\
&& \times \frac{\Gamma(2(-h_2)+1)}
{\Gamma(2(-h_2)+k+m+2)}
\, {z}_{12}^{m+1}\, \frac{1}{4!}\,
\frac{1}{2!}\, \ep^{ABCDIJKL}\, \ep_{EFGHIJKL}
{\pa}_{{z}_2}^m \,
{\cal O}_{+1}^{(-\frac{k}{2}+1+h_2),GH}(z_2,\bar{z}_2)+\cdots,
\nonu \\
&& \bar{\Phi}_{+\fft12}^{(-\frac{k}{2}+\fft54),ABC}(z_1,\bar{z}_1) \,
{\cal O}_{DE,-1}^{(h_2)}(z_2,\bar{z}_2)=-
\frac{\bar{\kappa}_{+\fft12,-1,-\fft32}}{2}\, \frac{1}{\bar{z}_{12}} \,
\sum_{m=0}^{-k-\fft32} \, \frac{(-1)^{-k-m-\fft32}}{m!\,
(-k-m-\fft32)!}
\nonu \\
&& \times \frac{\Gamma(2(-h_2)+1)}
{\Gamma(2(-h_2)+k+m+\tf52)}
\, {z}_{12}^{m+1}\, 3!\, 
\de^{[A}_{D}\,
{\pa}_{{z}_2}^m \,
{\cal O}_{+\fft32}^{(-\frac{k}{2}+\fft54+h_2),B}(z_2,\bar{z}_2) \,
\de^{C]}_{E}+\cdots,
\nonu \\
&& \bar{\Phi}_{+1}^{(-\frac{k}{2}+\fft32),AB}(z_1,\bar{z}_1) \,
{\cal O}_{CD,-1}^{(h_2)}(z_2,\bar{z}_2)=-
\frac{\bar{\kappa}_{+1,-1,-2}}{2}\, \frac{1}{\bar{z}_{12}} \,
\sum_{m=0}^{-k-2} \, \frac{(-1)^{-k-m-2}}{m!\,
(-k-m-2)!}
\nonu \\
&& \times \frac{\Gamma(2(-h_2)+1)}
{\Gamma(2(-h_2)+k+m+3)}
\, {z}_{12}^{m+1}\,
\de^{AB}_{CD}\,
{\pa}_{{z}_2}^m \,
{\cal O}_{+2}^{(-\frac{k}{2}+\fft32+h_2)}(z_2,\bar{z}_2)+\cdots,
\nonu \\
&& \bar{\Phi}_{0}^{(-\frac{k}{2}+1),ABCD}(z_1,\bar{z}_1) \,
{\cal O}_{EFG,-\fft12}^{(h_2)}(z_2,\bar{z}_2)=-
\frac{\bar{\kappa}_{0,-\fft12,-\fft32}}{2}\, \frac{1}{\bar{z}_{12}} \,
\sum_{m=0}^{-k-1} \, \frac{(-1)^{-k-m-1}}{m!\,
(-k-m-1)!}
\nonu \\
&& \times \frac{\Gamma(2(\fft12-h_2)+1)}
{\Gamma(2(\fft12-h_2)+k+m+2)}
\, {z}_{12}^{m+1}\, \frac{1}{4!}\,
\ep^{ABCDIJKL}\,\ep_{EFGHIJKL}
{\pa}_{{z}_2}^m \,
{\cal O}_{+\fft32}^{(-\frac{k}{2}+1+h_2),H}(z_2,\bar{z}_2)+\cdots,
\nonu \\
&& \bar{\Phi}_{+\fft12}^{(-\frac{k}{2}+\fft54),ABC}(z_1,\bar{z}_1) \,
{\cal O}_{DEF,-\fft12}^{(h_2)}(z_2,\bar{z}_2)=-
\frac{\bar{\kappa}_{+\fft12,-\fft12,-2}}{2}\, \frac{1}{\bar{z}_{12}} \,
\sum_{m=0}^{-k-\fft32} \, \frac{(-1)^{-k-m-\fft32}}{m!\,
(-k-m-\fft32)!}
\nonu \\
&& \times \frac{\Gamma(2(\fft12-h_2)+1)}
{\Gamma(2(\fft12-h_2)+k+m+\tf52)}
\, {z}_{12}^{m+1}\,
\de^{ABC}_{DEF}\,
{\pa}_{{z}_2}^m \,
{\cal O}_{+2}^{(-\frac{k}{2}+\fft54+h_2)}(z_2,\bar{z}_2)+\cdots.
\label{othertwenty}
\eea
The right-hand sides of the
tenth, the eleventh, the sixteenth, and the
nineteenth relations
of (\ref{othertwenty}) are different from those
in (\ref{twenty}).

\subsection{
 The first twenty-five (anti)commutator relations of (\ref{HOcomm1})
}

The (anti)commutators corresponding to
(\ref{HOcomm1}) are obtained
\bea
&& \bigg[ \hat{\bar{\Phi}}_{-2,n}^{(-\frac{k}{2})},
{\cal O}_{-2}^{(h_2)}(z,\bar{z}) \bigg]  =  -
\frac{\bar{\kappa}_{-2,-2,+2}}{2} \,
\sum_{m=0}^{-k+1} \, \frac{(-1)^{-\frac{k}{2}+n+1}\,
(m+1)}{ (-k-m+1)!}
\, \frac{\Gamma(2(-1-h_2)+1)}{\Gamma(2(-1-h_2)+k+m)}
\nonu \\
&& \times
\frac{ {z}^{m+\frac{k}{2}+n}}{(\frac{2-k}{2}-n)! \,
(m+\frac{k}{2}+n)!}
\, {\pa}_{z}^{m}\,
{\cal O}^{(-\frac{k}{2}+h_2)}_{-2}(z,\bar{z}),
\nonu \\
&& \bigg[ \hat{\bar{\Phi}}_{-2,n}^{(-\frac{k}{2})},
{\cal O}_{A,-\fft32}^{(h_2)}(z,\bar{z}) \bigg]  =  -
\frac{\bar{\kappa}_{-2,-\fft32,+\fft32}}{2} \,
\sum_{m=0}^{-k+1} \, \frac{(-1)^{-\frac{k}{2}+n+1}\,
(m+1)}{ (-k-m+1)!}
\, \frac{\Gamma(2(-\fft12-h_2)+1)}{\Gamma(2(-\fft12-h_2)+k+m)}
\nonu \\
&& \times
\frac{ {z}^{m+\frac{k}{2}+n}}{(\frac{2-k}{2}-n)! \,
(m+\frac{k}{2}+n)!}
\, {\pa}_{z}^{m}\,
{\cal O}^{(-\frac{k}{2}+h_2)}_{A,-\fft32}(z,\bar{z}),
\nonu \\
&& \bigg[ \hat{\bar{\Phi}}_{-2,n}^{(-\frac{k}{2})},
{\cal O}_{AB,-1}^{(h_2)}(z,\bar{z}) \bigg]  =  -
\frac{\bar{\kappa}_{-2,-1,+1}}{2} \,
\sum_{m=0}^{-k+1} \, \frac{(-1)^{-\frac{k}{2}+n+1}\,
(m+1)}{ (-k-m+1)!}
\, \frac{\Gamma(2(-h_2)+1)}{\Gamma(2(-h_2)+k+m)}
\nonu \\
&& \times
\frac{ {z}^{m+\frac{k}{2}+n}}{(\frac{2-k}{2}-n)! \,
(m+\frac{k}{2}+n)!}
\, {\pa}_{z}^{m}\,
{\cal O}^{(-\frac{k}{2}+h_2)}_{AB,-1}(z,\bar{z}),
\nonu \\
&& \bigg[ \hat{\bar{\Phi}}_{-2,n}^{(-\frac{k}{2})},
{\cal O}_{ABC,-\fft12}^{(h_2)}(z,\bar{z}) \bigg]  =  -
\frac{\bar{\kappa}_{-2,-\fft12,+\fft12}}{2} \,
\sum_{m=0}^{-k+1} \, \frac{(-1)^{-\frac{k}{2}+n+1}\,
(m+1)}{ (-k-m+1)!}
\, \frac{\Gamma(2(\fft12-h_2)+1)}{\Gamma(2(\fft12-h_2)+k+m)}
\nonu \\
&& \times
\frac{ {z}^{m+\frac{k}{2}+n}}{(\frac{2-k}{2}-n)! \,
(m+\frac{k}{2}+n)!}
\, {\pa}_{z}^{m}\,
{\cal O}^{(-\frac{k}{2}+h_2)}_{ABC,-\fft12}(z,\bar{z}),
\nonu \\
&& \bigg[ \hat{\bar{\Phi}}_{-2,n}^{(-\frac{k}{2})},
{\cal O}_{0}^{(h_2),ABCD}(z,\bar{z}) \bigg]  =  -
\frac{\bar{\kappa}_{-2,0,0}}{2} \,
\sum_{m=0}^{-k+1} \, \frac{(-1)^{-\frac{k}{2}+n+1}\,
(m+1)}{ (-k-m+1)!}
\, \frac{\Gamma(2(1-h_2)+1)}{\Gamma(2(1-h_2)+k+m)}
\nonu \\
&& \times
\frac{ {z}^{m+\frac{k}{2}+n}}{(\frac{2-k}{2}-n)! \,
(m+\frac{k}{2}+n)!}
\, {\pa}_{z}^{m}\,
{\cal O}^{(-\frac{k}{2}+h_2),ABCD}_{0}(z,\bar{z}),
\nonu \\
&& \bigg[ \hat{\bar{\Phi}}_{-2,n}^{(-\frac{k}{2})},
{\cal O}_{+\fft12}^{(h_2),ABC}(z,\bar{z}) \bigg]  =  -
\frac{\bar{\kappa}_{-2,+\fft12,-\fft12}}{2} \,
\sum_{m=0}^{-k+1} \, \frac{(-1)^{-\frac{k}{2}+n+1}\,
(m+1)}{ (-k-m+1)!}
\, \frac{\Gamma(2(\fft32-h_2)+1)}{\Gamma(2(\fft32-h_2)+k+m)}
\nonu \\
&& \times
\frac{ {z}^{m+\frac{k}{2}+n}}{(\frac{2-k}{2}-n)! \,
(m+\frac{k}{2}+n)!}
\, {\pa}_{z}^{m}\,
{\cal O}^{(-\frac{k}{2}+h_2),ABC}_{+\fft12}(z,\bar{z}),
\nonu \\
&& \bigg[ \hat{\bar{\Phi}}_{-2,n}^{(-\frac{k}{2})},
{\cal O}_{+1}^{(h_2),AB}(z,\bar{z}) \bigg]  =  -
\frac{\bar{\kappa}_{-2,+1,-1}}{2} \,
\sum_{m=0}^{-k+1} \, \frac{(-1)^{-\frac{k}{2}+n+1}\,
(m+1)}{ (-k-m+1)!}
\, \frac{\Gamma(2(2-h_2)+1)}{\Gamma(2(2-h_2)+k+m)}
\nonu \\
&& \times
\frac{ {z}^{m+\frac{k}{2}+n}}{(\frac{2-k}{2}-n)! \,
(m+\frac{k}{2}+n)!}
\, {\pa}_{z}^{m}\,
{\cal O}^{(-\frac{k}{2}+h_2),AB}_{+1}(z,\bar{z}),
\nonu \\
&& \bigg[ \hat{\bar{\Phi}}_{-2,n}^{(-\frac{k}{2})},
{\cal O}_{+\fft32}^{(h_2),A}(z,\bar{z}) \bigg]  =  -
\frac{\bar{\kappa}_{-2,+\fft32,-\fft32}}{2} \,
\sum_{m=0}^{-k+1} \, \frac{(-1)^{-\frac{k}{2}+n+1}\,
(m+1)}{ (-k-m+1)!}
\, \frac{\Gamma(2(\fft52-h_2)+1)}{\Gamma(2(\fft52-h_2)+k+m)}
\nonu \\
&& \times
\frac{ {z}^{m+\frac{k}{2}+n}}{(\frac{2-k}{2}-n)! \,
(m+\frac{k}{2}+n)!}
\, {\pa}_{z}^{m}\,
{\cal O}^{(-\frac{k}{2}+h_2),A}_{+\fft32}(z,\bar{z}),
\nonu \\
&& \bigg[ \hat{\bar{\Phi}}_{-2,n}^{(-\frac{k}{2})},
{\cal O}_{+2}^{(h_2)}(z,\bar{z}) \bigg]  =  -
\frac{\bar{\kappa}_{-2,+2,-2}}{2} \,
\sum_{m=0}^{-k+1} \, \frac{(-1)^{-\frac{k}{2}+n+1}\,
(m+1)}{ (-k-m+1)!}
\, \frac{\Gamma(2(3-h_2)+1)}{\Gamma(2(3-h_2)+k+m)}
\nonu \\
&& \times
\frac{ {z}^{m+\frac{k}{2}+n}}{(\frac{2-k}{2}-n)! \,
(m+\frac{k}{2}+n)!}
\, {\pa}_{z}^{m}\,
{\cal O}^{(-\frac{k}{2}+h_2)}_{+2}(z,\bar{z}),
\nonu \\
&& \bigg\{ \hat{\bar{\Phi}}_{A,-\fft32,n}^{(-\frac{k}{2}+\fft14)},
{\cal O}_{B,-\fft32}^{(h_2)}(z,\bar{z}) \bigg\}  =  -
\frac{\bar{\kappa}_{-\fft32,-\fft32,+1}}{2} \,
\sum_{m=0}^{-k+\fft12} \, \frac{(-1)^{-\frac{k}{2}+n+\fft34}\,
(m+1)}{ (-k-m+\fft12)!}
\nonu \\
&& \times
\frac{\Gamma(2(-\fft12-h_2)+1)}{\Gamma(2(-\fft12-h_2)+k+m+
\fft12)}
\,
\frac{ {z}^{m+\frac{k}{2}+n+\fft14}}{(\frac{\fft32-k}{2}-n)! \,
(m+\frac{k}{2}+n+\fft14)!}
\, {\pa}_{z}^{m}\,
{\cal O}^{(-\frac{k}{2}+\fft14+h_2)}_{AB,-1}(z,\bar{z}),
\nonu \\
&& \bigg[ \hat{\bar{\Phi}}_{A,-\fft32,n}^{(-\frac{k}{2}+\fft14)},
{\cal O}_{BC,-1}^{(h_2)}(z,\bar{z}) \bigg]  =  -
\frac{\bar{\kappa}_{-\fft32,-1,+\fft12}}{2} \,
\sum_{m=0}^{-k+\fft12} \, \frac{(-1)^{-\frac{k}{2}+n+\fft34}\,
(m+1)}{ (-k-m+\fft12)!}
\nonu \\
&& \times
\frac{\Gamma(2(-h_2)+1)}{\Gamma(2(-h_2)+k+m+
\fft12)}
\,
\frac{ {z}^{m+\frac{k}{2}+n+\fft14}}{(\frac{\fft32-k}{2}-n)! \,
(m+\frac{k}{2}+n+\fft14)!}
\, {\pa}_{z}^{m}\,
{\cal O}^{(-\frac{k}{2}+\fft14+h_2)}_{ABC,-\fft12}(z,\bar{z}),
\nonu \\
&& \bigg\{ \hat{\bar{\Phi}}_{A,-\fft32,n}^{(-\frac{k}{2}+\fft14)},
{\cal O}_{BCD,-\fft12}^{(h_2)}(z,\bar{z}) \bigg\}  =  -
\frac{\bar{\kappa}_{-\fft32,-\fft12,0}}{2} \,
\sum_{m=0}^{-k+\fft12} \, \frac{(-1)^{-\frac{k}{2}+n+\fft34}\,
(m+1)}{ (-k-m+\fft12)!}
\nonu \\
&& \times
\frac{\Gamma(2(\fft12-h_2)+1)}{\Gamma(2(\fft12-h_2)+k+m+
\fft12)}
\,
\frac{ {z}^{m+\frac{k}{2}+n+\fft14}}{(\frac{\fft32-k}{2}-n)! \,
  (m+\frac{k}{2}+n+\fft14)!}\nonu \\
&& \times \frac{1}{4!}\, \ep_{ABCDEFGH}
\, {\pa}_{z}^{m}\,
{\cal O}^{(-\frac{k}{2}+\fft14+h_2),EFGH}_{0}(z,\bar{z}),
\nonu \\
&& \bigg[ \hat{\bar{\Phi}}_{A,-\fft32,n}^{(-\frac{k}{2}+\fft14)},
{\cal O}_{0}^{(h_2),BCDE}(z,\bar{z}) \bigg]  =  -
\frac{\bar{\kappa}_{-\fft32,0,-\fft12}}{2} \,
\sum_{m=0}^{-k+\fft12} \, \frac{(-1)^{-\frac{k}{2}+n+\fft34}\,
(m+1)}{ (-k-m+\fft12)!}
\nonu \\
&& \times
\frac{\Gamma(2(1-h_2)+1)}{\Gamma(2(1-h_2)+k+m+
\fft12)}
\,
\frac{ {z}^{m+\frac{k}{2}+n+\fft14}}{(\frac{\fft32-k}{2}-n)! \,
  (m+\frac{k}{2}+n+\fft14)!}
\frac{1}{4!}\,
\frac{1}{3!}\, \ep^{BCDEIJKL}\, \ep_{AFGHIJKL} 
\nonu \\
&& \times {\pa}_{z}^{m}\,
{\cal O}^{(-\frac{k}{2}+\fft14+h_2),FGH}_{+\ft12}(z,\bar{z}),
\nonu \\
&& \bigg\{ \hat{\bar{\Phi}}_{A,-\fft32,n}^{(-\frac{k}{2}+\fft14)},
{\cal O}_{+\fft12}^{(h_2),BCD}(z,\bar{z}) \bigg\}  =  -
\frac{\bar{\kappa}_{-\fft32,+\fft12,-1}}{2} \,
\sum_{m=0}^{-k+\fft12} \, \frac{(-1)^{-\frac{k}{2}+n+\fft34}\,
(m+1)}{ (-k-m+\fft12)!}
\nonu \\
&& \times
\frac{\Gamma(2(\fft32-h_2)+1)}{\Gamma(2(\fft32-h_2)+k+m+
\fft12)}
\,
\frac{ {z}^{m+\frac{k}{2}+n+\fft14}}{(\frac{\fft32-k}{2}-n)! \,
(m+\frac{k}{2}+n+\fft14)!}
3\, \de^{[B}_{A}     \,
{\pa}_{z}^{m}\,
{\cal O}^{(-\frac{k}{2}+\fft14+h_2),CD]}_{+1}(z,\bar{z}),
\nonu \\
&& \bigg[ \hat{\bar{\Phi}}_{A,-\fft32,n}^{(-\frac{k}{2}+\fft14)},
{\cal O}_{+1}^{(h_2),BC}(z,\bar{z}) \bigg]  =  -
\frac{\bar{\kappa}_{-\fft32,+1,-\fft32}}{2} \,
\sum_{m=0}^{-k+\fft12} \, \frac{(-1)^{-\frac{k}{2}+n+\fft34}\,
(m+1)}{ (-k-m+\fft12)!}
\nonu \\
&& \times
\frac{\Gamma(2(2-h_2)+1)}{\Gamma(2(2-h_2)+k+m+
\fft12)}
\,
\frac{ {z}^{m+\frac{k}{2}+n+\fft14}}{(\frac{\fft32-k}{2}-n)! \,
(m+\frac{k}{2}+n+\fft14)!}
2!\, \de^{[B}_{A}     \,
{\pa}_{z}^{m}\,
{\cal O}^{(-\frac{k}{2}+\fft14+h_2),C]}_{+\fft32}(z,\bar{z}),
\nonu \\
&& \bigg\{ \hat{\bar{\Phi}}_{A,-\fft32,n}^{(-\frac{k}{2}+\fft14)},
{\cal O}_{+\fft32}^{(h_2),B}(z,\bar{z}) \bigg\}  =  -
\frac{\bar{\kappa}_{-\fft32,+\fft32,-2}}{2} \,
\sum_{m=0}^{-k+\fft12} \, \frac{(-1)^{-\frac{k}{2}+n+\fft34}\,
(m+1)}{ (-k-m+\fft12)!}
\nonu \\
&& \times
\frac{\Gamma(2(\fft52-h_2)+1)}{\Gamma(2(\fft52-h_2)+k+m+
\fft12)}
\,
\frac{ {z}^{m+\frac{k}{2}+n+\fft14}}{(\frac{\fft32-k}{2}-n)! \,
(m+\frac{k}{2}+n+\fft14)!}
\, \de^{B}_{A}     \,
{\pa}_{z}^{m}\,
{\cal O}^{(-\frac{k}{2}+\fft14+h_2)}_{+2}(z,\bar{z}),
\nonu \\
&& \bigg[ \hat{\bar{\Phi}}_{AB,-1,n}^{(-\frac{k}{2}+\fft12)},
{\cal O}_{CD,-1}^{(h_2)}(z,\bar{z}) \bigg]  =  -
\frac{\bar{\kappa}_{-1,-1,0}}{2} \,
\sum_{m=0}^{-k} \, \frac{(-1)^{-\frac{k}{2}+n+\fft12}\,
(m+1)}{ (-k-m)!}
\nonu \\
&& \times
\frac{\Gamma(2(-h_2)+1)}{\Gamma(2(-h_2)+k+m+1)}
\,
\frac{ {z}^{m+\frac{k}{2}+n+\fft12}}{(\frac{1-k}{2}-n)! \,
(m+\frac{k}{2}+n+\fft12)!}
\nonu \\
&& \times \frac{1}{4!}\, \ep_{ABCDEFGH}\,
{\pa}_{z}^{m}\,
{\cal O}^{(-\frac{k}{2}+\fft12+h_2),EFGH}_{0}(z,\bar{z}),
\nonu \\
&& \bigg[ \hat{\bar{\Phi}}_{AB,-1,n}^{(-\frac{k}{2}+\fft12)},
{\cal O}_{CDE,-\fft12}^{(h_2)}(z,\bar{z}) \bigg]  =  -
\frac{\bar{\kappa}_{-1,-\fft12,-\fft12}}{2} \,
\sum_{m=0}^{-k} \, \frac{(-1)^{-\frac{k}{2}+n+\fft12}\,
(m+1)}{ (-k-m)!}
\nonu \\
&& \times
\frac{\Gamma(2(\fft12-h_2)+1)}{\Gamma(2(\fft12-h_2)+k+m+1)}
\,
\frac{ {z}^{m+\frac{k}{2}+n+\fft12}}{(\frac{1-k}{2}-n)! \,
(m+\frac{k}{2}+n+\fft12)!}
\,
\frac{1}{3!}\, \ep_{ABCDEFGH}\nonu \\
&& \times
{\pa}_{z}^{m}\,
{\cal O}^{(-\frac{k}{2}+\fft12+h_2),FGH}_{+\fft12}(z,\bar{z}),
\nonu \\
&& \bigg[ \hat{\bar{\Phi}}_{AB,-1,n}^{(-\frac{k}{2}+\fft12)},
{\cal O}_{0}^{(h_2),CDEF}(z,\bar{z}) \bigg]  =  -
\frac{\bar{\kappa}_{-1,0,-1}}{2} \,
\sum_{m=0}^{-k} \, \frac{(-1)^{-\frac{k}{2}+n+\fft12}\,
(m+1)}{ (-k-m)!}
\nonu \\
&& \times
\frac{\Gamma(2(1-h_2)+1)}{\Gamma(2(1-h_2)+k+m+1)}
\,
\frac{ {z}^{m+\frac{k}{2}+n+\fft12}}{(\frac{1-k}{2}-n)! \,
  (m+\frac{k}{2}+n+\fft12)!}
\nonu \\
&& \times 
\frac{1}{4!}\, 
\frac{1}{2!}\, \ep^{CDEFIJKL}\,
\ep_{ABGHIJKL}\,
{\pa}_{z}^{m}\,
{\cal O}^{(-\frac{k}{2}+\fft12+h_2),GH}_{+1}(z,\bar{z}),
\nonu \\
&& \bigg[ \hat{\bar{\Phi}}_{AB,-1,n}^{(-\frac{k}{2}+\fft12)},
{\cal O}_{+\fft12}^{(h_2),CDE}(z,\bar{z}) \bigg]  =  -
\frac{\bar{\kappa}_{-1,+\fft12,-\fft32}}{2} \,
\sum_{m=0}^{-k} \, \frac{(-1)^{-\frac{k}{2}+n+\fft12}\,
(m+1)}{ (-k-m)!}
\nonu \\
&& \times
\frac{\Gamma(2(\fft32-h_2)+1)}{\Gamma(2(\fft32-h_2)+k+m+1)}
\,
\frac{ {z}^{m+\frac{k}{2}+n+\fft12}}{(\frac{1-k}{2}-n)! \,
(m+\frac{k}{2}+n+\fft12)!}
\,
3!\, \de^{[C}_{A} \, 
{\pa}_{z}^{m}\,
{\cal O}^{(-\frac{k}{2}+\fft12+h_2),D}_{+\fft32}(z,\bar{z})
\, \de^{E]}_{B},
\nonu \\
&& \bigg[ \hat{\bar{\Phi}}_{AB,-1,n}^{(-\frac{k}{2}+\fft12)},
{\cal O}_{+1}^{(h_2),CD}(z,\bar{z}) \bigg]  =  -
\frac{\bar{\kappa}_{-1,+1,-2}}{2} \,
\sum_{m=0}^{-k} \, \frac{(-1)^{-\frac{k}{2}+n+\fft12}\,
(m+1)}{ (-k-m)!}
\nonu \\
&& \times
\frac{\Gamma(2(2-h_2)+1)}{\Gamma(2(2-h_2)+k+m+1)}
\,
\frac{ {z}^{m+\frac{k}{2}+n+\fft12}}{(\frac{1-k}{2}-n)! \,
(m+\frac{k}{2}+n+\fft12)!}
\,
\de^{CD}_{AB} \, 
{\pa}_{z}^{m}\,
{\cal O}^{(-\frac{k}{2}+\fft12+h_2)}_{+2}(z,\bar{z}),
\nonu \\
&& \bigg\{ \hat{\bar{\Phi}}_{ABC,-\fft12,n}^{(-\frac{k}{2}+\fft34)},
{\cal O}_{DEF,-\fft12}^{(h_2)}(z,\bar{z}) \bigg\}  =  -
\frac{\bar{\kappa}_{-\fft12,-\fft12,-1}}{2} \,
\sum_{m=0}^{-k-\fft12} \, \frac{(-1)^{-\frac{k}{2}+n+\fft14}\,
(m+1)}{ (-k-m-\fft12)!}
\nonu \\
&& \times
\frac{\Gamma(2(\fft12-h_2)+1)}{\Gamma(2(\fft12-h_2)+k+m+\fft32)}
\,
\frac{ {z}^{m+\frac{k}{2}+n+\fft34}}{(\frac{\fft12-k}{2}-n)! \,
(m+\frac{k}{2}+n+\fft34)!}
\, \frac{1}{2!}\, \ep_{ABCDEFGH}
\nonu \\
&& \times
{\pa}_{z}^{m}\,
{\cal O}^{(-\frac{k}{2}+\fft34+h_2),GH}_{+1}(z,\bar{z}),
\nonu \\
&& \bigg[ \hat{\bar{\Phi}}_{ABC,-\fft12,n}^{(-\frac{k}{2}+\fft34)},
{\cal O}_{0}^{(h_2),DEFG}(z,\bar{z}) \bigg]  =  -
\frac{\bar{\kappa}_{-\fft12,0,-\fft32}}{2} \,
\sum_{m=0}^{-k-\fft12} \, \frac{(-1)^{-\frac{k}{2}+n+\fft14}\,
(m+1)}{ (-k-m-\fft12)!}
\nonu \\
&& \times
\frac{\Gamma(2(1-h_2)+1)}{\Gamma(2(1-h_2)+k+m+\fft32)}
\,
\frac{ {z}^{m+\frac{k}{2}+n+\fft34}}{(\frac{\fft12-k}{2}-n)! \,
(m+\frac{k}{2}+n+\fft34)!}
\, \frac{1}{4!} \, \ep^{DEFGIJKL} \, \ep_{ABCHIJKL}
\nonu \\
&& \times
{\pa}_{z}^{m}\,
{\cal O}^{(-\frac{k}{2}+\fft34+h_2),H}_{+\fft32}(z,\bar{z}),
\nonu \\
&& \bigg\{ \hat{\bar{\Phi}}_{ABC,-\fft12,n}^{(-\frac{k}{2}+\fft34)},
{\cal O}_{+\fft12}^{(h_2),DEF}(z,\bar{z}) \bigg\}  =  -
\frac{\bar{\kappa}_{-\fft12,+\fft12,-2}}{2} \,
\sum_{m=0}^{-k-\fft12} \, \frac{(-1)^{-\frac{k}{2}+n+\fft14}\,
(m+1)}{ (-k-m-\fft12)!}
\nonu \\
&& \times
\frac{\Gamma(2(\fft32-h_2)+1)}{\Gamma(2(\fft32-h_2)+k+m+\fft32)}
\,
\frac{ {z}^{m+\frac{k}{2}+n+\fft34}}{(\frac{\fft12-k}{2}-n)! \,
(m+\frac{k}{2}+n+\fft34)!}
\, \de^{DEF}_{ABC}
\,
{\pa}_{z}^{m}\,
{\cal O}^{(-\frac{k}{2}+\fft34+h_2)}_{+2}(z,\bar{z}),
\nonu \\
&& \bigg[ \hat{\bar{\Phi}}_{0,n}^{(-\frac{k}{2}+1),ABCD},
{\cal O}_{0}^{(h_2),EFGH}(z,\bar{z}) \bigg]  =  -
\frac{\bar{\kappa}_{0,0,-2}}{2} \,
\sum_{m=0}^{-k-1} \, \frac{(-1)^{-\frac{k}{2}+n}\,
(m+1)}{ (-k-m-1)!}
\label{anticomm25other}
\\
&& \times
\frac{\Gamma(2(1-h_2)+1)}{\Gamma(2(1-h_2)+k+m+2)}
\,
\frac{ {z}^{m+\frac{k}{2}+n+1}}{(\frac{-k}{2}-n)! \,
(m+\frac{k}{2}+n+1)!}
\, \ep^{ABCDEFGH}
\,
{\pa}_{z}^{m}\,
{\cal O}^{(-\frac{k}{2}+1+h_2)}_{+2}(z,\bar{z}).
\nonu
\eea


\subsection{
 The last twenty (anti)commutator relations of (\ref{HOcomm1})
}

The (anti)commutators corresponding to
(\ref{othertwenty}) are obtained
\bea
&& \bigg[ \hat{\bar{\Phi}}_{A,-\fft32,n}^{(-\frac{k}{2}+\fft14)},
{\cal O}_{-2}^{(h_2)}(z,\bar{z}) \bigg]  =  -
\frac{\bar{\kappa}_{-\fft32,-2,+\fft32}}{2} \,
\sum_{m=0}^{-k+\fft12} \, \frac{(-1)^{-\frac{k}{2}+n+\fft34}\,
(m+1)}{ (-k-m+\fft12)!}
\nonu \\
&& \times
\frac{\Gamma(2(-1-h_2)+1)}{\Gamma(2(-1-h_2)+k+m+
\fft12)}
\,
\frac{ {z}^{m+\frac{k}{2}+n+\fft14}}{(\frac{\fft32-k}{2}-n)! \,
(m+\frac{k}{2}+n+\fft14)!}   \,
{\pa}_{z}^{m}\,
{\cal O}^{(-\frac{k}{2}+\fft14+h_2)}_{A,-\fft32}(z,\bar{z}),
\nonu \\
&& \bigg[ \hat{\bar{\Phi}}_{AB,-1,n}^{(-\frac{k}{2}+\fft12)},
{\cal O}_{-2}^{(h_2)}(z,\bar{z}) \bigg]  =  -
\frac{\bar{\kappa}_{-1,-2,+1}}{2} \,
\sum_{m=0}^{-k} \, \frac{(-1)^{-\frac{k}{2}+n+\fft12}\,
(m+1)}{ (-k-m)!}
\nonu \\
&& \times
\frac{\Gamma(2(-1-h_2)+1)}{\Gamma(2(-1-h_2)+k+m+1)}
\,
\frac{ {z}^{m+\frac{k}{2}+n+\fft12}}{(\frac{1-k}{2}-n)! \,
(m+\frac{k}{2}+n+\fft12)!}   \,
{\pa}_{z}^{m}\,
{\cal O}^{(-\frac{k}{2}+\fft12+h_2)}_{AB,-1}(z,\bar{z}),
\nonu \\
&& \bigg[ \hat{\bar{\Phi}}_{ABC,-\fft12,n}^{(-\frac{k}{2}+\fft34)},
{\cal O}_{-2}^{(h_2)}(z,\bar{z}) \bigg]  =  -
\frac{\bar{\kappa}_{-\fft12,-2,+\fft12}}{2} \,
\sum_{m=0}^{-k-\fft12} \, \frac{(-1)^{-\frac{k}{2}+n+\fft14}\,
(m+1)}{ (-k-m-\fft12)!}
\nonu \\
&& \times
\frac{\Gamma(2(-1-h_2)+1)}{\Gamma(2(-1-h_2)+k+m+\fft32)}
\,
\frac{ {z}^{m+\frac{k}{2}+n+\fft34}}{(\frac{\fft12-k}{2}-n)! \,
(m+\frac{k}{2}+n+\fft34)!}   \,
{\pa}_{z}^{m}\,
{\cal O}^{(-\frac{k}{2}+\fft34+h_2)}_{ABC,-\fft12}(z,\bar{z}),
\nonu \\
&& \bigg[ \hat{\bar{\Phi}}_{0,n}^{(-\frac{k}{2}+1),ABCD},
{\cal O}_{-2}^{(h_2)}(z,\bar{z}) \bigg]  =  -
\frac{\bar{\kappa}_{0,-2,0}}{2} \,
\sum_{m=0}^{-k-1} \, \frac{(-1)^{-\frac{k}{2}+n}\,
(m+1)}{ (-k-m-1)!}
\nonu \\
&& \times
\frac{\Gamma(2(-1-h_2)+1)}{\Gamma(2(-1-h_2)+k+m+2)}
\,
\frac{ {z}^{m+\frac{k}{2}+n+1}}{(\frac{-k}{2}-n)! \,
(m+\frac{k}{2}+n+1)!}   \,
{\pa}_{z}^{m}\,
{\cal O}^{(-\frac{k}{2}+1+h_2),ABCD}_{0}(z,\bar{z}),
\nonu \\
&& \bigg[ \hat{\bar{\Phi}}_{+\fft12,n}^{(-\frac{k}{2}+\fft54),ABC},
{\cal O}_{-2}^{(h_2)}(z,\bar{z}) \bigg]  =  -
\frac{\bar{\kappa}_{+\fft12,-2,-\fft12}}{2} \,
\sum_{m=0}^{-k-\fft32} \, \frac{(-1)^{-\frac{k}{2}+n-\fft14}\,
(m+1)}{ (-k-m-\fft32)!}
\nonu \\
&& \times
\frac{\Gamma(2(-1-h_2)+1)}{\Gamma(2(-1-h_2)+k+m+\fft52)}
\,
\frac{ {z}^{m+\frac{k}{2}+n+\fft54}}{(\frac{-\fft12-k}{2}-n)! \,
(m+\frac{k}{2}+n+\fft54)!}   \,
{\pa}_{z}^{m}\,
{\cal O}^{(-\frac{k}{2}+\fft54+h_2),ABC}_{+\fft12}(z,\bar{z}),
\nonu \\
&& \bigg[ \hat{\bar{\Phi}}_{+1,n}^{(-\frac{k}{2}+\fft32),AB},
{\cal O}_{-2}^{(h_2)}(z,\bar{z}) \bigg]  =  -
\frac{\bar{\kappa}_{+1,-2,-1}}{2} \,
\sum_{m=0}^{-k-2} \, \frac{(-1)^{-\frac{k}{2}+n-\fft12}\,
(m+1)}{ (-k-m-2)!}
\nonu \\
&& \times
\frac{\Gamma(2(-1-h_2)+1)}{\Gamma(2(-1-h_2)+k+m+3)}
\,
\frac{ {z}^{m+\frac{k}{2}+n+\fft32}}{(\frac{-1-k}{2}-n)! \,
(m+\frac{k}{2}+n+\fft32)!}   \,
{\pa}_{z}^{m}\,
{\cal O}^{(-\frac{k}{2}+\fft32+h_2),AB}_{+1}(z,\bar{z}),
\nonu \\
&& \bigg[ \hat{\bar{\Phi}}_{+\fft32,n}^{(-\frac{k}{2}+\fft74),A},
{\cal O}_{-2}^{(h_2)}(z,\bar{z}) \bigg]  =  -
\frac{\bar{\kappa}_{+\fft32,-2,-\fft32}}{2} \,
\sum_{m=0}^{-k-\fft52} \, \frac{(-1)^{-\frac{k}{2}+n-\fft34}\,
(m+1)}{ (-k-m-\fft52)!}
\nonu \\
&& \times
\frac{\Gamma(2(-1-h_2)+1)}{\Gamma(2(-1-h_2)+k+m+\fft72)}
\,
\frac{ {z}^{m+\frac{k}{2}+n+\fft74}}{(\frac{-\fft32-k}{2}-n)! \,
(m+\frac{k}{2}+n+\fft74)!}   \,
{\pa}_{z}^{m}\,
{\cal O}^{(-\frac{k}{2}+\fft74+h_2),A}_{+\fft32}(z,\bar{z}),
\nonu \\
&& \bigg[ \hat{\bar{\Phi}}_{+2,n}^{(-\frac{k}{2}+2)},
{\cal O}_{-2}^{(h_2)}(z,\bar{z}) \bigg]  =  -
\frac{\bar{\kappa}_{+2,-2,-2}}{2} \,
\sum_{m=0}^{-k-3} \, \frac{(-1)^{-\frac{k}{2}+n-1}\,
(m+1)}{ (-k-m-3)!}
\nonu \\
&& \times
\frac{\Gamma(2(-1-h_2)+1)}{\Gamma(2(-1-h_2)+k+m+4)}
\,
\frac{ {z}^{m+\frac{k}{2}+n+2}}{(\frac{-2-k}{2}-n)! \,
(m+\frac{k}{2}+n+2)!}   \,
{\pa}_{z}^{m}\,
{\cal O}^{(-\frac{k}{2}+2+h_2)}_{+2}(z,\bar{z}),
\nonu \\
&& \bigg[ \hat{\bar{\Phi}}_{AB,-1,n}^{(-\frac{k}{2}+\fft12)},
{\cal O}_{C,-\fft32}^{(h_2)}(z,\bar{z}) \bigg]  =  -
\frac{\bar{\kappa}_{-1,-\fft32,+\fft12}}{2} \,
\sum_{m=0}^{-k} \, \frac{(-1)^{-\frac{k}{2}+n+\fft12}\,
(m+1)}{ (-k-m)!}
\nonu \\
&& \times
\frac{\Gamma(2(-\fft12-h_2)+1)}{\Gamma(2(-\fft12-h_2)+k+m+1)}
\,
\frac{ {z}^{m+\frac{k}{2}+n+\fft12}}{(\frac{1-k}{2}-n)! \,
(m+\frac{k}{2}+n+\fft12)!}
\, {\pa}_{z}^{m}\,
{\cal O}^{(-\frac{k}{2}+\fft12+h_2)}_{ABC,-\fft12}(z,\bar{z}),
\nonu \\
&& \bigg\{ \hat{\bar{\Phi}}_{ABC,-\fft12,n}^{(-\frac{k}{2}+\fft34)},
{\cal O}_{D,-\fft32}^{(h_2)}(z,\bar{z}) \bigg\}  =  -
\frac{\bar{\kappa}_{-\fft12,-\fft32,0}}{2} \,
\sum_{m=0}^{-k-\fft12} \, \frac{(-1)^{-\frac{k}{2}+n+\fft14}\,
(m+1)}{ (-k-m-\fft12)!}
\nonu \\
&& \times
\frac{\Gamma(2(-\fft12-h_2)+1)}{\Gamma(2(-\fft12-h_2)+k+m+\tf32)}
\,
\frac{ {z}^{m+\frac{k}{2}+n+\fft34}}{(\frac{\fft12-k}{2}-n)! \,
  (m+\frac{k}{2}+n+\fft34)!}
\nonu \\
&& \times \frac{1}{4!}\, \ep_{ABCDEFGH}
\, {\pa}_{z}^{m}\,
{\cal O}^{(-\frac{k}{2}+\fft34+h_2),EFGH}_{0}(z,\bar{z}),
\nonu \\
&& \bigg[ \hat{\bar{\Phi}}_{0,n}^{(-\frac{k}{2}+1),ABCD},
{\cal O}_{E,-\fft32}^{(h_2)}(z,\bar{z}) \bigg]  =  -
\frac{\bar{\kappa}_{0,-\fft32,-\fft12}}{2} \,
\sum_{m=0}^{-k-1} \, \frac{(-1)^{-\frac{k}{2}+n}\,
(m+1)}{ (-k-m-1)!}
\nonu \\
&& \times
\frac{\Gamma(2(-\fft12-h_2)+1)}{\Gamma(2(-\fft12-h_2)+k+m+2)}
\,
\frac{ {z}^{m+\frac{k}{2}+n+1}}{(\frac{-k}{2}-n)! \,
  (m+\frac{k}{2}+n+1)!}
\frac{1}{4!}
\, \frac{1}{3!}\, \ep^{ABCDIJKL} \, \ep_{EFGHIJKL}  \nonu \\
&& \times {\pa}_{z}^{m}\,
{\cal O}^{(-\frac{k}{2}+1+h_2),FGH}_{+\fft12}(z,\bar{z}),
\nonu \\
&& \bigg\{ \hat{\bar{\Phi}}_{+\fft12,n}^{(-\frac{k}{2}+\fft54),ABC},
{\cal O}_{D,-\fft32}^{(h_2)}(z,\bar{z}) \bigg\}  =  -
\frac{\bar{\kappa}_{+\fft12,-\fft32,-1}}{2} \,
\sum_{m=0}^{-k-\fft32} \, \frac{(-1)^{-\frac{k}{2}+n-\fft14}\,
(m+1)}{ (-k-m-\fft32)!}
\nonu \\
&& \times
\frac{\Gamma(2(-\fft12-h_2)+1)}{\Gamma(2(-\fft12-h_2)+k+m+\fft52)}
\,
\frac{ {z}^{m+\frac{k}{2}+n+\fft54}}{(\frac{-\fft12-k}{2}-n)! \,
(m+\frac{k}{2}+n+\fft54)!}
\, 3 \,   {\pa}_{z}^{m}\,
{\cal O}^{(-\frac{k}{2}+\fft54+h_2),[AB}_{+1}(z,\bar{z}) \,
\de^{C]}_{D},
\nonu \\
&& \bigg[ \hat{\bar{\Phi}}_{+1,n}^{(-\frac{k}{2}+\fft32),AB},
{\cal O}_{C,-\fft32}^{(h_2)}(z,\bar{z}) \bigg]  =  -
\frac{\bar{\kappa}_{+1,-\fft32,-\fft32}}{2} \,
\sum_{m=0}^{-k-2} \, \frac{(-1)^{-\frac{k}{2}+n-\fft12}\,
(m+1)}{ (-k-m-2)!}
\nonu \\
&& \times
\frac{\Gamma(2(-\fft12-h_2)+1)}{\Gamma(2(-\fft12-h_2)+k+m+3)}
\,
\frac{ {z}^{m+\frac{k}{2}+n+\fft32}}{(\frac{-1-k}{2}-n)! \,
(m+\frac{k}{2}+n+\fft32)!}
\, 2 \,  \de^{[A}_{C}\,  {\pa}_{z}^{m}\,
{\cal O}^{(-\frac{k}{2}+\fft32+h_2),B]}_{+\fft32}(z,\bar{z}),
\nonu \\
&& \bigg\{ \hat{\bar{\Phi}}_{+\fft32,n}^{(-\frac{k}{2}+\fft74),A},
{\cal O}_{B,-\fft32}^{(h_2)}(z,\bar{z}) \bigg\}  =  -
\frac{\bar{\kappa}_{+\fft32,-\fft32,-2}}{2} \,
\sum_{m=0}^{-k-\fft52} \, \frac{(-1)^{-\frac{k}{2}+n-\fft34}\,
(m+1)}{ (-k-m-\fft52)!}
\nonu \\
&& \times
\frac{\Gamma(2(-\fft12-h_2)+1)}{\Gamma(2(-\fft12-h_2)+k+m+\fft72)}
\,
\frac{ {z}^{m+\frac{k}{2}+n+\fft74}}{(\frac{-\fft32-k}{2}-n)! \,
(m+\frac{k}{2}+n+\fft74)!}
\,  \de^{A}_{B}\,  {\pa}_{z}^{m}\,
{\cal O}^{(-\frac{k}{2}+\fft74+h_2)}_{+2}(z,\bar{z}),
\nonu \\
&& \bigg[ \hat{\bar{\Phi}}_{ABC,-\fft12,n}^{(-\frac{k}{2}+\fft34)},
{\cal O}_{CD,-1}^{(h_2)}(z,\bar{z}) \bigg]  =  -
\frac{\bar{\kappa}_{-\fft12,-1,-\fft12}}{2} \,
\sum_{m=0}^{-k-\fft12} \, \frac{(-1)^{-\frac{k}{2}+n+\fft14}\,
(m+1)}{ (-k-m-\fft12)!}
\nonu \\
&& \times
\frac{\Gamma(2(-h_2)+1)}{\Gamma(2(-h_2)+k+m+\fft32)}
\,
\frac{ {z}^{m+\frac{k}{2}+n+\fft34}}{(\frac{\fft12-k}{2}-n)! \,
(m+\frac{k}{2}+n+\fft34)!}
\, \frac{1}{3!}\, \ep_{ABCDEFGH} \nonu \\
&& \times 
{\pa}_{z}^{m}\,
{\cal O}^{(-\frac{k}{2}+\fft34+h_2),FGH}_{+\fft12}(z,\bar{z}),
\nonu \\
&& \bigg[ \hat{\bar{\Phi}}_{0,n}^{(-\frac{k}{2}+1),ABCD},
{\cal O}_{EF,-1}^{(h_2)}(z,\bar{z}) \bigg]  =  -
\frac{\bar{\kappa}_{0,-1,-1}}{2} \,
\sum_{m=0}^{-k-1} \, \frac{(-1)^{-\frac{k}{2}+n}\,
(m+1)}{ (-k-m-1)!}
\nonu \\
&& \times
\frac{\Gamma(2(-h_2)+1)}{\Gamma(2(-h_2)+k+m+2)}
\,
\frac{ {z}^{m+\frac{k}{2}+n+1}}{(\frac{-k}{2}-n)! \,
(m+\frac{k}{2}+n+1)!}
\, \frac{1}{4!}\, \frac{1}{2!}\, \ep^{ABCDIJKL}\,
\ep_{EFGHIJKL} \nonu \\
&& \times 
{\pa}_{z}^{m}\,
{\cal O}^{(-\frac{k}{2}+1+h_2),GH}_{+1}(z,\bar{z}),
\nonu \\
&& \bigg[ \hat{\bar{\Phi}}_{+\fft12,n}^{(-\frac{k}{2}+\fft54),ABC},
{\cal O}_{DE,-1}^{(h_2)}(z,\bar{z}) \bigg]  =  -
\frac{\bar{\kappa}_{+\fft12,-1,-\fft32}}{2} \,
\sum_{m=0}^{-k-\fft32} \, \frac{(-1)^{-\frac{k}{2}+n-\fft14}\,
(m+1)}{ (-k-m-\fft32)!}
\nonu \\
&& \times
\frac{\Gamma(2(-h_2)+1)}{\Gamma(2(-h_2)+k+m+\fft52)}
\,
\frac{ {z}^{m+\frac{k}{2}+n+\fft54}}{(\frac{-\fft12-k}{2}-n)! \,
(m+\frac{k}{2}+n+\fft54)!}
3!\, \de^{[A}_{D}
\,
{\pa}_{z}^{m}\,
{\cal O}^{(-\frac{k}{2}+\fft54+h_2),B}_{+\fft32}(z,\bar{z}) \, \de^{C]}_{E},
\nonu \\
&& \bigg[ \hat{\bar{\Phi}}_{+1,n}^{(-\frac{k}{2}+\fft32),AB},
{\cal O}_{CD,-1}^{(h_2)}(z,\bar{z}) \bigg]  =  -
\frac{\bar{\kappa}_{+1,-1,-2}}{2} \,
\sum_{m=0}^{-k-2} \, \frac{(-1)^{-\frac{k}{2}+n-\fft12}\,
(m+1)}{ (-k-m-2)!}
\nonu \\
&& \times
\frac{\Gamma(2(-h_2)+1)}{\Gamma(2(-h_2)+k+m+3)}
\,
\frac{ {z}^{m+\frac{k}{2}+n+\fft32}}{(\frac{-1-k}{2}-n)! \,
(m+\frac{k}{2}+n+\fft32)!}
\,  \de^{AB}_{CD}
\,
{\pa}_{z}^{m}\,
{\cal O}^{(-\frac{k}{2}+\fft32+h_2)}_{+2}(z,\bar{z}) \, \de^{E}_{C]},
\nonu \\
&& \bigg[ \hat{\bar{\Phi}}_{0,n}^{(-\frac{k}{2}+1),ABCD},
{\cal O}_{EFG,-\fft12}^{(h_2)}(z,\bar{z}) \bigg]  =  -
\frac{\bar{\kappa}_{0,-\fft12,-\fft32}}{2} \,
\sum_{m=0}^{-k-1} \, \frac{(-1)^{-\frac{k}{2}+n}\,
(m+1)}{ (-k-m-1)!}
\nonu \\
&& \times
\frac{\Gamma(2(\tf12-h_2)+1)}{\Gamma(2(\tf12-h_2)+k+m+2)}
\,
\frac{ {z}^{m+\frac{k}{2}+n+1}}{(\frac{-k}{2}-n)! \,
(m+\frac{k}{2}+n+1)!}
\, \frac{1}{4!} \ep^{ABCDIJKL} \, \ep_{EFGHIJKL} \nonu \\
&& \times 
{\pa}_{z}^{m}\,
{\cal O}^{(-\frac{k}{2}+1+h_2),H}_{+\fft32}(z,\bar{z}),
\nonu \\
&& \bigg\{ \hat{\bar{\Phi}}_{+\fft12,n}^{(-\frac{k}{2}+\fft54),ABC},
{\cal O}_{DEF,-\fft12}^{(h_2)}(z,\bar{z}) \bigg\}  =  -
\frac{\bar{\kappa}_{+\fft12,-\fft12,-2}}{2} \,
\sum_{m=0}^{-k-\fft32} \, \frac{(-1)^{-\frac{k}{2}+n-\fft14}\,
(m+1)}{ (-k-m-\fft32)!}
\label{anticomm20other}
\\
&& \times
\frac{\Gamma(2(\fft12-h_2)+1)}{\Gamma(2(\fft12-h_2)+k+m+\fft52)}
\,
\frac{ {z}^{m+\frac{k}{2}+n+\fft54}}{(\frac{-\fft12-k}{2}-n)! \,
(m+\frac{k}{2}+n+\fft54)!}
\, \de^{ABC}_{DEF}
\,
{\pa}_{z}^{m}\,
{\cal O}^{(-\frac{k}{2}+\fft54+h_2)}_{+2}(z,\bar{z}).
\nonu
\eea
As before,
due to the structure of $SU(8)$ indices,  the right-hand sides of the
tenth, the eleventh, the sixteenth, and the
nineteenth relations
of (\ref{anticomm20other}) are different from those
in (\ref{anticomm20}).

\subsection{
The last twenty OPE relations of (\ref{HOOPEshadow1}) 
from shadow transform
}

The remaining twenty OPEs corresponding to (\ref{HOOPEshadow1})
can be described as
\bea
&& \tilde{\bar{\Phi}}_{A,-\fft32}^{(-\frac{k}{2}+\fft14)}(z_1,\bar{z}_1) \,
{\cal O}_{-2}^{(h_2)}(z_2,\bar{z}_2)=
\frac{\bar{\kappa}_{-\frac{3}{2},-2,+\fft32}}{2}\,
\frac{\pi \, (-1)^{k+\fft32}}{(\fft52-k)!}
\frac{1}{\bar{z}_{12}^{\fft12-k}} \,
\sum_{m=0}^{-k+\fft12} \,
(m+1)
\nonu \\
&& \times \frac{\Gamma(2(-1-h_2)+1)}
{\Gamma(2(-1-h_2)+k+m+\fft12)}
\, z_{12}^{k-\fft32+m}\, {\pa}_{{z}_2}^m \,
{\cal O}_{A,-\fft32}^{(-\frac{k}{2}+\fft14+h_2)}(z_2,\bar{z}_2)+\cdots,
\nonu \\
&& \tilde{\bar{\Phi}}_{AB,-1}^{(-\frac{k}{2}+\fft12)}(z_1,\bar{z}_1) \,
{\cal O}_{-2}^{(h_2)}(z_2,\bar{z}_2)=
\frac{\bar{\kappa}_{-1,-2,+1}}{2}\,
\frac{\pi \, (-1)^{k+1}}{(2-k)!}
\frac{1}{\bar{z}_{12}^{1-k}} \,
\sum_{m=0}^{-k} \,
(m+1)
\nonu \\
&& \times \frac{\Gamma(2(-1-h_2)+1)}
{\Gamma(2(-1-h_2)+k+m+1)}
\, z_{12}^{k-1+m}\,
{\pa}_{{z}_2}^m \,
{\cal O}_{AB,-1}^{(-\frac{k}{2}+\fft12+h_2)}(z_2,\bar{z}_2)+\cdots,
\nonu \\
&& \tilde{\bar{\Phi}}_{ABC,-\fft12}^{(-\frac{k}{2}+\fft34)}(z_1,\bar{z}_1) \,
{\cal O}_{-2}^{(h_2)}(z_2,\bar{z}_2)=
\frac{\bar{\kappa}_{-\fft12,-2,+\fft12}}{2}\,
\frac{\pi \, (-1)^{k+\fft12}}{(\fft32-k)!}
\frac{1}{\bar{z}_{12}^{\fft32-k}} \,
\sum_{m=0}^{-k-\fft12} \,
(m+1)
\nonu \\
&& \times \frac{\Gamma(2(-1-h_2)+1)}
{\Gamma(2(-1-h_2)+k+m+\tf32)}
\, z_{12}^{k-\fft12+m}\,
{\pa}_{{z}_2}^m \,
{\cal O}_{ABC,-\fft12}^{(-\frac{k}{2}+\fft34+h_2)}(z_2,\bar{z}_2)+\cdots,
\nonu \\
&& \tilde{\bar{\Phi}}_{0}^{(-\frac{k}{2}+1),ABCD}(z_1,\bar{z}_1) \,
{\cal O}_{-2}^{(h_2)}(z_2,\bar{z}_2)=
\frac{\bar{\kappa}_{0,-2,0}}{2}\,
\frac{\pi \, (-1)^{k}}{(1-k)!}
\frac{1}{\bar{z}_{12}^{2-k}} \,
\sum_{m=0}^{-k-1} \,
(m+1)
\nonu \\
&& \times \frac{\Gamma(2(-1-h_2)+1)}
{\Gamma(2(-1-h_2)+k+m+2)}
\, z_{12}^{k+m}\,
{\pa}_{{z}_2}^m \,
{\cal O}_{0}^{(-\frac{k}{2}+1+h_2),ABCD}(z_2,\bar{z}_2)+\cdots,
\nonu \\
&& \tilde{\bar{\Phi}}_{+\fft12}^{(-\frac{k}{2}+\fft54),ABC}(z_1,\bar{z}_1) \,
{\cal O}_{-2}^{(h_2)}(z_2,\bar{z}_2)=
\frac{\bar{\kappa}_{+\fft12,-2,-\fft12}}{2}\,
\frac{\pi \, (-1)^{k-\fft12}}{(\fft12-k)!}
\frac{1}{\bar{z}_{12}^{\fft52-k}} \,
\sum_{m=0}^{-k-\fft32} \,
(m+1)
\nonu \\
&& \times \frac{\Gamma(2(-1-h_2)+1)}
{\Gamma(2(-1-h_2)+k+m+\tf52)}
\, z_{12}^{k+\fft12+m}\,
{\pa}_{{z}_2}^m \,
{\cal O}_{+\fft12}^{(-\frac{k}{2}+\fft54+h_2),ABC}(z_2,\bar{z}_2)+\cdots,
\nonu \\
&& \tilde{\bar{\Phi}}_{+1}^{(-\frac{k}{2}+\fft32),AB}(z_1,\bar{z}_1) \,
{\cal O}_{-2}^{(h_2)}(z_2,\bar{z}_2)=
\frac{\bar{\kappa}_{+1,-2,-1}}{2}\,
\frac{\pi \, (-1)^{k-1}}{(-k)!}
\frac{1}{\bar{z}_{12}^{3-k}} \,
\sum_{m=0}^{-k-2} \,
(m+1)
\nonu \\
&& \times \frac{\Gamma(2(-1-h_2)+1)}
{\Gamma(2(-1-h_2)+k+m+3)}
\, z_{12}^{k+1+m}\,
{\pa}_{{z}_2}^m \,
{\cal O}_{+1}^{(-\frac{k}{2}+\fft32+h_2),AB}(z_2,\bar{z}_2)+\cdots,
\nonu \\
&& \tilde{\bar{\Phi}}_{+\fft32}^{(-\frac{k}{2}+\fft74),A}(z_1,\bar{z}_1) \,
{\cal O}_{-2}^{(h_2)}(z_2,\bar{z}_2)=
\frac{\bar{\kappa}_{+\fft32,-2,-\fft32}}{2}\,
\frac{\pi \, (-1)^{k-\fft32}}{(-\fft12-k)!}
\frac{1}{\bar{z}_{12}^{\fft72-k}} \,
\sum_{m=0}^{-k-\fft52} \,
(m+1)
\nonu \\
&& \times \frac{\Gamma(2(-1-h_2)+1)}
{\Gamma(2(-1-h_2)+k+m+\tf72)}
\, z_{12}^{k+\fft32+m}\,
{\pa}_{{z}_2}^m \,
{\cal O}_{+\fft32}^{(-\frac{k}{2}+\fft74+h_2),A}(z_2,\bar{z}_2)+\cdots,
\nonu \\
&& \tilde{\bar{\Phi}}_{+2}^{(-\frac{k}{2}+2)}(z_1,\bar{z}_1) \,
{\cal O}_{-2}^{(h_2)}(z_2,\bar{z}_2)=
\frac{\bar{\kappa}_{+2,-2,-2}}{2}\,
\frac{\pi \, (-1)^{k}}{(-1-k)!}
\frac{1}{\bar{z}_{12}^{4-k}} \,
\sum_{m=0}^{-k-3} \,
(m+1)
\nonu \\
&& \times \frac{\Gamma(2(-1-h_2)+1)}
{\Gamma(2(-1-h_2)+k+m+4)}
\, z_{12}^{k+2+m}\,
{\pa}_{{z}_2}^m \,
{\cal O}_{+2}^{(-\frac{k}{2}+2+h_2)}(z_2,\bar{z}_2)+\cdots,
\nonu \\
&& \tilde{\bar{\Phi}}_{AB,-1}^{(-\frac{k}{2}+\fft12)}(z_1,\bar{z}_1) \,
{\cal O}_{C,-\fft32}^{(h_2)}(z_2,\bar{z}_2)=
\frac{\bar{\kappa}_{-1,-\fft32,+\fft12}}{2}\,
\frac{\pi \, (-1)^{k+1}}{(2-k)!}
\frac{1}{\bar{z}_{12}^{1-k}} \,
\sum_{m=0}^{-k} \,
(m+1)
\nonu \\
&& \times \frac{\Gamma(2(-\fft12-h_2)+1)}
{\Gamma(2(-\fft12-h_2)+k+m+1)}
\, z_{12}^{k-1+m}\,
{\pa}_{{z}_2}^m \,
{\cal O}_{ABC,-\fft12}^{(-\frac{k}{2}+\fft12+h_2)}(z_2,\bar{z}_2)+\cdots,
\nonu \\
&& \tilde{\bar{\Phi}}_{ABC,-\fft12}^{(-\frac{k}{2}+\fft34)}(z_1,\bar{z}_1) \,
{\cal O}_{D,-\fft32}^{(h_2)}(z_2,\bar{z}_2)=
\frac{\bar{\kappa}_{-\fft12,-\fft32,0}}{2}\,
\frac{\pi \, (-1)^{k+\fft12}}{(\fft32-k)!}
\frac{1}{\bar{z}_{12}^{\fft32-k}} \,
\sum_{m=0}^{-k-\fft12} \,
(m+1)
\nonu \\
&& \times \frac{\Gamma(2(-\fft12-h_2)+1)}
{\Gamma(2(-\fft12-h_2)+k+m+\tf32)}
\, z_{12}^{k-\fft12+m}\, \frac{1}{4!}\, \ep_{ABCDEFGH}\,
{\pa}_{{z}_2}^m \,
{\cal O}_{0}^{(-\frac{k}{2}+\fft34+h_2),EFGH}(z_2,\bar{z}_2)+\cdots,
\nonu \\
&& \tilde{\bar{\Phi}}_{0}^{(-\frac{k}{2}+1),ABCD}(z_1,\bar{z}_1) \,
{\cal O}_{E,-\fft32}^{(h_2)}(z_2,\bar{z}_2)=
\frac{\bar{\kappa}_{0,-\fft32,-\fft12}}{2}\,
\frac{\pi \, (-1)^{k}}{(1-k)!}
\frac{1}{\bar{z}_{12}^{2-k}} \,
\sum_{m=0}^{-k-1} \,
(m+1)
\nonu \\
&& \times \frac{\Gamma(2(-\fft12-h_2)+1)}
{\Gamma(2(-\fft12-h_2)+k+m+2)}
\, z_{12}^{k+m}\,\frac{1}{4!}\,
\frac{1}{3!}\, \ep^{ABCDIJKL}\, \ep_{EFGHIJKL}
{\pa}_{{z}_2}^m \,
{\cal O}_{+\fft12}^{(-\frac{k}{2}+1+h_2),FGH}(z_2,\bar{z}_2)
\nonu \\
&& +\cdots,
\nonu \\
&& \tilde{\bar{\Phi}}_{+\fft12}^{(-\frac{k}{2}+\fft54),ABC}(z_1,\bar{z}_1) \,
{\cal O}_{D,-\fft32}^{(h_2)}(z_2,\bar{z}_2)=
\frac{\bar{\kappa}_{+\fft12,-\fft32,-1}}{2}\,
\frac{\pi \, (-1)^{k-\fft12}}{(\fft12-k)!}
\frac{1}{\bar{z}_{12}^{\fft52-k}} \,
\sum_{m=0}^{-k-\fft32} \,
(m+1)
\nonu \\
&& \times \frac{\Gamma(2(-\fft12-h_2)+1)}
{\Gamma(2(-\fft12-h_2)+k+m+\tf52)}
\, z_{12}^{k+\fft12+m}\,
3\, 
{\pa}_{{z}_2}^m \,
{\cal O}_{+1}^{(-\frac{k}{2}+\fft54+h_2),[AB}(z_2,\bar{z}_2)\,
\de^{C]}_{D}+\cdots,
\nonu \\
&& \tilde{\bar{\Phi}}_{+1}^{(-\frac{k}{2}+\fft32),AB}(z_1,\bar{z}_1) \,
{\cal O}_{C,-\fft32}^{(h_2)}(z_2,\bar{z}_2)=
\frac{\bar{\kappa}_{+1,-\fft32,-\fft32}}{2}\,
\frac{\pi \, (-1)^{k-1}}{(-k)!}
\frac{1}{\bar{z}_{12}^{3-k}} \,
\sum_{m=0}^{-k-2} \,
(m+1)
\nonu \\
&& \times \frac{\Gamma(2(-\fft12-h_2)+1)}
{\Gamma(2(-\fft12-h_2)+k+m+3)}
\, z_{12}^{k+1+m}\,
2\, \de^{[A}_{C} 
{\pa}_{{z}_2}^m \,
{\cal O}_{+\fft32}^{(-\frac{k}{2}+\fft32+h_2),B]}(z_2,\bar{z}_2)+\cdots,
\nonu \\
&& \tilde{\bar{\Phi}}_{+\fft32}^{(-\frac{k}{2}+\fft74),A}(z_1,\bar{z}_1) \,
{\cal O}_{B,-\fft32}^{(h_2)}(z_2,\bar{z}_2)=
\frac{\bar{\kappa}_{+\fft32,-\fft32,-2}}{2}\,
\frac{\pi \, (-1)^{k-\fft32}}{(-\fft12-k)!}
\frac{1}{\bar{z}_{12}^{\fft72-k}} \,
\sum_{m=0}^{-k-\fft52} \,
(m+1)
\nonu \\
&& \times \frac{\Gamma(2(-\fft12-h_2)+1)}
{\Gamma(2(-\fft12-h_2)+k+m+\tf72)}
\, z_{12}^{k+\fft32+m}\,
\de^{A}_{B} 
{\pa}_{{z}_2}^m \,
{\cal O}_{+2}^{(-\frac{k}{2}+\fft74+h_2)}(z_2,\bar{z}_2)+\cdots,
\nonu \\
&& \tilde{\bar{\Phi}}_{ABC,-\fft12}^{(-\frac{k}{2}+\fft34)}(z_1,\bar{z}_1) \,
{\cal O}_{DE,-1}^{(h_2)}(z_2,\bar{z}_2)=
\frac{\bar{\kappa}_{-\fft12,-1,-\fft12}}{2}\,
\frac{\pi \, (-1)^{k+\fft12}}{(\fft32-k)!}
\frac{1}{\bar{z}_{12}^{\fft32-k}} \,
\sum_{m=0}^{-k-\fft12} \,
(m+1)
\nonu \\
&& \times \frac{\Gamma(2(-h_2)+1)}
{\Gamma(2(-h_2)+k+m+\tf32)}
\, z_{12}^{k-\fft12+m}\,
\frac{1}{3!}\, \ep_{ABCDEFGH}\,
{\pa}_{{z}_2}^m \,
{\cal O}_{+\fft12}^{(-\frac{k}{2}+\fft34+h_2),FGH}(z_2,\bar{z}_2)+\cdots,
\nonu \\
&& \tilde{\bar{\Phi}}_{0}^{(-\frac{k}{2}+1),ABCD}(z_1,\bar{z}_1) \,
{\cal O}_{EF,-1}^{(h_2)}(z_2,\bar{z}_2)=
\frac{\bar{\kappa}_{0,-1,-1}}{2}\,
\frac{\pi \, (-1)^{k}}{(1-k)!}
\frac{1}{\bar{z}_{12}^{2-k}} \,
\sum_{m=0}^{-k-1} \,
(m+1)
\nonu \\
&& \times \frac{\Gamma(2(-h_2)+1)}
{\Gamma(2(-h_2)+k+m+2)}
\, z_{12}^{k+m}\, \frac{1}{4!}\,
\frac{1}{2!}\, \ep^{ABCDIJKL}\, \ep_{EFGHIJKL}
{\pa}_{{z}_2}^m \,
{\cal O}_{+1}^{(-\frac{k}{2}+1+h_2),GH}(z_2,\bar{z}_2)+\cdots,
\nonu \\
&& \tilde{\bar{\Phi}}_{+\fft12}^{(-\frac{k}{2}+\fft54),ABC}(z_1,\bar{z}_1) \,
{\cal O}_{DE,-1}^{(h_2)}(z_2,\bar{z}_2)=
\frac{\bar{\kappa}_{+\fft12,-1,-\fft32}}{2}\,
\frac{\pi \, (-1)^{k-\fft12}}{(\fft12-k)!}
\frac{1}{\bar{z}_{12}^{\fft52-k}} \,
\sum_{m=0}^{-k-\fft32} \,
(m+1)
\nonu \\
&& \times \frac{\Gamma(2(-h_2)+1)}
{\Gamma(2(-h_2)+k+m+\tf52)}
\, z_{12}^{k+\fft12+m}\,
3!\,
\de^{[A}_{D}\,
{\pa}_{{z}_2}^m \,
{\cal O}_{+\fft32}^{(-\frac{k}{2}+\fft54+h_2),B}(z_2,\bar{z}_2) \,
\de^{C]}_{E}+\cdots,
\nonu \\
&& \tilde{\bar{\Phi}}_{+1}^{(-\frac{k}{2}+\fft32),AB}(z_1,\bar{z}_1) \,
{\cal O}_{CD,-1}^{(h_2)}(z_2,\bar{z}_2)=
\frac{\bar{\kappa}_{+1,-1,-2}}{2}\,
\frac{\pi \, (-1)^{k-1}}{(-k)!}
\frac{1}{\bar{z}_{12}^{3-k}} \,
\sum_{m=0}^{-k-2} \,
(m+1)
\nonu \\
&& \times \frac{\Gamma(2(-h_2)+1)}
{\Gamma(2(-h_2)+k+m+3)}
\, z_{12}^{k+1+m}\,
\de^{AB}_{CD}\,
{\pa}_{{z}_2}^m \,
{\cal O}_{+2}^{(-\frac{k}{2}+\fft32+h_2)}(z_2,\bar{z}_2)+\cdots,
\nonu \\
&& \tilde{\bar{\Phi}}_{0}^{(-\frac{k}{2}+1),ABCD}(z_1,\bar{z}_1) \,
{\cal O}_{EFG,-\fft12}^{(h_2)}(z_2,\bar{z}_2)=
\frac{\bar{\kappa}_{0,-\fft12,-\fft32}}{2}\,
\frac{\pi \, (-1)^{k}}{(1-k)!}
\frac{1}{\bar{z}_{12}^{2-k}} \,
\sum_{m=0}^{-k-1} \,
(m+1)
\nonu \\
&& \times \frac{\Gamma(2(\fft12-h_2)+1)}
{\Gamma(2(\fft12-h_2)+k+m+2)}
\, z_{12}^{k+m}\, \frac{1}{4!}\,
\ep^{ABCDIJKL}\,\ep_{EFGHIJKL}
{\pa}_{{z}_2}^m \,
{\cal O}_{+\fft32}^{(-\frac{k}{2}+1+h_2),H}(z_2,\bar{z}_2)+\cdots,
\nonu \\
&& \tilde{\bar{\Phi}}_{+\fft12}^{(-\frac{k}{2}+\fft54),ABC}(z_1,\bar{z}_1) \,
{\cal O}_{DEF,-\fft12}^{(h_2)}(z_2,\bar{z}_2)=
\frac{\bar{\kappa}_{+\fft12,-\fft12,-2}}{2}\,
\frac{\pi \, (-1)^{k-\fft12}}{(\fft12-k)!}
\frac{1}{\bar{z}_{12}^{\fft52-k}} \,
\sum_{m=0}^{-k-\fft32} \,
(m+1)
\nonu \\
&& \times \frac{\Gamma(2(\fft12-h_2)+1)}
{\Gamma(2(\fft12-h_2)+k+m+\tf52)}
\, z_{12}^{k+\fft12+m}\,
\de^{ABC}_{DEF}\,
{\pa}_{{z}_2}^m \,
{\cal O}_{+2}^{(-\frac{k}{2}+\fft54+h_2)}(z_2,\bar{z}_2)+\cdots.
\label{othertwentysection6}
\eea
Due to the structure of $SU(8)$ indices,  the right-hand sides of the
tenth, the eleventh, the sixteenth, and the
nineteenth relations
of (\ref{othertwentysection6}) are different from those
in (\ref{twentysection6}).



\begin{thebibliography}{99}

\bibitem{BvM}
H.~Bondi, M.~G.~J.~van der Burg and A.~W.~K.~Metzner,
``Gravitational waves in general relativity. 7. Waves from axisymmetric isolated systems,''
Proc. Roy. Soc. Lond. A \textbf{269}, 21-52 (1962)
doi:10.1098/rspa.1962.0161

\bibitem{Sachs}
R.~K.~Sachs,
``Gravitational waves in general relativity. 8. Waves in asymptotically flat space-times,''
Proc. Roy. Soc. Lond. A \textbf{270}, 103-126 (1962)
doi:10.1098/rspa.1962.0206

\bibitem{HLMS}
T.~He, V.~Lysov, P.~Mitra and A.~Strominger,
``BMS supertranslations and Weinberg{\textquoteright}s soft graviton theorem,''
JHEP \textbf{05}, 151 (2015)
doi:10.1007/JHEP05(2015)151
[arXiv:1401.7026 [hep-th]].

\bibitem{Weinberg}
S.~Weinberg,
``Infrared photons and gravitons,''
Phys. Rev. \textbf{140}, B516-B524 (1965)
doi:10.1103/PhysRev.140.B516

\bibitem{BT}
G.~Barnich and C.~Troessaert,
``Symmetries of asymptotically flat 4 dimensional spacetimes at null infinity revisited,''
Phys. Rev. Lett. \textbf{105}, 111103 (2010)
doi:10.1103/PhysRevLett.105.111103
[arXiv:0909.2617 [gr-qc]].

\bibitem{KLPS}
D.~Kapec, V.~Lysov, S.~Pasterski and A.~Strominger,
``Semiclassical Virasoro symmetry of the quantum gravity $ \mathcal{S}$-matrix,''
JHEP \textbf{08}, 058 (2014)
doi:10.1007/JHEP08(2014)058
[arXiv:1406.3312 [hep-th]].

\bibitem{CS}
F.~Cachazo and A.~Strominger,
``Evidence for a New Soft Graviton Theorem,''
[arXiv:1404.4091 [hep-th]].

\bibitem{CL}
M.~Campiglia and A.~Laddha,
``Sub-subleading soft gravitons and large diffeomorphisms,''
JHEP \textbf{01}, 036 (2017)
doi:10.1007/JHEP01(2017)036
[arXiv:1608.00685 [gr-qc]].

\bibitem{FSTZ1912}
A.~Fotopoulos, S.~Stieberger, T.~R.~Taylor and B.~Zhu,
``Extended BMS Algebra of Celestial CFT,''
JHEP \textbf{03}, 130 (2020)
doi:10.1007/JHEP03(2020)130
[arXiv:1912.10973 [hep-th]].

\bibitem{Barnich}
G.~Barnich,
``Centrally extended BMS4 Lie algebroid,''
JHEP \textbf{06}, 007 (2017)
doi:10.1007/JHEP06(2017)007
[arXiv:1703.08704 [hep-th]].

\bibitem{FSTZ}
A.~Fotopoulos, S.~Stieberger, T.~R.~Taylor and B.~Zhu,
``Extended Super BMS Algebra of Celestial CFT,''
JHEP \textbf{09}, 198 (2020)
doi:10.1007/JHEP09(2020)198
[arXiv:2007.03785 [hep-th]].

\bibitem{AGS}
M.~A.~Awada, G.~W.~Gibbons and W.~T.~Shaw,
``CONFORMAL SUPERGRAVITY, TWISTORS AND THE SUPER BMS GROUP,''
Annals Phys. \textbf{171}, 52 (1986)
doi:10.1016/S0003-4916(86)80023-9

\bibitem{TZ}
T.~R.~Taylor and B.~Zhu,
``Celestial Supersymmetry,''
JHEP \textbf{06}, 210 (2023)
doi:10.1007/JHEP06(2023)210
[arXiv:2302.12830 [hep-th]].

\bibitem{Ahn2202}
C.~Ahn,
``A deformed supersymmetric $w_{1+\infty }$ symmetry in the celestial conformal field theory,''
Eur. Phys. J. C \textbf{82}, no.7, 630 (2022)
doi:10.1140/epjc/s10052-022-10582-9
[arXiv:2202.02949 [hep-th]].

\bibitem{Ahn2203}
C.~Ahn,
``The $ \mathcal{N}  = 2$ supersymmetric w$_{1+\infty}$
symmetry in the two-dimensional SYK models,''
JHEP \textbf{05}, 115 (2022)
doi:10.1007/JHEP05(2022)115
[arXiv:2203.03105 [hep-th]].

\bibitem{AK2309}
C.~Ahn and M.~H.~Kim,
``The $ \mathcal{N}  = 2, 4$ supersymmetric linear
W$_{\infty}$[{\ensuremath{\lambda}}] algebras
for generic {\ensuremath{\lambda}} parameter,''
JHEP \textbf{02}, 006 (2024)
doi:10.1007/JHEP02(2024)006
[arXiv:2309.01537 [hep-th]].

\bibitem{AK2407}
C.~Ahn and M.~H.~Kim,
``A supersymmetric extension of w$_{1+\infty}$
algebra in the celestial holography,''
JHEP \textbf{09}, 081 (2024)
doi:10.1007/JHEP09(2024)081
[arXiv:2407.05601 [hep-th]].

\bibitem{AK2501}
C.~Ahn and M.~H.~Kim,
``A supersymmetric $w_{1+\infty }$ symmetry,
the extended supergravity and the celestial holography,''
Eur. Phys. J. C \textbf{85}, no.8, 834 (2025)
doi:10.1140/epjc/s10052-025-14479-1
[arXiv:2501.11471 [hep-th]].

\bibitem{2212-2}
N.~Banerjee, T.~Rahnuma and R.~K.~Singh,
``Asymptotic symmetry algebra of N=8 supergravity,''
Phys. Rev. D \textbf{109}, no.4, 046010 (2024)
doi:10.1103/PhysRevD.109.046010
[arXiv:2212.12133 [hep-th]].

\bibitem{Tropper}
A.~Tropper,
``Symmetries of the Celestial Supersphere,''
[arXiv:2412.13113 [hep-th]].

\bibitem{GHPS}
A.~Guevara, E.~Himwich, M.~Pate and A.~Strominger,
``Holographic symmetry algebras for gauge theory and gravity,''
JHEP \textbf{11}, 152 (2021)
doi:10.1007/JHEP11(2021)152
[arXiv:2103.03961 [hep-th]].

\bibitem{Strominger}
A.~Strominger,
``$w_{1+\infty}$ Algebra and the Celestial Sphere: Infinite Towers of
Soft Graviton, Photon, and Gluon Symmetries,''
Phys. Rev. Lett. \textbf{127}, no.22, 221601 (2021)
doi:10.1103/PhysRevLett.127.221601
[arXiv:2105.14346 [hep-th]].

\bibitem{HPS}
E.~Himwich, M.~Pate and K.~Singh,
``Celestial operator product expansions and w$_{1+\infty}$
symmetry for all spins,''
JHEP \textbf{01}, 080 (2022)
doi:10.1007/JHEP01(2022)080
[arXiv:2108.07763 [hep-th]].

\bibitem{Strominger1703}
A.~Strominger,
``Lectures on the Infrared Structure of Gravity and Gauge Theory,''
Princeton University Press, 2018,
ISBN 978-0-691-17973-5
[arXiv:1703.05448 [hep-th]].

\bibitem{Raclariu}
A.~M.~Raclariu,
``Lectures on Celestial Holography,''
[arXiv:2107.02075 [hep-th]].

\bibitem{Pasterski}
S.~Pasterski,
``Lectures on celestial amplitudes,''
Eur. Phys. J. C \textbf{81}, no.12, 1062 (2021)
doi:10.1140/epjc/s10052-021-09846-7
[arXiv:2108.04801 [hep-th]].

\bibitem{PPR}
S.~Pasterski, M.~Pate and A.~M.~Raclariu,
``Celestial Holography,''
[arXiv:2111.11392 [hep-th]].

\bibitem{Donnay}
L.~Donnay,
``Celestial holography: An asymptotic symmetry perspective,''
Phys. Rept. \textbf{1073}, 1-41 (2024)
doi:10.1016/j.physrep.2024.04.003
[arXiv:2310.12922 [hep-th]].

\bibitem{Ruzziconi}
R.~Ruzziconi,
``Carrollian physics and holography,''
Phys. Rept. \textbf{1182}, 1-87 (2026)
doi:10.1016/j.physrep.2026.03.005
[arXiv:2602.02644 [hep-th]].

\bibitem{Zhu}
B.~Zhu,
``Topics in Celestial holography: A bottom-up perspective,''
Phys. Rept. \textbf{1200}, 1-46 (2026)
doi:10.1016/j.physrep.2026.08.001
[arXiv:2606.24285 [hep-th]].

\bibitem{dF}
B.~de Wit and D.~Z.~Freedman,
``On SO(8) Extended Supergravity,''
Nucl. Phys. B \textbf{130}, 105-113 (1977)
doi:10.1016/0550-3213(77)90395-9

\bibitem{AK2509}
C.~Ahn and M.~H.~Kim,
``Toward A celestial soft symmetry algebra in the $ \mathcal{N}=8 $ supergravity,''
JHEP \textbf{05}, 222 (2026)
doi:10.1007/JHEP05(2026)222
[arXiv:2509.06328 [hep-th]].

\bibitem{KS}
P.~Kravchuk and D.~Simmons-Duffin,
``Light-ray operators in conformal field theory,''
JHEP \textbf{11}, 102 (2018)
doi:10.1007/JHEP11(2018)102
[arXiv:1805.00098 [hep-th]].

\bibitem{CP}
M.~Calkins and M.~Pate,
``Multi-particle Celestial Operator Product Expansions from the Boundary,''
[arXiv:2601.04329 [hep-th]].

\bibitem{BHP}
A.~Ball, Y.~Hu and S.~Pasterski,
``Multicollinear singularities in celestial CFT,''
JHEP \textbf{02}, 219 (2024)
doi:10.1007/JHEP02(2024)219
[arXiv:2309.16602 [hep-th]].

\bibitem{Jiang}
H.~Jiang,
``Holographic chiral algebra: supersymmetry, infinite Ward identities, and EFTs,''
JHEP \textbf{01}, 113 (2022)
doi:10.1007/JHEP01(2022)113
[arXiv:2108.08799 [hep-th]].

\bibitem{2212-1}
N.~Banerjee, T.~Rahnuma and R.~K.~Singh,
``Soft and collinear limits in $ \mathcal{N} $ = 8 supergravity using double copy formalism,''
JHEP \textbf{04}, 126 (2023)
doi:10.1007/JHEP04(2023)126
[arXiv:2212.11480 [hep-th]].

\bibitem{CFT}
R.~Blumenhagen and E.~Plauschinn,
``Introduction to conformal field theory: with applications to String theory,''
Lect. Notes Phys. \textbf{779}, 1-256 (2009)
doi:10.1007/978-3-642-00450-6

\bibitem{HP}
E.~Himwich and M.~Pate,
``w$_{1+\infty}$ in 4D gravitational scattering,''
JHEP \textbf{07}, 180 (2024)
doi:10.1007/JHEP07(2024)180
[arXiv:2312.08597 [hep-th]].

\bibitem{Ahn2607}
C.~Ahn,
``Multi-Particle Contributions to the Celestial Algebra in the
${\cal N}=8$ Supergravity,''
[arXiv:2607.04611 [hep-th]].

\bibitem{Thielemans}
K.~Thielemans,
``A Mathematica package for computing operator product expansions,''
Int. J. Mod. Phys. C \textbf{2}, 787-798 (1991)
doi:10.1142/S0129183191001001

\bibitem{AMS}
T.~Adamo, L.~Mason and A.~Sharma,
``Celestial amplitudes and conformal soft theorems,''
Class. Quant. Grav. \textbf{36}, no.20, 205018 (2019)
doi:10.1088/1361-6382/ab42ce
[arXiv:1905.09224 [hep-th]].

\bibitem{Puhm}
A.~Puhm,
``Conformally Soft Theorem in Gravity,''
JHEP \textbf{09}, 130 (2020)
doi:10.1007/JHEP09(2020)130
[arXiv:1905.09799 [hep-th]].

\bibitem{Guevara}
A.~Guevara,
``Notes on Conformal Soft Theorems and Recursion Relations in Gravity,''
[arXiv:1906.07810 [hep-th]].

\bibitem{Ahn2111}
C.~Ahn,
``Towards a supersymmetric w1+{\ensuremath{\infty}} symmetry in the celestial conformal field theory,''
Phys. Rev. D \textbf{105}, no.8, 086028 (2022)
doi:10.1103/PhysRevD.105.086028
[arXiv:2111.04268 [hep-th]].

\bibitem{MRSV}
J.~Mago, L.~Ren, A.~Y.~Srikant and A.~Volovich,
``Deformed $w_{1+\infty}$ Algebras in the Celestial CFT,''
SIGMA \textbf{19}, 044 (2023)
doi:10.3842/SIGMA.2023.044
[arXiv:2111.11356 [hep-th]].

\bibitem{Liu}
Z.~W.~Liu,
``Soft theorems in maximally supersymmetric theories,''
Eur. Phys. J. C \textbf{75}, no.3, 105 (2015)
doi:10.1140/epjc/s10052-015-3304-1
[arXiv:1410.1616 [hep-th]].

\bibitem{KMRS}
D.~Kapec, P.~Mitra, A.~M.~Raclariu and A.~Strominger,
``2D Stress Tensor for 4D Gravity,''
Phys. Rev. Lett. \textbf{119}, no.12, 121601 (2017)
doi:10.1103/PhysRevLett.119.121601
[arXiv:1609.00282 [hep-th]].

\bibitem{BP}
S.~Banerjee and S.~Pasterski,
``Revisiting the shadow stress tensor in celestial CFT,''
JHEP \textbf{04}, 118 (2023)
doi:10.1007/JHEP04(2023)118
[arXiv:2212.00257 [hep-th]].

\bibitem{PRSY}
M.~Pate, A.~M.~Raclariu, A.~Strominger and E.~Y.~Yuan,
``Celestial operator products of gluons and gravitons,''
Rev. Math. Phys. \textbf{33}, no.09, 2140003 (2021)
doi:10.1142/S0129055X21400031
[arXiv:1910.07424 [hep-th]].

\bibitem{BEF}
M.~Bianchi, H.~Elvang and D.~Z.~Freedman,
``Generating Tree Amplitudes in N=4 SYM and N = 8 SG,''
JHEP \textbf{09}, 063 (2008)
doi:10.1088/1126-6708/2008/09/063
[arXiv:0805.0757 [hep-th]].

\bibitem{KLT}
H.~Kawai, D.~C.~Lewellen and S.~H.~H.~Tye,
``A Relation Between Tree Amplitudes of Closed and Open Strings,''
Nucl. Phys. B \textbf{269}, 1-23 (1986)
doi:10.1016/0550-3213(86)90362-7

\bibitem{EJN}
H.~Elvang, C.~R.~T.~Jones and S.~G.~Naculich,
``Soft Photon and Graviton Theorems in Effective Field Theory,''
Phys. Rev. Lett. \textbf{118}, no.23, 231601 (2017)
doi:10.1103/PhysRevLett.118.231601
[arXiv:1611.07534 [hep-th]].

\bibitem{BSYV}
A.~Ball, M.~Spradlin, A.~Yelleshpur Srikant and A.~Volovich,
``Supersymmetry and the celestial Jacobi identity,''
JHEP \textbf{04}, 099 (2024)
doi:10.1007/JHEP04(2024)099
[arXiv:2311.01364 [hep-th]].

\bibitem{mathematica}
Wolfram Research, Inc., Mathematica, Version 14.3, Champaign, IL
(2025).

\end{thebibliography}
\end{document}